\documentclass[sotoncolour]{uosthesis}      % Use the Thesis Style with custom link colour
\graphicspath{{../Figures/}}   % Location of your graphics files
\usepackage{bibentry}          % Use bibentry for prepublished works
\nobibliography*               % Use bibentry for prepublished works
\usepackage{cite}
\usepackage{attrib}            % Use the attrib package for quotations
\usepackage{float}
\usepackage{multirow}
\usepackage{wasysym}
\usepackage{mathtools}
\usepackage{colortbl}
\usepackage[compat=1.1.0]{tikz-feynman}
\tikzfeynmanset{warn luatex=false}
\usepackage[pdfpagemode={UseOutlines},bookmarks=true,bookmarksopen=true,
bookmarksopenlevel=0,bookmarksnumbered=true,hypertexnames=true,
colorlinks,linkcolor={linkBlue},citecolor={linkBlue},urlcolor={linkBlue},
pdfstartview={FitV},unicode,breaklinks=true,backref=page]{hyperref}
\backrefsetup{verbose=true}
\hypersetup{colorlinks=true}   % Set to false for black/white printing
\hypersetup{pdfstartpage=1}    % Set first page to skip copyright declaration
\renewcommand*{\backref}[1]{}
\renewcommand*{\backrefalt}[4]{[{\tiny%
    \ifcase #1 Not cited.%
          \or Cited on page~#2.%
          \else Cited on pages #2.%
    \fi%
    }]}

\newcommand{\BibTeX}{{\rm B\kern-.05em{\sc i\kern-.025em b}\kern-.08em T\kern-.1667em\lower.7ex\hbox{E}\kern-.125emX}}

\newcounter{address}
\newcounter{alg}

\def\nn{\nonumber\\ }

\def\gcg{{\overline g_{3}}}

\def\tc{{\overline \theta}}

\def\sc{{\overline s}}

\def\ckin{c_{H,\text{kin}}}

\def\one{\ensuremath{\mathbf{1}}}
\def\two{\ensuremath{\mathbf{2}}}
\def\three{\ensuremath{\mathbf{3}}}
\def\threeS{\ensuremath{\mathbf{\bar 3}}}
\def\GGM{\ensuremath{SU(5)}}
\def\555{\ensuremath{SU(5)^3}}
\def\five{\ensuremath{\mathbf{5}}}
\def\fiveS{\ensuremath{\mathbf{\bar 5}}}
\def\eight{\ensuremath{\mathbf{8}}}
\def\ten{\ensuremath{\mathbf{10}}}
\def\tenS{\ensuremath{\mathbf{\overline{10}}}}
\def\fifteen{\ensuremath{\mathbf{15}}}
\def\fifteenS{\ensuremath{\mathbf{\overline{15}}}}
\def\24{\ensuremath{\mathbf{24}}}
\def\tfive{\ensuremath{\mathbf{35}}}
\def\tfiveS{\ensuremath{\mathbf{\overline{35}}}}
\def\forty{\ensuremath{\mathbf{40}}}
\def\fortyS{\ensuremath{\mathbf{\overline{40}}}}
\def\ffive{\ensuremath{\mathbf{45}}}
\def\ffiveS{\ensuremath{\mathbf{\overline{45}}}}
\def\SU15#1{\ensuremath{\mathbf{#1}}}

\renewcommand{\O}{\mathcal{O}}
\newcommand{\op}[3]{\O^{#2,#3}_{#1}}

\newcommand{\hc}{\mathrm{h.c.}}

\renewenvironment{quote}
  {\begin{trivlist} \setlength\leftskip{5.3cm} \setlength\rightskip{0pt}
   \item\relax}
  {\end{trivlist}}

\faculty     {\MakeUppercase{Faculty of Engineering and Physical sciences}}
\FACULTY     {\MakeUppercase{\facname}}
\department  {School of Physics and Astronomy}
\DEPARTMENT  {\MakeUppercase{\deptname}}
\GROUP       {\MakeUppercase{\groupname}}
\title      {Theories of flavour from the Planck scale to the electroweak scale}
\authors    {Mario Fern\'andez Navarro} % Use of Soton Email unadvised, use ORCiD instead.
\orcidid    {0000-0002-8796-0172}
\addresses  {\groupname\\\deptname\\\univname}
\date       {March 2024}
\subject    {}
\begin{document}
%% ------------------ FRONT MATTER ORGANISATION -------------------
\pagenumbering{gobble} % removes page number
\newgeometry{
            %left=1.25in,
            %right=1.25in,
            %left=1.5in,
            %right=0.6in,
            hmarginratio=1:2,
            textwidth=146.5mm,
            top=0.6in,
            bottom=0.8in,
            headheight=20pt,
            headsep=0.25in,
            foot=9pt,
            footskip=0.3in,
            bindingoffset=0.5in,
            %bindingoffset=25mm,
            includeheadfoot}
%\copyrightDeclaration{} % !!! Comment this line when printing the hardcopy !!!
%\raggedright                  %% Set the style to Left justification, remove for fill justification
                              %% Must be done after copyrightDeclaration
\restoregeometry
\frontmatter
\maketitle
\begin{abstract}
\noindent The flavour puzzle remains as one of the most intriguing enigmas of
particle physics. \textit{A priori}, there is no apparent reason for
the existence of three identical families of fundamental fermions
in Nature. Moreover, the high number of free parameters in the flavour
sector, along with
their particular hierarchical patterns, suggest the existence
of new physics that provide a dynamical explanation for the flavour
structure of the Standard Model: such a theory describing the complicated flavour
sector in terms of simple and natural principles is called
a \textit{theory of flavour}.

In this thesis, we propose and study theories of flavour which generically
hint to a multi-scale origin of flavour. First we explore the idea
of fermiophobic models, where the fermion mass hierarchies and the
smallness of quark mixing are explained via the mechanism of messenger
dominance. The general idea is that the chiral fermions of the Standard Model
are uncharged under (part of) an extended gauge symmetry, which also
forbids the presence of Yukawa couplings for chiral fermions. These
are generated effectively due to the presence of hierarchically heavy
messengers, including vector-like fermions and/or heavy Higgs doublets.
The heavy messengers may also induce effective couplings for the chiral
fermions to TeV-scale gauge bosons associated to the spontaneous
breaking of the extended symmetry, leading to a predictive phenomenology
connected to the origin of flavour hierarchies.

First we apply this idea to an extension of the Standard Model by
a $U(1)'$ local abelian factor, where we seek to provide a significant
contribution to the anomalous magnetic moment of the muon via exchange of
a heavy $Z'\sim(\mathbf{1,1},0)$ boson, and we find an interesting
correlation with a suppression of the Higgs decay to two photons.
Then we apply the same idea to a twin Pati-Salam
symmetry which provides a TeV-scale
vector leptoquark $U_{1}\sim(\mathbf{3,1},2/3)$ that can explain
the so-called $B$-physics anomalies. We find that this model can
be tested due to the correlated enhancement of key low-energy observables,
and also via direct production of the new heavy degrees of freedom
at the LHC.
\newpage
\thispagestyle{empty}
Secondly, we leave behind the ideas of messenger dominance and fermiophobic models to
study the possibility that the Standard Model originates
from a non-universal gauge theory in the ultraviolet. We argue that one of the most simple
ways to achieve this is by assigning a separate gauge hypercharge
to each fermion family at high energies, spontaneously broken down
to the usual weak hypercharge which is the diagonal subgroup. This
simple framework denoted as ``tri-hypercharge'' avoids the family replication of the Standard Model, and could
be the first step towards a deep non-universal gauge structure in
the ultraviolet. If the Higgs doublet(s) only carries third family hypercharge,
then only third family renormalisable Yukawa couplings are allowed
by the gauge symmetry. However, non-renormalisable Yukawa couplings for
the light families
may be induced by the high scale scalar fields which break the three
hypercharges down to the usual weak hypercharge, providing an explanation
for fermion mass hierarchies and the smallness of quark mixing.
Interestingly, in order to explain neutrino mixing, it is useful to
introduce right-handed neutrinos which carry non-zero hypercharges
(although their sum must vanish), which then turn out to get Majorana masses
at the lowest scale of symmetry breaking, that could be as low as
a few TeV. In fact, we find that the model has a rich phenomenology
via $Z'$ bosons if the new physics scales are relatively low: from flavour-violating observables
to LHC physics and electroweak precision observables.

Finally, we propose a gauge unified origin for gauge non-universal
frameworks such as the aforementioned tri-hypercharge theory.
The model consists on assigning a separate $SU(5)$ group
to each fermion family. However, assuming that the three $SU(5)$
groups are related by a cyclic permutation symmetry $\mathbb{Z}_{3}$, then
the model is described by a single gauge coupling in the ultraviolet, despite
$SU(5)^{3}$ being a non-simple group. First, we show a general
$SU(5)^{3}$ ``tri-unification'' framework where gauge non-universal
theories of flavour may be embedded, and secondly we construct
a minimal tri-hypercharge example which can account for all the quark
and lepton (including neutrino) masses and mixing parameters, with
the five gauge couplings of the tri-hypercharge group unifying at
the GUT scale into a single gauge coupling associated to the cyclic
$SU(5)^{3}$ group, and we study the implications for the stability of the proton
in such a setup.
\end{abstract}

\newpage
\dedicatory{To those who, despite all the sacrifices, never lost the joy of learning more about physics \\ \, \\ \, \\ We owe all progress to them}

\pdfbookmark[0]{\contentsname}{toc} % Add pdf bookmark for contents page for navigation
\tableofcontents

\chapter*{Acknowledgements\protect\markboth{Acknowledgements}{Acknowledgements}} \label{Chapter:Acknowledgements}
\addcontentsline{toc}{chapter}{Acknowledgements}
\begin{quote}
  ``Yes, I began my journey alone, and I ended it alone. \\ 
    But that does not mean that I walked alone.''
  \begin{flushright}
  \hfill \hfill - Brandon Sanderson, \textit{Oathbringer}
  \par\end{flushright}
\end{quote}

\noindent This is most likely the most difficult part to write of the whole
thesis. Ever since I started doing ``research'', back in late 2018
during the final year of my physics degree,~several PhD thesis
have passed through my hands, shared with me by colleagues and supervisors
as useful introductory texts into specific topics. I cannot help but
say that the first thing I would always read were the Acknowledgements:
here, for a few lines of text, you can connect with the emotions of
a young researcher who struggled to push a PhD thesis forward for
several years, and finally made it to this section. Writing these lines
is a task that I have been putting off and, for various reasons, I
know it will be impossible to write it without getting emotional.
Well, this is finally my turn: here we go.

First and foremost, I have to start by expressing my most sincere
(and endless) gratitude to my supervisor, guide, mentor, boss and
also colleague, Steve King. I believe that I always carried inside
myself the joy of learning more about physics, the enthusiasm for
discovering the unknown and the optimism of the model builder,
but I know that Steve is the one who was able to enlighten all
this latent feelings inside me. I lost the hope once, on all of this,
but Steve was able to drive me back into the right way, and I will
never forget that conversation. Your time has been invaluable to me,
and I will always be thankful for all the fruitful discussions and
for the exciting work we did together. And of course, thanks for giving
me the opportunity to work with you back in 2020 when I was a student
who knew nothing, and thanks for putting me forward to give a plenary
talk at Moriond (where I learnt how to ski). I have become the physicist
I am today thanks to Steve.

My heart-felt gratitude also goes to all my collaborators for their
unwavering support and for their endless patience: Marzia, Valerie,
Avelino and Miguel.

From Marzia I learnt many things, including how important it is to
do precise calculations (otherwise we are no better than model builders)
and to understand the Standard Model in order to discuss possible
new physics. I really appreciate your time and advice when I was in
the difficult period of postdoc applications (and offers). I feel
that Marzia is a person who always remembers, protects and fights
for her friends and collaborators, and I feel blessed to find myself
in this select group.

I thank Valerie for welcoming me at CERN-TH and for introducing me
into the amazing environment of CERN-Cosmo. I feel that I am in the
company of a world-class leading expert in the field of early Universe
cosmology, where I think we may learn a lot about fundamental physics
in the coming future. She always had good advice when I needed it
and the answer to many of my fundamental, out-of-context questions.
Valerie is also the creator of the 4$\pi$ factor theorem in scientific
writing (``always multiply your intended deadline for paper submission
by a 4$\pi$ factor''). I express my most sincere gratitude to her,
for the time spent at CERN, for offering me to collaborate with her and for
giving me a reference letter when I most needed it.

I also owe many thanks to Avelino. He was my first contact into the
world of model building, which I really wanted to join. Avelino made
a lot of work in order to prepare an excellent FPU application back
in 2020, so that I could join the group at IFIC and start a PhD with
him, work which in the end was unpaid because I couldn't join the
group. I have been looking forward to collaborate with you since
then, and I am happy that the opportunity finally came to us recently!

And of course I cannot forget about Miguel. I still remember
your kind smile at Invisibles22 when you first approached to my poster,
a poster about flavour physics and anomalies nonetheless, and we ended
up talking about neutrinos, cosmology, research and life. This was
the beginning of a friendship that I hope will last forever. Thank
you as well for your kind welcome at CERN and for showing me the most
beautiful, hidden places of Geneva. I couldn't have made my postdoc
applications without all your help, guidance and support during the
most uncertain times. You always had time for me when I needed it,
including that warm phone call from Naples when I was deciding between
my postdoc offers (this I will never forget). I hope that eventually
I can learn from you how to be such an excellent, organised physicist, so
that I can have a proper work-life balance and stop working during the
weekends!

Obviously I cannot escape these lines without dedicating a paragraph to Xavi 
(and honestly I could write a full chapter). Everyday I am thankful that I was
assigned the secondment in Padova so that we could met. I sincerely
admit that I couldn't have finished this PhD without you. You appeared
like a gift fallen from heaven when I needed help with Madgraph and package-X on the more technical side, and on the more personal side
I needed someone around who would listen to the ramblings about my
research. I hope one day to know half the physics that you know. Beyond all of this, I am happy for every conversation we
had about the open ``problems'' and ``puzzles'' in physics, and
about life in general. You were also there when the difficulties of
the academic life hit me the hardest (I will always remember that
call from Glasgow when I was lost in flat-hunting), and I admit that all the uncertainties and all
the problems were easier to handle because you were always there.
I know our friendship goes beyond physics and our PhDs, and I hope
I have here a friend for life. The future right now is uncertain,
but I hope you will eventually find in yourself the enthusiasm~for discovering
new physics that I found while doing this PhD, which is as yours as
it is mine.

I also want to dedicate a few lines to Nikolai, who was sharing the
office with me at Southampton for the last two years. I am happy for
every conversation we had and for every time you gave me some chat
to disconnect from work. I remember those nights when we stayed working
in the office beyond 10 pm and 11 pm, which would have been much more
difficult if you had not been around. Even though we work in very
different topics, you were always kind to listen to my maunderings about
the problems in my research, and you were there to help me the best
you could. Having you around was also great to learn more about lattice
QCD, although I admit that sometimes I didn't understand everything
you say, but this is only because I have the small mind of a phenomenologist!
Even though now we part ways to start new postdocs in different places,
I hope more opportunities to meet and catch up will arise in the future.

Even if it may seem that the life of a researcher, and by extension
the life of a PhD student, can be a lonely enterprise, I must admit
fortunately it was not. I was lucky to share the journey with many
fellow students and researchers who made the whole task of doing a
PhD much easier. Among them I count all my fellow students at Southampton:
Alessandro, Arran, Ben, Mauricio, Giorgio, Michele, Rajnandini, Dalius,
Vlad, Jacan, Jack, Bowen, Giovanna, Shubhani... I had the best time as well
during my two months at CERN thanks to Stefan, Pepe, Salva and Virgile,
who shared all the lunches and ping pong sessions with me, plus I also thank
all the nice, more senior members of CERN-TH who contributed to a very welcoming
and friendly environment to discuss about physics, including Nick, Fabrizio,
Andreas, Tobias... and thanks to Alba as well for the nice party tours
at night around Geneva. 

Before CERN, I found a second home in Padova that I will always miss,
within the very warm group of local PhD students who became my italian family: Federico, Sofia, Clemens,
Gabriele Levati, Gabriele Perna, Federica, Fabio, Anna, Giulio, Leonardo,
Sara, Alfredo, Pietro... and of course, last but not least, \textit{il grandissimo} Stefano Di Noi.
Thank you guys for the amazing months that we spent together in Padova.
I also thank Stefano Rigolin for the nice welcome to Padova and Luca
Di Luzio for his time, patience and feedback back from when I was
working on 4321 models.

I am thankful as well to all the nice people of the HIDDeN family.
In particular, thanks to Jaime, Joao, Arturo, María, Salva, Virgile, Francesco,
Patrick, Gioacchino, Federica, Valentina, Giacomo and Luca, for making
the best time possible at any school or conference. I keep all the
adventures in my heart. Thanks also to Silvia, Olga, Joerg, Rebeca and Enrico
for keeping everything under control, and more generally thanks to
HIDDeN for the excellent conditions of my PhD and for the amazing secondments in Padova and CERN,
which made me grow both as a physicist and as a person.

The people of the AHEP group in Valencia also deserve my most sincere
gratitude: they have always remembered me and supported me even though
we parted ways three years ago in 2020. I am especially thankful to
Sergio for always trusting me, for always remembering me and for doing the effort of keeping in touch.
I am thankful to Pablo Martínez as well, who was there when I needed him the most.

I don't
have words to express how thankful I am to José Ignacio Illana, who
lead my first steps into research back in 2018 and 2019 when I was finishing
my physics degree in Granada: I would not be the physicist I am today without
you, thanks José.

I am also thankful to the people of the Glasgow PPT group, which will
be my new home for the next few years. In particular, thanks Sophie and Christoph for
trusting me for working with you, and thanks as well for the support
with housing and flat-hunting and for the nice conversations about
physics and life in academia that we had already (hopefully the first of many!). I express
my gratitude to Ben Allanach and Stefano Moretti as well, for being my reviewers
despite the significant extension of this manuscript, I am sure we will
have nice discussions during the viva!

I met many nice people in conferences and schools as well, among them I
am the most thankful to Daniel Naredo, Nico Gubernari, Simone, Jonathan, Renato, Miguel
Levy, Jorge Terol and Javier Lizana for the nice sessions of beers after the talks,
and for the great discussions about physics and life that came with
them.

We are now getting close to the end, where I want to express my gratitude to my family. Despite
all the complications and sacrifices of the last three years, I
thank my parents, Rosa and Miguel, for always supporting my decisions,
even though some of them were difficult. The life of a physicist
may not be what they were expecting for me, but they have always been
there when I needed someone to hold my back, and all of this would be
impossible without them. I am also thankful to my aunts, uncles and
cousins who, despite having a growing family to take care of,
always remember that cousin who travelled far away in order to pursue
his dream of being a physicist. I don't forget my best friends from Málaga,
Álvaro, Andrés and Nacho, who have been there since I have memory.
Each day with you guys is a blessing, and even though I have travelled far
away and visited many countries, the best place I have ever been to is in
Málaga with you, having a beer and talking about anything.

Finally, I am lovingly indebted to Sonalee, who joined my life when
I started this thesis three years ago. Thank you for so many wonderful
moments and for making me smile. Thank you, for always having time
to listen to me, for your patience and support in my darkest moments.
I am sorry that the circumstances may have not been the best, and
I am sorry for having to work on the weekends, for having to travel
so much and for having to move to a different city, away from you.
I am sorry that physics takes so much from me that I have little left
to give, but I am thankful to you for sharing this path, difficult
as it is, with me. 

While working on this thesis I found that the Universe and its fundamental
laws are an exciting study, which goes beyond our small role as tiny
humans living finite lives. I discovered something that is worth doing
for a living, something that will make me feel happier the day I die.
This is why I do what I do: because I want to be there when something
new about Nature is revealed to us. And I hope I can continue sharing
this journey with all of you who make it possible. From the deepest
of my heart, thank you for not letting me walk alone.

\newpage
\thispagestyle{empty}
%\,
%\newpage
%\thispagestyle{empty}
%\dedicatory{}

%\,
%\newpage
%\clearpage{\thispagestyle{empty}\cleardoublepage}

%\newpage
%\dedicatory{To my friend Xavi,\\ who shared this journey with me, \\ \, \\ \, \\ and to my parents, Rosa and Miguel,\\ who made it possible}

%% ---------- AUTHORSHIP DECLARATION/ ACKNOW. / DEDICATORY ----------
%% Either include citations like below (as many as required spaced with commas or 'and').
%% \bibentry command must be used here with prepublished papers
\authorshipdeclaration{\cite{FernandezNavarro:2021sfb},  \cite{FernandezNavarro:2022gst}, \cite{FernandezNavarro:2023rhv}, \cite{Bordone:2023ybl} and \cite{FernandezNavarro:2023hrf} in peer-reviewed journals, \cite{FernandezNavarro:2023lgk} and \cite{FernandezNavarro:2023ykw} as conference proceedings.}
%% Or state no citations like below
%% \authorshipdeclaration{}
%% -----------------------

\listoffigures
\listoftables
%% The List of listings does not, by default, appear in the ToC, so....
%\addtotoc{Listings}
%\lstlistoflistings
%\listofaddmaterial
%\addtolom{Material Name e.g Map}
%\addtolom{Material Name e.g CD}
%\addtolom{Test Material}
%%Lightweight Definitions and Abbreviations see package:nomencl for alternative
%% Include if relevant to discipline
\listofsymbols{ll}{2HDM & Two Higgs Doublet Model\\BP & Benchmark Point\\BSM & Beyond the Standard Model\\CKM & Cabibbo–Kobayashi–Maskawa\\CL & Confidence Level\\CLFV & Charged Lepton Flavour Violation\\$CP$ & Charge conjugation Parity\\DM & Dark Matter\\EDM & Electric Dipole Moment\\EFT & Effective Field Theory\\Eq. & Equation\\EW & Electroweak\\EWPOs & Electroweak Precision Observables\\FCC & Future Circular Collider\\FCCCs & Flavour Changing Charged Currents\\FCNCs & Flavour Changing Neutral Currents\\Fig. & Figure\\FN & Froggatt-Nielsen\\GR & General Relativity\\GIM & Glashow–Iliopoulos–Maiani\\GUT & Grand Unified Theory\\HL-LHC & High-Luminosity Large Hadron Collider\\IR & Infrared\\LEFT & Low-energy Effective Field Theory\\LEP & Large Electron-Positron collider\\LFU & Lepton Flavour Universality\\LFUV & Lepton Flavour Universality Violation\\LFV & Lepton Flavour Violation\\LH & Left-handed\\LHC & Large Hadron Collider\\MFV & Minimal Flavour Violation\\$\overline{\mathrm{MS}}$ & Modified Minimal Subtraction (a renormalisation scheme)\\MSSM & Minimal Supersymmetric Standard Model\\NLO & Next-to-Leading-Order\\NNLO & Next-to-Next-to-Leading-Order\\NP & New Physics\\PMNS & Pontecorvo–Maki–Nakagawa–Sakata\\PS & Pati-Salam\\QCD & Quantum Chromodynamics\\QED & Quantum Electrodynamics\\QFT & Quantum Field Theory\\Ref. & Reference\\RGE & Renormalisation Group Evolution\\RH & Right-handed\\SM & Standard Model\\SMEFT & Standard Model Effective Field Theory\\SSB & Spontaneous Symmetry Breaking\\SUSY & Supersymmetry\\TH & Tri-hypercharge\\UFO & Universal FeynRules Output\\UV & Ultraviolet\\VEV & Vacuum Expectation Value\\VL & Vector-like\\WC & Wilson Coefficient\\WIMPs & Weakly Interacting Massive Particles}
\mainmatter
%% ------------------ MAIN MATTER (CONTENT) --------------------
\hypersetup{
linkcolor=linkBlue,
urlcolor=linkBlue,
citecolor=linkBlue,
linktocpage=true,
}
\interfootnotelinepenalty=10000
\allowdisplaybreaks

\chapter*{Preface\protect\markboth{Preface}{Preface}} \label{Chapter:Preface}
\addcontentsline{toc}{chapter}{Preface}

\begin{quote}
  ``The purpose of a storyteller is not to tell you how \\ to think, but to give you questions to think upon.''
  \begin{flushright}
  \hfill \hfill $-$ Brandon Sanderson, \textit{The Way of Kings} 
  \par\end{flushright}
\end{quote}

\noindent Physics thrives on crisis\footnote{Quoting the great sentence by Steven Weinberg in \cite{Weinberg:1988cp}.}. This is probably the most important lesson
that one can extract from the history of (particle) physics. During
the last few hundred years, we have witnessed how every apparent failure
of the contemporaneous theories has led to abrupt developments in
our understanding of Nature. The Standard Model of particle physics
is no exception: in spite of its remarkable success, it leaves several
open problems and puzzles that very likely hint to a path towards
a more fundamental understanding of Nature. Despite my own experience
forces me to be humble, this is the reason for writing this thesis:
the honest hope that I will be able to learn something about Nature
that has not yet been revealed to us. 

Among the various open questions of the Standard Model (SM), the flavour
puzzle may be one of the most perverse. In contrast to other
shortcomings of the SM, the flavour puzzle does not
point to any particular energy scale for the new dynamics that might
be behind the origin of flavour. Most of the proposed theories rely
on \textit{a priori} undetermined flavour scales, which may be \textit{anywhere
from the Planck scale to the electroweak scale}, giving its name to
this thesis. 

However, this is no reason to be negative: the origin of flavour may
still be around the corner, waiting to be discovered in particle physics experiments that test the origin of flavour \textit{from the
bottom-up}. In fact, one of the findings of this thesis is that flavour
might originate from several new physics scales that may cover several
orders of magnitude from the Planck scale to the electroweak scale,
suggesting a possible multi-scale origin of flavour. If the lower layer
of new physics is low enough, we might be seeing its first signals
in the form of anomalies in low-energy flavour observables that hint
to new flavour-specific interactions, or that suggest the breaking of accidental flavour
symmetries of the SM such as lepton flavour universality.
This thesis is therefore motivated by the spirit of model building,
in the sense that I humbly believe that I can provide something new
and significant to the already vast set of theories that try to explain
the unknown. Moreover, I remain optimistic that my theories may be
tested in the current or upcoming generation of particle physics experiments,
and this motivates the lengthy phenomenological analyses that you
will find along the following chapters.

In Chapter~\ref{chap:Chapter1} I will perform a somewhat lengthy introduction to
the Standard Model, focusing on its open questions and on the particular
role of flavour. Among other topics, I will discuss the full set of free parameters of
the SM, I will introduce the type I seesaw mechanism
as a possible origin of neutrino masses and I will discuss
the approximate flavour symmetries of the SM, which are of great importance
for flavour model building. I will also motivate here the need for
a theory of flavour, and I will introduce the \textit{a priori} undetermined
flavour scales of a theory of flavour in the illustrative example
of Froggatt-Nielsen models.

In Chapter~\ref{chap:2} I will introduce the LEFT and the SMEFT, which
are very useful effective field theories for phenomenological analyses
of heavy new physics. I will discuss key flavour observables which
show experimental anomalies, hinting to a consistent departure from the SM emerging
from an underlying theory of flavour. I will highlight the significance of these anomalies
and the emerging puzzles regarding theory prediction or experimental determination
that put the BSM interpretation of these observables under question.
I will also introduce a significant set of observables which are so
far not anomalous, but are correlated to the anomalous observables
in well-motivated BSM scenarios, hence being important for testing
the proposed models. Regarding the so-called $B$-physics anomalies, I will
highlight an emerging BSM scenario that is preferred by current data,
which will be later on realised by the explicit theory of flavour
discussed in Chapter~\ref{Chapter:TwinPS}.

In Chapter~\ref{Chapter:Fermiophobic} I will study a class of local $U(1)'$ extensions of the SM,
where chiral fermions are uncharged under the $U(1)'$ but an exotic
family of vector-like fermions is not, providing effective $Z'$ couplings
for chiral fermions via mixing. This feature gives the name \textit{fermiophobic}
to this class of models. I will show how a simplified fermiophobic
framework can explain the $(g-2)_{\mu}$ anomaly via the basic idea of chiral
enhancement, correlating the enhancement of $(g-2)_{\mu}$ with a
suppression of Higgs diphoton decay. Afterwards, I will study a theory
of flavour with fermiophobic $Z'$ that can explain the origin of
the SM flavour structure via the mechanism of messenger dominance.
This theory connects the origin of Yukawa couplings in the SM with
the origin of effective $Z'$ couplings for chiral fermions, hence
connecting the origin of flavour hierarchies in the SM with the low-energy
phenomenology of the model. We conclude that ultimately the $(g-2)_{\mu}$
anomaly cannot be explained in the context of this theory of flavour
due to a correlated enhancement of $\mathcal{B}(\tau\rightarrow\mu\gamma)$.
Nevertheless, we find that the flavour structure of the model allows
for the $Z'$ boson to be as light as 1 TeV, within the reach of current
particle physics experiments.

In Chapter~\ref{Chapter:TwinPS} I will study a twin Pati-Salam theory of flavour which
contains a TeV scale vector leptoquark $U_{1}\sim(\mathbf{3,1},2/3)$.
The model features a fermiophobic framework as well, where both the
effective Yukawa couplings for chiral fermions and their effective
$U_{1}$ couplings originate again from mixing with heavy vector-like
fermions. The mechanism of messenger dominance plays a fundamental
role here to simultaneously explain the fermion mass hierarchies and
deliver the flavour structure required to explain the $B$-physics
anomalies. I will show that one vector-like fermion family is not
enough to achieve such flavour structure, but indeed three vector-like
fermion families are required. In this case, the model predicts a
plethora of low-energy signals in flavour observables, several of
them fundamentally related to the origin of fermion mass hierarchies
and mixing. The model can also be tested via direct searches of the
new heavy degrees of freedom at the LHC, including the vector-like
fermions, the $U_{1}$ leptoquark, and a coloron $g'\sim(\mathbf{8,1},0)$
and $Z'$ gauge bosons.

In Chapter~\ref{Chapter:Tri-hypercharge} I will study the possibility that the Standard Model
originates from a non-universal gauge theory in the ultraviolet. I will argue that one of the most simple
ways to achieve this is by assigning a separate gauge hypercharge
to each fermion family at high energies, broken down to the usual
weak hypercharge which is the diagonal subgroup. This simple framework
avoids the family replication of the SM, and could be the first step
towards a deep non-universal gauge structure in the UV. If the Higgs
doublet(s) only carry third family hypercharge, then only third family
renormalisable Yukawa couplings are allowed by the gauge symmetry. However, non-renormalisable
Yukawa couplings for the light families may be induced by the high scale scalar fields
which break the three hypercharges down to the SM hypercharge, providing
an explanation for fermion mass hierarchies and the smallness of CKM
quark mixing. I will show that in order to explain neutrino mixing,
it is useful to introduce right-handed neutrinos which carry non-zero
hypercharges (although their sum must vanish), which then turn out
to get Majorana masses at the lowest scale of symmetry breaking, that could
be as low as a few TeV. Indeed, I will motivate that the model has
a rich phenomenology via $Z'$ bosons if the flavour scales are relatively low: from flavour-violating observables to LHC physics and electroweak precision observables.

In Chapter~\ref{Chapter:Tri-unification} I will propose a gauge unified origin for gauge non-universal
frameworks such as the aforementioned tri-hypercharge theory.
The model consists on assigning a separate $SU(5)$ group
to each fermion family. However, assuming that the three $SU(5)$
groups are related by a cyclic permutation symmetry $\mathbb{Z}_{3}$, then
the model is described by a single gauge coupling in the UV, despite
$SU(5)^{3}$ being a non-simple group. In this
manner, $SU(5)^{3}$ ``tri-unification'' reconciles the idea of gauge non-universality
with the idea of gauge coupling unification, opening the possibility to build consistent
non-universal descriptions of Nature that are valid all the way up to the GUT scale. First, I will show a general
$SU(5)^{3}$ tri-unification framework where gauge non-universal
theories of flavour may be embedded, and secondly I will construct
a minimal tri-hypercharge example which can account for all the quark
and lepton (including neutrino) masses and mixing parameters, with
the five gauge couplings of the tri-hypercharge group unifying at
the GUT scale into a single gauge coupling associated to the cyclic
$SU(5)^{3}$ group. I will study the implications for the stability of the proton
in such a setup.

Finally, in Chapter~\ref{Chapter: Conclusions} I will summarise my own findings and the main results extracted from each chapter. I will also motivate next possible steps in my research and give my own view about the future ahead of us in particle physics. I hope the reader finds interesting the particular approaches
to the flavour puzzle discussed here, along with all the
related phenomenology and discovery prospects. And on top of everything, I hope
the reader enjoys as much as I did when learning and thinking about
the work included in this thesis.
% Vista preliminar cuerpo

\chapter{Flavour in the Standard Model and beyond}\label{chap:Chapter1}

\begin{quote}
  ``The standard theory may survive as a part of the ultimate theory, or it may turn out to be fundamentally wrong. In either case, it will have been an important way-station, and the next theory will have to be better.''
  \begin{flushright}
  \hfill \hfill $-$ Sheldon L. Glashow
  \par\end{flushright}
\end{quote}

\noindent In this chapter we provide an introduction to the Standard Model
(SM) of particle physics, with special attention to the flavour sector
and its particular properties (for a more complete review see e.g.~\cite{langacker2009standard,Avelino:2016,Illana:2022hab,Pich:2012sx}).
The SM is a mathematical model of Nature built upon fundamental principles.
It provides a simple framework in which the different fundamental
components of matter and their interactions can be understood. At
the end of the chapter, we will show that despite its remarkable success
in describing the vast majority of experimental data, the SM cannot
be the ultimate theory of Nature: it leaves several experimental and
theoretical questions unanswered, motivating us to go beyond.

\section{The basics of the Standard Model: Fundamental Principles and particle
content}

The SM was built to understand the strong, weak and electromagnetic
interactions observed in Nature, along with the fundamental matter
components which experience such interactions at the quantum level.
As such, the SM is a quantum field theory based on the principles
of locality, causality and renormalisability. In the SM, the concept
of symmetry plays a central role. The SM postulates that Nature is
invariant under the spacetime (Poincaré) symmetries and under the
so-called SM gauge (local) symmetry group
\begin{equation}
SU(3)_{c}\times SU(2)_{L}\times U(1)_{Y}\,.\label{eq:SM_symmetry}
\end{equation}
Each of the components of the SM gauge group is associated with a
quantum number to be preserved in Nature. $SU(3)_{c}$ is associated
to the so-called color, while $U(1)_{Y}$ is associated to the so-called
hypercharge. Remarkably, $SU(2)_{L}$ acts over fermions as the left-handed
chirality of the Lorentz group\footnote{However, the $SU(2)_{L}$ gauge symmetry of the SM should not be confused with the $SU(2)_{L}$ symmetry of the Lorentz group, as they are fundamentally different. An example of this is the Higgs doublet, which transforms as a doublet under the SM $SU(2)_{L}$ but is a spin-0 scalar under the Lorentz symmetry.} (subgroup of the full Poincaré spacetime
symmetry). This will have important
implications for the fermion content of the theory. Electric charge
$U(1)_{Q}$ is contained within the so-called \textit{electroweak}
symmetry group $SU(2)_{L}\times U(1)_{Y}$. 
\begin{table}
\begin{centering}
\begin{tabular}{lccc}
\toprule 
Field & $SU(3)_{c}$ & $SU(2)_{L}$ & $U(1)_{Y}$\tabularnewline
\midrule
\midrule 
$Q_{L1}=\left(\begin{array}{c}
u_{L1}\\
d_{L1}
\end{array}\right)$ & $\mathrm{\mathbf{3}}$ & $\mathbf{2}$ & $1/6$\tabularnewline
$u_{R1}$ & $\mathrm{\mathbf{3}}$ & $\mathbf{1}$ & $2/3$\tabularnewline
$d_{R1}$ & $\mathrm{\mathbf{3}}$ & $\mathbf{1}$ & $-1/3$\tabularnewline
$L_{L1}=\left(\begin{array}{c}
\nu_{L1}\\
e_{L1}
\end{array}\right)$ & $\mathbf{1}$ & $\mathbf{2}$ & $-1/2$\tabularnewline
$e_{R1}$ & $\mathbf{1}$ & $\mathbf{1}$ & $-1$\tabularnewline
\midrule 
$Q_{L2}=\left(\begin{array}{c}
u_{L2}\\
d_{L2}
\end{array}\right)$ & $\mathrm{\mathbf{3}}$ & $\mathbf{2}$ & $1/6$\tabularnewline
$u_{R2}$ & $\mathrm{\mathbf{3}}$ & $\mathbf{1}$ & $2/3$\tabularnewline
$d_{R2}$ & $\mathrm{\mathbf{3}}$ & $\mathbf{1}$ & $-1/3$\tabularnewline
$L_{L2}=\left(\begin{array}{c}
\nu_{L2}\\
e_{L2}
\end{array}\right)$ & $\mathbf{1}$ & $\mathbf{2}$ & $-1/2$\tabularnewline
$e_{R2}$ & $\mathbf{1}$ & $\mathbf{1}$ & $-1$\tabularnewline
\midrule 
$Q_{L3}=\left(\begin{array}{c}
u_{L3}\\
d_{L3}
\end{array}\right)$ & $\mathrm{\mathbf{3}}$ & $\mathbf{2}$ & $1/6$\tabularnewline
$u_{R3}$ & $\mathrm{\mathbf{3}}$ & $\mathbf{1}$ & $2/3$\tabularnewline
$d_{R3}$ & $\mathrm{\mathbf{3}}$ & $\mathbf{1}$ & $-1/3$\tabularnewline
$L_{L3}=\left(\begin{array}{c}
\nu_{L3}\\
e_{L3}
\end{array}\right)$ & $\mathbf{1}$ & $\mathbf{2}$ & $-1/2$\tabularnewline
$e_{R3}$ & $\mathbf{1}$ & $\mathbf{1}$ & $-1$\tabularnewline
\bottomrule
\end{tabular}
\par\end{centering}
\caption[Transformation properties of fundamental fermions under the Standard Model
gauge group]{The fundamental fermions of the SM and their transformation properties
under the SM gauge group. We show explicitly the two components of
the $SU(2)_{L}$ doublets, while colour indices are implicit.\label{tab:SM_fermion_content}}

\end{table}

The SM further postulates the existence of fundamental
spin-1/2 fermions, understood as Dirac four-component spinor fields $\psi(x)=\psi(x)_{L}\oplus\psi(x)_{R}$
containing both the left-handed and right-handed\footnote{With the exception of the right-handed chirality of neutrinos, as discussed
later in this section.} chiralities of the Lorentz group, which are related via the parity
symmetry. The left-handed and right-handed components of a Dirac four-spinor
can be extracted as two-component Weyl spinors via the chiral projectors
\begin{equation}
\psi_{L}\equiv P_{L}\psi=\frac{1-\gamma_{5}}{2}\psi\,,\qquad\psi_{R}\equiv P_{R}\psi=\frac{1+\gamma_{5}}{2}\psi\,,
\end{equation}
where $\gamma_{5}\equiv i\gamma^{0}\gamma^{1}\gamma^{2}\gamma^{3}$
and $\gamma^{\mu}$ are the gamma matrices (also called Dirac matrices), with $\mu=0,1,2,3$ being
a Lorentz index.

The fundamental fermions undergo particular transformation properties
under the SM gauge group, as given in Table~\ref{tab:SM_fermion_content}.
Each fundamental fermion comes in three copies, with identical transformation
rules under the SM gauge group. Remarkably, the left-handed components
of fundamental fermions transform as doublets under $SU(2)_{L}$,
while the right-handed components transform as singlets,
\begin{flalign}
Q_{Li}^{a}=\left(\begin{array}{c}
u_{Li}^{a}\\
d_{Li}^{a}
\end{array}\right), & \quad u_{Ri}^{a},\quad d_{Ri}^{a},\label{eq:fermions_SU(2)}\\
L_{Li}=\left(\begin{array}{c}
\nu_{Li}\\
e_{Li}
\end{array}\right), & \quad e_{Ri}.\nonumber 
\end{flalign}
In this manner, parity is explicitly violated within the SM, and we
say that the SM is a chiral theory\footnote{The fundamental reason for the SM to be a chiral theory is a mystery,
related to the question of why Nature has chosen the particular symmetry
group in Eq.~(\ref{eq:SM_symmetry}) and not any other.}. As anticipated before, fundamental fermions come in three identical
copies that we denote as flavours, such that $i=1,2,3$. The fermions
in the first line of Eq.~(\ref{eq:fermions_SU(2)}) all carry colour
$a=r,g,b$ (red, green, blue), in such a way that
they can be arranged as a triplet under $SU(3)_{c}$,
\begin{equation}
q=\left(\begin{array}{c}
\textcolor{red}{q^{r}}\\
\textcolor{DarkGreen}{q^{g}}\\
\textcolor{blue}{q^{b}}
\end{array}\right)\,,
\end{equation}
where $q=u_{Li}\,,d_{Li}\,,u_{Ri}\,,d_{Ri}$ since $SU(3)_{c}$ does
not discriminate by chirality nor flavour. These ``colourful'' fermions
are denoted as \textit{quarks}. In this manner, the quark content
of the SM consists of three \textit{up-type} quarks $u_{L,Ri}^{a}$
and three \textit{down-type} quarks $d_{L,Ri}^{a}$, all carrying
colour but discriminated by their different hypercharge, plus their
left-handed components are arranged in $SU(2)_{L}$ doublets as per
the first line in Eq.~(\ref{eq:fermions_SU(2)}). The remaining fermions
are called \textit{leptons}, and they transform as singlets under $SU(3)_{c}$,
hence carrying no color. We discriminate between \textit{charged
leptons $e_{L,Ri}$} and \textit{neutrinos} $\nu_{Li}$. The latter
will be shown to carry zero electric charge, in contrast with the
former. In this manner, we have identified the different \textit{charged
sectors} of the SM fermions, each of them including three identical
flavours. Remarkably, we have chosen not to introduce the right-handed
counterpart of the neutrino fields: it would transform as a complete
singlet under the SM, $\nu_{R}\sim(\mathbf{1},\mathbf{1},0)$, therefore
it would not experience any gauge interaction and would be an unphysical
field so far. We will come back to the discussion of right-handed neutrinos
when we introduce the origin of fermion masses in the SM.

Having discussed the transformation properties of fundamental fermions
under the SM gauge group, we proceed to explicitly include for completeness
the gauge transformations of a generic fermion $\psi(x)$ under each
component of the SM gauge group,
\begin{equation}
U(1)_{Y}:\quad\psi(x)\rightarrow\mathrm{exp}\left[-iY_{\psi}\theta_{Y}(x)\right]\psi(x)\,,\label{eq:U(1)_transformation}
\end{equation}
\begin{equation}
SU(2)_{L}:\quad\psi(x)\rightarrow\mathrm{exp}\left[-iT^{a}\theta_{L}^{a}(x)\right]\psi(x)\,,
\end{equation}
\begin{equation}
SU(3)_{c}:\quad\psi(x)\rightarrow\mathrm{exp}\left[-it^{\alpha}\theta_{c}^{\alpha}(x)\right]\psi(x)\,,\label{eq:SU(3)_transformation}
\end{equation}
where repeated indices are always taken as summed. $Y$ is the generator
of the hypercharge $U(1)_{Y}$ group, such that $Y_{\psi}$ is the
hypercharge of the $\psi$ fermion. $T^{a}$ with $a=1,2,3$ are the
generators of $SU(2)_{L}$, while $t^{\alpha}$ with $\alpha=1,...,8$
are the generators of $SU(3)_{c}$. When acting upon a doublet representation
of $SU(2)_{L}$, $T^{a}=\sigma^{a}/2$ where $\sigma^{a}$ are the
Pauli matrices, while when acting upon singlets $T^{a}=0$. This way,
the two different components of a $SU(2)_{L}$ doublet can be discriminated
by their third component of weak isospin $T_{3}$: by convention,
up-quarks and neutrinos carry $T_{3}=1/2$, while down-quarks and
charged leptons carry $T_{3}=-1/2$. For the case of $SU(3)_{c}$,
the triplet representations transform with $t^{\alpha}=\lambda^{\alpha}/2$,
where $\lambda^{\alpha}$ are the Gell-Mann matrices, while singlet
representations transform as $t^{\alpha}=0$.

The particular choice of fermion transformation properties and hypercharge
assignments under the SM gauge group is not arbitrary. We will see that the 
transformation properties of the SM fermions are motivated by experimental evidence
(such as the fact that only quarks experience the strong interaction), 
plus the requirement to cancel gauge anomalies, which is crucial for any consistent quantum field theory.
Any other choice would be incompatible with experimental data, or require the addition of extra, 
unseen fermions, being again incompatible with experimental evidence.
Nevertheless, there is no fundamental reason to have three copies
for each fermion (flavours), since gauge anomaly cancellation occurs
in the SM for each individual family. The most simple choice would
be to have only one family formed by one up-quark, down-quark, charged
lepton and neutrino. However, the experimental data tell us that three
families do exist, with identical transformation properties under
the SM gauge group.

Postulating the existence of a gauge symmetry naturally predicts the
existence of the so-called gauge bosons: physical, fundamental and
massless particles with spin-1, transforming as four-vectors $V_{\mu}$
($\mu=0,1,2,3$) under the Lorentz group. Gauge bosons act as mediators
of interactions among spin-1/2 fermions. The gauge bosons are ultimately
associated to the generators of the gauge group and live in the adjoint
representation, as we shall see. Therefore, in the SM we have 8 massless
gluons $G_{\mu}^{\alpha}\sim(\mathbf{8},\mathbf{1},0)$ which mediate the strong interactions,
with $\alpha=1,...,8$, associated to $SU(3)_{c}$. We also have three
gauge bosons $W_{\mu}^{a}\sim(\mathbf{1},\mathbf{3},0)$ with $a=1,2,3$ associated to $SU(2)_{L}$,
and one gauge boson $B_{\mu}\sim(\mathbf{1},\mathbf{1},0)$ associated to $U(1)_{Y}$.

\section{Gauge sector \label{sec:The-gauge-sector}}

Having introduced the gauge symmetry of the SM, the fundamental fermions
and their transformation rules, now we are led to write a Lagrangian
in order to study the dynamics of this system. We require our Lagrangian
to preserve the Poincaré symmetry of spacetime plus the gauge symmetry
of the SM. Following common practice, we start by introducing kinetic
terms for all fermion fields and gauge fields, so they can propagate
through spacetime. The resulting Lagrangian is
\begin{equation}
\mathcal{L}_{\mathrm{gauge}}=-\frac{1}{4}G_{\mu\nu}^{\alpha}G^{\mu\nu\alpha}-\frac{1}{4}W_{\mu\nu}^{a}W^{\mu\nu a}-\frac{1}{4}B_{\mu\nu}B^{\mu\nu}+i\sum_{f}\overline{\psi}_{f}\gamma^{\mu}\partial_{\mu}\psi_{f}\,,\label{eq:Gauge_Lagrangian_1}
\end{equation}
where $\overline{\psi}\equiv\psi^{\dagger}\gamma^{0}$ is the adjoint Dirac spinor. The sum over fermions
runs for all the different fields in Table~\ref{tab:SM_fermion_content}.
The so-called strength tensors are given by 
\begin{equation}
G_{\mu\nu}^{\alpha}=\partial_{\mu}G_{\nu}^{\alpha}-\partial_{\nu}G_{\mu}^{\alpha}+g_{s}f_{\alpha\beta\rho}G_{\mu}^{\beta}G_{\nu}^{\rho}\,,
\end{equation}
\begin{equation}
W_{\mu\nu}^{a}=\partial_{\mu}W_{\nu}^{a}-\partial_{\nu}W_{\mu}^{a}+g_{L}\epsilon_{abc}W_{\mu}^{b}W_{\nu}^{c}\,,
\end{equation}
\begin{equation}
B_{\mu\nu}=\partial_{\mu}B_{\nu}-\partial_{\nu}B_{\mu}\,,
\end{equation}
where $g_{s}$ and $g_{L}$ are the coupling constants of $SU(3)_{c}$
and $SU(2)_{L}$, respectively. The commutator between generators
of a Lie group provides the structure constants\footnote{We remind the reader that the vector space of generators of a Lie
group, with the commutator between generators as inner product, provides
the so-called Lie algebra of the given group.}. For $SU(3)_{c}$, $[t^{\alpha},t^{\beta}]=if_{\alpha\beta\rho}t^{\rho}$
provides the antisymmetric structure constants $f_{\alpha\beta\rho}$.
For $SU(2)_{L}$, $[T^{a},T^{b}]=i\epsilon_{abc}T^{c}$ where $\epsilon_{abc}$
is the completely antisymmetric three-index tensor.

One can check that the strength tensor of the abelian factor is gauge invariant.
In contrast, the strength tensors of the non-abelian factors are not gauge invariant, in this case
only the product of two strength tensors contracted over all indices is gauge invariant\footnote{Remember
the infinitesimal gauge transformations of gauge boson fields
$V_{\mu}^{a}\rightarrow V_{\mu}^{a}-\frac{1}{g}\partial_{\mu}\alpha^{a}(x)-f_{abc}V_{\mu}^{c}\alpha^{b}(x)$
for $SU(N)$ gauge bosons and $V_{\mu}\rightarrow V_{\mu}-\frac{1}{g}\partial_{\mu}\alpha(x)$
for $U(1)$ gauge bosons, where $\alpha^{(a)}(x)\ll1$ are arbitrary
functions of spacetime.}. This ensures 
that the kinetic gauge terms preserve the SM gauge symmetry. However, one
can check that the fermion kinetic terms introduced in Eq.~(\ref{eq:Gauge_Lagrangian_1})
are not gauge invariant by simply applying the transformations in
Eqs.~(\ref{eq:U(1)_transformation}-\ref{eq:SU(3)_transformation}).
We can obtain gauge invariant fermion kinetic terms by replacing the
derivative in Eq.~(\ref{eq:Gauge_Lagrangian_1}) by a covariant derivative,
\begin{equation}
D_{\mu}=\partial_{\mu}-ig_{s}t^{\alpha}G_{\mu}^{\alpha}-ig_{L}T^{a}W_{\mu}^{a}-ig_{Y}YB_{\mu}\,.
\end{equation}
The covariant derivative transforms as an adjoint of the complete gauge group,
and therefore the gauge boson fields live in the adjoint representation.

One can check that the resulting Lagrangian,
\begin{equation}
\mathcal{L}_{\mathrm{gauge}}=-\frac{1}{4}G_{\mu\nu}^{\alpha}G^{\mu\nu\alpha}-\frac{1}{4}W_{\mu\nu}^{a}W^{\mu\nu a}-\frac{1}{4}B_{\mu\nu}B^{\mu\nu}+i\sum_{f}\overline{\psi}_{f}\gamma^{\mu}D_{\mu}\psi_{f}\,,\label{eq:Gauge_Lagrangian_2}
\end{equation}
is invariant under the SM gauge group. As a consequence of imposing
gauge invariance, the Lagrangian now contains interaction terms coupling
fermions with gauge bosons. For example, after expanding the covariant
derivative, one can find a term $g_{Y}\overline{\psi}\gamma^{\mu}B_{\mu}\psi$
in the Lagrangian, coupling fermions to the hypercharge gauge boson.
This is the beauty of the gauge principle: just by imposing gauge
symmetry arguments, we can reproduce complicated interactions between
fermions and gauge bosons observed in Nature. Since all the SM families have the same
gauge symmetry transformations, the interactions among the SM fermions
and gauge bosons are \textit{universal in flavour}. In other words, all the
fermion flavours have exactly the same interactions with the gauge
bosons. This fact translates into the appearance of a global flavour
symmetry (exact in the gauge sector at the classical level) as
\begin{equation}
U(3)^{5}=U(3)_{Q}\times U(3)_{L}\times U(3)_{u}\times U(3)_{d}\times U(3)_{e}\,.\label{eq:Flavour_Symmetry}
\end{equation}
This flavour symmetry of the SM is denoted as \textit{accidental},
because it is not imposed nor postulated, but simply arises as a consequence
of having three identical families of fermions.

Due to the non-abelian nature of the $SU(3)_{c}$ and the $SU(2)_{L}$
symmetries, the kinetic terms for the corresponding gauge fields also
generate self-interactions. These terms play a crucial role in the
explanation of several key features of the strong interactions, such
as confinement and asymptotic freedom \cite{Gross:1973id,Politzer:1973fx,Coleman:1973sx,Fritzsch:1973pi,Politzer:1974fr}.
The strong interactions mediated by gluons lead to composite subatomic
particles made by quarks, that we denote as hadrons. Hadrons consist
of baryons, made of three quarks, and mesons, made of a quark-antiquark
pair. Some remarkable examples of baryons are the proton and neutron,
which interact as well via the strong force in order to build the
atomic nuclei. A dedicated review of the strong interactions is beyond
the scope of this thesis, but we refer the interested reader to Ref.~\cite{Altarelli2017}.

The Lagrangian included so far would be enough to describe a classical
theory. The quantisation of gauge fields involves a number of subtleties,
related to the fact that the quanta (gauge bosons) are massless spin-1
particles with just two degrees of freedom embedded in a vector field
that has four. The usual solution consists in modifying the Lagrangian
by adding a gauge-fixing term, $\mathcal{L}=\mathcal{L}_{\mathrm{gauge}}+\mathcal{L}_{\mathrm{GF}}$,
in order to get rid of the extra degrees of freedom that would lead
to an inconsistent quantum field theory. The quantisation of non-abelian
gauge theories, like $SU(3)_{c}$ and $SU(2)_{L}$, also requires
the addition of auxiliary anticommuting spin-0 fields, usually denoted as \textit{ghost}
fields. However, these fields are unphysical, as they never appear
as external fields but only in internal lines of Feynman diagrams,
and are just required for consistency of the quantum field theory
\cite{Faddeev:1967fc}. It can be shown that a quantum field theory
with these ingredients is renormalisable (see e.g. \cite{peskin1995introduction}).
This means that the ultraviolet divergences appearing from quantum
corrections (\textit{loops}) can be absorbed by an appropriate redefinition
of the parameters and fields in the classical Lagrangian. In this
manner, all physical observables can be reliably calculated.

All things considered, so far we have constructed a consistent and
renormalisable gauge theory described by the following Lagrangian
\begin{equation}
\mathcal{L}_{\mathrm{SM}}=\mathcal{L}_{\mathrm{gauge}}+\mathcal{L}_{\mathrm{GF}}+\mathcal{L}_{\mathrm{ghosts}}\,,
\end{equation}
where massless gauge bosons interact with massless fermions. Notice
that the fact that the SM is a chiral theory, i.e.~that left-handed
and right-handed fermions transform differently under the gauge group,
implies that conventional Dirac mass terms violate gauge invariance.
Therefore, all fermions should indeed be massless. In a similar manner, all
gauge bosons associated to unbroken gauge symmetries are massless.
This description of the SM introduced so far, based only on fundamental
principles, gauge symmetry and fundamental fermions, is however not
realised in Nature. Quarks need to become massive in order to
provide the observed spectrum of hadrons, plus charged lepton masses
have been measured to be non-zero with good precision \cite{PDG:2022ynf}.
Moreover, data suggest that the electroweak part of the SM, $SU(2)_{L}\times U(1)_{Y}$,
is somehow broken down to electric charge $U(1)_{Q}$, and as a consequence
of such a process, linear combinations of the $W_{\mu}^{a}$ and $B_{\mu}$
bosons become massive, leading to the mediators of the weak force
$W_{\mu}^{\pm}$ and $Z_{\mu}$. The massless photon is recovered
as the linear combination of $W_{\mu}^{3}$ and $B_{\mu}$ associated
to the unbroken $U(1)_{Q}$. In the next two sections, we introduce
the mechanism responsible for breaking the electroweak symmetry, leading
to the weak and electromagnetic interactions, along with the masses
of (most) fundamental fermions. This will involve the addition of extra terms
to the renormalisable Lagrangian, leading to
\begin{equation}
\mathcal{L}_{\mathrm{SM}}=\mathcal{L}_{\mathrm{gauge}}+\mathcal{L}_{\mathrm{GF}}+\mathcal{L}_{\mathrm{ghosts}}+\mathcal{L}_{\mathrm{scalar}}-\mathcal{L}_{\mathrm{Yukawa}}\,.\label{eq:SM_Lagrangian}
\end{equation}

\section{Scalar sector and electroweak symmetry breaking} \label{sec:ScalarSectorandEWSSB}

In the previous sections, we have built a consistent and renormalisable
quantum field theory based on the SM gauge symmetry. However, such a
theory only contains massless gauge bosons, making it impossible to
explain the weak interactions observed in Nature. Nevertheless, in
1964, Peter Higgs, François Englert and Robert Brout proposed a mechanism
that allows to \textit{spontaneously} break a gauge symmetry, giving
mass to the gauge bosons associated to the broken generators \cite{Higgs:1964pj,Englert:1964et}.
This mechanism was later applied by Weinberg and Salam \cite{Salam:1964ry,Weinberg:1967tq,Salam:1968rm} to the electroweak
part of the SM originally proposed by Glashow \cite{Glashow:1961tr}, leading to its spontaneous
symmetry breaking down to $U(1)_{Q}$. The idea consists in constructing
a theory with a fully symmetric Lagrangian, but whose vacuum is not
invariant under the symmetry. This can be achieved in a rather minimal
way by introducing in the theory a complex spin-0 field, transforming
as a scalar under the Lorentz group. This so-called Higgs field will
develop a vacuum expectation value (VEV) that will trigger the spontaneous
breaking of the electroweak symmetry.

Since we need to break $SU(2)_{L}$, the Higgs should transform non-trivially
under it. The most simple choice is to postulate that the Higgs field
transforms as a doublet under $SU(2)_{L}$. In a similar manner, we
want to break $U(1)_{Y}$, so the Higgs field must carry non-zero
hypercharge. Finally, it should carry no color in order to preserve $SU(3)_{c}$,
which describes the strong interaction. With these considerations, a particularly appealing choice
is $H\sim(\mathrm{\mathbf{1},\mathbf{2},1/2)}$. We can parameterise such 
a Higgs doublet as
\begin{equation}
H=\left(\begin{array}{c}
\sigma\\
\phi
\end{array}\right)\,.
\end{equation}
Interestingly, the component $\phi$ of the doublet carries $T_{3}=-1/2$ (and $Y=1/2$), meaning that
if such component develops a VEV, then the electroweak symmetry is spontaneously broken
but a gauge $U(1)$ associated to the linear combination $T_{3}+Y$ remains unbroken.
Indeed, such a gauge $U(1)$ can be associated to the electromagnetic interactions or
quantum electrodynamics (QED), giving rise to the massless photon.
The most general renormalisable Lagrangian for the Higgs
doublet can be written as
\begin{equation}
\mathcal{L}_{\mathrm{scalar}}=\left(D_{\mu}H\right)^{\dagger}\left(D^{\mu}H\right)-V(H)\,,\label{eq:L_Scalar}
\end{equation}
where the first term is the Higgs kinetic term. Just like in the fermion
sector, the covariant derivative has been introduced to render the
kinetic term gauge invariant. When applied to the Higgs doublet, the
covariant derivative reads
\begin{equation}
D_{\mu}H=\left(\partial_{\mu}-ig_{L}\frac{\sigma^{a}}{2}W_{\mu}^{a}-ig_{Y}\frac{1}{2}B_{\mu}\right)\,.
\end{equation}
On the other hand, gauge invariance and renormalisability also allow
the inclusion of a scalar potential with the form
\begin{equation}
V(H)=m_{H}^{2}H^{\dagger}H+\lambda\left(H^{\dagger}H\right)^{2}\,,\label{eq:Scalar_Potential}
\end{equation}
containing a mass term for the Higgs doublet and a self-interaction
term.

\begin{figure}
\begin{centering}
\includegraphics[scale=1.2]{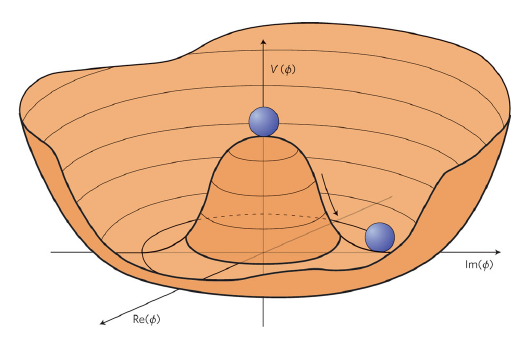}
\par\end{centering}
\caption[Higgs scalar potential]{Higgs scalar potential for the case $m_{H}^{2}<0$ in terms of the
neutral component of the Higgs doublet, $\phi$. Figure taken from
\cite{Ellis:2015tba}.\label{fig:SSB_potential}}
\end{figure}

The vacuum state of the theory can be obtained by minimising the scalar
potential 
\begin{equation}
\left.\frac{\partial V}{\partial H}\right|_{H=\langle H\rangle}=m_{H}^{2}\frac{v_{\mathrm{SM}}}{\sqrt{2}}+\lambda\frac{v_{\mathrm{SM}}^{3}}{\sqrt{2}}=0\,,\label{eq:Minimum_Condition}
\end{equation}
where
\begin{equation}
\langle H\rangle=\frac{1}{\sqrt{2}}\left(\begin{array}{c}
0\\
v_{\mathrm{SM}}
\end{array}\right)
\end{equation}
is the VEV of the Higgs doublet, and we can take $v_{\mathrm{SM}}$ to be real and positive without loss of generality. It is straightforward to check that
for $m_{H}^{2}>0$ , the condition in Eq.~(\ref{eq:Minimum_Condition})
provides just one real minimum at $v_{\mathrm{SM}}=0$. However, for $m_{H}^{2}<0$,
the potential has two real and positive extrema
\begin{equation}
v_{\mathrm{SM}}=0\,,\qquad v_{\mathrm{SM}}=\sqrt{-\frac{m_{H}^{2}}{\lambda}}\,.
\end{equation}
If $\lambda<0$, the potential is not bounded from below, making it
impossible to define the ground state. Instead, if we impose $\lambda>0$,
then $v_{\mathrm{SM}}=0$ corresponds to a local maximum while $v_{\mathrm{SM}}=\sqrt{-m_{H}^{2}/\lambda}>0$
corresponds to the global minimum of the scalar potential, that we
identify with the ground state. As depicted in Fig.~\ref{fig:SSB_potential},
there is an infinite set of degenerate configurations for the neutral
component $\phi$ that lead to the same vacua. Choosing one of those
vacua triggers the spontaneous breaking of the electroweak gauge symmetry,
\begin{equation}
SU(2)_{L}\times U(1)_{Y}\rightarrow U(1)_{Q}\,,
\end{equation}
where, as anticipated before, the unbroken $U(1)$ is associated to
the electromagnetic force, in such a way that
only electrically charged fermions carry non-zero charges under $U(1)_{Q}$. We
can study which generators of the subgroup $SU(2)_{L}\times U(1)_{Y}$
are left unbroken by applying the generators to the vacuum $\langle H\rangle$
and verifying whether it is left invariant or not. The vacuum is left
invariant by a generator $G$ if 
\begin{equation}
e^{i\alpha G}\langle H\rangle=\langle H\rangle\,,
\end{equation}
which, for an infinitesimal transformation ($\alpha\ll1$) leads to
\begin{equation}
e^{i\alpha G}\langle H\rangle\simeq\left(1+i\alpha G\right)\langle H\rangle\,,
\end{equation}
which implies $G\langle H\rangle=0$. In this case, we say that \textquotedblleft $G$
annihilates the vacuum\textquotedblright . We can now apply all four
generators of the electroweak gauge group to the vacuum. We find
\begin{flalign}
T_{1}\langle H\rangle=\frac{\sigma_{1}}{2}\langle H\rangle=\frac{1}{2}\left(\begin{array}{cc}
0 & 1\\
1 & 0
\end{array}\right)\left(\begin{array}{c}
0\\
\frac{v_{\mathrm{SM}}}{\sqrt{2}}
\end{array}\right)=\frac{1}{2}\left(\begin{array}{c}
\frac{v_{\mathrm{SM}}}{\sqrt{2}}\\
0
\end{array}\right)\neq\left(\begin{array}{c}
0\\
0
\end{array}\right)\text{ Broken}\,,\\
T_{2}\langle H\rangle=\frac{\sigma_{2}}{2}\langle H\rangle=\frac{1}{2}\left(\begin{array}{cc}
0 & -i\\
i & 0
\end{array}\right)\left(\begin{array}{c}
0\\
\frac{v_{\mathrm{SM}}}{\sqrt{2}}
\end{array}\right)=-\frac{i}{2}\left(\begin{array}{c}
\frac{v_{\mathrm{SM}}}{\sqrt{2}}\\
0
\end{array}\right)\neq\left(\begin{array}{c}
0\\
0
\end{array}\right)\text{ Broken}\,,\\
T_{3}\langle H\rangle=\frac{\sigma_{3}}{2}\langle H\rangle=\frac{1}{2}\left(\begin{array}{cc}
1 & 0\\
0 & -1
\end{array}\right)\left(\begin{array}{c}
0\\
\frac{v_{\mathrm{SM}}}{\sqrt{2}}
\end{array}\right)=-\frac{1}{2}\left(\begin{array}{c}
0\\
\frac{v_{\mathrm{SM}}}{\sqrt{2}}
\end{array}\right)\neq\left(\begin{array}{c}
0\\
0
\end{array}\right)\text{ Broken}\,,\\
Y\langle H\rangle=Y_{H}\langle H\rangle=\frac{1}{2}\langle H\rangle=\frac{1}{2}\left(\begin{array}{c}
0\\
\frac{v_{\mathrm{SM}}}{\sqrt{2}}
\end{array}\right)\neq\left(\begin{array}{c}
0\\
0
\end{array}\right)\text{ Broken}\,.
\end{flalign}
In this manner, we confirm that the electric charge operator $Q$ is associated to the
combination of generators of $SU(2)_{L}\times U(1)_{Y}$ that leaves
the vacuum invariant,
\begin{equation}
Q\langle H\rangle=\left(T_{3}+Y\right)\langle H\rangle=\left(\frac{\sigma_{3}}{2}+Y_{H}\right)\langle H\rangle=\frac{1}{2}\left(\begin{array}{cc}
1 & 0\\
0 & 0
\end{array}\right)\left(\begin{array}{c}
0\\
\frac{v_{\mathrm{SM}}}{\sqrt{2}}
\end{array}\right)=\left(\begin{array}{c}
0\\
0
\end{array}\right)\text{ Unbroken}\,.
\end{equation}
Therefore, as anticipated before, the electric charge operator
generating $U(1)_{Q}$ is given by
\begin{equation}
Q=T_{3}+Y\,,\label{eq:electric_charge}
\end{equation}
which immediately allows to extract the electric charges of fundamental
fermions as per Table~\ref{tab:SM_fermion_Qcharge}.
\begin{table}
\begin{centering}
\begin{tabular}{lcccc}
\toprule 
Field & $u_{i}$ & $d_{i}$ & $\nu_{i}$ & $e_{i}$\tabularnewline
\midrule 
$U(1)_{Q}$ & $2/3$ & $-1/3$ & 0 & $-1$\tabularnewline
\bottomrule
\end{tabular}
\par\end{centering}
\caption[Electric charges of fundamental fermions]{Electric charges of fundamental fermions, related to weak isospin
and hypercharge via Eq.~(\ref{eq:electric_charge}). $i=1,2,3$ is
the flavour index.\label{tab:SM_fermion_Qcharge}}
\end{table}

We find that electric charge is indeed unbroken after SSB. This is
a consequence of choosing the neutral component of the Higgs doublet
as the one that develops the VEV.

In the following, we will show how gauge bosons associated to (combinations
of) the broken generators of $SU(2)_{L}\times U(1)_{Y}$ get a mass.
When expanding the covariant derivative applied to the vacuum of the
Higgs doublet, we obtain
\begin{align}
{\displaystyle {\displaystyle D_{\mu}\langle H\rangle}} & {\displaystyle {\displaystyle =\left(\partial_{\mu}-i\frac{g_{L}}{2}\left[\begin{array}{cc}
W_{\mu}^{3} & \sqrt{2}W_{\mu}\\
\sqrt{2}W_{\mu}^{\dagger} & -W_{\mu}^{3}
\end{array}\right]-i\frac{g_{Y}}{2}B_{\mu}\right)\left(\begin{array}{c}
0\\
v_{\mathrm{SM}}/\sqrt{2}
\end{array}\right)}}\\
 & {\displaystyle {\displaystyle =\frac{v_{\mathrm{SM}}}{\sqrt{2}}\left(\begin{array}{c}
-i\frac{g_{L}}{2}\sqrt{2}W_{\mu}\\
i\frac{g_{L}}{2}W_{\mu}^{3}-i\frac{g_{Y}}{2}B_{\mu}
\end{array}\right)\,,}}\nonumber 
\end{align}
where we have defined $W_{\mu}=(W_{\mu}^{1}-iW_{\mu}^{2})/\sqrt{2}$.
With this, we can expand the kinetic term of the Higgs doublet in
Eq.~(\ref{eq:L_Scalar}), obtaining
\begin{equation}
(D_{\mu}\langle H\rangle)^{\dagger}D^{\mu}\langle H\rangle=\frac{g_{L}^{2}v_{\mathrm{SM}}^{2}}{4}W_{\mu}W^{\mu}+\frac{v_{\mathrm{SM}}^{2}}{2}\frac{1}{4}\left(\begin{array}{cc}
B_{\mu} & W_{\mu}^{3}\end{array}\right)\left(\begin{array}{lc@{}}
g_{Y}^{2} & -g_{Y}g_{L}\\
-g_{Y}g_{L} & g_{L}^{2}
\end{array}\right)\left(\begin{array}{c}
B^{\mu}\\
W^{3\mu}
\end{array}\right)\,.\label{eq:Mass_Terms_SMbosons}
\end{equation}
It directly follows that the gauge bosons $W_{\mu}$ and $W_{\mu}^{\dagger}$,
which we associate with the charged bosons $W_{\mu}^{+}$ and $W_{\mu}^{-}$,
get a mass term after SSB. The gauge bosons $B_{\mu}$ and $W_{\mu}^{3}$
also get mass terms, indicating a mass mixing. In order to obtain
the mass eigenstates we need to diagonalise the off-diagonal matrix
above, obtaining
\begin{equation}
(D_{\mu}\langle H\rangle)^{\dagger}D^{\mu}\langle H\rangle=\frac{g_{L}^{2}v_{\mathrm{SM}}^{2}}{4}W_{\mu}^{\dagger}W^{\mu}+\frac{v_{\mathrm{SM}}^{2}}{2}\frac{1}{4}\left(\begin{array}{cc}
A_{\mu} & Z_{\mu}\end{array}\right)\left(\begin{array}{lc@{}}
0 & 0\\
0 & (g_{L}^{2}+g_{Y}^{2})
\end{array}\right)\left(\begin{array}{c}
A^{\mu}\\
Z^{\mu}
\end{array}\right)\,,\label{eq:Higgs_Kinetic_physical}
\end{equation}
where
\begin{equation}
\left(\begin{array}{c}
A_{\mu}\\
Z_{\mu}
\end{array}\right)=\left(\begin{array}{cc}
\cos\theta_{W} & \sin\theta_{W}\\
-\sin\theta_{W} & \cos\theta_{W}
\end{array}\right)\left(\begin{array}{c}
B_{\mu}\\
W_{\mu}^{3}
\end{array}\right)=\left(\begin{array}{c}
\cos\theta_{W}B_{\mu}+\sin\theta_{W}W_{\mu}^{3}\\
-\sin\theta_{W}B_{\mu}+\cos\theta_{W}W_{\mu}^{3}
\end{array}\right)\,,
\end{equation}
where the mixing angle is commonly denoted as the ``weak mixing angle''
or the ``Weinberg\footnote{Although it appeared for the first time in the 1961 paper
by Glashow \cite{Glashow:1961tr}.} angle'', and is given by
\begin{equation}
\sin\theta_{W}=\frac{g_{Y}}{\sqrt{g_{Y}^{2}+g_{L}^{2}}}\,.
\end{equation}
In conclusion, expanding the kinetic term of the Higgs doublet reveals
that the $W_{\mu}^{\pm}$ bosons and the $Z_{\mu}$ boson, given as
linear combinations of the original gauge bosons of the massless theory, become
massive gauge bosons after SSB. Their mass can be readily extracted
from Eq.~(\ref{eq:Higgs_Kinetic_physical}) as
\begin{equation}
M_{W}=\frac{1}{2}g_{L}v_{\mathrm{SM}}\,,\qquad M_{Z}=\frac{v_{\mathrm{SM}}}{2}\sqrt{g_{L}^{2}+g_{Y}^{2}}=\frac{1}{2\cos\theta_{W}}g_{L}v_{\mathrm{SM}}\,.
\end{equation}
Instead, the $A_{\mu}$ boson (also known as the photon) remains massless, associated to the
unbroken generator of $U(1)_{Q}$ with gauge coupling
\begin{equation}
e=\frac{g_{L}g_{Y}}{\sqrt{g_{L}^{2}+g_{Y}^{2}}}\,,\label{eq:QED_coupling}
\end{equation}
which is commonly expressed as $\alpha_{\mathrm{EM}}=e^{2}/(4\pi)$  and denoted as the \textit{fine-structure constant}.
We can see that the masses of
the $W_{\mu}^{\pm}$ and the $Z_{\mu}$ bosons are not independent but instead
they follow the tree-level relation
\begin{equation}
\rho\equiv\frac{M_{W}^{2}}{M_{Z}^{2}\cos^{2}\theta_{W}}=1\,.\label{eq:rho_parameter}
\end{equation}
The tree-level relation in Eq.~(\ref{eq:rho_parameter}) is a consequence
of the so-called \textit{custodial symmetry}. The Higgs potential
$V(H)$ is invariant not only under the SM symmetry, but it is also
accidentally invariant under a global $SO(4)\cong SU(2)_{L}\times SU(2)_{R}$
transformation. This can be easily seen by parameterising the Higgs
doublet in terms of a four-dimensional vector of real scalar fields,
$\vec{H}=(H_{1},H_{2},H_{3},H_{4})$, and noting that one has for
the terms in the potential
\begin{equation}
(H^{\dagger}H)=\vec{H}\cdot\vec{H}\,,
\end{equation}
which is invariant under four-dimensional rotations. After SSB, the
accidental symmetry $SU(2)_{L}\times SU(2)_{R}$ gets spontaneously
broken down to the diagonal subgroup. This symmetry is the reason
for the tree-level relation among the gauge boson masses given in
Eq.~(\ref{eq:rho_parameter}). However, the custodial symmetry is
not a true accidental symmetry of the full Lagrangian, and gets explicitly
broken by the gauging of the $U(1)_{Y}$ symmetry (and also by the
different Yukawa couplings of up-quarks and down-quarks, to be introduced
in Sections~\ref{sec:Yukawa-sector} and \ref{subsec:Parameters-of-the-SM}).
Thanks to the smallness of the hypercharge gauge coupling and the smallness of
most Yukawa couplings, these breaking effects are small and custodial symmetry remains 
as a good approximate symmetry. A consequence of this is the fact that the
$W_{\mu}^{\pm}$ and $Z_{\mu}$ bosons have similar masses.
Another consequence is that one obtains just small departures from $\rho=1$ at higher
orders in perturbation theory. Since these corrections are small,
extensions of the SM that explicitly break the custodial symmetry,
altering the tree-level relation in Eq.~(\ref{eq:rho_parameter}),
typically receive very strong bounds from the experimental measurements
of the $\rho$ parameter, which has been tested to
be in good agreement with the SM prediction \cite{PDG:2022ynf}.

A careful reader might have noticed that after SSB, our physical system
seems to have more degrees of freedom than before SSB. Indeed, massless
gauge bosons carry two degrees of freedom, while massive gauge bosons
carry three. What is the origin of these extra degrees of freedom,
associated to the now massive $W_{\mu}^{\pm}$ and $Z_{\mu}$ bosons?
The Goldstone theorem \cite{Goldstone:1961eq} states that the spontaneous
breaking of a symmetry leads to the appearance of a set of massless
bosons, one for each of the broken generators. These are the so-called
Goldstone bosons. Interestingly, when the broken symmetry is a gauge
symmetry, the Goldstone bosons ``get eaten'' by the gauge fields
and become their longitudinal components, providing
enough degrees of freedom for a massive gauge boson. This can be easily
seen if we parameterise the Higgs doublet in the following way
\begin{equation}
{\displaystyle {\displaystyle H=e^{i\frac{\varphi^{a}\sigma^{a}}{2v_{\mathrm{SM}}}}\left(\begin{array}{c}
0\\
\frac{1}{\sqrt{2}}(v_{\mathrm{SM}}+h)
\end{array}\right)\,.}}\label{eq:Unitary_gauge}
\end{equation}
where the Goldstone bosons, $\varphi^{a}$ ($a=1,2,3$), related to
excitations along the minima of the scalar potential, are isolated
in the exponential factor. It is clear that the Goldstone bosons are
no longer physical: we can use the local $SU(2)_{L}$ gauge invariance
to rotate away the Goldstone bosons, which effectively consists in
setting $\varphi^{a}=0$ in Eq.~(\ref{eq:Unitary_gauge}). This rotation
corresponds to the choice of a specific gauge that receives the name
of \textit{unitary gauge}. In this gauge, the Higgs doublet then reads
\begin{equation}
{\displaystyle H=\frac{1}{\sqrt{2}}\left(\begin{array}{c}
0\\
v_{\mathrm{SM}}+h
\end{array}\right)\,.}\label{eq:Unitary_gauge-1}
\end{equation}
In this manner, the unphysical Goldstone bosons had been rotated away,
their three physical degrees of freedom absorbed by the now massive
$W_{\mu}^{\pm}$ and $Z_{\mu}$ gauge bosons. Nevertheless, we notice
the existence of a physical scalar field associated to the radial
excitation of the Higgs doublet, $h$. If we expand the kinetic term
in Eq.~(\ref{eq:L_Scalar}) with the Higgs doublet in the
unitary gauge, then
we find that the field $h$ couples at tree-level to the massive gauge
bosons $W_{\mu}^{\pm}$ and $Z_{\mu}$, with couplings proportional
to their squared masses, but it does not couple to the massless photon.
Indeed, the neutral scalar $h$ is a key prediction of the model:
the \textit{Higgs boson}.

Finally, if we write the scalar potential of Eq.~(\ref{eq:Scalar_Potential})
in terms of the Higgs boson via Eq.~(\ref{eq:Unitary_gauge-1}),
we obtain
\begin{equation}
V(h)=\frac{1}{2}m_{h}^{2}h^{2}+\frac{m_{h}^{2}}{2v_{\mathrm{SM}}}h^{3}+\frac{m_{h}^{2}}{8v_{\mathrm{SM}}^{2}}h^{4}-\frac{m_{h}^{2}v_{\mathrm{SM}}^{2}}{8}\,,\label{eq:Higgs_boson_potential}
\end{equation}
which reveals that the Higgs boson acquires a mass proportional to
its VEV, $m_{h}^{2}=2\lambda v_{\mathrm{SM}}^{2}$, as well as cubic
and quartic self-interactions that are proportional to its mass. The
Higgs boson predicted by the SM was discovered at the LHC in 2012
\cite{ATLAS:2012yve,CMS:2012qbp}, and its mass is now a well known
quantity $m_{h}=125.25\pm0.17\,\mathrm{GeV}$ \cite{PDG:2022ynf}. In this manner, the
Higgs self-couplings are completely fixed in the SM at tree-level
in terms of its mass. However, testing this key prediction of the
SM is difficult: The HL-LHC will only probe order one departures from
the SM prediction \cite{Mlynarikova:2023bvx}, and pushing the energy frontier further will be
required in order to gain one order of magnitude in precision \cite{Durieux:2022hbu}. Finally, the
constant term in Eq.~(\ref{eq:Higgs_boson_potential}) contributes to
the energy of the vacuum, so it becomes a contribution to the cosmological
constant of the Universe. However, this
contribution turns out to be 56 orders of magnitude larger (and with
the opposite sign) than the observed value of the cosmological constant,
which is required to be introduced in the equations of General Relativity
in order to explain the accelerated expansion of our Universe. This
implies that there should be an incredibly fine-tuned cancellation
between $-m_{h}^{2}v_{\mathrm{SM}}^{2}/8$ and a bare vacuum energy
parameter that can be introduced in the Lagrangian, leading to the
worst fine-tuning problem of the SM \cite{Weinberg:1988cp,Adler:1995vd,Sola:2013gha}.

\section{Yukawa sector \label{sec:Yukawa-sector}}

We have now built the Standard Model as a consistent QFT based on
gauge symmetry, where the electroweak part of the SM group is spontaneously
broken via the Higgs mechanism down to QED, and the $W_{\mu}^{\pm}$
and $Z_{\mu}$ bosons become massive and mediate the weak interactions.
In this section, we shall explain how most of the fundamental fermions
of the SM also become massive via the Higgs mechanism (except for
neutrinos that will remain massless). The particular representations
of fundamental fermions and the Higgs doublet under the SM gauge group
allow to introduce the following terms in the SM Lagrangian,
\begin{equation}
\mathcal{L}_{\mathrm{Yukawa}}=y_{ij}^{u}\overline{Q}_{Li}\tilde{H}u_{Rj}+y_{ij}^{d}\overline{Q}_{Li}Hd_{Rj}+y_{ij}^{e}\overline{L}_{Li}He_{Rj}+\mathrm{h.c.}\,,
\end{equation}
where $\tilde{H}=i\sigma^{2}H^{*}$ is the conjugate of $H$ with well-defined transformations
(doublet of $SU(2)_{L}$ with $Y=-1/2$), $i,j=1,2,3$ are flavour indices and $y_{ij}^{u,d,e}$
are completely generic $3\times3$ complex matrices. The interaction
terms above, and usually also the matrices $y_{ij}^{u,d,e}$ containing
the coupling constants, are denoted as the \textit{Yukawa couplings}
of the SM. After the Higgs doublet develops a VEV, the Yukawa couplings
above provide mass terms for charged fermions along with couplings
to the Higgs boson. Working in the unitary gauge as introduced in Eq.~(\ref{eq:Unitary_gauge-1}),
we obtain
\begin{equation}
\mathcal{L}_{\mathrm{Yukawa}}=\frac{1}{\sqrt{2}}(v_{\mathrm{SM}}+h)\left[y_{ij}^{u}\bar{u}_{Li}u_{Rj}+y_{ij}^{d}\bar{d}_{Li}d_{Rj}+y_{ij}^{e}\bar{e}_{Li}e_{Rj}\right]+\mathrm{h.c.}\label{eq:Yukawa_matrices_SM}
\end{equation}
Remarkably, neutrinos do not obtain mass terms nor couplings to the
Higgs boson, so they remain massless: this is a consequence of the
absent of right-handed neutrinos in the SM. On the other hand, it
is definitely remarkable that the same mechanism that gives masses to
the gauge bosons (SSB), also gives masses to the charged fermions.
Now, in general the three Yukawa matrices are not diagonal, but rather
completely generic $3\times3$ complex matrices. In order to obtain
mass eigenstates and eigenvalues, the Yukawa matrices in Eq.~(\ref{eq:Yukawa_matrices_SM})
must be brought to diagonal form. Since all mass terms in $\mathcal{L}_{\mathrm{Yukawa}}$
are of Dirac type, this must be done by means of biunitary transformations:
Given a general complex matrix $\mathcal{M}$, there exist two unitary
matrices $V$ and $V'$ (verifying $VV^{\dagger}=V^{\dagger}V=\mathbb{I}$
and $V'V'^{\dagger}=V'^{\dagger}V'=\mathbb{I}$) such that
\begin{equation}
V^{\dagger}\mathcal{M}V'=\tilde{\mathcal{M}}\,,
\end{equation}
where $\tilde{\mathcal{M}}$ is diagonal. This general result can
be applied to the Yukawa matrices of Eq.~(\ref{eq:Yukawa_matrices_SM}).
From a physical point of view, independent transformations rotate
the left-handed and right-handed spinors from the interaction basis to the
mass basis, obtaining
\begin{flalign}
\mathcal{L}_{\mathrm{Yukawa}} & =\frac{1}{\sqrt{2}}(v_{\mathrm{SM}}+h)\left[\bar{u}_{L}V_{u_{L}}V^{\dagger}_{u_{L}}y^{u}V_{u_{R}}V^{\dagger}_{u_{R}}u_{Rj}+\bar{d}_{L}V_{d_{L}}V^{\dagger}_{d_{L}}y^{d}V_{d_{R}}V^{\dagger}_{d_{R}}d_{Rj}\right.\label{eq:SM_Yukawas_diagonal}\\
 & \left.+\bar{e}_{L}V_{e_{L}}V^{\dagger}_{e_{L}}y^{e}V_{e_{R}}V^{\dagger}_{e_{R}}e_{Rj}\right]+\mathrm{h.c.}\,,\nonumber 
\end{flalign}
where rather than including a large number of flavour indices, we
have defined three-component fermion vectors containing the three
flavours as $f_{L}=(f_{L1},f_{L2},f_{L3})^{\mathrm{T}}$ and $f_{R}=(f_{R1},f_{R2},f_{R3})^{\mathrm{T}}$,
where $f=u,d,e$. Thanks to the unitary property of the rotation matrices,
one can check that $\mathcal{L}_{\mathrm{Yukawa}}$ remains invariant,
but written in the form of Eq.~(\ref{eq:SM_Yukawas_diagonal}) one
can identify the diagonal Yukawa matrices as
\begin{equation}
V^{\dagger}_{u_{L}}y^{u}V_{u_{R}}=\mathrm{diag}(y_{u},y_{c},y_{t})\,,
\end{equation}
\begin{equation}
V^{\dagger}_{d_{L}}y^{d}V_{d_{R}}=\mathrm{diag}(y_{d},y_{s},y_{b})\,,
\end{equation}
\begin{equation}
V^{\dagger}_{e_{L}}y^{e}V_{e_{R}}=\mathrm{diag}(y_{e},y_{\mu},y_{\tau})\,,
\end{equation}
along with the mass eigenstates
\begin{equation}
\left(\begin{array}{c}
u_{L}\\
c_{L}\\
t_{L}
\end{array}\right)=V^{\dagger}_{u_{L}}\left(\begin{array}{c}
u_{L1}\\
u_{L2}\\
u_{L3}
\end{array}\right)\,,\qquad\left(\begin{array}{c}
u_{R}\\
c_{R}\\
t_{R}
\end{array}\right)=V^{\dagger}_{u_{R}}\left(\begin{array}{c}
u_{R1}\\
u_{R2}\\
u_{R3}
\end{array}\right)\,,
\end{equation}
\begin{equation}
\left(\begin{array}{c}
d_{L}\\
s_{L}\\
b_{L}
\end{array}\right)=V^{\dagger}_{d_{L}}\left(\begin{array}{c}
d_{L1}\\
d_{L2}\\
d_{L3}
\end{array}\right)\,,\qquad\left(\begin{array}{c}
d_{R}\\
s_{R}\\
b_{R}
\end{array}\right)=V^{\dagger}_{d_{R}}\left(\begin{array}{c}
d_{R1}\\
d_{R2}\\
d_{R3}
\end{array}\right)\,,
\end{equation}
\begin{equation}
\left(\begin{array}{c}
e_{L}\\
\mu_{L}\\
\tau_{L}
\end{array}\right)=V^{\dagger}_{e_{L}}\left(\begin{array}{c}
e_{L1}\\
e_{L2}\\
e_{L3}
\end{array}\right)\,,\qquad\left(\begin{array}{c}
e_{R}\\
\mu_{R}\\
\tau_{R}
\end{array}\right)=V^{\dagger}_{e_{R}}\left(\begin{array}{c}
e_{R1}\\
e_{R2}\\
e_{R3}
\end{array}\right)\,.
\end{equation}
One can notice that the different flavours of mass eigenstates enjoy
a dedicated notation, as they receive particular names by convention.
The first family of charged fermions is formed by the up-quark, the
down-quark and the electron. The second family is formed by the charm-quark,
the strange-quark and the muon. Finally, the third family is formed
by the top-quark, the bottom-quark and the tau. The three flavours
of neutrinos, massless in the SM, remain aligned with the original interaction eigenstates and are given names associated
to the flavours of charged leptons: electron neutrino, muon neutrino
and tau neutrino. 

After diagonalising the Yukawa matrices, the mass eigenstates of charged
fermions obtain a well-defined mass given by
\begin{equation}
m_{\alpha}=y_{\alpha}\frac{v_{\mathrm{SM}}}{\sqrt{2}}\,.
\end{equation}

The masses of charged leptons can be measured in the experiment, while
the masses of quarks can be estimated by QCD methods from the masses
of hadrons. The Higgs VEV $v_{\mathrm{SM}}$ can also be extracted from
the experiment such that the physical Yukawa couplings $y_{\alpha}$
can be univocally determined. The interactions of the Higgs boson
with charged fermions are proportional to the physical Yukawa couplings
$y_{\alpha}$, which are generally different for each fermion. Given
that the gauge sector is flavour universal, the Yukawa couplings are
the only source of \textit{flavour physics} in the SM. This is translated
to the fact that the Yukawa couplings break the accidental $U(3)^{5}$ flavour
symmetry of the SM introduced in Eq.~(\ref{eq:Flavour_Symmetry}).
More about this will be discussed in Section~\ref{sec:Flavour-symmetries}.

Particle physics experiments are sensitive to the physical mass eigenstates
of fermions, therefore it is important to write the SM gauge interactions
in the mass basis. For the particular case of the weak interactions
mediated by the $W_{\mu}^{\pm}$ bosons, we will see that fermion
mixing originated from the Yukawa couplings plays a fundamental role.

\section{Weak interactions and CKM mixing: flavour violation}

Having completed the SM with the scalar and Yukawa sectors, now we
are ready to write the weak interactions for fundamental fermions
as the final step of our description of the SM. We are interested
on the terms of the covariant derivative related to the physical bosons
$W_{\mu}^{\pm}$, $Z_{\mu}$ and $A_{\mu}$,
\begin{equation}
\mathcal{L}_{\mathrm{gauge}}\supset\left(\frac{g_{L}}{\sqrt{2}}J_{\mu}^{+}W^{\mu+}+\mathrm{h.c.}\right)+\left(g_{L}\cos\theta_{W}J_{\mu}^{3}-g_{Y}\sin\theta_{W}J_{\mu}^{Y}\right)Z^{\mu}+eJ_{\mu}^{\mathrm{em}}A^{\mu}\,,
\end{equation}
where
\begin{flalign}
 & J_{\mu}^{+}=\bar{u}_{L}\gamma_{\mu}d_{L}+\bar{\nu}_{L}\gamma_{\mu}e_{L}\,,\\
 & J_{\mu}^{3}=\frac{1}{2}\bar{u}_{L}\gamma_{\mu}u_{L}-\frac{1}{2}\bar{d}_{L}\gamma_{\mu}d_{L}+\frac{1}{2}\bar{\nu}_{L}\gamma_{\mu}\nu_{L}-\frac{1}{2}\bar{e}_{L}\gamma_{\mu}e_{L}\,,\\
 & J_{\mu}^{Y}=\frac{1}{6}\bar{u}_{L}\gamma_{\mu}u_{L}+\frac{1}{6}\bar{d}_{L}\gamma_{\mu}d_{L}-\frac{1}{2}\bar{\nu}_{L}\gamma_{\mu}\nu_{L}-\frac{1}{2}\bar{e}_{L}\gamma_{\mu}e_{L}\\
 & \quad\quad+\frac{2}{3}\bar{u}_{R}\gamma_{\mu}u_{R}-\frac{1}{3}\bar{d}_{R}\gamma_{\mu}d_{R}-\bar{e}_{R}\gamma_{\mu}e_{R}\,,\nonumber \\
 & J_{\mu}^{\mathrm{em}}=\frac{2}{3}\left(\bar{u}_{L}\gamma_{\mu}u_{L}+\bar{u}_{R}\gamma_{\mu}u_{R}\right)-\frac{1}{3}\left(\bar{d}_{L}\gamma_{\mu}d_{L}+\bar{d}_{R}\gamma_{\mu}d_{R}\right)-\left(\bar{e}_{L}\gamma_{\mu}e_{L}+\bar{e}_{R}\gamma_{\mu}e_{R}\right)\,.
\end{flalign}
and $e=g_{L}\sin\theta_{W}=g_{Y}\cos\theta_{W}$ is the gauge coupling
of QED, univocally determined by $g_{L}$ and $g_{Y}$ via Eq.~(\ref{eq:QED_coupling}). The equations
above are written in terms of interaction eigenstates, defined as
three-component vectors containing the three flavours as $f_{L}=(f_{L1},f_{L2},f_{L3})^{\mathrm{T}}$
and $f_{R}=(f_{R1},f_{R2},f_{R3})^{\mathrm{T}}$, where $f=u,d,\nu,e$.
We can notice already that QED interactions and the neutral currents mediated
by $Z_{\mu}$ are flavour universal: the rotations $V_{f_{L}}$ and
$V_{f_{R}}$ cancel by unitarity for every term in the Lagrangian. 
The fundamental reason for $Z_{\mu}$ interactions being flavour universal is that
all mass eigenstates of a given electric charge originate from the
same $SU(2)_{L}\times U(1)_{Y}$ representations, otherwise $Z_{\mu}$
would mediate flavour transitions at tree-level. For QED (and QCD),
the fundamental reason is the flavour universality of the kinetic
terms in the canonical basis, which imposes the universality of the
interactions related to the unbroken symmetries. Remarkably, the SM was not
proposed as a minimal theory to explain the observed phenomena,
but rather as a non-minimal model that predicted the neutral currents mediated
by the $Z_{\mu}$ boson as a smoking-gun signal. Weak neutral currents were
discovered in 1973 at the Gargamelle experiment of CERN \cite{Haidt:2015bgg}.

We now focus on the charged currents mediated by $W_{\mu}^{\pm}$,
and we rotate fermion states from the interaction basis to the mass
basis
\begin{flalign}
\mathcal{L}_{\mathrm{cc}} & =\frac{g_{L}}{\sqrt{2}}\left(\bar{u}_{L}\gamma_{\mu}d_{L}+\bar{\nu}_{L}\gamma_{\mu}e_{L}\right)W^{\mu+}+\mathrm{h.c.}\label{eq:Charged_Currents}\\
 & =\frac{g_{L}}{\sqrt{2}}\left(\hat{\bar{u}}_{L}V_{u_{L}}V^{\dagger}_{d_{L}}\gamma_{\mu}\hat{d}_{L}+\bar{\nu}_{L}V^{\dagger}_{e_{L}}\gamma_{\mu}\hat{e}_{L}\right)W^{\mu+}+\mathrm{h.c.}\,,\nonumber 
\end{flalign}
where $\hat{f}$ denotes mass eigenstates. We notice that the charged currents are sensitive to the product of
left-handed quark mixing matrices, that is denoted as the \textit{Cabbibo-Kobayashi-Maskawa
(CKM) matrix} \cite{Cabibbo:1963yz,Kobayashi:1973fv},
\begin{equation}
V_{\mathrm{CKM}}=V^{\dagger}_{u_{L}}V_{d_{L}}\,.\label{eq:CKM_matrix}
\end{equation}
On the other hand, since neutrinos are massless in the SM, the rotation
matrix $V_{e_{L}}$ is unphysical in the SM, as it can be absorbed
by redefining the neutrino states.

In contrast, the neutral currents mediated by $Z_{\mu}$, along with the
QED and QCD interactions, all couple fermions with their conjugate
counterparts, such that the rotation matrices cancel by unitarity. Therefore,
the right-handed rotation matrices $V_{u_{R},d_{R},e_{R}}$, along with
$V_{e_{L}}$, are all unphysical in the SM: no interaction is sensitive
to them. As a consequence, these interactions are flavour universal
and do not mediate flavour transitions at tree-level. In the lepton
sector, this is denoted as lepton flavour universality (LFU) in the
SM: gauge interactions do not discriminate between the different lepton
flavours, up to small corrections given by the different masses of charged leptons.

The rotation matrices $V_{u_{L}}$ and $V_{d_{L}}$ are not physical
themselves, but rather their product $V_{\mathrm{CKM}}$ as per Eq.~(\ref{eq:CKM_matrix})
is physical. This way, the charged currents provide tree-level flavour
transitions proportional to the off-diagonal elements of the CKM matrix,
leading to flavour physics. Remarkably, in the absence of Yukawa couplings,
fermion mixing would vanish and the charged currents would not mediate
family transitions. As anticipated, Yukawa couplings in the SM are
the only source of flavour physics. Moreover, we shall see that the
CKM matrix contains one physical complex phase, meaning that weak interactions
break the so-called $CP$ symmetry that relates particles and antiparticles.

As a final remark, when one goes beyond the SM, it is possible that
the new interactions are sensitive to each of the individual rotation matrices that connect
the interaction and mass basis, such that they become physical. This
can happen even in an effective field theory framework such as the
SMEFT \cite{Grzadkowski:2010es}.

\section{The flavour puzzle}

\subsection{Parameters of the SM: mass hierarchies and quark mixing\label{subsec:Parameters-of-the-SM}}

The Higgs VEV $v_{\mathrm{SM}}$ can be extracted from the Fermi constant
$G_{F}=1.1663788(6)\times10^{-5}\;\mathrm{GeV}^{-2}$ \cite{PDG:2022ynf},
which has been accurately measured from the process of muon decay
into an electron and two neutrinos. The resulting value is
\begin{equation}
v_{\mathrm{SM}}=\left(\sqrt{2}G_{F}\right)^{-1/2}=246.21964(6)\;\mathrm{GeV\,,}
\end{equation}
\begin{figure}
\begin{centering}
\includegraphics[scale=0.22]{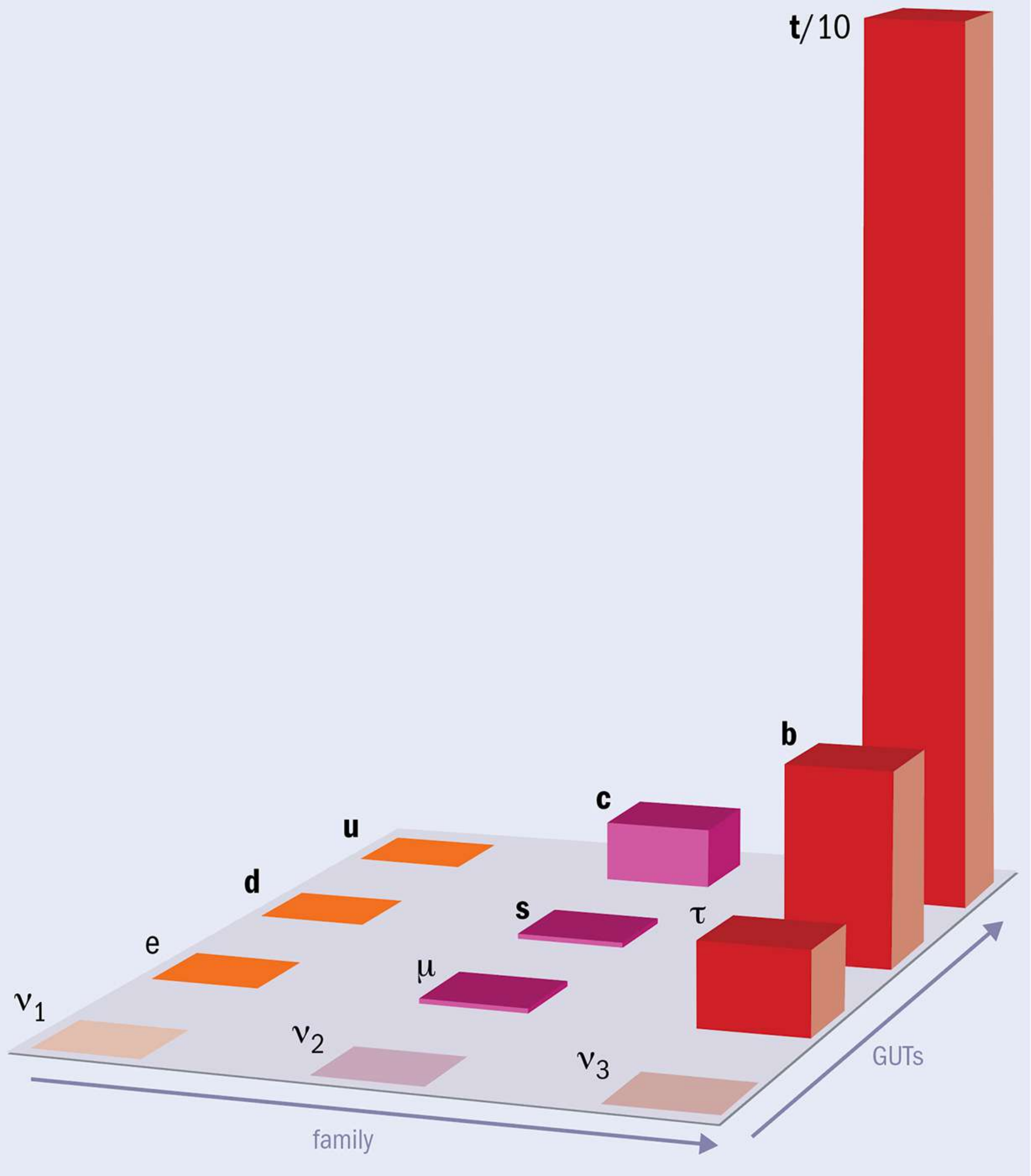}
\par\end{centering}
\caption[Fermion mass hierarchies]{Fermion mass hierarchies between the different families and charge
sectors. \label{fig:Fermion-mass-hierarchies}}
\end{figure}
which sets the energy scale for the spontaneous breaking of the electroweak symmetry. The Yukawa interactions
of the Higgs doublet with charged fermions introduce most of the free
parameters of the SM: 3 charged lepton masses, 6 quark masses, 3 mixing
angles for the CKM matrix and the $CP$-violating phase. The masses of charged leptons
can be directly measured in the experiment, while the masses of quarks can
be estimated by QCD methods from the masses of hadrons (which can
be measured in the experiment as well). The determination of these parameters provides a hierarchical spectrum of fermion masses. At low energies, the masses of charged fermions
can be approximately described in terms of the Wolfenstein parameter
$\lambda\simeq0.225$ as \cite{PDG:2022ynf}
\allowdisplaybreaks[0]
\begin{alignat}{2}
 & m_{t}\sim\frac{v_{\mathrm{SM}}}{\sqrt{2}}\,,\qquad & m_{c}\sim\lambda^{3.3}\frac{v_{\mathrm{SM}}}{\sqrt{2}}\,,\qquad & m_{u}\sim\lambda^{7.5}\frac{v_{\mathrm{SM}}}{\sqrt{2}}\,,\label{eq:up_masses}\\
 & m_{b}\sim\lambda^{2.5}\frac{v_{\mathrm{SM}}}{\sqrt{2}}\,,\qquad & m_{s}\sim\lambda^{5.0}\frac{v_{\mathrm{SM}}}{\sqrt{2}}\,,\qquad & m_{d}\sim\lambda^{7.0}\frac{v_{\mathrm{SM}}}{\sqrt{2}}\,,\\
 & m_{\tau}\sim\lambda^{3.0}\frac{v_{\mathrm{SM}}}{\sqrt{2}}\,,\qquad & m_{\mu}\sim\lambda^{4.9}\frac{v_{\mathrm{SM}}}{\sqrt{2}}\,,\qquad & m_{e}\sim\lambda^{8.4}\frac{v_{\mathrm{SM}}}{\sqrt{2}}\,.\label{eq:lepton_masses}
\end{alignat}
\allowdisplaybreaks
Although the specific values of fermion masses depend on the energy
scale, as they experience RGE running, the mass ratios of fermions
from different families but in the same charged sector are constant.
For example, $m_{c}/m_{t}\sim\lambda^{3.3}$ is RGE-independent,
and similar ratios can be constructed for each charged sector. It
is clear this way that the third family is the heaviest, while the
first and second families are hierarchically lighter, with the first
family being lighter than the second, as highlighted in Fig.~\ref{fig:Fermion-mass-hierarchies}.
These hierarchies are difficult to understand due to the fact that the three fermion families are identical objects under the gauge symmetry.
There also exist hierarchies between charged sectors, i.e.~among fermions that transform differently under the gauge group: 
\begin{itemize}
\item In the third family, top is heavier than bottom and tau, while bottom
and tau have a mass of a similar order of magnitude.
\item In the second family, charm is heavier than strange and muon, while
strange and muon have a mass of a similar order of magnitude.
\item In contrast, in the first family the down-quark is the heaviest, while
the up-quark is heavier than the electron.
\end{itemize}
The hierarchies between charged sectors are, however, affected by
RGE running.
We introduce the specific numerical values of low-energy fermion masses
in Table~\ref{tab:Fermion_masses}.
\begin{table}
\begin{centering}
\begin{tabular}{cccccc}
\toprule 
\multicolumn{2}{c}{up-quarks} & \multicolumn{2}{c}{down-quarks} & \multicolumn{2}{c}{charged leptons}\tabularnewline
\midrule 
$m_{t}$ & $172.69\pm0.30\:\mathrm{GeV}$ & $m_{b}$ & $4.18_{-0.02}^{+0.03}\:\mathrm{GeV}$ & $m_{\tau}$ & $1.77686\pm0.12\:\mathrm{GeV}$\tabularnewline
\midrule 
$m_{c}$ & $1.27\pm0.02\:\mathrm{GeV}$ & $m_{s}$ & $93.4_{-3.4}^{+8.6}\:\mathrm{MeV}$ & $m_{\mu}$ & $105.6583755(23)\:\mathrm{MeV}$\tabularnewline
\midrule 
$m_{u}$ & $2.16_{-0.26}^{+0.49}\:\mathrm{MeV}$ & $m_{d}$ & $4.67_{-0.17}^{+0.48}\:\mathrm{MeV}$ & $m_{e}$ & $0.51099895000(15)\:\mathrm{MeV}$\tabularnewline
\bottomrule
\end{tabular}
\par\end{centering}
\caption[Numerical values for charged fermion masses]{Numerical values for charged fermion masses in the SM \cite{PDG:2022ynf}.
The $u$-, $d$-, and $s$-quark masses are the $\overline{\mathrm{MS}}$
masses at the scale $\mu=2\:\mathrm{GeV}$. The $c$- and $b$-quark
masses are the $\overline{\mathrm{MS}}$ masses renormalised at their
$\overline{\mathrm{MS}}$ mass scale, i.e. $\mu=m_{c}\,,m_{b}$ respectively.
The $t$-quark mass is extracted from event kinematics. Charged lepton
masses are measured in low-energy experiments. \label{tab:Fermion_masses}}

\end{table}

Quark mixing is accounted by the CKM matrix, which is a $3\times3$
complex and unitary matrix. In all generality, such a matrix would
contain 3 moduli and 6 phases, however 5 of them can be reabsorbed
by quark field redefinitions,
\begin{equation}
u_{L}^{i}\rightarrow e^{i\alpha_{i}}u_{L}^{i}\,,\qquad d_{L}^{j}\rightarrow e^{i\beta_{j}}d_{L}^{j}\,,\qquad V_{\mathrm{CKM}}^{ij}\rightarrow V_{\mathrm{CKM}}^{ij}e^{i(\alpha_{i}-\beta_{j})}\,,
\end{equation}
such that only one phase remains physical. In this manner, the CKM
matrix can be parameterised by three moduli associated to mixing angles,
and one $CP$-violating phase. Of the many possible conventions, a
standard choice has become \cite{Chau:1984fp}
\begin{flalign}
V_{\mathrm{CKM}} & =\left(\begin{array}{ccc}
1 & 0 & 0\\
0 & c_{23} & s_{23}\\
0 & -s_{23} & c_{23}
\end{array}\right)\left(\begin{array}{ccc}
c_{13} & 0 & s_{13}e^{-i\delta_{\mathrm{CKM}}}\\
0 & 1 & 0\\
-s_{13}e^{i\delta_{\mathrm{CKM}}} & 0 & c_{13}
\end{array}\right)\left(\begin{array}{ccc}
c_{12} & s_{12} & 0\\
-s_{12} & c_{12} & 0\\
0 & 0 & 1
\end{array}\right)\label{eq:CKM_sines}\\
 & =\left(\begin{array}{ccc}
c_{12}c_{13} & s_{12}c_{13} & s_{13}e^{-i\delta_{\mathrm{CKM}}}\\
-s_{12}c_{13}-c_{12}s_{23}s_{13}e^{i\delta_{\mathrm{CKM}}} & c_{12}c_{23}-s_{12}s_{23}s_{13}e^{i\delta_{\mathrm{CKM}}} & s_{23}c_{13}\\
s_{12}s_{23}-c_{12}c_{23}s_{13}e^{i\delta_{\mathrm{CKM}}} & -c_{12}s_{23}-s_{12}c_{23}s_{13}e^{i\delta_{\mathrm{CKM}}} & c_{23}c_{13}
\end{array}\right)\,,\nonumber 
\end{flalign}
where $s_{ij}\equiv\sin\theta_{ij}$, $c_{ij}\equiv\cos\theta_{ij}$
and $\delta_{\mathrm{CKM}}$ is the phase responsible for all $CP$-violating
phenomena in flavour-changing processes in the SM. The violation of
the $CP$ symmetry in the charged currents leads to different interactions
for particles and antiparticles. $CP$ violation was experimentally
discovered in neutral kaon decays in 1964 \cite{Christenson:1964fg},
and observed in recent years in $B$ meson decays \cite{BaBar:2001pki}
and in the charm sector \cite{LHCb:2019hro}.

The angles $\theta_{ij}$ can be chosen to lie in the first quadrant,
so that $s_{ij},c_{ij}\geq0$. A global fit to the mixing angles and
$CP$-violating phase in Eq.~(\ref{eq:CKM_sines}) reveals \cite{PDG:2022ynf}
\begin{align}
 & \sin\theta_{12}=0.22500\pm0.00067\,, &  & \sin\theta_{13}=0.00369\pm0.00011\,,\\
 & \sin\theta_{23}=0.04182_{-0.00074}^{+0.00085}\,, &  & \delta_{\mathrm{CKM}}=1.144\pm0.027\,.
\end{align}
 It follows that the CKM matrix is hierarchical, described by $s_{13}\ll s_{23}\ll s_{12}\ll1$
where the largest angle $\theta_{12}$ is traditionally known as the
\textit{Cabibbo angle} \cite{Cabibbo:1963yz}, and the $CP$-violating
phase is of $\mathcal{O}(1)$. To exhibit more clearly the hierarchy
of the CKM matrix, it is convenient to adopt the Wolfenstein parameterisation,
where we define \cite{PDG:2022ynf}
\begin{equation}
s_{12}\equiv\lambda\,,\qquad s_{23}\equiv A\lambda^{2}\,,\qquad s_{13}e^{i\delta_{\mathrm{CKM}}}\equiv A\lambda^{3}(\rho+i\eta)\,,
\end{equation}
where the Wolfenstein parameter $\lambda=s_{12}\simeq0.225$ is associated
to the Cabibbo angle, and is also used to parameterise charged fermion
masses as per Eqs.~(\ref{eq:up_masses}-\ref{eq:lepton_masses}).
We now write the CKM matrix in the Wolfenstein parameterisation up
to $\mathcal{O}(\lambda^{4})$, obtaining
\begin{equation}
V_{\mathrm{CKM}}=\left(\begin{array}{ccc}
1-\lambda^{2}/2 & \lambda & A\lambda^{3}(\rho-i\eta)\\
-\lambda & 1-\lambda^{2}/2 & A\lambda^{2}\\
A\lambda^{3}(1-\rho-i\eta) & -A\lambda^{2} & 1
\end{array}\right)+\mathcal{O}(\lambda^{4})\,.
\end{equation}
A global fit of recent data to the Wolfenstein parameters (assuming
the unitarity constraints of a three generation CKM matrix) reveals
\cite{PDG:2022ynf}
\begin{align}
 & \lambda=0.22500\pm0.00067\,, &  & A=0.826_{-0.015}^{+0.018}\,,\\
 & \rho=0.159\pm0.010\,, &  & \eta=0.348\pm0.010\,.\nonumber 
\end{align}
Finally, it is also common to denote the different elements of the
CKM matrix by the quark transitions associated to them in the charged
currents,
\begin{equation}
V_{\mathrm{CKM}}=\left(\begin{array}{ccc}
V_{ud} & V_{us} & V_{ub}\\
V_{cd} & V_{cs} & V_{cb}\\
V_{td} & V_{ts} & V_{tb}
\end{array}\right)\,.\label{eq:CKM_elements}
\end{equation}
We also introduce the Jarlskog invariant \cite{Jarlskog:1985ht,PDG:2022ynf}
\begin{flalign}
J & =\mathrm{Im}(V_{us}V_{cb}V_{ub}^{*}V_{cs}^{*})=c_{12}c_{13}^{2}c_{23}s_{12}s_{13}s_{23}\sin\delta_{\mathrm{CKM}}\\
 & =(3.08_{-0.13}^{+0.15})\times10^{-5}\,,\nonumber 
\end{flalign}
which is a quantitative measure of $CP$ violation in weak interactions
independent of conventions. In the parameterisation of Eq.~(\ref{eq:CKM_elements}),
we highlight key off-diagonal elements associated to the mixing angles,
and their scaling with $\lambda$ as
\begin{equation}
V_{us}\sim s_{12}\sim\lambda\,,\qquad V_{cb}\sim s_{23}\sim\lambda^{2}\,,\qquad V_{ub}\sim s_{13}\sim\lambda^{3}\,.
\end{equation}
The specific numerical values for the magnitude of each CKM element
$\left|V_{\alpha\beta}\right|$ are \cite{PDG:2022ynf}
\begin{equation}
\left|V_{\mathrm{CKM}}\right|=\left(\begin{array}{ccc}
0.97373(31) & 0.2243(8) & 0.00382(20)\\
0.221(4) & 0.975(6) & 0.0408(14)\\
0.0086(2) & 0.0415(9) & 1.014(29)
\end{array}\right)\,,
\end{equation}
where for $V_{cb}$ and $V_{ub}$ we show the results obtained from
exclusive meson decays\footnote{We note that there exists tension between the inclusive and exclusive
determinations of $V_{cb}$ and $V_{ub}$, and also a small deficit
in first row unitarity of the CKM matrix. This is briefly discussed
in Section~\ref{sec:Other-open-questions-SM}.}. We note here that the CKM elements are commonly used as input in
flavour physics calculations. However, if NP afflict the observables
used in the experimental determination of the CKM elements, then this
NP infection would be translated to flavour physics computations,
making the results unreliable. Alternative approaches to circumvent
this issue have been proposed in the literature \cite{Buras:2022qip}. 

Finally, the remaining parameters of the SM include the gauge couplings
$g_{s}$, $g_{L}$ and $g_{Y}$. $g_{s}$ can be extracted via QCD
methods as long as a confinement scale is provided \cite{DelDebbio:2021ryq}, which is usually
given via the pion decay constant $f_{\pi}=130.19(89)\;\mathrm{MeV}$ or the $\Omega^{-}$ baryon mass, $m_{\Omega^{-}}=1672.45(29)\;\mathrm{MeV}$, \cite{DelDebbio:2021ryq,PDG:2022ynf}. The couplings $g_{L}$ and $g_{Y}$
are commonly extracted from the mass of the $Z$ boson, $M_{Z}=91.1880(21)\;\mathrm{GeV}$,
and from the fine-structure constant, $\alpha^{-1}_{\mathrm{EM}}(M_{Z})=137.035999084(21)$ \cite{PDG:2022ynf},
obtaining (using tree-level expressions)
\begin{equation}
g_{s}(M_{Z})=1.217(5)\,,\qquad g_{L}(M_{Z})=0.64842(3)\,,\qquad g_{Y}(M_{Z})=0.35804(3)\,.\label{eq:gauge_couplings}
\end{equation} 
For a more precise calculation, see the values computed at NNLO
in Ref.~\cite{Buttazzo:2013uya}. The smallness of the gauge
couplings $g_{L}$ and $g_{Y}$ allows to perform reliable perturbative
calculations in quantum field theory for QED and weak interactions
at low energies. This is not the case for $g_{s}$, which is larger
and grows via RGE at low energies due to asymptotic freedom. It is
also common to express the gauge couplings as $\alpha_{i}=g_{i}^{2}/4\pi$,
such that \cite{PDG:2022ynf}
\begin{equation}
\alpha_{s}(M_{Z})=0.1179(9)\,,\qquad\alpha_{L}(M_{Z})=0.033458(3)\,,\qquad\alpha_{Y}(M_{Z})=0.0102012(17)\,. \label{eq:alpha_couplings}
\end{equation}

In principle, one may add a gauge invariant topological term for each of $SU(3)_{c}$, $SU(2)_{L}$ and $U(1)_{Y}$
to the Lagrangian of the gauge sector (\ref{eq:Gauge_Lagrangian_2}), leading to three extra parameters.
The so-called $\theta_{\mathrm{QCD}}$ and $\theta_{SU(2)_{L}}$ may be observable due to non-perturbative instanton
effects, while $\theta_{U(1)_{Y}}$ is unphysical \cite{FileviezPerez:2004hn,Shifman:2017lkj}. Given that $SU(2)_{L}$ acts chirally on SM fermions, it is possible to absorb $\theta_{SU(2)_{L}}$ via a chiral rotation associated to an anomalous (but classically conserved) $U(1)$ (say e.g.~$U(1)_{B+L}$), meaning it is unphysical as well\footnote{Although $\theta_{SU(2)_{L}}$ becomes
observable (but its effects are generally very suppressed) in specific BSM scenarios where both baryon and lepton number are broken \cite{FileviezPerez:2014xju}.
An example of this are gauge unified frameworks such as $SU(5)$ spontaneously broken to the SM, where all group factors get the same $\theta$ parameter
and $\theta_{SU(2)_{L}}$ becomes observable via $\Delta B=\Delta L=\pm1$ processes (further suppressed by the 
scale of grand unification) \cite{Shifman:2017lkj}.}. In contrast, since QCD
acts vector-like over SM quarks, $\theta_{\mathrm{QCD}}$ cannot be rotated away and is responsible for 
$CP$ violation in QCD interactions. However, from experimental constraints on the neutron EDM one derives\footnote{Formally speaking, only the linear combination $\overline{\theta}_{\mathrm{QCD}}=\theta_{\mathrm{QCD}}+\mathrm{Arg[}\mathrm{Det}(Y_{u}Y_{d})]$
is invariant under quark chiral rotations, and hence physically observable.} $|\overline{\theta}_{\mathrm{QCD}}|<10^{-10}$
\cite{Dragos:2019oxn}. We denote such apparent (and unexplained) absence of $CP$ violation in the
QCD sector as the \textit{strong $CP$ puzzle}, which will be briefly
introduced in Section~\ref{sec:Other-open-questions-SM}.

The last parameter of the SM is the mass of the Higgs boson $m_{h}=125.25\pm0.17\,\mathrm{GeV}$
\cite{PDG:2022ynf}. All things considered, the SM contains 19 parameters:
the three gauge couplings and $\overline{\theta}_{\mathrm{QCD}}$,
nine fermion masses and four quark mixing parameters, the Higgs VEV and
the mass of the Higgs boson. This is already a rather large set of parameters
for a fundamental theory. Going beyond the SM, the parameter counting
is enlarged when massive neutrinos are considered.

\subsection{The origin of neutrino masses and mixing \label{subsec:The-origin-of-neutrinos}}

In the decade of the 1990s, a series of neutrino experiments showed
conspicous discrepancies with the neutrino flux predictions of the
SM. In 1998, the Super-Kamiokande collaboration discovered evidence
that neutrinos do oscillate in flavour \cite{Super-Kamiokande:1998kpq,Kajita:1998bw,Kajita:2006cy},
explaining the discrepancies in the observed fluxes of atmospheric
neutrinos. In 2002, the SNO collaboration further confirmed the evidence
for neutrino oscillations, reporting that solar neutrinos do also
oscillate in flavour \cite{SNO:2002tuh}. The neutrino flavour oscillations
can only be realised in Nature if there exists neutrino mixing, which
can only happen if at least two neutrinos are massive. Neutrino oscillations
have become a confirmed discrepancy between the experiment and the
SM, which cannot account for the existence of massive neutrinos without
being extended.

Going beyond the SM, the most simple way to obtain a mass term for
neutrinos is to introduce right-handed neutrinos that transform as
complete singlets under the SM, $N_{Rj}\sim(\mathbf{1},\mathbf{1},0)$.
Given that for all the other fermions both chiralities are present
in Nature, it seems natural to think that right-handed neutrinos do
also exists. With this addition, one can write Yukawa couplings between
neutrinos and the Higgs doublet in a similar manner as for charged
fermions. As a toy example, let us consider that only one left-handed
neutrino flavour $\nu_{L}$ exists in Nature, transforming as a doublet
under $SU(2)_{L}$ along with his left-handed charged lepton partner
$L_{L}=(\nu_{L},e_{L})^{\mathrm{T}}$. Now let us introduce one right-handed
neutrino $N_{R}\sim(\mathbf{1},\mathbf{1},0)$. We are now
free to write the following gauge invariant coupling in our Lagrangian,
\begin{equation}
\mathcal{L}_{\nu}\supset y^{\nu}\bar{L}_{L}\widetilde{H}N_{R}+\mathrm{h.c.}
\end{equation}
However, the fact that $N_{R}$ is a complete singlet under the SM
allows to introduce a bare mass term in the Lagrangian that preserves
gauge invariance,
\begin{equation}
\mathcal{L}_{\nu}= y^{\nu}\bar{L}_{L}\widetilde{H}N_{R}+\frac{1}{2}M_{R}\bar{N}^{C}_{R}N_{R}+\mathrm{h.c.}\,,\label{eq:neutrino_lagrangian}
\end{equation}
where $N_{R}^{C}=C\bar{N}_{R}^{\mathrm{T}}$, with $C=i\gamma^{0}\gamma^{2}$
denoting the charge conjugation operator. This new term is called a
Majorana mass: it violates all $U(1)$ charges by two units, including lepton
number. In this manner, only SM singlet fermions can receive a Majorana
mass, and are denoted Majorana fermions. In principle, $M_{R}$ is
not subject to any symmetry and could take any value. One possible
choice is to take $M_{R}=0$. In this case, the neutrino becomes massive
via a Dirac mass term once the Higgs doublet gets a VEV, in the usual
way. However, in order to obtain a very tiny neutrino mass $m_{\nu}\sim\mathcal{O}(0.05\,\mathrm{eV})$,
as suggested by data, one would need a very small Yukawa coupling
as $y^{\nu}\sim10^{-12}$. This process can be generalised to three
neutrino generations, and having complete freedom over the Yukawa
couplings $y_{ij}^{\nu}$, one could fit all neutrino oscillations
data with Dirac neutrinos. However, this mechanism does not seem very
natural: the flavour sector of the SM becomes even more puzzling
with incredibly tiny Yukawa couplings and Majorana masses set to zero by
hand (or by imposing the conservation of lepton number). A more appealing solution is to consider that $M_{R}$ is non-zero
and then extract the physical masses of neutrinos. One can write all the
couplings in Eq.~(\ref{eq:neutrino_lagrangian}) in matrix form as
\begin{equation}
\mathcal{L}_{\nu}=\frac{1}{2}\left(\begin{array}{cc}
\bar{\nu}_{L} & \bar{N}^{C}_{R}\end{array}\right)\left(\begin{array}{cc}
0 & m_{D}\\
m_{D} & M_{R}
\end{array}\right)\left(\begin{array}{c}
\nu_{L}^{C}\\
N_{R}
\end{array}\right)+\mathrm{h.c.}
\end{equation}
where we have exchanged the Higgs doublet by its VEV and expanded
the product of $SU(2)_{L}$ doublets, such that $m_{D}=y^{\nu}v_{\mathrm{SM}}/\sqrt{2}$.
Since the right-handed neutrino is an electroweak singlet, the Majorana
mass of the right-handed neutrino $M_{R}$ may be orders of magnitude
above the electroweak scale. It follows that $m_{D}\ll M_{R}$,
and in this case we can extract in all generality the smaller eigenvalue
of the mass matrix above as
\begin{equation}
m_{\nu}\simeq\frac{m_{D}^{2}}{M_{R}}\,.\label{eq:seesaw_formula}
\end{equation}
Assuming that $y^{\nu}$ is of $\mathcal{O}(1)$, which is the natural
expectation, we find that $M_{R}\approx\mathcal{O}(10^{15}\,\mathrm{GeV})$
is required in order to obtain $m_{\nu}\sim\mathcal{O}(0.05\,\mathrm{eV})$.
This is known as the\textit{ seesaw mechanism}: the left-handed neutrino
is very light because its right-handed counterpart is very heavy. More exactly, 
the left-handed neutrino gets a tiny effective Majorana mass through mixing with the heavy Majorana neutrino, propitiated via the Higgs Yukawa coupling. 
Since Majorana particles are their own antiparticle, active
neutrinos being Majorana particles leads to key phenomenological predictions such
as neutrinoless double beta decay \cite{Jones:2021cga}, which are
currently being tested. It is also intriguing the fact that the mass
scale of right-handed neutrinos is close to the scale of Grand Unification
Theories, where the gauge couplings of the SM may unify into a single
gauge coupling (see a longer discussion in Section~\ref{sec:Other-open-questions-SM}).
So far we have only discussed a simplified example of the so-called
\textit{type I seesaw mechanism}, firstly introduced in 1977 by Peter Minkowsky
\cite{Minkowski:1977sc}, although many more versions had been proposed since
then, including low scale variations (see e.g.~\cite{Mohapatra:1986bd})
where the right-handed neutrinos may be much lighter. Many more seesaw
models beyond type I are available in the literature, containing different
types of BSM particle content, most notably the type II seesaw containing
a scalar triplet \cite{Schechter:1980gr} and the type III seesaw
containing a fermion triplet \cite{Foot:1988aq}. Given that the right-handed
neutrinos of the traditional type I seesaw mechanism are much heavier than the
electroweak scale, we may integrate them out to parameterise their
low-energy effect in terms of an effective, non-renormalisable operator
with energy dimension 5,
\begin{equation}
\frac{c}{\Lambda}(\bar{L}_{L}^{C}\widetilde{H})(L_{L}\widetilde{H})+\mathrm{h.c.}\,,\label{eq:Weinberg}
\end{equation}
denoted as the Weinberg operator \cite{Weinberg:1979sa},
which violates lepton number explicitly in two units. The heavy cut-off
scale is associated with $M_{R}$, while $c=(y^{\nu})^{2}$. A similar
operator arises from other neutrino mass models in the literature
that generate Majorana masses for the active neutrinos, just with different matching conditions
for $\Lambda$ and $c$. This way, such an operator can be incorporated
into the SM in order to account for Majorana neutrino masses, while
remaining agnostic to the details of the neutrino mass mechanism in
the UV.

The type I seesaw mechanism can be easily generalised to three generations
of neutrinos. If we introduce two right-handed neutrinos, then two
neutrinos will become massive and one will remain massless, which
is compatible with current data. The new Lagrangian is given by 
\begin{equation}
\mathcal{L}_{\nu}= y_{ij}^{\nu}\bar{L}_{Li}\widetilde{H}N_{Rj}+\frac{1}{2}M_{Rjk}\bar{N}^{C}_{Rj}N_{Rk}+\mathrm{h.c.}\,,\label{eq:neutrino_lagrangian-1}
\end{equation}
where $i=1,2,3$, $j,k=1,2$. Note that since $N_{R1}$ and $N_{R2}$ share the same quantum numbers, they are related by a global $U(2)$ symmetry. We can take advantage of this symmetry to rotate away the mixing term $M_{R12}$ without loss of generality (the consequence is simply an unphysical redefinition of the couplings $y^{\nu}_{ij}$ and $M_{Rjj}$ with respect to the original basis). Then we can write the Lagrangian in matrix form as
\begin{equation}
\mathcal{L}_{\nu}=\frac{1}{2}\left(\begin{array}{cc}
\bar{\nu}_{L} & \bar{N}^{C}_{R}\end{array}\right)\left(\begin{array}{cc}
0 & m_{D}\\
m_{D}^{\mathrm{T}} & M_{R}
\end{array}\right)\left(\begin{array}{c}
\nu_{L}^{C}\\
N_{R}
\end{array}\right)+\mathrm{h.c.}\,,
\end{equation}
where now $\nu_{L}\equiv(\nu_{L1},\nu_{L2},\nu_{L3})^{\mathrm{T}}$ contains the
three left-handed neutrino interaction eigenstates, $N_{R}\equiv(N_{R1},N_{R2})^{\mathrm{T}}$
contains the two right-handed neutrinos, and we have defined the following
matrices in flavour space
\begin{equation}
m_{D}=y_{ij}^{\nu}\frac{v_{\mathrm{SM}}}{\sqrt{2}}\,,
\end{equation}
\begin{equation}
M_{R}=\left(\begin{array}{cc}
M_{R11} & 0\\
0 & M_{R22}
\end{array}\right)\,.
\end{equation}
Provided that $m_{D}\ll M_{R}$, the seesaw formula in Eq.~(\ref{eq:seesaw_formula})
can be generalised to a matrix product
\begin{equation}
m_{\nu}\simeq m_{D}M_{R}^{-1}m_{D}^{\mathrm{T}}\,.\label{eq:seesaw_formula-1}
\end{equation}
After applying the seesaw formula above, we obtain $m_{\nu}$ as a
non-trivial and symmetric $3\times3$ matrix full of $y_{ij}^{\nu}$ couplings of $\mathcal{O}(1)$ that we can use to fit neutrino oscillations data. The precise
values of the neutrino masses are still unknown, however their squared
mass splittings, $\Delta m_{ij}^{2}=m_{i}^{2}-m_{j}^{2}$, can be extracted from oscillations data.
The mass ordering of neutrino masses is unknown as well: it
could be normal with $\Delta m_{31}^{2}>0$, or inverted with $\Delta m_{31}^{2}<0$,
although the former is currently preferred by data. Indeed, if one
neutrino is massless, then normal ordering reveals that the mass of
the heaviest neutrino is $m_{3}=\sqrt{\Delta m_{31}^{2}}\approx0.05\;\mathrm{eV}$,
as anticipated before. 

In analogy with the CKM matrix\footnote{We note that the usual conventions in the literature for the definition of the CKM and PMNS are opposite regarding the ordering of the weak isospin states, i.e. $V_{\mathrm{CKM}}=V^{\dagger}_{u_{L}}V_{d_{L}}$ while $V_{\mathrm{PMNS}}=V^{\dagger}_{e_{L}}V_{\nu_{L}}$, despite the fact that up-quarks and neutrinos share the same weak isosping.} in Eq.~(\ref{eq:CKM_matrix}), one
can define a mixing matrix $V_{\mathrm{PMNS}}$ describing lepton
mixing in terms of left-handed rotations as,
\begin{equation}
V_{\mathrm{PMNS}}=V^{\dagger}_{e_{L}}V_{\nu_{L}}\,,
\end{equation}
where $V_{\nu_{L}}$ is obtained from diagonalising $m_{\nu}$ as $V_{\nu_{L}} ^{\mathrm{T}}m_{\nu}V_{\nu_{L}}=\mathrm{diag}(m_{1},m_{2},m_{3})$, while $V_{e_{L}}$ is obtained
from diagonalising the charged lepton Yukawa couplings. In analogy
with the quark sector, due to PMNS mixing charged currents mediate
tree-level flavour-changing transitions in the lepton sector, that
would violate lepton flavour universality. However, such transitions
only exist because neutrinos are massive, being very suppressed due
to the very tiny neutrino masses. Notice that a non-trivial $V_{\mathrm{PMNS}}$
can be obtained via both neutrino mixing or charged lepton mixing:
only the product $V_{\mathrm{PMNS}}=V_{e_{L}}^{\dagger}V_{\nu_{L}}$
is physical with the known interactions.

\begin{table}
\begin{centering}
\begin{tabular}{ccc}
\toprule 
 & Normal Ordering & Inverted Ordering\tabularnewline
\midrule
\midrule 
$\sin\theta_{12}$ & $0.550\pm0.011$ & $0.550\pm0.011$\tabularnewline
\midrule 
$\sin\theta_{23}$ & $0.756\pm0.012$ & $0.760\pm0.011$\tabularnewline
\midrule 
$\sin\theta_{13}$ & $0.1492\pm0.0020$ & $0.1491\pm0.0019$\tabularnewline
\midrule 
$\delta_{\mathrm{PMNS}}/\pi$ & $1.29_{-0.16}^{+0.19}$ & $1.53_{-0.16}^{+0.12}$\tabularnewline
\midrule 
${\displaystyle \frac{\Delta m_{21}^{2}}{10^{-5}\mathrm{eV^{2}}}}$ & $7.41_{-0.20}^{+0.21}$ & $7.41_{-0.20}^{+0.21}$\tabularnewline
\midrule 
${\displaystyle \frac{\Delta m_{31}^{2}}{10^{-3}\mathrm{eV^{2}}}}$ & $+2.507_{-0.027}^{+0.026}$ & $-2.486_{-0.028}^{+0.025}$\tabularnewline
\bottomrule
\end{tabular}
\par\end{centering}
\caption[Numerical values for neutrino oscillation parameters]{Numerical values for neutrino oscillation parameters taken from the
global fit \cite{Gonzalez-Garcia:2021dve}. Uncertainties in the mixing
angles have been symmetrised and assumed to be Gaussian distributed.
\label{tab:Neutrino_Data}}
\end{table}

However, it can be shown that due to their large mass splittings,
charged leptons in weak interactions are always produced as states
with well-defined mass, since any admixture of $e$, $\mu$ and $\tau$
mass eigenstates is always produced incoherently or cannot maintain coherence
over macroscopic distances \cite{Akhmedov:2007fk} (except at extremely
high energies, not accessible to current experiments). In this manner,
from the observational point of view, the PMNS matrix is carried over
to the neutrino state, which is produced as a coherent admixture of neutrino
mass eigenstates $\left|\nu_{i}\right\rangle $ (where $i=1,2,3$),
such that the produced neutrino is $\left|\nu_{\alpha}\right\rangle =V_{\mathrm{PMNS}}^{\alpha i}\left|\nu_{i}\right\rangle $.
The states $\left|\nu_{\alpha}\right\rangle $ are denoted by convention
as ``interaction eigenstates''\footnote{Sometimes also denoted as ``flavour eigenstates''.},
associated to the mass eigenstates of the charged leptons, that we
can detect in the experiment (such that $\alpha=e,\,\mu,\,\tau$). By
solving the Schrödinger equation, one can check that the produced
neutrino evolves into an admixture of interaction eigenstates. Finally,
at the detection point, the wave function collapses into a well-defined
interaction eigenstate and the associated charged lepton is detected.
Therefore, the interaction state at the detection point can be different
from the interaction state originally produced. This is the phenomenon
of neutrino oscillations\footnote{For a review of neutrino oscillations as a quantum mechanical phenomenon,
see e.g.~\cite{Akhmedov:2019iyt}. }.

Just like the CKM matrix, $V_{\mathrm{PMNS}}$ is described by three
mixing angles and six $CP$-violating phases. If neutrinos are Dirac
particles, then only one $CP$ phase $\delta_{\mathrm{PMNS}}$ is
physical, which has not yet been measured with enough precision. Instead,
if neutrinos are Majorana, one has less freedom to absorb complex phases
via field redefinitions, such that $V_{\mathrm{PMNS}}$ contains three
physical $CP$-violating phases. Therefore, the flavour sector now
contains 26 parameters if neutrinos are Dirac, or 28 if they are Majorana.
Neutrino mixing angles can be parameterised as \cite{deSalas:2020pgw,Gonzalez-Garcia:2021dve}
\begin{alignat}{2}
 & \tan\theta_{23}\sim1\,,\qquad & \tan\theta_{12}\sim\frac{1}{\sqrt{2}}\,,\qquad & \theta_{13}\sim\frac{\lambda}{\sqrt{2}}\,,
\end{alignat}
while the exact numerical values are given in Table~\ref{tab:Neutrino_Data},
which shows also the measured values of the mass splittings $\Delta m_{ij}^{2}=m_{i}^{2}-m_{j}^{2}$.

Oscillation experiments with atmospheric neutrinos are particularly
sensitive to the angle $\theta_{23}$, which is commonly denoted as the
\textit{atmospheric} angle. In a similar manner, $\theta_{12}$ is
denoted as the \textit{solar} angle and $\theta_{13}$ is denoted
as the \textit{reactor} angle. In contrast with quark mixing angles,
neutrino mixing angles are large and seemingly anarchic, with the smaller angle
$\theta_{13}$ being of the same order as the Cabibbo angle. The reason
of why neutrino mixing is so different from quark mixing is unknown,
maybe suggesting a deeper understanding of both in a UV theory that
goes beyond the simplified seesaw mechanism introduced here. Such
a theory could explain the very complicated flavour sector of the
SM+neutrinos in terms of simple and natural principles: this is called
\textit{a theory of flavour}.

\section{Other open questions in the SM\label{sec:Other-open-questions-SM}}

Beyond the flavour puzzle, there are several hints for an extended
framework beyond the SM from both the theoretical and the experimental
side:
\begin{itemize}
\item \textbf{Quantum gravity:} The SM, being a quantum field theory, explains
all the fundamental interactions observed in Nature except for gravity.
General Relativity is a successful theory of gravity at the classical
level, but the SM \textit{must} be extended in order to include a
theory of quantum gravity, which would be crucial to understand physical
phenomena like the singularity of black holes or the (possible) singularity at
the beginning of our Universe. However, attempts to quantise gravity
via the traditional methods of quantum field theory have failed so
far. Since quantum gravity effects are only expected to become manifest
around the Planck scale $M_{\mathrm{Planck}}\sim10^{19}\;\mathrm{GeV}$,
the SM can be understood as an effective field theory of Nature that provides
a good description of low energy physics, well below the Planck scale.
\item \textbf{Dark Matter}: According to cosmological observations (described
in the framework of General Relativity), the SM only accounts for
25\% of matter in our Universe, where we understand matter as massive
particles experiencing the gravity force. The remaining 75\% of matter
could correspond to BSM particles, new colorless matter that does
not interact electromagnetically, the so-called \textit{Dark Matter}.
A well-motivated candidate for Dark Matter were WIMPs, particles
with masses around the EW scale interacting via SM-weak-like interactions,
although they have not been detected so far. Axions are a well-motivated
candidate as well, also connected with the strong $CP$ puzzle. 
However, notice that Dark Matter is a problem related to our understanding
of gravity in the Universe: since we believe that a further theory
of gravity beyond GR exists, at least to account for quantum gravity,
it is possible that such a theory can properly describe the observed
Universe without the need of DM. In this case, DM would be an artifact
of GR being not the final theory of gravity in Nature. It is also 
possible that Dark Matter is made by BSM particles which however do not interact with
the SM particles at all, hence making it difficult to test their properties.
\item \textbf{Accelerated expansion of the Universe}: An Universe consisting
of SM matter and cold Dark Matter is expected to experience a decelerated
expansion. However, cosmological observations show that the Universe
is experiencing an accelerated expansion. This can only be accounted
for within the equations of General Relativity by adding a cosmological
constant term, which is associated to the energy of vacuum itself.
The SM does provide a contribution to the vacuum energy (see the end of
Section~\ref{sec:ScalarSectorandEWSSB}), however this contribution turns out
to be 56 orders of magnitude larger (and with the opposite sign) than the observed
value of the cosmological constant. This implies that there should be an incredibly
\textit{fine-tuned} cancellation between the SM contribution and a
bare vacuum energy parameter that can be introduced in the Lagrangian,
leading to the largest fine-tuning problem of the SM \cite{Weinberg:1988cp,Adler:1995vd,Sola:2013gha}.
In the absence of an explanation for the accelerated expansion of the
Universe, the cosmological constant is associated with an unknown
\textit{Dark Energy} component that constitutes roughly 70\% of the
energy density in our Universe.
\item \textbf{Horizon and flatness problems of the standard cosmology:}
Cosmology aims to describe the history of the Universe in terms of
the known theories of physics. General Relativity describes the evolution
of the Universe and their different components (SM matter, Dark Matter,
radiation and Dark Energy), while the SM describes the interactions
among particles that are crucial to understand the early Universe,
when all particles were interacting in a thermal plasma. The so-called
$\Lambda\mathrm{CDM}$ model provides an overall successful description
of most cosmological observations. However, it fails to provide a proper
explanation for the observed isotropy and homogeneity of the CMB \cite{Planck:2018vyg}.
For example, photons from the last scattering surface coming from
different directions were not in casual contact in the past, yet they
show the same temperature today. This is known as the \textit{horizon
problem}. Cosmological observations also suggest that the Universe
is geometrically flat \cite{Planck:2018vyg} (the metric of the spatial
sections is close to that of the euclidian plane $\mathbb{R}^{3}$).
However, General Relativity predicts that if the Universe is flat
now, it should have been incredibly flatter in the very early Universe,
implying a large fine-tuning of the energy density of the Universe
during the very first instants of time. This is known as the \textit{flatness problem}.
Both problems cannot be understood within the $\Lambda\mathrm{CDM}$
framework based on the SM. A well-motivated explanation is the \textit{inflationary
model}, which requires the addition of a new scalar, the \textit{inflaton}
field, which couples to SM particles. However, the inflationary model
has not yet been experimentally confirmed.
\item \textbf{Matter-antimatter asymmetry:} The SM interactions are $CP$
invariant, meaning that matter and antimatter experience the same
interactions, with the exception of the $CP$-violating phase in the
CKM matrix that breaks $CP$ in the charged currents mediated by $W_{\mu}^{\pm}$.
However, this amount of $CP$ violation is not enough to explain that
our observed Universe is mostly made of matter. Assuming that in the early Universe
matter and antimatter were initially produced in a similar amount (as inflationary
models generally suggest \cite{Nanopoulos:1979gx}),
then in order to explain the matter-antimatter asymmetry of our Universe the SM has to be
extended. It is possible however that the starting conditions of our Universe as 
a dynamic system included more matter than antimatter, or that the matter-antimatter asymmetry is
generated at the end of inflation through the decay of the inflaton: this only deepens into our lack of
understanding about the origin of the Universe and the origin of matter, calling for a more
fundamental theory that can describe the very first instants of time (when quantum gravity effects
were non-negligible), in order to shed light over the initial conditions of the Universe.
\item \textbf{Hierarchy puzzle of the Higgs mass:} The Higgs field responsible
for electroweak symmetry breaking in the SM is quadratically sensitive to NP
scales. Notice that this is not the case for fermions, whose masses
are protected by chiral symmetry \cite{tHooft:1979rat,Klett:2022iga}, or the $W^{\pm}$ and $Z$ bosons, whose masses are
protected by electroweak symmetry. In the presence of NP, the Higgs
mass would receive radiative corrections proportional to the new energy
scale, i.e. $m_{h}^{2}\rightarrow m_{h}^{2}+\delta m_{h}^{2}$ where
\begin{equation}
\delta m_{h}^{2}\sim\frac{\Lambda_{\mathrm{NP}}^{2}}{16\pi^{2}}\,,
\end{equation}
if the NP couple to the Higgs (otherwise the corrections are still
quadratically divergent but they carry further loop suppression). We
have introduced $\Lambda_{\mathrm{NP}}$ as the NP scale where new
degrees of freedom become manifest. Given that we expect new degrees
of freedom to account for neutrino masses, one could consider $\Lambda_{\mathrm{NP}}\sim M_{\mathrm{seesaw}}\sim10^{15}\;\mathrm{GeV}$.
In this manner, one would expect the Higgs to be very heavy due to
the large corrections provided by $M_{\mathrm{seesaw}}$ \cite{Vissani:1997ys,Clarke:2015gwa}. In order to preserve its mass at the electroweak scale,
one should introduce a large cancellation between the bare mass $m_{h}^{2}$
and the large correction $\delta m_{h}^{2}$. This is called \textit{fine-tuning}.
The situation is even worse if one considers NP degrees of freedom
at the Planck scale $M_{\mathrm{Planck}}\sim10^{19}\:\mathrm{GeV}$,
where quantum gravity becomes manifest, and an even larger fine-tuning
is needed. In other words, the Higgs hierarchy puzzle is also the
question of why the electroweak scale and the Planck scale are so far from each other:
a puzzle of hierarchies. The Higgs hierarchy puzzle could be solved
by invoking new physics that screen the Higgs mass from heavier NP
scales. Examples of this are Supersymmetry \cite{Martin:1997ns,Haber:2017aci} (a
symmetry imposing that each fermion has a boson partner and viceversa)
and composite Higgs models \cite{Csaki:2018muy} (the Higgs boson is not an elementary
particle, but a bound state of some more
fundamental, strongly-interacting fermions). Ideally, these NP should
be manifest close to the electroweak scale in order to ameliorate the fine-tuning of the
Higgs mass. Given that flavour observables are sensitive to NP far
above the TeV scale, it is remarkable that no signals of NP
addressing the hierarchy puzzle have been found so far. Is there any
suppression mechanism of FCNCs in the UV that prevents NP from showing
up? This is known as the \textit{NP flavour puzzle}, and motivates
that NP might approximately preserve some flavour symmetry in order
to suppress FCNCs (see Section~\ref{sec:Flavour-symmetries}).
\item \textbf{Strong $CP$ puzzle:} In principle, QCD could violate $CP$
invariance. The QCD Lagrangian allows for the addition of a topological  gauge invariant
term accounting for $CP$ violation, parameterised via the free parameter
$\overline{\theta}_{\mathrm{QCD}}$. However, from non-observation
of the neutron EDM, we know that $|\overline{\theta}_{\mathrm{QCD}}|<10^{-10}$
\cite{Dragos:2019oxn}, meaning that either $CP$ is conserved by
the strong force or its violation is extremely small. Given that $CP$
is largely violated in the quark sector by the weak interaction, explaining
the non-observation of strong $CP$ violation in QCD is challenging.
Ultimately, the puzzle is related to why $\theta_{\mathrm{QCD}}$
is so small when the other source of $CP$ violation in the SM, $\delta_{\mathrm{CKM}}$,
is of $\mathcal{O}(1)$: another puzzle of hierarchies. A well-motivated
solution is the existence of an axion field, as a pseudo-Goldstone
boson from a spontaneously broken Peccei-Quinn global symmetry. Such
a particle would naturally ensure $\overline{\theta}_{\mathrm{QCD}}=0$ \cite{DiLuzio:2020wdo},
and it could also behave as Dark Matter in our Universe \cite{Adams:2022pbo}.
\item \textbf{Experimental anomalies:} Despite the remarkable success of
the SM in explaining the vast majority of particle physics data, some observables remain
in tension with the SM predictions. An illustrative compilation of selected experimental anomalies can be found in Fig.~\ref{fig:Anomalies_Compilation}, for a review see e.g.~\cite{Crivellin:2023gky}.
However, no single tension is significant enough to claim the discovery
of NP. We discuss the flavour-related anomalies in detail in Chapter~\ref{chap:2}.
For the moment, we just highlight the anomalous magnetic moment
of the muon $(g-2)_{\mu}$, which is in 5.1$\sigma$ tension\footnote{We discuss $(g-2)_{\mu}$ in detail in Section~\ref{subsec:g-2}. However, we note
already that different SM predictions for the hadronic vacuum polarisation,
that enters into the determination of $(g-2)_{\mu}$, are in tension.
Namely, the predictions from lattice QCD do not agree with the data driven
predictions obtained from $e^{+}e^{-}\rightarrow\mathrm{hadrons}$
data.} with
the SM \cite{Muong-2:2021ojo,Muong-2:2023cdq}. The $R_{D^{(*)}}$ ratios are in
3.3$\sigma$ tension with the SM \cite{HFLAV:2022wzx}, suggesting the breaking
of lepton flavour universality in $B$-meson decays. The recent data in $\mathcal{B}(B^{+}\rightarrow K^{+}\nu \bar{\nu})$ obtained by Belle II is in $2.8\sigma$ discrepancy with the SM as well \cite{BelleIIEPS:2023}. There is also a
substantial tension in $b\rightarrow s\mu\mu$ data, although these
observables are afflicted by large QCD uncertainties \cite{Alguero:2023jeh}.
The CDF collaboration reported a very anomalous measurement of the
$W_{\mu}^{\pm}$ boson mass \cite{CDF:2022hxs}, although this measurement
is in tension with current and previous data by LHC, LEP and Tevatron.
The Cabibbo angle anomaly \cite{Cirigliano:2022yyo} is related to
a deficit of unitarity in the first row of the CKM matrix, and the
determinations of $V_{ub}$ and $V_{cb}$ via exclusive and inclusive
meson decays do not agree with each other \cite{PDG:2022ynf}. There
are several hints for new resonances at the LHC, mainly in di-photon
channels at 95 GeV \cite{CMS:2023yay,ATLAS-CONF-2023-035} (this one
is supported by mild excesses in both ATLAS and CMS data, plus an
old excess at LEP), 151 GeV \cite{ATLAS:2021jbf} and 670 GeV \cite{ATLAS:2021uiz}.
Finally, there are mild deviations in electroweak precision observables
(EWPOs) that worsen the global fit of the SM to electroweak data, for example
the $2\sigma$ tensions in the forward-backward asymmetry of $Z\rightarrow\bar{b}b$
and the asymmetry observable in $Z\rightarrow e^{+}e^{-}$ \cite{ALEPH:2005ab}.
\begin{figure}[t]
%\begin{centering}
\includegraphics[scale=0.72]{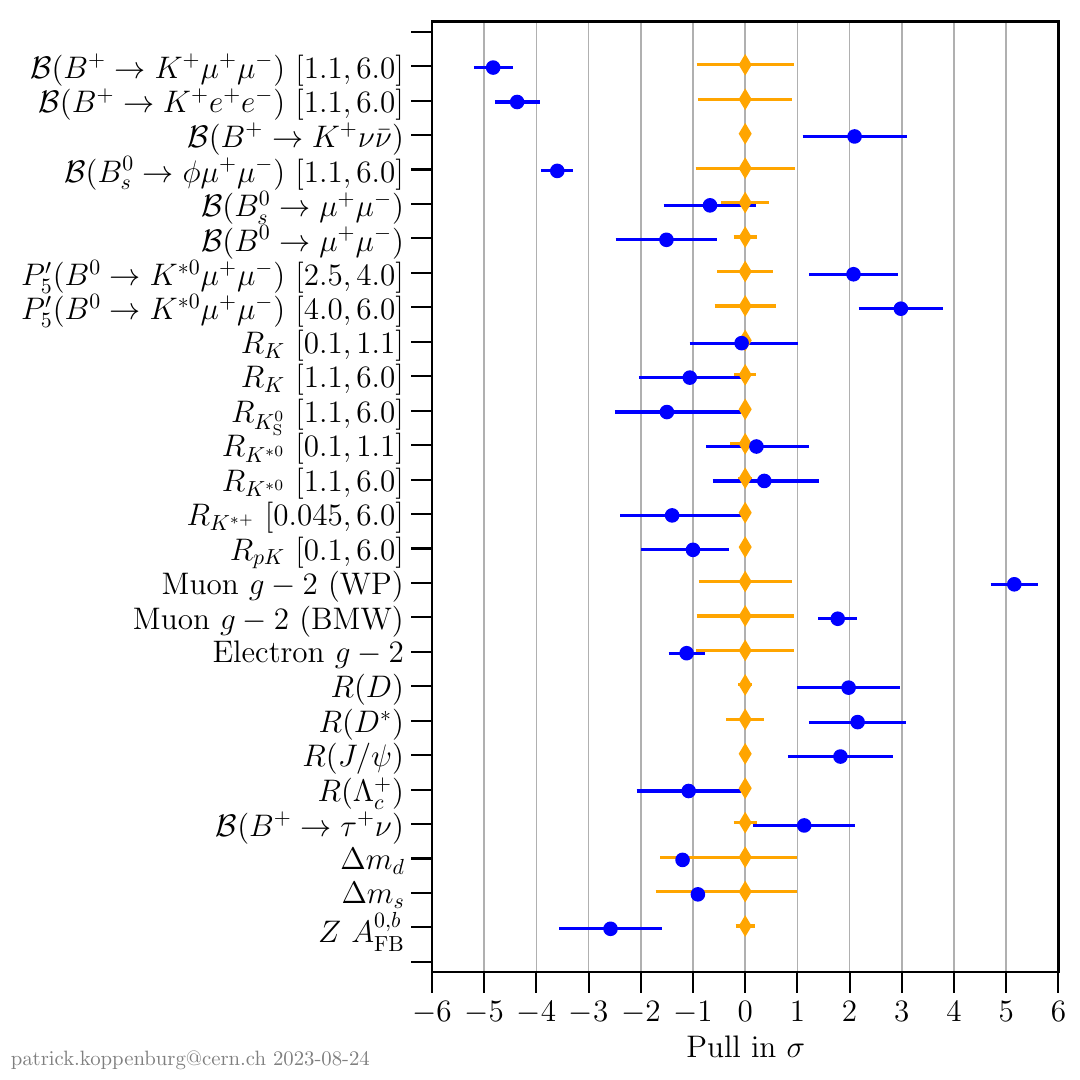}
%\par\end{centering}
\caption[Compilation of experimental anomalies in low-energy observables and their pull in standard deviations]{Compilation of experimental anomalies in low-energy observables and their pull in standard deviations. Experimental measurements are shown in blue, the SM predictions are shown in orange. Figure available in the website of Patrick Koppenburg: \href{https://www.nikhef.nl/~pkoppenb/anomalies.html}{https://www.nikhef.nl/~pkoppenb/anomalies.html}.
\label{fig:Anomalies_Compilation}}
\end{figure}
\item \textbf{Hubble tension:} Beyond the above anomalies in particle physics experiments,
there exists a long-lasting discrepancy in the different cosmological determinations of the Hubble constant,
$H_{0}$, which measures the current expansion rate of our Universe \cite{Weinberg:2013agg,DiValentino:2021izs}. More accurately, the early time determinations
(e.g.~from the CMB) are in roughly $5\sigma$ tension with the late time determinations obtained from observing
stars such as superonovae and cepheids. This hints to a possible failure of the cosmological model $\Lambda\mathrm{CDM}$,
which nevertheless provides a very successful description of many other observables. Given that $\Lambda\mathrm{CDM}$
is supported on SM physics, it is possible that the solution to the long-lasting Hubble tension may be due to 
the presence of BSM physics.
\end{itemize}

\section{Gauge unification and flavour} \label{sec:GaugeAndFlavour}

Beyond the aforementioned open questions of the SM, charge quantisation,
namely why the proton and the electron have equal but opposite electric
charges despite being apparently very different in nature, has been
a mystery since the early days of quantum mechanics. With the advent
of the SM, this question has escalated to explain the particular quantum
numbers and transformation properties of known elementary particles
under the SM gauge group, and the relative strength of their interactions. 

\begin{figure}[t]
\begin{centering}
\includegraphics[scale=0.85]{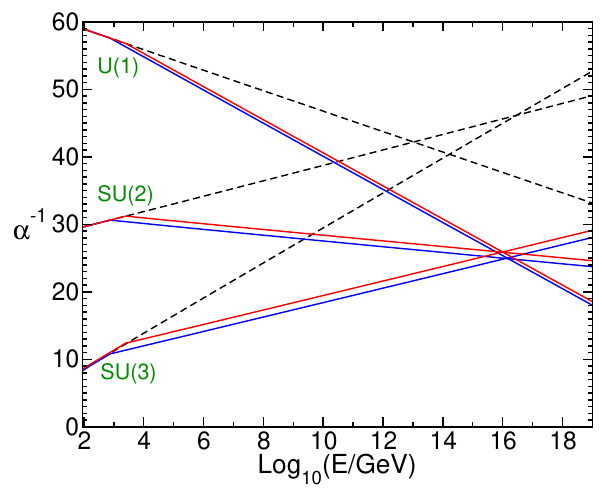}
\par\end{centering}
\caption[RGE of the inverse gauge couplings $\alpha_{a}^{-1}=4\pi/g_{a}^{2}$
in the SM and the MSSM]{Two-loop RGE of the inverse gauge couplings $\alpha_{a}^{-1}=4\pi/g_{a}^{2}$
in the SM (dashed lines) and the MSSM (solid lines). In the MSSM case,
the sparticle masses are treated as a common threshold varied between
750 GeV and 2.5 TeV, and $\alpha_{3}(M_{Z})$ is varied between 0.117
and 0.120. Figure taken from \cite{Martin:1997ns}.\label{fig:GaugeUnification}}

\end{figure}

With great elegance and simplicity, a Grand Unified Theory (GUT) postulates
that the plethora of different charges and interactions of the SM
is just a low-energy manifestation of a deeper underlying unity. Therefore,
in the ultraviolet, the separate parts of the SM gauge group are merged
into an unique gauge symmetry, and SM fermions are unified into representations
under the GUT gauge group. GUTs rely on the assumption that SM gauge
couplings evolve in the UV and join each other, via RGE running. In
the SM they approach but do not quite unify, although
they unify in the Minimal Supersymmetric SM (MSSM) at the very high scale
$M_{\mathrm{GUT}}\sim10^{16}\:\mathrm{GeV}$ \cite{Amaldi:1991cn}, see Fig.~\ref{fig:GaugeUnification}. This scale is suspiciously close to the heavy scale
$M_{\mathrm{seesaw}}\sim10^{15}\:\mathrm{GeV}$ of right-handed neutrinos
in the type I seesaw mechanism.

The minimal\footnote{The other simple rank 4 algebras beyond $SU(5)$ do not work since they
do not have complex representations.} gauge group where the SM forces may unify is $SU(5)$
\cite{Georgi:1974sy}, which we introduce here as it
will be greatly discussed in Chapter~\ref{Chapter:Tri-unification} of this thesis. One can check
that $SU(5)$ contains the generators of $SU(3)$ and $SU(2)$ as
\begin{equation}
t^{\alpha}=\left(\begin{array}{c|c}
\frac{1}{2}\lambda^{\alpha} & 0\\
\hline 0 & 0
\end{array}\right)\,,\qquad T^{a}=\left(\begin{array}{c|c}
0 & 0\\
\hline 0 & \frac{1}{2}\sigma^{a}
\end{array}\right)\,,
\end{equation}
where $\alpha=1,...,8$ and $a=1,2,3$, with $\lambda^{\alpha}$ and
$\sigma^{a}$ being the Gell-Mann and Pauli matrices, respectively.
The only remaining generator of $SU(5)$ commuting with $t^{\alpha}$
and $T^{a}$ is
\begin{equation}
\mathcal{T}_{Y}=\mathcal{N}\mathrm{diag}(-1/3,-1/3,-1/3,1/2,1/2)\equiv\mathcal{N}Y\,,\label{eq:hypercharge_SU(5)}
\end{equation}
which we associate to SM hypercharge up to a normalisation factor.
The normalisation factor is chosen by convention so that all $SU(5)$ generators
satisfy $\mathrm{Tr}(\mathcal{T}_{A}\mathcal{T}_{B})=\frac{1}{2}\delta_{AB}$,
where $A,B=1,...,24$. Therefore, one can check that $\mathrm{Tr}(\mathcal{T}_{Y}\mathcal{T}_{Y})=\frac{1}{2}$
if $\mathcal{N}=\sqrt{\frac{3}{5}}$. This normalisation factor is commonly absorbed in a redefinition of the
coupling strength $g_{Y}$.

The adjoint representation of
$SU(5)$ may be decomposed under the SM as
\begin{equation}
\mathbf{24}=(\mathbf{8},\mathbf{1})_{0}\oplus(\mathbf{1},\mathbf{3})_{0}\oplus(\mathbf{1},\mathbf{1})_{0}\oplus(\mathbf{3},\mathbf{2})_{-5/6}\oplus(\mathbf{\overline{3}},\mathbf{\overline{2}})_{5/6}\,.
\end{equation}
Beyond gluons and the electroweak gauge bosons, we also find 12 exotic
gauge bosons commonly denoted as $X\sim(\mathbf{3},\mathbf{2})_{-5/6}$
and $X^{*}\sim(\mathbf{\overline{3}},\mathbf{\overline{2}})_{5/6}$\footnote{We note that $\mathbf{\overline{2}}=\mathbf{2}$ because the fundamental
representation of $SU(2)$ is pseudo-real, but commonly we will still show the bar in $\mathbf{\overline{2}}$ for the sake of clarity.}, which are
associated to the remaining generators of $SU(5)$. The $X^{(*)}$
gauge bosons become massive after spontaneous breaking of the $SU(5)$
group down to the SM, which usually proceeds via a fundamental scalar
in the adjoint representation getting a VEV in the SM singlet
component.

The chiral fermions of the SM do also unify into representations of
$SU(5)$. For a given family of SM chiral fermions described by two-component
Weyl spinors, the lepton doublet and the $CP$-conjugate down-quark singlet
are combined into a $\mathbf{\overline{5}}$, and the quark doublet
is combined with the $CP$-conjugate up-quark and lepton singlets into
a $\mathbf{10}$\footnote{For this reason, this convention where all fermions are described
by two-component left-handed Weyl spinors ($CP$-conjugate right-handed
fermions are actually left-handed) is very convenient for model building
studies, in contrast with the convention of four-component left-right
Dirac spinors which is more common for phenomenological studies. Both
conventions are described in Appendix~\ref{app:2-component_notation}.},
\begin{equation}
\mathbf{\overline{5}}_{i}=\left(\begin{array}{c}
\textcolor{red}{d_{r}^{c}}\\
\textcolor{DarkGreen}{d_{g}^{c}}\\
\textcolor{blue}{d_{b}^{c}}\\
\nu\\
e
\end{array}\right)_{i}\,,\qquad\mathbf{10}_{i}=\frac{1}{\sqrt{2}}\left(\begin{array}{ccccc}
0 & \textcolor{blue}{u_{b}^{c}} & \textcolor{DarkGreen}{-u_{g}^{c}} & \textcolor{red}{u_{r}} & \textcolor{red}{d_{r}}\\
-\textcolor{blue}{u_{b}^{c}} & 0 & \textcolor{red}{u_{r}^{c}} & \textcolor{DarkGreen}{u_{g}} & \textcolor{DarkGreen}{d_{g}}\\
\textcolor{DarkGreen}{u_{g}^{c}} & -\textcolor{red}{u_{r}^{c}} & 0 & \textcolor{blue}{u_{b}} & \textcolor{blue}{d_{b}}\\
-\textcolor{red}{u_{r}} & -\textcolor{DarkGreen}{u_{g}} & -\textcolor{blue}{u_{b}} & 0 & e^{c}\\
-\textcolor{red}{d_{r}} & -\textcolor{DarkGreen}{d_{g}} & -\textcolor{blue}{d_{b}} & -e^{c} & 0
\end{array}\right)_{i}\,,\label{eq:fermions_SU(5)}
\end{equation}
where $i=1,2,3$ denote the three fermion families of the SM. It is
clear then than in the $SU(5)$ grand unified model, the three fermion
families remain as three identical copies under the gauge symmetry.
Remarkably, the hypercharges of chiral fermions are no longer
seemingly arbitrary values (constrained by the cancellation of gauge anomalies), but are rather associated to a discretely valued
and traceless generator of $SU(5)$ as given in Eq.~\eqref{eq:hypercharge_SU(5)}. As a consequence, the generator of the residual group $U(1)_{Q}$ is given as a linear combination
of generators of $SU(5)$, namely $Q=T_{3}+\mathcal{T}_{Y}$, therefore being discretely valued and traceless as well. Given the fermion representations
in Eq.~\eqref{eq:fermions_SU(5)}, this enforces the charge of the down quark to be 1/3 of the charge of the electron, and the charge of the up quark to be 2/3 the charge of the positron, explaining the observed quantisation of electric charge.

The Higgs doublet of the SM is embedded into a $\mathbf{5}$ representation
that allows to write Yukawa couplings as
\begin{equation}
\mathcal{L}_{\mathrm{Yukawa}}=Y_{ij}^{u}\mathbf{10}_{i}\mathbf{10}_{j}\mathbf{5}_{H}+Y_{ij}^{de}\mathbf{10}_{i}\mathbf{\overline{5}}_{j}\mathbf{\overline{5}}_{H}+\mathrm{h.c.}
\end{equation}
Important consequences are that the up-quark Yukawa matrix is symmetric,
while the down-quark and charged lepton Yukawa couplings both emerge
from $Y_{ij}^{de}$ and verify $Y^{d}=(Y^{e})^{\dagger}$. This means
that down quarks and charged leptons are predicted to have the same
mass at $M_{\mathrm{GUT}}\sim10^{16}\:\mathrm{GeV}$ where $SU(5)$
is spontaneously broken. However, this unification of Yukawa couplings
is so far inconsistent with experimental data and further model building
needs to be done \cite{Georgi:1979df}, plus Yukawa couplings remain hierarchical for the different families
as in the SM, leaving the flavour puzzle unanswered or even worsened.
Moreover, note that the $SU(5)$ theory says nothing about neutrino
masses, since right-handed neutrinos remain singlets as in the SM.

The fact that some quarks and leptons are unified in the same representations
means that the $X^{(*)}$ gauge bosons can transform quarks into leptons
and viceversa, giving them the name of \textit{leptoquarks}. In particular,
this leads to the striking prediction of proton decay, which is in
tension with current data unless the GUT symmetry is spontaneously
broken at very high scales, compatible with $M_{\mathrm{GUT}}\sim10^{16}\:\mathrm{GeV}$ \cite{Super-Kamiokande:2016exg}
(see a more complete treatment of gauge boson-mediated proton decay
in Section~\ref{subsec:ProtonDecay}). Notice that colour-triplet Higgs scalars contained
in $\mathbf{5}_{H}$ commonly mediate proton decay as well, requiring
their masses to be very heavy. This is in conflict with the fact that
the SM Higgs doublet contained in $\mathbf{5}_{H}$ lives at the electroweak
scale, since one would naturally expect fields within the same multiplet
to have masses at the same scale. This inconsistency is denoted as
the doublet-triplet splitting problem.

Beyond $SU(5)$, the next simple unification framework is based on
the gauge group $SO(10)$\footnote{Notice that the Lie group with $\mathbf{16}$-dimensional spinor representations
is formally $\mathrm{Spin}(10)$, and by $SO(10)$ we refer to its
Lie algebra.} \cite{Fritzsch:1974nn,Georgi:1974my}, which is rank 5. Here all 15 chiral
fermions of a given family\footnote{Indeed even in $SO(10)$ one still needs three fermion representations,
one for each SM fermion family. Naively, all SM+3$\nu_{R}$ fermions
(48 Weyl spinors) can be unified into the fundamental representation
of $SU(48)$ without introducing exotic fermions (beware of gauge
anomalies)~\cite{Fonseca:2015aoa,Allanach:2021bfe}. However, with
increasingly big unification groups the possibilities for embedding
the SM in them grow in a seemingly exponential way \cite{Fonseca:2015aoa}.
For this reason, one might argue that models based on very large gauge
groups are not as attractive as those based on smaller ones: they
contain many subgroups, therefore a significant tuning of the scalar
sector parameters would likely be needed in order to have the correct
symmetry breaking.} are unified into a single $\mathbf{16}$-dimensional spinor representation, including one SM singlet
that is associated to a right-handed neutrino. The SM Higgs doublet
is then embedded into a $\mathbf{10}$ representation, which allows
to write a common Yukawa coupling for all fermions within the same
family. This is phenomenologically unsuccessful and requires the addition
of further Higgs bosons and model building. The $SO(10)$ group may
be broken down to the SM in various steps, which include $SU(5)$ and the Pati-Salam
gauge group $SU(4)_{c}\times SU(2)_{L}\times SU(2)_{R}$ \cite{Pati:1974yy}.
Despite not being a grand unified theory, the latter predicts the quantisation of hypercharge
as well via $Y=(B-L)/2+T_{3R}$\footnote{Note that baryon number minus lepton number $B-L$ is associated to a discretely valued and traceless generator of $SU(4)_{c}$ (up to a 1/2 normalisation factor).}, the restoration
of parity and the unification of quarks and leptons within the same
representations, with leptons being the ``fourth colour'' (including
right-handed neutrinos),
\begin{flalign}
 & \psi_{i}(\mathbf{4,2,1})=\left(\begin{array}{cccc}
\textcolor{red}{u_{r}} & \textcolor{DarkGreen}{u_{g}} & \textcolor{blue}{u_{b}} & \nu\\
\textcolor{red}{d_{r}} & \textcolor{DarkGreen}{d_{g}} & \textcolor{blue}{d_{b}} & e
\end{array}\right)_{i}\equiv\left(Q_{i},L_{i}\right)\,,\\
 & \psi_{j}^{c}(\mathbf{\overline{4},1,\overline{2}})=\left(\begin{array}{cccc}
\textcolor{red}{u_{r}^{c}} & \textcolor{DarkGreen}{u_{g}^{c}} & \textcolor{blue}{u_{b}^{c}} & \nu^{c}\\
\textcolor{red}{d_{r}^{c}} & \textcolor{DarkGreen}{d_{g}^{c}} & \textcolor{blue}{d_{b}^{c}} & e^{c}
\end{array}\right)_{j}\equiv\left(u_{j}^{c},d_{j}^{c},\nu_{j}^{c},e_{j}^{c}\right)\,,
\end{flalign}
where $i,j=1,2,3$ are flavour indices. The SM Higgs doublet is embedded as $H\sim(\mathbf{1},\mathbf{2},\mathbf{2})$
which again provides a common Yukawa coupling for all fermions within
the same family as in $SO(10)$. As shown, the correct description
of fermion masses and mixings is usually a challenge for the GUT program
and for many BSM theories. In Chapters~\ref{Chapter:TwinPS} and \ref{Chapter:Tri-unification} we will propose models
based on modifications of Pati-Salam and $SU(5)$ respectively, that
are able to describe fermion masses and mixings successfully along
with explaining their hierarchical patterns.

\section{Flavour symmetries of the SM and beyond\label{sec:Flavour-symmetries}}

Flavour dynamics in the SM are encoded as processes or parameters
that break the flavour symmetry $U(3)^{5}$. The flavour symmetry may
be of great importance in order to understand the precise structure
of the Yukawa couplings, which in the SM are just free parameters,
and might provide some insights about possible underlying flavour dynamics,
which may be connected to the origin of flavour in the SM. Therefore,
understanding $U(3)^{5}$ and its possible breaking by NP is fundamental
for flavour model building.

\subsection{\texorpdfstring{$U(3)^{5}$: the accidental symmetries of the SM}{U(3)5: the accidental symmetries of the SM}}

We have built the SM as a theory where each fundamental fermion comes in three flavours
that transform in the same way under the gauge group, the only
difference between fermion flavours being their masses. As mentioned
in Section~\ref{sec:The-gauge-sector}, this fact translates into
the appearance of an approximate, global flavour symmetry
\begin{equation}
U(3)^{5}=U(3)_{Q}\times U(3)_{L}\times U(3)_{u}\times U(3)_{d}\times U(3)_{e}\,.\label{eq:Flavour_Symmetry-1}
\end{equation}
Given that $U(3)^{5}\cong SU(3)^{5}\times U(1)^{5}$, we can replace the
flavour symmetry by
\begin{equation}
U(3)^{5}\cong SU(3)_{q}^{3}\times SU(3)_{\ell}^{2}\times U(1)^{5}\,,\label{eq:flavour_symmetry-2}
\end{equation}
where
\begin{equation}
SU(3)_{q}^{3}=SU(3)_{Q}\times SU(3)_{u}\times SU(3)_{d}\,,
\end{equation}
\begin{equation}
SU(3)_{\ell}^{2}=SU(3)_{L}\times SU(3)_{e}\,,
\end{equation}
\begin{equation}
U(1)^{5}=U(1)_{B}\times U(1)_{L}\times U(1)_{Y}\times U(1)_{\mathrm{PQ}}\times U(1)_{e_{R}}\,.
\end{equation}
Out of the five $U(1)$ charges, we identify baryon number ($B$),
total lepton number ($L$), SM hypercharge (which is gauged), the
Peccei-Quinn ($\mathrm{PQ}$) symmetry whereby the Higgs doublet and the $d_{Ri}$, $e_{Ri}$ fields
have opposite charges, and finally $U(1)_{e_{R}}$ corresponds to a global phase of $e_{Ri}$ only.

The Yukawa couplings of the SM break the flavour symmetry in Eq.~(\ref{eq:Flavour_Symmetry-1}). However, one can always rephase all quark fields by the same phase, which is associated to a global $U(1)$ in the quark sector that remains unbroken by the Yukawa couplings. This is the abelian symmetry associated to baryon number. The fact that there are no right-handed neutrinos in the SM implies that the three charged lepton masses are the only physical parameters of the lepton sector, associated to the diagonal entries of the lepton Yukawa matrix. We are always free to rephase the three diagonal charged lepton bilinears by three independent phases, associated to the three global abelian symmetries of the three lepton family numbers. Therefore, we conclude that the global flavour symmetry of the SM is broken by the Yukawa couplings down to (not displaying the gauged hypercharge which of course remains unbroken)
\begin{equation}
U(3)^{5}\rightarrow U(1)_{B}\times U(1)_{L_{e}}\times U(1)_{L_{\mu}}\times U(1)_{L_{\tau}}\,,\label{eq:anomalous_symmetries}
\end{equation}
where the lepton family numbers, $L_{e}$, $L_{\mu}$ and $L_{\tau}$, contain the total lepton number $L$ as the diagonal subgroup.
These accidental symmetries have fundamental consequences: the proton
is stable in the SM, and lepton flavour violating processes
such as $\mu\rightarrow e\gamma$ are forbidden. However,
given that these symmetries are accidental, nothing forbids us to
write operators with energy dimension higher than four (non-renormalisable)
which break these symmetries. Two examples are the following,
\begin{equation}
\frac{c_{ij}}{\Lambda}(\bar{L}_{Li}^{C}\widetilde{H})(L_{Lj}\widetilde{H})+\mathrm{h.c.}\,,\label{eq:Weinberg-1}
\end{equation}
\begin{equation}
\frac{c_{ijkl}}{\Lambda^{2}}(\bar{Q}_{Li}^{C}Q_{Lj})(\bar{Q}_{Lk}^{C}L_{l})\,.\label{eq:B_violating}
\end{equation}
The careful reader may have noticed the Weinberg operator already
introduced in Eq.~(\ref{eq:Weinberg}) (generalised to three generations
in Eq.~(\ref{eq:Weinberg-1})), that generates Majorana neutrino
masses and violates lepton number in two units. The operator in Eq.~(\ref{eq:B_violating})
violates both baryon number and lepton number in one unit, and can mediate proton decay for particular
flavour indices. $B$-violating operators such that (\ref{eq:B_violating})
are predicted by GUTs, as mentioned in Section~\ref{sec:GaugeAndFlavour}.
Indeed, the proton has been observed to be extremely stable \cite{Super-Kamiokande:2016exg},
setting strong bounds on the cut-off scale of the operators (\ref{eq:B_violating}).

The global symmetries in Eq.~(\ref{eq:anomalous_symmetries}) are
anomalous, meaning that they are only preserved at the classical level.
In fact, they are explicitly broken by non-perturbative quantum effects
such as the sphaleron process \cite{Manton:1983nd}. Even though this symmetry
breaking effects are negligible at low energies (or equivalently at low
temperatures), they may play an important role in the early Universe
when temperatures were very high. For example, they could play a
role in the origin of the matter-antimatter asymmetry (see a brief
discussion in Section~\ref{sec:Other-open-questions-SM}), or when
one aims for the construction of an extended gauge sector. Out of the
four remaining $U(1)$ factors, only the subgroup corresponding to one of
the combinations of lepton family numbers $U(1)_{L_{\alpha}-L_{\beta}}$ ($\alpha,\beta=e,\mu,\tau$
with $\alpha\neq\beta$) remains anomaly-free in the SM \cite{Foot:1990mn}.
Interestingly, if one includes three right-handed neutrinos, singlets
under the SM gauge group (which could account for neutrino masses
as described in Section~\ref{subsec:The-origin-of-neutrinos}), the
global accidental anomaly-free symmetry of the theory is extended
to \cite{Araki:2012ip}
\begin{equation}
U(1)_{B-L}\times U(1)_{L_{e}-L_{\mu}}\times U(1)_{L_{\mu}-L_{\tau}}\,.
\end{equation}
These accidental, anomaly-free symmetries will get either partially
or totally broken once we include all the details from the neutrino
sector. For instance, we know from the measured values of the PMNS
matrix that the $U(1)_{L_{e}-L_{\mu}}\times U(1)_{L_{\mu}-L_{\tau}}$ symmetry is
broken. On the other hand, the $U(1)_{B-L}$ symmetry would be broken
if neutrinos are Majorana particles.

Even though the Yukawa couplings indeed break the flavour symmetry
as per Eq.~(\ref{eq:anomalous_symmetries}), the smallness of the
Yukawa couplings in the SM leads to a small breaking
of the flavour symmetry. Notice that if the Yukawa couplings were
absent, the flavour symmetry would remain unbroken and the SM would
be a flavour conserving theory. This has important consequences for
flavour physics in the SM: flavour-changing charged currents (FCCCs)
are suppressed by the smallness of the CKM mixing, and flavour-changing
neutral currents (FCNCs) only appear at loop-level and are suppressed
as well by small CKM elements. Another extra suppression enters in
FCNCs: notice that if all quarks in a given sector (up or down) were
degenerate, the flavour-changing $W^{\pm}$-couplings in Eq.~(\ref{eq:Charged_Currents})
would vanish. A consequence of this is the fact that FCNCs in the down
(up) sector are proportional to mass-squared differences between the
quarks of the up (down) sector. For FCNCs that involve only quarks
of the first two generations, this leads to a strong suppression factor
related to the light quark masses, and known as \textit{Glashow-Iliopoulos-Maiani
(GIM) suppression} \cite{Glashow:1970gm}. This extra suppression
factor allowed Glashow, Iliopoulos and Maiani to explain kaon mixing
and predict the existence of the charm quark, giving an upper bound
for its mass. 

Instead, going beyond the SM, FCNCs can occur at tree-level, with
negligible suppression. However, the large suppression of FCNCs predicted
by the SM turns out to be very well realised in Nature: rare decays
of mesons and meson-antimeson mixing have been tested to agree well
with the SM. Given that there are good reasons to expect new physics
beyond the SM near the TeV scale (see Section~\ref{sec:Other-open-questions-SM}), why do we see no deviations in flavour physics observables, which
are sensitive to very high NP scales?

In the following we provide a few examples of flavour symmetries proposed
to dictate the flavour structure of NP in order to make it compatible with current bounds
from flavour observables.

\subsection{Minimal Flavour Violation}

\begin{figure}[t]
\begin{centering}
\includegraphics[scale=0.95]{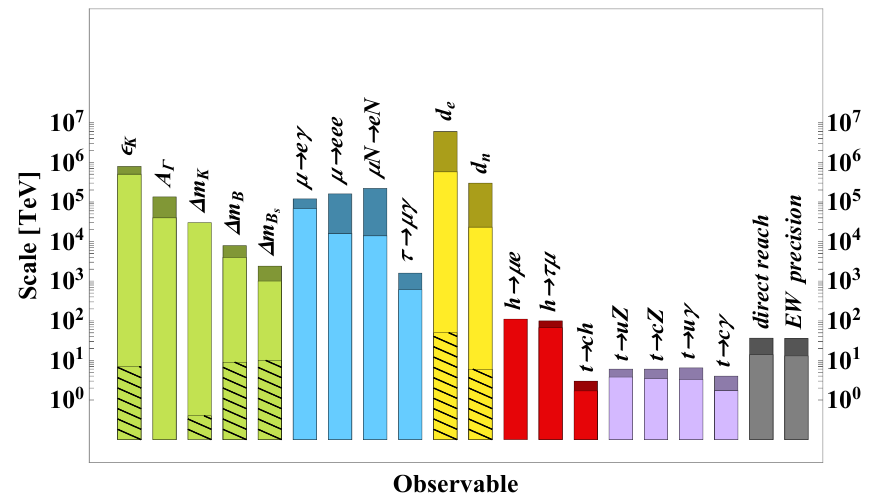}
\par\end{centering}
\caption[Reach in new physics scale from flavour observables]{Reach in new physics scale of present and future facilities, from
generic dimension-six operators. The colour coding of observables is:
green for mesons, blue for leptons, yellow for EDMs, red for Higgs
flavoured couplings and purple for the top quark. The grey columns
illustrate the reach of direct flavour-blind searches and EW precision
measurements. The operator coefficients are taken to be either \ensuremath{\sim}
1 (plain coloured columns) or suppressed by MFV factors (hatch filled
surfaces). Light (dark) colours correspond to present data (mid-term
prospects, including HL-LHC, Belle II, MEG II, Mu3e, Mu2e, COMET,
ACME, PIK and SNS). Figure taken from \cite{EuropeanStrategyforParticlePhysicsPreparatoryGroup:2019qin}.
\label{fig:Reach-in-NP_flavour}}

\end{figure}

A possibility to accommodate relatively light NP with flavour physics observables
arises from the understanding of $U(3)^{5}$ and its breaking in the
SM: if the SM Yukawa couplings remain as the only breaking source
of $U(3)^{5}$, this guarantees that low-energy flavour changing processes
deviate only very little from the SM predictions. In this scenario,
denoted as \textit{Minimal Flavour Violation} (MFV) \cite{DAmbrosio:2002vsn},
flavour transitions mediated by NP enjoy a similar suppression as they do in the
SM. This ameliorates the strong flavour bounds over the scale of NP, as shown
in Fig.~\ref{fig:Reach-in-NP_flavour}, allowing the possibility of
TeV-scale NP as suggested by the Higgs hierarchy puzzle.

Let us now formulate this principle in a formal way. Given that in the SM the flavour symmetry $U(3)^{5}$
is only broken by the Yukawa couplings, we can formally recover flavour invariance
by promoting the Yukawa matrices to dimensionless auxiliary fields, \textit{spurions},
transforming under $SU(3)_{q}^{3}\times SU(3)_{\ell}^{2}$
as
\begin{equation}
Y_{u}\sim(\mathbf{3},\mathbf{\overline{3}},\mathbf{1})_{SU(3)_{q}^{3}}\,,\qquad Y_{d}\sim(\mathbf{3},\mathbf{1},\mathbf{\overline{3}})_{SU(3)_{q}^{3}}\,,\qquad Y_{e}\sim(\mathbf{3},\mathbf{1})_{SU(3)_{\ell}^{2}}\,.
\end{equation}
In this manner, the Yukawa interactions
\begin{equation}
\mathcal{L}_{\mathrm{Yukawa}}=Y_{u}\overline{Q}_{L}\tilde{H}u_{R}+Y_{d}\overline{Q}_{L}Hd_{R}+Y_{e}\overline{L}_{L}He_{R}+\mathrm{h.c.}
\end{equation}
now preserve the flavour symmetry, with quarks and charged leptons transforming as triplets under their
corresponding $SU(3)$ factors. Having the freedom of
the $SU(3)_{q}^{3}\times SU(3)_{\ell}^{2}$ symmetry, we can use unitary
transformations to rotate to the particular basis
\begin{equation}
Y_{u}=V_{\mathrm{CKM}}^{\dagger}\lambda_{u}\,,\qquad Y_{d}=\lambda_{d}\,,\qquad Y_{e}=\lambda_{e}\,,
\end{equation}
where $\lambda_{u,d,e}$ are diagonal matrices,
\begin{equation}
\lambda_{u}=\mathrm{diag}(y_{u},y_{c},y_{t})\,,\qquad\lambda_{d}=\mathrm{diag}(y_{d},y_{s},y_{b})\,,\qquad\lambda_{e}=\mathrm{diag}(y_{e},y_{\mu},y_{\tau})\,.
\end{equation}
As an example, we may apply the MFV hypothesis to effective operators
describing NP effects in an EFT framework. We define that our effective
field theory satisfies the criterion of MFV if all higher-dimensional
operators, constructed from SM and Yukawa spurions, are invariant under
$CP$ and (formally) under the flavour group $U(3)^{5}$. In other words,
MFV requires that the dynamics of flavour violation are completely
determined by the structure of the ordinary Yukawa couplings. This
requirement translates into the fact that all relevant flavour non-diagonal
operators in the EFT are proportional to powers of $Y_{u}Y_{u}^{\dagger}$.
Since the SM Yukawa couplings are small for all fermions except for the top quark,
one obtains $(Y_{u}Y_{u}^{\dagger})_{ij}\approx y_{t}^{2}V_{\mathrm{CKM}}^{ti*}V_{\mathrm{CKM}}^{tj}$.
For example, for a four-quark operator violating fermion number by
two units ($\Delta F=2$), like the operators which contribute to
meson-antimeson mixing processes, one obtains
\begin{equation}
\mathcal{O}_{ijkl}=\left(\overline{Q}_{Li}(\lambda_{\mathrm{FC}})_{ij}\gamma_{\mu}Q_{Lj}\right)\left(\overline{Q}_{Lk}(\lambda_{\mathrm{FC}})_{kl}\gamma^{\mu}Q_{Ll}\right)\,,
\end{equation}
where we have defined
\begin{equation}
(\lambda_{\mathrm{FC}})_{ij}=\left\{ \begin{array}{c}
(Y_{u}Y_{u}^{\dagger})_{ij}\approx y_{t}^{2}V_{\mathrm{CKM}}^{ti*}V_{\mathrm{CKM}}^{tj}\qquad i\neq j\,,\\
0\qquad\qquad\qquad\qquad\qquad\qquad\;\, i=j\,.
\end{array}\right.
\end{equation}
It is clear now that the factor $(\lambda_{\mathrm{FC}})_{ij}$ provides
CKM suppression for the flavour-violating operators as in the SM. This
would explain the absence of NP signals in the flavour observables
of the quark sector, allowing relatively light NP as long as their flavour structure
is SM-like. The MFV hypothesis can also be extended to the lepton
sector. However, since the mechanism responsible for neutrino masses
is unknown at present, there is no unique way to introduce the MFV
principle in the lepton sector. For the realisation of MFV in a scenario
inspired in the type I seesaw mechanism, see e.g.~\cite{Cirigliano:2005ck}.

The MFV prescription imposes that the flavour structure of NP is SM-like.
From the point of view of a theory of flavour, this means that the
new flavour dynamics addressing the flavour puzzle are very heavy,
leaving as a low-energy remnant the flavour structure of the SM and
relatively light MFV NP. In this sense, the MFV hypothesis suggests that
the \textit{a priori} unknown scales of the theory of flavour ($\Lambda_{F}$
and $\left\langle \phi_{F}\right\rangle $) are very heavy, possibly
close to the seesaw scale $M_{\mathrm{seesaw}}\approx10^{15}\;\mathrm{GeV}$
or to the GUT scale $M_{\mathrm{GUT}}\approx10^{16}\;\mathrm{GeV}$, where
we definitely expect BSM physics to manifest. This is an interesting
idea: NP following the flavour structure of the SM may manifest
at relatively low energies, as required in order to solve the Higgs
hierarchy puzzle and other open questions of the SM. In contrast, the new flavour
dynamics from the theory of flavour, that would break explicitly $U(3)^{5}$,
would manifest at much higher scales.

\subsection{\texorpdfstring{$U(2)^{5}$}{U(2)5} and a multi-scale origin of flavour} \label{subsec:U(2)5}

Although the MFV hypothesis is successful in suppressing flavour-violating
NP effects, it also predicts large flavour-universal NP effects. Notice
that the Higgs boson has its largest fermion coupling with the top-quark,
and NP addressing the Higgs hierarchy puzzle are usually coupled
to both. If we would introduce TeV-scale NP coupled to the top-quark
and the Higgs boson, the MFV prescription would also impose couplings
to the light quark generations. However, TeV-scale NP coupled to
light quark generations are strongly constrained by direct searches
at the LHC, see e.g.~the direct reach band in Fig.~\ref{fig:Reach-in-NP_flavour}
or the CMS summary plots in Ref.~\cite{CMS:EXO}. This usually pushes the scale of
MFV NP above 10 TeV, worsening the fine-tuning of the Higgs mass.

In contrast with the MFV prescription, one may consider NP that
dominantly couple to the third family. This is interesting from the
point of view of a theory of flavour: just like the third family Yukawa
couplings are the largest, flavour dynamics connected to the origin of flavour
might be dominantly coupled to the third family. Remarkably,
direct LHC bounds over NP that predominantly couple to the third family
are much weaker than those over flavour-universal NP.

\begin{figure}
\begin{centering}
\includegraphics[scale=0.6]{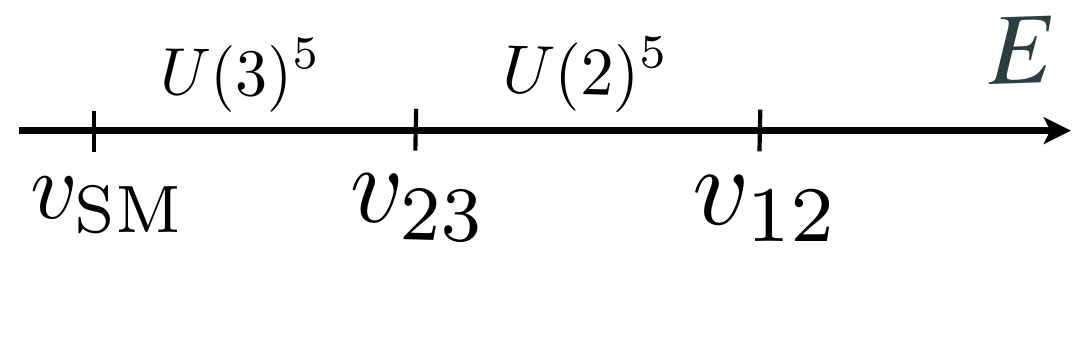}
\par\end{centering}
\caption[Multi-scale picture for a theory of flavour]{Multi-scale picture for a theory of flavour: $v_{23}$ denotes the
low scale where $U(2)^{5}$ is approximately preserved, and the hierarchy
$m_{2}/m_{3}$ is explained. $v_{12}$ denotes the higher scale where
$U(2)^{5}$ is explicitly broken and the hierarchy $m_{1}/m_{2}$
is explained.\label{fig:Multiscale-picture-flavour}}
\end{figure}

We can formally establish this hypothesis by imposing that in the
theory of flavour, only third family Yukawa couplings
are allowed at renormalisable level,
\begin{equation}
\mathcal{L}_{\mathrm{Yukawa}}=y_{t}\overline{Q}_{L3}\tilde{H}u_{R3}+y_{b}\overline{Q}_{L3}Hd_{R3}+y_{\tau}\overline{L}_{L3}He_{R3}\,.
\end{equation}
The Yukawa couplings above break the usual $U(3)^{5}$ flavour symmetry
of the SM down to \cite{Agashe:2005hk,Barbieri:2011ci,Isidori:2012ts,Barbieri:2012uh}
\begin{equation}
U(2)^{5}=U(2)_{Q}\times U(2)_{L}\times U(2)_{u}\times U(2)_{d}\times U(2)_{e}\,.\label{eq:U(2)_5}
\end{equation}
Given that the Yukawa couplings of first and second family fermions
are very small compared to the third family, the SM with massless
first and second family fermions is a good first order description
of the SM fermion spectrum. In that case, the $U(2)^{5}$ symmetry would be
exactly preserved at the classical level. In reality, the small Yukawa
couplings of first and second family fermions provide small
breaking effects of $U(2)^{5}$ that we can parameterise via the spurion
formalism. In the Yukawa matrices of the quark sector, we introduce
a spurion transforming as $V_{Q}\sim(\mathbf{2},\mathbf{1},\mathbf{1})$
under $U(2)_{Q}\times U(2)_{u}\times U(2)_{d}$. Similarly, we introduce
$\Delta Y_{u}\sim(\mathbf{2},\mathbf{\overline{2}},\mathbf{1})$ and
$\Delta Y_{d}\sim(\mathbf{2},\mathbf{1},\mathbf{\overline{2}})$,
obtaining
\begin{equation}
\mathcal{L}_{\mathrm{Yukawa}}^{q}=\left(\begin{array}{cc}
\Delta Y_{u} & V_{Q}\\
0 & y_{t}
\end{array}\right)\overline{Q}_{L}\tilde{H}u_{R}+\left(\begin{array}{cc}
\Delta Y_{d} & V_{Q}\\
0 & y_{b}
\end{array}\right)\overline{Q}_{L}Hd_{R}+\mathrm{h.c.}\label{eq:U(2)_5_Yukawas}
\end{equation}
The spurion $V_{Q}$ parameterises the small breaking of $U(2)^{5}$ that
provides $V_{cb}$ and $V_{ub}$. The $U(2)^{5}$ hypothesis cannot
distinguish between them, but this is not a bad
approximation given that $V_{cb}\approx\lambda^{2}$ and $V_{ub}\approx\lambda^{3}$.
In a similar manner, the $U(2)^{5}$ hypothesis cannot distinguish
between first and second family masses: further dynamics in the UV
must provide the splitting of the spurions $\Delta Y_{u}$, $\Delta Y_{d}$
and $V_{Q}$ into first and second family contributions. This is an
interesting idea: light NP at the TeV-scale may approximately preserve
$U(2)^{5}$ and explain the smallness of $V_{cb}$ and $V_{ub}$,
along with the mass hierarchies between light families and the
third family, i.e.~$m_{2}/m_{3}$, while $m_{1}/m_{2}$, the difference
(and alignment) between $V_{cb}$ and $V_{ub}$ and the Cabibbo angle
are explained in the UV via heavier dynamics that explicitly break $U(2)^{5}$. This naturally leads to the
idea of a multi-scale picture behind the origin of flavour: the theory of flavour could
be a non-universal theory broken in (at least) two hierarchical steps
down to the SM \cite{Craig:2011yk}, as in Fig.~\ref{fig:Multiscale-picture-flavour}.
The approximate $U(2)^{5}$ flavour symmetry may arise accidentally if the theory
of flavour is based on a non-universal gauge group \cite{Craig:2011yk,Panico:2016ull,Bordone:2017bld,Allwicher:2020esa,Fuentes-Martin:2020pww,Barbieri:2021wrc,Fuentes-Martin:2022xnb,Davighi:2022bqf,Davighi:2022fer,Davighi:2023iks,FernandezNavarro:2023rhv,Davighi:2023evx,FernandezNavarro:2023hrf,Chiang:2009kb,Davighi:2023xqn,Capdevila:2024gki,Fuentes-Martin:2024fpx}.

However, we notice that the $U(2)^{5}$ hypothesis does not give
any information about the flavour hierarchies between different charged
sectors, e.g.~$m_{b}/m_{t}$, which are left to be reproduced via
small dimensionless coefficients. This is unsatisfactory
given that some hierarchies like $m_{b}/m_{t}$ are significant,
suggesting the need to incorporate a dynamical mechanism to generate
these hierarchies in theories of flavour based on $U(2)^{5}$. In a similar
manner, the $U(2)^{5}$ hypothesis does not give any predictions about
the alignment of the CKM matrix.

From a phenomenological point of view, the $U(2)^{5}$ hypothesis
dictates that NP may couple non-universally to the third family but universally to the
lighter generations. In particular, couplings to light generations
can be absent in the interaction basis, only generated via fermion
mixing. This latter case provides an efficient suppression of the production
of NP degrees of freedom at colliders, relaxing bounds from LHC, while
large NP in the third family (that could address the Higgs hierarchy
puzzle) are allowed. If NP couplings to light generations are present,
$U(2)^{5}$ protects from the appearance of the most dangerous FCNCs at tree-level, since they involve flavour
transitions between the first and second families\footnote{In the complete theory of flavour, heavier NP would eventually break
explicitly $U(2)^{5}$ to explain the origin of the flavour structure that involves
first and second family fermions. Such NP may potentially mediate
the most dangerous 1-2 FCNCs, however they would be heavier in a multi-scale
picture of flavour, allowing to pass the stringent bounds from flavour
observables.}. Flavour transitions between the (left-handed) third family and light families
are allowed, but they carry the CKM suppression of $V_{cb}$ and $V_{ub}$.
Flavour transitions between right-handed fermions, if mediated by
NP approximately preserving $U(2)^{5}$, are naturally suppressed
with respect to left-handed flavour transitions. This is a very desirable
feature, given the tight bounds over FCNCs involving right-handed fermions (see Section~\ref{subsec:BsMixing}).

Finally, we remark that in a realistic
framework, the spurions in Eq.~(\ref{eq:U(2)_5_Yukawas}) would be
promoted to physical scalars that provide non-renormalisable operators
of the form 
\begin{equation}
\mathcal{L}_{\mathrm{Yukawa}}^{q,d=5}=c_{i3}^{u}\frac{V_{Q}}{\Lambda_{Q}}\overline{Q}_{Li}\tilde{H}u_{R3}+c_{ij}^{u}\frac{\Delta Y_{u}}{\Lambda_{u}}\overline{Q}_{Li}\tilde{H}u_{Rj}+c_{i3}^{d}\frac{V_{Q}}{\Lambda_{Q}}\overline{Q}_{Li}Hd_{R3}+c_{ij}^{d}\frac{\Delta Y_{d}}{\Lambda_{d}}\overline{Q}_{Li}Hd_{Rj}+\mathrm{h.c.}
\end{equation}
such that 
\begin{equation}
{\displaystyle {\displaystyle \mathcal{L}_{\mathrm{Yukawa}}^{q}={\displaystyle \left({\displaystyle \begin{array}{cc}
{\displaystyle c_{ij}^{u}\frac{\Delta Y_{u}}{\Lambda_{u}}} & {\displaystyle c_{i3}^{u}\frac{V_{Q}}{\Lambda_{Q}}}\\
{\displaystyle 0} & {\displaystyle y_{t}}
\end{array}}\right)}\overline{Q}_{L}\tilde{H}u_{R}+{\displaystyle \left({\displaystyle \begin{array}{cc}
{\displaystyle c_{ij}^{d}\frac{\Delta Y_{d}}{\Lambda_{d}}} & {\displaystyle c_{i3}^{d}\frac{V_{Q}}{\Lambda_{Q}}}\\
{\displaystyle 0} & {\displaystyle y_{b}}
\end{array}}\right)}\overline{Q}_{L}Hd_{R}}+\mathrm{h.c.}}
\end{equation}
where $i,j=1,2$. If the physical scalars $V_{Q}$, $\Delta Y_{u}$ and $\Delta Y_{d}$ develop a VEV spontaneously (minimally) breaking the global $U(2)^{5}$ symmetry, then the effective Yukawa couplings involving light families
are naturally suppressed by the heavy scales $\Lambda_{Q,u,d}$.
The coefficients $c_{23}^{u,d}$, $c_{22}^{u,d}$ and $c_{12}^{u,d}$
could naturally be $\mathcal{O}(1)$, while
$y_{b}$ and $c_{11}^{u,d}$ need to be small in order to reproduce
the remaining flavour hierarchies that $U(2)^{5}$ cannot explain. 

A similar formalism can be applied to the charged lepton sector. However,
the $U(2)^{5}$ hypothesis, if extended to the neutrino sector, would
naively predict a third family neutrino much heavier than the others,
with small mixing. This is at odds with neutrino oscillation data
that suggests large and seemingly anarchic neutrino mixing, therefore requiring
to introduce an extra mechanism in the theory of flavour in order to
account for a proper description of neutrino masses and mixing.

\section{Towards a theory of flavour: from the Planck scale to the electroweak
scale} \label{subsec:FromPlanckToEW}

The complicated flavour sector of the SM leaves several questions unanswered:
\begin{itemize}
\item \textbf{Why three families of fermions, transforming as identical copies
under the SM gauge group?} One may argue that we need at least three quark families to have $CP$-violating
phases in the CKM matrix \cite{Kobayashi:1973fv}, and we need less than nine quark families
to preserve the asymptotic freedom of QCD \cite{Gross:1973id}. But this is equivalent to arguing that
all experimental data suggests the existence of three families of
fermions, see e.g.~the invisible decay width of the $Z$ boson \cite{PDG:2022ynf}. It
is clear that these \textit{a posteriori} explanations are unsatisfactory
as they do not provide any fundamental principle to understand why
Nature has chosen the number three.
\item \textbf{Why the three identical families of fermions interact so differently
with the Higgs, leading to a hierarchical pattern of charged fermion
masses and CKM mixing?}
\item \textbf{What is the origin of the very tiny neutrino masses and lepton
mixing?}
\item \textbf{Why quark mixing and lepton mixing are so different, namely
why the CKM matrix is almost diagonal while the PMNS matrix is seemingly
anarchic?}
\end{itemize}
We highlight that the gauge couplings of the SM in Eq.~(\ref{eq:gauge_couplings})
are not far from $\mathcal{O}(1)$. Even the tree-level mass of the Higgs boson is $\mathcal{O}(v_{\mathrm{SM}})$,
such that the quartic coupling $\lambda$ of the scalar potential (remember
$m_{h}=\sqrt{\text{\ensuremath{2\lambda}}}v_{\mathrm{SM}}$) is not
much smaller than $\mathcal{O}(1)$. This is what one would expect
from a fundamental theory based on \textit{naturalness} arguments:
the free parameters take arbitrary values of the same order of magnitude,
and any hierarchy or cancellation is explained in terms of dynamical
mechanisms. There is no apparent reason for the parameters of a fundamental
theory to greatly differ from each other, since they all appear in
a similar way as free parameters of the renormalisable Lagrangian.
Nevertheless, the flavour sector apparently is not guided by naturalness
principles: rather than being of the same order, both the charged
fermion masses and the CKM mixing angles follow hierarchical patterns.
The high number of free parameters in the flavour sector, maybe too
many for a fundamental theory of Nature, along with their particular
hierarchical patterns, may be hinting at the existence of new physics
that provide a dynamical explanation for the flavour structure of the
SM: such a theory describing the complicated flavour sector of the
SM in terms of simple and natural principles is called a \textit{theory
of flavour}.

The lack of understanding of the flavour sector of the SM has classically
being denoted as the \textit{flavour puzzle}. We stress here that
this is not a \textit{problem} of the SM, which works perfectly well
with the input of the flavour parameters, but rather a puzzle of Nature
for us to identify the dynamical mechanism behind these parameters in terms of
our mathematical models. After the discovery of neutrino masses, the
flavour sector is enlarged with extra flavour parameters accounting
for neutrino masses and mixing, which now do become a problem of the
SM: the inability of the SM to account for the observed physical phenomenon
of neutrino oscillations, making the flavour puzzle difficult
to ignore.

This thesis is devoted to the flavour puzzle: the development and
study of new models to understand the origin of flavour in the SM,
along with their phenomenology and discovery prospects. Flavour physics
phenomenology will also play a central role, as it would not be unreasonable
that a theory of flavour beyond the SM leads to new flavour specific
interactions connected to the dynamical mechanisms behind the origin
of flavour.

Given that most of the Yukawa couplings are much smaller than $\mathcal{O}(1)$,
a good starting point to model them in a more natural way would be
to assume that Yukawa couplings in the SM are effective remnants of
a UV theory. In this manner, they would carry the natural suppression
of non-renormalisable operators. For example, let us consider that 
a symmetry beyond the SM forbids the SM Yukawa couplings at renormalisable
level\footnote{We note that the flavour puzzle can be addressed without invoking
new symmetries, e.g.~in extra dimensional frameworks (see e.g.~\cite{Arkani-Hamed:1999ylh,Dvali:2000ha}),
however we will not consider this approach in this thesis. For a review
about different approaches to the flavour puzzle, see e.g.~\cite{Feruglio:2015jfa}.}.
One example would be the global $U(1)_{\mathrm{FN}}$ flavour
symmetry\footnote{Note that spontaneously broken discrete flavour symmetries are also possible, for a review of mass matrices in such a framework see \cite{Leurer:1992wg,Leurer:1993gy}.} of Froggatt-Nielsen (FN) models \cite{Froggatt:1978nt},
which is spontaneously broken by the VEV of a SM scalar singlet $\phi_{F}$,
that we denote as the \textit{flavon} field. For the moment, we shall remain
agnostic to the particular UV theory, but we will assume that the
flavon field $\phi_{F}$ transforms in the appropriate way under the
BSM symmetry, allowing to write dimension-5 operators as
\begin{equation}
\mathcal{L}_{\mathrm{Yukawa}}^{d=5}=\frac{c}{\Lambda_{F}}\phi_{F}\overline{\Psi}_{L}\overset{(\sim)}{H}\psi_{R}+\mathrm{h.c.}\,,\label{eq:Eff_Yukawa}
\end{equation}
where $\Psi_{L}$ denotes a generic $SU(2)_{L}$ fermion doublet of
the SM, $\psi_{R}$ denotes the accompanying $SU(2)_{L}$ fermion
singlet, and $H$ couples to down-quarks and charged leptons while $\widetilde{H}$ couples
to up-quarks. The heavy scale $\Lambda_{F}$ is the cut-off of the
EFT, associated to new degrees of freedom present in the UV theory.
When the flavon field $\phi_{F}$ develops a VEV spontaneously breaking the BSM
symmetry, the operator in Eq.~(\ref{eq:Eff_Yukawa}) provides an
\textit{effective Yukawa coupling }as
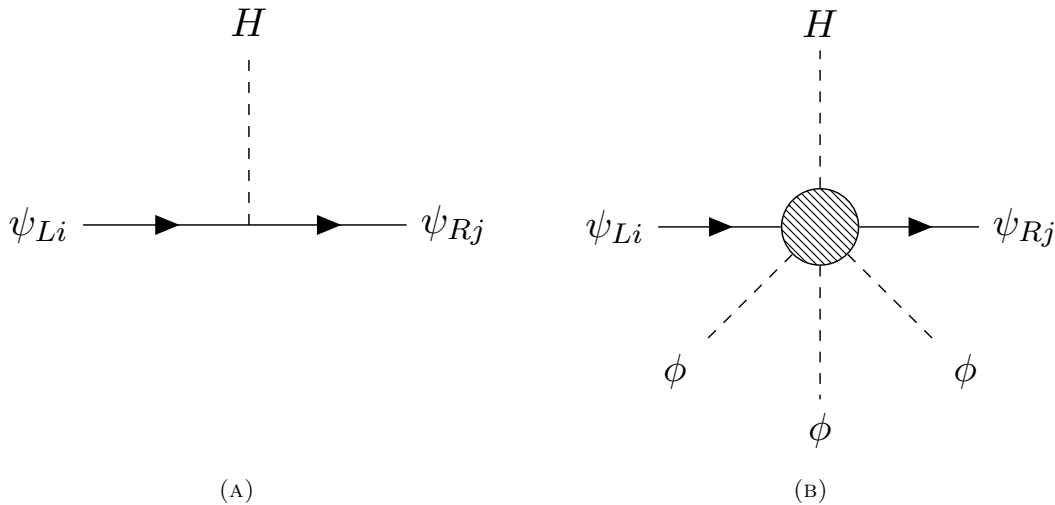
\begin{figure}[t]
\subfloat[]{\begin{centering}
\resizebox{0.284\textheight}{!}{ \begin{tikzpicture}
	\begin{feynman}
		\vertex (a) {\(\psi_{Li}\)};
		\vertex [right=20mm of a] (b);
		\vertex [right=of b] (c) {\(\psi_{Rj}\)};
		\vertex [above=16.8mm of b] (f1) {\(H\)};
        \vertex [below left=20mm of b] (g1) {};
		\vertex [below=20mm of b] (g2) {};
		\vertex [below right=20mm of b] (g3) {};
		\diagram* {
			(a) -- [fermion] (b) -- [fermion] (c),
			(b) -- [scalar] (f1),
	};
	\end{feynman}
\end{tikzpicture}} 
\par\end{centering}
}$\quad$\subfloat[]{\begin{centering}
\resizebox{0.28\textheight}{!}{ \begin{tikzpicture}
	\begin{feynman}
		\vertex (a) {\(\psi_{Li}\)};
		\vertex [right=20mm of a, blob] (b) {};
		\vertex [right=20mm of b] (c) {\(\psi_{Rj}\)};
		\vertex [above=20mm of b] (f1) {\(H\)};
		\vertex [below left=20mm of b] (g1) {\(\phi\)};
		\vertex [below=20mm of b] (g2) {\(\phi\)};
		\vertex [below right=20mm of b] (g3) {\(\phi\)};
		\diagram* {
			(a) -- [fermion] (b),
            (b) -- [fermion] (c),
			(b) -- [scalar] (f1),
			(b) -- [scalar] (g1),
			(b) -- [scalar] (g2),
			(b) -- [scalar] (g3),
	};
	\end{feynman}
\end{tikzpicture}} 
\par\end{centering}
}\caption[Effective Yukawa couplings]{ \textbf{\textit{Left:}} Renormalisable Yukawa couplings in the SM.
\textbf{\textit{Right:}} Effective Yukawa couplings arising from non-renormalisable
operators containing insertions of flavon fields $\phi$. \label{fig:Eff_Yukawas}}
\end{figure}
\begin{equation}
\mathcal{L}_{\mathrm{Yukawa}}=c\frac{\left\langle \phi_{F}\right\rangle }{\Lambda_{F}}\overline{\Psi}_{L}\overset{(\sim)}{H}\psi_{R}+\mathrm{h.c.}\label{eq:Effective_Yukawa}
\end{equation}
In contrast with the SM, now the Yukawa couplings are obtained from
non-renormalisable operators built with the insertion of the flavon
field, as in Fig.~\ref{fig:Eff_Yukawas}. Assuming $\left\langle \phi_{F}\right\rangle \ll\Lambda_{F}$,
the effective Yukawa coupling in Eq.~(\ref{eq:Effective_Yukawa}) is
naturally suppressed from unity, with the dimensionless coefficient
$c$ being naturally of $\mathcal{O}(1)$. At the level of our EFT
framework, we have the freedom to assign a numerical value to the
ratio $\left\langle \phi_{F}\right\rangle/\Lambda_{F}$.
A convenient choice is 
\begin{equation}
\frac{\left\langle \phi_{F}\right\rangle }{\Lambda_{F}}\simeq\lambda\,,\label{eq:VEV_Lambda_Assignments}
\end{equation}
where $\lambda\simeq0.225$ is the Wolfenstein parameter already introduced
in Section~\ref{subsec:Parameters-of-the-SM}. In the Froggatt-Nielsen
framework, all SM fermions and $\phi_{F}$ may carry charge assignments
under the global $U(1)_{\mathrm{FN}}$ symmetry. By convention, we 
assign -1 FN charge to $\phi_{F}$ without loss of generality. In this manner, the effective
Yukawa couplings of SM fermions may be given by
\begin{flalign}
\mathcal{L}_{\mathrm{Yukawa}} & =c_{ij}^{u}\left(\frac{\left\langle \phi_{F}\right\rangle }{\Lambda_{F}}\right)^{\left|q_{\mathrm{FN}}(\overline{Q}_{Li})+q_{\mathrm{FN}}(u_{Rj})+q_{\mathrm{FN}}(\tilde{H})\right|}\overline{Q}_{Li}\widetilde{H}u_{Rj}\label{eq:Effective_Yukawa-1}\\
 & +c_{ij}^{d}\left(\frac{\left\langle \phi_{F}\right\rangle }{\Lambda_{F}}\right)^{\left|q_{\mathrm{FN}}(\overline{Q}_{Li})+q_{\mathrm{FN}}(d_{Rj})+q_{\mathrm{FN}}(H)\right|}\overline{Q}_{Li}Hd_{Rj}\\
 & +c_{ij}^{e}\left(\frac{\left\langle \phi_{F}\right\rangle }{\Lambda_{F}}\right)^{\left|q_{\mathrm{FN}}(\overline{L}_{Li})+q_{\mathrm{FN}}(e_{Rj})+q_{\mathrm{FN}}(H)\right|}\overline{L}_{Li}He_{Rj}+\mathrm{h.c.}\,,
\end{flalign}
where $i,j=1,2,3$. With the assignment of Eq.~(\ref{eq:VEV_Lambda_Assignments}),
each entry in the effective Yukawa matrices carries an individual
suppression via powers of the Wolfenstein parameter, $\lambda^{\alpha}$,
where $\alpha$ is determined in terms of the FN charge assignments
of SM fermions and the Higgs doublet. As a concrete example, we take
the following set of charges:
\begin{flalign}
 & q_{\mathrm{FN}}(\overline{Q}_{Li})=(3,2,0)\,, &  & q_{\mathrm{FN}}(u_{Ri})=(4,1,0)\,, &  & \,\\
 & q_{\mathrm{FN}}(d_{Ri})=(4,3,2)\,, &  & q_{\mathrm{FN}}(\overline{L}_{Li})=(5,3,0)\,, &  & \,\\
 & q_{\mathrm{FN}}(e_{Ri})=(3,2,2)\,, &  & q_{\mathrm{FN}}(H)=0\,. &  & \,
\end{flalign}
This choice leads to the following parametric suppression of the
Yukawa couplings:
\begin{flalign}
\mathcal{L}_{\mathrm{Yukawa}} & =\begin{pmatrix}\overline{Q}_{L1} & \overline{Q}_{L2} & \overline{Q}_{L3}\end{pmatrix}\left(\begin{array}{ccc}
\lambda^{7} & \lambda^{4} & \lambda^{3}\\
\lambda^{6} & \lambda^{3} & \lambda^{2}\\
\lambda^{4} & \lambda & 1
\end{array}\right)\begin{pmatrix}u_{R1}\\
u_{R2}\\
u_{R3}
\end{pmatrix}\widetilde{H} \label{eq:FNtextureUp}\\
 & +\begin{pmatrix}\overline{Q}_{L1} & \overline{Q}_{L2} & \overline{Q}_{L3}\end{pmatrix}\left(\begin{array}{ccc}
\lambda^{7} & \lambda^{6} & \lambda^{5}\\
\lambda^{6} & \lambda^{5} & \lambda^{4}\\
\lambda^{4} & \lambda^{3} & \lambda^{2}
\end{array}\right)\begin{pmatrix}d_{R1}\\
d_{R2}\\
d_{R3}
\end{pmatrix}H\\
 & +\begin{pmatrix}\overline{L}_{L1} & \overline{L}_{L2} & \overline{L}_{L3}\end{pmatrix}\left(\begin{array}{ccc}
\lambda^{8} & \lambda^{7} & \lambda^{7}\\
\lambda^{6} & \lambda^{5} & \lambda^{5}\\
\lambda^{3} & \lambda^{2} & \lambda^{2}
\end{array}\right)\begin{pmatrix}e_{R1}\\
e_{R2}\\
e_{R3}
\end{pmatrix}H+\mathrm{h.c.}\,, \label{eq:FNtextureLeptons}
\end{flalign}
where we have omitted the dimensionless coefficients $c_{ij}^{u,d,e}$,
naturally expected to be of $\mathcal{O}(1)$. The Yukawa matrices
above are hierarchical and approximately diagonal (the off-diagonal
entries are small), therefore, in good approximation, the physical
Yukawa couplings obtained after the diagonalisation will scale
with powers of $\lambda$ as the diagonal entries above. In this manner,
one can check that our simplified FN model performs a good description
of charged fermion masses.

Fermion mixing is approximately given by the ratios of
off-diagonal over diagonal entries, upper off-diagonal entries for left-handed mixing
and lower off-diagonal entries for right-handed mixing. In this manner, we obtain $V_{cb}\sim\lambda^{2}$,
$V_{ub}\sim\lambda^{3}$ and $V_{us}\approx\lambda^{4}/\lambda^{3}\sim\lambda$
for the CKM mixing. Remarkably, our toy FN model predicts non-vanishing
charged lepton mixing, although too small to account for PMNS mixing,
that must therefore come dominantly from the neutrino sector. The
model also predicts significant right-handed fermion mixing, such as
$s_{23}^{u_{R}}\sim\lambda$. In this manner, the Yukawa matrices in Eqs.~(\ref{eq:FNtextureUp}-\ref{eq:FNtextureLeptons}) are an example of Yukawa \textit{textures} broadly compatible with current data. Other examples include texture zeros,
where one or more of the entries in the effective mass matrices may be filled with zeros. For a systematic study of possible textures 
for the charged fermion and neutrino mass matrices, we refer the reader to Refs.~\cite{Gupta:2012fsl,Ludl:2014axa,Gupta:2015iku}.

We notice that the description of the flavour sector does not point
to any particular scale for $\left\langle \phi_{F}\right\rangle $
and $\Lambda_{F}$: the explanation of the SM flavour structure is
successful as long as the ratio $\left\langle \phi_{F}\right\rangle /\Lambda_{F}$
is held fixed, but the independent scales of flavour $\left\langle \phi_{F}\right\rangle $
and $\Lambda_{F}$ may be anywhere \textit{from the Planck scale to
the electroweak scale}.

This is not just a feature of the FN mechanism, but common to most
theories of flavour based on BSM symmetries where Yukawa couplings are explained via effective
operators. In particular, this applies to all the theories of flavour
that will be explored in this thesis.

In the FN setup, the tree-level exchange of a radial mode of $\phi_{F}$
provides NP contributions to meson mixing observables that are compatible
with current data only if the mass of $\phi_{F}$ is above $\sim10^{5}\;\mathrm{TeV}$
\cite{Altmannshofer:2022aml,Cornella:2023zme} (note that this is the approximate bound
that recent studies of successful FN models reveal, despite the obvious dependence on the charge assignments of fermions
under the FN symmetry), thus setting the NP
scales of the model far above our current reach for direct detection.
Going beyond the FN setup, one may find different
arguments to fix the \textit{a priori} undetermined flavour scales:
\begin{itemize}
\item Motivated by gauge coupling unification, one may suggest that the
very heavy scale where the gauge sector gets simplified and described
by a single gauge coupling is also the scale where the new dynamics
that explain and simplify the flavour sector become manifest \cite{Barbieri:1996ww,King:2003rf,King:2005bj,Dimou:2015yng,Belyaev:2018vkl,Bernigaud:2018qky,Ekstedt:2020gaj,Fonseca:2020tgx,Fu:2022lrn}.
This hypothesis is also supported by the heavy scale for the origin
of neutrino masses suggested by the seesaw mechanism, and in good
agreement with the prescription of Minimal Flavour Violation \cite{Dimou:2015yng}.
In particular, in recent years it has been noted the possibility that
modular forms, motivated by string theory, could play
an important role to explain the flavour sector \cite{Altarelli:2005yx,deAdelhartToorop:2011re,Feruglio:2017spp,deMedeirosVarzielas:2019cyj,King:2020qaj,Kobayashi:2023zzc}.
Modular symmetries can be incorporated to GUTs in order to build elegant
theories of flavour at very high energy scales \cite{deAnda:2018ecu,Ding:2021zbg,King:2021fhl,Chen:2021zty,Ding:2021eva}.
\item New flavour dynamics addressing the flavour puzzle may leave its
imprints in flavour observables, which are sensitive to scales far
above the TeV. In this direction, experimental anomalies in observables
that suggest new flavour-specific interactions \cite{Muong-2:2021ojo,Muong-2:2023cdq},
or the breaking of lepton flavour universality \cite{BaBar:2012obs},
could be indirect signals of a theory of flavour. This suggests that the
NP flavour scales may be closer to the electroweak scale, within
the range for detection in current experiments. In particular, TeV-scale
leptoquarks or $Z'$ bosons would mediate new interactions to
explain the flavour anomalies, with the flavour structure of their
couplings to fermions dictated by the theory of flavour and connected to the origin of Yukawa couplings in the SM \cite{deMedeirosVarzielas:2015yxm,Hiller:2017bzc,King:2017anf,King:2018fcg,deMedeirosVarzielas:2018bcy,Grinstein:2018fgb,deMedeirosVarzielas:2019lgb,DeMedeirosVarzielas:2019nob,King:2020mau,FernandezNavarro:2021sfb,King:2021jeo,FernandezNavarro:2022gst}.
\item Another interesting possibility is that the theory of flavour consists
of multiple NP scales, that may cover a wide range of
energy \textit{from the Planck scale to the electroweak scale}.
This is realised in multi-scale theories of flavour \cite{Craig:2011yk,Panico:2016ull,Bordone:2017bld,King:2018fcg,Allwicher:2020esa,Fuentes-Martin:2020pww,Barbieri:2021wrc,King:2021jeo,Fuentes-Martin:2022xnb,FernandezNavarro:2022gst,Davighi:2022fer,Davighi:2022bqf,Davighi:2023iks,FernandezNavarro:2023rhv,Davighi:2023evx,FernandezNavarro:2023hrf,Chiang:2009kb,Davighi:2023xqn,Capdevila:2024gki,Fuentes-Martin:2024fpx},
where a first layer of NP explains the flavour hierarchies $m_{2}/m_{3}$
and the smallness of CKM mixing, while a second layer explains the
flavour hierarchy $m_{1}/m_{2}$ and the Cabibbo angle. The lower
layer of NP also offers the opportunity to connect the theory with
the flavour anomalies \cite{Bordone:2017bld,Bordone:2018nbg,King:2021jeo,Fuentes-Martin:2022xnb,FernandezNavarro:2022gst,Davighi:2022bqf},
while the higher layer may provide a gauge unified framework \cite{FernandezNavarro:2023hrf}.
The explanation of neutrino masses and PMNS mixing could be incorporated
at low energies via a low scale seesaw mechanism \cite{FernandezNavarro:2023rhv,FernandezNavarro:2023hrf,Fuentes-Martin:2020pww},
or at very high energies as a new step in the multi-scale picture \cite{King:2021jeo}.
Some examples of multi-scale theories of flavour predict an approximate
$U(2)^{5}$ flavour symmetry \cite{Craig:2011yk,Panico:2016ull,Bordone:2017bld,Allwicher:2020esa,Fuentes-Martin:2020pww,Barbieri:2021wrc,Fuentes-Martin:2022xnb,Davighi:2022fer,Davighi:2022bqf,Davighi:2023iks,FernandezNavarro:2023rhv,Davighi:2023evx,FernandezNavarro:2023hrf,Chiang:2009kb,Davighi:2023xqn,Capdevila:2024gki},
but their exist other alternatives such as \cite{King:2021jeo,FernandezNavarro:2022gst,King:2018fcg}
based on the idea of messenger dominance \cite{Ferretti:2006df}.
Remarkably, the various steps of symmetry breaking may offer the
opportunity to connect the theory with other subjects
like quark-lepton unification, the origin of matter-antimatter asymmetry \cite{Shu:2006mm} or the unification of electroweak
and flavour symmetry \cite{Davighi:2022fer,Davighi:2022bqf},
and might be tested via cosmological observations of the different
phase transitions in the early Universe associated to the several
steps of symmetry breaking \cite{Greljo:2019xan}.
\end{itemize}
In this thesis, we propose and explore theories of flavour of the
last kind, which may be connected to other open problems of the SM via the different layers of the multi-scale picture. Moreover, these theories enjoy a rich phenomenology and have
the potential to be discovered in the current or next generation of
particle physics experiments. In Chapter~\ref{Chapter:Fermiophobic} we discuss a class of fermiophobic $U(1)'$ extensions of the SM, where the flavour structure of the SM is explained via the mechanism of messenger dominance \cite{Ferretti:2006df}, and we seek for an enhancement of the anomalous magnetic moment of the muon. In Chapter~\ref{Chapter:TwinPS}, we explore a twin Pati-Salam theory of flavour also based on messenger dominance, where the origin of flavour hierarchies is connected to the effective couplings of a TeV-scale vector leptoquark $U_{1}\sim(\mathbf{3,1},2/3)$ that explains the so-called $B$-physics anomalies. In Chapter~\ref{Chapter:Tri-hypercharge} we propose a gauge non-universal embedding of the SM in which a separate weak hypercharge is assigned to each fermion family. If the Higgs doublet(s) only carries
third family hypercharge, then the third family is naturally heavier and flavour hierarchies arise naturally after the spontaneous breaking of the
tri-hypercharge group. Finally, in Chapter~\ref{Chapter:Tri-unification} we show how gauge non-universal frameworks like the tri-hypercharge theory, among others, may emerge from a gauge unified group containing one separate $SU(5)$ for each family, where the three $SU(5)$ groups are related by a cyclic permutation symmetry that ensures a single gauge coupling at the GUT scale and the unification of all SM fermions into a single representation.

\chapter{Testing a theory of flavour: EFT formalism and flavour observables} \label{chap:2}

\begin{quote}
  ``Soon I knew the craft of experimental physics was \\ beyond me - it was the sublime quality of patience - \\ patience in accumulating data, patience with \\ recalcitrant equipment - which I sadly lacked.''
  \begin{flushright}
  \hfill \hfill $-$ Abdus Salam
  \par\end{flushright}
\end{quote}

\noindent In order to test a new physics model, such as a theory of flavour,
it is usually convenient to integrate out the new heavy degrees of
freedom to obtain the low energy Effective Field Theory (EFT) of
the model. In this context, we will introduce the Standard Model Effective
Field Theory (SMEFT) that extends the SM via non-renormalisable operators
which capture the NP effects originated by heavy physics above the electroweak scale. We will also
introduce the Low Energy Effective Field Theory (LEFT) that contains the
effective Lagrangian below the electroweak scale, and is useful to
study NP contributions to low-energy observables. Finally, we will
introduce and discuss particular flavour observables that suggest
the presence of NP, and are well motivated from the point of view
of a theory of flavour. 

\section{The Standard Model Effective Field Theory}

The SMEFT is the effective field theory that contains the SM Lagrangian
(\ref{eq:SM_Lagrangian}) plus all possible higher dimensional operators
invariant under the SM gauge symmetry (\ref{eq:SM_symmetry}). In
this thesis, we will only consider operators in the SMEFT up to dimension
six. In fact, a few SMEFT operators have already been introduced in
Chapter~\ref{chap:Chapter1}. One example is the dimension five Weinberg
operator (\ref{eq:Weinberg}) that violates lepton number explicitly
in two units, and provides Majorana masses for active neutrinos. Remarkably,
the Weinberg operator is the only dimension five operator that one
can write with SM fields. Another example is the dimension six operator
in Eq.~(\ref{eq:B_violating}), which breaks both lepton number and
baryon number in one unit and can mediate the decay of nucleons.

We introduce the SMEFT Lagrangian as 
\begin{equation}
\mathcal{L}_{\mathrm{SMEFT}}^{d\leq6}=\mathcal{L}^{d\leq4}_{\mathrm{SM}}+\mathcal{L}^{d=5}_{\mathrm{Weinberg}}-\sum_{i}\frac{1}{\Lambda_{i}^{2}}C_{i}(\mu)Q_{i}\,,\label{eq:SMEFT_Lagrangian}
\end{equation}
Although in Eq.~(\ref{eq:SMEFT_Lagrangian}) we have included one
high cut-off scale for every operator, in principle each operator
can contain different contributions from different UV models associated
to different NP scales. Notice also that the SMEFT Wilson coefficients
$C_{i}(\mu)$ depend on the energy scale $\mu$, and hence experience
renormalisation group evolution (RGE) effects. We will take the latter
into account by using dedicated software such as \texttt{DsixTools 2.1} \cite{Fuentes-Martin:2020zaz}.

In Appendix~\ref{app:SMEFT_Operators}, Table~\ref{tab:smeft6ops}
\cite{Grzadkowski:2010es}, we list all dimension six SMEFT operators
conserving baryon and lepton number, while the dimension six operators
that violate baryon and lepton number are listed in Table~\ref{tab:smeft6baryonops}.
We highlight Higgs-bifermion operators (class-3) and purely Higgs
operators (class-7), which are useful to study contributions to electroweak
precision observables, while the baryon number violating operators
in Table~\ref{tab:smeft6baryonops} are useful to study nucleon decay.
Finally, we also highlight the set of baryon number conserving
four-fermion operators (class-8), which are relevant for the study
of flavour observables.

When referring to effective operators through this chapter, we denote lepton flavour
indices as $\alpha=e,\mu,\tau$, and quark flavour indices as $i=1,2,3$, in such a way that
greek indices denote lepton flavours while latin indices denote quark flavours.

Finally, we comment that a more general EFT than the SMEFT do exists, the so-called 
Higgs Effective Field Theory (HEFT). SMEFT contains one Higgs doublet field as prescripted in the SM,
however scenarios where the observed Higgs boson does not belong to an elementary
exact $SU(2)_{L}$ doublet are still allowed within the current
experimental accuracy. Those may be described by the HEFT where the Higgs boson is treated as a gauge singlet
and the Goldstone bosons are treated separately. However, in this thesis all the UV models proposed contain
at least one exact Higgs doublet performing EW symmetry breaking, with the canonical SM being a low scale limit in
all cases, therefore all the NP effects will be well captured by the SMEFT framework.

\section{The Low Energy Effective Field Theory}

The LEFT is the effective field theory that describes low energy scales
$\mu\ll M_{Z}$, at which the electroweak gauge invariance is broken
and the remaining gauge symmetry is $SU(3)_{c}\times U(1)_{Q}$. The
effective Lagrangian containing operators up to dimension six is given
by
\begin{equation}
\mathcal{L}_{\mathrm{LEFT}}^{d\leq6}=\mathcal{L}_{\mathrm{QED}}+\mathcal{L}_{\mathrm{QCD}}-\frac{4G_{F}}{\sqrt{2}}\sum_{i}C_{i}(\mu)\mathcal{O}_{i}\,,\label{eq:LEFT_Lagrangian}
\end{equation}
where $-4G_{F}/\sqrt{2}$ is a conventional normalisation factor that
allows to easily compare the strength of the NP effect with that of
the weak interactions, both generally contributing to the Wilson coefficients
$C_{i}(\mu)$. Throughout this thesis, we will commonly consider $\mu=m_{b}$
as our low-energy scale, since we will consider several observables
related to $B$-meson physics.

In Appendix~\ref{app:LEFT_Operators}, Table~\ref{tab:oplist1}
\cite{Jenkins:2017jig}, we list baryon and lepton number conserving
operators up to dimension six, while in Table~\ref{tab:oplist2}
we list baryon and/or lepton number violating operators up to dimension
six. We highlight the four-fermion operators in Table~\ref{tab:oplist1},
distinguished by their different chiralities, which will play an important
role for the study of flavour observables.

We provide tree-level matching conditions
between the operators in the SMEFT and the LEFT \cite{Jenkins:2017jig} in Appendix~\ref{app:Matching}.
Notice that particular operators in the LEFT do not get
any contributions from the SMEFT at tree-level, therefore they can
only get suppressed contributions (e.g.~via RGE) from dimension
six NP operators preserving $SU(2)_{L}\times U(1)_{Y}$.

\section{Flavour observables}

In the following, we discuss key flavour observables that are in tension
with the SM and might be connected to a possible theory of flavour.
We denote these observables as \textit{anomalies}. We also discuss
several observables that get modified as well in NP scenarios that
explain the anomalies. These observables offer the possibility to
test and discriminate between the different NP explanations.

\subsection{\texorpdfstring{$R_{K^{(*)}}$ and $b\rightarrow s\mu\mu$}{RK(*) and bsmumu}
\label{subsec:bsll}}

Lepton flavour universality (LFU) is a key prediction of the SM: all
lepton flavours experience gauge interactions in the same way, up to corrections
related to the different masses of charged leptons\footnote{These corrections are most relevant for processes involving
$\tau$ charged leptons, which are heavier.}. After the discovery of neutrino oscillations, we know that LFU is
not an exact symmetry of Nature. However, the breaking effects of
LFU via lepton mixing are suppressed by the very tiny neutrino masses,
being generally unobservable with current experimental precision.
In this manner, the observation of LFU breaking in low-energy processes
would be a clear indication of new physics.

Although purely leptonic observables so far show no significant hints of violation
of LFU (see e.g.~LFU in $\tau$ decays in Section~\ref{subsec:Universality-in-tau}),
semileptonic observables are also sensitive to the breaking of LFU.
In contrast with purely leptonic modes, they can be afflicted by substantial
QCD uncertainties, however it is possible to build very clean observables in terms of
ratios of semileptonic processes. In this direction, the $R_{K^{(*)}}$
ratios were proposed
\begin{equation}
R_{K^{(*)}}=\frac{\mathcal{B}(B\rightarrow K^{(*)}\mu^{+}\mu^{-})}{\mathcal{B}(B\rightarrow K^{(*)}e^{+}e^{-})}\,.
\end{equation}
Within the SM, lepton universality predicts $R_{K^{(*)}}=1$ for $q^{2}\,\epsilon \,[1.1,6]\:\mathrm{GeV}^{2}$, where
$q^{2}$ denotes the dilepton invariant-mass squared, up to
corrections of order 1\% \cite{Bordone:2016gaq} due to the different
masses of muons and electrons. Notice that the $b\rightarrow s\mu\mu$
transition is a FCNC in the SM, therefore being generally loop suppressed,
GIM suppressed and CKM suppressed. This strong suppression makes these
processes particularly sensitive to new physics.

It turns out that the experimental measurements of the $R_{K^{(*)}}$
ratios showed deviations from the SM for almost eight years. In particular,
the $R_{K}$ ratio alone reached a 3.1$\sigma$ tension with the
SM in the LHCb update of 2021 \cite{LHCb:2021trn}, and several deviations
sitting at the $2\sigma$ level showed up in $R_{K^{*}}$ \cite{LHCb:2017avl}
and other LFU ratios involving kaons. These measurements suggested
the presence of NP contributions interfering with the SM contribution and
mainly coupled to muons, leading to $R_{K^{(*)}}<1$.

This pattern was supported by other semileptonic observables, including
$\mathcal{B}(B\rightarrow K^{(*)}\mu^{+}\mu^{-})$, $\mathcal{B}(B_{s}\rightarrow\phi\mu^{+}\mu^{-})$
and the angular observable $P'_{5}$, all of them afflicted however
by significant hadronic uncertainties \cite{Gubernari:2022hxn}. Remarkably, the very clean leptonic
decay $\mathcal{B}(B_{s}\rightarrow\mu^{+}\mu^{-})$ was also in good
agreement with the muon deficit observed in $R_{K^{(*)}}$ \cite{LHCb:2021vsc}.
This consistent set of anomalies was easy to accommodate in the context of a theory
of flavour: new dynamics connected to the origin of the Yukawa couplings
$y_{e}\ll y_{\mu}$ might as well couple preferentially to muons. In
this direction, we proposed a simplified phenomenological model based
on a fermiophobic $Z'$ boson \cite{FernandezNavarro:2021sfb} where
effective couplings to SM fermions were obtained via mixing with a
fourth family of vector-like fermions. This simplified model could
not only explain $R_{K^{(*)}}$ but also $(g-2)_{\mu}$ simultaneously,
and was motivated by a theory of flavour with fermiophobic $Z'$ \cite{King:2018fcg}
already proposed in the literature to explain $R_{K^{(*)}}$. Later
on, we considered a complete theory of flavour based on a twin Pati-Salam
gauge group that contains a TeV scale $U_{1}\sim(\mathbf{3},\mathbf{1},2/3)$
vector leptoquark \cite{King:2021jeo}. This theory could potentially
explain $R_{K^{(*)}}$ via $U_{1}$ exchange, along with the $R_{D^{(*)}}$
anomalies which also suggest a consistent breaking of LFU (see Section~\ref{subsec:bctaunu}).
We concluded that such a theory could not explain the anomalies in
its minimal version, but with extra
model building efforts we showed that the theory was able to simultaneously
explain both $R_{K^{(*)}}$ and $R_{D^{(*)}}$ while remaining compatible
with all experimental data \cite{FernandezNavarro:2022gst}. The anomalous measurements of $R_{K^{(*)}}$
that motivated these efforts were \cite{LHCb:2017avl,LHCb:2021trn}
\begin{flalign}
R_{K}^{[1.1,6]} & =\frac{\mathcal{B}\left(B\rightarrow K\mu^{+}\mu^{-}\right)}{\mathcal{B}\left(B\rightarrow Ke^{+}e^{-}\right)}=0.846_{-0.041}^{+0.044}\,,\label{eq:RK_old}\\
R_{K^{*}}^{[1.1,6]} & =\frac{\mathcal{B}\left(B\rightarrow K^{*}\mu^{+}\mu^{-}\right)}{\mathcal{B}\left(B\rightarrow K^{*}e^{+}e^{-}\right)}=0.69_{-0.12}^{+0.16}\,,\nonumber 
\end{flalign}
where $q^{2}\,\epsilon \,[1.1,6]\:\mathrm{GeV}^{2}$ denotes the dilepton
invariant-mass squared. As of 2021, the global average of the theoretically
clean observable $\mathcal{B}(B_{s}\rightarrow\mu^{+}\mu^{-})$ was (see e.g.~\cite{Geng:2021nhg})
\begin{equation}
\mathcal{B}(B_{s}\rightarrow\mu^{+}\mu^{-})=(2.8\pm0.3)\times10^{-9}\,,\label{eq:Bsmumu_old}
\end{equation}
to be compared with the SM prediction $\mathcal{B}(B_{s}\rightarrow\mu^{+}\mu^{-})_{\mathrm{SM}}=(3.67\pm0.15)\times10^{-9}$
\cite{Bobeth:2013uxa}. In order to describe these measurements at
the level of the LEFT, it is convenient to define a new basis of LEFT
operators beyond the San Diego basis \cite{Jenkins:2017jig} that
discriminates operators by the chirality of quarks and by the vector-like
or vector-axial components of muons, leading to the following effective
Lagrangian: 
\begin{equation}
\begin{aligned}\mathcal{L}_{b\rightarrow s\mu\mu}=\frac{4G_{F}}{\sqrt{2}}V_{tb}V_{ts}^{*}\frac{\alpha_{\mathrm{\mathrm{EM}}}}{4\pi} & \left[(C_{9}^{\mathrm{SM}}+C_{9}^{\mu\mu})\mathcal{O}_{9}^{\mu\mu}+(C_{10}^{\mathrm{SM}}+C_{10}^{\mu\mu})\mathcal{O}_{10}^{\mu\mu}+\mathrm{h.c.}\right]\end{aligned}
\,,\label{eq:L_bsmumu}
\end{equation}
where
\begin{align}
\mathcal{O}_{9}^{\mu\mu}= & \left(\bar{s}\gamma_{\mu}P_{L}b\right)\left(\bar{\mu}\gamma^{\mu}\mu\right)\,,\label{eq:O9mumu}\\
\mathcal{O}_{10}^{\mu\mu}= & \left(\bar{s}\gamma_{\mu}P_{L}b\right)\left(\bar{\mu}\gamma^{\mu}\gamma_{5}\mu\right)\,.\label{eq:O10mumu}
\end{align}
We have omitted semileptonic scalar
and tensor operators from Eq.~(\ref{eq:L_bsmumu}). The former provide a chiral enhancement of $\mathcal{B}(B_{s}\rightarrow\mu^{+}\mu^{-})$
which is at odds with current data, and the latter are not generated
at tree-level from dimension six SMEFT operators. Similarly, primed operators
$\mathcal{O}'^{\mu\mu}_{9}$ and $\mathcal{O}'^{\mu\mu}_{10}$ are
obtained by exchanging $P_{L}$ by $P_{R}$ in Eqs.~(\ref{eq:O9mumu})
and (\ref{eq:O10mumu}), however these operators involving right-handed
quarks are as well disfavoured by current data. Notice also the different
normalisation factor of Eq.~(\ref{eq:L_bsmumu}) with respect to
Eq.~(\ref{eq:LEFT_Lagrangian}), highlighting the CKM suppression
of the $b\rightarrow s\mu\mu$ transition. For simplicity, in Eq.~(\ref{eq:L_bsmumu})
we suppressed the scale dependence of the Wilson coefficients, that
have to be evaluated at $\mu=m_{b}$.

\begin{figure}
\subfloat[\label{fig:CleanFit_2021}]{\includegraphics[scale=0.5]{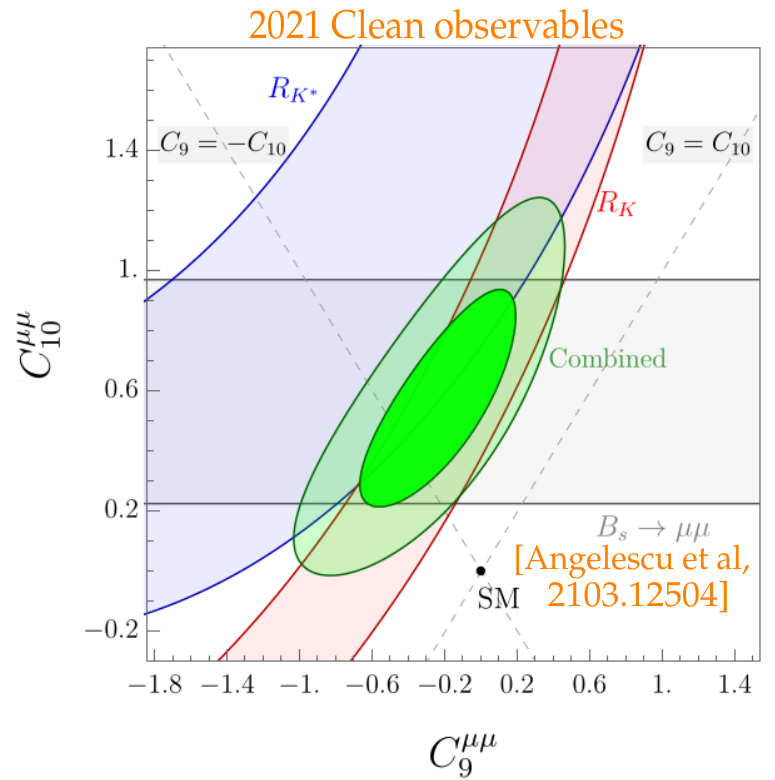}

}$\quad$\subfloat[\label{fig:CleanFit_2023}]{\includegraphics[scale=0.42]{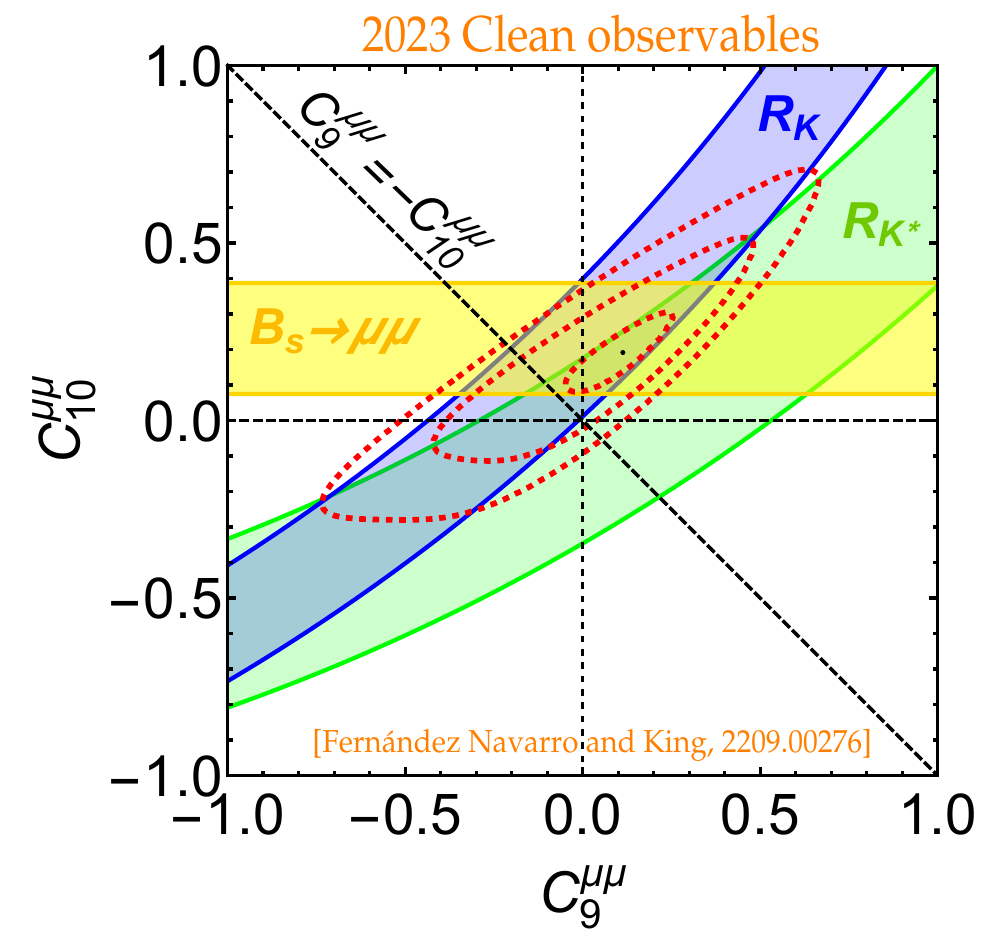}

}

\caption[Model independent analysis of $b\rightarrow s\mu\mu$ theoretically
clean observables]{ Allowed regions in the plane $C_{9}^{\mu\mu}$ vs $C_{10}^{\mu\mu}$
to $1\sigma$ accuracy derived by using 2021 (left) and 2023 (right)
data on $R_{K}$, $R_{K^{*}}$ and $\mathcal{B}(B_{s}\rightarrow\mu^{+}\mu^{-})$.
The green (left) and red (right) contours denote the $1\sigma$, $2\sigma$
and $3\sigma$ regions of the $\chi^{2}$ fit. The left plot is taken
from \cite{Angelescu:2021lln}. The right plot is our own work (originally
presented in \cite{FernandezNavarro:2022gst}), and the black dot
denotes the best fit point with $\Delta\chi^{2}/\mathrm{dof}\approx0.56$.\label{fig:C9vsC10}}

\end{figure}

The Wilson coefficients $C_{9}^{\mathrm{SM}}=4.27$ and $C_{10}^{\mathrm{SM}}=-4.17$
\cite{Bruggisser:2021duo} encode the SM contributions, while $C_{9}^{\mu\mu}$
and $C_{10}^{\mu\mu}$ are associated to NP. Performing a combined
$\chi^{2}$ fit of the observables $R_{K^{(*)}}$ and $\mathcal{B}(B_{s}\rightarrow\mu^{+}\mu^{-})$
one obtains the parameter space of $C_{9}^{\mu\mu}$ and $C_{10}^{\mu\mu}$
preferred by NP, as can be seen in Fig.~\ref{fig:CleanFit_2021}.
As of 2021, scenarios involving only $C_{9}^{\mu\mu}$ or $C_{10}^{\mu\mu}$
could describe the experimental data up to $2\sigma$ precision, while
left-handed NP $C_{9}^{\mu\mu}=-C_{10}^{\mu\mu}$ were in excellent
agreement with experimental data and preferred over the SM hypothesis
by more than $4\sigma$. 

In late 2022, a reanalysis of the $R_{K^{(*)}}$ ratios by LHCb
revealed that backgrounds in the electron channel had been misidentified
in all previous analyses. After this systematic effect was taken into
account, the collaboration updated the $R_{K^{(*)}}$ ratios as\cite{LHCb:2022qnv}
\begin{flalign}
R_{K}^{[1.1,6]} & =\frac{\mathcal{B}\left(B\rightarrow K\mu^{+}\mu^{-}\right)}{\mathcal{B}\left(B\rightarrow Ke^{+}e^{-}\right)}=0.949_{-0.046}^{+0.047}\,,\label{eq:RK}\\
R_{K^{*}}^{[1.1,6]} & =\frac{\mathcal{B}\left(B\rightarrow K^{*}\mu^{+}\mu^{-}\right)}{\mathcal{B}\left(B\rightarrow K^{*}e^{+}e^{-}\right)}=1.027_{-0.073}^{+0.077}\,,\nonumber 
\end{flalign}
with correlation factor $\rho=-0.017$. This way, the $R_{K^{(*)}}$
ratios are now in good agreement with SM lepton universality, although
some space for NP is still left. The CMS collaboration presented
a precise new measurement of $\mathcal{B}(B_{s}\rightarrow\mu^{+}\mu^{-})$
in 2022 as well \cite{CMS:2022mgd}, which is in good agreement with the SM. When
combined with the existing measurements by LHCb and ATLAS, the global
average now reads (see e.g.~\cite{Allanach:2022iod,Greljo:2022jac})
\begin{equation}
\mathcal{B}(B_{s}\rightarrow\mu^{+}\mu^{-})=(3.28\pm0.26)\times10^{-9}\,,
\end{equation}
which is in good agreement with the SM prediction $\mathcal{B}\left(B_{s}\rightarrow\mu^{+}\mu^{-}\right)_{\mathrm{SM}}=(3.67\pm0.15)\times10^{-9}$,
albeit leaving some space for NP. The expressions for the observables
of interest $R_{K}^{[1.1,6]}$, $R_{K^{*}}^{[1.1,6]}$ and $\mathcal{B}(B_{s}\rightarrow\mu^{+}\mu^{-})$
in terms of the Wilson coefficients $C_{9}^{\mu\mu}$ and $C_{10}^{\mu\mu}$ are
\cite{Becirevic:2016zri} (we do not include expressions for the lower
$q^{2}$ interval for $R_{K^{(*)}}$ where NP contributions are suppressed)
\begin{equation}
R_{K}^{[1.1,6]}=R_{K,\mathrm{SM}}^{[1.1,6]}\frac{1+0.24\mathrm{Re}(C_{9}^{\mu\mu})-0.26\mathrm{Re}(C_{10}^{\mu\mu})+0.03(\left|C_{9}^{\mu\mu}\right|^{2}+\left|C_{10}^{\mu\mu}\right|^{2})}{1+0.24\mathrm{Re}(C_{9}^{ee})-0.26\mathrm{Re}(C_{10}^{ee})+0.03(\left|C_{9}^{ee}\right|^{2}+\left|C_{10}^{ee}\right|^{2})}\,,\label{eq:RKth}
\end{equation}
\begin{equation}
R_{K^{*}}^{[1.1,6]}=R_{K^{*},\mathrm{SM}}^{[1.1,6]}\frac{1+0.18\mathrm{Re}(C_{9}^{\mu\mu})-0.29\mathrm{Re}(C_{10}^{\mu\mu})+0.03(\left|C_{9}^{\mu\mu}\right|^{2}+\left|C_{10}^{\mu\mu}\right|^{2})}{1+0.18\mathrm{Re}(C_{9}^{ee})-0.29\mathrm{Re}(C_{10}^{ee})+0.03(\left|C_{9}^{ee}\right|^{2}+\left|C_{10}^{ee}\right|^{2})}\,,\label{eq:RKstarth}
\end{equation}
\begin{equation}
\mathcal{B}\left(B_{s}\rightarrow\mu^{+}\mu^{-}\right)=\mathcal{B}\left(B_{s}\rightarrow\mu^{+}\mu^{-}\right)_{\mathrm{SM}}\left|1+\frac{C_{10}^{\mu\mu}}{C_{10}^{\mathrm{SM}}}\right|^{2}\,.\label{eq:Bs_mumu}
\end{equation}

In Fig.~\ref{fig:CleanFit_2023} we show the parameter space in the
plane ($C_{9}^{\mu\mu}$, $C_{10}^{\mu\mu}$) preferred by the 2023
$R_{K^{(*)}}$ ratios and the 2023 average of $\mathcal{B}(B_{s}\rightarrow\mu^{+}\mu^{-})$.
We also display the result of a combined $\chi^{2}$ fit to the three
observables as the red ellipses, denoting 1$\sigma$, $2\sigma$ and
$3\sigma$ intervals. Our results show that a small but non-zero value
of $C_{10}^{\mu\mu}$ is still preferred by $\mathcal{B}(B_{s}\rightarrow\mu^{+}\mu^{-})$.
On the other hand, $C_{9}^{\mu\mu}$ is compatible with zero, but
small positive and negative values are still allowed by the new $R_{K^{(*)}}$
ratios at $1\sigma$.

In particular, we highlight that left-handed NP $C_{9}^{\mu\mu}=-C_{10}^{\mu\mu}$
are not far away from the $1\sigma$ region, and our 1-dimensional
fit reveals
\begin{equation}
C_{9}^{\mu\mu}=-C_{10}^{\mu\mu}=[-0.0111,-0.1425]\quad(1\ensuremath{\sigma})\,,\label{eq:C9_-C10}
\end{equation}
with a best fit value of $C_{9}^{\mu\mu}=-C_{10}^{\mu\mu}=-0.0725$
with $\Delta\chi^{2}/\mathrm{dof}\approx0.58$. Although left-handed
NP are still allowed by the new data, the WCs are much smaller than
those preferred by 2021 data (see Fig.~\ref{fig:CleanFit_2021}).

Finally, we highlight that although the anomalies in $R_{K^{(*)}}$
have disappeared, strong hints for NP still remain in $b\rightarrow s\mu\mu$
observables such as $\mathcal{B}(B\rightarrow K^{(*)}\mu^{+}\mu^{-})$,
$\mathcal{B}(B_{s}\rightarrow\phi\mu^{+}\mu^{-})$ and the angular
observable $P'_{5}$. However, the significance of these anomalies
depends on assumptions about the unknown QCD uncertainties that affect
these observables. Although some analyses suggest that the tension
in $\mathcal{B}(B\rightarrow K^{(*)}\mu^{+}\mu^{-})$ can reach the
$4\sigma$ level (see e.g.~\cite{Gubernari:2022hxn}), these claims
should be taken with care. The remaining anomalies in $b\rightarrow s\mu\mu$
data could be explained by a lepton universal contribution to the
operator $C_{9}$, provided that the NP effect is $C_{9}\approx1$
\cite{Alguero:2023jeh} (roughly one fourth of the SM $C_{9}$). We
denote such contribution as $C_{9}^{U}$. It could be generated via
a $Z'$ boson with flavour universal couplings to leptons, or it 
could also be generated via RGE effects provided by leptoquarks that
couple preferentially to third family fermions, such as those proposed
to address $R_{D^{(*)}}$ (see Section~\ref{subsec:bctaunu}). Remarkably,
the scenario most preferred by current data involves a large $C_{9}^{U}$
plus a small LFU-violating left-handed contribution $C_{9}^{\mu\mu}=-C_{10}^{\mu\mu}$
\cite{Alguero:2023jeh},
\begin{equation}
C_{9}^{U}=-1.10_{-0.19}^{+0.17}\quad(\mathrm{LFU})\,,\qquad C_{9}^{\mu\mu}=-C_{10}^{\mu\mu}=-0.08_{-0.06}^{+0.07}\quad(\mathrm{LFUV})\,.
\end{equation}
We will see that this scenario is very well motivated from the point
of view of a theory of flavour, as it could arise from leptoquarks
with hierarchical couplings to charged leptons, following the pattern
of SM Yukawa couplings $y_{e}\ll y_{\mu}\ll y_{\tau}$.

\subsection{\texorpdfstring{$R_{D^{(*)}}$ anomalies and their interpretation in a theory of flavour}{RD(*) anomalies and their interpretation in a theory of flavour}\label{subsec:bctaunu}}

Even though the $R_{K^{(*)}}$ ratios are now in good agreement with
the SM, strong hints for the breaking of LFU are still present in
$b\rightarrow c\ell\nu$ transitions. Here one can construct relatively
clean LFU ratios of $B$ mesons decaying to $D^{(*)}$ mesons and
a lepton-neutrino pair. Particularly interesting are the $R_{D^{(*)}}$
ratios,
\begin{equation}
R_{D^{(*)}}=\left.\frac{\mathcal{B}(B\rightarrow D^{(*)}\tau\bar{\nu})}{\mathcal{B}(B\rightarrow D^{(*)}\ell\bar{\nu})}\right|_{\ell=e,\,\mu}\,,
\end{equation}
which test the universality of the decays into taus with respect to
the decays into light charged leptons. In the following, we display
the arithmetic average of existing SM predictions given in \cite{HFLAV:2022wzx},
\begin{equation}
R_{D}^{\mathrm{SM}}=0.298\pm0.004\,,\qquad R_{D^{*}}^{\mathrm{SM}}=0.254\pm0.005\,.
\end{equation}
In contrast with the $R_{K^{(*)}}$ ratios, the SM prediction of $R_{D^{(*)}}$
is smaller than unity due to the large mass of tau with respect to
the light charged leptons. The uncertainties sit at the level of a
few per cent (slightly larger than those of the $R_{K^{(*)}}$ ratios), due
to uncertainties in the lattice QCD determinations of hadronic form
factors \cite{FermilabLattice:2021cdg,Harrison:2023dzh,Aoki:2023qpa}.
It is important to mention that, despite the recent progress, the
lattice QCD results for the $B\rightarrow D^{*}$ form factors show
tensions among each other and with experimental data, hence requiring
further investigation.
\begin{figure}
\begin{centering}
\includegraphics[scale=0.72]{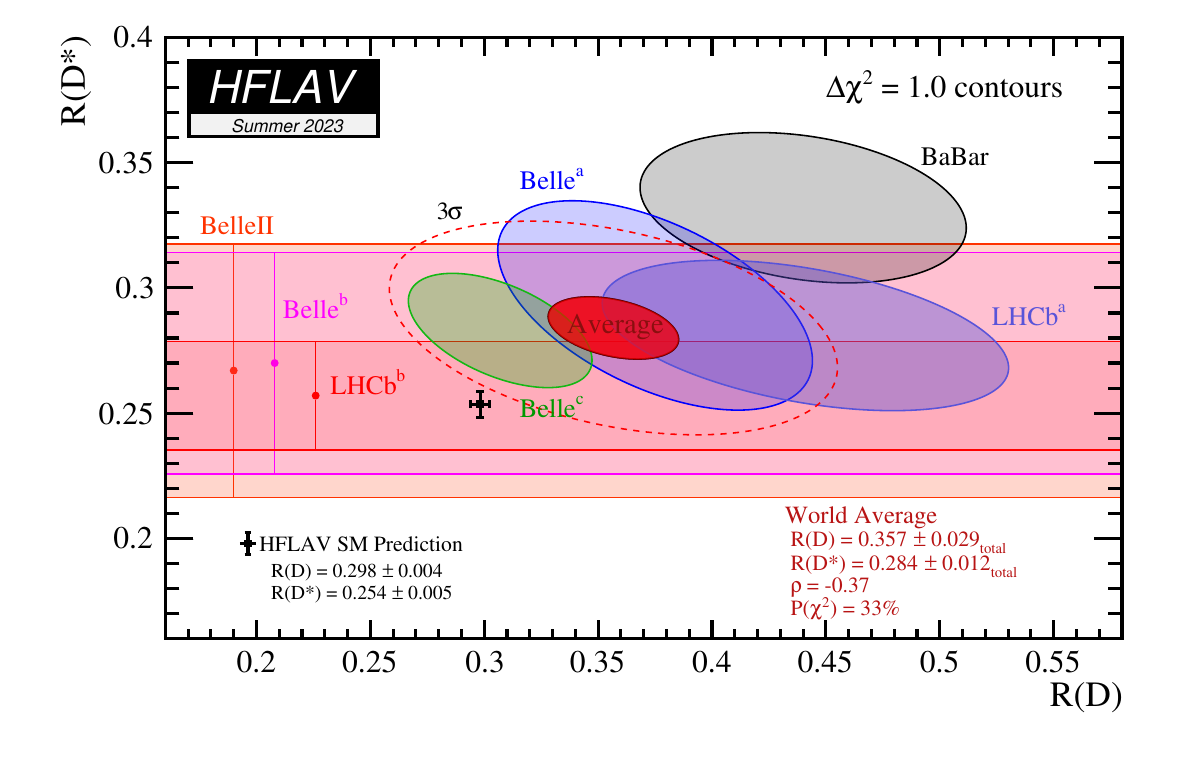}
\par\end{centering}
\caption[Experimental picture of the $R_{D^{(*)}}$ anomalies as of Summer
2023]{ Experimental picture of the $R_{D^{(*)}}$ anomalies as of Summer
2023, along with the HFLAV average SM prediction. The horizontal bands
are $1\sigma$ (68\% CL) bands, whereas the solid ellipses correspond
to $\Delta\chi^{2}=1$ (61\% CL) contours. In particular, the solid
red ellipse corresponds to the $\Delta\chi^{2}=1$ (61\% CL) contour
for the global average, while the dashed red ellipse corresponds to
the $3\sigma$ contour for the global average. Figure taken from
the \href{https://hflav-eos.web.cern.ch/hflav-eos/semi/summer23/html/RDsDsstar/RDRDs.html}{website of HFLAV}. \label{fig:RD_measurements}}

\end{figure}

The $R_{D^{(*)}}$ ratios were firstly measured by the BaBar collaboration
in 2012 \cite{BaBar:2012obs,BaBar:2013mob}, who presented experimental
values of $R_{D}$ and $R_{D^{*}}$ larger than the SM prediction
by about 2-$3\sigma$. Since then, new measurements by the Belle collaboration
\cite{Belle:2015qfa,Belle:2016dyj,Belle:2017ilt,Belle:2019rba} and
by the LHCb collaboration \cite{LHCb:2023cjr,LHCb:2023zxo} have shown
small deviations from the SM, consistent with the BaBar measurements,
but no singular measurement has been significant enough to
claim evidence for new physics. Nevertheless, the global picture shows
a consistent pattern of deviations where $R_{D}$ and $R_{D^{*}}$ are both larger
than the SM predictions. As of Summer 2023, the global average by HFLAV
of all official measurements of the $R_{D^{(*)}}$ ratios reads \cite{HFLAV:2022wzx}
\begin{equation}
R_{D}^{\mathrm{\mathrm{HFLAV}}}=0.357\pm0.029\,,\qquad R_{D^{*}}^{\mathrm{HFLAV}}=0.284\pm0.013\,.
\end{equation}
These values exceed the SM predictions given before by $2\sigma$
and $2.2\sigma$, respectively. Considering the $R_{D}$-$R_{D^{*}}$
correlation of $\rho=-0.37$, the resulting combined deviation from
the SM is at the level of $3.3\sigma$, as shown in Fig.~\ref{fig:RD_measurements}.
We note that in the near future, we expect
experimental input from the Belle II\footnote{We note here that the Belle II collaboration presented their first
preliminary measurement of $R_{D^{*}}$ simultaneously at the Lepton-Photon
2023 and SUSY 2023 conferences, being compatible with the existing
pattern of deviations, although this measurement is not significant
enough to extract any further conclusions.} collaboration to confirm or exclude the existing pattern of deviations.

Other $b\rightarrow c\ell\nu$ observables beyond $R_{D^{(*)}}$ have
been measured, namely $R_{J/\psi}$ \cite{LHCb:2017vlu} and $R_{\Lambda_{b}}$
\cite{LHCb:2022piu}, which are affected by larger experimental uncertainties.
They are consistent with the current deviations in $R_{D^{(*)}}$,
but they are not yet precise enough to shed light on the $b\rightarrow c\ell\nu$
LFU puzzle.

In order to describe the $b\to c\tau\bar{\nu}$ transition at the
level of the LEFT, we introduce the following effective operators, 
\begin{flalign}
\mathcal{L}_{b\rightarrow c\tau\nu} & =-\frac{4G_{F}}{\sqrt{2}}V_{cb}\left[\left(1+\left[C_{\nu edu}^{V,LL}\right]^{\tau\tau32*}\right)\left[\mathcal{O}_{\nu edu}^{V,LL}\right]^{\tau\tau32\dagger}\right.\label{eq:btoctaunu_low}\\
 & \left.+\left[C_{\nu edu}^{S,RL}\right]^{\tau\tau32*}\left[\mathcal{O}_{\nu edu}^{S,RL}\right]^{\tau\tau32\dagger}+\mathrm{h.c.}\right] \,,\nonumber 
\end{flalign}
where the Wilson coefficients are at the $m_{b}$ scale. In the effective
Lagrangian of Eq.~(\ref{eq:btoctaunu_low}), we have omitted operators
which will not be relevant for the NP models presented in this thesis.
The $R_{D^{(*)}}$ ratios are described in terms of the NP Wilson
coefficients as, 
\begin{flalign}
R_{D}=R_{D}^{\mathrm{SM}} & \left[\left|1+\left[C_{\nu edu}^{V,LL}\right]^{\tau\tau32*}\right|^{2}+1.5\mathrm{Re}\left\{ \left(1+\left[C_{\nu edu}^{V,LL}\right]^{\tau\tau32*}\right)\left[C_{\nu edu}^{S,RL}\right]^{\tau\tau32*}\right\} \right.\label{eq:RD}\\
 & \left.+1.03\left|\left[C_{\nu edu}^{S,RL}\right]^{\tau\tau32*}\right|^{2}\right]\,,\nonumber 
\end{flalign}
\begin{flalign}
R_{D^{*}}=R_{D^{*}}^{\mathrm{SM}} & \left[\left|1+\left[C_{\nu edu}^{V,LL}\right]^{\tau\tau32*}\right|^{2}+0.12\mathrm{Re}\left\{ \left(1+\left[C_{\nu edu}^{V,LL}\right]^{\tau\tau32*}\right)\left[C_{\nu edu}^{S,RL}\right]^{\tau\tau32*}\right\} \right.\label{eq:RD*}\\
 & \left.+0.04\left|\left[C_{\nu edu}^{S,RL}\right]^{\tau\tau32*}\right|^{2}\right]\,,\nonumber 
\end{flalign}
where the numerical coefficients are obtained from integrating over
the full kinematical distributions for the $B\rightarrow D^{(*)}$
semileptonic decay \cite{Murgui:2019czp,Mandal:2020htr}. Notice that
the vector operator $\left[\mathcal{O}_{\nu edu}^{V,LL}\right]^{\tau\tau32\dagger}$
predicts that both $R_{D}$ and $R_{D^{*}}$ are similarly modified
by NP, such that their deviations from the SM follow $\Delta R_{D}=\Delta R_{D^{*}}$,
where a fit to $b\to c\tau\bar{\nu}$ data prefers $\left[C_{\nu edu}^{V,LL}\right]^{\tau\tau32*}=0.08\pm0.02$
\cite{Iguro:2022yzr}. Instead, the scalar operator $\left[\mathcal{O}_{\nu edu}^{S,RL}\right]^{\tau\tau32\dagger}$
predicts a larger enhancement of $R_{D}$ with respect to $R_{D^{*}}$,
such that $\Delta R_{D}>\Delta R_{D^{*}}$, where a fit to $b\to c\tau\bar{\nu}$
data prefers $\left[C_{\nu edu}^{S,RL}\right]^{\tau\tau32*}=0.17\pm0.05$
\cite{Iguro:2022yzr}. The preferred NP candidates to generate these
operators and explain the $R_{D^{(*)}}$ anomalies are vector and/or
scalar leptoquarks, which generally avoid tree-level contributions
to the most dangerous $\Delta F=2$ processes.

If confirmed, the $R_{D^{(*)}}$ anomalies suggest the existence of
NP dominantly interacting with taus rather than with light charged
leptons, leading to a 10-20\% enhancement over the SM prediction in
the tau channel. Notice that the $b\rightarrow c\ell\nu$ transition
is a tree-level charged current in the SM. Therefore, relevant NP
contributions to the $R_{D^{(*)}}$ ratios very likely have to be
tree-level and associated to the TeV scale, otherwise the NP effect
would be completely screened by the SM contribution. This is in contrast
with the $R_{K^{(*)}}$ ratios, which are sensitive to heavier NP scales
due to loop, GIM and CKM suppressions.

The $R_{D^{(*)}}$ ratios and the still anomalous $b\rightarrow s\mu\mu$
data are commonly denoted as the ``$B$-anomalies''. During the
time when the $R_{K^{(*)}}$ ratios showed tensions with the SM, deviations
in both LFU ratios $R_{D^{(*)}}$ and $R_{K^{(*)}}$ were consistently
understood in the framework of a theory of flavour. New dynamics connected
to the origin of the SM flavour structure might very well couple hierarchically
to SM fermions, following the behaviour of SM Yukawa couplings. Given
that in the SM $y_{e}\ll y_{\mu}\ll y_{\tau}$, then if the new dynamics follow
the same hierarchical pattern, it would be natural to see a large
effect in the $R_{D^{(*)}}$ ratio, followed by a hierarchically smaller effect
that modifies the $R_{K^{(*)}}$ ratios. From the point of view of
a theory of flavour, the fact that the $R_{K^{(*)}}$ ratios are now
consistent with the SM only means that the new dynamics coupling to
muons are smaller than we expected. In other words, if the theory
of flavour predicts generic NP couplings $\beta_{\mu}\ll\beta_{\tau}$,
then the new data on $R_{K^{(*)}}$ only means that $\beta_{\mu}$
is smaller than we expected, but still completely consistent with the expected hierarchical
pattern. However, if $\beta_{\mu}$ is indeed
connected to the origin of $y_{\mu}$, then eventually a deviation
in $R_{K^{(*)}}$ from the SM prediction should be seen with further
precision. Model building in this direction will be considered in
Chapter~\ref{Chapter:TwinPS} via a theory of flavour containing a $U_{1}\sim(\mathbf{3,1},2/3)$
vector leptoquark.

\subsection{\texorpdfstring{$(g-2)_{\mu}$ anomaly}{(g-2)mu anomaly} \label{subsec:g-2}}

\begin{figure}[t]
\begin{centering}
\includegraphics[scale=0.7]{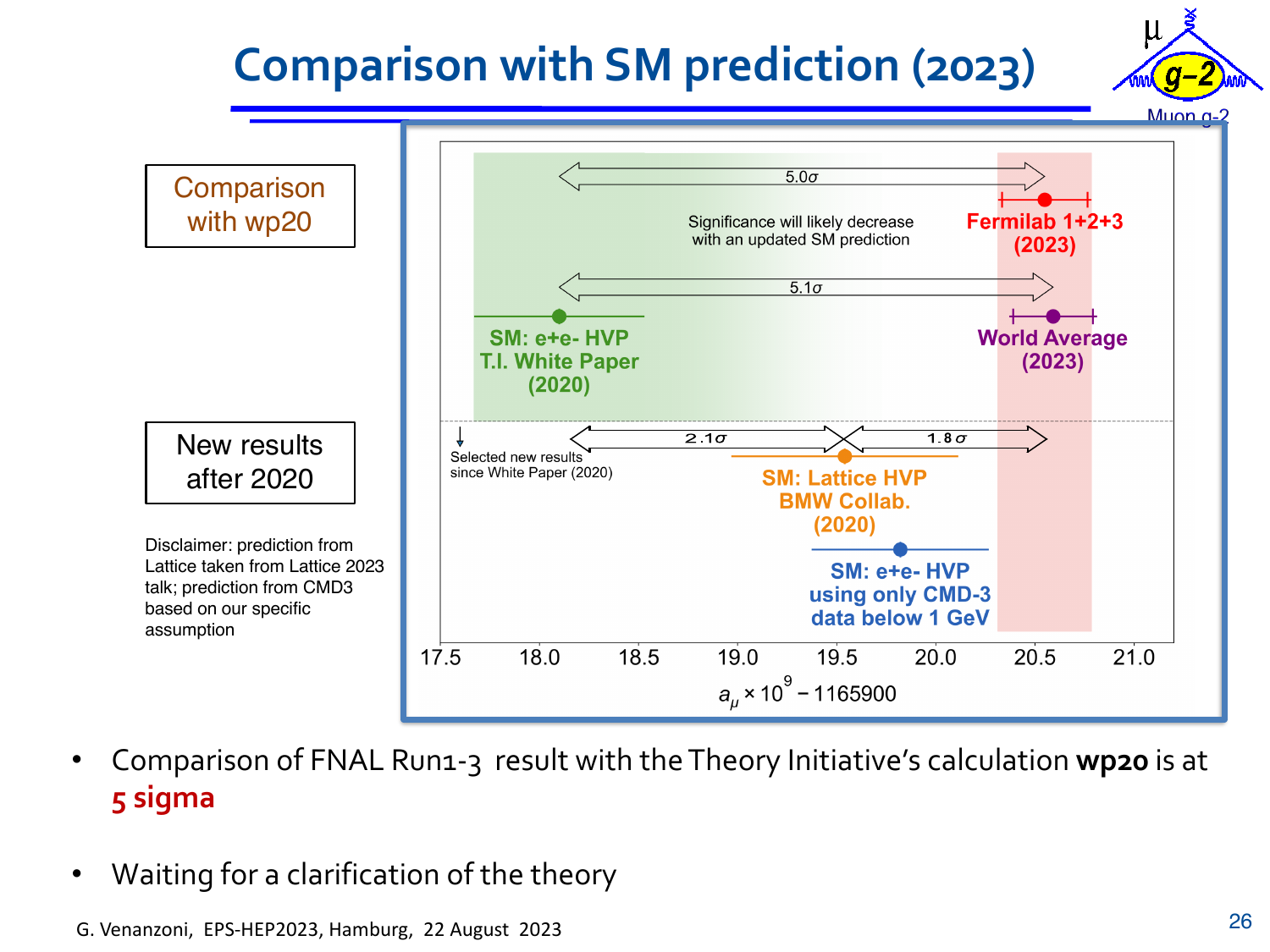}\caption[Tensions in $(g-2)_{\mu}$ data, theory and experiment]{ Current picture of $a_{\mu}\equiv(g-2)_{\mu}/2$ data, including
the data driven SM prediction (White Paper), the data driven prediction considering only the recent CMD-3 data, the BMW lattice prediction,
and the experimental measurements including FNAL and the current world average (see more in
the main text). Figure taken from the slides of \protect\href{https://indico.desy.de/event/34916/contributions/150287/attachments/84171/111449/gv_eps220823_s_pdf.pdf}{Graziano Venanzoni for the EPS-HEP2023 conference}.
\label{fig:g-2_data}}
\par\end{centering}
\end{figure}

Independent of the anomalies in $B$-physics
data, there also exists a possible discrepancy with the SM prediction
in the experimentally measured anomalous magnetic moment $a=(g-2)/2$
of the muon. The long-lasting non-compliance of $a_{\mu}$ with the
SM was first observed by the Brookhaven E821 experiment at BNL \cite{Muong-2:2006rrc}.
More recently, this discrepancy has been confirmed by the FNAL
experiment \cite{Muong-2:2021ojo,Muong-2:2023cdq},
\begin{equation}
\Delta a^{\mathrm{R}}_{\mu}=a_{\mu}^{\mathrm{exp}}-a^{\mathrm{SM,R}}_{\mu}=(249\pm48)\times10^{-11}\,,\label{eq:g-2_mu}
\end{equation}
a result $5.1\sigma$ larger than the SM prediction obtained by the muon g-2 theory initiative \cite{Aoyama:2020ynm}. However, this SM prediction is based on data from $e^{+}e^{-}\rightarrow\mathrm{hadrons}$ \cite{Colangelo:2018mtw,Davier:2019can,Keshavarzi:2019abf},
and does not include the lattice QCD results by the BMW collaboration
for the hadronic vacuum polarisation \cite{Borsanyi:2020mff}, which
reduce the tension to the $1.8\sigma$ level,
\begin{equation}
\Delta a_{\mu}^{\mathrm{BMW}}=a_{\mu}^{\mathrm{exp}}-a_{\mu}^{\mathrm{SM,BMW}}=(105\pm59)\times10^{-11}\,.
\end{equation}
Note that the results of BMW have been confirmed by other lattice
collaborations \cite{Ce:2022kxy,ExtendedTwistedMass:2022jpw,Blum:2023qou},
but only in the so-called ``intermediate window'' \cite{RBC:2018dos},
which represents only a third of the total contribution to the hadronic
vacuum polarisation. Apparently, the hadronic vacuum polarisation predicted by
BMW also worsens the SM fit to EWPOs \cite{Crivellin:2020zul}. The
situation becomes even more puzzling if we consider the most recent
measurement of $e^{+}e^{-}\rightarrow\mathrm{hadrons}$ by the CMD-3 collaboration \cite{CMD-3:2023alj},
which would render the data driven prediction closer to the measurement, however the recent results obtained by CMD-3 are in conflict with more than 20 years of data from $e^{+}e^{-}$ experiments. This complicated situation regarding the theory prediction of $(g-2)_{\mu}$ is illustrated in Fig.~\ref{fig:g-2_data}.

While we wait for new data and theory improvement to establish a clear
picture, it is interesting to study the BSM interpretation of the
result in Eq.~(\ref{eq:g-2_mu}). Moreover, although the BMW prediction is still in rough agreement
with the experimental value, with the increasing precision in the
experimental measurement is possible that a small tension at the $2\sigma$
level emerges in the near future.

Any BSM contribution to $(g-2)_{\mu}$
involves both left-handed and right-handed muons, hence relying on
a chirality flip, which can be provided by the muon Yukawa coupling
in the SM. Given the smallness of $y_{\mu}$ in the SM, only light
NP such as a sub-GeV $Z'$ \cite{Pospelov:2008zw} can explain the anomaly in this manner.
However, such NP are difficult to connect to a theory of flavour,
plus the parameter space is becoming very constrained (see e.g.~Ref.~\cite{Athron:2021iuf}
for a review of new physics in $(g-2)_{\mu}$). Because the deviation
from the SM prediction is as large as its electroweak contribution, heavy NP
at or above the TeV scale must possess an enhancement factor. This
can be provided via the mechanism of chiral enhancement, meaning that
the chirality flip does not originate from the small muon Yukawa coupling
but rather from a larger coupling of other particles to the SM Higgs.
Models of this kind include the MSSM, where chiral enhancement is
connected to $\mathrm{tan}\beta$ \cite{Everett:2001tq}, models with
generic new scalars and fermions \cite{Czarnecki:2001pv,Kannike:2011ng,Kowalska:2017iqv,Crivellin:2018qmi,Crivellin:2021rbq,Hernandez:2021tii},
and also models with the scalar leptoquarks $S_{1}\sim(\mathbf{\overline{3},1,}1/3)$ and/or $R_{2}\sim(\mathbf{3,2,}7/6)$
\cite{Djouadi:1989md,Cheung:2001ip,Bauer:2015knc,ColuccioLeskow:2016dox,Crivellin:2020tsz}
which provide a $m_{t}/m_{\mu}$ chiral enhancement.

Another interesting class of models involves the addition of extra
vector-like fermions charged under a new $U(1)'$ gauge group. The
SM fermions remain uncharged under $U(1)'$, such that the massive
$Z'$ boson does not couple directly to SM fermions, only through
possible mixing between vector-like fermions and SM fermions. This feature
gives the name of ``fermiophobic'' to this class of models \cite{Raby:2017igl,Belanger:2015nma,King:2017anf,Falkowski:2018dsl,CarcamoHernandez:2019ydc,Kawamura:2019rth,Kawamura:2019hxp,FernandezNavarro:2021sfb}, which were able to connect $(g-2)_{\mu}$
with $R_{K^{(*)}}$ in the past \cite{FernandezNavarro:2021sfb}, and will be
further discussed in Chapter~\ref{Chapter:Fermiophobic}. We will see that such models can
be connected as well with the origin of Yukawa couplings in the SM
\cite{King:2018fcg}, providing a connection between the $(g-2)_{\mu}$
anomaly and the origin of the flavour structure of the SM.

\subsection{Meson mixing} \label{subsec:BsMixing}

\begin{figure}
\begin{centering}
\includegraphics[scale=1.03]{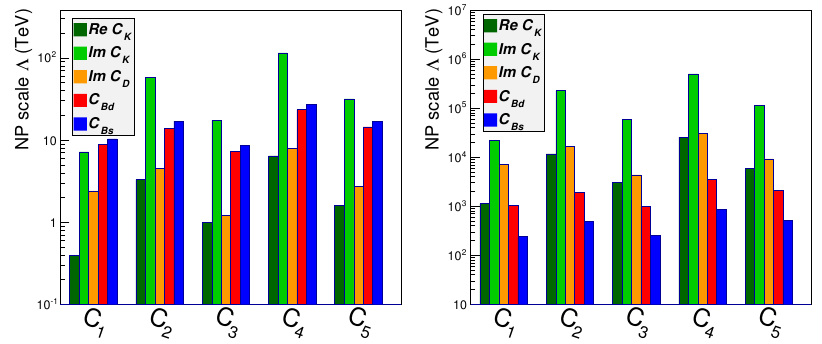}\caption[NP bounds from meson mixing observables]{ Summary of the 95\% CL lower bound on the NP scale $\Lambda$ for
strongly-interacting NP in the $U(2)^{5}$ scenario (left) and in
the flavour anarchic NP scenario (right). Results from all the neutral
meson systems are shown. Figure taken from \cite{Bona:2016bvr}.\label{fig:MesonMixing_Bounds}}
\par\end{centering}
\end{figure}

In the SM, charged current weak interactions provide meson-antimeson
transitions for neutral mesons, which are commonly denoted as $\Delta F=2$
processes because they change fermion flavour in two units\footnote{This is in contrast with semileptonic processes, which change flavour
in one unit $\Delta F=1$.}. This mixing provides a misalignment between meson flavour and mass
eigenstates, that was firstly observed throught the oscillations of
neutral kaons \cite{Good:1961ik}. The meson mixing process is highly loop, CKM
and GIM suppressed in the SM, being sensitive to very high NP scales.
Given that no significant deviation from the SM has been found so
far in meson mixing observables, they set very strong bounds over
the scale of NP contributions. The effective Lagrangian to describe
meson-antimeson mixing contains the following 4-quark operators
\begin{equation}
\mathcal{L}_{\Delta F=2}=\sum_{i=1}^{5}C_{i}^{qq'}\mathcal{Q}_{i}^{qq'}+\sum_{i=1}^{5}\tilde{C}_{i}^{qq'}\tilde{\mathcal{Q}}_{i}^{qq'}\,,\label{eq:MesonMixing_operators}
\end{equation}
where
\begin{flalign}
\mathcal{Q}_{1}^{qq'}= & (\bar{q}_{L}^{\alpha}\gamma_{\mu}q'^{\alpha}_{L}(\bar{q}_{L}\gamma^{\mu}q'^{\beta}_{L})\,,\\
\mathcal{Q}_{2}^{qq'}= & (\bar{q}_{R}^{\alpha}q'^{\alpha}_{L})(\bar{q}_{R}^{\beta}q'^{\beta}_{L})\,,\\
\mathcal{Q}_{3}^{qq'}= & (\bar{q}_{R}^{\alpha}q'^{\beta}_{L})(\bar{q}_{R}^{\beta}q'^{\alpha}_{L})\,,\\
\mathcal{Q}_{4}^{qq'}= & (\bar{q}_{R}^{\alpha}q'^{\alpha}_{L})(\bar{q}_{L}^{\beta}q'^{\beta}_{R})\,,\\
\mathcal{Q}_{5}^{qq'}= & (\bar{q}_{R}^{\alpha}q'^{\beta}_{L})(\bar{q}_{L}^{\beta}q'^{\alpha}_{R})\,,
\end{flalign}
where $\alpha$ and $\beta$ are colour indices, $q$ and $q'$ refer
to the two different quark flavours in the neutral mesonic system, and the
tilde operators are obtained by replacing $L\longleftrightarrow R$
everywhere. Note that some of the operators in the basis above are
not included in the San Diego basis of the LEFT, however they match
into operators of the San Diego basis via Fierz rearrangements. In
Fig.~\ref{fig:MesonMixing_Bounds}, we show the very strong bounds from meson mixing observables
over the Wilson coefficients in Eq.~(\ref{eq:MesonMixing_operators}),
under the assumption of anarchic flavour structure (right panel) and
$U(2)^{5}$ flavour structure (left panel) for the NP contributions (see Section~\ref{subsec:U(2)5}).
The figures highlight the high reach of flavour observables over NP
scales, but the fact that such reach is significantly reduced in flavour
symmetry frameworks, such as MFV and $U(2)^{5}$, gives hints about the possible
flavour structure of NP. In contrast, in the case of anarchic
flavour structure, the bounds can reach scales as high as $10^{6}\;\mathrm{TeV}$
for the imaginary part of the Wilson coefficients of scalar operators.
Remarkably, the largest bounds come from $K-\bar{K}$ mixing
observables, followed by $D-\bar{D}$, $B_{d}-\bar{B}_{d}$
and $B_{s}-\bar{B}_{s}$ in decreasing order \cite{UTfit:2007eik,Isidori:2014rba}.

\subsubsection*{$B_{s}-\bar{B}_{s}$ mixing}

For the NP models presented in this thesis, $B_{s}-\bar{B}_{s}$
is particularly interesting because is sensitive to models featuring
2-3 flavour transitions. The most relevant observable is the mass
difference $\Delta M_{s}$, which controls the frequency of the $B_{s}-\bar{B}_{s}$
oscillations. The experimental value is known very precisely, see
for example the most recent HFLAV average \cite{HFLAV:2022wzx}, which
is dominated by the updated measurement by LHCb \cite{LHCb:2021moh}.
However, the SM prediction historically suffered from larger uncertainties,
and we need a precise knowledge of the SM contribution in order to
quantify the impact of possible contributions from new physics. The
theoretical determination of $\Delta M_{s}$ is limited by our understanding
of non-perturbative matrix elements of dimension six operators. The
matrix elements can be determined with lattice simulations or sum
rules. As discussed in Ref.~\cite{DiLuzio:2019jyq}, the 2019 FLAG
average \cite{FlavourLatticeAveragingGroup:2019iem} is dominated
by the lattice results \cite{FermilabLattice:2016ipl,Boyle:2018knm,Dowdall:2019bea},
and suffers from an uncertainty just below 10\% with the central value
being $1.8\sigma$ above the experiment,
\begin{equation}
\Delta M_{s}^{\mathrm{FLAG'19}}=\left(1.13_{-0.09}^{+0.07}\right)\Delta M_{s}^{\mathrm{exp}}\,.
\end{equation}
If one considers the value above as the SM prediction for $\Delta M_{s}$,
then NP models with positive contributions to $\Delta M_{s}$ (which
is the common case for $Z'$ and leptoquark models suggested to explain
the $B$-anomalies), have very small room to be compatible with the
experimental value at the $2\sigma$ level. Instead,
\begin{equation}
\Delta M_{s}^{\mathrm{Average'19}}=\left(1.04_{-0.07}^{+0.04}\right)\Delta M_{s}^{\mathrm{exp}}\,,
\end{equation}
was computed in \cite{DiLuzio:2019jyq} as a weighted average of both
the FLAG'19 average \cite{FlavourLatticeAveragingGroup:2019iem} and
sum rule results \cite{Kirk:2017juj,Grozin:2016uqy,King:2019lal}.
The weighted average shows better agreement with experiment, and
a reduction of the total uncertainty (see the further discussion in
\cite{DiLuzio:2019jyq}). The Average'19 result for $\Delta M_{s}$
leaves some room for positive NP contributions at the $2\sigma$ level.
We extract an upper bound over the NP contribution by considering
the lower limit of the $2\sigma$ range, $\Delta M_{s}^{\mathrm{SM}}\approx0.9\Delta M_{s}^{\mathrm{exp}}$,
hence
\begin{equation}
\frac{\Delta M_{s}^{\mathrm{SM}}+\Delta M_{s}^{\mathrm{NP}}}{\Delta M_{s}^{\mathrm{exp}}}\approx0.9\frac{\Delta M_{s}^{\mathrm{SM}}+\Delta M_{s}^{\mathrm{NP}}}{\Delta M_{s}^{\mathrm{SM}}}\approx1\Rightarrow\Delta M_{s}^{\mathrm{NP}}\lesssim0.11\Delta M_{s}^{\mathrm{SM}}\,.\label{eq:DeltaMs_bound}
\end{equation}
In other words, $\Delta M_{s}^{\mathrm{Average'19}}$ allows for roughly
a 10\% positive NP correction over the SM value. This is in line with
the 10\% criteria commonly considered in the literature, which are
possibly motivated by $\Delta M_{s}^{\mathrm{Average'19}}$ as well.
As a specific NP example, the bound in Eq.~(\ref{eq:DeltaMs_bound})
translates directly over the Wilson coefficient $\left.C_{1}^{bs}\right|_{\mathrm{NP}}$.
Let us normalise the effective operator as
\begin{equation}
\mathcal{L}_{bs}=-\frac{C_{\text{1}}^{bs}}{2}\mathcal{Q}_{1}^{bs}\label{eq:MesonMixing_operators2}
\end{equation}
The bound over $\delta(\Delta M_{s})$ then translates to a bound over $\left.C_{1}^{bs}\right|_{\mathrm{NP}}$
as
\begin{equation}
\delta(\Delta M_{s})\equiv\frac{\Delta M_{s}-\Delta M_{s}^{\mathrm{SM}}}{\Delta M_{s}^{\mathrm{SM}}}=\left|1+\frac{\left.C_{1}^{bs}\right|_{\mathrm{NP}}}{\left.C_{1}^{bs}\right|_{\mathrm{SM}}}\right|-1=\frac{\left.C_{1}^{bs}\right|_{\mathrm{NP}}}{\left.C_{1}^{bs}\right|_{\mathrm{SM}}}\apprle0.11\,,\label{eq:delta_DeltaMs}
\end{equation}
where in the second step we have assumed real and positive Wilson
coefficients. The SM contribution to the Wilson coefficient reads
\begin{equation}
\left.C_{1}^{bs}\right|_{\mathrm{SM}}=\frac{G_{F}^{2}m_{W}^{2}}{2\pi^{2}}\left(V_{tb}^{*}V_{ts}\right)^{2}S_{0}(x_{t})\,,
\end{equation}
with $S_{0}(x_{t})=2.37$ \cite{Buchalla:1995vs}. This way, we obtain
the numerical bound 
\begin{equation}
\left.C_{1}^{bs}\right|_{\mathrm{NP}}\lesssim\frac{1}{\left(225\,\mathrm{TeV}\right)^{2}}\,,\label{eq:bound_Bsmixing}
\end{equation}
which is in good agreement with the $(220\,\mathrm{TeV)^{-2}}$ bound obtained
in \cite{DiLuzio:2019jyq} from $\Delta M_{s}^{\mathrm{Average'19}}$.
Instead, if we consider $\Delta M_{s}^{\mathrm{FLAG'19}}$ as the
SM prediction, the resulting bound is
\begin{equation}
\Delta M_{s}^{\mathrm{NP}}\lesssim0.0526\,\Delta M_{s}^{\mathrm{SM}}\Rightarrow\left.C_{1}^{bs}\right|_{\mathrm{NP}}\lesssim\frac{1}{\left(330\,\mathrm{TeV}\right)^{2}}\,,\label{eq:bound_Bsmixing-1}
\end{equation}
which is again in good agreement with the bound presented in \cite{DiLuzio:2019jyq}.

We finally note that NP mediators proposed to address the $B$-anomalies
usually receive strong constraints from $\Delta M_{s}$. In particular,
$Z'$ bosons and other neutral mediators like colour octets usually contribute at tree-level
to $\Delta M_{s}$, and particular suppression mechanisms are sometimes
implemented in order to ameliorate the bounds. On the other hand,
leptoquark mediators contribute to $\Delta M_{s}$ at 1-loop, enjoying
a natural suppression. Because of this, leptoquarks are the preferred
NP explanation of the $R_{D^{(*)}}$ anomalies, where a NP scale not
far above the TeV is required.

\subsection{\texorpdfstring{$b\rightarrow s\tau\tau$}{bstautau}} \label{subsec:bstautau}

We treat the $b\rightarrow s\tau\tau$ transition separately from the muon and
electron channels discussed in Section~\ref{subsec:bsll}, because
the tau channel remains much more unexplored and allows for larger
NP contributions. We introduce the effective Lagrangian
\begin{equation}
\begin{aligned}\mathcal{L}_{b\rightarrow s\tau\tau}=\frac{4G_{F}}{\sqrt{2}}V_{tb}V_{ts}^{*}\frac{\alpha_{\mathrm{\mathrm{EM}}}}{4\pi} & \bigg[(C_{9}^{\mathrm{SM}}+C_{9}^{\tau\tau})\mathcal{O}_{9}^{\tau\tau}+(C_{10}^{\mathrm{SM}}+C_{10}^{\tau\tau})\mathcal{O}_{10}^{\tau\tau}\\
 & +C_{S}^{\tau\tau}\mathcal{O}_{S}^{\tau\tau}+C_{P}^{\tau\tau}\mathcal{O}_{P}^{\tau\tau}\bigg]+\mathrm{h.c.}\,,
\end{aligned}
\label{eq:bstautau}
\end{equation}
where
\begin{align}
\mathcal{O}_{9}^{\tau\tau}= & \left(\bar{s}\gamma_{\mu}P_{L}b\right)\left(\bar{\tau}\gamma^{\mu}\tau\right)\,, & \mathcal{O}_{S}^{\tau\tau}= & \left(\bar{s}P_{R}b\right)\left(\bar{\tau}\tau\right)\,,\\
\mathcal{O}_{10}^{\tau\tau}= & \left(\bar{s}\gamma_{\mu}P_{L}b\right)\left(\bar{\tau}\gamma^{\mu}\gamma_{5}\tau\right)\,, & \mathcal{O}_{P}^{\tau\tau}= & \left(\bar{s}P_{R}b\right)\left(\bar{\tau}\gamma_{5}\tau\right)\,,
\end{align}
where for simplicity we suppressed the scale dependence of the Wilson
coefficients, which are at the scale $\mu=m_{b}$. With these definitions,
the expressions for the observables of interest in the $b\rightarrow s\tau\tau$
transition read \cite{Becirevic:2016zri} 
\begin{flalign}
\mathcal{B}(B_{s}\rightarrow\tau\tau) & =\mathcal{B}(B_{s}\rightarrow\tau\tau)_{\mathrm{SM}}\bigg\{\left|1+\frac{C_{10}^{\tau\tau}}{C_{10}^{\mathrm{SM}}}+\frac{C_{P}^{\tau\tau}}{C_{10}^{\mathrm{SM}}}\frac{M_{B_{s}}^{2}}{2m_{\tau}\left(m_{b}+m_{s}\right)}\right|^{2}\nonumber \\
 & +\left(1-\frac{4m_{\tau}^{2}}{M_{B_{s}}^{2}}\right)\left|\frac{C_{S}^{\tau\tau}}{C_{10}^{\mathrm{SM}}}\frac{M_{B_{s}}^{2}}{2m_{\tau}\left(m_{b}+m_{s}\right)}\right|^{2}\bigg\}\label{eq:Bs_tautau}\\
\mathcal{B}(B^{+}\rightarrow K^{+}\tau\tau) & =10^{-9}\bigg(2.2\left|C_{9}^{\tau\tau}+C_{9}^{\mathrm{SM}}\right|^{2}+6.0\left|C_{10}^{\tau\tau}+C_{10}^{\mathrm{SM}}\right|^{2}+8.3\left|C_{S}^{\tau\tau}\right|^{2}\nonumber \\
 & +8.9\left|C_{P}^{\tau\tau}\right|^{2}+4.8\mathrm{Re}[C_{S}^{\tau\tau}(C_{9}^{\tau\tau}+C_{9}^{\mathrm{SM}})^{*}]+5.9\mathrm{Re}[C_{P}^{\tau\tau}(C_{10}^{\tau\tau}+C_{10}^{\mathrm{SM}})^{*}]\bigg)\,, \label{eq:B_Ktautau}
\end{flalign}
where we use $\mathcal{B}(B_{s}\rightarrow\tau\tau)_{\mathrm{SM}}=\left(7.73\pm0.49\right)\times10^{-7}$
\cite{Bobeth:2013uxa}. The numerical values for the NP contributions
to $B^{+}\to K^{+}\tau\tau$ decays are taken from \cite{Bouchard:2013mia}.

It is interesting to introduce the matching between the operators
in the basis of Eq.~(\ref{eq:bstautau}) and the San Diego LEFT basis.
In particular, the operators $C_{9}^{\tau\tau}$ and $C_{10}^{\tau\tau}$
are related to $[C_{ed}^{V,LL}]^{\tau\tau23}$ and $[C_{de}^{V,LR}]^{23\tau\tau}$
via
\begin{flalign}
C_{9}^{\tau\tau}=\frac{2\pi}{\alpha_{\mathrm{\mathrm{EM}}}V_{ts}^{*}V_{tb}}\left([C_{ed}^{V,LL}]^{\tau\tau23}+[C_{de}^{V,LR}]^{23\tau\tau}\right)\,,\label{eq:C9C10_CedV}\\
C_{10}^{\tau\tau}=\frac{2\pi}{\alpha_{\mathrm{\mathrm{EM}}}V_{ts}^{*}V_{tb}}\left([C_{de}^{V,LR}]^{23\tau\tau}-[C_{ed}^{V,LL}]^{\tau\tau23}\right)\,,\nonumber 
\end{flalign}
while the operators $[C_{ed}^{S,LL}]^{\tau\tau23}$ and $[C_{ed}^{S,LR}]^{\tau\tau23}$
are related to $C_{S}^{\tau\tau}$ and $C_{P}^{\tau\tau}$ via
\begin{flalign}
C_{S}^{\tau\tau}=\frac{2\pi}{\alpha_{\mathrm{\mathrm{EM}}}V_{ts}^{*}V_{tb}}\left([C_{ed}^{S,RR}]^{\tau\tau23}-[C_{ed}^{S,LR}]^{\tau\tau23}\right)\,,\label{eq:CSCP_CedS}\\
C_{P}^{\tau\tau}=\frac{2\pi}{\alpha_{\mathrm{\mathrm{EM}}}V_{ts}^{*}V_{tb}}\left([C_{ed}^{S,RR}]^{\tau\tau23}+[C_{ed}^{S,LR}]^{\tau\tau23}\right)\,.\nonumber
\end{flalign}
Even though the current experimental bounds over $\mathcal{B}(B_{s}\rightarrow\tau\tau)$
and $\mathcal{B}(B\rightarrow K\tau\tau)$ are weak, specific NP models
that address the $R_{D^{(*)}}$ anomalies predict a large enhancement of these
processes that might be testable in the near future. The enhancement
is larger in models featuring scalar operators, in particular for
$\mathcal{B}(B_{s}\rightarrow\tau\tau)$, since these provide a significant
chiral enhancement. Current bounds and future projections given by
the experimental collaborations are given in Table~\ref{tab:Bounds_bstautau}.
\begin{table}[t]
\begin{centering}
\begin{tabular}{ccc}
\toprule 
\multirow{3}{*}{$\mathcal{B}(B_{s}\rightarrow\tau\tau)$} & LHCb ($36\,\mathrm{fb}^{-1}$) (current) & $<6.8\times10^{-3}$ (95\% CL) \cite{LHCb:2017myy}\tabularnewline
\cmidrule{2-3} \cmidrule{3-3} 
 & LHCb (50 $\mathrm{fb}^{-1}$) & $<1.3\times10^{-3}$ (95\% CL) \cite{LHCb:2018roe}\tabularnewline
\cmidrule{2-3} \cmidrule{3-3} 
 & LHCb (300 $\mathrm{fb}^{-1}$) & $<5\times10^{-4}$ (95\% CL) \cite{LHCb:2018roe}\tabularnewline
\midrule
\midrule 
\multirow{3}{*}{$\mathcal{B}(B^{+}\rightarrow K^{+}\tau\tau)$} & BaBar (424 $\mathrm{fb}^{-1}$) (current) & $<2.25\times10^{-3}$ (90\% CL) \cite{BaBar:2016wgb}\tabularnewline
\cmidrule{2-3} \cmidrule{3-3} 
 & Belle II (5 $\mathrm{ab}^{-1}$) & $<6.5\times10^{-5}$ (95\% CL) \cite{Belle-II:2018jsg}\tabularnewline
\cmidrule{2-3} \cmidrule{3-3} 
 & Belle II (50 $\mathrm{ab}^{-1}$) & $<2.0\times10^{-5}$ (95\% CL) \cite{Belle-II:2018jsg}\tabularnewline
\bottomrule
\end{tabular}
\par\end{centering}
\caption[Current and projected bounds for $\mathcal{B}(B_{s}\rightarrow\tau\tau)$
and $\mathcal{B}(B^{+}\rightarrow K^{+}\tau\tau)$ as given by the
experimental collaborations]{ Current and projected bounds for $\mathcal{B}(B_{s}\rightarrow\tau\tau)$
and $\mathcal{B}(B^{+}\rightarrow K^{+}\tau\tau)$ as given by the
experimental collaborations. \label{tab:Bounds_bstautau}}
\end{table}

Beyond the direct bounds obtained by the experimental collaborations,
in \cite{Bordone:2023ybl} we studied alternative flavour observables
that have the potential to provide indirect bounds over $b\rightarrow s\tau\tau$.
We present these results in the following section, along with their application to the
well motivated example of the vector leptoquark $U_{1}\sim(\mathbf{3},\mathbf{1},2/3)$.

\subsection{\texorpdfstring{$\tau_{B_{s}}/\tau_{B_d}$: the $U_{1}\sim(\mathbf{3,1},2/3)$ vector leptoquark example}{tauBs/tauBd: the U1(3,1,2/3) vector leptoquark example}} \label{subsec:LifetimeRationU1}

Beyond the direct bounds obtained by the experimental collaborations
over $b\rightarrow s\tau\tau$ observables, in \cite{Bordone:2023ybl}
we studied alternative flavour observables that have the potential
to provide indirect bounds over $b\rightarrow s\tau\tau$. Our study
highlights that NP effects in $b\rightarrow s\tau\tau$ operators
also affect the lifetime ratio of $B_{s}$ and $B_{d}$ mesons. If
we assume no NP effects in the $B_{d}$ lifetime, we have
\begin{equation}
\frac{\tau_{B_{s}}}{\tau_{B_{d}}}=\left(\frac{\tau_{B_{s}}}{\tau_{B_{d}}}\right)_{\mathrm{SM}}\left(1+\frac{\Gamma(B_{s}\rightarrow\tau\tau)_{\mathrm{NP}}}{\Gamma(B_{s})_{\mathrm{SM}}}\right)^{-1}\,,\label{eq:LifetimeRatio_NP}
\end{equation}
where we define $\Gamma(B_{s}\rightarrow\tau\tau)_{\mathrm{NP}}=\Gamma(B_{s}\rightarrow\tau\tau)_{\mathrm{total}}-\Gamma(B_{s}\rightarrow\tau\tau)_{\mathrm{SM}}$,
which encodes the NP contribution to the partial decay width. The
expression for $\Gamma(B_{s}\rightarrow\tau\tau)_{\mathrm{total}}$
can be extracted from Eq.~(\ref{eq:Bs_tautau}). The SM prediction
for the lifetime ratio can be found in Ref.~\cite{Lenz:2022rbq},
and it depends on non-perturbative parameters in the Heavy Quark Expansion
as well as on the size of $SU(3)_{f}$ breaking between the $B_{s}$
and the $B_{d}$ system. We employ the central values and errors for
the expectation values of the next-to-leading power matrix element
in the $B_{d}$ field from \cite{Bordone:2021oof}. Concerning the
size of $SU(3)_{f}$ breaking, estimates using Heavy-Quark Effective
Theory relations \cite{Bordone:2022qez,Lenz:2022rbq} and preliminary
lattice QCD estimations \cite{Gambino:2017vkx,Gambino:2017dfa} are
affected by large errors. To be very conservative, we use the central
values from \cite{Bordone:2022qez} and assign 100\% errors. With
this, we obtain: 
\begin{align}
\left(\frac{\tau_{B_{s}}}{\tau_{B_{d}}}\right)_{\mathrm{SM}}=1.02\pm0.02\,, &  & %1.0279\pm0.0113\,,
\Gamma(B_{s})_{\mathrm{SM}}=0.597_{-0.069}^{+0.106}\,\mathrm{ps^{-1}}\,,\label{eq:LifetimeRatio_SM}
\end{align}
that has to be compared with the current experimental HFLAV average
\cite{HFLAV:2022wzx}, 
\begin{equation}
\left(\frac{\tau_{B_{s}}}{\tau_{B_{d}}}\right)_{\mathrm{HFLAV\,2022}}=1.001\pm0.004\,.
\end{equation}
At the current status, we find good agreement between the SM predictions
and the lifetime average, albeit with large uncertainties due to the
unknown $SU(3)_{f}$ breaking.

We note that in the literature, it has been discussed the impact of
using the values from a different set of non-perturbative parameters
in the lifetime ratio \cite{Bernlochner:2022ucr,Lenz:2022rbq}, which
yield to a large shift. However, it has to be noticed that the values
for these parameters change a lot depending on whether higher dimensional
operators are considered or not, hinting at non-trivial correlations.
This is not observed in \cite{Bordone:2021oof}, that we adopt as
our reference. We can now extract an indirect limit over NP contributions
to $\mathcal{B}(B_{s}\to\tau\tau)$ from the lifetime ratio. Using
Eq.~(\ref{eq:LifetimeRatio_NP}), we obtain 
\begin{equation}
\mathcal{B}(B_{s}\rightarrow\tau\tau)_{\mathrm{NP}}<5.5\times10^{-2}\,,
\end{equation}
at the 95\% CL, which has to be compared with the direct bound from
LHCb \cite{LHCb:2017myy}, namely $\mathcal{B}(B_{s}\rightarrow\tau\tau)_{\mathrm{NP}}<6.8\times10^{-3}$.
Currently, the direct bound over $\mathcal{B}(B_{s}\rightarrow\tau\tau)_{\mathrm{NP}}$
obtained by the LHCb collaboration is 40\% better than the indirect
bound obtained from the lifetime ratio.
\begin{table}[t]
\begin{centering}
\resizebox{0.92\textwidth}{!}{
\begin{tabular}{|c|c|c|c|c|}
\cline{2-5} \cline{3-5} \cline{4-5} \cline{5-5} 
\multicolumn{1}{c|}{} &  & \multicolumn{3}{c|}{$\mathcal{B}(B_{s}\rightarrow\tau\tau)_{\mathrm{NP}}$}\tabularnewline
\hline 
Assumptions  & Input  & $\tau_{B_{s}}/\tau_{B_{d}}$  & LHCb ($50\,\mathrm{fb}^{-1}$)  & LHCb (300 $\mathrm{fb}^{-1}$)\tabularnewline
\hline 
\multirow{3}{*}{H1} & \multirow{2}{*}{$\left(\frac{\tau_{B_{s}}}{\tau_{B_{d}}}\right)_{\mathrm{SM}}=1.020(5)$} & \multirow{4}{*}{$1.7(6)\cdot10^{-2}$} &  & \multicolumn{1}{c|}{}\tabularnewline
 &  &  &  & \tabularnewline
\cline{2-2} 
 & $\left(\frac{\tau_{B_{s}}}{\tau_{B_{d}}}\right)_{\mathrm{exp}}=1.001(1)$  &  & \multirow{3}{*}{$<1.3\cdot10^{-3}$} & \multirow{3}{*}{$<5\cdot10^{-4}$}\tabularnewline
\cline{1-3} \cline{2-3} \cline{3-3} 
\multirow{3}{*}{H2} & \multirow{2}{*}{$\left(\frac{\tau_{B_{s}}}{\tau_{B_{d}}}\right)_{\mathrm{SM}}=1.001(1)$} & \multirow{4}{*}{$<2.6\cdot10^{-3}$} &  & \tabularnewline
 &  &  &  & \tabularnewline
\cline{2-2} 
 & $\left(\frac{\tau_{B_{s}}}{\tau_{B_{d}}}\right)_{\mathrm{exp}}=1.001(1)$  &  &  & \tabularnewline
\hline 
\end{tabular}}
\par\end{centering}
\caption[Projected bounds at 95\% CL for $\mathcal{B}(B_{s}\rightarrow\tau\tau)_{\mathrm{NP}}$
obtained from the lifetime ratio $\tau_{B_{s}}/\tau_{B_{d}}$ confronted
against the projected bounds from LHCb]{Projected bounds at 95\% CL for $\mathcal{B}(B_{s}\rightarrow\tau\tau)_{\mathrm{NP}}$
obtained from the lifetime ratio $\tau_{B_{s}}/\tau_{B_{d}}$ are
confronted against the projected bounds from LHCb \cite{LHCb:2018roe}.
For the projections in the lifetime ratio, we assume that the uncertainties
will reduce to 1 per mille in the experiment. We display different
results under two different hypothesis for the SM prediction: that
the central value will remain as the current one and the $SU(3)_{F}$
breaking parameters could be measured to a 10\% precision (H1), and
that the central value will shift to match the experiment and there
is no $SU(3)_{F}$ breaking up to the per mille level (H2).}
\label{tab:Lifetime_Projected} 
\end{table}

We then repeat this comparison with the projected sensitivities. The
results are shown in Table~\ref{tab:Lifetime_Projected}. For the
experimental measurement, we explore the possibility that the error
will reduce to 1 per mille. For the SM prediction, we explore two
hypotheses corresponding to either no change in the central value
or a substantial reduction of it, towards a strong indication of small
$SU(3)_{f}$ breaking. In hypothesis H1, we assume that the $SU(3)_{f}$
breaking parameters could be measured to a 10\% precision, as possible
in the foreseeable future using lattice QCD, but retaining the current
central values, while in H2 we impose no $SU(3)_{f}$ breaking up
to the per mill level. The LHCb collaboration provides two expected
upper bounds for $\mathcal{B}(B_{s}\to\tau\tau)$: a first projection
is based on a luminosity of 50 $\mathrm{fb^{-1}}$, which in contrast
to the expectations in \cite{LHCb:2018roe} will be reached only after
2032. The second upper bound from the LHCb collaboration is based
on an expected luminosity of 300 $\mathrm{fb^{-1}}$, which with respect
to the expectations in \cite{LHCb:2018roe}, will be reached only
after 2041. This shows that improved measurements and predictions
of the lifetime ratios have the potential of improving the current
bound on $\mathcal{B}(B_{s}\to\tau\tau)$, while waiting for LHCb
to collect the necessary statistics to obtain even more stringent
bounds. This motivates extra efforts from both the theoretical and
experimental communities to investigate $\tau_{B_{s}}/\tau_{B_{d}}$
as a potential channel to constrain NP effects.

\subsubsection*{Application to the vector leptoquark $U_{1}\sim(\mathbf{3},\mathbf{1},2/3)$}

The $U_{1}\sim(\mathbf{3},\mathbf{1},2/3)$ vector leptoquark is a
well motivated mediator to explain the $R_{D}$ and $R_{D^{*}}$ anomalies
\cite{Cornella:2019hct,Cornella:2021sby,Greljo:2018tuh,Calibbi:2017qbu,Blanke:2018sro,DiLuzio:2017vat,DiLuzio:2018zxy,Aebischer:2022oqe}.
A gauge $U_{1}$ leptoquark is predicted by the Pati-Salam group \cite{Pati:1974yy},
that provides a natural connection with quark-lepton unification.
Moreover, explanations of the $R_{D^{(*)}}$ anomalies via exchange
of the $U_{1}$ vector leptoquark had been shown to be naturally connected
with the origin of flavour hierarchies and the flavour structure of
the SM \cite{Bordone:2017bld,Bordone:2018nbg,Allwicher:2020esa,Fuentes-Martin:2020pww,Fuentes-Martin:2022xnb,Davighi:2022bqf,King:2021jeo,FernandezNavarro:2022gst}.
In this direction, we will provide in Chapter~\ref{Chapter:TwinPS} of this thesis a
theory of flavour containing a TeV scale $U_{1}$ leptoquark that can
explain the $B$-anomalies \cite{FernandezNavarro:2022gst}. Remarkably,
the contributions of the $U_{1}$ vector leptoquark to $R_{D^{(*)}}$
are correlated to an enhancement of $b\rightarrow s\tau\tau$, hence
potentially undergoing the constraints from $\tau_{B_{s}}/\tau_{B_{d}}$
described before.
\begin{figure}[t]
\begin{centering}
\includegraphics[scale=0.60]{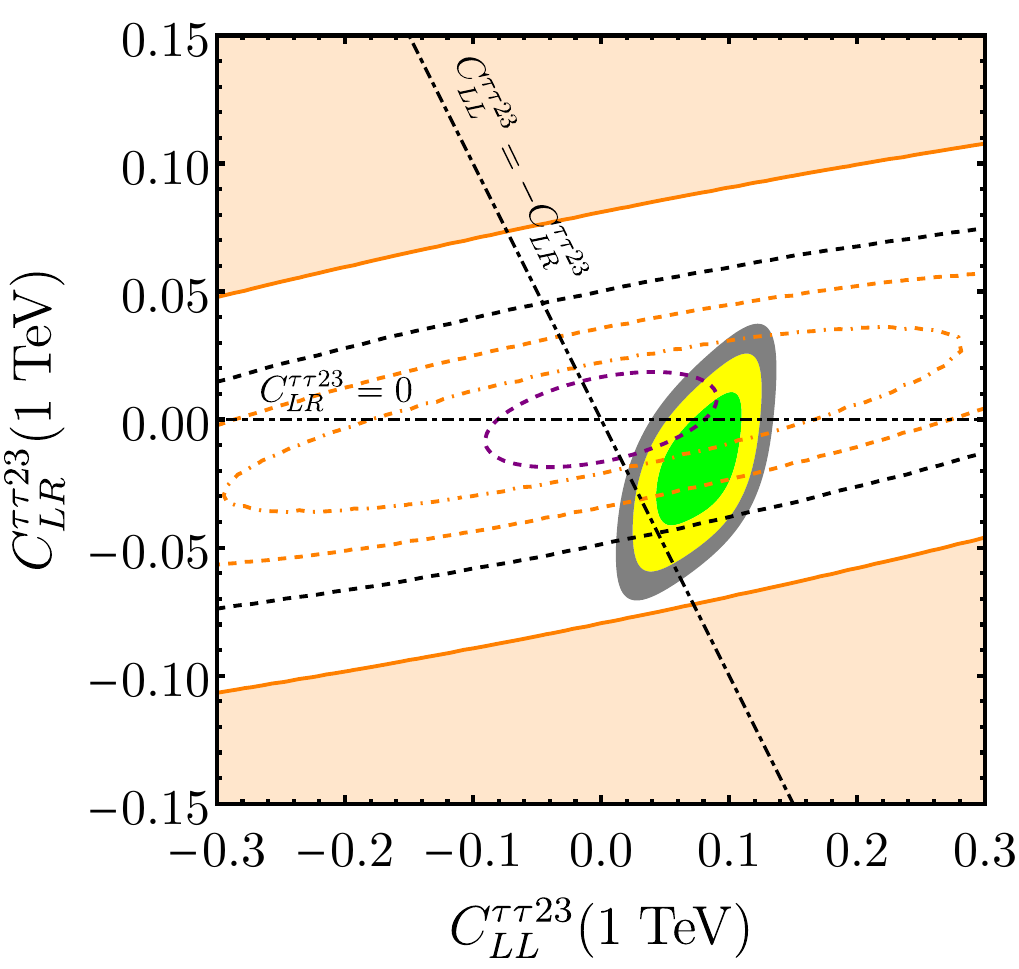}
\par\end{centering}
\caption[Model independent analysis of $U_{1}$ vector leptoquark: $R_{D^{(*)}}$
and $b\rightarrow s\tau\tau$]{Parameter space of Wilson coefficients motivated by the $U_{1}$
vector leptoquark explanation of $R_{D^{(*)}}$ (see main text). The
green, yellow and grey regions represent the $1\sigma$, $2\sigma$
and $3\sigma$ regions preferred by $R_{D^{(*)}}$, respectively.
Orange contours represent the direct bounds from $\mathcal{B}(B_{s}\rightarrow\tau\tau)$,
while the black contour represents the indirect bound obtained from
$\tau_{B_{s}}/\tau_{B_{d}}$ and the purple contour represents the
direct bounds from $\mathcal{B}(B^{+}\rightarrow K^{+}\tau\tau)$.
Solid (dashed) contours denote current (projected) 95\% CL exclusions,
except for the two projections for $\mathcal{B}(B_{s}\rightarrow\tau\tau)$
by LHCb: 50 $\mathrm{fb}^{-1}$ (orange dashed) and 300 $\mathrm{fb}^{-1}$
(orange dash-dotted). Dash-dotted black lines represent two interesting
benchmark scenarios motivated in the main text.}
\label{Fig:U1_LifetimeRatios}
\end{figure}

At an effective scale $\Lambda$ higher than the electroweak scale,
the $U_{1}$ interactions are well described in the context of the
SMEFT as:
\begin{equation}
\mathcal{L}_{\mathrm{SMEFT}}^{U_{1}}\supset-\frac{1}{\Lambda^{2}}\left[\frac{C_{LL}^{\alpha\beta ij}}{2}\left(Q_{\ell q}^{(1)}+Q_{\ell q}^{(3)}\right)^{\alpha\beta ij}-\left(2C_{LR}^{\alpha\beta ij}\left(Q_{\ell edq}^{\dagger}\right)^{\alpha\beta ij}+\mathrm{h.c.}\right)\right]\,.
\end{equation}
The matching between the relevant LEFT and SMEFT Wilson coefficients
reads: 
\begin{align}
 & \left[C_{\nu edu}^{V,LL}\right]^{\tau\tau32*}(m_{b})=\eta_{V}^{\tau\nu}C_{LL}^{\tau\tau23}(\Lambda)\frac{v^{2}}{2V_{cb}\Lambda^{2}}\,,\label{eq:U1_ops1}\\
 & \left[C_{\nu edu}^{S,RL}\right]^{\tau\tau32*}(m_{b})=-\eta_{S}^{\tau\nu}2C_{LR}^{\tau\tau23}(\Lambda)\frac{v^{2}}{2V_{cb}\Lambda^{2}}\,,\nonumber \\
 & \left[C_{ed}^{V,LL}\right]^{\tau\tau23}(m_{b})=\eta_{V}^{\tau\tau}C_{LL}^{\tau\tau23}(\Lambda)\frac{v^{2}}{2\Lambda^{2}}\,,\label{eq:U1_ops3}\\
 & \left[C_{ed}^{S,LR}\right]^{\tau\tau23}(m_{b})=-\eta_{S}^{\tau\tau}2C_{LR}^{\tau\tau23}(\Lambda)\frac{v^{2}}{2\Lambda^{2}}\,,\nonumber 
\end{align}
where the factors $\eta_{i}^{\tau\tau}$ and $\eta_{i}^{\tau\nu}$
encode the running from the high scale $\Lambda=1\,\mathrm{TeV}$
and are evaluated with \texttt{DsixTools 2.1} \cite{Fuentes-Martin:2020zaz}, obtaining
$\eta_{V}^{\tau\tau}\simeq0.96$, $\eta_{S}^{\tau\tau}\simeq1.57$
$\eta_{V}^{\tau\nu}\simeq1.03$ and $\eta_{S}^{\tau\nu}\simeq1.64$.
The operators in Eq.~(\ref{eq:U1_ops3}) match into $C_{9}^{\tau\tau}=-C_{10}^{\tau\tau}$
and $C_{S}^{\tau\tau}=-C_{P}^{\tau\tau}$ via Eq.~(\ref{eq:C9C10_CedV})
and Eq.~(\ref{eq:CSCP_CedS}), respectively. The presence of the
scalar operator $\left[C_{ed}^{S,LR}\right]^{\tau\tau23}$, which ultimately
provides $C_{S}^{\tau\tau}=-C_{P}^{\tau\tau}$, delivers a chirally
enhanced contribution to $\mathcal{B}(B_{s}\rightarrow\tau\tau)$
connected to the size of $C_{LR}^{\tau\tau23}$. If $C_{LR}^{\tau\tau23}=0$,
then $\mathcal{B}(B_{s}\rightarrow\tau\tau)$ is still substantially
enhanced by the presence of $C_{9}^{\tau\tau}=-C_{10}^{\tau\tau}$,
but chiral enhancement is lost.

In Fig.~\ref{Fig:U1_LifetimeRatios} we explore the parameter space
of SMEFT Wilson coefficients in the model, highlighting two particularly
motivated benchmark scenarios. The case $C_{LL}^{\tau\tau23}=-C_{LR}^{\tau\tau23}$
is a good benchmark for 4321 models featuring TeV scale third family
quark-lepton unification \cite{Bordone:2017bld,Cornella:2019hct,Cornella:2021sby,Greljo:2018tuh,Allwicher:2020esa,Fuentes-Martin:2020pww,Fuentes-Martin:2022xnb,Davighi:2022bqf},
while the case $C_{LR}^{\tau\tau23}=0$ is a good benchmark for the
flavour universal (fermiophobic) 4321 model \cite{DiLuzio:2017vat,DiLuzio:2018zxy,King:2021jeo,FernandezNavarro:2022gst},
including the twin Pati-Salam theory of flavour \cite{FernandezNavarro:2022gst}
introduced in Chapter~\ref{Chapter:TwinPS} of this thesis. 

Given that $U_{1}$ is a vector leptoquark, the leading contribution
to $\Delta M_{s}$ arising at 1-loop depends on the specific UV completion.
For the well-motivated case of 4321 models, the contribution to $\Delta M_{s}$
is dominated by a vector-like lepton running in the loop, and the
most stringent constraints can be avoided as long as the mass of the
vector-like lepton is around or below the TeV scale \cite{DiLuzio:2018zxy,Cornella:2021sby,FernandezNavarro:2022gst}.
Similarly, the $U_{1}$ leptoquark also avoids tree-level contributions
to $b\rightarrow s\nu\nu$ transitions. In this manner, the model
is able to address $R_{D^{(*)}}$ and the enhancement of $\mathcal{B}(B_{s}\rightarrow\tau\tau)$
becomes a key prediction.

Due to chiral enhancement, $\mathcal{B}(B_{s}\rightarrow\tau\tau)$
is particularly sensitive to scenarios with large $\left|C_{LR}^{\tau\tau23}\right|$,
but current direct bounds from LHCb cannot yet test the preferred
region by the benchmark case $C_{LL}^{\tau\tau23}=-C_{LR}^{\tau\tau23}$.
Remarkably, in the near future we expect the indirect bound from the
lifetime ratio $\tau_{B_{s}}/\tau_{B_{d}}$ \cite{Bordone:2023ybl} to constrain a significant region
of the parameter space preferred by $C_{LL}^{\tau\tau23}=-C_{LR}^{\tau\tau23}$,
while the parameter space preferred by $C_{LR}^{\tau\tau23}=0$ is
expected to remain unconstrained. In the longer term, projected direct
measurements of $\mathcal{B}(B_{s}\rightarrow\tau\tau)$ and $\mathcal{B}(B^{+}\rightarrow K^{+}\tau\tau)$
have the potential to test most of the parameter space preferred
by $R_{D}$ and $R_{D^{*}}$.

In conclusion, the lifetime ratio $\tau_{B_{s}}/\tau_{B_{d}}$ could
be able to discriminate between different $U_{1}$ models explaining
$R_{D^{(*)}}$ in the near future, setting strong constraints over
models featuring TeV scale quark-lepton unification \cite{Bordone:2017bld,Cornella:2019hct,Cornella:2021sby,Greljo:2018tuh,Allwicher:2020esa,Fuentes-Martin:2020pww,Fuentes-Martin:2022xnb,Davighi:2022bqf},
while the twin Pati-Salam model \cite{FernandezNavarro:2022gst} is expected to remain as the only
viable theory of flavour containing a $U_{1}$ leptoquark that explains
the $R_{D^{(*)}}$ anomalies.

\subsection{\texorpdfstring{$b\rightarrow s\nu\nu$}{bsnunu}} \label{subsec:bsnunu}

The process $b\rightarrow s\nu\nu$ is correlated to the enhancement
of $b\rightarrow c\tau\nu$ in particular NP scenarios via $SU(2)_{L}$ invariance.
Given the significant bounds over $b\rightarrow s\nu\nu$ enhancement,
this transition allows to discriminate between different NP models
proposed to address the $R_{D^{(*)}}$ anomalies. We define the relevant
Lagrangian to describe $b\rightarrow s\nu\nu$ transitions as
\begin{equation}
\mathcal{L}_{b\rightarrow s\nu\nu}=\frac{4G_{F}}{\sqrt{2}}V_{tb}V_{ts}^{*}\left(C_{\nu,\mathrm{NP}}^{\alpha\beta}+C_{\nu,\mathrm{SM}}\right)\left(\bar{s}_{L}\gamma_{\mu}b_{L}\right)\left(\bar{\nu}^{\alpha}_{L}\gamma^{\mu}\nu^{\beta}_{L}\right)+\mathrm{h.c.}\,,\label{eq:B_Knunu_Effective}
\end{equation}
plus the primed operators that involve the exchange $L\rightarrow R$
in the quark fields, however these operators are highly constrained
by data so we do not consider them. The universal SM contribution
reads
\begin{equation}
C_{\nu,\mathrm{SM}}=-\frac{\alpha_{L}}{2\pi}X_{t}\,,
\end{equation}
where $X_{t}=1.48\pm0.01$ \cite{Buchalla:1998ba}, and $\alpha_{L}=g_{L}^{2}/(4\pi)$
as defined in Eq.~(\ref{eq:alpha_couplings}). In the context of models
explaining $R_{D^{(*)}}$, we expect the NP to couple mostly to the
third family (this structure is also well motivated from the point
of view of a theory of flavour, as we shall see), therefore we will
assume that all $C_{\nu,\mathrm{NP}}^{\alpha\beta}$ are negligible
except for $C_{\nu,\mathrm{NP}}^{\tau\tau}$. It is interesting to
study the deviation from the SM due to the NP effects. In this direction,
we define
\begin{equation}
\delta\mathcal{B}(B\rightarrow K^{(*)}\nu\bar{\nu})=\frac{\mathcal{B}(B\rightarrow K^{(*)}\nu\bar{\nu})}{\mathcal{B}(B\rightarrow K^{(*)}\nu\bar{\nu})_{\mathrm{SM}}}-1\approx\frac{1}{3}\left|\frac{C_{\nu,\mathrm{NP}}^{\tau\tau}+C_{\nu\nu}^{\mathrm{SM}}}{C_{\nu\nu}^{\mathrm{SM}}}\right|^{2}-\frac{1}{3}\,,\label{eq:BtoK_nunu}
\end{equation}
where current data sets the following bounds \cite{BaBar:2013npw,Belle:2017oht}
\begin{equation}
\delta\mathcal{B}(B\rightarrow K^{(*)}\nu\bar{\nu})<2.6\;(1.7)\;(90\%\;\mathrm{CL})\,.
\end{equation}
We note however that in Summer 2023 the Belle II collaboration presented
the following measurement \cite{BelleIIEPS:2023}
\begin{equation}
\delta\mathcal{B}(B^{+}\rightarrow K^{+}\nu\bar{\nu})=3.8\pm1.5\,,
\end{equation}
which shows a $2.8\sigma$ tension with the SM prediction. When combined
with the previous measurements by Belle and BaBar \cite{BaBar:2013npw,Belle:2017oht} the significance
is reduced to the $2.2\sigma$ level.
In any case, current data implies that $b\rightarrow s\nu\nu$ cannot be enhanced
much above the SM prediction, while in the previous sections we have
shown that $b\rightarrow s\tau\tau$ is enhanced several orders of
magnitude above the SM prediction in models that explain $R_{D^{(*)}}$.
Therefore, $b\rightarrow s\nu\nu$ sets important constraints to NP
models where $b\rightarrow s\nu\nu$ and $b\rightarrow c\tau\nu$
are strongly correlated. This is the case for the $S_{3}\sim(\mathbf{\overline{3},3,}1/3)$
scalar leptoquark\footnote{Notice that in the $S_{1}\sim(\mathbf{\overline{3},1,}1/3)$ leptoquark model,
the explanation of $R_{D^{(*)}}$ via $\left[C_{\nu edu}^{V,LL}\right]^{\tau\tau32*}$
receives constraints from $b\rightarrow s\nu\nu$ as well, but these constraints
can be avoided if $R_{D^{(*)}}$ are explained via scalar and tensor
operators (which involve right-handed couplings) \cite{Angelescu:2021lln}.}. Instead, the $U_{1}\sim(\mathbf{3},\mathbf{1},2/3)$ model avoids
tree-level contributions to $b\rightarrow s\nu\nu$, but contributions
at 1-loop can be relevant (although they depend on the specific UV
completion, see e.g.~Section~\ref{subsec:Signals-in-rare-processes}).

Remarkably, the Belle II collaboration is expected to measure $\mathcal{B}(B\rightarrow K^{(*)}\nu\bar{\nu})$
up to 10\% of the SM value \cite{Belle-II:2018jsg}, hence confirming
the anomaly hinted in \cite{BelleIIEPS:2023} and testing most of the remaining NP models
which contribute to $b\rightarrow s\nu\nu$ either at tree-level or
1-loop.

\subsection{Purely leptonic CLFV processes} \label{subsec:leptonic_CLFV}

Processes involving violation of family lepton number via charged
lepton transitions are strongly constrained by data. In order to study
CLFV decays of the type $e_{\beta}^{-}\rightarrow e_{\alpha}^{-}\gamma$
and $e_{\beta}^{-}\rightarrow e_{\alpha}^{-}e_{\alpha}^{+}e_{\alpha}^{-}$,
where $\alpha\neq\beta$, we define the following effective Lagrangian
built from operators in the San Diego basis of the LEFT,
\begin{flalign}
\mathcal{L}_{\mathrm{leptonic\,LFV}}= & -\frac{4G_{F}}{\sqrt{2}}\left[m_{\beta}\left[C_{e\gamma}\right]^{\alpha\beta}\bar{e}_{L}^{\alpha}\sigma^{\mu\nu}e_{R}^{\beta}F_{\mu\nu}+m_{\beta}\left[C_{e\gamma}\right]^{\beta\alpha}\bar{e}_{L}^{\beta}\sigma^{\mu\nu}e_{R}^{\alpha}F_{\mu\nu}\right.\\
 & +\left[C_{ee}^{V,LL}\right]^{\alpha\beta\alpha\alpha}(\bar{e}_{L}^{\alpha}\gamma_{\mu}e_{L}^{\beta})(\bar{e}_{L}^{\alpha}\gamma^{\mu}e_{L}^{\alpha})+\left[C_{ee}^{V,LR}\right]^{\alpha\beta\alpha\alpha}(\bar{e}_{L}^{\alpha}\gamma_{\mu}e_{L}^{\beta})(\bar{e}_{R}^{\alpha}\gamma^{\mu}e_{R}^{\alpha})\nonumber \\
 & +\left[C_{ee}^{V,LR}\right]^{\alpha\alpha\alpha\beta}(\bar{e}_{L}^{\alpha}\gamma_{\mu}e_{L}^{\alpha})(\bar{e}_{R}^{\alpha}\gamma^{\mu}e_{R}^{\beta})+\left[C_{ee}^{V,RR}\right]^{\alpha\beta\alpha\alpha}(\bar{e}_{R}^{\alpha}\gamma_{\mu}e_{R}^{\beta})(\bar{e}_{R}^{\alpha}\gamma^{\mu}e_{R}^{\alpha})\nonumber \\
 & \left.+\left[C_{ee}^{S,RR}\right]^{\alpha\beta\alpha\alpha}(\bar{e}_{L}^{\alpha}e_{R}^{\beta})(\bar{e}_{L}^{\alpha}e_{R}^{\alpha})+\left[C_{ee}^{S,RR}\right]^{\beta\alpha\alpha\alpha}(\bar{e}_{L}^{\beta}e_{R}^{\alpha})(\bar{e}_{L}^{\alpha}e_{R}^{\alpha})\right]+\mathrm{h.c.}\,,\nonumber 
\end{flalign}
where all Wilson coefficients are dimensionless, including those of the dipole operators.
In terms of this basis of effective operators, the branching fraction of the
$e_{\beta}^{-}\rightarrow e_{\alpha}^{-}\gamma$ decay is given by
(neglecting small corrections proportional to the lightest lepton
mass) \cite{Kuno:1999jp}
\begin{equation}
\mathcal{B}(e_{\beta}^{-}\rightarrow e_{\alpha}^{-}\gamma)=\left(\frac{4 G_{F}}{\sqrt{2}}\right)^{2}\alpha_{\mathrm{EM}}m^{5}_{\beta}\tau_{\beta}\left(\left|\left[C_{e\gamma}\right]^{\alpha\beta}\right|^{2}+\left|\left[C_{e\gamma}\right]^{\beta\alpha}\right|^{2}\right)\,,\label{eq:tau_mu_photon}
\end{equation}
where $\tau_{\beta}$ is the lifetime of the $\beta=e,\mu,\tau$ charged lepton. All the processes of the form $e_{\beta}\rightarrow e_{\alpha}\gamma$
, namely $\mu\rightarrow e\gamma$, $\tau\rightarrow\mu\gamma$ and
$\tau\rightarrow e\gamma$ are predicted by well motivated BSM models,
and this has motivated an intensive search by the experimental collaborations.
In the absence of any NP signal so far, strong bounds are set over
the various branching fractions. The bound over the $\mu\rightarrow e\gamma$
process is particularly strong as \cite{PDG:2022ynf}
\begin{equation}
\mathcal{B}(\mu^{-}\rightarrow e^{-}\gamma)<3.1\times10^{-13}\;(90\%\;\mathrm{CL})\,,
\end{equation}
having the largest reach in NP scale out of all LFV processes, and
having a NP reach comparable to that of $K-\bar{K}$ mixing observables, as shown in Fig.~\ref{fig:Reach-in-NP_flavour}.
The processes involving tau lepton decays remain so far less constrained
as \cite{PDG:2022ynf}
\begin{equation}
\mathcal{B}(\tau^{-}\rightarrow\mu^{-}\gamma)<4.2\times10^{-8}\;(90\%\;\mathrm{CL})\,,
\end{equation}
\begin{equation}
\mathcal{B}(\tau^{-}\rightarrow e^{-}\gamma)<3.3\times10^{-8}\;(90\%\;\mathrm{CL})\,.
\end{equation}
These processes are interesting because they test scenarios with NP mostly
coupled to the third family that predict 2-3 and 1-3 charged lepton
mixing. A well motivated example are theories of flavour based on
the $U(2)^{5}$ flavour symmetry (see e.g.~\cite{FernandezNavarro:2023rhv}).

In order to describe the $e_{\beta}^{-}\rightarrow e_{\alpha}^{-}e_{\alpha}^{+}e_{\alpha}^{-}$
branching fractions, it is useful to introduce auxiliary variables,
\begin{flalign}
 & C_{1}=\frac{\left|\left[C_{ee}^{S,RR}\right]^{\alpha\beta\alpha\alpha}\right|^{2}}{16}+\left|\left[C_{ee}^{V,RR}\right]^{\alpha\beta\alpha\alpha}\right|^{2}\,,\\
 & C_{2}=\frac{\left|\left[C_{ee}^{S,RR}\right]^{\beta\alpha\alpha\alpha}\right|^{2}}{16}+\left|\left[C_{ee}^{V,LL}\right]^{\alpha\beta\alpha\alpha}\right|^{2}\,,\\
 & C_{3}=\left|\left[C_{ee}^{V,LR}\right]^{\alpha\alpha\alpha\beta}\right|^{2}\,,\\
 & C_{4}=\left|\left[C_{ee}^{V,LR}\right]^{\alpha\beta\alpha\alpha}\right|^{2}\,,\\
 & C_{5}=\left|e\left[C_{e\gamma}\right]^{\alpha\beta}\right|^{2}\,,\\
 & C_{6}=\left|e\left[C_{e\gamma}\right]^{\beta\alpha}\right|^{2}\,,\\
 & C_{7}=\mathrm{Re}\left[e\left[C_{e\gamma}\right]^{\alpha\beta}\left(\left[C_{ee}^{V,LL}\right]^{\alpha\beta\alpha\alpha}\right)^{*}\right]\,,\\
 & C_{8}=\mathrm{Re}\left[e\left[C_{e\gamma}\right]^{\beta\alpha}\left(\left[C_{ee}^{V,RR}\right]^{\alpha\beta\alpha\alpha}\right)^{*}\right]\,,\\
 & C_{9}=\mathrm{Re}\left[e\left[C_{e\gamma}\right]^{\alpha\beta}\left(\left[C_{ee}^{V,LR}\right]^{\alpha\beta\alpha\alpha}\right)^{*}\right]\,,\\
 & C_{10}=\mathrm{Re}\left[e\left[C_{e\gamma}\right]^{\beta\alpha}\left(\left[C_{ee}^{V,LR}\right]^{\alpha\alpha\alpha\beta}\right)^{*}\right]\,,
\end{flalign}
where $e=\sqrt{4\pi\alpha_{\mathrm{EM}}}$ is the QED gauge
coupling (also associated to the elementary electric charge). Having
defined these auxiliary variables, the branching fraction of the $e_{\beta}^{-}\rightarrow e_{\alpha}^{-}e_{\alpha}^{+}e_{\alpha}^{-}$
process is given by (neglecting small corrections proportional to the lightest
lepton mass) \cite{Kuno:1999jp}
\begin{flalign}
\mathcal{B}(e_{\beta}^{-}\rightarrow e_{\alpha}^{-}e_{\alpha}^{+}e_{\alpha}^{-})= & 2(C_{1}+C_{2})+(C_{3}+C_{4})+32\left[\log\left(\frac{m_{\beta}^{2}}{m_{\alpha}^{2}}\right)-\frac{11}{4}\right](C_{5}+C_{6})\\
 & +16(C_{7}+C_{8})+8(C_{9}+C_{10})\,.\nonumber \label{eq:tau_3mu}
\end{flalign}
Notice the significant enhancement factor $\log(m_{\beta}^{2}/m_{\alpha}^{2})$
that enters the branching fraction when the dipole operators are present (although it becomes negligible if dipoles are generated at 1-loop level).
The various processes $e_{\beta}^{-}\rightarrow e_{\alpha}^{-}e_{\alpha}^{+}e_{\alpha}^{-}$,
namely $\mu\rightarrow3e$, $\tau\rightarrow3\mu$ and $\tau\rightarrow3e$
are also well motivated by BSM models, and have been searched by the experimental
collaborations with no positive signal so far. The bounds read \cite{PDG:2022ynf},
\begin{flalign}
\qquad\qquad & \mathcal{B}(\mu^{-}\rightarrow e^{-}e^{+}e^{-})<1.0\times10^{-12} & \, & (90\%\;\mathrm{CL})\,, & \,\\
\qquad\qquad & \mathcal{B}(\tau^{-}\rightarrow\mu^{-}\mu^{+}\mu^{-})<2.1\times10^{-8} & \, & (90\%\;\mathrm{CL})\,, & \,\\
\qquad\qquad & \mathcal{B}(\tau^{-}\rightarrow e^{-}e^{+}e^{-})<2.7\times10^{-8} & \, & (90\%\;\mathrm{CL})\,. & 
\end{flalign}
Again the bound over the muon decay process is the strongest, just
below $\mathcal{B}(\mu^{-}\rightarrow e^{-}\gamma)$, while the bounds
over the tau decay processes are weaker. Before concluding, we notice
that current experiments are seeking to improve the current bounds
over the CLFV processes outlined in this section. We first highlight
the future bounds over $\mu\rightarrow e\gamma$ and $\mu\rightarrow3e$
projected by the MEG II and Mu3e collaborations (highlighting the
impressive bound projected by Mu3e) \cite{MEGII:2018kmf,Blondel:2013ia}
\begin{flalign}
\qquad\qquad & \mathcal{B}(\mu^{-}\rightarrow e^{-}\gamma)<6\times10^{-14}\qquad & \, & (90\%\;\mathrm{CL})\,, & \,\\
\qquad\qquad & \mathcal{B}(\mu^{-}\rightarrow e^{-}e^{+}e^{-})<10^{-16}\qquad & \, & (90\%\;\mathrm{CL})\,. & \,
\end{flalign}
Bounds over the processes involving tau decays are expected to be
improved by the Belle~II collaboration, which provides the following
projections \cite{Belle-II:2018jsg}
\begin{flalign}
\qquad\qquad & \mathcal{B}(\tau^{-}\rightarrow\mu^{-}\gamma)<4\times10^{-10} & \, & (90\%\;\mathrm{CL})\,, & \,\\
\qquad\qquad & \mathcal{B}(\tau^{-}\rightarrow\mu^{-}\mu^{+}\mu^{-})<3\times10^{-10} & \, & (90\%\;\mathrm{CL})\,, & \,\\
\qquad\qquad & \mathcal{B}(\tau^{-}\rightarrow e^{-}\gamma)<10^{-9} & \, & (90\%\;\mathrm{CL})\,, & \,\\
\qquad\qquad & \mathcal{B}(\tau^{-}\rightarrow e^{-}e^{+}e^{-})<5\times10^{-10} & \, & (90\%\;\mathrm{CL})\,. & \,
\end{flalign}
The impressive efforts by the experimental collaborations will allow
to test very well motivated BSM scenarios, including specific theories
of flavour presented in this thesis.

\subsection{Semileptonic CLFV processes} 
\label{subsec:Semileptonic-CLFV-processes}

Violation of family lepton number via charged lepton transitions is
also possible in semileptonic processes. We are particularly interested
in $b\rightarrow se_{\alpha}e_{\beta}$ processes, where $\alpha\neq\beta$,
which are predicted in several NP scenarios addressing the anomalies
in $R_{D^{(*)}}$ and/or $b\rightarrow s\mu\mu$ data. We define the
effective Lagrangian as
\begin{equation}
\begin{aligned}\mathcal{L}_{b\rightarrow se_{\alpha}e_{\beta}}=\frac{4G_{F}}{\sqrt{2}}V_{tb}V_{ts}^{*}\frac{\alpha_{\mathrm{\mathrm{EM}}}}{4\pi} & \bigg[C_{9}^{\alpha\beta}\mathcal{O}_{9}^{\alpha\beta}+C_{10}^{\alpha\beta}\mathcal{O}_{10}^{\alpha\beta}\\
 & +C_{S}^{\alpha\beta}\mathcal{O}_{S}^{\alpha\beta}+C_{P}^{\alpha\beta}\mathcal{O}_{P}^{\alpha\beta}\bigg]+\mathrm{h.c.}\,,
\end{aligned}
\label{eq:bsl1l2}
\end{equation}
where
\begin{align}
\mathcal{O}_{9}^{\alpha\beta}= & \left(\bar{s}\gamma_{\mu}P_{L}b\right)\left(\bar{e}^{\alpha}\gamma^{\mu}e^{\beta}\right)\,, & \mathcal{O}_{S}^{\alpha\beta}= & \left(\bar{s}P_{R}b\right)\left(\bar{e}^{\alpha}e^{\beta}\right)\,,\\
\mathcal{O}_{10}^{\alpha\beta}= & \left(\bar{s}\gamma_{\mu}P_{L}b\right)\left(\bar{e}^{\alpha}\gamma^{\mu}\gamma_{5}e^{\beta}\right)\,, & \mathcal{O}_{P}^{\alpha\beta}= & \left(\bar{s}P_{R}b\right)\left(\bar{e}^{\alpha}\gamma_{5}e^{\beta}\right)\,,
\end{align}
plus primed operators which are obtained by exchanging $L\longleftrightarrow R$
in the quark bilineals. Having defined the relevant set of effective
operators, the branching fraction of the process $B_{s}\rightarrow e_{\alpha}e_{\beta}$
is given by \cite{Becirevic:2016zri}
\begin{flalign}
\mathcal{B}(B_{s}  & \rightarrow e_{\alpha}e_{\beta})=\frac{\tau_{B_{s}}}{64\pi^{3}}\frac{\alpha_{\mathrm{EM}}^{2}G_{F}^{2}}{m_{B_{s}}^{3}}f_{B_{s}}^{2}\left|V_{tb}V_{ts}^{*}\right|^{2}\sqrt{f(m_{B_{s}},m_{\alpha},m_{\beta})}\label{eq:Bs_tau_mu}\\
 & \times\left\{ [m_{B_{s}}^{2}-(m_{\alpha}+m_{\beta})^{2}]\cdot\left|(C_{9}^{\alpha\beta}-C'^{\alpha\beta}_{9})(m_{\alpha}-m_{\beta})+(C_{S}^{\alpha\beta}-C'^{\alpha\beta}_{S})\frac{m_{B_{s}}^{2}}{m_{b}+m_{s}}\right|^{2}\right.\nonumber \\
 & \left.+[m_{B_{s}}^{2}-(m_{\alpha}-m_{\beta})^{2}]\cdot\left|(C_{10}^{\alpha\beta}-C'^{\alpha\beta}_{10})(m_{\alpha}+m_{\beta})+(C_{P}^{\alpha\beta}-C'^{\alpha\beta}_{P})\frac{m_{B_{s}}^{2}}{m_{b}+m_{s}}\right|^{2}\right\} \nonumber
\end{flalign}
where $f(a,b,c)=[a^{2}-(b-c)^{2}][a^{2}-(b+c)^{2}]$, and we use $f_{B_{s}}=230.3\pm1.3\;\mathrm{MeV}$
\cite{FLAG:2021npn}, $\tau_{B_{s}}=1.515\pm0.005\;\mathrm{ps}$\cite{PDG:2022ynf}
and $M_{B_{s}}=5366.92\pm0.10\;\mathrm{MeV}$ \cite{PDG:2022ynf}
as input values.

Assuming only vector operators, the branching fraction of the processes
$B\rightarrow K^{(*)}e_{\alpha}e_{\beta}$ are given by \cite{Becirevic:2016zri}
\begin{flalign}
\mathcal{B}\left(B\rightarrow K^{(*)}e_{\alpha}e_{\beta}\right)=10^{-9} & \left(a_{K^{(*)}}^{\alpha\beta}\left|C_{9}^{\alpha\beta}+C'^{\alpha\beta}_{9}\right|^{2}+b_{K^{(*)}}^{\alpha\beta}\left|C_{10}^{\alpha\beta}+C'^{\alpha\beta}_{10}\right|^{2}\right.\label{eq:BKLFV1}\\
 & \left.+c_{K^{(*)}}^{\alpha\beta}\left|C_{9}^{\alpha\beta}-C'^{\alpha\beta}_{9}\right|^{2}+d_{K^{(*)}}^{\alpha\beta}\left|C_{10}^{\alpha\beta}-C'^{\alpha\beta}_{10}\right|^{2}\right)\,,\nonumber
\end{flalign}
while if we assume only scalar operators, the expression for the branching
fractions is given by \cite{Becirevic:2016zri}
\begin{flalign}
\mathcal{B}\left(B\rightarrow K^{(*)}e_{\alpha}e_{\beta}\right)=10^{-9} & \left(e_{K^{(*)}}^{\alpha\beta}\left|C_{S}^{\alpha\beta}+C'^{\alpha\beta}_{S}\right|^{2}+f_{K^{(*)}}^{\alpha\beta}\left|C_{P}^{\alpha\beta}+C'^{\alpha\beta}_{P}\right|^{2}\right.\label{eq:BKLFV2}\\
 & \left.+g_{K^{(*)}}^{\alpha\beta}\left|C_{S}^{\alpha\beta}-C'^{\alpha\beta}_{S}\right|^{2}+h_{K^{(*)}}^{\alpha\beta}\left|C_{P}^{\alpha\beta}-C'^{\alpha\beta}_{P}\right|^{2}\right)\,.\nonumber
\end{flalign}
The numerical values for the factors $a_{K^{(*)}}^{\alpha\beta}-h_{K^{(*)}}^{\alpha\beta}$
are given in Table~\ref{tab:CoefficientsbsLFV} for the various processes.
We will not consider the scenario with both vector and scalar Wilson
coefficients because it is not predicted by any of the NP models proposed
in this thesis, but we refer the interested reader to \cite{Becirevic:2016zri}.
\begin{table}
\begin{centering}
\subfloat[]{
\global\long\def\arraystretch{1.5}%
 \centering %
\begin{tabular}{|c|cccc|cccc|}
\hline 
$\alpha\beta$ & $a_{K^{\ast}}^{\alpha\beta}$ & $b_{K^{\ast}}^{\alpha\beta}$ & $c_{K^{\ast}}^{\alpha\beta}$ & $d_{K^{\ast}}^{\alpha\beta}$ & $a_{K}^{\alpha\beta}$ & $b_{K}^{\alpha\beta}$ & $c_{K}^{\alpha\beta}$ & $d_{K}^{\alpha\beta}$\tabularnewline
\hline 
\hline 
$e\mu$ & $7.8(9)$ & $7.8(9)$ & $34(6)$ & $34(6)$ & $20(2)$ & $20(2)$ & $0$ & $0$\tabularnewline
$e\tau$ & $3.8(4)$ & $3.9(4)$ & $18(2)$ & $18(2)$ & $12.7(9)$ & $12.7(9)$ & $0$ & $0$\tabularnewline
$\mu\tau$ & $4.1(5)$ & $3.6(4)$ & $18(2)$ & $17(2)$ & $12.5(1.0)$ & $12.9(9)$ & $0$ & $0$\tabularnewline
\hline 
\end{tabular}}
\par \end{centering}
\begin{centering}
\subfloat[]{
\global\long\def\arraystretch{1.5}%
 \centering %
\begin{tabular}{|c|cccc|cccc|}
\hline 
$\alpha\beta$ & $e_{K^{\ast}}^{\alpha\beta}$ & $f_{K^{\ast}}^{\alpha\beta}$ & $g_{K^{\ast}}^{\alpha\beta}$ & $h_{K^{\ast}}^{\alpha\beta}$ & $e_{K}^{\alpha\beta}$ & $f_{K}^{\alpha\beta}$ & $g_{K}^{\alpha\beta}$ & $h_{K}^{\alpha\beta}$\tabularnewline
\hline 
\hline 
$e\mu$ & $0$ & $0$ & $12(1)$ & $12(1)$ & $26.2(4)$ & $26.2(4)$ & $0$ & $0$\tabularnewline
$e\tau$ & $0$ & $0$ & $5.5(6)$ & $5.5(6)$ & $15.0(2)$ & $15.0(2)$ & $0$ & $0$\tabularnewline
$\mu\tau$ & $0$ & $0$ & $5.2(6)$ & $5.8(7)$ & $14.4(2)$ & $15.5(2)$ & $0$ & $0$\tabularnewline
\hline 
\end{tabular}}
\par \end{centering}
\caption[Numerical coefficients for $\mathcal{B}(B\rightarrow K^{(*)}e_{\alpha}e_{\beta})$]{Values for the multiplicative factors defined in Eqs.~(\ref{eq:BKLFV1})
and (\ref{eq:BKLFV2}). The quoted uncertainties are at the $1\sigma$
level. Table taken from \cite{Becirevic:2016zri}. \label{tab:CoefficientsbsLFV} }

\end{table}

Out of the different $b\rightarrow se_{\alpha}e_{\beta}$ transitions,
processes of the form $b\rightarrow s\tau\mu$ are very interesting
because they receive contributions in particular theories of flavour
proposed in this thesis, see more in Chapter~\ref{Chapter:TwinPS}. Current bounds over these
processes exclude branching fractions larger than $10^{-5}$
\cite{LHCb:2019ujz,BaBar:2012azg},
\begin{equation}
\mathcal{B}\left(B_{s}\rightarrow\tau^{\pm}\mu^{\mp}\right)<3.4\times10^{-5}\;(90\%\;\mathrm{CL})\,,
\end{equation}
\begin{equation}
\mathcal{B}\left(B^{+}\rightarrow K^{+}\tau^{\pm}\mu^{\mp}\right)<2.25\times10^{-5}\;(90\%\;\mathrm{CL})\,.
\end{equation}
Remarkably, the LHCb collaboration is expected to improve the bounds
over the processes above by one order of magnitude after Upgrade II
\cite{LHCb:2018roe}
\begin{equation}
\mathcal{B}\left(B_{s}\rightarrow\tau^{\pm}\mu^{\mp}\right)<3\times10^{-6}\;(90\%\;\mathrm{CL})\,,
\end{equation}
\begin{equation}
\mathcal{B}\left(B^{+}\rightarrow K^{+}\tau^{\pm}\mu^{\mp}\right)<10^{-6}\;(90\%\;\mathrm{CL})\,.
\end{equation}

The hadronic tau decay $\tau^{\pm}\rightarrow\mu^{\pm}\phi$ can be
modified as well in models addressing the $B$-anomalies. We include
its branching fraction in the presence of left-handed NP currents
\cite{Pich:2013lsa},
\begin{equation}
\mathcal{B}\left(\tau^{\pm}\rightarrow\mu^{\pm}\phi\right)=\frac{\tau_{\tau}G_{F}^{2}f_{\phi}^{2}m_{\tau}^{3}}{16\pi}\left(1-\frac{m_{\phi}^{2}}{m_{\tau}^{2}}\right)^{2}\left(1+2\frac{m_{\phi}^{2}}{m_{\tau}^{2}}\right)\left|\left[C_{ed}^{V,LL}\right]^{\mu\tau22}\right|^{2}\,,\label{eq:tau_muphi}
\end{equation}
where $\tau_{\tau}=(290.3\pm0.5)\times10^{-15}\;s$, $f_{\phi}\simeq225\,\mathrm{MeV}$
and $m_{\phi}^{2}/m_{\tau}^{2}\approx0.33$ \cite{PDG:2022ynf}. The current
bound is $\mathcal{B}\left(\tau\rightarrow\mu\phi\right)<8.4\times10^{-8}\;(90\%\;\mathrm{CL})$
\cite{Belle:2011ogy}, which is expected to be improved by the Belle II collaboration to $\mathcal{B}\left(\tau\rightarrow\mu\phi\right)<2\times10^{-9}\;(90\%\;\mathrm{CL})$
\cite{Belle-II:2018jsg}.

Before concluding, we highlight the process $K_{L}\rightarrow\mu e$
that provides strong constraints over the vector leptoquark $U_{1}\sim(\mathbf{3},\mathbf{1},2/3)$
of the Pati-Salam model \cite{Pati:1974yy}. We introduce the effective
Lagrangian
\begin{equation}
\begin{aligned}\mathcal{L}_{s\rightarrow d\mu e}=-\frac{G_{F}}{\sqrt{2}}V_{ud}V_{us}^{*} & \left[C_{7V}^{\mu e}\left(\bar{s}\gamma_{\mu}P_{L}d\right)\left(\bar{\mu}\gamma^{\mu}e\right)+C_{7A}^{\mu e}\left(\bar{s}\gamma_{\mu}P_{L}d\right)\left(\bar{\mu}\gamma^{\mu}\gamma_{5}e\right)\right]\,,\end{aligned}
\label{eq:bsl1l2-1}
\end{equation}
where the Wilson coefficients $C_{7V}^{\mu e}$ and $C_{7A}^{\mu e}$,
equivalent to the $B$-physics coefficients $C_{9}^{\alpha\beta}$
and $C_{10}^{\alpha\beta}$ but for the kaon system, are related to
operators from the San Diego basis via
\begin{align}
C_{7V}^{\mu e}=\frac{2}{V_{ub}V_{us}^{*}}\left([C_{ed}^{V,LL}]^{\mu e21}+[C_{de}^{V,LR}]^{21\mu e}\right)\,, &  & C_{7A}^{\mu e}=\frac{2}{V_{ub}V_{us}^{*}}\left([C_{de}^{V,LR}]^{21\mu e}-[C_{ed}^{V,LL}]^{\mu e21}\right).\label{eq:C9C10_CedV-1}
\end{align}
The branching fraction of the $K_{L}\rightarrow\mu e$ process is
given by \cite{Crivellin:2016vjc}
\begin{equation}
\mathcal{B}\left(K_{L}\rightarrow\mu^{\pm}e^{\mp}\right)=\frac{\tau_{K_{L}}f_{K}^{2}m_{\mu}^{2}m_{K^{0}}}{64\pi}G_{F}^{2}\left|V_{ub}V_{us}^{*}\right|^{2}\left(1-\frac{m_{\mu}^{2}}{m_{K}^{2}}\right)^{2}\left(\left|C_{7V}^{\mu e}\right|^{2}+\left|C_{7A}^{\mu e}\right|^{2}\right)\,,\label{eq:KL_mue}
\end{equation}
where $m_{K^{0}}=497.611\pm0.013\;\mathrm{MeV}$, $f_{K}=155.7\pm0.3\;\mathrm{MeV}$,
and $\tau_{K_{L}}=(5.116\pm0.021)\times10^{-8}\;s$ \cite{PDG:2022ynf}.
The very strong bound over $\mathcal{B}\left(K_{L}\rightarrow\mu^{\pm}e^{\mp}\right)$
was obtained by the BNL collaboration \cite{BNL:1998apv}
\begin{equation}
\mathcal{B}\left(K_{L}\rightarrow\mu^{\pm}e^{\mp}\right)<4.7\times10^{-12}\;(90\%\;\mathrm{CL})\,.
\end{equation}
which naively pushes the breaking scale of the traditional Pati-Salam group \cite{Pati:1974yy} above the PeV \cite{Valencia:1994cj}. In this manner, realistic model building based on low-scale implementations of the Pati-Salam gauge
group faces the challenge of ameliorating this constraint, which can
be achieved e.g.~if the $U_{1}\sim(\mathbf{3},\mathbf{1},2/3)$ vector
leptoquark is mostly coupled to the third family.

\subsection{\texorpdfstring{Universality in $\tau$ decays}{Universality in tau decays}\label{subsec:Universality-in-tau}}

As discussed in Section~\ref{subsec:bctaunu}, the $R_{D^{(*)}}$
anomalies suggest the breaking of LFU in semileptonic $b\rightarrow c\ell\nu$
processes, hinting at NP interactions that discriminate between the
tau and light charged leptons. In this direction, it seems sensible
to ask whether the decays of taus into light charged leptons are sensitive
to this kind of NP. 

By comparing the measured decay widths of leptonic or hadronic
tau decays which only differ by the flavour of the final charged lepton, one can test experimentally
that the $W^{\pm}$ interaction is indeed universal to good approximation, i.e.~that $g_{e}=g_{\mu}=g_{e}$,
where $g_{\alpha}$ denotes the couplings of leptons to the $W^{\pm}$ boson.
In this manner, tau decay rates provide a powerful test of lepton flavor
universality via the ratios \cite{Pich:2013lsa}
\begin{equation}
\left(\frac{g_{\tau}}{g_{\mu(e)}}\right)_{\ell}=\left[\frac{\mathcal{B}(\tau\rightarrow e(\mu)\nu\bar{\nu})/\mathcal{B}(\tau\rightarrow e(\mu)\nu\bar{\nu})_{\mathrm{SM}}}{\mathcal{B}(\mu\rightarrow e\nu\bar{\nu})/\mathcal{B}(\mu\rightarrow e\nu\bar{\nu})_{\mathrm{SM}}}\right]^{\frac{1}{2}}\,.
\end{equation}
One can build similar ratios to test LFU in tau decays to light mesons
and one neutrino,
\begin{equation}
\left(\frac{g_{\tau}}{g_{\mu}}\right)_{\pi}=\left[\frac{\mathcal{B}(\tau\rightarrow\pi\nu)/\mathcal{B}(\tau\rightarrow\pi\nu)_{\mathrm{SM}}}{\mathcal{B}(\pi\rightarrow\mu\bar{\nu})/\mathcal{B}(\pi\rightarrow\mu\bar{\nu})_{\mathrm{SM}}}\right]^{\frac{1}{2}}\,,
\end{equation}
\begin{equation}
\left(\frac{g_{\tau}}{g_{\mu}}\right)_{K}=\left[\frac{\mathcal{B}(\tau\rightarrow K\nu)/\mathcal{B}(\tau\rightarrow K\nu)_{\mathrm{SM}}}{\mathcal{B}(K\rightarrow\mu\bar{\nu})/\mathcal{B}(K\rightarrow\mu\bar{\nu})_{\mathrm{SM}}}\right]^{\frac{1}{2}}\,,
\end{equation}
where we have not included the hadronic $(g_{\tau}/g_{e})$ ratios
because the meson decays to electrons are strongly helicity suppressed,
leading to less precise experimental measurements. NP contributions
to the three ratios above can be described via the following effective
Lagrangian (containing operators from the San Diego basis)
\begin{flalign}
\mathcal{L}_{\tau,\,\mathrm{LFU}}=-\frac{4G_{F}}{\sqrt{2}} & \left[\left[C_{\nu e}^{V,LL}\right]^{\alpha\beta\rho\lambda}(\bar{\nu}_{L}^{\alpha}\gamma_{\mu}\nu_{L}^{\beta})(\bar{e}_{L}^{\rho}\gamma^{\mu}e_{L}^{\lambda})\right.\\
 & \left.+\sum_{\alpha}\left(\delta_{\alpha3}V_{ud}^{*}+\left[C_{\nu edu}^{V,LL}\right]^{\alpha\beta11}\right)(\bar{\nu}_{L}^{\alpha}\gamma_{\mu}e_{L}^{\beta})(\bar{d}_{L}\gamma^{\mu}u_{L})\right.\nonumber \\
 & \left.+\sum_{\alpha}\left(\delta_{\alpha3}V_{us}^{*}+\left[C_{\nu edu}^{V,LL}\right]^{\alpha\beta21}\right)(\bar{\nu}_{L}^{\alpha}\gamma_{\mu}e_{L}^{\beta})(\bar{s}_{L}\gamma^{\mu}u_{L})\right]+\mathrm{h.c.}\,,\nonumber 
\end{flalign}
where we have neglected operators containing right-handed fields, plus scalar
and tensor Lorentz structures, as they are all strongly constrained
by current data. We have also included the SM contributions for the
semileptonic Wilson coefficients. We find the following theoretical
predictions for the LFU ratios in terms of the Wilson coefficients
above,
\begin{equation}
{\displaystyle \left(\frac{g_{\tau}}{g_{\mu(e)}}\right)_{\ell}=\left[\frac{\sum_{\alpha\beta}\left(\delta_{\alpha\tau}\delta_{\beta e(\mu)}+\left|\left[C_{\nu e}^{V,LL}\right]^{\alpha\beta e(\mu)\tau}\right|^{2}\right)}{\sum_{\alpha\beta}\left(\delta_{\alpha\mu}\delta_{\beta e}+\left|\left[C_{\nu e}^{V,LL}\right]^{\alpha\beta e\mu}\right|^{2}\right)}\right]^{\frac{1}{2}}}\,,
\end{equation}
\begin{equation}
{\displaystyle \left(\frac{g_{\tau}}{g_{\mu}}\right)_{\pi}=\left[\frac{\sum_{\alpha}\left(\delta_{\alpha\tau}V_{ud}^{*}+\left|\left[C_{\nu edu}^{V,LL}\right]^{\alpha\tau11}\right|^{2}\right)}{\sum_{\alpha}\left(\delta_{\alpha\mu}V_{ud}+\left|\left[C_{\nu edu}^{V,LL}\right]^{\alpha\mu11}\right|^{2}\right)}\right]^{\frac{1}{2}}}\,,
\end{equation}
\begin{equation}
{\displaystyle \left(\frac{g_{\tau}}{g_{\mu}}\right)_{K}=\left[\frac{\sum_{\alpha}\left(\delta_{\alpha\tau}V_{us}^{*}+\left|\left[C_{\nu edu}^{V,LL}\right]^{\alpha\tau21}\right|^{2}\right)}{\sum_{\alpha}\left(\delta_{\alpha\mu}V_{us}+\left|\left[C_{\nu edu}^{V,LL}\right]^{\alpha\mu21}\right|^{2}\right)}\right]^{\frac{1}{2}}}\,.
\end{equation}
Current data over the various LFU ratios in tau decays is in good
agreement with the SM. By averaging the three $(g_{\tau}/g_{\mu})$ ratios,
the observed bound is \cite{HFLAV:2022wzx}
\begin{equation}
\left(\frac{g_{\tau}}{g_{\mu}}\right)_{\ell+\pi+K}=1.0003\pm0.0014\,.
\end{equation}
Therefore, given the $R_{D^{(*)}}$ anomalies suggesting NP that discriminate
between taus and light charged leptons, one could ask why the $(g_{\tau}/g_{\mu(e)})$
ratios show no deviations so far. Regarding the hadronic ratios, the
NP invoked to address the anomalies usually have very small couplings to light quarks, therefore
suppressing the contributions to $(g_{\tau}/g_{\mu})_{\pi+K}$. Regarding
the purely leptonic ratios, the point is that the NP proposed to address
$R_{D^{(*)}}$, such as leptoquark mediators, provide semileptonic
operators at tree-level but purely leptonic operators only arise at
1-loop, therefore suppressing the contributions to $(g_{\tau}/g_{\mu})_{\ell}$.
It is also worth mentioning that the operator $\left[\mathcal{O}_{\nu edu}^{V,LL}\right]^{\tau\tau33\dagger}$that
is well motivated in models that address $R_{D^{(*)}}$ mixes via
RGE running into $\left[C_{\nu e}^{V,LL}\right]^{\tau\tau e(\mu)\tau}$.
Therefore, the ratios $(g_{\tau}/g_{\mu})_{\ell+\pi+K}$ can still
provide significant bounds over models addressing $R_{D^{(*)}}$,
and very likely some deviation in $(g_{\tau}/g_{\mu(e)})$ should
be seen in the future with more experimental precision, if indeed the $R_{D^{(*)}}$
anomalies are due to NP.

\subsection{Proton decay \label{subsec:ProtonDecay}}

Proton decay is a crucial prediction of Grand Unified Theories, which
we will explore in Chapter~\ref{Chapter:Tri-unification} of this thesis by using the tools outlined
here. We will focus on the golden channel $p\rightarrow e^{+}\pi^{0}$
which drives the phenomenology in many well-motivated GUTs. We focus
as well in the contributions mediated by the superheavy gauge bosons
arising after spontaneous breaking of the GUT group. When the heavy
leptoquarks are integrated out, we obtain dimensions six operators
violating both baryon and lepton number in one unit, but preserving
$B-L$. SMEFT operators of this type are listed in Table~\ref{tab:smeft6baryonops} for the
Warsaw basis, although here we shall choose to work in a different
basis more common in proton decay studies\footnote{Note that here we are actually working with 4-component Dirac spinors
but removing the chiral left-right notation, and $C$ denotes charge
conjugation.},
\begin{flalign}
 & \mathcal{O}_{I}^{d=6}=\Lambda_{1}^{2}\epsilon^{\alpha\beta\lambda}\epsilon^{ab}(\overline{u_{\alpha i}^{C}}\gamma^{\mu}Q_{\beta ai})(\overline{e_{j}^{C}}\gamma_{\mu}Q_{\lambda bj})\,, & \mathcal{O}_{II}^{d=6}=\Lambda_{1}^{2}\epsilon^{\alpha\beta\lambda}\epsilon^{ab}(\overline{u_{\alpha i}^{C}}\gamma^{\mu}Q_{\beta ai})(\overline{d_{\lambda j}^{C}}\gamma_{\mu}L_{bj})\,,\label{eq:operatorsX}\\
 & \mathcal{O}_{III}^{d=6}=\Lambda_{2}^{2}\epsilon^{\alpha\beta\lambda}\epsilon^{ab}(\overline{d_{\alpha i}^{C}}\gamma^{\mu}Q_{\beta bi})(\overline{u_{\lambda j}^{C}}\gamma_{\mu}L_{aj})\,, & \mathcal{O}_{IV}^{d=6}=\Lambda_{2}^{2}\epsilon^{\alpha\beta\lambda}\epsilon^{ab}(\overline{d_{\alpha i}^{C}}\gamma^{\mu}Q_{\beta bi})(\overline{\nu_{j}^{C}}\gamma_{\mu}Q_{\lambda aj})\,,\label{eq:operatorsXprime}
\end{flalign}
where $\alpha,\beta,\lambda=1,2,3$ are colour indices, $a,b=1,2$
are $SU(2)_{L}$ indices and $i,j=1,2,3$ are flavour indices. The
effective operators $\mathcal{O}_{I}^{d=6}$ and $\mathcal{O}_{II}^{d=6}$
are generated when we integrate out the $X\sim(\mathbf{3},\mathbf{2})_{-5/6}$
leptoquarks. This is the case of theories based on the gauge group
$SU(5)$. In contrast, the effective operators $\mathcal{O}_{III}^{d=6}$
and $\mathcal{O}_{IV}^{d=6}$ are generated when we integrate out
the $X'\sim(\mathbf{3},\mathbf{2})_{1/6}$ leptoquarks. This is the
case of flipped $SU(5)$ theories \cite{DeRujula:1980qc,Barr:1981qv}, while in $SO(10)$ models both
$X$ and $X'$ are present. In this manner, $\Lambda_{1}=g_{\mathrm{GUT}}/(\sqrt{2}M_{X})$
and $\Lambda_{2}=g_{\mathrm{GUT}}/(\sqrt{2}M_{X'})$, where $M_{X},M_{X'}\sim M_{\mathrm{GUT}}$
are the masses of the superheavy gauge bosons and $g_{\mathrm{GUT}}$
is the single gauge coupling of the theory at the GUT scale.

The operators in Eqs.~(\ref{eq:operatorsX}) and (\ref{eq:operatorsXprime})
are written in the interaction basis. In the mass basis, the relevant
effective operators leading to the $p\rightarrow e^{+}\pi^{0}$ decay
are expressed as \cite{FileviezPerez:2004hn}
\begin{equation}
\mathcal{O}_{L}^{d=6}=C_{L}\epsilon^{\alpha\beta\lambda}(\overline{u_{\alpha}^{C}}\gamma^{\mu}u_{\beta})(\overline{e^{C}}\gamma_{\mu}d_{\lambda})\,, \label{eq:operator_OL}
\end{equation}
\begin{equation}
\mathcal{O}_{R}^{d=6}=C_{R}\epsilon^{\alpha\beta\lambda}(\overline{u_{\alpha}^{C}}\gamma^{\mu}u_{\beta})(\overline{d_{\lambda}^{C}}\gamma_{\mu}e)\,, \label{eq:operator_OR}
\end{equation}
where the Wilson coefficients are given by
\begin{equation}
C_{L}=\Lambda_{1}^{2}\left[(V_{u^{c}}^{\dagger}V_{u})^{11}(V_{e^{c}}^{\dagger}V_{d})^{11}+(V_{u^{c}}^{\dagger}V_{u}V_{\mathrm{CKM}})^{11}(V_{e^{c}}^{\dagger}V_{d}V_{\mathrm{CKM}}^{\dagger})^{11}\right]\,, \label{eq:proton_CL}
\end{equation}
\begin{equation}
C_{R}=\Lambda_{1}^{2}(V_{u^{c}}^{\dagger}V_{u})^{11}(V_{d^{c}}^{\dagger}V_{e})^{11}+\Lambda_{2}^{2}(V_{d^{c}}^{\dagger}V_{d}V_{\mathrm{CKM}}^{\dagger})^{11}(V_{u^{c}}^{\dagger}V_{u}V_{\mathrm{CKM}}V_{d^{c}}V_{d}^{\dagger}V_{d^{c}}^{\dagger}V_{e})^{11}\,, \label{eq:proton_CR}
\end{equation}
where the $V_{\psi^{(c)}}$ matrices refer to fermion mixing in the
$\psi^{(c)}$ sector. The partial decay width of the $p\to e^{+}\pi^{0}$
process is then given by \cite{Nath:2006ut,Chakrabortty:2019fov}
\begin{flalign}
\Gamma(p\rightarrow e^{+}\pi^{0})= & \frac{m_{p}}{8}\pi\left(1-\frac{m_{\pi^{0}}^{2}}{m_{p}^{2}}\right)^{2}A_{L}^{2}\frac{\alpha_{\mathrm{GUT}}^{2}}{M_{\mathrm{GUT}}^{4}}\times\left[A_{SL}^{2}\left|C_{L}\right|^{2}\left|\langle\pi^{0}|(ud)_{L}u_{L}|p\rangle\right|^{2}\right.\label{eq:proton_decay_width}\\
 & \left.+A_{SR}^{2}\left|C_{R}\right|^{2}\left|\langle\pi^{0}|(ud)_{R}u_{L}|p\rangle\right|^{2}\right]\,,\nonumber 
\end{flalign}
where $\alpha_{\mathrm{GUT}}=g_{\mathrm{GUT}}^{2}/4\pi$, $A_{L}\approx1.247$
accounts for the QCD RGE from the $M_{Z}$ scale to $m_{p}$ \cite{Nath:2006ut},
and $A_{SL(R)}$ accounts for the short-distance RGE from the GUT
scale to $M_{Z}$, given by 
\begin{equation}
A_{SL(R)}=\prod_{A}^{M_{Z}\leq M_{A}\leq M_{\mathrm{GUT}}}\prod_{i}\left[\frac{\alpha_{i}(M_{A+1})}{\alpha_{i}(M_{A})}\right]^{\frac{\gamma_{iL(R)}}{b_{i}}}\,,
\end{equation}
where $b_{i}$ and $\gamma_{iL(R)}$ denote the $\beta$-function
coefficients and the anomalous dimensions respectively, computed for
the various intermediate scales $M_{A}$ that may contain the given
model. The $\gamma_{iL(R)}$ are computed as loop corrections to the
effective operators $\mathcal{O}_{L}^{d=6}$ and $\mathcal{O}_{R}^{d=6}$
(vertex corrections and the self-energy corrections), which can be
done by following the algorithm in Appendix A of Ref.~\cite{Chakrabortty:2019fov}.
Therefore, most model dependence is carried by the $A_{SL(R)}$ factors
and by the $C_{L}$ and $C_{R}$ coefficients.

The following form factors \cite{Aoki:2017puj} 
\begin{equation}
\langle\pi^{0}|(ud)_{L}u_{L}|p\rangle=0.134(5)(16)\,\mathrm{GeV^{2}}\,,
\end{equation}
\begin{equation}
\langle\pi^{0}|(ud)_{R}u_{L}|p\rangle=-0.131(4)(13)\,\mathrm{GeV^{2}}\,,
\end{equation}
correspond to the $\mathcal{O}_{L}^{d=6}$ and $\mathcal{O}_{R}^{d=6}$ operators,
respectively, and the errors (shown in the parenthesis) denote statistical
and systematic uncertainties. The lifetime of the proton is finally
computed as $\tau_{p}\simeq1/\Gamma(p\rightarrow e^{+}\pi^{0})$.
In many cases, the following estimation
\begin{equation}
\tau_{p}\approx\frac{M_{\mathrm{GUT}}^{4}}{\alpha_{\mathrm{GUT}}^{2}m_{p}^{5}}\,,
\end{equation}
provides a very good approximation to the more accurate results obtained
via Eq.~(\ref{eq:proton_decay_width}), as the technicalities of the full calculation usually
lead to just $\mathcal{O}(1)$ variations.

\subsection{\texorpdfstring{Connection between $R_{D^{(*)}}$ and $b\rightarrow s\ell\ell$}{Connection between RD* and bsll}} \label{subsec:ConnectionRDbsll}

In Section~\ref{subsec:bctaunu} we have sketched the connection
between $R_{D^{(*)}}$ and $R_{K^{(*)}}$ that becomes manifest in
a theory of flavour. However, the connection between the $R_{D^{(*)}}$
anomalies and $b\rightarrow s\ell\ell$ goes beyond $R_{K^{(*)}}$
\cite{Capdevila:2017iqn}. Let us assume that the $R_{D^{(*)}}$ anomalies
are dominantly explained via the NP operator $\left[\mathcal{O}_{\nu edu}^{V,LL}\right]^{\tau\tau32\dagger}$
in Eq.~(\ref{eq:btoctaunu_low}), which involves only left-handed
fermions. At the level of the SMEFT, semileptonic decays involving
only left-handed quarks and leptons are described by the two $SU(2)_{L}$
invariant operators $[Q_{\ell q}^{(1)}]^{\tau\tau23}$ and $[Q_{\ell q}^{(3)}]^{\tau\tau23}$.
At low energies, these operators provide $\left[\mathcal{O}_{\nu edu}^{V,LL}\right]^{\tau\tau32\dagger}$
plus $\left[\mathcal{O}_{ed}^{V,LL}\right]^{\tau\tau23}$ and $\left[\mathcal{O}_{\nu d}^{V,LL}\right]^{\tau\tau23}$. The latter leads to dangerous contributions to $b\rightarrow s\nu\nu$,
unless $[C_{\ell q}^{(1)}]^{\tau\tau23}=[C_{\ell q}^{(3)}]^{\tau\tau23}$. This
is a prominent SMEFT scenario which predicts $\left[C_{\nu edu}^{V,LL}\right]^{\tau\tau32*}\approx\left[C_{ed}^{V,LL}\right]^{\tau\tau23}$,
that then matches into a large contribution to $C_{9}^{\tau\tau}=-C_{10}^{\tau\tau}$
correlated to $\left[C_{\nu edu}^{V,LL}\right]^{\tau\tau32*}$. In other
words, this scenario correlates an enhancement of the $R_{D^{(*)}}$
ratios to a significant enhancement of $b\rightarrow s\tau\tau$.
\begin{figure}[t]
\begin{centering}
\subfloat[\label{fig:Off-shell penguin_EFT}]{\noindent \begin{centering}
\resizebox{.4\textwidth}{!}{
\begin{tikzpicture}
	\begin{feynman}
		\vertex (a) {\(b_{L}\)};
		\vertex [crossed dot, right=20mm of a] (b) {};
		\vertex [below=16mm of b] (c);
		\vertex [below=16mm of c] (f);
		\vertex [below left=16mm of f] (f1) {\(\ell^{+}\)};
		\vertex [below right=16mm of f] (f2) {\(\ell^{-}\)};
		\vertex [right=20mm of b] (d) {\(s_{L}\)};
		\diagram* {
			(a) -- [fermion] (b) -- [fermion] (d),
			(b) -- [fermion, half right, edge label'={\(\tau_{L}\)}, inner sep=2pt] (c) -- [boson, blue, edge label'=\(\gamma\)] (f),
			(f1) -- [fermion] (f) -- [fermion] (f2),
			(c) -- [fermion, half right, edge label'={\(\tau_{L}\)}, inner sep=4pt] (b),
	};
	\end{feynman}
\end{tikzpicture}}

\par\end{centering}
}$\quad$\subfloat[\label{fig:RD_bsll_Correlation}]{\includegraphics[scale=0.027]{FiguresEFT/C9vsC10_Alguero5}

}
\par\end{centering}
\caption[{Connection between $R_{D^{(*)}}$ and $b\rightarrow s\ell\ell$ in
the SMEFT scenario $[C_{\ell q}^{(1)}]^{\tau\tau23}=[C_{\ell q}^{(3)}]^{\tau\tau23}$}]{\textbf{\textit{Left:}} Off-shell photon penguin diagram mixing
$\left[\mathcal{O}_{ed}^{V,LL}\right]^{\tau\tau32}$ into a lepton universal
contribution to $\mathcal{O}_{9}^{\ell\ell}$. The crossed dot indicates the
insertion of a 4-fermion operator. \textbf{\textit{Right:}} Parameter
space of $C_{9}^{U}$ and LFUV $C_{9}^{\mu\mu}=-C_{10}^{\mu\mu}$
in the particular SMEFT scenario $[C_{\ell q}^{(1)}]^{\tau\tau23}=[C_{\ell q}^{(3)}]^{\tau\tau23}$
that correlates $R_{D^{(*)}}$ with $b\rightarrow s\ell\ell$ (see
main text). The blue region is preferred by $R_{D^{(*)}}$ at $2\sigma$,
while the yellow region is preferred by the updated, SM-like $R_{K^{(*)}}$
ratios at $2\sigma$. The green region is preferred by the global
fit of $b\rightarrow s\mu\mu$ (contours denote $1\sigma$, $2\sigma$, $3\sigma$
respectively), and the contours get modified to the red ellipses if
$R_{D^{(*)}}$ are included in the global fit. Plot taken from \cite{Alguero:2023jeh}. }
\end{figure}

Due to RGE effects, the large Wilson coefficients $C_{9}^{\tau\tau}=-C_{10}^{\tau\tau}$
then mix into a universal contribution to $\mathcal{O}_{9}^{\ell\ell}$,
with the leading diagram being an off-shell photon penguin involving
the insertion of $\left[\mathcal{O}_{ed}^{V,LL}\right]^{\tau\tau23}$ (or
equivalently $\mathcal{O}_{9}^{\tau\tau}$ and $\mathcal{O}_{10}^{\tau\tau}$
with the relation $C_{9}^{\tau\tau}=-C_{10}^{\tau\tau}$), see Fig.~\ref{fig:Off-shell penguin_EFT}.
This provides a sizable $C_{9}^{U}$, which is known to provide an
excellent fit to the anomalies in $b\rightarrow s\mu\mu$ data without
entering in conflict with the SM-like $R_{K^{(*)}}$ ratios. Therefore,
this scenario provides a correlation between an enhancement of the
$R_{D^{(*)}}$ ratios and $b\rightarrow s\ell\ell$, which greatly
improves the global fit to the latter \cite{Alguero:2022wkd,Alguero:2023jeh}.
The scenario $[C_{\ell q}^{(1)}]^{\tau\tau23}=[C_{\ell q}^{(3)}]^{\tau\tau23}$
can be obtained by integrating out a heavy singlet vector leptoquark
$U_{1}\sim(\mathbf{3,1},2/3)$ \cite{Crivellin:2018yvo}, or alternatively
several copies of scalar leptoquarks $R_{2}\sim(\mathbf{3,2,}7/6)$\cite{Crivellin:2022mff}
or $S_{1}+S_{3}$ \cite{Crivellin:2017zlb,Buttazzo:2017ixm}.

In Fig.~\ref{fig:RD_bsll_Correlation} it can be seen that the
region of parameter space where $R_{D^{(*)}}$ is enhanced to explain
the experimental values agrees nicely with the region preferred by
$b\rightarrow s\mu\mu$ data. Remarkably, the inclusion of a small
LFUV contribution $C_{9}^{\mu\mu}=-C_{10}^{\mu\mu}$, which contributes
to $R_{K^{(*)}}$ without being in conflict with the recent update
by LHCb, further improves the global fit to all existing data. 

In conclusion, low-energy data on semileptonic $B$-decays suggests
the existence of a large contribution to $\left[\mathcal{O}_{\nu edu}^{V,LL}\right]^{\tau\tau32\dagger}$
which simultaneously explains $R_{D^{(*)}}$ and the anomalous $b\rightarrow s\mu\mu$
data, plus a small LFUV contribution $C_{9}^{\mu\mu}=-C_{10}^{\mu\mu}$
that slightly diminish $R_{K^{(*)}}$ without being in conflict with
current data. In Chapter~\ref{Chapter:TwinPS}, we will show that this scenario arises
naturally from a theory of flavour involving two copies of the Pati-Salam
gauge group \cite{FernandezNavarro:2022gst}.
%% ----------------------------------------------------------------
%% Fermiophobic.tex
%% ---------------------------------------------------------------- 
\chapter{\texorpdfstring{Fermiophobic $Z'$ models}{Fermiophobic Z' models}} \label{Chapter:Fermiophobic}

\begin{quote}
    ``The most important step a man can take. It's not the first one, is it? 
      It's the next one. Always the next step, Dalinar.''
  \begin{flushright}
  \hfill \hfill $-$ Brandon Sanderson, \textit{Oathbringer}
  \par\end{flushright}
\end{quote}

\noindent In this chapter we discuss a class of well-motivated extensions of
the SM that contain a $Z'$ boson which does not couple directly to
SM fermions, only through mixing via heavy vector-like fermions. This
mixing controls the size of the $Z'$ couplings to SM fermions, which
can then be large or very small in a natural way. Such a setup provides
a framework where flavour anomalies can be explained with $Z'$ masses
ranging from a few GeV to a few TeV, and it can be connected to
a natural origin of the SM Yukawa couplings and the flavour hierarchies.

\section{Introduction}

The 2021 updates of $R_{K}$ \cite{LHCb:2021trn} (see Section~\ref{subsec:bsll})
and $(g-2)_{\mu}$ \cite{Muong-2:2021ojo} (see Section~\ref{subsec:g-2})
by LHCb and FNAL, respectively, increased the mounting evidence for
new physics preferentially coupled to muons, suggesting the breaking
of lepton flavour universality. This motivated model building efforts
to understand the so-called \textit{flavour anomalies} in terms of
extensions of the SM. Beyond the well-motivated case of the $U_{1}\sim(\mathbf{3,1,}2/3)$
vector leptoquark (see Chapter~\ref{Chapter:TwinPS}), scalar leptoquarks
or $Z'\sim(\mathbf{1,1,}0)$ bosons were promising candidates for
the explanation of the $R_{K^{(*)}}$ anomalies, see e.g.~the dedicated study of Ref.~\cite{Allanach:2022iod}.

Scalar leptoquarks could just be added by hand to the SM Lagrangian,
providing a renormalisable theory. This setup, however, is not very
predictive: a large number of possible leptoquark couplings are allowed
in the renormalisable Lagrangian, not restricted by any principle,
but rather the model builder usually assumes that only the minimal
set of couplings required for the flavour anomalies are non-zero.
Moreover, such a scalar leptoquark apparently does not give any hints
about the possible solution to long-standing puzzles in fundamental
physics, but actually it does the opposite: the flavour sector is
enlarged by an extra number of leptoquark couplings with arbitrary
values motivated by phenomenology, worsening the flavour puzzle, and
the hierarchy puzzle is enlarged with the mass of another relatively
light fundamental scalar that is quadratically sensitive to NP corrections.

In contrast to scalar leptoquarks, $Z'$ bosons cannot be added
by hand to the SM Lagrangian if one seeks for a renormalisable theory.
Instead, massive $Z'$ bosons generally arise from extra $U(1)$ gauge
groups, spontaneously broken by the VEV of a scalar(s) singlet $\phi$,
such that $M_{Z'}\sim g'\left\langle \phi\right\rangle $. Moreover,
the couplings of the $Z'$ to chiral fermions are given by their charges
under the extra $U(1)$, constrained by the requirement of cancelling
gauge anomalies. This provides a more predictive framework than that
of scalar leptoquarks, featuring an extension of the SM gauge group
that might be connected to more fundamental questions like the origin
of flavour.

However, $Z'$ bosons explaining the $R_{K^{(*)}}$ anomalies contribute
at tree-level to $B_{s}-\bar{B}_{s}$ meson mixing, requiring the
$bsZ'$ coupling to be rather small, while the $\mu\mu Z'$ coupling
then need to be rather large to explain $R_{K^{(*)}}$. Beyond opening
some questions about the naturalness of the framework, these conditions
might be difficult to achieve within the stringent constraints from
gauge anomaly cancellation \cite{Davighi:2021oel,Bonilla:2017lsq,Alonso:2017uky,Chun:2018ibr,Allanach:2020kss,Allanach:2018lvl,Allanach:2019iiy,Allanach:2018vjg,Allanach:2020zna}.
All these concerns are solved if the SM fermions are not charged under
the extra $U(1)$, but only vector-like fermions are charged, which
then mix with the SM fermions via Yukawa couplings provided by the
new scalar singlet $\phi$. This is called a \textit{fermiophobic} $Z'$
model \cite{King:2017anf,Falkowski:2018dsl}. The setup is naturally
anomaly-free and allows effective couplings to SM fermions that are
controlled by the mixing angles, connected to ratios of the form $\left\langle \phi\right\rangle /M_{i}$
where $M_{i}$ generically denotes the masses of the vector-like
fermions. The case $\left\langle \phi\right\rangle \ll M_{i}$ provides
naturally small mixing angles, hence the effective $Z'$ couplings
are naturally small as well, while the case $\left\langle \phi\right\rangle /M_{i}\sim1$
provides mixing angles and $Z'$ couplings not much smaller than $\mathcal{O}(1)$.

The situation of the $(g-2)_{\mu}$ anomaly is not very different:
both scalar leptoquarks and $Z'$ bosons can explain the anomaly (see
also models with just new scalar content and/or vector-like fermions
\cite{Arnan:2019uhr,Crivellin:2018qmi,Crivellin:2021rbq,Hernandez:2021tii}).
The $Z'$ models can largely be classified depending on the origin
of the chirality flip that occurs in the loop. The traditional $Z'$
model involves a chirality flip on the muon line, requiring the $Z'$
boson to be light, below the GeV range \cite{Pospelov:2008zw}. Another
option involves $Z'$ models that include $\tau\mu Z'$ couplings,
such that a chirally enhanced contribution to $(g-2)_{\mu}$ is obtained
via the ratio $m_{\tau}/m_{\mu}$ \cite{Ma:2001md}. However, a much
larger chiral enhancement is required for heavy $Z'$ explanations
of $(g-2)_{\mu}$. Fermiophobic $Z'$ models provide a very efficient
solution: a heavy vector-like lepton of the fermiophobic model gets
a Yukawa coupling to the SM Higgs doublet. This coupling provides
a chiral enhancement of $(g-2)_{\mu}$ via the ratio $M_{4}^{C}/m_{\mu}$,
with the ``chiral mass'' of the vector-like lepton $M_{4}^{C}$
naturally of the order of the top mass. The vector-like lepton mixes
with muons, such that the $Z'$ boson mediates the new chirally enhanced
contribution to $(g-2)_{\mu}$\cite{Belanger:2015nma,CarcamoHernandez:2019ydc,CarcamoHernandez:2019xkb,Allanach:2015gkd,Raby:2017igl,Kawamura:2019rth,Kawamura:2019hxp,Kawamura:2021ygg}.
\begin{figure}[t]

\includegraphics[scale=0.75]{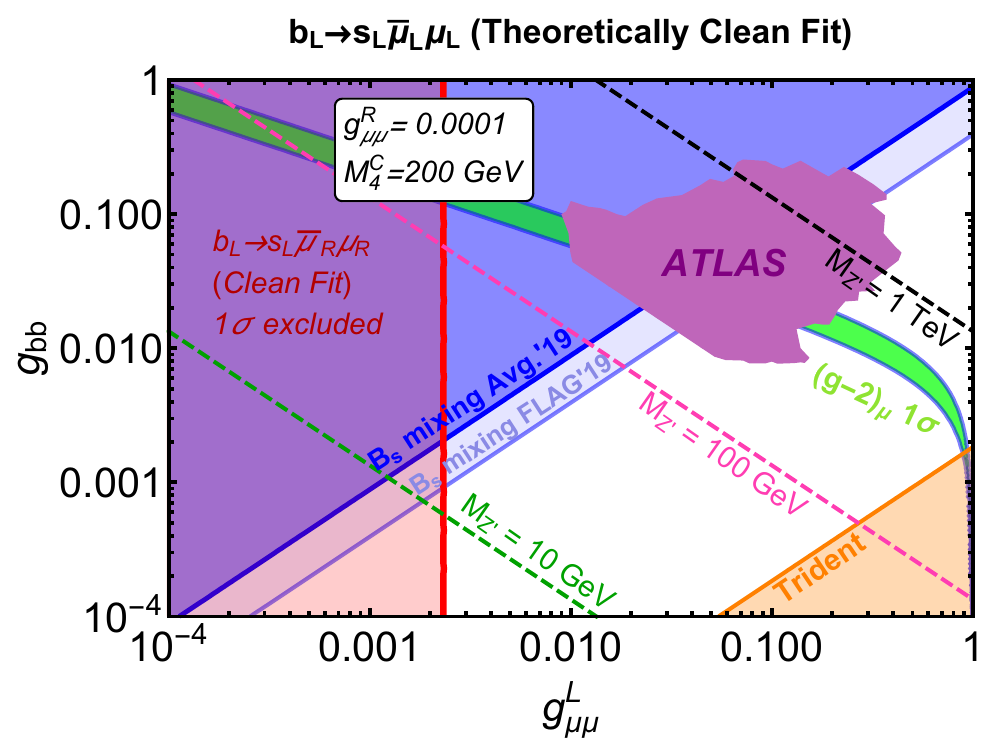}

\caption[Legacy plot with simultaneous explanation of the 2021 $R_{K^{(*)}}$ and
$(g-2)_{\mu}$ anomalies in the fermiophobic $Z'$ model]{Legacy plot with simultaneous explanation of the 2021 $R_{K^{(*)}}$ and
$(g-2)_{\mu}$ anomalies in the fermiophobic $Z'$ model of Ref.~\cite{FernandezNavarro:2021sfb}.
All points in the plot are made compatible with the central values
of the 2021 $R_{K^{(*)}}$ anomalies via a varying $M_{Z'}$ mass.
Shaded regions are excluded except for the green region, which is
preferred by the $(g-2)_{\mu}$ anomaly at 1$\sigma$.\label{fig:Legacy-plot-explaining}}
\end{figure}

This way, the class of fermiophobic $Z'$ models provided a well-motivated
solution to the $R_{K^{(*)}}$ and $(g-2)_{\mu}$ anomalies separately.
In Ref.~\cite{FernandezNavarro:2021sfb} we presented a novel study
showing that both anomalies could be addressed simultaneously in a
fermiophobic $Z'$ framework with a minimal number of couplings, as
shown in Fig~\ref{fig:Legacy-plot-explaining}. Moreover, fermiophobic
$Z'$ models can also be naturally connected to the origin of Yukawa
couplings in the SM \cite{King:2018fcg}. The basic idea is replacing
the SM Higgs doublet by a pair of Higgs doublets charged under the
new $U(1)$, while SM fermions remain uncharged, in such a way that
SM Yukawa couplings for SM fermions are forbidden but generated
effectively through the presence of heavy messengers, including the vector-like fermions. The same
mechanism generates effective $Z'$ couplings for SM fermions, hence
connecting the origin of Yukawa couplings and flavour hierarchies
with the origin of the low-energy flavour anomalies. In this manner,
the flavour anomalies fix the usually undetermined NP scales of the
theory of flavour to the TeV, within the reach of current experiments.
Notice that this is not the case for similar constructions based on
scalar leptoquarks \cite{deMedeirosVarzielas:2018bcy,DeMedeirosVarzielas:2019nob},
where the mass of the leptoquark needs to be at the TeV scale but
the scales of the theory of flavour may be anywhere \textit{from the
Planck scale to the electroweak scale}. We shall see that both the origin of flavour hierarchies
and hierarchical $Z'$ couplings are explained via the hierarchical masses of heavy messengers, known as the mechanism of \textit{messenger dominance} \cite{Ferretti:2006df}.
Despite the elegance of the
theory of flavour with fermiophobic $Z'$, it was never shown whether
this model could actually account simultaneously for fermion masses
and for both flavour anomalies $R_{K^{(*)}}$ and $(g-2)_{\mu}$ while remaining compatible with all known data.

In spite of the vanishing of the $R_{K^{(*)}}$ anomalies that motivated
our previous work \cite{FernandezNavarro:2021sfb}, $(g-2)_{\mu}$
remains as a possible hint for new physics. As shown in Section~\ref{subsec:g-2},
even in the case that the BMW computation is capturing well the SM
prediction, a small tension of almost $2\sigma$ with the increasingly
precise experimental measurement may arise. Therefore, in Section~\ref{sec:Simplified_Fermiophobic}
of this chapter we will present a minimal version of the fermiophobic
$Z'$ model that can explain the $(g-2)_{\mu}$ anomaly. In Section~\ref{sec:Theory-of-flavour_Fermiophobic}
we will present a theory of flavour based on the fermiophobic $Z'$
model, and study whether the $(g-2)_{\mu}$ anomaly can be explained
in this framework while being compatible with all constraints. Finally,
we will assess the experimental constraints over the theory of flavour
independently of the $(g-2)_{\mu}$ anomaly, to extract lower bounds
over the undetermined NP scales of the flavour model.

\section{\texorpdfstring{Fermiophobic $Z'$ model for the $(g-2)_{\mu}$ anomaly}{Fermiophobic Z' model for the (g-2)mu anomaly}\label{sec:Simplified_Fermiophobic}}

\subsection{The model}

The fermiophobic model features a $U(1)'$ gauge group under which
the chiral fermions of the SM are uncharged, as shown in Table~\ref{tab:The-field-content_simplified}.
The model includes a SM singlet scalar $\phi$ that gets a VEV to
spontaneously break the $U(1)'$ symmetry, leading to a $Z'$ boson
with mass $M_{Z'}\sim g'\left\langle \phi\right\rangle $, where $g'$
is the gauge coupling of the $U(1)'$ group. The model also includes
a so-called ``4th family'' of vector-like leptons, including
doublets $L_{L4},$ $\widetilde{L}_{R4}$ and singlets $e_{L4}$,
$\widetilde{e}_{R4}$ with the same charge under $U(1)'$. The mass
terms for the various fermions are then given by the following renormalisable
Lagrangian
\begin{table}[!t]
\begin{centering}
\begin{tabular}{ccccc}
\toprule 
\multicolumn{1}{c}{Field} & \multicolumn{1}{c}{$SU(3)_{c}$} & \multicolumn{1}{c}{$SU(2)_{L}$} & \multicolumn{1}{c}{$U(1)_{Y}$} & $U(1)'$\tabularnewline
\midrule
\midrule 
$Q_{Li}$ & $\mathbf{3}$ & $\mathbf{2}$ & 1/6 & 0\tabularnewline
$u_{Ri}$ & $\mathbf{3}$ & $\mathbf{1}$ & 2/3 & 0\tabularnewline
$d_{Ri}$ & $\mathbf{3}$ & $\mathbf{1}$ & -1/3 & 0\tabularnewline
$L_{Li}$ & $\mathbf{1}$ & $\mathbf{2}$ & -1/2 & 0\tabularnewline
$e_{Ri}$ & $\mathbf{1}$ & $\mathbf{1}$ & -1 & 0\tabularnewline
\midrule 
$L_{L4},\widetilde{L}_{R4}$ & $\mathbf{1}$ & $\mathbf{2}$ & -1/2 & 1\tabularnewline
$\tilde{e}_{L4},e_{R4}$ & $\mathbf{1}$ & $\mathbf{1}$ & -1 & 1\tabularnewline
\midrule 
$\phi$ & $\mathbf{1}$ & $\mathbf{1}$ & 0 & -1\tabularnewline
\midrule
$H$ & $\mathbf{1}$ & $\mathbf{2}$ & 1/2 & 0\tabularnewline
\bottomrule
\end{tabular}
\par\end{centering}
\caption[Field content of the simplified $Z'$ model]{Particle assignments under the $SU(3)_{c}\times SU(2)_{L}\times U(1)_{Y}\times U(1)'$
gauge symmetry, with $i=1,2,3$. \label{tab:The-field-content_simplified}}
\end{table}
\begin{align}
\begin{aligned}\mathcal{L}^{\mathrm{ren}} & =y_{ij}^{u}\overline{Q}_{Li}\widetilde{H}u_{Rj}+y_{ij}^{d}\overline{Q}_{Li}Hd_{Rj}+y_{ij}^{e}\overline{L}_{Li}He_{Rj} & {}\\
{} & +y_{4}^{e}\overline{L}_{L4}He_{R4}+\widetilde{y}_{4}^{e}\overline{\widetilde{e}}_{L4}H^{\dagger}\tilde{L}_{R4} & {}\\
{} & +x_{i}^{L}\phi\overline{L}_{Li}\widetilde{L}_{R4}+x_{i}^{e}\phi\overline{\widetilde{e}}_{L4}e_{Ri} & {}\\
{} & +M_{4}^{L}\overline{L}_{L4}\widetilde{L}_{R4}+M_{4}^{e}\overline{\widetilde{e}}_{L4}e_{R4}+\mathrm{h.c.}\,, & {}
\end{aligned}
\label{eq:fermoiphobic_lagrangian}
\end{align}
where $\widetilde{H}=i\sigma_{2}H^{\dagger}$ and $i=1,2,3$. Out
of the various terms in the Lagrangian above, those in the first line
provide Yukawa couplings for chiral fermions in the usual way, leading
to the known masses of chiral fermions after the Higgs doublet gets
a VEV. In contrast, the vector-like leptons get mass from two different
sources:
\begin{itemize}
\item Firstly, from the arbitrary vector-like masses $M_{4}^{L}$ and $M_{4}^{e}$.
\item Secondly, from Yukawa terms involving the SM Higgs doublet, i.e.~$y_{4}^{e}\overline{L}_{L4}He_{R4}$
and $\widetilde{y}_{4}^{e}\overline{\widetilde{e}}_{L4}\widetilde{H}\tilde{L}_{R4}$,
which get promoted to fourth family mass terms $M_{4}^{C}$ and $\widetilde{M}_{4}^{C}$
relating the vector-like doublet and singlet once the Higgs doublet acquires
a VEV,
\begin{equation}
M_{4}^{C}=y_{4}^{e}\frac{v_{\mathrm{SM}}}{\sqrt{2}}\,,
\end{equation}
\begin{equation}
\widetilde{M}_{4}^{C}=\widetilde{y}_{4}^{e}\frac{v_{\mathrm{SM}}}{\sqrt{2}}\,.\label{eq:Tilde_M4C}
\end{equation}
We shall denote these mass terms as ``chiral masses'' to make a distinction
with the arbitrary vector-like mass terms. Perturbation theory naively requires
$y\apprle\sqrt{4\pi}\approx3.5$ for generic Yukawa couplings in the
renormalisable Lagrangian, therefore the chiral masses above naturally
live at the electroweak scale, setting an effective bound $M_{4}^{C}\apprle600\;\mathrm{GeV}$
in order to preserve the perturbativity of the model.
\end{itemize}
Bounds over vector-like leptons generally require their physical masses
to be larger than 200 GeV \cite{Dermisek:2014qca}, therefore without loss of generality we
work in the regime $m_{\mu}\ll M_{E_{1}},M_{E_{2}}$, where $M_{E_{1}}$
and $M_{E_{2}}$ are the physical masses of the fourth family (vector-like)
leptons. In this regime, the physical masses $M_{E_{1}}$ and $M_{E_{2}}$
are obtained by diagonalising the following mass matrix,
\begin{equation}
\left(
\global\long\def\arraystretch{0.7}%
\begin{array}{@{}llc@{}}
 & \multicolumn{1}{c@{}}{\widetilde{L}_{R4}} & e_{R4}\\
\cmidrule(l){2-3}\left.\overline{L}_{L4}\right| & M_{4}^{L} & M_{4}^{C}\\
\:\left.\overline{\widetilde{e}}_{L4}\right| & \widetilde{M}_{4}^{C} & M_{4}^{e}
\end{array}\right)\,,\label{eq:Physical_Masses_VL}
\end{equation}
via two unitary rotations that we generally parameterise by the mixing
angles $\sin\theta_{L}^{E}\equiv s_{L}^{E}$ and $\sin\theta_{R}^{E}\equiv s_{R}^{E}$.
We shall see that the presence of the chiral masses, especially $M_{4}^{C}$,
is of fundamental importance to explain the $(g-2)_{\mu}$ anomaly.
We anticipate that the terms in the third line of Eq.~(\ref{eq:fermoiphobic_lagrangian})
mix chiral lepton doublets $L_{Li}$ with the partner doublet $L_{L4}$
and the chiral lepton singlets $e_{Ri}$ with the partner singlet $e_{R4}$,
leading to effective $Z'$ couplings for the SM chiral leptons as
discussed in the next subsection. The presence of chiral masses
for the fourth family leptons in the Lagrangian also leads to mixing between the
conjugate (tilde) leptons $\widetilde{L}_{R4}$ and $\widetilde{e}_{L4}$
and the chiral leptons of the SM, however in the regime $v_{\mathrm{SM}}\ll M_{4}^{L,e}$
where we shall (mostly) work, this mixing will be suppressed by the
small angles $s_{L}^{E}$ and $s_{R}^{E}$, such that it can be neglected.

\subsection{\texorpdfstring{Fermion mixing and effective $Z'$ couplings}{Fermion mixing and effective Z' couplings}}

For the sake of simplicity, we assume a minimal mixing framework which
provides the minimal set of couplings needed to address the $(g-2)_{\mu}$
anomaly\footnote{For a complete framework also connected with the origin of the SM
flavour structure see the theory of flavour with fermiophobic $Z'$
in Section~\ref{sec:Theory-of-flavour_Fermiophobic}.}. This requires that the vector-like leptons only mix with $L_{L2}$
and with $e_{R2}$. Such a minimal mixing framework could be enforced for example by a $Z_{2}$
discrete symmetry under which only vector-like leptons and $L_{L2}$,
$e_{R2}$ are not even. We also assume $v_{\mathrm{SM}}\ll M_{4}^{L,e}$
such that the physical masses of vector-like leptons are well approximated
by $M_{E_{1}}\approx M_{4}^{L}$ and $M_{E_{2}}\approx M_{4}^{e}$.
In this regime, for the purpose of extracting the effective $Z'$
couplings we can neglect the mixing angles $s_{L}^{E}$ and $s_{R}^{E}$
introduced in the previous subsection. We shall see that this approximation
holds for all the parameter space relevant for $(g-2)_{\mu}$.

In the mass insertion approximation where $\left\langle \phi\right\rangle \ll M_{4}^{L},M_{4}^{e}$,
the mixing angles are well approximated by $s_{24}^{L}\sim x_{2}^{L}\left\langle \phi\right\rangle /M_{4}^{L}$
and $s_{24}^{e}\sim x_{2}^{L}\left\langle \phi\right\rangle /M_{4}^{e}$,
see also the diagrams in Fig.~\ref{fig:Z_couplings_mass_insertion}.
However, it might be necessary to go beyond the mass insertion approximation,
where $s_{24}^{L},s_{24}^{e}\ll1$, since the explanation of the $(g-2)_{\mu}$
anomaly might require larger mixing angles, which are naturally obtained
in the regime $\left\langle \phi\right\rangle /M_{4}^{L,e}\sim1$.
In this case, the mass insertion approximation breaks and we need
to work in a large mixing angle formalism (see Appendix~\ref{app:mixing_angle_formalism}).
We always work in the regime where $m_{\mu}\ll\left\langle \phi\right\rangle $,
i.e.~we do not consider $Z'$ masses lighter\footnote{We shall see that this choice is also motivated by the strong bounds
on kinetic mixing for light $Z'$ bosons with masses below 1 GeV.} than 1 GeV. In this regime, the mixing induced by the couplings in the third
line of Eq.~(\ref{eq:fermoiphobic_lagrangian}) is well captured
by the following mixing angles
\begin{equation}
s_{24}^{L}=\frac{x_{2}^{L}\left\langle \phi\right\rangle }{\sqrt{\left(x_{2}^{L}\left\langle \phi\right\rangle \right)^{2}+\left(M_{4}^{L}\right)^{2}}}\,,\qquad c_{24}^{L}=\frac{M_{4}^{L}}{\sqrt{\left(x_{2}^{L}\left\langle \phi\right\rangle \right)^{2}+\left(M_{4}^{L}\right)^{2}}}\,,
\end{equation}
\begin{equation}
s_{24}^{e}=\frac{x_{2}^{e}\left\langle \phi\right\rangle }{\sqrt{\left(x_{2}^{e}\left\langle \phi\right\rangle \right)^{2}+\left(M_{4}^{e}\right)^{2}}}\,,\qquad c_{24}^{e}=\frac{M_{4}^{e}}{\sqrt{\left(x_{2}^{e}\left\langle \phi\right\rangle \right)^{2}+\left(M_{4}^{e}\right)^{2}}}\,,
\end{equation}
where $s_{24}^{L,e}\equiv\sin\theta_{24}^{L,e}$, $c_{24}^{L,e}\equiv\cos\theta_{24}^{L,e}$.
Assuming that we have freedom over the various parameters $\left\langle \phi\right\rangle $,
$x_{2}^{L,e}$ and $M_{4}^{L,e}$, we have complete freedom over the
numerical values of the mixing angles $s_{24}^{L,e}$. Notice that
this mixing modifies the expression for the muon mass as $m_{\mu}\approx y_{\mu}c_{24}^{L}c_{24}^{e}v_{\mathrm{SM}}/\sqrt{2}$.
However, due to the smallness of the muon Yukawa coupling in the SM,
it is possible to recover the experimental value of the muon mass
as long as $c_{24}^{L}c_{24}^{e}\gtrsim0.0006$, which only excludes
extremely large mixing angles with sine close to unity.
\begin{figure}[t]
\subfloat[]{\begin{centering}
\begin{tikzpicture}
	\begin{feynman}
		\vertex (a);
		\vertex [right=13mm of a] (b);
		\vertex [right=11mm of b] (c) [label={ [xshift=0.1cm, yshift=0.1cm] \small $M^{L}_{4}$}];
		\vertex [right=11mm of c] (d);
		\vertex [right=11mm of d] (e) [label={ [xshift=0.1cm, yshift=0.1cm] \small $M^{L}_{4}$}];
		\vertex [right=11mm of e] (f);
		\vertex [right=11mm of f] (g);
		\vertex [above=14mm of b] (f1) {\(\phi\)};
		\vertex [above=14mm of d] (f2) {\(Z'\)};
		\vertex [above=14mm of f] (f3) {\(\phi\)};
		\diagram* {
			(a) -- [fermion, edge label'=\(L_{L2}\)] (b) -- [charged scalar] (f1),
			(b) -- [edge label'=\(\widetilde{L}_{R4}\)] (c),
			(c) -- [edge label'=\(L_{L4}\), inner sep=6pt, insertion=0] (d) -- [boson, blue] (f2),
			(d) -- [edge label'=\(L_{L4}\), inner sep=6pt] (e),
			(e) -- [edge label'=\(\widetilde{L}_{R4}\), insertion=0] (f) -- [charged scalar] (f3),
			(f) -- [fermion, edge label'=\(L_{L2}\)] (g),
	};
	\end{feynman}
\end{tikzpicture}
\par\end{centering}
}$\quad$\subfloat[]{\begin{centering}
\begin{tikzpicture}
	\begin{feynman}
		\vertex (a);
		\vertex [right=13mm of a] (b);
		\vertex [right=11mm of b] (c) [label={ [xshift=0.1cm, yshift=0.1cm] \small $M^{e}_{4}$}];
		\vertex [right=11mm of c] (d);
		\vertex [right=11mm of d] (e) [label={ [xshift=0.1cm, yshift=0.1cm] \small $M^{e}_{4}$}];
		\vertex [right=11mm of e] (f);
		\vertex [right=11mm of f] (g);
		\vertex [above=14mm of b] (f1) {\(\phi\)};
		\vertex [above=14mm of d] (f2) {\(Z'\)};
		\vertex [above=14mm of f] (f3) {\(\phi\)};
		\diagram* {
			(a) -- [fermion, edge label'=\(e_{R2}\), inner sep=6pt] (b) -- [charged scalar] (f1),
			(b) -- [edge label'=\(\widetilde{e}_{L4}\)] (c),
			(c) -- [edge label'=\(e_{R4}\), inner sep=6pt, insertion=0] (d) -- [boson, blue] (f2),
			(d) -- [edge label'=\(e_{R4}\), inner sep=6pt] (e),
			(e) -- [edge label'=\(\widetilde{e}_{L4}\), insertion=0] (f) -- [charged scalar] (f3),
			(f) -- [fermion, edge label'=\(e_{R2}\), inner sep=6pt] (g),
	};
	\end{feynman}
\end{tikzpicture}
\par\end{centering}
}

\caption[Diagrams in the fermiophobic $Z'$ model which lead to the effective
$Z'$ couplings in the mass insertion approximation]{Diagrams in the model which lead to the effective $Z'$ couplings
in the mass insertion approximation. \label{fig:Z_couplings_mass_insertion}}
\end{figure}

Before any mixing, the $Z'$ boson only couples to the vector-like
leptons as
\begin{equation}
\mathcal{L}_{Z'}\supset g'\left(\overline{L}_{L4}\gamma^{\mu}L_{L4}+\overline{e}_{R4}\gamma^{\mu}e_{R4}+\overline{\widetilde{L}}_{R4}\gamma^{\mu}\widetilde{L}_{R4}+\overline{\widetilde{e}}_{L4}\gamma^{\mu}\widetilde{e}_{L4}\right)Z'_{\mu}\,.
\end{equation}
We then define the following matrices in lepton flavour space
\begin{equation}
D_{L,e}=\mathrm{diag}(0,0,0,1)\,,
\end{equation}
such that 
\begin{equation}
\mathcal{L}_{Z'}=g'\left(\overline{L}_{L}\gamma^{\mu}D{}_{L}L_{L}+\overline{e}_{R}\gamma^{\mu}D{}_{e}e_{R}+\overline{\widetilde{L}}_{R4}\gamma^{\mu}\widetilde{L}_{R4}+\overline{\widetilde{e}}_{L4}\gamma^{\mu}\widetilde{e}_{L4}\right)Z'_{\mu}\,,
\end{equation}
where we have defined $L_{L}$ and $e_{R}$ as 4-component vectors
containing the leptonic interaction eigenstates.

Now we define the mixing transformations
\begin{equation}
V_{24}^{L,e}=\left(\begin{array}{cccc}
1 & 0 & 0 & 0\\
0 & c_{24}^{L,e} & 0 & s_{24}^{L,e}\\
0 & 0 & 1 & 0\\
0 & -s_{24}^{L,e} & 0 & c_{24}^{L,e}
\end{array}\right)\,.
\end{equation}
Therefore, after the mixing, the $Z'$ couplings read
\begin{equation}
\mathcal{L}_{Z'}=g'\left(\overline{L'}_{L}\gamma^{\mu}D'_{L}L'_{L4}+\overline{e'}_{R}\gamma^{\mu}D'_{e}e'_{R}+\overline{\widetilde{L}}_{R4}\gamma^{\mu}\widetilde{L}_{R4}+\overline{\widetilde{e}}_{L4}\gamma^{\mu}\widetilde{e}_{L4}\right)Z'_{\mu}\,,\label{eq:ZpCouplings}
\end{equation}
where
\begin{equation}
D'_{L,e}=\left(V_{24}^{L,e}\right)^{\dagger}D_{L,e}V_{24}^{L,e}=\left(\begin{array}{cccc}
1 & 0 & 0 & 0\\
0 & \left(s_{24}^{L,e}\right)^{2} & 0 & -s_{24}^{L,e}c_{24}^{L,e}\\
0 & 0 & 1 & 0\\
0 & -s_{24}^{L,e}c_{24}^{L,e} & 0 & \left(c_{24}^{L,e}\right)^{2}
\end{array}\right)\,.
\end{equation}
where $L'_{L}$ and $e'_{R}$ are the 4-component vectors containing
the leptonic mass eigenstates defined as
\begin{equation}
L'_{L}=V_{24}^{L,e}L_{L}\,,\qquad e'_{R}=V_{24}^{L,e}e_{R}\,.
\end{equation}
From Eq.~(\ref{eq:ZpCouplings}), it is clear that the $Z'$ boson
now has effective couplings to muon pairs,
\begin{flalign}
\mathcal{L}_{Z'} & \supset\left[g_{\mu\mu}^{L}\overline{L'}_{L2}\gamma^{\mu}L'_{L2}+g_{\mu\mu}^{R}\overline{\mu}_{R}\gamma^{\mu}\mu{}_{R}+g_{EE}^{L}\overline{L'}_{L4}\gamma^{\mu}L'_{L4}+g_{EE}^{R}\overline{e'}_{R4}\gamma^{\mu}e'_{R4}\right.\label{eq:Z'_couplings}\\
 & \left.+\left(g_{\mu E}^{L}\overline{L'}_{L2}\gamma^{\mu}L'_{L4}+g_{\mu E}^{R}\overline{\mu}_{R}\gamma^{\mu}e'_{R4}+\mathrm{h.c.}\right)\right]Z'_{\mu}\nonumber 
\end{flalign}
where
\begin{equation}
g_{\mu\mu}^{L,R}=g'\left(s_{24}^{L,e}\right)^{2}\,,\qquad g_{EE}^{L,R}=g'\left(c_{24}^{L,e}\right)^{2}\,,
\end{equation}
\begin{equation}
g_{\mu E}^{L,R}=-g's_{24}^{L,e}c_{24}^{L,e}\,.\label{eq:gLmuE}
\end{equation}
As we shall see, the effective couplings $g_{\mu E}^{L,R}$
are of crucial importance for the explanation of the $(g-2)_{\mu}$
anomaly. Notice that the tilde leptons $\widetilde{L}_{R4}$ and $\widetilde{e}_{L4}$
do not mix with SM leptons in the regime $v_{\mathrm{SM}}\ll M_{4}^{L,e}$ where we can neglect the
mixing angles $s_{L}^{E}$ and $s_{R}^{E}$.

\subsection{Higgs diphoton decay}

After electroweak symmetry breaking, the Yukawa terms in Eq.~(\ref{eq:fermoiphobic_lagrangian})
involving the SM Higgs field and the vector-like leptons give rise
to the chiral masses $M_{4}^{C}$ and $\widetilde{M}_{4}^{C}$, which
will be crucial for accommodating $(g-2)_{\mu}$ with the experimental
measurements. On the other hand, the chiral masses are also expected
to give an extra contribution to the decay of the SM Higgs to two
photons, a key process that played a major role in the discovery of
the Higgs boson at the LHC. In the following, we check whether the
current data on Higgs diphoton decay can set any constraints over
the chiral masses.

Firstly, within the SM, chiral fermions (Fig.~\ref{fig:h2photons SMfermions})
and $W^{\pm}$ bosons (Figs.~\ref{fig:h2photons Wa}, \ref{fig:h2photons Wb})
contribute to the decay channel $h^{0}\rightarrow\gamma\gamma$ \cite{Gunion:1989we}
\begin{equation}
\Gamma(h^{0}\rightarrow\gamma\gamma)_{\mathrm{SM}}=\frac{\alpha_{\mathrm{EM}}^{2}m_{h}^{3}}{256\pi^{3}v_{\mathrm{SM}}^{2}}\left|F_{1}(\tau_{W})+\sum_{f\,\epsilon\,\mathrm{SM}}N_{cf}Q_{f}^{2}F_{1/2}(\tau_{f})\right|^{2}\,,
\end{equation}
where $N_{cf}=1\text{\,(leptons)},\,3\text{ (quarks)}$ and $Q_{f}$
is the electromagnetic charge of the fermion $f$, with the loop
functions defined as
\begin{equation}
F_{1}(\tau)=2+3\tau+3\tau\left(2-\tau\right)f(\tau)\,,
\end{equation}
\begin{equation}
F_{1/2}(\tau)=-2\tau\left[1+\left(1-\tau\right)f(\tau)\right]\,,
\end{equation}
where
\begin{equation}
\tau_{i}=4m_{i}^{2}/m_{h}^{2}
\end{equation}
and
\begin{equation}
f(\tau)=\left\{ \begin{array}{c}
{\displaystyle \left[\arcsin\left(1/\sqrt{\tau}\right)\right]^{2}\,,\quad\text{if }\quad\tau\geq1\,,}\\
\,\\
{\displaystyle -\frac{1}{4}\left[\ln\left(\frac{1+\sqrt{1-\tau}}{1-\sqrt{1-\tau}}\right)-i\pi\right]^{2}\,,\quad\text{if }\quad\tau<1\,.}
\end{array}\right.
\end{equation}
Note here that for large $\tau$, $F_{1/2}\rightarrow-4/3$. The dominant
contribution to $\Gamma(h^{0}\rightarrow\gamma\gamma)_{\mathrm{SM}}$
is the contribution of the $W^{\pm}$ bosons,
\begin{equation}
F_{1}(\tau_{W})\approx8.33\,,
\end{equation}
and it interferes destructively with the top-quark loop that dominates
the contribution from SM chiral fermions,
\begin{equation}
N_{ct}Q_{t}^{2}F_{1/2}(\tau_{t})\approx-1.84\,.
\end{equation}
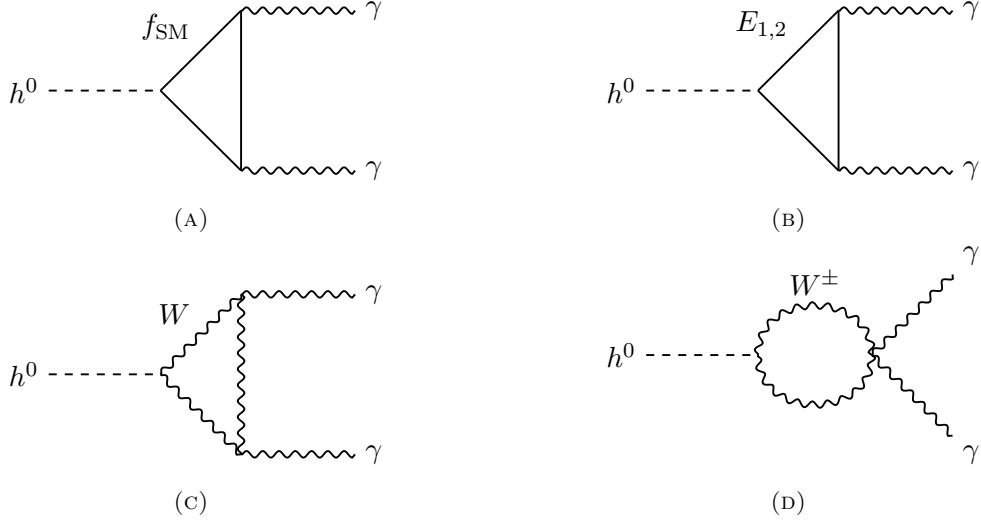
\begin{figure}[t]
\subfloat[\label{fig:h2photons SMfermions}]{\begin{centering}
\begin{tikzpicture}
	\begin{feynman}
		\vertex (a) {\(h^{0}\)};
		\vertex [right=18mm of a] (b);
		\vertex [above right=of b] (f1);
		\vertex [below right=of b] (g1);
		\vertex [right= of f1] (f2) {\(\gamma\)};
		\vertex [right= of g1] (g2) {\(\gamma\)};
		\diagram* {
			(a) -- [scalar, line width=0.25mm] (b) -- [plain, edge label=\(f_{\mathrm{SM}}\), line width=0.25mm] (f1),
			(b) --  [plain, line width=0.25mm] (g1),
			(f1) -- [boson, line width=0.25mm] (f2),
			(g1) -- [boson, line width=0.25mm] (g2),
			(g1) -- [plain, line width=0.25mm] (f1),
	};
	\end{feynman}
\end{tikzpicture}
\par\end{centering}
}$\qquad$$\qquad\qquad$\subfloat[\label{fig:h2photons VLfermion}]{\begin{centering}
\begin{tikzpicture}
	\begin{feynman}
		\vertex (a) {\(h^{0}\)};
		\vertex [right=18mm of a] (b);
		\vertex [above right=of b] (f1);
		\vertex [below right=of b] (g1);
		\vertex [right= of f1] (f2) {\(\gamma\)};
		\vertex [right= of g1] (g2) {\(\gamma\)};
		\diagram* {
			(a) -- [scalar, line width=0.25mm] (b) -- [plain, edge label=\(E_{1,2}\), line width=0.25mm] (f1),
			(b) --  [plain, line width=0.25mm] (g1),
			(f1) -- [boson, line width=0.25mm] (f2),
			(g1) -- [boson, line width=0.25mm] (g2),
			(g1) -- [plain, line width=0.25mm] (f1),
	};
	\end{feynman}
\end{tikzpicture}
\par\end{centering}
}

\subfloat[\label{fig:h2photons Wa}]{\begin{centering}
\begin{tikzpicture}
	\begin{feynman}
		\vertex (a) {\(h^{0}\)};
		\vertex [right=18mm of a] (b);
		\vertex [above right=of b] (f1);
		\vertex [below right=of b] (g1);
		\vertex [right= of f1] (f2) {\(\gamma\)};
		\vertex [right= of g1] (g2) {\(\gamma\)};
		\diagram* {
			(a) -- [scalar, line width=0.25mm] (b) -- [boson, edge label=\(W\), line width=0.25mm] (f1),
			(b) --  [boson, line width=0.25mm] (g1),
			(f1) -- [boson, line width=0.25mm] (f2),
			(g1) -- [boson, line width=0.25mm] (g2),
			(g1) -- [boson, line width=0.25mm] (f1),
	};
	\end{feynman}
\end{tikzpicture}
\par\end{centering}
}$\qquad\qquad\qquad$\subfloat[\label{fig:h2photons Wb}]{\begin{centering}
\begin{tikzpicture}
	\begin{feynman}
		\vertex (a) {\(h^{0}\)};
		\vertex [right=18mm of a] (b);
		\vertex [right=of b] (c);
		\vertex [above right=of c] (f1) {\(\gamma\)};
		\vertex [below right=of c] (g1) {\(\gamma\)};
		\diagram* {
			(a) -- [scalar, line width=0.25mm] (b) -- [boson, half left, edge label=\(W^{\pm}\), line width=0.25mm] (c),
			(b) --  [boson, half right, line width=0.25mm] (c),
			(c) -- [boson, line width=0.25mm] (f1),
			(c) -- [boson, line width=0.25mm] (g1),
	};
	\end{feynman}
\end{tikzpicture}
\par\end{centering}
}\caption[Higgs diphoton decay in the fermiophobic $Z'$ model]{Diagrams contributing to the Higgs diphoton decay, $h^{0}\rightarrow\gamma\gamma$,
where $f_{\mathrm{SM}}=u_{i},d_{i},e_{i}$, $i=1,2,3$ and $E_{1,2}$
denotes the ``4th family'' (vector-like) leptons. \label{fig: h2photons}}
\end{figure}
Now we add the contributions of the vector-like leptons (Fig.~\ref{fig:h2photons VLfermion})
with physical masses $M_{E_{1}}$ and $M_{E_{2}}$. The couplings
of the physical fourth family leptons to the SM Higgs boson are obtained
by rotating the Yukawa couplings in the second line of Eq.~(\ref{eq:fermoiphobic_lagrangian})
by the same transformations that diagonalise the matrix in Eq.~(\ref{eq:Physical_Masses_VL}),
parameterised by the mixing angles $s_{L,R}^{E}$. We obtain \cite{Joglekar:2012vc,Kearney:2012zi,Bizot:2015zaa},
\begin{flalign}
\Gamma(h^{0}\rightarrow\gamma\gamma)= & \frac{\alpha_{\mathrm{EM}}^{2}m_{h}^{3}}{256\pi^{3}v^{2}_{\mathrm{SM}}}\left|F_{1}(\tau_{W})+\sum_{f\,\epsilon\,\mathrm{SM}}N_{cf}Q_{f}^{2}F_{1/2}(\tau_{f})\right.\label{eq:Higgs_Diphoton}\\
 & \left.+\frac{M_{4}^{C}c_{L}^{E}c_{R}^{E}+\widetilde{M}_{4}^{C}s_{L}^{E}s_{R}^{E}}{M_{E_{1}}}F_{1/2}\left(\tau_{E_{1}}\right)+\frac{M_{4}^{C}s_{L}^{E}s_{R}^{E}+\widetilde{M}_{4}^{C}c_{L}^{E}c_{R}^{E}}{M_{E_{2}}}F_{1/2}\left(\tau_{E_{2}}\right)\right|^{2}\,. \nonumber
\end{flalign}
We can see that the new contributions proportional to the chiral masses
are suppressed by the physical masses of the fourth family (vector-like)
leptons $M_{E_{1,2}}$. In the regime $v_{\mathrm{SM}}\ll M_{4}^{L,e}$,
we obtain $M_{E_{1}}\approx M_{4}^{L}$ and $M_{E_{2}}\approx M_{4}^{e}$,
along with $s_{L,R}^{E}\ll1$, such that $\Gamma(h^{0}\rightarrow\gamma\gamma)$
is given by the more simple form
\begin{flalign}
\Gamma(h^{0}\rightarrow\gamma\gamma)\approx & \frac{\alpha_{\mathrm{EM}}^{2}m_{h}^{3}}{256\pi^{3}v^{2}_{\mathrm{SM}}}\left|F_{1}(\tau_{W})+\sum_{f\,\epsilon\,\mathrm{SM}}N_{cf}Q_{f}^{2}F_{1/2}(\tau_{f})\right.\label{eq:Higgs_diphoton_2}\\
 & \left.+\frac{M_{4}^{C}}{M_{4}^{L}}F_{1/2}\left(\tau_{L_{4}}\right)+\frac{\widetilde{M}_{4}^{C}}{M_{4}^{e}}F_{1/2}\left(\tau_{e_{4}}\right)\right|^{2}\,.\nonumber 
\end{flalign}
Although in the phenomenological analysis we work with the exact expression
in Eq.~(\ref{eq:Higgs_Diphoton}), we find that all the parameter
space motivated by the $(g-2)_{\mu}$ anomaly is within the regime
$v_{\mathrm{SM}}\ll M_{4}^{L,e}$, well described by the equation
above. Therefore, the new contributions to $h^{0}\rightarrow\gamma\gamma$
mediated by the fourth family leptons are suppressed by the heavy vector-like
masses $M_{4}^{L,e}$. Moreover, these new contributions generally
interfere destructively with the most sizable contribution of the
$W^{\pm}$ bosons, decreasing $\Gamma(h^{0}\rightarrow\gamma\gamma)$.
Let us now compare with the experimental results for the $h^{0}$
signal strength in the $h^{0}\rightarrow\gamma\gamma$ channel,
\begin{equation}
R_{\gamma\gamma}=\frac{\Gamma(h^{0}\rightarrow\gamma\gamma)}{\Gamma(h^{0}\rightarrow\gamma\gamma)_{\mathrm{SM}}}\,,
\end{equation}
where current data reveals $R_{\gamma\gamma}^{\mathrm{PDG,\,2023}}=1.10\pm0.07$
\cite{PDG:2022ynf}, and future projections show that ATLAS and CMS
could reduce the uncertainties down to the per cent level after the
high luminosity phase of the LHC \cite{CMS:2018qgz,ATLAS:2018jlh}.

Finally, we comment that the vector-like leptons do not only contribute
at 1-loop level to the $h^{0}\rightarrow\gamma\gamma$ decay, but they
also modify the $h^{0}\rightarrow Z\gamma$ mode which has been recently
observed at the LHC \cite{CMS:2023mku}. However, in our scenario where we consider vector-like
leptons with similar quantum numbers as the SM leptons, notice that
the $Z$ couplings are suppressed with respect to electromagnetic
couplings (roughly by $1-s_{\theta_{W}}^{2}\simeq0.08$). Moreover,
it has been shown that the doublet and the singlet fermions have opposite
trends in terms of the interference pattern with the SM amplitudes \cite{Carena:2012xa}.
Therefore, the two effects tend to cancel each other and $h^{0}\rightarrow Z\gamma$
remains SM-like to an excellent approximation. We note however that
this strong suppression does not arise in models where the vector-like
leptons carry exotic hypercharges \cite{Altmannshofer:2013zba}.

\subsection{\texorpdfstring{$(g-2)_{\mu}$ anomaly}{(g-2)mu anomaly}} \label{subsec:g-2_simplified}

As discussed in Section~\ref{subsec:g-2}, the current picture of
$(g-2)_{\mu}$ is very puzzling. The SM prediction based on data from
$e^{+}e^{-}\rightarrow\mathrm{hadrons}$ \cite{Aoyama:2020ynm} is
in 5.1$\sigma$ tension with the most recent experimental measurement
by FNAL \cite{Muong-2:2023cdq}. Numerically, we obtain (see Section~\ref{subsec:g-2})
\begin{equation}
\Delta a_{\mu}^{\mathrm{R}}=a_{\mu}^{\mathrm{exp}}-a_{\mu}^{\mathrm{SM,R}}=(249\pm48)\times10^{-11}\,,\label{eq:g-2_Rpaper}
\end{equation}
where $a=(g-2)/2$. In contrast, the SM prediction by the BMW lattice
collaboration is in agreement with the experiment at the 1.8$\sigma$
level,
\begin{equation}
\Delta a_{\mu}^{\mathrm{BMW}}=a_{\mu}^{\mathrm{exp}}-a_{\mu}^{\mathrm{SM,BMW}}=(105\pm59)\times10^{-11}\,.\label{eq:g-2_BMWpaper}
\end{equation}
While we wait for new data and theory improvement to establish a clear
picture, it is interesting to study the BSM interpretation of the
results above. Although the BMW prediction is still in rough agreement
with the experimental value, with the increasing precision in the
experimental measurement is possible that a small tension at the $2\sigma$
level is established in the near future.

In our model, in the regime $v_{\mathrm{SM}}\ll M_{4}^{L,e}$ where
we can neglect the $s_{L,R}^{E}$ mixing, the NP contribution to $(g-2)_{\mu}$
is given as \cite{Jegerlehner:2009ry,Dermisek:2013gta}
\begin{equation}
\begin{aligned}\Delta a_{\mu} & =-\frac{m_{\mu}^{2}}{8\pi^{2}M_{Z'}^{2}}\left[\vphantom{\frac{A^{C}}{A_{\mu}}}\right.\left(\left|g_{\mu\mu}^{L}\right|^{2}+\left|g_{\mu\mu}^{R}\right|^{2}\right)F(m_{\mu}^{2}/M_{Z'}^{2})+\left(\left|g_{\mu E}^{L}\right|^{2}+\left|g_{\mu E}^{R}\right|^{2}\right)F(M_{E_{1}}^{2}/M_{Z'}^{2}) & {}\\
{} & {\displaystyle \left.+\mathrm{Re}\left[g_{\mu\mu}^{L}\left(g_{\mu\mu}^{R}\right)^{*}\right]G(m_{\mu}^{2}/M_{Z'}^{2})+\mathrm{Re}\left[g_{\mu E}^{L}\left(g_{\mu E}^{R}\right)^{*}\right]\frac{M_{4}^{C}}{m_{\mu}}G(M_{E_{1}}^{2}/M_{Z'}^{2})\right]\,,} & {}
\end{aligned}
\label{eq:g-2}
\end{equation}
where $M_{E_{1}}\approx M_{4}^{L}$ is the physical mass of the vector-like
lepton mostly aligned with the fourth family doublets. The loop functions
are given by
\begin{figure}
\begin{centering}
\begin{tikzpicture}
	\begin{feynman}
		\vertex (a) {\(L_{L2}\)};
		\vertex [right=18mm of a] (b);
		\vertex [right=25mm of b] (c);
		\vertex [right=of c] (d) {\(e_{R2}\)};
		\vertex [above right=18mm of b] (f1);
		\vertex [above=of f1] (f2) {\(H\)};
		\diagram* {
			(a) -- [fermion] (b) -- [boson, blue, edge label'=$Z'$] (c) -- [fermion] (d),
			(b) -- [fermion, edge label=\(L_{L4}\)] (f1) -- [scalar] (f2),
			(f1) -- [fermion, edge label=\(e_{R4}\)] (c),
	};
	\end{feynman}
\end{tikzpicture}
\par\end{centering}
\caption[Dominant and chirally enhanced contribution to $(g-2)_{\mu}$ in the
fermiophobic $Z'$ model]{Dominant and chirally enhanced contribution to $(g-2)_{\mu}$ in
the fermiophobic $Z'$ model. The Higgs insertion provides a chiral
mass $M_{4}^{C}$ for the fourth family (vector-like) lepton after electroweak symmetry
breaking, such that the low-energy process involves the emission of
a photon (not shown). \label{fig:g-2_Diagram} }
\end{figure}
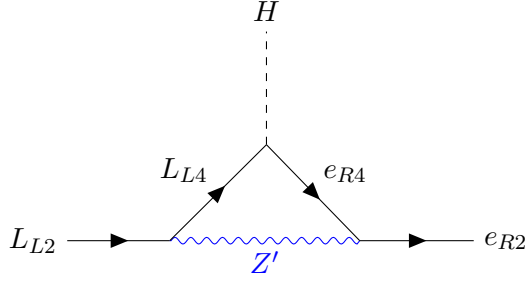
\begin{equation}
F(x)=\frac{5x^{4}-14x^{3}+39x^{2}-38x-18x^{2}\log x+8}{12(1-x)^{4}}\,,
\end{equation}
\begin{equation}
G(x)=\frac{x^{3}+3x-6x\log x-4}{2(1-x)^{3}}\,.
\end{equation}
Since the loop functions satisfy $G(x)<0$ and $F(x)>0$, the contributions
proportional to $G(x)$ and $F(x)$ in Eq.~(\ref{eq:g-2}) interfere
negatively. However, for a chiral mass $M_{4}^{C}$ of $\mathcal{O}(v_{\mathrm{SM}})$,
as naturally expected, the term proportional to $M_{4}^{C}$ in Eq.~(\ref{eq:g-2})
is dominant and positive due to $G(x)<0$, matching the required sign
to explain the deviations in Eqs.~(\ref{eq:g-2_Rpaper}) and (\ref{eq:g-2_BMWpaper}).
This dominant contribution arises from the diagram in Fig.~\ref{fig:g-2_Diagram},
and given that naturally $M_{4}^{C}\gg m_{\mu}$, we can approximate
$\Delta a_{\mu}$ as
\begin{equation}
\Delta a_{\mu}\simeq-\frac{m_{\mu}^{2}}{8\pi^{2}M_{Z'}^{2}}\mathrm{Re}\left[g_{\mu E}^{L}\left(g_{\mu E}^{R}\right)^{*}\right]\frac{M_{4}^{C}}{m_{\mu}}G\left((M_{4}^{L}/M_{Z'})^{2}\right)\,.
\end{equation}
We see that the relevant couplings to address the $(g-2)_{\mu}$ anomaly
are $g_{\mu E}^{L,R}$, which are connected to the mixing angles of
the model via Eq.~(\ref{eq:gLmuE}). For the parameter space relevant
for $(g-2)_{\mu}$ we find $M_{4}^{L}\apprge M_{Z'}$. In this case,
we have $\left|G(x)\right|\approx\mathcal{O}(1)$, and we can neglect
the order 1 factor of the loop function in order to extract $M_{Z'}$
as a function of the relevant parameters of the model and $\Delta a_{\mu}$,
obtaining
\begin{equation}
M_{Z'}^{2}\simeq\frac{m_{\mu}^{2}}{8\pi^{2}\Delta a_{\mu}}\left(g'\right)^{2}\left(s_{34}^{L}\right)^{2}\left(s_{34}^{e}\right)^{2}\frac{M_{4}^{C}}{m_{\mu}}\,.\label{eq:FixingZpMass}
\end{equation}

\begin{figure}[t]
\begin{centering}
\subfloat[]{
\begin{tikzpicture}
	\begin{feynman}
		\vertex (f1) {\(\nu_{\mu L}\)};
		\vertex [above right=20mm of f1] (a);
		\vertex [above left=of a] (f2) {\(\nu_{\mu L}\)};
		\vertex [right=of a] (b);
		\vertex [above right=of b] (g1) {\(\mu_{L,R}\)};
		\vertex [below right=of b] (g2) {\(\mu_{L,R}\)};
		\diagram* {
			(f1) -- [fermion] (a) -- [fermion] (f2),
			(a) -- [boson, blue, edge label'=\(Z'\), inner sep=6pt] (b),
			(b) -- [fermion] (g1),
			(b) -- [anti fermion] (g2),
	};
	\end{feynman}
\end{tikzpicture}

}$\qquad$\subfloat[]{\noindent \begin{centering}
\begin{tikzpicture}
	\begin{feynman}
		\vertex (a) {\(Z\)};
		\vertex [right=20mm of a] (b);
		\vertex [above right=of b] (f1);
		\vertex [below right=of b] (g1);
		\vertex [right=of f1] (f2) {\(\mu_{L(R)}\)};
		\vertex [right=of g1] (g2) {\(\mu_{L(R)}\)};
		\diagram* {
			(a) -- [boson] (b) -- [fermion,  edge label=\(L_{L4}\,(e_{R4})\),  inner sep=2pt] (f1) -- [fermion] (f2),
			(b) -- [anti fermion,   edge label'=\(L_{L4}\,(e_{R4})\),  inner sep=2pt] (g1) -- [boson, blue, edge label'=\(Z'\), inner sep=4pt] (f1),
			(g1) -- [anti fermion] (g2),
	};
	\end{feynman}
\end{tikzpicture}
\par\end{centering}
}
\par\end{centering}
\caption[NP contributions to neutrino trident and $Z\rightarrow\mu\mu$ in
the fermiophobic $Z'$ model]{$Z'$ exchange diagrams contributing to neutrino trident production
(left) and 1-loop contribution to $Z\rightarrow\mu\mu$ involving
the $Z'$ and the heavy vector-like leptons (right). \label{fig:trident_Zmumu}}
\end{figure}
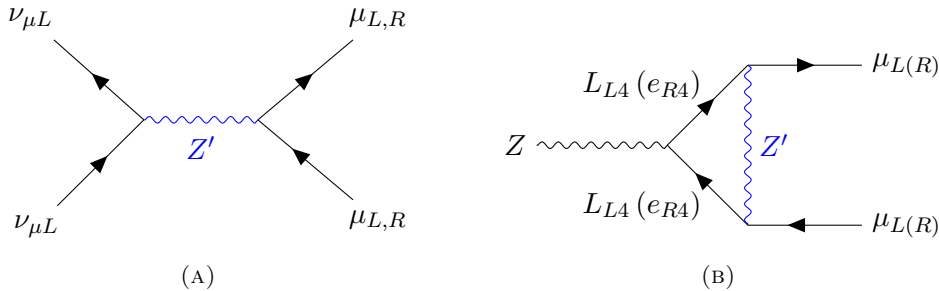

\begin{figure}[t]

\includegraphics[scale=0.75]{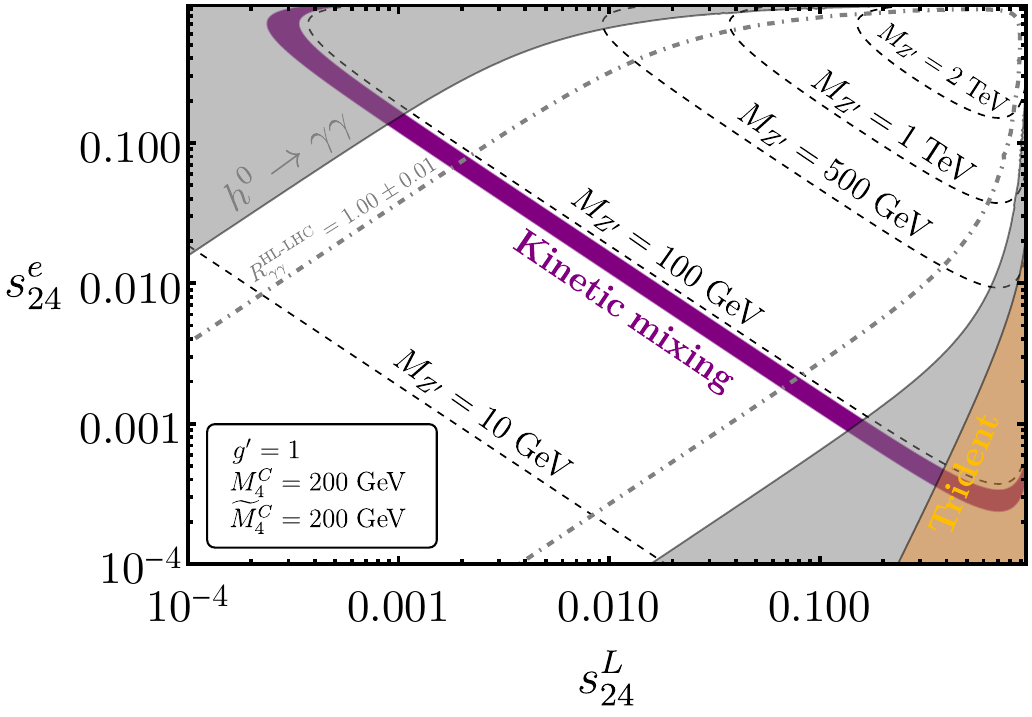}

\caption[Parameter space in the simplified fermiophobic $Z'$ model]{All the parameter space shown in the plot is compatible with the
central value of $\Delta a_{\mu}^{\mathrm{R}}$ . $M_{Z'}$ varies through
the parameter space because it is related to the mixing angles and
to $\Delta a_{\mu}^{\mathrm{R}}$ via Eq.~(\ref{eq:FixingZpMass}).
Shaded regions are excluded at 95\% CL, while the dot-dashed line
shows the projection for $R_{\gamma\gamma}$ that might be reached
by ATLAS and CMS after HL-LHC \cite{CMS:2018qgz,ATLAS:2018jlh}. \label{fig:g-2_simplified}}
\end{figure}

In Fig.~\ref{fig:g-2_simplified} we study the parameter space of
$s_{34}^{L}$ and $s_{34}^{e}$ while fixing $M_{Z'}$ via the relation
(\ref{eq:FixingZpMass}) above, where we take the central value of $\Delta a_{\mu}^{\mathrm{R}}$.
We also take $g'\sim1$ for simplicity, as a natural benchmark value.
In this manner, all the points in Fig.~\ref{fig:g-2_simplified}
address the central value of the $5\sigma$ $(g-2)_{\mu}$ anomaly
with respect to $e^{+}e^{-}\rightarrow\mathrm{hadrons}$ data. We also
highlight the parameter space excluded by the neutrino\footnote{Note that muon neutrinos also get effective $Z'$ couplings in our
model, along with muons.} trident production $\nu_{\mu}\gamma^{*}\rightarrow\nu_{\mu}\mu^{+}\mu^{-}$
\cite{CHARM-II:1990dvf,CCFR:1991lpl,NuTeV:1998khj,Altmannshofer:2014pba} (see also Fig.~\ref{fig:trident_Zmumu}),
\begin{equation}
\frac{\sigma_{\mathrm{SM+NP}}}{\sigma_{\mathrm{SM}}}=1+8\frac{g_{\mu\mu}^{L}}{g_{L}^{2}}\frac{M_{W}^{2}}{M_{Z'}^{2}}\frac{(1+4s_{W}^{2})(g_{\mu\mu}^{L}+g_{\mu\mu}^{R})+(g_{\mu\mu}^{L}-g_{\mu\mu}^{R})}{(1+4s_{W}^{2})^{2}+1}\apprle0.83\pm0.18\,.
\end{equation}
We notice another constraint arising from LHC measurements of the
$Z$ decays to four muons, with the second muon pair produced in the
SM via a virtual photon \cite{CMS:2012bw,ATLAS:2014jlg} and in our
model via the $Z'$, $pp\rightarrow Z\rightarrow4\mu$. This process
sets constraints in the region $5\,\mathrm{GeV}\lesssim M_{Z'}\apprle70\,\mathrm{GeV}$,
however we note that these constraints mostly overlap with neutrino
trident production \cite{Falkowski:2018dsl,Altmannshofer:2014cfa,Altmannshofer:2014pba,Altmannshofer:2016jzy},
and hence we neglect it. We further notice a 1-loop contribution to the precisely measured
$Z\rightarrow\mu\mu$ decay with the $Z'$ and the vector-like leptons
running in the loop, as shown in Fig.~\ref{fig:trident_Zmumu}. We find the following modifications of the $Z$
couplings due to the $Z'$ loop:
\begin{equation}
\frac{g_{\mu_{L}\mu_{L}}^{Z}}{g_{e_{L}e_{L}}^{Z}}\simeq1+\frac{(g_{\mu E}^{L})^{2}}{16\pi^{2}}\mathcal{K}(M_{Z}^{2}/M_{Z'}^{2})\,,\label{eq:Zboson_coupling}
\end{equation}
plus a similar modification of the right-handed coupling obtained by setting $L\rightarrow R$ everywhere. The loop function $\mathcal{K}$ is given by \cite{Haisch:2011up}
\begin{equation}
\mathcal{K}(x)=-\frac{4+7x}{2x}+\frac{2+3x}{x}\log x-\frac{2(1+x)^{2}}{x^{2}}\left[\log x\log(1+x)+\mathrm{Li}_{2}(-x)\right]\,,
\end{equation}
where $\mathrm{Li}_{2}(x)=-\int_{0}^{x}dt\log(1-t)$ is the di-logarithm.
In Eq.~(\ref{eq:Zboson_coupling}) we use the electron $Z$ couplings
as a convenient normalisation, as they are not affected by NP. We
find that LEP measurements \cite{ALEPH:2005ab} allow for per mille deviations from
unity in the ratios of Eq.~(\ref{eq:Zboson_coupling}), however in
our model the $Z'$ loop is suppressed by small mixing angles and/or
by a heavy $Z'$ mass, leading to no constraints at 95\% CL over the
relevant parameter space shown in Fig.~\ref{fig:g-2_simplified}.

Another relevant constraint originates from kinetic mixing between
the abelian groups $U(1)_{Y}$ and $U(1)'$. The vector-like leptons
in the model are charged under both $U(1)$s, generating
kinetic mixing at 1-loop as $\epsilon\sim3g_{Y}g'\log(M_{E}/\mu)(32\pi^{2})$,
where $\mu$ is a renormalisation scale. Typically this provides $\epsilon\sim0.01$.
Kinetic mixing of this size is bounded in the sub-GeV region of $M_{Z'}$ \cite{Hook:2010tw}, which
we hence neglect in our analysis, and also around the $Z$-pole mass,
which we exhibit as the purple region excluded in Fig.~\ref{fig:g-2_simplified}.

As shown in Fig.~\ref{fig:g-2_simplified}, Higgs diphoton decay
constrains large mixing angles when the $Z'$ boson is relatively
light. The reason is that large mixing angles require $\left\langle \phi\right\rangle /M_{4}^{L,e}\sim1$
(remember $M_{Z'}\sim\left\langle \phi\right\rangle $), such that
if $M_{Z'}$ is light then $M_{4}^{L,e}$ are also light and the NP
contributions to $h^{0}\rightarrow\gamma\gamma$ in Eq.~(\ref{eq:Higgs_diphoton_2})
are no longer suppressed. In contrast, if $M_{Z'}$ is heavy then
$\left\langle \phi\right\rangle /M_{4}^{L,e}\sim1$ can be achieved
with $M_{4}^{L,e}\gg v_{\mathrm{SM}}$ and the NP contributions to
Higgs diphoton decay remain suppressed. Notice that for all the parameter
space in Fig.~\ref{fig:g-2_simplified} at least one of the mixing
angles is small, except for the region of heavy $Z'$. This implies
that at least one of the vector-like masses $M_{4}^{L}$ and $M_{4}^{e}$ (or
both) is heavy, such that at least one of the angles $s_{L,R}^{E}$
that diagonalise Eq.~(\ref{eq:Physical_Masses_VL}) is small. This
allows us to neglect the presence of the conjugate (tilde) fermions
$\widetilde{L}_{R4}$ and $\widetilde{e}_{L4}$ in the phenomenological
analysis, as anticipated at the beginning of this section, because
their $Z'$ couplings to SM fermions are further suppressed by
the small mixing angles $s_{L,R}^{E}$. 

All in all, we show in Fig.~\ref{fig:g-2_simplified} that our model
can address the $\Delta a_{\mu}^{\mathrm{R}}$ anomaly for $Z'$ masses
ranging from the GeV scale to the few TeV range. The nature of the
sine and cosine functions does not allow effective $Z'$ couplings
larger than 1, which requires $M_{Z'}\apprle3.6\;\mathrm{TeV}$ in
order to fit the central values of $\Delta a_{\mu}^{\mathrm{R}}$,
but one can always take $g'$ larger than 1 or increase $M_{4}^{C}$
above its natural values $M_{4}^{C}\approx200\;\mathrm{GeV}$ up to
the perturbative limit $M_{4}^{C}\approx600\,\mathrm{GeV}$ if one
seeks to explain $\Delta a_{\mu}^{\mathrm{R}}$ with even heavier
$Z'$ masses. In this case, the bounds from Higgs diphoton decay should
remain similar because even though we increase $M_{4}^{C}$, a heavier
$Z'$ is needed as well and therefore heavier $M_{4}^{L,e}$ are also
required to obtain large mixing angles. Of course if we consider the
BMW prediction, $\Delta a_{\mu}^{\mathrm{BMW}}$, we can further improve
the agreement with the experiment and we have even more freedom in
the parameter space due to the smaller shift required with respect
to $\Delta a_{\mu}^{\mathrm{R}}$. The HL-LHC projections by ATLAS
and CMS over Higgs diphoton decay show that the uncertainties in $R_{\gamma\gamma}$
could be reduced down to the per cent level \cite{CMS:2018qgz,ATLAS:2018jlh},
testing a significant region of parameter space in the model as shown
by the dot-dashed line in Fig.~\ref{fig:g-2_simplified} (assuming
that the central values match the SM prediction). We find that the
whole parameter space of the model that can explain the central value
of $\Delta a_{\mu}^{\mathrm{R}}$ can be tested if $h^{0}\rightarrow\gamma\gamma$
is measured to the few per mille precision.

\section{\texorpdfstring{Theory of flavour with fermiophobic $Z'$ boson}{Theory of flavour with fermiophobic Z' boson}\label{sec:Theory-of-flavour_Fermiophobic}}

In the previous section we introduced a simplified fermiophobic $Z'$
model that can address the $(g-2)_{\mu}$ anomaly. In the following
we will show how this simple construction may emerge from a complete
theory of flavour, where the same mixing that generates the effective
$Z'$ couplings will generate effective Yukawa couplings for the SM
fermions, connecting the origin of the flavour hierarchies with the
origin of flavour anomalies.

\subsection{The renormalisable Lagrangian for charged fermions \label{subsec:The-renormalisable-Lagrangian}}

In contrast with the simplified model introduced in the previous section,
here we consider a complete ``fourth'' family of vector-like fermions
(including a vector-like neutrino) charged under $U(1)'$, while chiral
fermions remain uncharged. Notice that $SU(2)_{L}$ doublet and singlet
fermions now carry opposite charges under $U(1)'$, as shown in Table~\ref{tab:The_field_content_2}. We preserve the
SM singlet scalar $\phi$ that gets a VEV to spontaneously break the
$U(1)'$ symmetry, leading to a $Z'$ boson with mass $M_{Z'}\sim g'\left\langle \phi\right\rangle $,
where $g'$ is the gauge coupling of the $U(1)'$ group. However,
we exchange the SM-like Higgs of the simplified model by a pair of
Higgs doublets equally charged under $U(1)'$. The fact that the Higgs
doublets are charged under $U(1)'$ while the chiral fermions are
not forbids SM-like Yukawa couplings, in contrast with the simplified
model presented in the previous section. The renormalisable mass terms involving
charged fermions are then given by
\begin{table}
\begin{centering}
\begin{tabular}{ccccc}
\toprule 
\multicolumn{1}{c}{Field} & \multicolumn{1}{c}{$SU(3)_{c}$} & \multicolumn{1}{c}{$SU(2)_{L}$} & \multicolumn{1}{c}{$U(1)_{Y}$} & $U(1)'$\tabularnewline
\midrule
\midrule 
$Q_{Li}$ & $\mathbf{3}$ & $\mathbf{2}$ & 1/6 & 0\tabularnewline
$u_{Ri}$ & $\mathbf{3}$ & $\mathbf{1}$ & 2/3 & 0\tabularnewline
$d_{Ri}$ & $\mathbf{3}$ & $\mathbf{1}$ & -1/3 & 0\tabularnewline
$L_{Li}$ & $\mathbf{1}$ & $\mathbf{2}$ & -1/2 & 0\tabularnewline
$e_{Ri}$ & $\mathbf{1}$ & $\mathbf{1}$ & -1 & 0\tabularnewline
\midrule 
$Q_{L4},\widetilde{Q}_{R4}$ & $\mathbf{3}$ & $\mathbf{2}$ & 1/6 & 1\tabularnewline
$\tilde{u}_{L4},u_{R4}$ & $\mathbf{3}$ & $\mathbf{1}$ & 2/3 & -1\tabularnewline
$\tilde{d}_{L4},d_{R4}$ & $\mathbf{3}$ & $\mathbf{1}$ & -1/3 & -1\tabularnewline
$L_{L4},\widetilde{L}_{R4}$ & $\mathbf{1}$ & $\mathbf{2}$ & -1/2 & 1\tabularnewline
$\tilde{e}_{L4},e_{R4}$ & $\mathbf{1}$ & $\mathbf{1}$ & -1 & -1\tabularnewline
$\tilde{\nu}_{L4},\nu_{R4}$ & $\mathbf{1}$ & $\mathbf{1}$ & 0 & -1\tabularnewline
\midrule 
$\phi$ & $\mathbf{1}$ & $\mathbf{1}$ & 0 & -1\tabularnewline
\midrule 
$H_{u}$ & $\mathbf{1}$ & $\mathbf{2}$ & -1/2 & 1\tabularnewline
$H_{d}$ & $\mathbf{1}$ & $\mathbf{2}$ & 1/2 & 1\tabularnewline
\bottomrule
\end{tabular}
\par\end{centering}
\caption[Field content in the theory of flavour with fermiophobic $Z'$ ]{Particle assignments under the $SU(3)_{c}\times SU(2)_{L}\times U(1)_{Y}\times U(1)'$
gauge symmetry in the theory of flavour with fermiophobic $Z'$ ,
where $i=1,2,3$. \label{tab:The_field_content_2}}

\end{table}
\begin{align}
\begin{aligned}\mathcal{L}_{\mathrm{cf}}^{\mathrm{ren}} & =y_{i4}^{u}H_{u}\overline{Q}_{Li}u_{R4}+y_{i4}^{d}H_{d}\overline{Q}_{Li}d_{R4}+y_{i4}^{e}H_{d}\overline{L}_{Li}e_{R4} & {}\\
{} & +y_{4i}^{u}H_{u}\overline{Q}_{L4}u_{Ri}+y_{4i}^{d}H_{d}\overline{Q}_{L4}d_{Ri}+y_{4i}^{e}H_{d}\overline{L}_{L4}e_{Ri} & {}\\
{} & +x_{i}^{Q}\phi\overline{Q}_{Li}\widetilde{Q}_{R4}+x_{i}^{L}\phi\overline{L}_{Li}\widetilde{L}_{R4}+x_{i}^{u}\phi\overline{\widetilde{u}}_{L4}u_{Ri}+x_{i}^{d}\phi\overline{\widetilde{d}}_{L4}d_{Ri}+x_{i}^{e}\phi\overline{\widetilde{e}}_{L4}e_{Ri} & {}\\
{} & +M_{4}^{Q}\overline{Q}_{L4}\widetilde{Q}_{R4}+M_{4}^{L}\overline{L}_{L4}\widetilde{L}_{R4}+M_{4}^{u}\overline{\widetilde{u}}_{L4}u_{R4}+M_{4}^{d}\overline{\widetilde{d}}_{L4}d_{R4}+M_{4}^{e}\overline{\widetilde{e}}_{L4}e_{R4}+\mathrm{h.c.}\,, & {}
\end{aligned}
\label{eq:fermoiphobic_lagrangian-1}
\end{align}
where $i=1,2,3$. Notice that the fact that both Higgs doublets are
equally charged under $U(1)'$ while carrying opposite hypercharges
not only forbids the SM-like Yukawa couplings, but also enforces a
natural type II two Higgs doublet model (2HDM) that forbids tree-level
FCNCs mediated by the Higgs doublets. The terms in the fourth line
of the Lagrangian above are the arbitrary vector-like masses of the
fourth family fermions. The terms in the third line provide mixing
between chiral and vector-like fermions mediated by the singlet scalar
$\phi$ when it acquires a VEV. Once chiral and vector-like
fermions mix, the terms in the first and second lines provide effective
Yukawa couplings for chiral fermions as $y\sim\left\langle \phi\right\rangle /M$,
which are able to explain the origin of naturally small Yukawa couplings
in the SM. As usual for a theory of flavour, the NP scales $\left\langle \phi\right\rangle $
and $M$ may be anywhere \textit{from the Planck scale to the electroweak
scale}, as long as the ratios $\left\langle \phi\right\rangle /M$
are held fixed. However, flavour anomalies like $(g-2)_{\mu}$ might
suggest that these NP scales are actually close to the TeV, as we
shall consider here. In fact, the $Z'$ boson only couples originally
to vector-like fermions, but obtains effective couplings to chiral
fermions via the same mixing that provides the effective Yukawa couplings,
leading to a predictive phenomenology as we shall see.

\subsection{Neutrino masses and the type Ib seesaw mechanism\label{subsec:Neutrino-masses-and-type.Ib}}

The renormalisable Lagrangian for neutrinos includes the mass terms
\begin{equation}
\mathcal{L}_{\mathrm{\nu}}^{\mathrm{ren}}=y_{i4}^{\nu}H_{u}\overline{L}_{Li}\nu_{R4}+\widetilde{y}_{i4}^{\nu}\widetilde{H}_{d}\overline{L}_{Li}\widetilde{\nu}_{L4}^{C}+M_{4}^{\nu}\overline{\widetilde{\nu}}_{L4}\nu_{R4}+\mathrm{h.c.}\label{eq:Ib_Lagrangian}
\end{equation}
where $C$ denotes charge conjugation. Notice the unusual Yukawa coupling
as the second term above, breaking lepton number and involving the
second Higgs $H_{d}$, hinting that this is a different version of
the usual type I seesaw mechanism. Indeed, one may have noticed that
the choice of two Higgs doublets with opposite hypercharge but equal
charges under $U(1)'$ forbids the usual Weinberg operator $(\overline{L}_{L}^{C}H)(L_{L}H)$,
but allows for an alternative Weinberg operator involving the two
Higgs doublets
\begin{equation}
\mathcal{L}_{\mathrm{Weinberg}}=c_{ij}^{\nu}(\overline{L}_{Li}^{C}H_{u})(L_{Lj}\widetilde{H}_{d})+\mathrm{h.c.}
\end{equation}
By simply assuming $M_{4}^{\nu}\gg\left\langle H_{u,d}\right\rangle$
and integrating out the vector-like neutrino, one obtains the operator
above and the coefficients $c_{ij}^{\nu}$ as
\begin{equation}
c_{ij}^{\nu}=\frac{1}{M_{4}^{\nu}}\left(y_{i4}^{\nu}\widetilde{y}{}_{j4}^{\nu}+\widetilde{y}{}_{i4}^{\nu}y_{j4}^{\nu}\right)\,.
\end{equation}
After the Higgs doublets develop VEVs $\left\langle H_{u,d}\right\rangle =v_{u,d}/\sqrt{2}$,
the Weinberg operator provides an effective mass matrix for active
neutrinos as
\begin{equation}
m_{\nu}=\frac{v_{u}v_{d}}{2M_{4}^{\nu}}\left(y_{i4}^{\nu}\widetilde{y}_{j4}^{\nu}+\widetilde{y}{}_{i4}^{\nu}y_{j4}^{\nu}\right)\,.
\end{equation}
The Yukawa couplings above provide enough freedom to fit the observed
PMNS mixing and neutrino mass splittings \cite{deSalas:2020pgw,Gonzalez-Garcia:2021dve},
for both normal and inverted orderings. Of course if we arrange the
various mass terms of Eq.~(\ref{eq:Ib_Lagrangian}) in matrix formalism
as 
\begin{equation}
\mathcal{L}_{\mathrm{\nu}}^{\mathrm{ren}}=\frac{1}{2}\left(\begin{array}{ccc}
\overline{L}_{Li} & \overline{\nu}_{R4}^{C} & \overline{\widetilde{\nu}}_{L4}\end{array}\right)M^{\nu}\left(\begin{array}{c}
L_{Li}^{C}\\
\nu_{R4}\\
\widetilde{\nu}_{L4}^{C}
\end{array}\right)+\mathrm{h.c.}\,,
\end{equation}
where
\begin{equation}
M^{\nu}=\left(
\global\long\def\arraystretch{1.3}%
\begin{array}{@{}lccccc@{}}
 & L_{L1}^{C} & L_{L2}^{C} & L_{L3}^{C} & \nu_{R4} & \widetilde{\nu}_{L4}^{C}\\
\cmidrule(l){2-6}\overline{L}_{L1} & 0 & 0 & 0 & y_{14}^{\nu}H_{u} & \widetilde{y}{}_{14}^{\nu}\widetilde{H}_{d}\\
\overline{L}_{L2} & 0 & 0 & 0 & y_{24}^{\nu}H_{u} & \widetilde{y}{}_{24}^{\nu}\widetilde{H}_{d}\\
\overline{L}_{L3} & 0 & 0 & 0 & y_{34}^{\nu}H_{u} & \widetilde{y}{}_{34}^{\nu}\widetilde{H}_{d}\\
\overline{\nu}_{R4}^{C} & y_{14}^{\nu}H_{u} & y_{24}^{\nu}H_{u} & y_{34}^{\nu}H_{u} & 0 & M_{4}^{\nu}\\
\overline{\widetilde{\nu}}_{L4} & \widetilde{y}{}_{14}^{\nu}\widetilde{H}_{d} & \widetilde{y}{}_{24}^{\nu}\widetilde{H}_{d} & \widetilde{y}{}_{34}^{\nu}\widetilde{H}_{d} & M_{4}^{\nu} & 0
\end{array}\right)\equiv\left(\begin{array}{cc}
0_{\mathrm{3\times3}} & m_{D\,3\times2}\\
m_{D\,2\times3}^{\mathrm{T}} & M_{N\,2\times2}
\end{array}\right)\,,\label{eq: mass matrix}
\end{equation}
and now we apply the seesaw formula assuming $M_{4}^{\nu}\gg\left\langle H_{u,d}\right\rangle $,
then we obtain the same result as
\begin{equation}
m_{\nu}\simeq m_{D}M_{N}^{-1}m_{D}^{\mathrm{T}}=\frac{v_{u}v_{d}}{2M_{4}^{\nu}}\left(y_{i4}^{\nu}\widetilde{y}_{j4}^{\nu}+\widetilde{y}{}_{i4}^{\nu}y_{j4}^{\nu}\right)\,.
\end{equation}
This mechanism is denoted as \textit{type Ib} seesaw mechanism \cite{Hernandez-Garcia:2019uof}.
Note that in Ref.~\cite{Hernandez-Garcia:2019uof} it was considered the possibility that the couplings
$\widetilde{y}_{i4}^{\nu}$ are small, as they originate from an operator breaking lepton number. This scenario
leads to a low scale seesaw with potentially large violations of unitarity of the leptonic mixing matrix.
However, in this chapter we refrain to discuss further the neutrino sector,
and instead we focus in the following on the origin of charged fermion
masses and the associated $Z'$ phenomenology.

\subsection{Effective quark Yukawa couplings and messenger dominance \label{subsec:Messenger_Dominance}}

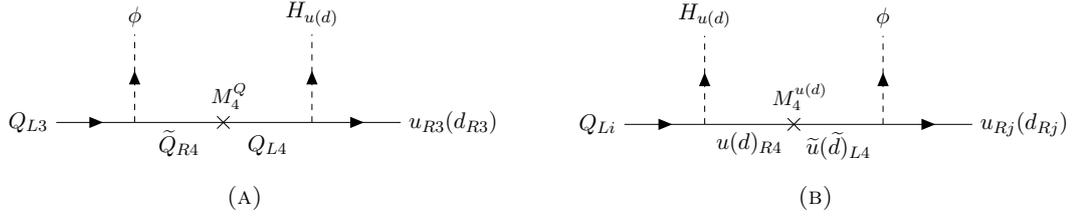
\begin{figure}[t]
\subfloat[]{\begin{centering}
\resizebox{.46\textwidth}{!}{
\begin{tikzpicture}
	\begin{feynman}
		\vertex (a) {\(Q_{L3}\)};
		\vertex [right=18mm of a] (b);
		\vertex [right=of b] (c) [label={ [xshift=0.1cm, yshift=0.1cm] \small $M^{Q}_{4}$}];
		\vertex [right=of c] (d);
		\vertex [right=of d] (e) {\(u_{R3}(d_{R3})\)};
		\vertex [above=of b] (f1) {\(\phi\)};
		\vertex [above=of d] (f2) {\(H_{u(d)}\)};
		\diagram* {
			(a) -- [fermion] (b) -- [charged scalar] (f1),
			(b) -- [edge label'=\(\widetilde{Q}_{R4}\)] (c),
			(c) -- [edge label'=\(Q_{L4}\), inner sep=6pt, insertion=0] (d) -- [charged scalar] (f2),
			(d) -- [fermion] (e),
	};
	\end{feynman}
\end{tikzpicture}}
\par\end{centering}
}$\quad$\subfloat[]{\begin{centering}
\resizebox{.46\textwidth}{!}{
\begin{tikzpicture}
	\begin{feynman}
		\vertex (a) {\(Q_{Li}\)};
		\vertex [right=18mm of a] (b);
		\vertex [right=of b] (c) [label={ [xshift=0.1cm, yshift=0.1cm] \small $M^{u(d)}_{4}$}];
		\vertex [right=of c] (d);
		\vertex [right=of d] (e) {\(u_{Rj}(d_{Rj})\)};
		\vertex [above=of b] (f1) {\(H_{u(d)}\)};
		\vertex [above=of d] (f2) {\(\phi\)};
		\diagram* {
			(a) -- [fermion] (b) -- [charged scalar] (f1),
			(b) -- [edge label'=\(u(d)_{R4}\)] (c),
			(c) -- [edge label'=\(\widetilde{u}(\widetilde{d})_{L4}\), insertion=0] (d) -- [charged scalar] (f2),
			(d) -- [fermion] (e),
	};
	\end{feynman}
\end{tikzpicture}}
\par\end{centering}
}\caption[Diagrams in the fermiophobic $Z'$ model which lead to the effective
Yukawa couplings for second and third family fermions in the mass
insertion approximation]{ Diagrams in the model which lead to the effective Yukawa couplings
for second and third family fermions ($i,j=2,3$) in the mass insertion
approximation. \label{fig: mass_insertion_4thVL}}
\end{figure}

We may arrange the plethora of quark mass terms in Eq.~(\ref{eq:fermoiphobic_lagrangian-1})
into a more convenient matrix formalism as
\begin{equation}
\mathcal{L}_{q}^{\mathrm{ren}}=\left(\begin{array}{ccc}
\overline{Q}_{Li} & \overline{Q}_{L4} & \overline{\widetilde{u}}_{L4}\end{array}\right)M^{u}\left(\begin{array}{c}
u_{Ri}\\
u_{R4}\\
\widetilde{Q}_{R4}
\end{array}\right)+\left(\begin{array}{ccc}
\overline{Q}_{Li} & \overline{Q}_{L4} & \overline{\widetilde{d}}_{L4}\end{array}\right)M^{d}\left(\begin{array}{c}
d_{Ri}\\
d_{R4}\\
\widetilde{Q}_{R4}
\end{array}\right)+\mathrm{h.c.}
\end{equation}
where $i=1,2,3$ and 
\begin{equation}
M^{u}=\left(
\global\long\def\arraystretch{1.3}%
\begin{array}{@{}llcccc@{}}
 & \multicolumn{1}{c@{}}{u_{R1}} & u_{R2} & u_{R3} & u_{R4} & \widetilde{Q}_{R4}\\
\cmidrule(l){2-6}\left.\overline{Q}_{L1}\right| & 0 & 0 & 0 & y_{14}^{u}H_{u} & x_{1}^{Q}\phi\\
\left.\overline{Q}_{L2}\right| & 0 & 0 & 0 & y_{24}^{u}H_{u} & x_{2}^{Q}\phi\\
\left.\overline{Q}_{L3}\right| & 0 & 0 & 0 & y_{34}^{u}H_{u} & x_{3}^{Q}\phi\\
\left.\overline{Q}_{L4}\right| & y_{41}^{u}H_{u} & y_{42}^{u}H_{u} & y_{43}^{u}H_{u} & 0 & M_{4}^{Q}\\
\left.\overline{\widetilde{u}}_{L4}\right| & x_{1}^{u}\phi & x_{2}^{u}\phi & x_{3}^{u}\phi & M_{4}^{u} & 0
\end{array}\right)\,,\label{eq: convenient basis-1}
\end{equation}
with $M^{d}$ given by replacing $u \rightarrow d$ everywhere in the matrix above. Notice that since the upper
$3\times3$ block of Eq.~(\ref{eq: convenient basis-1})
contains zeros, the rotations of the first three families are so far
unphysical. As a consequence, the $Z'$ couplings to the first three
families remain zero under such rotations. Therefore, we are free
to rotate the first three families as we wish, in other words we are
free to choose a convenient basis to start the diagonalisation process.
For example, we are allowed to rotate $Q_{L1}$ and $Q_{L3}$ to set
$x_{1}^{Q}$ to zero and then rotate $Q_{L2}$ and $Q_{L3}$ to set
$x_{2}^{Q}$ to zero. We can apply similar rotations to $u_{R1}$
and $u_{R2}$ to set $y_{41}^{u}$ to zero, and then we rotate $u_{R2}$
and $u_{R3}$ to set $y_{42}^{u}$ to zero. We can repeat these rotations
in the down sector to set $y_{41}^{d}$ and $y_{42}^{d}$ to zero.
We can rotate as well $Q_{L1}$ and $Q_{L2}$ to set $y_{14}^{u}$
to zero (note that in general $y_{14}^{d}\neq0$ since the quark doublet
rotations have all already been used up), and a similar rotation goes
also for $u_{R1}$ and $u_{R2}$ to switch off $x_{1}^{u}$. Finally,
we rotate $d_{R1}$ and $d_{R2}$ to switch off $x_{1}^{d}$. In
this basis, the matrices $M^{u}$ and $M^{d}$ become respectively
\begin{equation}
M^{u}=\left(
\global\long\def\arraystretch{1.3}%
\begin{array}{@{}llcccc@{}}
 & \multicolumn{1}{c@{}}{u_{R1}} & u_{R2} & u_{R3} & u_{R4} & \widetilde{Q}_{R4}\\
\cmidrule(l){2-6}\left.\overline{Q}_{L1}\right| & 0 & 0 & 0 & 0 & 0\\
\left.\overline{Q}_{L2}\right| & 0 & 0 & 0 & y_{24}^{u}H_{u} & 0\\
\left.\overline{Q}_{L3}\right| & 0 & 0 & 0 & y_{34}^{u}H_{u} & x_{3}^{Q}\phi\\
\left.\overline{Q}_{L4}\right| & 0 & 0 & y_{43}^{u}H_{u} & 0 & M_{4}^{Q}\\
\left.\overline{\widetilde{u}}_{L4}\right| & 0 & x_{2}^{u}\phi & x_{3}^{u}\phi & M_{4}^{u} & 0
\end{array}\right)\,,\label{eq: convenient basis_u}
\end{equation}
\begin{equation}
M^{d}=\left(
\global\long\def\arraystretch{1.3}%
\begin{array}{@{}llcccc@{}}
 & \multicolumn{1}{c@{}}{d_{R1}} & d_{R2} & d_{R3} & d_{R4} & \widetilde{Q}_{R4}\\
\cmidrule(l){2-6}\left.\overline{Q}_{L1}\right| & 0 & 0 & 0 & y_{14}^{d}H_{d} & 0\\
\left.\overline{Q}_{L2}\right| & 0 & 0 & 0 & y_{24}^{d}H_{d} & 0\\
\left.\overline{Q}_{L3}\right| & 0 & 0 & 0 & y_{34}^{d}H_{d} & x_{3}^{Q}\phi\\
\left.\overline{Q}_{L4}\right| & 0 & 0 & y_{43}^{d}H_{d} & 0 & M_{4}^{Q}\\
\left.\overline{\widetilde{d}}_{L4}\right| & 0 & x_{2}^{d}\phi & x_{3}^{d}\phi & M_{4}^{d} & 0
\end{array}\right)\,.\label{eq: convenient basis_d}
\end{equation}
There are several distinct mass scales in
the mass matrices above: the VEVs of the Higgs doublets $\left\langle H_{u,d}\right\rangle $, the VEV of the scalar singlet $\left\langle \phi\right\rangle $ and the vector-like fourth family mass terms $M_{4}^{Q,u,d}$. Assuming the latter are much heavier than all the VEVs, we
may integrate out the fourth family to generate effective Yukawa couplings for chiral
quarks, as in the diagrams of Fig.~\ref{fig: mass_insertion_4thVL}. This is denoted as the \textit{mass insertion approximation}, which provides the following effective Yukawa couplings for chiral quarks (see Appendix~\ref{app:mixing_angle_formalism})
\begin{equation}
m_{\mathrm{eff}}^{u}\simeq y_{ij}^{u}\left\langle H_{u}\right\rangle =\left(\begin{array}{ccc}
0 & 0 & 0\\
0 & 0 & 0\\
0 & 0 & x_{3}^{Q}y_{43}^{u}
\end{array}\right)\frac{\left\langle \phi\right\rangle }{M_{4}^{Q}}\left\langle H_{u}\right\rangle +\left(\begin{array}{ccc}
0 & 0 & 0\\
0 & y_{24}^{u}x_{2}^{u} & y_{24}^{u}x_{3}^{u}\\
0 & y_{34}^{u}x_{2}^{u} & y_{34}^{u}x_{3}^{u}
\end{array}\right)\frac{\left\langle \phi\right\rangle }{M_{4}^{u}}\left\langle H_{u}\right\rangle \,,\label{eq:Effective_up}
\end{equation}
\begin{equation}
m_{\mathrm{eff}}^{d}\simeq y_{ij}^{d}\left\langle H_{d}\right\rangle =\left(\begin{array}{ccc}
0 & 0 & 0\\
0 & 0 & 0\\
0 & 0 & x_{3}^{Q}y_{43}^{d}
\end{array}\right)\frac{\left\langle \phi\right\rangle }{M_{4}^{Q}}\left\langle H_{d}\right\rangle +\left(\begin{array}{ccc}
0 & y_{14}^{d}x_{2}^{d} & y_{14}^{d}x_{3}^{d}\\
0 & y_{24}^{d}x_{2}^{d} & y_{24}^{d}x_{3}^{d}\\
0 & y_{34}^{d}x_{2}^{d} & y_{34}^{d}x_{3}^{d}
\end{array}\right)\frac{\left\langle \phi\right\rangle }{M_{4}^{d}}\left\langle H_{d}\right\rangle \,.\label{eq:Effective_down}
\end{equation}
By imposing the \textit{dominance} of the doublet messenger fermions over the
singlet messenger fermions \cite{Ferretti:2006df}, i.e.
\begin{equation}
M_{4}^{Q}\ll M_{4}^{u,d}\,,\label{eq:messenger_dominance_1}
\end{equation}
then the first matrices in Eqs.~(\ref{eq:Effective_up}) and (\ref{eq:Effective_down})
will dominate over the second ones, explaining the heaviness of third
family quarks and the smallness of the CKM elements $V_{cb}$ and
$V_{ub}$. In contrast, the Cabibbo angle is naturally larger and
connected to the ratio $y_{14}^{d}/y_{24}^{d}$. Notice that in order to reproduce
the smaller mass hierarchy in the down sector, we expect $M^{u}_{4}<M^{d}_{4}$
such that $V_{cb}$ and $V_{ub}$ originate mostly from down-quark
mixing. In addition, notice that the effective Yukawa matrices in
Eqs.~(\ref{eq:Effective_up}) and (\ref{eq:Effective_down}) consist
of the sum of two rank 1 matrices, so the first family quarks are
massless so far. Indeed, first family quark masses are protected by
an accidental $U(1)^{3}$ flavour symmetry, that could be minimally
broken in the UV to generate the tiny masses of the up-quark and down-quark,
suggesting a multi-scale origin of flavour as discussed in Section~\ref{subsec:FromPlanckToEW}.
A simple way to achieve this is to introduce a heavy Higgs messenger
uncharged under $U(1)'$ as $h(\mathbf{1,2,}1/2,0)$, which does
not get a VEV. The addition of such field allows to write new terms
in the Lagrangian as
\begin{flalign}
\mathcal{L}_{h}\supset & \;Y_{ij}^{u}\overline{Q}_{Li}\widetilde{h}u_{Rj}+Y_{ij}^{d}\overline{Q}_{Li}hd_{Rj}+Y_{ij}^{e}\overline{L}_{Li}he_{Rj}\\
 & +\frac{1}{2}M_{h}^{2}hh^{\dagger}+f^{u}h\phi H_{u}+f^{d}h^{\dagger}\phi H_{d}+\mathrm{h.c.}\,,\nonumber 
\end{flalign}
where the couplings $f^{u,d}$ carry mass dimensions, and we generally
assume $f^{u,d}\sim M_{h}$. Assuming $\left\langle H_{u,d}\right\rangle \ll\left\langle \phi\right\rangle \ll M_{h}$,
we are able to obtain effective Yukawa couplings for chiral
quarks in the mass insertion approximation via the diagrams in Fig.~\ref{fig: mass_insertion_newHiggs},
\begin{figure}[t]
\subfloat[]{\begin{centering}
\resizebox{.46\textwidth}{!}{
\begin{tikzpicture}
	\begin{feynman}
		\vertex (a) {\(Q_{Li}\)};
		\vertex [right=20mm of a] (b);
		\vertex [right=of b] (c) {\(u_{Rj}\)};
		\vertex [above=of b] (f1) [label={ [xshift=0.5cm, yshift=-0.3cm] \small $M_{h}$}];
		\vertex [above=of f1] (f2);
		\vertex [above left=of f2] (g1) {\(\phi\)};
		\vertex [above right=of f2] (g2) {\(H_{u}\)};
		\diagram* {
			(a) -- [fermion] (b) -- [fermion] (c),
			(b) -- [scalar, edge label'=\(h\)] (f1),
			(f1) -- [scalar, edge label'=\(h\),inner sep=6pt, insertion=0] (f2),
			(f2) -- [scalar] (g1),
			(f2) -- [scalar] (g2),
	};
	\end{feynman}
\end{tikzpicture}}
\par\end{centering}
}$\quad$\subfloat[]{\begin{centering}
\resizebox{.46\textwidth}{!}{
\begin{tikzpicture}
	\begin{feynman}
		\vertex (a) {\(Q_{Li}\)};
		\vertex [right=20mm of a] (b);
		\vertex [right=of b] (c) {\(d_{Rj}\)};
		\vertex [above=of b] (f1) [label={ [xshift=0.5cm, yshift=-0.3cm] \small $M_{h}$}];
		\vertex [above=of f1] (f2);
		\vertex [above left=of f2] (g1) {\(\phi\)};
		\vertex [above right=of f2] (g2) {\(H_{d}\)};
		\diagram* {
			(a) -- [fermion] (b) -- [fermion] (c),
			(b) -- [scalar, edge label'=\(h\)] (f1),
			(f1) -- [scalar, edge label'=\(h\),inner sep=6pt, insertion=0] (f2),
			(f2) -- [scalar] (g1),
			(f2) -- [scalar] (g2),
	};
	\end{feynman}
\end{tikzpicture}}
\par\end{centering}
}\caption[Diagrams in the fermiophobic $Z'$ model which lead to the masses
of the up-quark and down-quark in the mass insertion approximation]{ Diagrams in the model which lead to the masses of the up-quark
and down-quark in the mass insertion approximation. \label{fig: mass_insertion_newHiggs}}
\end{figure}
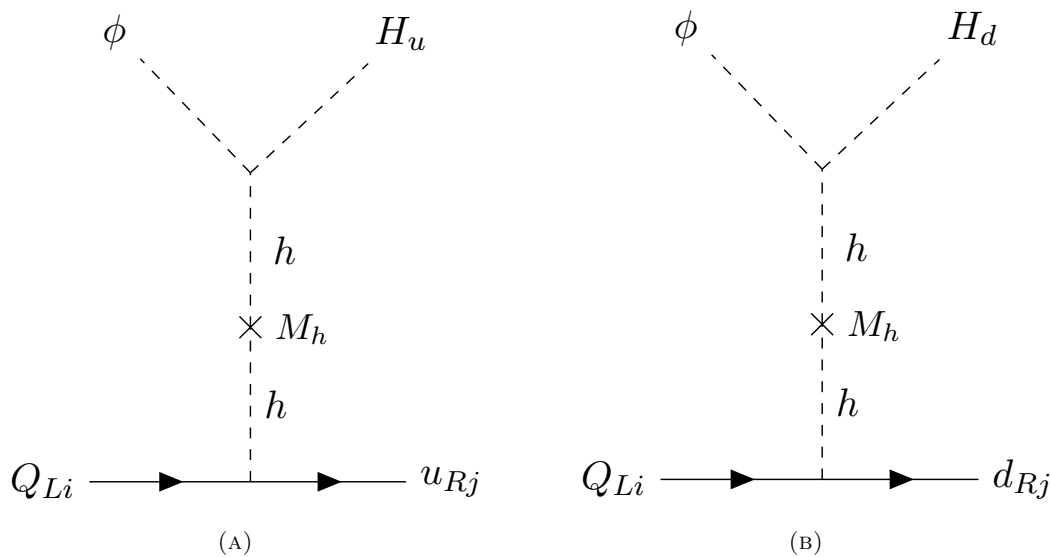
\begin{equation}
Y'^{u}_{ij}=Y_{ij}^{u}f^{u}\frac{\left\langle \phi\right\rangle }{M_{h}^{2}}\,,\label{eq:effective_up_h}
\end{equation}
\begin{equation}
Y'^{d}_{ij}=Y_{ij}^{d}f^{d}\frac{\left\langle \phi\right\rangle }{M_{h}^{2}}\,,\label{eq:effective_down_h}
\end{equation}
which need to be added to the effective mass matrices in Eqs.~(\ref{eq:Effective_up})
and (\ref{eq:Effective_down}). The new Higgs needs to be much heavier
than the vector-like quarks in order to explain the smallness of first
family quark masses, i.e.~the \textit{messenger dominance} of Eq.~(\ref{eq:messenger_dominance_1})
is extended to
\begin{equation}
M_{4}^{Q}\ll M_{4}^{u}<M_{4}^{d}\ll M_{h}\,.\label{eq:messenger_dominance_2}
\end{equation}
Although Eqs.~(\ref{eq:effective_up_h}) and (\ref{eq:effective_down_h})
generate $3\times3$ effective Yukawa matrices, the messenger dominance
above implies that the matrices in Eqs.~(\ref{eq:Effective_up})
and (\ref{eq:Effective_down}) dominate, explaining not only second
and third family quark masses but also CKM mixing originated mostly
from the down sector. The matrices in Eqs.~(\ref{eq:effective_up_h})
and (\ref{eq:effective_down_h}) suppressed by the smaller ratio $\left\langle \phi\right\rangle /M_{h}$
provide masses for the up-quark and down-quark, along with tiny 1-2
and 1-3 right-handed mixing in both the up and down sectors. Tiny
1-2 and 1-3 left-handed mixing in the up sector is negligible with
respect to the larger down-quark mixing, giving a negligible contribution
to the CKM matrix. Given that the masses of the up-quark and down-quark
are very similar, the Yukawa couplings $Y_{11}^{u}$ and $Y_{11}^{d}$
can fit both masses with $\mathcal{O}(1)$ values.

Finally, notice that since the new Higgs is not charged under $U(1)'$,
no effective $Z'$ couplings for first family quarks are generated,
hence the first family can be ignored for phenomenological purposes.

\subsection{Effective quark Yukawa couplings beyond the mass insertion approximation
\label{subsec:Effective-quark-Yukawa}}

The mass insertion approximation discussed in the previous subsection
works generally very well due to the smallness of Yukawa couplings
in the SM, and is very useful for illustrative purposes. However,
the mass insertion approximation breaks for the top Yukawa coupling
which requires $\left\langle \phi\right\rangle /M_{4}^{Q}\sim1$.
Therefore, for the top-quark we need to go beyond and work in a large
mixing angle formalism, as described in Appendix~\ref{app:mixing_angle_formalism}.
Assuming $\left\langle H_{u,d}\right\rangle \ll\left\langle \phi\right\rangle ,M_{4}^{Q,u,d}$
we can block-diagonalise the matrices in Eqs.~(\ref{eq: convenient basis_u}-\ref{eq: convenient basis_d})
via the transformations
\begin{equation}
V_{34}^{Q,u,d}=\left(\begin{array}{cccc}
1 & 0 & 0 & 0\\
0 & 1 & 0 & 0\\
0 & 0 & c_{34}^{Q,u,d} & s_{34}^{Q,u,d}\\
0 & 0 & -s_{34}^{Qu,d} & c_{34}^{Q,u,d}
\end{array}\right)\,,\quad V_{24}^{u,d}=\left(\begin{array}{cccc}
1 & 0 & 0 & 0\\
0 & c_{24}^{u,d} & 0 & s_{24}^{u,d}\\
0 & 0 & 1 & 0\\
0 & -s_{24}^{u,d} & 0 & c_{24}^{u,d}
\end{array}\right)\,,\label{eq:rotations_quarks}
\end{equation}
where
\begin{flalign}
 & s_{34}^{Q}=\frac{x_{3}^{Q}\left\langle \phi\right\rangle }{\sqrt{\left(x_{3}^{Q}\left\langle \phi\right\rangle \right)^{2}+\left(M_{4}^{Q}\right)^{2}}}\,, & s_{24}^{u,d}=\frac{x_{2}^{u,d}\left\langle \phi\right\rangle }{\sqrt{\left(x_{2}^{u,d}\left\langle \phi\right\rangle \right)^{2}+\left(M_{4}^{u,d}\right)^{2}}}\,,\label{eq:mixing_quarks}\\
 & s_{34}^{u,d}=\frac{x_{3}^{u,d}\left\langle \phi\right\rangle }{\sqrt{\left(x_{2}^{u,d}\left\langle \phi\right\rangle \right)^{2}+\left(x_{3}^{u,d}\left\langle \phi\right\rangle \right)^{2}+\left(M_{4}^{u,d}\right)^{2}}}\,.\nonumber 
\end{flalign}
The 3-4 and 2-4 rotations above introduce zeros in the fifth rows and
columns of Eqs.~(\ref{eq: convenient basis_u}-\ref{eq: convenient basis_d}), by absorbing the $x_{i}^{Q,u,d}\left\langle \phi\right\rangle $
factors into a redefinition of the vector-like masses as
\begin{flalign}
 & \hat{M}_{4}^{Q}=\sqrt{\left(x_{3}^{Q}\left\langle \phi\right\rangle \right)^{2}+\left(M_{4}^{Q}\right)^{2}}\,,\\
 & \hat{M}_{4}^{u,d}=\sqrt{\left(x_{2}^{u,d}\left\langle \phi\right\rangle \right)^{2}+\left(x_{3}^{u,d}\left\langle \phi\right\rangle \right)^{2}+\left(M_{4}^{u,d}\right)^{2}}\,,
\end{flalign}
which are indeed the physical masses of the vector-like fermions valid
for arbitrary $\left\langle \phi\right\rangle $ and $M_{4}^{Q,u,d}$
(as long as $\left\langle H_{u,d}\right\rangle \ll\left\langle \phi\right\rangle ,M_{4}^{Q,u,d}$).
Therefore, in the basis (primed) obtained after applying the rotations
in Eq.~(\ref{eq:rotations_quarks}) the fourth family is effectively
decoupled, and the total mass matrices are
\begin{equation}
M^{u}=\left(
\global\long\def\arraystretch{1.3}%
\begin{array}{@{}llcccc@{}}
 & \multicolumn{1}{c@{}}{u'_{R1}} & u'_{R2} & u'_{R3} & u'_{R4} & \widetilde{Q'}_{R4}\\
\cmidrule(l){2-6}\left.\overline{Q'}_{L1}\right| &  &  &  &  & 0\\
\left.\overline{Q'}_{L2}\right| &  &  &  &  & 0\\
\left.\overline{Q'}_{L3}\right| &  &  & y'^{u}_{\alpha\beta}H_{u} &  & 0\\
\left.\overline{Q'}_{L4}\right| &  &  &  &  & \hat{M}_{4}^{Q}\\
\left.\overline{\widetilde{u'}}_{L4}\right| & 0 & 0 & 0 & \hat{M}_{4}^{u} & 0
\end{array}\right)\,,\label{eq: convenient basis-3}
\end{equation}
with $M^{d}$ given by replacing $u \rightarrow d$ everywhere in the matrix above.
The effective Yukawa couplings $y'^{u,d}_{\alpha\beta}$ are given by
\begin{equation}
y'^{u}_{\alpha\beta}=\left(V_{34}^{Q}\right)^{\dagger}\left(\begin{array}{cccc}
0 & 0 & 0 & 0\\
0 & 0 & 0 & y_{24}^{u}H_{u}\\
0 & 0 & 0 & y_{34}^{u}H_{u}\\
0 & 0 & y_{43}^{u}H_{u} & 0
\end{array}\right)V_{24}^{u}V_{34}^{u}\,,
\end{equation}
\begin{equation}
y'^{d}_{\alpha\beta}=\left(V_{34}^{Q}\right)^{\dagger}\left(\begin{array}{cccc}
0 & 0 & 0 & y_{14}^{d}H_{d}\\
0 & 0 & 0 & y_{24}^{d}H_{d}\\
0 & 0 & 0 & y_{34}^{d}H_{d}\\
0 & 0 & y_{43}^{d}H_{d} & 0
\end{array}\right)V_{24}^{d}V_{34}^{d}\,.
\end{equation}
After performing the algebra we obtain,
\begin{equation}
m_{\mathrm{eff}}^{u}=y'^{u}_{ij}\left\langle H_{u}\right\rangle =\left(\begin{array}{ccc}
0 & 0 & 0\\
0 & 0 & 0\\
0 & 0 & y_{43}^{u}c_{34}^{u}s_{34}^{Q}
\end{array}\right)\left\langle H_{u}\right\rangle +\left(\begin{array}{ccc}
0 & 0 & 0\\
0 & y_{24}^{u}s_{24}^{u} & y_{24}^{u}c_{24}^{u}s_{34}^{u}\\
0 & y_{34}^{u}c_{34}^{Q}s_{24}^{u} & y_{34}^{u}c_{34}^{Q}c_{24}^{u}s_{34}^{u}
\end{array}\right)\left\langle H_{u}\right\rangle \,,
\end{equation}
\begin{equation}
m_{\mathrm{eff}}^{d}=y'^{d}_{ij}\left\langle H_{d}\right\rangle =\left(\begin{array}{ccc}
0 & 0 & 0\\
0 & 0 & 0\\
0 & 0 & y_{43}^{d}c_{34}^{d}s_{34}^{Q}
\end{array}\right)\left\langle H_{d}\right\rangle +\left(\begin{array}{ccc}
0 & y_{14}^{d}s_{24}^{d} & y_{14}^{d}c_{24}^{d}s_{34}^{d}\\
0 & y_{24}^{d}s_{24}^{d} & y_{24}^{d}c_{24}^{d}s_{34}^{d}\\
0 & y_{34}^{d}c_{34}^{Q}s_{24}^{d} & y_{34}^{d}c_{34}^{Q}c_{24}^{d}s_{34}^{d}
\end{array}\right)\left\langle H_{d}\right\rangle \,,
\end{equation}
where, if we take the limit $\left\langle \phi\right\rangle \ll M_{4}^{Q,u,d}$
in Eq.~(\ref{eq:mixing_quarks}) to obtain $s_{i4}^{Q,u,d}\approx x_{i}^{Q,u,d}\left\langle \phi\right\rangle /M_{4}^{Q,u,d}$,
then the results in the mass insertion approximation as shown in the previous
subsection are recovered. However, we shall continue working
in the large mixing angle formalism to account for $\left\langle \phi\right\rangle /M_{4}^{Q}\sim1$
as required to address the large top Yukawa coupling.

Since the angles $s_{34}^{u,d}$ need to be small in order
to explain the smallness of $V_{cb}$, we are left with $s_{34}^{Q}$,
the Higgs VEVs $\left\langle H_{u,d}\right\rangle $ and the couplings
$y_{43}^{u,d}$ to fit the top and bottom masses. The hierarchy $m_{b}/m_{t}$
could be explained either by having an unnaturally small $y_{43}^{d}\sim0.01$
coupling, or by assuming that the ratio of Higgs VEVs $\tan\beta=v_{u}/v_{d}$
is large. We follow this second path, taking a large 3-4 mixing $s_{34}^{Q}\sim1/\sqrt{2}$
in order to explain the large top Yukawa. By taking $\tan\beta=v_{u}/v_{d}\approx30$
we have $y_{43}^{d}\approx1$ and $y_{43}^{u}\approx\sqrt{2}\approx1.4$,
which are reasonable values. Notice that $s_{34}^{Q}\sim1/\sqrt{2}$
requires $M_{4}^{Q}\sim\left\langle \phi\right\rangle $, hinting
to the existence of a relatively light vector-like
quark, within the reach of the LHC.

The smallness of $m_{c}$ and $m_{s}$ can be naturally explained
by taking small mixing angles $s_{24}^{u,d}$. In particular, we find
$s_{24}^{u}\sim0.01$ and $s_{24}^{d}\sim0.02$ to deliver a good
fit of $m_{c}$ and $m_{s}$. In this manner, we expect $s_{34}^{u,d}$
of similar order because they live at the same scale. Therefore, given
that only $s_{34}^{Q}$ is large while all the other angles are small,
we can apply the limit $\left\langle \phi\right\rangle \ll M_{4}^{u,d}$
to recover the expressions in the mass insertion approximation as
\begin{flalign}
m_{\mathrm{eff}}^{u}=y'^{u}_{ij}\left\langle H_{u}\right\rangle  & =\left(\begin{array}{ccc}
0 & 0 & 0\\
0 & 0 & 0\\
0 & 0 & y_{43}^{u}c_{34}^{u}s_{34}^{Q}
\end{array}\right)\left\langle H_{u}\right\rangle +\left(\begin{array}{ccc}
0 & 0 & 0\\
0 & y_{24}^{u}x_{2}^{u} & y_{24}^{u}x_{3}^{u}\\
0 & y_{34}^{u}c_{34}^{Q}x_{2}^{u} & y_{34}^{u}c_{34}^{Q}x_{3}^{u}
\end{array}\right)\frac{\left\langle \phi\right\rangle }{M_{4}^{d}}\left\langle H_{u}\right\rangle \\
 & +\left(\begin{array}{ccc}
Y_{11}^{u} & Y_{12}^{u} & Y_{13}^{u}\\
Y_{21}^{u} & Y_{22}^{u} & Y_{23}^{u}\\
Y_{31}^{u}c_{34}^{Q} & Y_{32}^{u}c_{34}^{Q} & Y_{33}^{u}c_{34}^{Q}
\end{array}\right)f^{u}\frac{\left\langle \phi\right\rangle }{M_{h}^{2}}\left\langle H_{u}\right\rangle \,,\nonumber 
\end{flalign}
\begin{flalign}
m_{\mathrm{eff}}^{d}=y'^{d}_{ij}\left\langle H_{d}\right\rangle  & =\left(\begin{array}{ccc}
0 & 0 & 0\\
0 & 0 & 0\\
0 & 0 & y_{43}^{d}c_{34}^{d}s_{34}^{Q}
\end{array}\right)\left\langle H_{d}\right\rangle +\left(\begin{array}{ccc}
0 & y_{14}^{d}x_{2}^{d} & y_{14}^{d}x_{3}^{d}\\
0 & y_{24}^{d}x_{2}^{d} & y_{24}^{d}x_{3}^{d}\\
0 & y_{34}^{d}c_{34}^{Q}x_{2}^{d} & y_{34}^{d}c_{34}^{Q}x_{3}^{d}
\end{array}\right)\frac{\left\langle \phi\right\rangle }{M_{4}^{d}}\left\langle H_{d}\right\rangle \\
 & +\left(\begin{array}{ccc}
Y_{11}^{d} & Y_{12}^{d} & Y_{13}^{d}\\
Y_{21}^{d} & Y_{22}^{d} & Y_{23}^{d}\\
Y_{31}^{d}c_{34}^{Q} & Y_{32}^{d}c_{34}^{Q} & Y_{33}^{d}c_{34}^{Q}
\end{array}\right)f^{d}\frac{\left\langle \phi\right\rangle }{M_{h}^{2}}\left\langle H_{d}\right\rangle \,,\nonumber 
\end{flalign}
where we have also included the contributions from the heavy Higgs
messengers that provide the mass of the up-quark and down-quark, plus
negligible up-quark mixing and right-handed down-quark mixing.

\subsection{Effective charged lepton Yukawa couplings \label{subsec:Effective-charged-lepton}}

Having discussed in detail the origin of quark mass hierarchies and
mixing, we now build the charged lepton sector in a similar way. We
construct the full charged lepton mass matrix as
\begin{equation}
\mathcal{L}_{q}^{\mathrm{ren}}=\left(\begin{array}{ccc}
\overline{L}_{Li} & \overline{L}_{L4} & \overline{\widetilde{e}}_{L4}\end{array}\right)M^{e}\left(\begin{array}{c}
e_{Ri}\\
e_{R4}\\
\widetilde{L}_{R4}
\end{array}\right)+\mathrm{h.c.}\,,
\end{equation}
where
\begin{equation}
M^{e}=\left(
\global\long\def\arraystretch{1.3}%
\begin{array}{@{}llcccc@{}}
 & \multicolumn{1}{c@{}}{e_{R1}} & e_{R2} & e_{R3} & e_{R4} & \widetilde{L}_{R4}\\
\cmidrule(l){2-6}\left.\overline{L}_{L1}\right| & 0 & 0 & 0 & 0 & 0\\
\left.\overline{L}_{L2}\right| & 0 & 0 & 0 & y_{24}^{e}H_{d} & 0\\
\left.\overline{L}_{L3}\right| & 0 & 0 & 0 & y_{34}^{e}H_{d} & x_{3}^{L}\phi\\
\left.\overline{L}_{L4}\right| & 0 & 0 & y_{43}^{e}H_{d} & 0 & M_{4}^{L}\\
\left.\overline{\widetilde{e}}_{L4}\right| & 0 & x_{2}^{e}\phi & x_{3}^{e}\phi & M_{4}^{e} & 0
\end{array}\right)\label{eq:mass_matrix}
\end{equation}
where we have already taking advantage of the upper $3\times3$ zeros
to perform suitable rotations of the first family fields that introduce
extra zeros in the full mass matrix without loss generality, as we
did in the quark sector (see Eqs.~(\ref{eq: convenient basis_u}-\ref{eq: convenient basis_d}) and
the discussion therein). Here we work directly in the large mixing
angle formalism as in Section~\ref{subsec:Effective-quark-Yukawa},
because we are seeking for large mixing angles in the charged lepton sector
that will contribute to large $Z'$ couplings for $(g-2)_{\mu}$.
Assuming $\left\langle H_{d}\right\rangle \ll\left\langle \phi\right\rangle ,M_{4}^{L,e}$
we can block-diagonalise the matrix in Eq.~(\ref{eq:mass_matrix})
via the transformations
\begin{equation}
V_{34}^{L,e}=\left(\begin{array}{cccc}
1 & 0 & 0 & 0\\
0 & 1 & 0 & 0\\
0 & 0 & c_{34}^{L,e} & s_{34}^{L,e}\\
0 & 0 & -s_{34}^{L,e} & c_{34}^{L,e}
\end{array}\right)\,,\qquad V_{24}^{e}=\left(\begin{array}{cccc}
1 & 0 & 0 & 0\\
0 & c_{24}^{e} & 0 & s_{24}^{e}\\
0 & 0 & 1 & 0\\
0 & -s_{24}^{e} & 0 & c_{24}^{e}
\end{array}\right)\,,\label{eq:rotations_leptons}
\end{equation}
where
\begin{flalign}
 & s_{34}^{L}=\frac{x_{3}^{L}\left\langle \phi\right\rangle }{\sqrt{\left(x_{3}^{L}\left\langle \phi\right\rangle \right)^{2}+\left(M_{4}^{L}\right)^{2}}}\,, & s_{24}^{e}=\frac{x_{2}^{e}\left\langle \phi\right\rangle }{\sqrt{\left(x_{2}^{e}\left\langle \phi\right\rangle \right)^{2}+\left(M_{4}^{e}\right)^{2}}}\,,\\
 & s_{34}^{e}=\frac{x_{3}^{e}\left\langle \phi\right\rangle }{\sqrt{\left(x_{2}^{e}\left\langle \phi\right\rangle \right)^{2}+\left(x_{3}^{e}\left\langle \phi\right\rangle \right)^{2}+\left(M_{4}^{e}\right)^{2}}}\,.\nonumber 
\end{flalign}
The 3-4 and 2-4 rotations above introduce zeros in the fifth row and
column of Eq.~(\ref{eq:mass_matrix}), by absorbing the $x_{i}^{L,e}\left\langle \phi\right\rangle $
factors into a redefinition of the vector-like masses as
\begin{flalign}
 & \hat{M}_{4}^{L}=\sqrt{\left(x_{3}^{L}\left\langle \phi\right\rangle \right)^{2}+\left(M_{4}^{L}\right)^{2}}\,,\\
 & \hat{M}_{4}^{e}=\sqrt{\left(x_{2}^{e}\left\langle \phi\right\rangle \right)^{2}+\left(x_{3}^{e}\left\langle \phi\right\rangle \right)^{2}+\left(M_{4}^{e}\right)^{2}}\,,
\end{flalign}
in complete analogy with the situation in the quark sector. Therefore,
in the basis (primed) obtained after applying the rotations in Eq.~(\ref{eq:rotations_leptons})
the fourth family is effectively decoupled, and the total mass matrix
is similar to the up-quark mass matrix in Eq.~(\ref{eq: convenient basis-3}).
The effective Yukawa couplings can then be readily extracted as
\begin{equation}
m_{\mathrm{eff}}^{e}=y'^{e}_{ij}\left\langle H_{d}\right\rangle =\left(\begin{array}{ccc}
0 & 0 & 0\\
0 & 0 & 0\\
0 & 0 & y_{43}^{e}c_{34}^{e}s_{34}^{L}
\end{array}\right)\left\langle H_{d}\right\rangle +\left(\begin{array}{ccc}
0 & 0 & 0\\
0 & y_{24}^{e}s_{24}^{e} & y_{24}^{e}c_{24}^{e}s_{34}^{e}\\
0 & y_{34}^{e}c_{34}^{L}s_{24}^{e} & y_{34}^{e}c_{34}^{L}c_{24}^{e}s_{34}^{e}
\end{array}\right)\left\langle H_{d}\right\rangle \,.
\end{equation}
Notice that the electron remains massless so far, however it can get
a mass at the scale of the up-quark and down-quark masses via the
heavy Higgs $h$, in complete analogy with the quark sector as discussed
in Eqs.~(\ref{eq:effective_up_h}-\ref{eq:effective_down_h}) and
paragraphs therein. However, in order to explain the smallness of
the electron mass with respect to the masses of first family quarks,
we would need a small dimensionless coupling at the 10\% level. This could
be avoided by adding another Higgs $h'$ a bit heavier than $h$ which
generates the smaller mass of the electron, without generating any
effective couplings to the $Z'$ boson, such that the related phenomenology
is dominated by the 2-3 sector. Therefore, in the following we ignore
the first family and proceed with our discussion of the 2-3 sector.

Given that we have enforced $\tan\beta=v_{u}/v_{d}\approx30$ in our
discussion of the quark sector, we need $s_{34}^{L}\approx0.3$ to
explain the tau mass, and $s_{24}^{e}\approx0.02$ to explain the
muon mass. Therefore, the charged fermion mass hierarchies are explained
by the messenger dominance $M_{4}^{L}\ll M_{4}^{e}$, which then predicts
that $s_{34}^{e}$ is of a similar order than $s_{24}^{e}$. By taking
$s_{34}^{e}\approx0.02$ and assuming $y_{43}^{e}\approx y_{24}^{e}\approx1$,
we obtain a significant 2-3 left-handed charged lepton mixing as $s_{23}^{e_{L}}\approx s_{34}^{e}/s_{34}^{L}\approx0.07$.
Similarly, by taking $y_{34}^{e}\approx1$ we obtain significant 2-3
right-handed charged lepton mixing as $s_{23}^{e_{R}}\approx s_{24}^{e}/s_{34}^{L}\approx0.07$.
This non-vanishing 2-3 mixing will be very relevant for the low-energy
phenomenology. On the one hand, it will generate effective couplings
to left-handed muons necessary to give a contribution to $(g-2)_{\mu}$,
which would otherwise vanish, and on the other hand it will induce lepton flavour-violating
transitions such as $\tau\rightarrow\mu\gamma$.

\subsection{\texorpdfstring{Effective $Z'$ couplings}{Effective Z' couplings}}

Since all chiral fermions are uncharged under $U(1)'$, in the interaction
basis the massive $Z'$ boson only couples to the fourth family vector-like
fermions, 
\begin{equation}
\mathcal{L}_{Z'}\supset g'\left(\overline{Q}_{L}D_{Q}\gamma^{\mu}Q_{L}+\overline{u}_{R}D_{u}\gamma^{\mu}u_{R}+\overline{d}_{R}D_{d}\gamma^{\mu}d_{R}+\overline{L}_{L}D_{L}\gamma^{\mu}L_{L}+\overline{\ell}_{R}D_{\ell}\gamma^{\mu}\ell_{R}\right)Z'_{\mu}\,,
\end{equation}
where we are ignoring couplings to the conjugate (tilde) fermions,
because so far they do not mix with chiral fermions and so are not
relevant for the low-energy phenomenology. We are also ignoring couplings
to vector-like neutrinos, assuming that they are very heavy and decoupled
from low-energy phenomenology, as suggested by the type Ib seesaw
mechanism implemented in Section~\ref{subsec:Neutrino-masses-and-type.Ib}.
We have defined 4-component vectors including the four families (3
chiral families plus the fourth vector-like), and the $D_{\psi}$
matrices are defined as
\begin{equation}
D_{Q}=D_{L}=\mathrm{diag}(0,0,0,1)\,,
\end{equation}
\begin{equation}
D_{u}=D_{d}=D_{e}=\mathrm{diag}(0,0,0,-1)\,.
\end{equation}
The same rotations that diagonalise the full mass matrices in Sections~\ref{subsec:Effective-quark-Yukawa}
and \ref{subsec:Effective-charged-lepton} need to be applied now
in order to go to the basis of mass eigenstates. For illustrative
purposes, one can see the diagrams in Fig.~\ref{fig:Z_couplings_mass_insertion}
that provided effective $Z'$ couplings in the mass insertion approximation
for the simplified model. In contrast, here we shall work in the more
general large mixing angle formalism, even though most of the mixing
angles turn out to be small. Up to the effects of chiral fermion mixing,
we obtain
\begin{equation}
(V_{34}^{Q,L})^{\dagger}D_{Q,L}V_{34}^{Q,L}=\left(\begin{array}{cccc}
0 & 0 & 0 & 0\\
0 & 0 & 0 & 0\\
0 & 0 & (s_{34}^{Q,L})^{2} & -c_{34}^{Q,L}s_{34}^{Q,L}\\
0 & 0 & -c_{34}^{L}s_{34}^{Q,L} & (c_{34}^{Q,L})^{2}
\end{array}\right)\,,
\end{equation}
\begin{equation}
(V_{24}^{u,d,e}V_{34}^{u,d,e})^{\dagger}D_{u,d,e}V_{24}^{u,d,e}V_{34}^{u,d,e}=-\left(\begin{array}{cccc}
0 & 0 & 0 & 0\\
0 & (s_{24}^{e_{R}})^{2} & s_{24}^{e_{R}}s_{34}^{e_{R}} & c_{34}^{e_{R}}c_{24}^{e_{R}}s_{24}^{e_{R}}\\
0 & s_{24}^{e_{R}}s_{34}^{e_{R}} & (c_{24}^{e_{R}}s_{34}^{e_{R}})^{2} & -c_{34}^{e_{R}}(c_{24}^{e_{R}})^{2}s_{34}^{e_{R}}\\
0 & c_{34}^{e_{R}}c_{24}^{e_{R}}s_{24}^{e_{R}} & -c_{34}^{e_{R}}(c_{24}^{e_{R}})^{2}s_{34}^{e_{R}} & (c_{24}^{e_{R}}c_{34}^{e_{R}})^{2}
\end{array}\right).
\end{equation}
Notice that the same mixing that provides the effective Yukawa couplings
also leads to effective $Z'$ couplings for chiral fermions. In the
left-handed sector only the third family (and fourth) couples to the
$Z'$ so far, but effective couplings to the second family are introduced
when 2-3 chiral fermion mixing is considered, splitting the $SU(2)_{L}$ doublets
and leading to
\begin{flalign}
 & (V_{34}^{Q}V_{23}^{u_{L},d_{L}})^{\dagger}D_{u_{L},d_{L}}V_{34}^{Q}V_{23}^{u_{L},d_{L}}\\
 & =\left(\begin{array}{cccc}
0 & 0 & 0 & 0\\
0 & (s_{34}^{Q}s_{23}^{u_{L},d_{L}})^{2} & -(s_{34}^{Q})^{2}c_{23}^{u_{L},d_{L}}s_{23}^{u_{L},d_{L}} & c_{34}^{Q}s_{34}^{Q}s_{23}^{u_{L},d_{L}}\\
0 & -(s_{34}^{Q})^{2}c_{23}^{u_{L},d_{L}}s_{23}^{u_{L},d_{L}} & (s_{34}^{Q}c_{23}^{u_{L},d_{L}})^{2} & -c_{34}^{Q}s_{34}^{Q}c_{23}^{u_{L},d_{L}}\\
0 & c_{34}^{Q}s_{34}^{Q}s_{23}^{u_{L},d_{L}} & -c_{34}^{Q}s_{34}^{Q}c_{23}^{u_{L},d_{L}} & (c_{34}^{Q})^{2}
\end{array}\right)\,,\nonumber 
\end{flalign}
\begin{flalign}
 & (V_{34}^{L}V_{23}^{e_{L}})^{\dagger}D_{e_{L}}V_{34}^{L}V_{23}^{e_{L}}\\
 & =\left(\begin{array}{cccc}
0 & 0 & 0 & 0\\
0 & (s_{34}^{L}s_{23}^{e_{L}})^{2} & -(s_{34}^{L})^{2}c_{23}^{e_{L}}s_{23}^{e_{L}} & c_{34}^{L}s_{34}^{L}s_{23}^{e_{L}}\\
0 & -(s_{34}^{L})^{2}c_{23}^{e_{L}}s_{23}^{e_{L}} & (s_{34}^{L}c_{23}^{e_{L}})^{2} & -c_{34}^{L}s_{34}^{L}c_{23}^{e_{L}}\\
0 & c_{34}^{L}s_{34}^{L}s_{23}^{e_{L}} & -c_{34}^{L}s_{34}^{L}c_{23}^{e_{L}} & (c_{34}^{L})^{2}
\end{array}\right)\,.\nonumber 
\end{flalign}
The origin of first family masses is connected to the much heavier
Higgs messengers that are uncharged under $U(1)'$, providing no effective
$Z'$ couplings to first family fermions. These would arise via chiral
fermion mixing, but will be further suppressed by powers of small
mixing angles so that we can safely neglect them. When we include
2-3 chiral fermion mixing for right-handed chiral fermions, the expressions
for the $D_{u_{R},d_{R},e_{R}}$ matrices become very lengthy. Since
only the charged lepton sector is relevant for $(g-2)_{\mu}$, we
shall include below the explicit $Z'$ couplings for charged leptons
as
\begin{flalign}
 & g_{\mu\mu}^{R}=-g'(s_{24}^{e_{R}}c_{23}^{e_{R}}-c_{24}^{e_{R}}s_{34}^{e_{R}}s_{23}^{e_{R}})^{2}\,,\\
 & g_{\tau\tau}^{R}=-g'(s_{24}^{e_{R}}s_{23}^{e_{R}}+c_{24}^{e_{R}}s_{34}^{e_{R}}c_{23}^{e_{R}})^{2}\,,\\
 & g_{\mu\tau}^{R}=g'(s_{24}^{e_{R}}c_{23}^{e_{R}}-c_{24}^{e_{R}}s_{34}^{e_{R}}s_{23}^{e_{R}})(s_{24}^{e_{R}}s_{23}^{e_{R}}+c_{24}^{e_{R}}s_{34}^{e_{R}}c_{23}^{e_{R}})\,,\\
 & g_{\mu E}^{R}=g'c_{34}^{e_{R}}c_{24}^{e_{R}}(c_{23}^{e_{R}}s_{24}^{e_{R}}-c_{24}^{e_{R}}s_{34}^{e_{R}}s_{23}^{e_{R}})\,,\\
 & g_{\tau E}^{R}=g'c_{34}^{e_{R}}c_{24}^{e_{R}}(s_{24}^{e_{R}}s_{23}^{e_{R}}+c_{24}^{e_{R}}c_{23}^{e_{R}}s_{34}^{e_{R}})\,,\\
 & g_{EE}^{R}=-g'(c_{24}^{e_{R}}c_{34}^{e_{R}})^{2}\,,
\end{flalign}
defined as in the Lagrangian of Eq.~(\ref{eq:Z'_couplings}). In
the next section we shall study the phenomenology of the model, with
particular attention to the lepton sector and $(g-2)_{\mu}$.

\section{Phenomenology of the flavour model}

\subsection{\texorpdfstring{$B_{s}-\bar{B}_{s}$ mixing}{Bs-Bsbar mixing}\label{subsec:MesonMixing_FermiophobicZp}}

The effective $Z'$ couplings of the flavour model in the previous
section naturally contribute to flavour-violating 2-3 transitions.
Assuming that the CKM mixing originates from the down sector as discussed
in Section~\ref{subsec:Messenger_Dominance}, and neglecting the
small $Z'$ couplings to right-handed quarks which are suppressed
by the heavier messengers, then the model predicts a significant tree-level
contribution to the operator $\mathcal{Q}_{1}^{bs}$ contributing
to $B_{s}-\bar{B}_{s}$ meson mixing (see Section~\ref{subsec:BsMixing}).
We obtain the bound
\begin{equation}
C_{1}^{bs}=\frac{\left(g_{bs}^{L}\right)^{2}}{M_{Z'}^{2}}\simeq\frac{\left(g'(s_{34}^{Q})^{2}V_{cb}\right)^{2}}{M_{Z'}^{2}}\apprle\frac{1}{(225\,\mathrm{TeV})^{2}}\,.
\end{equation}
The mixing angle $s_{34}^{Q}$ needs to be large in order to explain
the heaviness of the top mass. In particular, perturbativity of the
Yukawa coupling $y_{43}^{u}\apprle\sqrt{4\pi}$ imposes $s_{34}^{Q}\apprge0.285$,
which translates to the lower 95\% CL bound $M_{Z'}/g'\apprge750\:\mathrm{GeV}$
via $B_{s}-\bar{B}_{s}$ meson mixing. In contrast, if we take $y_{43}^{u}\approx\sqrt{2}$
then we require $s_{34}^{Q}\approx1/\sqrt{2}$ as discussed in Section~\ref{subsec:Effective-quark-Yukawa},
which translates into a bound $M_{Z'}/g'\apprge4.5\:\mathrm{TeV}$.

\subsection{\texorpdfstring{$B$-physics: $R_{K^{(*)}}$ and $B_{s}\rightarrow\mu\mu$}{B-physics: RK(*) and Bsmumu}}

In our model, the effective $Z'$ couplings are connected to the origin
of flavour hierarchies in the SM. Therefore, the effective $Z'$ couplings
break lepton flavour universality in a similar manner as the usual Yukawa
couplings do in the SM. Our $Z'$ is mostly coupled to third family left-handed
fermions, while couplings to light fermions are induced via small
chiral fermion mixing in the left-handed sector, or via small mixing
angles connected to the origin of second family masses in the right-handed
sector. The smallest $Z'$ couplings are then found for first family
fermions, suppressed by powers of small mixing angles.

This explicit breaking of lepton flavour universality was suggested
as a possible explanation of the $R_{K^{(*)}}$ anomalies \cite{King:2018fcg},
but it was soon concluded that providing a significant contribution
to the $R_{K^{(*)}}$ ratios was either in conflict with the combined
bounds from $B_{s}-\bar{B}_{s}$ meson mixing and $\tau\rightarrow3\mu$
\cite{King:2020fdl}, or required the vector-like leptons to mix predominantly
with muons rather than taus, disposing of the natural explanation
of charged lepton mass hierarchies \cite{King:2018fcg}. 

After the recent update by LHCb \cite{LHCb:2022qnv}, the $R_{K^{(*)}}$
ratios are now in agreement with the SM but leaving some space for
NP contributions. In our model, we can safely neglect very small contributions
to $\mathcal{O}_{9}^{ee}$ and $\mathcal{O}_{10}^{ee}$, and consider
only the contributions to the operators $\mathcal{O}_{9}^{\mu\mu}$
and $\mathcal{O}_{10}^{\mu\mu}$ obtained as (see Section~\ref{subsec:bsll}
for the definition of the operators)
\begin{equation}
C_{9}^{\mu\mu}=-\frac{2\pi}{\alpha_{\mathrm{EM}}(V_{tb}V_{ts})^{*}}\frac{\sqrt{2}}{4G_{F}}\frac{\left(g_{bs}^{L}g_{\mu\mu}^{L}+g_{bs}^{L}g_{\mu\mu}^{R}\right)}{M_{Z'}^{2}}\,,
\end{equation}
\begin{equation}
C_{10}^{\mu\mu}=-\frac{2\pi}{\alpha_{\mathrm{EM}}(V_{tb}V_{ts})^{*}}\frac{\sqrt{2}}{4G_{F}}\frac{\left(-g_{bs}^{L}g_{\mu\mu}^{L}+g_{bs}^{L}g_{\mu\mu}^{R}\right)}{M_{Z'}^{2}}\,.
\end{equation}
The $R_{K^{(*)}}$ ratios are then computed via Eqs.~(\ref{eq:RKth}-\ref{eq:RKstarth}).
The effective $Z'$ couplings above involving second family fermions
are generally suppressed by small mixing angles, in such a way that
for the typical configurations of mixing angles in the model (see
Sections~\ref{subsec:Effective-quark-Yukawa} and \ref{subsec:Effective-charged-lepton})
we obtain at 95\% CL $M_{Z'}/g'\apprge350\;\mathrm{GeV}$. The Wilson
coefficient $C_{10}^{\mu\mu}$ also contributes to $\mathcal{B}(B_{s}\rightarrow\mu\mu)$,
however we find the obtained bounds to be generally smaller than those
from $R_{K^{(*)}}$, obtaining for the typical benchmark of the model
$M_{Z'}/g'\apprge270\;\mathrm{GeV}$ at 95\% CL. Going beyond the
benchmark discussed in Sections~\ref{subsec:Effective-quark-Yukawa}
and \ref{subsec:Effective-charged-lepton}, as it might be required
by the $(g-2)_{\mu}$ anomaly, these bounds could become more significant
if we consider larger mixing angles in the charged lepton sector. In this manner,
we do not expect that our model can provide significant contributions to $b\rightarrow s\mu \mu$
without entering in conflict with the $R_{K^{(*)}}$ ratios.

Given that the $Z'$ boson is mostly coupled to third family fermions, we have studied the enhancement of $b\rightarrow s \tau \tau$ and $b\rightarrow s \nu \nu$. However, in both cases we find that any significant enhancement is in tension with bounds from $B_{s}-\bar{B}_{s}$ mixing over the $\bar{s}bZ'$ coupling. For example, we find that $\mathcal{B}(B\rightarrow K^{(*)}\nu \bar{\nu})$ can be enhanced just by a 10\% factor over the SM prediction (similar to the conclucion of \cite{Athron:2023hmz}), making impossible to provide a significant contribution that ameliorates the $2.8\sigma$ tension in the recent measurement by Belle II \cite{BelleIIEPS:2023}.

\subsection{Collider searches}

The $Z'$ boson is produced at the LHC mainly via the partonic process
$b\overline{b}\rightarrow Z'$, because couplings to light quarks
are heavily suppressed by powers of small mixing angles. Then it decays
mostly to third families fermions, and to a lesser extent to second
family fermions. Other modes such as $Z'\rightarrow W^{+}W^{-}$ and $Z'\rightarrow Zh$ arise from
$Z-Z'$ mixing. Assuming kinetic mixing is not generated at tree-level
(e.g.~if $U(1)_{Y}$ and $U(1)'$ originate from different semi-simple
groups), then $Z-Z'$ mixing may still be generated at 1-loop mediated by
$H_{u,d}$ and the VL fermions which are charged under both $U(1)$s (as discussed in Section~\ref{subsec:g-2_simplified}),
or via the VEVs of $H_{u,d}$ but carrying a suppression of $\sim M_{Z}^{2}/M_{Z'}^{2}$.
In either case, we find the modes $Z'\rightarrow WW$ and $Z'\rightarrow Zh$ to be sufficiently suppressed
and the bounds extracted from the experimental searches \cite{ATLAS:2017xel,CMS:2018ljc} are not
competitive with those obtained from the Drell-Yan modes.

We have prepared the UFO file of the model using
\texttt{FeynRules} \cite{Feynrules:2013bka}, and then we have computed
the $Z'$ production cross section from 13 TeV $pp$ collisions using
\texttt{Madgraph5} \cite{Madgraph:2014hca} with the default PDF \texttt{NNPDF23LO}.
We estimated analytically the decay width to fermion modes as
\begin{equation}
\mathrm{\Gamma}(Z'\rightarrow f_{\alpha}\overline{f}_{\beta})=\frac{N_{c}}{24\pi}M_{Z'}\left((g_{f_{\alpha}f_{\beta}}^{L})^{2}+(g_{f_{\alpha}f_{\beta}}^{R})^{2}\right)\,,
\end{equation}
where $N_{c}=\text{3 (quarks), 1 (leptons)}$. We assume that decays
into vector-like fermions are kinematically forbidden or suppressed,
and we treat the top-quark separately due to its larger mass,
\begin{equation}
\mathrm{\Gamma}(Z'\rightarrow t\overline{t})=\frac{1}{8\pi}M_{Z'}\left((g_{tt}^{L})^{2}+(g_{tt}^{R})^{2}\right)\left(1-\frac{m_{t}^{2}}{M_{Z'}^{2}}\right)\sqrt{\left(1-\frac{4m_{t}^{2}}{M_{Z'}^{2}}\right)}\,,
\end{equation}
and finally the branching fraction is estimated as $\mathcal{B}(Z'\rightarrow f_{\alpha}\overline{f}_{\beta})=\mathrm{\Gamma}(Z'\rightarrow f_{\alpha}\overline{f}_{\beta})/\mathrm{\Gamma}(Z'\rightarrow\mathrm{all})$.
In the natural benchmark for the flavour model we find the total decay width to be narrow $\Gamma_{Z'}/M_{Z'}\sim\mathcal{O}(1\%)$, as the $Z'$ couplings to light families are generally suppressed and decays to VL fermions are kinematically forbidden.
In the next section we will tune $s_{24}^{e_{R}}$ to be large in
order to get a larger coupling to muons, in this case we find $\Gamma_{Z'}/M_{Z'}\sim5\%$.

We compute the total cross section by applying the narrow width
approximation $\sigma(pp\rightarrow Z'\rightarrow f_{\alpha}\overline{f}_{\beta})\approx\sigma(pp\rightarrow Z')\mathcal{B}(Z'\rightarrow f_{\alpha}\overline{f}_{\beta})$.
We confront our results with the limits from the most recent light
dilepton resonance searches by ATLAS \cite{ATLAS:2019erb} and CMS
\cite{CMS:2021ctt} in order to obtain 95\% CL exclusion bounds. However,
given that the $Z'$ is mostly coupled to third family fermions, the
bounds from ditau \cite{ATLAS:2017eiz} and ditop \cite{ATLAS:2020lks}
searches generally dominate over the light dilepton searches. For
the benchmark values motivated in Sections~\ref{subsec:Effective-quark-Yukawa}
and \ref{subsec:Effective-charged-lepton} to explain the origin of
flavour hierarchies, we obtain that the most stringent bounds come
from ditau searches (despite the lesser integrated luminosity of this
analysis), leading to $M_{Z'}\apprge1.1\;\mathrm{TeV}$ for $g'\approx1$
at 95\% CL. However, as we shall see in the next subsection, if we
increase the mixing angles of the lepton sector in our seek for explaining
$(g-2)_{\mu}$, the light dilepton searches become more constraining
than ditau and ditop searches.

Finally we comment on vector-like fermion searches. In order to explain fermion mass hierarchies,
we expect most of the vector-like fermions to be much
heavier than the $Z'$ boson, and only $Q_{4}$ and $L_{4}$ may live at the TeV scale.
Current bounds on vector-like quark masses lie around 1 TeV, however the strongest
bounds are usually model dependent (see e.g.~\cite{CMS:2019eqb}). Therefore, while $L_{4}$ most likely escapes detection,
$Q_{4}$ is produced at the LHC via gluon fusion and may be observable
through its decays to third family quarks.

\subsection{\texorpdfstring{$(g-2)_{\mu}$ and $\mathcal{B}(\tau\rightarrow\mu\gamma)$}{(g-2)mu and B(taumugamma)}}

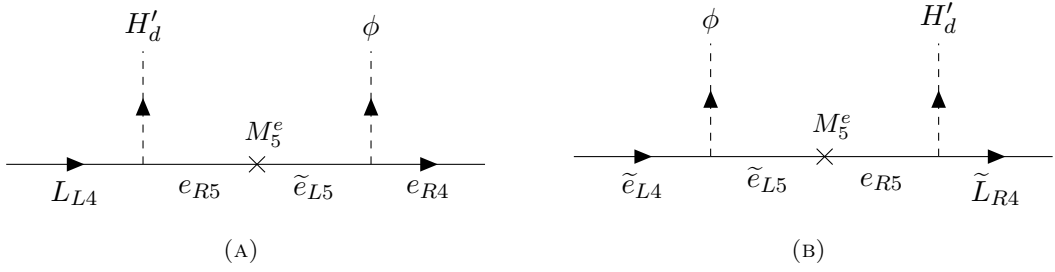
\begin{figure}
\subfloat[]{\noindent \begin{centering}
\begin{tikzpicture}
	\begin{feynman}
		\vertex (a);
		\vertex [right=18mm of a] (b);
		\vertex [right=of b] (c) [label={ [xshift=0.1cm, yshift=0.1cm] \small $M^{e}_{5}$}];
		\vertex [right=of c] (d);
		\vertex [right=of d] (e);
		\vertex [above=of b] (f1) {\(H'_{d}\)};
		\vertex [above=of d] (f2) {\(\phi\)};
		\diagram* {
			(a) -- [fermion, edge label'=\(L_{L4}\), inner sep=6pt] (b) -- [charged scalar] (f1),
			(b) -- [edge label'=\(e_{R5}\), inner sep=6pt] (c),
			(c) -- [edge label'=\(\widetilde{e}_{L5}\), insertion=0] (d) -- [charged scalar] (f2),
			(d) -- [fermion, edge label'=\(e_{R4}\), inner sep=6pt] (e),
	};
	\end{feynman}
\end{tikzpicture}
\par\end{centering}
}$\qquad$\subfloat[]{\noindent \begin{centering}
\begin{tikzpicture}
	\begin{feynman}
		\vertex (a);
		\vertex [right=18mm of a] (b);
		\vertex [right=of b] (c) [label={ [xshift=0.1cm, yshift=0.1cm] \small $M^{e}_{5}$}];
		\vertex [right=of c] (d);
		\vertex [right=of d] (e);
		\vertex [above=of b] (f1) {\(\phi\)};
		\vertex [above=of d] (f2) {\(H'_{d}\)};
		\diagram* {
			(a) -- [fermion, edge label'=\(\widetilde{e}_{L4}\), inner sep=6pt] (b) -- [charged scalar] (f1),
			(b) -- [edge label'=\(\widetilde{e}_{L5}\)] (c),
			(c) -- [edge label'=\(e_{R5}\), inner sep=6pt, insertion=0] (d) -- [charged scalar] (f2),
			(d) -- [fermion, edge label'=\(\widetilde{L}_{R4}\), inner sep=6pt] (e),
	};
	\end{feynman}
\end{tikzpicture}
\par\end{centering}
}

\caption[Effective Higgs Yukawa couplings for the fourth family vector-like
leptons in the theory of flavour with fermiophobic $Z'$]{Effective Higgs Yukawa couplings for the fourth family vector-like
leptons in the theory of flavour with fermiophobic $Z'$. \label{fig:Effective-Higgs-Yukawa_exotic}}
\end{figure}

As discussed in the simplified model, if we want to provide a significant
contribution to $(g-2)_{\mu}$ mediated by a relatively heavy $Z'$
boson, then we need vector-like leptons to couple to the SM Higgs in
order to provide a chiral enhancement. This was achieved in the simplified
model via a coupling of the form $\overline{L}_{L4}He_{R4}$, which
is forbidden in the theory of flavour due to the non-zero $U(1)'$
charges of the Higgs doublets. Nevertheless, we can obtain effective
couplings to $H_{d}$ via non-renormalisable operators of the form
$\phi\overline{L}_{L4}H_{d}e_{R4}$. Such operators may be provided
by adding an extra vector-like lepton singlet uncharged under $U(1)'$,
i.e.~$\widetilde{e}_{L5}$ and $e_{R5}$ with vector-like mass $M_{5}^{e}$.
After $\phi$ develops a VEV, we obtain effective couplings in the
mass insertion approximation as shown in Fig.~\ref{fig:Effective-Higgs-Yukawa_exotic}.
Going beyond the mass insertion approximation but preserving the assumption
$\left\langle H_{d}\right\rangle \ll\left\langle \phi\right\rangle $,
one of the effective couplings is explicitly given by
\begin{table}
\begin{centering}
\begin{tabular}{cccccc}
\toprule 
\multicolumn{1}{c}{Field} & \multicolumn{1}{c}{$SU(3)_{c}$} & \multicolumn{1}{c}{$SU(2)_{L}$} & \multicolumn{1}{c}{$U(1)_{Y}$} & $U(1)'$ & $Z_{2}$\tabularnewline
\midrule
$\tilde{e}_{L5},e_{R5}$ & $\mathbf{1}$ & $\mathbf{1}$ & -1 & 0 & (-)\tabularnewline
$H'_{d}$ & $\mathbf{1}$ & $\mathbf{2}$ & -1/2 & 1 & (-)\tabularnewline
\bottomrule
\end{tabular}
\par\end{centering}
\caption[New fields required to obtain a significant chiral mass for fourth
family vector-like leptons in the theory of flavour]{New fields required to obtain a significant chiral mass for fourth
family vector-like leptons in the theory of flavour. The rest of fields
in the model (shown in Table~\ref{tab:The_field_content_2}) are
even under $Z_{2}$. \label{tab:New-fields-required}}

\end{table}
\begin{equation}
y_{5}^{e}\frac{x_{5}^{e}\left\langle \phi\right\rangle }{\sqrt{(x_{5}^{e}\left\langle \phi\right\rangle )^{2}+(M_{5}^{e})^{2}}}\overline{L}_{L4}H_{d}e_{R4}\,.\label{eq:effective_Chirality_flip}
\end{equation}
However, given that we are working in the large $\tan\beta$ regime
in order to explain the $m_{b}/m_{t}$ hierarchy, we expect $\left\langle H_{d}\right\rangle \approx\mathcal{O}(\mathrm{GeV})$,
too small to provide a significant chiral enhancement. A cheap solution
consists in adding another Higgs doublet $H'_{d}\sim(\mathbf{1,2},1/2,1)$,
along with a $Z_{2}$ discrete symmetry that only discriminates $\widetilde{e}_{L5}$,
$e_{R5}$ and the new Higgs $H'_{d}$, such that these new particles
only play the role of providing the effective couplings in Fig.~\ref{fig:Effective-Higgs-Yukawa_exotic}
(see also Table~\ref{tab:New-fields-required}). We assume that the new
Higgs gets a VEV $\left\langle H'_{d}\right\rangle \approx\mathcal{O}(100\,\mathrm{GeV})$.
Then if $\left\langle \phi\right\rangle \sim M_{5}^{e}$, the effective
coupling in Eq.~(\ref{eq:effective_Chirality_flip}) is well approximated
by $y_{5}^{e}$, and we assume the usual perturbativity constraint
$y_{5}^{e}\apprle\sqrt{4\pi}$. Of course, with three Higgs doublets
we have $v_{\mathrm{SM}}^{2}=v_{u}^{2}+v_{d}^{2}+v'^{2}_{d}$, where
in our case $v_{d}\ll v_{u},v'_{d}$. Notice that we remain protected
from FCNCs in the Higgs sector thanks to the $U(1)'$ charge assignments
and the $Z_{2}$ discrete symmetry.

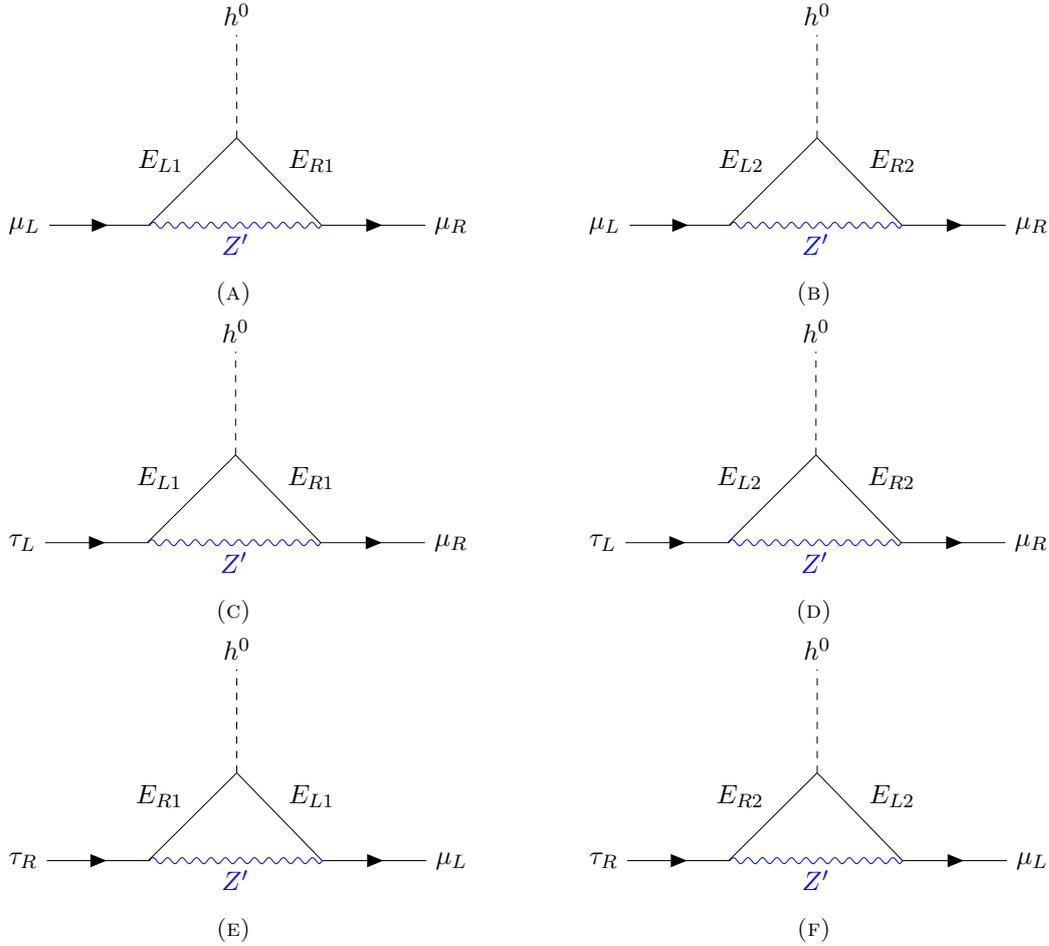
\begin{figure}[t]
\subfloat[]{\resizebox{.46\textwidth}{!}{
\begin{centering}
\begin{tikzpicture}
	\begin{feynman}
		\vertex (a) {\(\mu_{L}\)};
		\vertex [right=18mm of a] (b);
		\vertex [right=25mm of b] (c);
		\vertex [right=of c] (d) {\(\mu_{R}\)};
		\vertex [above right=18mm of b] (f1);
		\vertex [above=of f1] (f2) {\(h^{0}\)};
		\diagram* {
			(a) -- [fermion] (b) -- [boson, blue, edge label'=$Z'$] (c) -- [fermion] (d),
			(b) -- [edge label=\(E_{L1}\)] (f1) -- [scalar] (f2),
			(f1) -- [edge label=\(E_{R1}\)] (c),
	};
	\end{feynman}
\end{tikzpicture}
\par\end{centering}
}

}$\qquad$\subfloat[]{\resizebox{.46\textwidth}{!}{
\begin{centering}
\begin{tikzpicture}
	\begin{feynman}
		\vertex (a) {\(\mu_{L}\)};
		\vertex [right=18mm of a] (b);
		\vertex [right=25mm of b] (c);
		\vertex [right=of c] (d) {\(\mu_{R}\)};
		\vertex [above right=18mm of b] (f1);
		\vertex [above=of f1] (f2) {\(h^{0}\)};
		\diagram* {
			(a) -- [fermion] (b) -- [boson, blue, edge label'=$Z'$] (c) -- [fermion] (d),
			(b) -- [edge label=\(E_{L2}\)] (f1) -- [scalar] (f2),
			(f1) -- [edge label=\(E_{R2}\)] (c),
	};
	\end{feynman}
\end{tikzpicture}
\par\end{centering}
}

}

\subfloat[]{\resizebox{.46\textwidth}{!}{
\begin{centering}
\begin{tikzpicture}
	\begin{feynman}
		\vertex (a) {\(\tau_{L}\)};
		\vertex [right=18mm of a] (b);
		\vertex [right=25mm of b] (c);
		\vertex [right=of c] (d) {\(\mu_{R}\)};
		\vertex [above right=18mm of b] (f1);
		\vertex [above=of f1] (f2) {\(h^{0}\)};
		\diagram* {
			(a) -- [fermion] (b) -- [boson, blue, edge label'=$Z'$] (c) -- [fermion] (d),
			(b) -- [edge label=\(E_{L1}\)] (f1) -- [scalar] (f2),
			(f1) -- [edge label=\(E_{R1}\)] (c),
	};
	\end{feynman}
\end{tikzpicture}
\par\end{centering}
}

}$\qquad$\subfloat[]{\resizebox{.46\textwidth}{!}{
\begin{centering}
\begin{tikzpicture}
	\begin{feynman}
		\vertex (a) {\(\tau_{L}\)};
		\vertex [right=18mm of a] (b);
		\vertex [right=25mm of b] (c);
		\vertex [right=of c] (d) {\(\mu_{R}\)};
		\vertex [above right=18mm of b] (f1);
		\vertex [above=of f1] (f2) {\(h^{0}\)};
		\diagram* {
			(a) -- [fermion] (b) -- [boson, blue, edge label'=$Z'$] (c) -- [fermion] (d),
			(b) -- [edge label=\(E_{L2}\)] (f1) -- [scalar] (f2),
			(f1) -- [edge label=\(E_{R2}\)] (c),
	};
	\end{feynman}
\end{tikzpicture}
\par\end{centering}
}

}

\subfloat[]{\resizebox{.46\textwidth}{!}{
\begin{centering}
\begin{tikzpicture}
	\begin{feynman}
		\vertex (a) {\(\tau_{R}\)};
		\vertex [right=18mm of a] (b);
		\vertex [right=25mm of b] (c);
		\vertex [right=of c] (d) {\(\mu_{L}\)};
		\vertex [above right=18mm of b] (f1);
		\vertex [above=of f1] (f2) {\(h^{0}\)};
		\diagram* {
			(a) -- [fermion] (b) -- [boson, blue, edge label'=$Z'$] (c) -- [fermion] (d),
			(b) -- [edge label=\(E_{R1}\)] (f1) -- [scalar] (f2),
			(f1) -- [edge label=\(E_{L1}\)] (c),
	};
	\end{feynman}
\end{tikzpicture}
\par\end{centering}
}

}$\qquad$\subfloat[]{\resizebox{.46\textwidth}{!}{
\begin{centering}
\begin{tikzpicture}
	\begin{feynman}
		\vertex (a) {\(\tau_{R}\)};
		\vertex [right=18mm of a] (b);
		\vertex [right=25mm of b] (c);
		\vertex [right=of c] (d) {\(\mu_{L}\)};
		\vertex [above right=18mm of b] (f1);
		\vertex [above=of f1] (f2) {\(h^{0}\)};
		\diagram* {
			(a) -- [fermion] (b) -- [boson, blue, edge label'=$Z'$] (c) -- [fermion] (d),
			(b) -- [edge label=\(E_{R2}\)] (f1) -- [scalar] (f2),
			(f1) -- [edge label=\(E_{L2}\)] (c),
	};
	\end{feynman}
\end{tikzpicture}
\par\end{centering}
}

}

\caption[Leading contributions to $(g-2)_{\mu}$ and $\mathcal{B}(\tau\rightarrow\mu\gamma)$
in the theory of flavour with fermiophobic $Z'$]{Leading contributions to $(g-2)_{\mu}$ and $\mathcal{B}(\tau\rightarrow\mu\gamma)$
in the theory of flavour with fermiophobic $Z'$. The effective couplings of the vector-like fermions $E_{1,2}$ to the SM Higgs boson lead to a chiral enhancement of both $(g-2)_{\mu}$ and $\mathcal{B}(\tau\rightarrow\mu\gamma)$, and a photon is emitted (not shown).  \label{fig:Leading-contributions-to_g-2_taumugamma}}
\end{figure}

After the Higgs doublet $H'_{d}$ gets a VEV, the effective coupling
$\overline{L}_{L4}H'_{d}e_{R4}$ in Eq.~(\ref{eq:effective_Chirality_flip})
provides a chiral mass $M_{4}^{C}$ for the fourth family (vector-like) leptons. A similar
coupling for the conjugate fermions $\overline{\widetilde{e}}_{L4}H'_{d}\widetilde{L}_{R4}$
provides a chiral mass $\widetilde{M}_{4}^{C}$, reproducing the framework
of the simplified model in Section~\ref{sec:Simplified_Fermiophobic}.
As in the simplified model, the presence of chiral masses for fourth family (vector-like)
leptons leads to mixing between the conjugate leptons $\widetilde{L}_{R4}$
and $\widetilde{e}_{L4}$ and the chiral leptons of the SM. The dominant
contributions to $(g-2)_{\mu}$ are indeed proportional to the chiral
masses as (see also the top panels of Fig.~\ref{fig:Leading-contributions-to_g-2_taumugamma})
\begin{flalign}
\Delta a_{\mu}\simeq & -\frac{m_{\mu}^{2}}{8\pi^{2}M_{Z'}^{2}}\left[\mathrm{Re}\left[g_{\mu E_{1}}^{L}\left(g_{\mu E_{1}}^{R}\right)^{*}\right]G(M_{E_{1}}^{2}/M_{Z'}^{2})\frac{M_{4}^{C}c_{L}^{E}c_{R}^{E}+\widetilde{M}_{4}^{C}s_{L}^{E}s_{R}^{E}}{m_{\mu}}\right.\\
 & \left.+\mathrm{Re}\left[g_{\mu E_{2}}^{L}\left(g_{\mu E_{2}}^{R}\right)^{*}\right]G(M_{E_{2}}^{2}/M_{Z'}^{2})\frac{M_{4}^{C}s_{L}^{E}s_{R}^{E}+\widetilde{M}_{4}^{C}c_{L}^{E}c_{R}^{E}}{m_{\mu}}\right]\,.\nonumber 
\end{flalign}
Notice that the couplings $g_{\mu E_{2}}^{L,R}$ are proportional
to the mixing angles $s_{L,R}^{E}$ obtained after diagonalising Eq.~(\ref{eq:Physical_Masses_VL}).
For the simplified model discussed in Section~\ref{sec:Simplified_Fermiophobic},
these mixing angles where suppressed since all the relevant parameter
space was fulfilling $v_{\mathrm{SM}}\ll M_{4}^{L,e}$ for at least
one of the vector-like masses. However, notice that in the flavour
model the $Z'$ dominantly couples to third family left-handed fermions,
and couplings to right-handed fermions or second family left-handed
fermions are generally suppressed. This implies that in order to provide
a significant contribution to $(g-2)_{\mu}$, we will need to work
in a regime of large mixing angles where both $M_{4}^{L,e}\sim\left\langle \phi\right\rangle $
with $\left\langle \phi\right\rangle \sim1\;\mathrm{TeV}$. Notice
that smaller $\left\langle \phi\right\rangle $ would be in tension
with $B_{s}-\bar{B}_{s}$ meson mixing, while larger $\left\langle \phi\right\rangle $
would suppress the contribution to $(g-2)_{\mu}$. We need as well
both $M_{4}^{C}$ and $\widetilde{M}_{4}^{C}$ close to perturbativity
limits in order to obtain the largest contribution to $(g-2)_{\mu}$.
In this regime, we find that both $M_{4}^{L}$ and $M_{4}^{e}$ may be
not much larger than $M_{4}^{C}$ and $\widetilde{M}_{4}^{C}$, such
that the mixing angles $s_{L,R}^{E}$ may be large and the effective
couplings $g_{\mu E_{2}}^{L,R}$ cannot be neglected, contrary to
the case of the simplified model.

In contrast with the simplified model, due to the presence of $Z'$
couplings for both third and second family fermions, in the flavour
model a significant contribution to $\mathcal{B}(\tau\rightarrow\mu\gamma)$
arises, connected to the contribution to $(g-2)_{\mu}$ and to the
chiral masses $M_{4}^{C}$ and $\widetilde{M}_{4}^{C}$, namely (see
also central and bottom panels in Fig.~\ref{fig:Leading-contributions-to_g-2_taumugamma})
\begin{flalign}
\mathcal{B}(\tau\rightarrow\mu\gamma)\simeq & \frac{\alpha_{\mathrm{EM}}}{1024\pi^{4}}\frac{m_{\tau}^{5}}{\Gamma_{\tau}M_{Z'}^{4}}\left[\left|g_{\mu E_{1}}^{L}\left(g_{\tau E_{1}}^{R}\right)^{*}G(M_{E_{1}}^{2}/M_{Z'}^{2})\frac{M_{4}^{C}c_{L}^{E}c_{R}^{E}+\widetilde{M}_{4}^{C}s_{L}^{E}s_{R}^{E}}{m_{\mu}}\right.\right.\\
 & \left.+g_{\mu E_{2}}^{L}\left(g_{\tau E_{2}}^{R}\right)^{*}G(M_{E_{2}}^{2}/M_{Z'}^{2})\frac{M_{4}^{C}s_{L}^{E}s_{R}^{E}+\widetilde{M}_{4}^{C}c_{L}^{E}c_{R}^{E}}{m_{\mu}}\right|^{2}\nonumber \\
 & +\left|g_{\mu E_{1}}^{R}\left(g_{\tau E_{1}}^{L}\right)^{*}G(M_{E_{1}}^{2}/M_{Z'}^{2})\frac{M_{4}^{C}c_{L}^{E}c_{R}^{E}+\widetilde{M}_{4}^{C}s_{L}^{E}s_{R}^{E}}{m_{\mu}}\right.\nonumber \\
 & \left.\left.+g_{\mu E_{2}}^{R}\left(g_{\tau E_{2}}^{L}\right)^{*}G(M_{E_{2}}^{2}/M_{Z'}^{2})\frac{M_{4}^{C}s_{L}^{E}s_{R}^{E}+\widetilde{M}_{4}^{C}c_{L}^{E}c_{R}^{E}}{m_{\mu}}\right|^{2}\right]\,.\nonumber 
\end{flalign}
Contributions to related LFV processes such as $\tau\rightarrow3\mu$
also arise, however we find $\tau\rightarrow\mu\gamma$ to provide
the leading constraints in the regime of large chiral masses $M_{4}^{C}$
and $\widetilde{M}_{4}^{C}$ where we shall work.
\begin{figure}
\subfloat[\label{fig:Parameter-space_FlavourModel_g-2_a}]{\includegraphics[scale=0.38]{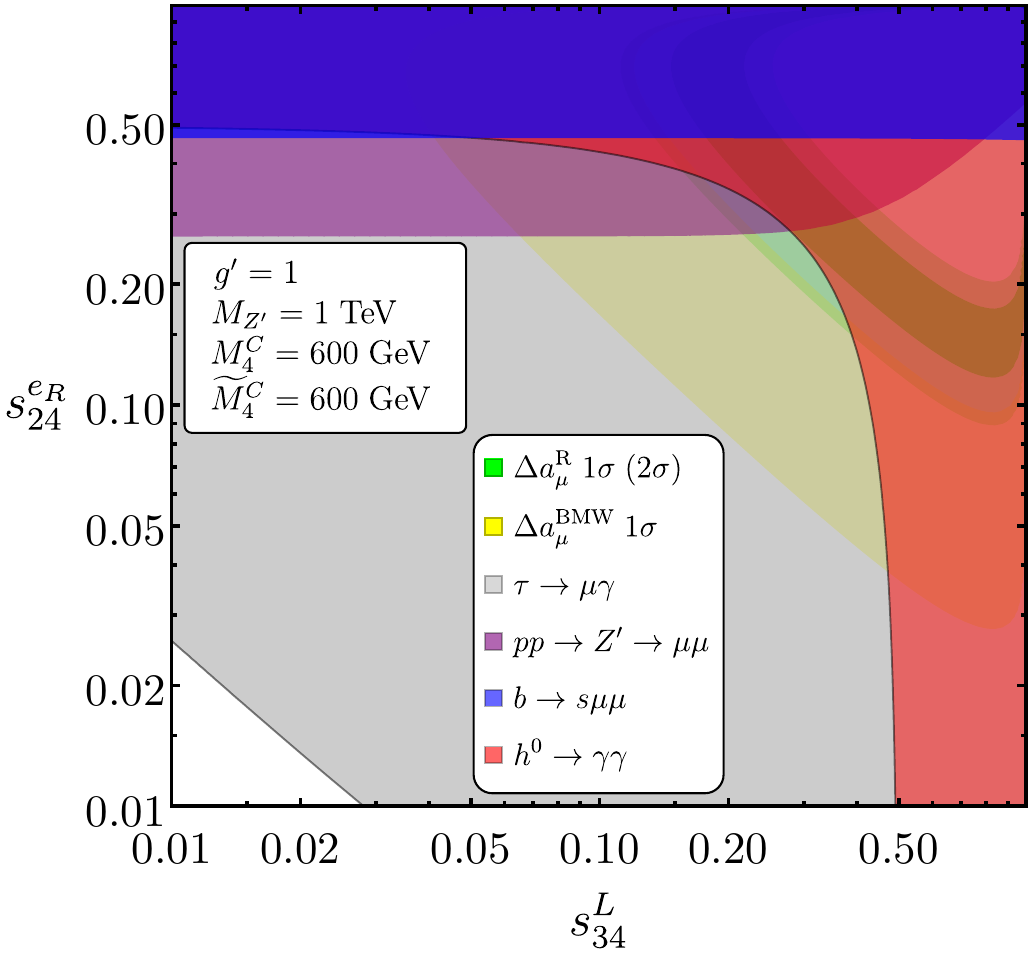}

}$\quad$\subfloat[\label{fig:Parameter-space_FlavourModel_g-2_b}]{\includegraphics[scale=0.38]{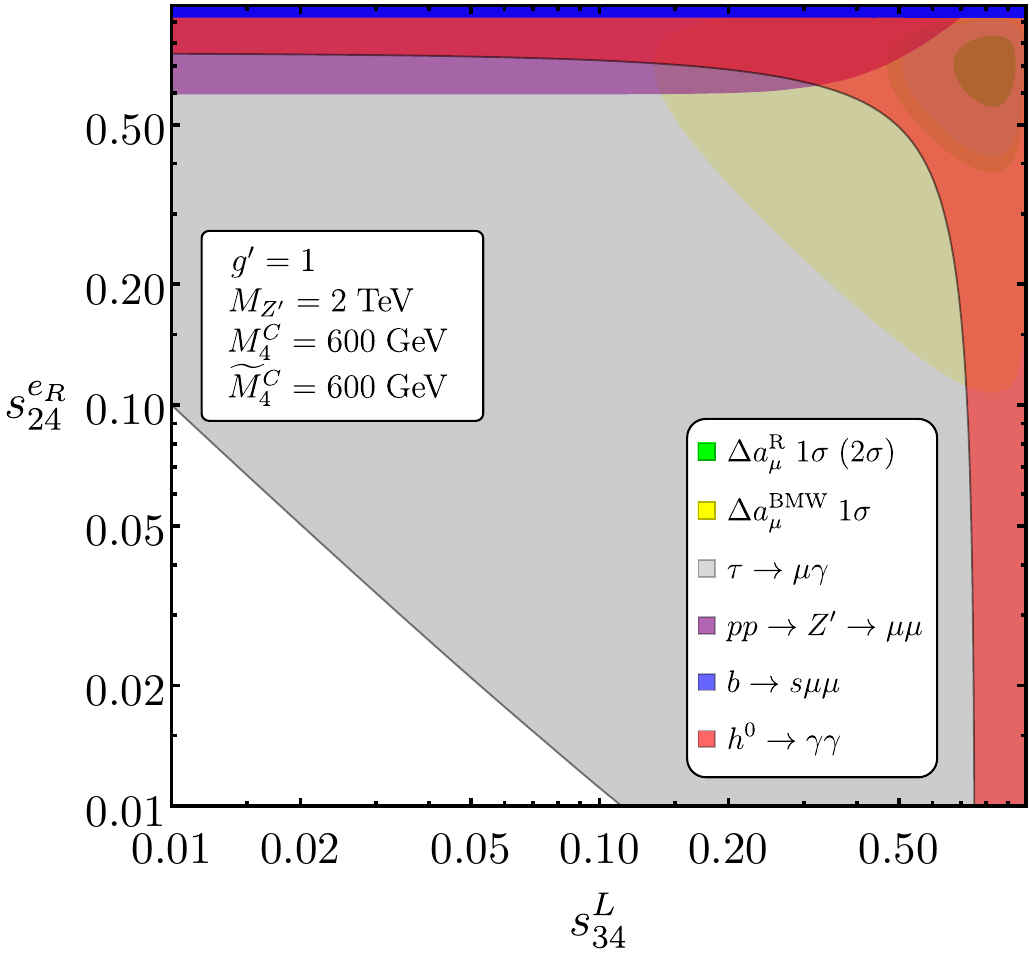}

}

\caption[Parameter space for $(g-2)_{\mu}$ in the theory of flavour with fermiophobic $Z'$ boson]{Parameter space $s_{34}^{L}$ vs $s_{24}^{e_{R}}$ for $M_{Z'}=1\;\mathrm{TeV}$
(left) and $M_{Z'}=2\;\mathrm{TeV}$ (right) with chiral masses close
to the perturbativity limit as shown in the panels. The green (lighter
green) region is preferred at 1$\sigma$ (2$\sigma$) by $\Delta a_{\mu}^{\mathrm{R}}$
($e^{+}e^{-}\rightarrow\mathrm{hadrons}$) while the yellow region
is preferred at $1\sigma$ by $\Delta a_{\mu}^{\mathrm{BMW}}$. The
rest of the shaded regions are excluded at 95\% CL, with the leading
constraint given by $\mathcal{B}(\tau\rightarrow\mu\gamma)$ as shown
by the grey-shaded region. \label{fig:Parameter-space_FlavourModel_g-2}}

\end{figure}

As shown in Fig.~\ref{fig:Parameter-space_FlavourModel_g-2}, the
large contribution to $\tau\rightarrow\mu\gamma$ is the dominant
constraint over the parameter space, ruling out the possibility of
having a significant contribution to $(g-2)_{\mu}$ in the theory
of flavour with fermiophobic $Z'$. Higgs diphoton decay also constrains
the region of parameter space preferred by $\Delta a_{\mu}^{\mathrm{R}}$
(the $5\sigma$ deviation with respect to $e^{+}e^{-}\rightarrow\mathrm{hadrons}$
data). Notice that reducing $M_{4}^{C}$ and $\widetilde{M}_{4}^{C}$
ameliorates the bounds from $\tau\rightarrow\mu\gamma$ but also reduces
the contribution to $(g-2)_{\mu}$. Note that in the parameter space motivated
by $(g-2)_{\mu}$, the leading LHC constraint is the dimuon Drell-Yan channel due 
to the large $s^{e_{R}}_{24}\approx 0.5$ required, while for its natural value $s^{e_{R}}_{24}\approx V_{cb}$
the ditau Drell-Yan channel is more competitive as mentioned.

The crucial difference with respect
to the simplified model is the presence of flavour-violating $\mu\tau$
couplings to the $Z'$ boson. The $(g-2)_{\mu}$ anomaly could be
addressed if the vector-like leptons mix dominantly with muons, but
this is not possible in the theory of flavour since we need at least
one of the vector-like leptons to mix with taus in order to obtain
their effective Yukawa coupling. Nevertheless, it is possible that with
further model building one could achieve a cancellation in $\tau\rightarrow\mu\gamma$
while preserving the enhancement of $(g-2)_{\mu}$, however this probably
involves the addition of more fields and symmetries, enlarging the
number of free parameters, making the model less predictive and disconnecting
the explanation of $(g-2)_{\mu}$ from the origin of flavour hierarchies,
going against the spirit of the model.

\subsection{Parameter space in the flavour model}

For the first time, we have considered the possibility of addressing
the $(g-2)_{\mu}$ anomaly within the framework of a complete theory
of flavour with a fermiophobic $Z'$ boson. Although our analysis
reveals that a significant contribution to $(g-2)_{\mu}$ is in tension
with $\mathcal{B}(\tau\rightarrow\mu\gamma)$, the dangerous contributions
to LFV processes are much reduced if we neglect the chiral masses
of vector-like fermions $M_{4}^{C}$ and $\widetilde{M}_{4}^{C}$,
which anyway required an extension of the original flavour model.
The model as presented in Section~\ref{sec:Theory-of-flavour_Fermiophobic}
remains a well-motivated possibility to explain the origin of the
SM flavour structure, and it is interesting to study the allowed parameters
and the experimental bounds over the NP scales of the model.
\begin{figure}
\subfloat[]{\includegraphics[scale=0.46]{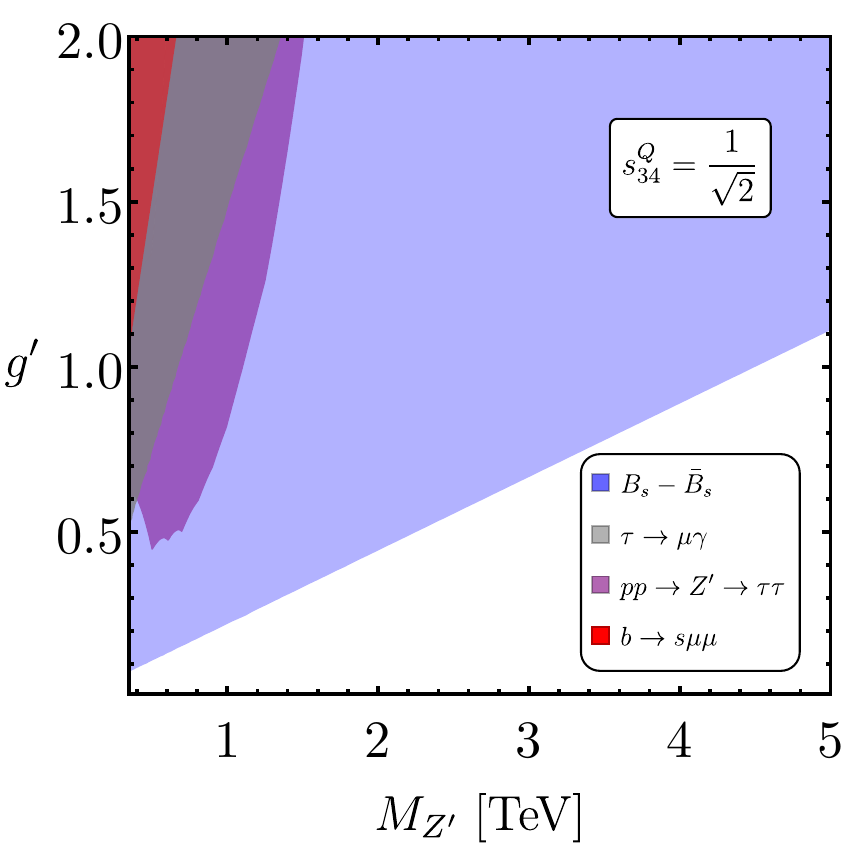}

}$\quad$\subfloat[]{\includegraphics[scale=0.46]{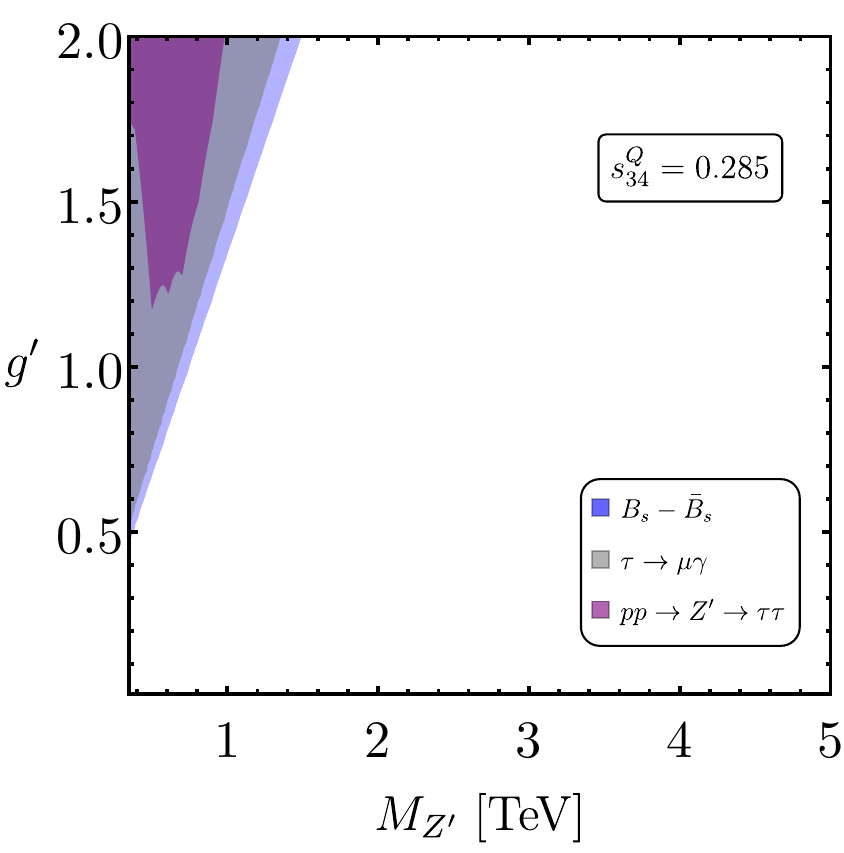}

}

\caption[Parameter space in the theory of flavour with fermiophobic $Z'$ without
$(g-2)_{\mu}$]{Parameter space $M_{Z'}$ vs $g'$ for $s_{34}^{Q}=1/\sqrt{2}$ (left) and
$s_{34}^{Q}=0.285$ (right). The shaded regions are excluded at 95\%
CL. \label{fig:Parameter-space_FlavourModel}}
\end{figure}

As motivated in Section~\ref{subsec:MesonMixing_FermiophobicZp},
the stronger bounds from $B_{s}-\bar{B}_{s}$ meson mixing are very
sensitive to the value of the mixing angle $s_{34}^{Q}$. As presented
in Section~\ref{subsec:Effective-quark-Yukawa}, if we assume $y_{43}^{u}\approx\sqrt{2}$,
then we expect $s_{34}^{Q}\approx1/\sqrt{2}$. In this case, $B_{s}-\bar{B}_{s}$
meson mixing leads to the most stringent bounds over $M_{Z'}/g'$ as shown
in Fig.~\ref{fig:Parameter-space_FlavourModel_g-2_a}, such that
for natural values of $g'$ we would expect to find $M_{Z'}$ above
4 TeV. We find $pp\rightarrow Z'\rightarrow\tau\tau$ to be the leading
LHC constraint over the parameter space, and $\tau\rightarrow\mu\gamma$
to be the leading signal in LFV processes. $pp\rightarrow Z'\rightarrow\mu\mu$
, $b\rightarrow s\mu\mu$ and $\tau\rightarrow3\mu$ are further suppressed
due to the small right-handed mixing angles or small 2-3 mixing.

If we push $y_{43}^{u}$ close to perturbativity bounds (i.e.~$y_{43}^{u}\approx\sqrt{4\pi}$),
then we can fit the top mass via $s_{34}^{Q}\approx0.285$ and reduce
the bounds from $B_{s}-\bar{B}_{s}$ mixing. As shown in Fig.~\ref{fig:Parameter-space_FlavourModel_g-2_b},
in this case the bounds from $pp\rightarrow Z'\rightarrow\tau\tau$
and $b\rightarrow s\mu\mu$ are also suppressed due to the smaller
$Z'$ couplings to bottom quarks, while the bounds from $\tau\rightarrow\mu\gamma$
remain the same. In this case, for natural values of $g'$ we could
have $M_{Z'}$ as low as 1 TeV.

We conclude that in any case, the flavour structure of the fermiophobic
$Z'$ flavour model allows for relatively low $Z'$ masses, within
the reach of current and upcoming experiments.

\section{Conclusions}

In this chapter we have studied fermiophobic $Z'$ models, which
were a well-motivated class of SM extensions to address flavour anomalies
in $R_{K^{(*)}}$ and $(g-2)_{\mu}$ \cite{FernandezNavarro:2021sfb}.
We have shown that a simplified fermiophobic $Z'$ model is still
able to explain large deviations in $(g-2)_{\mu}$ with $Z'$ masses
ranging from the GeV to the few TeV. The main idea is that the
$Z'$ boson only couples originally to vector-like fermions, which
then mix with chiral fermions providing effective $Z'$ couplings
that can be controlled by the size of the mixing angles. The vector-like
leptons obtain chiral masses at the electroweak scale via couplings
to the SM Higgs (although their physical mass is dominated by the
vector-like mass terms), which provide a chiral enhancement of $(g-2)_{\mu}$.
Interestingly, this enhancement is correlated to a suppression of
the Higgs decay to two photons, where further experimental precision
could test the full parameter space of the model.

Afterwards, we went beyond the simplified framework to propose a complete
theory of flavour containing a fermiophobic $Z'$ boson. The flavour
structure of the SM is explained via messenger dominance \cite{Ferretti:2006df},
such that third family fermion masses are obtained from mixing with
TeV scale vector-like fermion doublets while second family fermion
masses and small CKM mixing are obtained from mixing with heavier
vector-like fermion singlets. This mixing also provides effective
couplings for chiral fermions to the $Z'$ boson, leading to a predictive
phenomenology at low energies connected to the origin of flavour hierarchies.
First family fermion masses are explained via a heavy Higgs doublet
that gets an small effective VEV via mixing with the much lighter
Higgs doublets that perform electroweak symmetry breaking. The heavier
Higgs does not couple to the $Z'$ boson and therefore does not introduce
effective $Z'$ couplings for first family fermions, protecting the
model from the appearance of the most dangerous FCNCs. Finally, the
origin of tiny neutrino masses and PMNS mixing is addressed via the
type Ib seesaw mechanism \cite{Hernandez-Garcia:2019uof}.

Obtaining an effective chiral mass for vector-like leptons in the
flavour model requires the addition of an extra vector-like lepton
and a $Z_{2}$ discrete symmetry. However, we have found that in the
flavour model, the enhancement of $(g-2)_{\mu}$ is correlated as
well with a chiral enhancement of $\mathcal{B}(\tau\rightarrow\mu\gamma)$,
which renders impossible to obtain large contributions to $(g-2)_{\mu}$.
It is remarkable that the simplified model works fine for $(g-2)_{\mu}$
while the full flavour model fails, questioning the BSM interpretation of $(g-2)_{\mu}$
within a theory of flavour.

Finally, we have dropped the extra dynamics introduced to
obtain chiral masses for the vector-like leptons, in order to study
the parameter space of the flavour model without addressing $(g-2)_{\mu}$. We have found that the leading
constraint is generally $B_{s}-\bar{B}_{s}$ meson mixing, followed
by ditau searches at the LHC and $\mathcal{B}(\tau\rightarrow\mu\gamma)$.
We conclude that the flavour structure in the model allows for $Z'$
masses as low as 1 TeV for natural values of the gauge coupling, within
the reach of current and upcoming experiments.
%% ----------------------------------------------------------------
%% TwinPS.tex
%% ---------------------------------------------------------------- 
\chapter{Twin Pati-Salam theory of flavour} \label{Chapter:TwinPS}

\begin{quote}
  ``Make everything as simple as possible, but not simpler.''
  \begin{flushright}
  \hfill \hfill $-$ Albert Einstein
  \par\end{flushright}
\end{quote}

\noindent In this chapter, based on Refs.~\cite{FernandezNavarro:2022gst,FernandezNavarro:2023lgk,FernandezNavarro:2023ykw}, we introduce a theory of flavour consisting of two copies
of the Pati-Salam gauge group, broken in two steps down to the SM.
The last step of the symmetry breaking can be as low as the TeV scale,
giving rise to a $U_{1}\sim(\mathbf{3,1},2/3)$ vector leptoquark which
explains the $B$-anomalies. This model
will connect the origin of Yukawa couplings and flavour hierarchies
of the SM with the effective couplings of the vector leptoquark that
explain the $B$-anomalies.

\section{Introduction}

The picture of anomalies in $B$-meson decays, including the discrepancy
in the $R_{K^{(*)}}$ ratios that lasted until late 2022 (see Section~\ref{subsec:bsll}),
led to important model building efforts by the community during the
last eight years, in order to interpret these anomalies as a low-energy
signal of a consistent BSM model. Leptoquarks were identified as
excellent NP candidates for the $B$-anomalies, because
unlike other mediators they avoid dangerous contributions to $B_{s}-\bar{B}_{s}$
meson mixing observables at tree-level. Different scalar leptoquarks
were proposed to address either $R_{K^{(*)}}$ or $R_{D^{(*)}}$ separately
(see e.g.~\cite{Buttazzo:2017ixm,Becirevic:2017jtw,Crivellin:2017zlb,deMedeirosVarzielas:2018bcy,Angelescu:2021lln,Becirevic:2022tsj,Crivellin:2022mff}).
Interestingly, the vector leptoquark $U_{1}\sim(\boldsymbol{3},\boldsymbol{1},2/3)$
was identified as the only single mediator capable of addressing both
the $R_{K^{(*)}}$ and $R_{D^{(*)}}$ anomalies simultaneously \cite{Angelescu:2021lln}.
Given that both anomalies suggested the departure from lepton flavour universality, it was soon
realised that the $U_{1}$ explanation of the $B$-anomalies could
be connected to the origin of the flavour structure of the SM (see Section~\ref{subsec:bctaunu}).

However, the gauge nature of $U_{1}$ requires to specify a
UV completion that explains its origin. The original
ideas by Pati and Salam (PS) \cite{Pati:1974yy}, led to tensions
with unobserved processes such as $K_{L}\rightarrow\mu e$ (see Section~\ref{subsec:Semileptonic-CLFV-processes}).
Instead, an interesting proposal was firstly laid out in the Appendix
of \cite{Diaz:2017lit}, and more formally later in \cite{DiLuzio:2017vat},
following the idea introduced in \cite{Georgi:2016xhm} that colour
could appear as a diagonal subgroup of a larger $SU(3+N)\times SU(3)'$
local symmetry restored at high energies. The particular choice $N=1$
leads to the so-called ``4321'' gauge symmetry,
\begin{equation}
G_{4321}\equiv SU(4)\times SU(3)_{c}^{'}\times SU(2)_{L}\times U(1)_{Y'}\,,
\end{equation}
which can be broken at the TeV scale while satisfying the experimental
bounds \cite{DiLuzio:2017vat,DiLuzio:2018zxy,Cornella:2019hct,Cornella:2021sby},
provided that at least the first and second families of SM fermions
are singlets under $SU(4)$. This breaking leads to a rich gauge boson
spectrum at the TeV scale, containing the vector leptoquark $U_{1}$
along with a massive colour octet $g'\sim(\boldsymbol{8},\boldsymbol{1},0)$
and a massive $Z'\sim(\boldsymbol{1},\boldsymbol{1},0)$ with suppressed
couplings to light SM fermions. Vector-like fermions need to
be introduced in order to obtain effective couplings for (at least)
second family fermions to $U_{1}$. The model, though it is not minimal,
is very predictive and leads to a rich phenomenology in both low-energy
and high-$p_{T}$ searches. However, the flavour structure of the
model was rather ad-hoc, and it was hinted that the 4321 gauge group
could be the TeV scale effective field theory of a complete model
addressing more open questions of the SM. Given that both the $R_{K^{(*)}}$
and $R_{D^{(*)}}$ anomalies suggested a departure from LFU consistent
with the idea of a theory of flavour (see Section~\ref{subsec:bctaunu}),
it was soon realised that the $U_{1}$ explanation of the $B$-anomalies,
and in particular the 4321 model, could originate from a theory of
flavour.
\begin{figure}[t]
\begin{centering}
\includegraphics[scale=0.20]{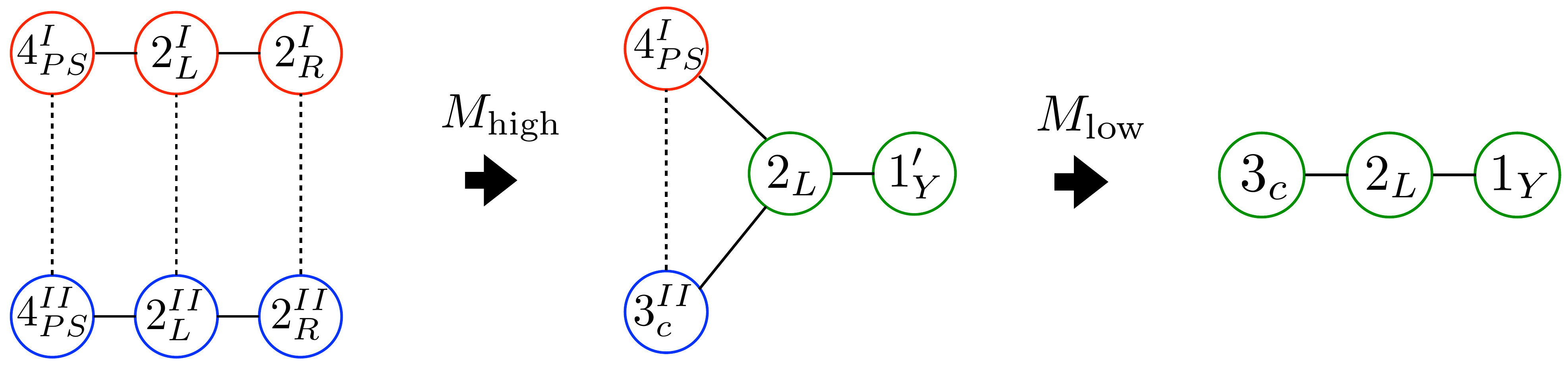}
\par\end{centering}
\caption[Symmetry breaking chain in the twin Pati-Salam model]{The model is based on two copies of the Pati-Salam gauge group $SU(4)_{PS}\times SU(2)_{L}\times SU(2)_{R}$.
The circles represent the gauge groups with the indicated symmetry
breaking. The twin Pati-Salam symmetry is broken down to the 4321
symmetry at high energies $M_{\mathrm{High}}\apprge1\,\mathrm{PeV}$,
then the 4321 group is further broken to the SM at the TeV scale $M_{\mathrm{low}}\sim\mathcal{O}(\mathrm{TeV}).$\label{fig:Model_Diagram}}
\end{figure}

The very first theories of flavour containing a TeV-scale $U_{1}$
to explain the $B$-anomalies \cite{Bordone:2017bld,Fuentes-Martin:2020pww,Fuentes-Martin:2022xnb,Davighi:2022bqf}
were based on the flavour deconstruction of the SM gauge group, predicting an
accidental $U(2)^{5}$ flavour symmetry (see Section~\ref{subsec:U(2)5}).
Instead, the twin Pati-Salam theory \cite{King:2021jeo} was proposed
from a completely different perspective. Unlike the alternative models
in the market, the twin PS model treats all three fermion families
in the same way, and does not require to perform a very aggressive
family decomposition of the SM (avoiding dangerous contributions
to electroweak precision observables). The basic idea is that all
three families of SM chiral fermions transform under one PS group,
while families of vector-like fermions, which are required in any
implementation of the 4321 model, transform under the other one. The
first PS group, broken at a high scale, provides Pati-Salam unification
of all SM quarks and leptons, while a fourth family of vector-like
fermions transforms under a second PS group, broken at the TeV scale
to the SM, as shown in Fig.~\ref{fig:Model_Diagram}. The full twin Pati-Salam
symmetry, together with the absence of a standard Higgs doublet,
forbids the usual Yukawa couplings for the SM fermions. Instead, effective
Yukawa couplings arise through the mixing between SM fermions and
vector-like partners. The same mixing leads to $U_{1}$ couplings
for SM fermions which could address the $B$-anomalies. This way,
$B$-anomalies and the flavour puzzle are dynamically and parametrically
connected. In this manner, the twin Pati-Salam model features a \textit{fermiophobic}
framework where both $B$-anomalies and flavour hierarchies are explained
via the mechanism of \textit{messenger dominance} \cite{Ferretti:2006df}, as in the $Z'$ model of Chapter~\ref{Chapter:Fermiophobic}.

The origin of second and third family masses and mixing is connected with the 
TeV scale dynamics and first PS breaking that address the $B$-anomalies, while the origin of first family masses
and mixing is connected to the heavier scale of second PS breaking. In this manner, the
twin Pati-Salam model is an example of a multi-scale origin of flavour as introduced in Section~\ref{subsec:FromPlanckToEW}, however without the need to family decompose the SM as all the alternative theories do. Furthermore, the twin PS model predicts dominantly left-handed
$U_{1}$ currents as preferred by the current picture of $B$-anomalies
\cite{Alguero:2023jeh,Aebischer:2022oqe}, while
the alternative proposals \cite{Bordone:2017bld,Allwicher:2020esa,Fuentes-Martin:2020pww,Fuentes-Martin:2022xnb,Davighi:2022bqf}
predict large couplings for right-handed third family fermions,
which lead to tight constraints from high-$p_{T}$ searches \cite{Aebischer:2022oqe}.

In this chapter we study the phenomenology of the simplified twin
PS model proposed in \cite{King:2021jeo} to address the $B$-anomalies, which turns out to be
incompatible with low-energy data. Afterwards, we perform further
model building and present an extended version of the model that can
explain the most updated picture of $B$-anomalies and address charged
fermion masses and mixings, while being compatible with all existing
data. The model \cite{FernandezNavarro:2022gst,FernandezNavarro:2023lgk,FernandezNavarro:2023ykw} was initially built to connect both $R_{K^{(*)}}$
and $R_{D^{(*)}}$, however we will show that the model is also compatible
with the 2022 updates of $R_{K^{(*)}}$ by LHCb, which are consistent with
the SM. Nevertheless, the connection is not lost: eventually some
deviations in $R_{K^{(*)}}$ should be seen with more precision if
our model is indeed realised in Nature at the TeV scale to explain
the $R_{D^{(*)}}$ anomalies.

The layout of the remainder of the chapter is as follows. In Section~\ref{subsec:Twin-Pati-Salam-Theory_1VL}
we introduce the simplified
twin Pati-Salam model as presented in \cite{King:2021jeo}, featuring
only one vector-like family coupled to $U_{1}$, and we show the origin of 
fermion masses and mixing in this model. In Section~\ref{subsec:EFT_model_Appendix}
we show the tree-level SMEFT matching of 4-fermion operators in the model,
relevant for phenomenological analyses. Then in Section~\ref{subsec:pheno_simplified}
we show that the simplified model is unable to explain $R_{D^{(*)}}$ while being
compatible with the stringent constraints from $B_{s}-\bar{B}_{s}$
mixing. Instead, in Section~\ref{sec:Twin-Pati-Salam-Theory_3VL}
we present a new, extended version of the twin Pati-Salam model including
three vector-like families and a discrete flavour symmetry, which
is successful to address the $B$-anomalies while being compatible with all the experimental constraints. The 
phenomenological analysis and the discussion of the results are shown in Section~\ref{subsec:Low-energy-phenomenology}, highlighting promising signals to test the model in low-energy observables
and high-$p_{T}$ searches, along with a study of the perturbativity
of the model. Section~\ref{sec:Comparison_models} includes a comparison
of our predictions with alternative models in the market. Finally, we conclude
the chapter in Section~\ref{sec:Conclusions}.

\section{Simplified twin Pati-Salam theory of flavour \label{subsec:Twin-Pati-Salam-Theory_1VL}}

\subsection{The High Energy Model}

In the traditional Pati-Salam theory \cite{Pati:1974yy}, the chiral
quarks and leptons are unified into $SU(4)_{PS}$ multiplets with
leptons as the fourth colour (red, green, blue, lepton),
\begin{flalign}
 & \psi_{i}(\mathbf{4,2,1})=\left(\begin{array}{cccc}
\textcolor{red}{u_{r}} & \textcolor{DarkGreen}{u_{g}} & \textcolor{blue}{u_{b}} & \nu\\
\textcolor{red}{d_{r}} & \textcolor{DarkGreen}{d_{g}} & \textcolor{blue}{d_{b}} & e
\end{array}\right)_{i}\equiv\left(Q_{i},L_{i}\right)\,,\\
 & \psi_{j}^{c}(\mathbf{\overline{4},1,\overline{2}})=\left(\begin{array}{cccc}
\textcolor{red}{u_{r}^{c}} & \textcolor{DarkGreen}{u_{g}^{c}} & \textcolor{blue}{u_{b}^{c}} & \nu^{c}\\
\textcolor{red}{d_{r}^{c}} & \textcolor{DarkGreen}{d_{g}^{c}} & \textcolor{blue}{d_{b}^{c}} & e^{c}
\end{array}\right)_{j}\equiv\left(u_{j}^{c},d_{j}^{c},\nu_{j}^{c},e_{j}^{c}\right)\,,
\end{flalign}
where $\psi_{i}$ contains the left-handed quarks and leptons while
$\psi_{j}^{c}$ contains the $CP$-conjugated right-handed quarks and
leptons (so that they become left-handed\footnote{The reader who is not familiar with this notation based on left-handed
2-component Weyl spinors can find the connection with the traditional
4-component, left-right notation in Appendix~\ref{app:2-component_notation}.}), and $i,j=1,2,3$ are family indices. We consider here two copies
of the Pati-Salam symmetry \cite{King:2021jeo},
\begin{equation}
G_{422}^{I}\times G_{422}^{II}=\left(SU(4)_{PS}^{I}\times SU(2)_{L}^{I}\times SU(2)_{R}^{I}\right)\times\left(SU(4)_{PS}^{II}\times SU(2)_{L}^{II}\times SU(2)_{R}^{II}\right)\,.\label{eq:TwinPS_symmetry}
\end{equation}

The matter content and the quantum numbers of each field are displayed
in Table~\ref{tab:Field_content_TwinPS}. The usual three families of chiral
fermions originate from the second PS group $G_{422}^{II}$,
broken at a high scale, and transform under Eq.~(\ref{eq:TwinPS_symmetry})
as
\begin{equation}
\psi_{1,2,3}(\mathbf{1,1,1;4,2,1})\,,\quad\psi_{1,2,3}^{c}(\mathbf{1,1,1;\overline{4},1,\overline{2}})\,.
\end{equation}
This simplified version of the theory includes only one vector-like
family of fermions originating from the first PS group, whose $SU(4)^{I}_{PS}$
is broken at the TeV scale, and transforms under Eq.~(\ref{eq:TwinPS_symmetry})
as
\begin{equation}
\psi_{4}(\mathbf{4,2,1;1,1,1})\,,\quad\overline{\psi_{4}}(\mathbf{\overline{4},\overline{2},1;1,1,1})\,,\quad\psi_{4}^{c}(\mathbf{\overline{4},1,\overline{2};1,1,1})\,,\quad\overline{\psi_{4}^{c}}(\mathbf{4,1,2;1,1,1})\,.
\end{equation}

On the other hand, according to the matter content in Table~\ref{tab:Field_content_TwinPS},
there are no standard Higgs fields which transform as ($\mathbf{1,\overline{2},2}$)
under $G_{422}^{II}$, hence the standard Yukawa couplings involving
the chiral fermions are forbidden by the twin PS symmetry. This is
what we call a \textit{fermiophobic model}, in complete analogy with the fermiophobic $Z'$ model discussed in Chapter~\ref{Chapter:Fermiophobic}. For the third and second
families, these will be generated effectively via mixing with the
fourth family of vector-like fermions, which only have quantum numbers
under the first PS group, $G_{422}^{I}$. This mixing is facilitated
by the non-standard Higgs scalar doublets contained in $\phi$, $\overline{\phi}$,
$H$, $\overline{H}$ in Table~\ref{tab:Field_content_TwinPS}, via
the couplings,
\begin{table}[t]
\begin{centering}
\begin{tabular}{lcccccc}
\toprule 
Field & $SU(4)_{PS}^{I}$ & $SU(2)_{L}^{I}$ & $SU(2)_{R}^{I}$ & $SU(4)_{PS}^{II}$ & $SU(2)_{L}^{II}$ & $SU(2)_{R}^{II}$\tabularnewline
\midrule
\midrule 
$\psi_{1,2,3}$ & $\mathbf{1}$ & $\mathbf{1}$ & $\mathbf{1}$ & $\mathbf{4}$ & $\mathbf{2}$ & $\mathbf{1}$\tabularnewline
$\psi_{1,2,3}^{c}$ & $\mathbf{1}$ & $\mathbf{1}$ & $\mathbf{1}$ & $\mathbf{\overline{4}}$ & $\mathbf{1}$ & $\mathbf{\overline{2}}$\tabularnewline
\midrule 
$\psi_{4}$ & $\mathbf{4}$ & $\mathbf{2}$ & $\mathbf{1}$ & $\mathbf{1}$ & $\mathbf{1}$ & $\mathbf{1}$\tabularnewline
$\overline{\psi_{4}}$ & $\mathbf{\overline{4}}$ & $\mathbf{\overline{2}}$ & $\mathbf{1}$ & $\mathbf{1}$ & $\mathbf{1}$ & $\mathbf{1}$\tabularnewline
$\psi_{4}^{c}$ & $\mathbf{\overline{4}}$ & $\mathbf{1}$ & $\mathbf{\overline{2}}$ & $\mathbf{1}$ & $\mathbf{1}$ & $\mathbf{1}$\tabularnewline
$\overline{\psi_{4}^{c}}$ & $\mathbf{4}$ & $\mathbf{1}$ & $\mathbf{2}$ & $\mathbf{1}$ & $\mathbf{1}$ & $\mathbf{1}$\tabularnewline
\midrule 
$\phi$ & $\mathbf{4}$ & $\mathbf{2}$ & $\mathbf{1}$ & $\mathbf{\overline{4}}$ & $\mathbf{\overline{2}}$ & $\mathbf{1}$\tabularnewline
$\overline{\phi}$ & $\mathbf{\overline{4}}$ & $\mathbf{1}$ & $\mathbf{\overline{2}}$ & $\mathbf{4}$ & $\mathbf{1}$ & $\mathbf{2}$\tabularnewline
\midrule
$H$ & $\mathbf{\overline{4}}$ & $\mathbf{\overline{2}}$ & $\mathbf{1}$ & $\mathbf{4}$ & $\mathbf{1}$ & $\mathbf{2}$\tabularnewline
$\overline{H}$ & $\mathbf{4}$ & $\mathbf{1}$ & $\mathbf{2}$ & $\mathbf{\overline{4}}$ & $\mathbf{\overline{2}}$ & $\mathbf{1}$\tabularnewline
\midrule
$H'$ & $\mathbf{1}$ & $\mathbf{1}$ & $\mathbf{1}$ & $\mathbf{4}$ & $\mathbf{1}$ & $\mathbf{2}$\tabularnewline
$\Phi$ & $\mathbf{1}$ & $\mathbf{2}$ & $\mathbf{1}$ & $\mathbf{1}$ & $\mathbf{\overline{2}}$ & $\mathbf{1}$\tabularnewline
$\overline{\Phi}$ & $\mathbf{1}$ & $\mathbf{1}$ & $\mathbf{\overline{2}}$ & $\mathbf{1}$ & $\mathbf{1}$ & $\mathbf{2}$\tabularnewline
\bottomrule
\end{tabular}
\par\end{centering}
\caption[Field content in the simplified twin PS model]{The field content under $G_{422}^{I}\times G_{422}^{II}$, see the main text for details. We do not include here extra content related to the origin of first family fermion masses and mixing, to be discussed in Section~\ref{subsec:firstFamily}. \label{tab:Field_content_TwinPS}}
\end{table}
\begin{equation}
\mathcal{L}_{\mathrm{mass}}^{ren}=y_{i4}^{\psi}\overline{H}\psi_{i}\psi_{4}^{c}+y_{4i}^{\psi}H\psi_{4}\psi_{i}^{c}+x_{i4}^{\psi}\phi\psi_{i}\overline{\psi_{4}}+x_{4i}^{\psi^{c}}\overline{\psi_{4}^{c}\phi}\psi_{i}^{c}+M_{4}^{\psi}\psi_{4}\overline{\psi_{4}}+M_{4}^{\psi^{c}}\psi_{4}^{c}\overline{\psi_{4}^{c}}\label{eq:Lren_4thVL-1}
\end{equation}
plus h.c., where $i=1,2,3$; $x,y$ are dimensionless coupling
constants and $M_{4}^{\psi,\psi^{c}}$ are the vector-like mass terms. These
couplings mix the chiral fermions with the vector-like fermions, and will be
responsible for generating effective Yukawa couplings for the second
and third families (the origin of first family Yukawa couplings is discussed
in Section~\ref{subsec:firstFamily} and involves the second PS group). Moreover, the same
mixing leads to effective couplings to TeV scale $SU(4)^{I}_{PS}$ gauge
bosons which violate lepton universality between the second and third
families, as we shall see.

\subsection{High scale symmetry breaking\label{subsec:High-scale-symmetry}}

The twin Pati-Salam symmetry displayed in Eq.~(\ref{eq:TwinPS_symmetry})
is spontaneously broken to the 4321 symmetry at the high scale
$M_{\mathrm{High}}\apprge1\,\mathrm{PeV}$ (the latter bound due to
the non-observation of $K_{L}\rightarrow\mu e$ \cite{BNL:1998apv,Valencia:1994cj}),
\begin{equation}
G_{422}^{I}\times G_{422}^{II}\rightarrow G_{4321}\equiv SU(4)_{PS}^{I}\times SU(3)_{c}^{II}\times SU(2)_{L}^{I+II}\times U(1)_{Y'}\,.
\end{equation}
We can think of this as a two part symmetry breaking:\\
(i) The two pairs of left-right groups break down to their diagonal
left-right subgroup, via the VEVs $\left\langle \Phi\right\rangle \sim v_{\Phi}$
and $\langle\overline{\Phi}\rangle\sim v_{\overline{\Phi}}$, leading
to the symmetry breaking,
\begin{equation}
SU(2)_{L}^{I}\times SU(2)_{L}^{II}\rightarrow SU(2)_{L}^{I+II}\,,\qquad SU(2)_{R}^{I}\times SU(2)_{R}^{II}\rightarrow SU(2)_{R}^{I+II}\,.
\end{equation}
Since the two $SU(4)_{PS}$ groups remain intact, the above symmetry
breaking corresponds to\footnote{We note that the mechanism for generating first family masses discussed in Section~\ref{subsec:firstFamily} suggests that the scale of 
$SU(2)_{L}^{I}\times SU(2)_{L}^{II}\rightarrow SU(2)_{L}^{I+II}$ breaking may be below the scale of $SU(4)^{II}_{PS}$ breaking. However, this has no implications for the conclusions of this chapter.}
\begin{equation}
G_{422}^{I}\times G_{422}^{II}\rightarrow G_{4422}\equiv SU(4)_{PS}^{I}\times SU(4)_{PS}^{II}\times SU(2)_{L}^{I+II}\times SU(2)_{R}^{I+II}\,.
\end{equation}
(ii) Then we assume that the second PS group is broken at a high scale
via the Higgs $H'$ in Table~\ref{tab:Field_content_TwinPS}, which
under $G_{4422}$ transforms as
\begin{equation}
H'(\mathbf{1,4,1,2})=\left(\begin{array}{cccc}
u_{H'}^{r} & u_{H'}^{b} & u_{H'}^{g} & \nu_{H'}\\
d_{H'}^{r} & d_{H'}^{b} & d_{H'}^{g} & e_{H'}
\end{array}\right)\,,
\end{equation}
and develops a VEV in its right-handed neutrino (neutral) component, $\left\langle \nu_{H'}\right\rangle \apprge1\,\mathrm{PeV}$, leading
to the symmetry breaking
\begin{equation}
G_{4422}\rightarrow G_{4321}\equiv SU(4)_{PS}^{I}\times SU(3)_{c}^{II}\times SU(2)_{L}^{I+II}\times U(1)_{Y'}\,,
\end{equation}
where $SU(4)_{PS}^{II}$ is broken to $SU(3)_{c}^{II}\times U(1)_{B-L}^{II}$
(at the level of fermion representations, chiral quarks and leptons
are split $\mathbf{4}^{II}\rightarrow(\mathbf{3},1/6)^{II}\oplus(\mathbf{1},-1/2)^{II}$),
while $SU(2)^{I+II}_{R}$ is broken to $U(1)^{I+II}_{T_{3R}}$ and the abelian generators
are broken to $U(1)_{Y'}$, where $Y'=T_{B-L}^{II}+T_{3R}^{I+II}$.
The broken generators of $SU(4)_{PS}^{II}$ are associated with PeV-scale
gauge bosons that will mediate processes at acceptable rates, beyond
the sensitivity of current experiments and colliders. Instead, the
further symmetry breaking of $G_{4321}$ will lead to a rich phenomenology
at the TeV scale, as we shall see. We anticipate that $SU(2)_{L}^{I+II}$
is already the $SU(2)_{L}$ of the SM gauge group, while SM color
and hypercharge are embedded in $SU(4)_{PS}^{I}\times SU(3)_{c}^{II}\times U(1)_{Y'}$. 

On the other hand, the Yukon scalars $\phi$ and $\overline{\phi}$
in Table~\ref{tab:Field_content_TwinPS}, responsible for mixing chiral and vector-like fermions, decompose under $G_{422}^{I}\times G_{422}^{II}\rightarrow G_{4422}\rightarrow G_{4321}$
as
\begin{equation}
\begin{array}{c}
\phi(\mathbf{4,2,1;\overline{4},\overline{2},1})\rightarrow\phi(\mathbf{4,\overline{4},1\oplus3,1})\rightarrow\phi_{3}(\mathbf{4,\overline{3},1\oplus3},-1/6)\oplus\phi_{1}(\mathbf{4,1,1\oplus3},1/2)\,,\\
\,\\
\overline{\phi}(\mathbf{\overline{4},1,\overline{2};4,1,2})\rightarrow\overline{\phi}(\mathbf{\overline{4},4,1,1\oplus3})\rightarrow\overline{\phi_{3}}(\mathbf{\overline{4},3,1},1/6)\oplus\overline{\phi_{1}}(\mathbf{\overline{4},1,1},-1/2)\,,
\end{array}\label{eq:phi_decomposition}
\end{equation}
plus extra $\overline{\phi_{3}}$ and $\overline{\phi_{1}}$ with
different values of $Y'$ associated to the breaking of the $SU(2)_{R}^{I+II}$
triplet, that we ignore because they do not couple to fermions. The
decomposition above is of phenomenological interest, as the Yukons
$\phi_{3}$, $\overline{\phi_{3}}$ will couple to quarks while $\phi_{1}$,
$\overline{\phi_{1}}$ will couple to leptons, allowing non-trivial
mixing between SM fermions and vector-like fermions. They will also lead to
a non-trivial breaking of $G_{4321}$ down to the SM.

The Higgs scalars $H$ and $\overline{H}$ in Table~\ref{tab:Field_content_TwinPS}
decompose under $G_{422}^{I}\times G_{422}^{II}\rightarrow G_{4321}$
as (we skip the $G_{4422}$ decomposition here for simplicity)
\begin{equation}
H(\mathbf{\overline{4},\overline{2},1;4,1,2})\rightarrow H_{t}(\mathbf{\overline{4},3,\overline{2}},2/3),\:H_{b}(\mathbf{\overline{4},3,\overline{2}},-1/3),\,H_{\tau}(\mathbf{\overline{4},1,\overline{2}},-1),\,H_{\nu_{\tau}}(\mathbf{\overline{4},1,\overline{2}},0)\,,\label{eq:H_Higgs}
\end{equation}
\begin{equation}
\overline{H}(\mathbf{4,1,2;\overline{4},\overline{2},1})\rightarrow H_{c}(\mathbf{4,\overline{3},\overline{2}},1/3),\:H_{s}(\mathbf{4,\overline{3},\overline{2}},-2/3),\,H_{\mu}(\mathbf{\overline{4},1,\overline{2}},0),\,H_{\nu_{\mu}}(\mathbf{\overline{4},1,\overline{2}},1)\,,\label{eq:Hbar_Higgs}
\end{equation}
where the notation anticipates that a separate personal Higgs doublet
contributes to each of the second and third family quark and lepton
masses, as we shall see. Models with multiple light Higgs doublets
face the phenomenological challenge of FCNCs arising from tree-level
exchange of the scalar doublets in the Higgs basis. Therefore we
assume that only one pair of Higgs doublets $H_{u}$ and $H_{d}$
are light, given by linear combinations of the personal Higgs,
\begin{equation}
\begin{array}{c}
H_{u}=\widetilde{\alpha}_{u}H_{t}+\widetilde{\beta}_{u}H_{c}+\widetilde{\gamma}_{u}H_{\nu_{\tau}}+\widetilde{\delta}_{u}H_{\nu_{\mu}}\,,\quad H_{d}=\widetilde{\alpha}_{d}H_{b}+\widetilde{\beta}_{d}H_{s}+\widetilde{\gamma}_{d}H_{\tau}+\widetilde{\delta}_{d}H_{\mu}\,,\end{array}\label{eq:Light_Higgses}
\end{equation}
where $\widetilde{\alpha}_{u,d}$, $\widetilde{\beta}_{u,d}$, $\widetilde{\gamma}_{u,d}$,
$\widetilde{\delta}_{u,d}$ are complex elements of two unitary Higgs
mixing matrices. The orthogonal linear combinations are assumed to
be very heavy, well above the TeV scale in order to sufficiently suppress
the FCNCs. We will further assume that only the light Higgs doublets
$H_{u}$ and $H_{d}$ get VEVs in order to perform EW symmetry breaking,
\begin{equation}
\left\langle H_{u}\right\rangle =v_{u},\quad\left\langle H_{d}\right\rangle =v_{d},\label{eq:VeVs_2HDM}
\end{equation}
while the heavy linear combinations do not, i.e.~we assume that in
the Higgs basis the linear combinations which do not get VEVs are
very heavy. The discussion of such Higgs potential is beyond the scope
of this work, for the interested reader a deeper discussion was made
in Section 3.4 of \cite{King:2021jeo}. We shall just anticipate that
the situation is familiar from $SO(10)$ models \cite{Fukuyama:2012rw},
where there are six Higgs doublets arising from the $\mathbf{10}$,
$\mathbf{120}$ and $\mathbf{\overline{126}}$ representations, denoted
as $H_{\mathbf{10}}$, $H_{\mathbf{120}}$ and $H_{\mathbf{\overline{126}}}$,
two from each, but below the $SO(10)$ breaking scale only two Higgs
doublets are assumed to be light, similar to $H_{u}$ and $H_{d}$
above in our case. Another example is the $\mathrm{PS^{3}}$ theory
of flavour \cite{Bordone:2017bld}, also proposed to address the $B$-anomalies,
where the Higgs $(\mathbf{15,1,1})_{3}$ and $(\mathbf{1,2,2})_{3}$
are assumed to give rise to a specific set of light doublets. In any
case, we shall invert the unitary transformations in Eq.~(\ref{eq:Light_Higgses})
to express each of the personal Higgs doublets in terms of the light
doublets $H_{u}$, $H_{d}$,
\begin{equation}
\begin{array}{c}
H_{t}=\alpha_{u}H_{u}+...\,,\quad H_{b}=\alpha_{d}H_{d}+...\,,\quad H_{\tau}=\gamma_{d}H_{d}+...\,,\quad H_{\nu_{\tau}}=\gamma_{u}H_{u}+...\,,\\
\,\\
H_{c}=\beta_{u}H_{u}+...\,,\quad H_{s}=\beta_{d}H_{d}+...\,,\quad H_{\mu}=\delta_{d}H_{d}+...\,,\quad\,H_{\nu_{\mu}}=\delta_{u}H_{u}+...\,, \label{eq:Higgs_matching}
\end{array}
\end{equation}
ignoring the heavy states indicated by dots. When the light Higgs
$H_{u}$ and $H_{d}$ gain their VEVs in Eq.~(\ref{eq:VeVs_2HDM}),
the personal Higgs in the original basis can be thought of as gaining
effective VEVs $\left\langle H_{t}\right\rangle =\alpha_{u}v_{u}$, etc... This
approach will be used in the next subsection when constructing the low energy
quark and lepton mass matrices.

\subsection{Effective Yukawa couplings and fermion masses for the second and
third family \label{subsec:Effective-Yukawa-couplings}}

We have already remarked that the usual Yukawa couplings involving
purely chiral fermions are absent in the twin PS model. In this subsection,
we show how they may be generated effectively via mixing with the
vector-like fermions. 

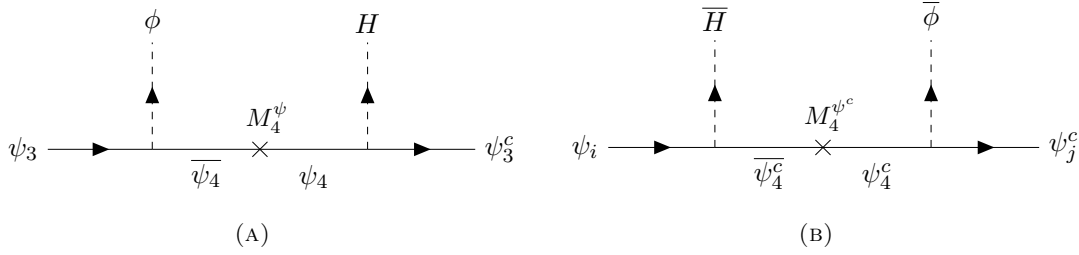
\begin{figure}[t]
\subfloat[]{\begin{centering}
\resizebox{.48\textwidth}{!}{
\begin{tikzpicture}
	\begin{feynman}
		\vertex (a) {\(\psi_{3}\)};
		\vertex [right=18mm of a] (b);
		\vertex [right=of b] (c) [label={ [xshift=0.1cm, yshift=0.1cm] \small $M^{\psi}_{4}$}];
		\vertex [right=of c] (d);
		\vertex [right=of d] (e) {\(\psi^{c}_{3}\)};
		\vertex [above=of b] (f1) {\(\phi\)};
		\vertex [above=of d] (f2) {\(H\)};
		\diagram* {
			(a) -- [fermion] (b) -- [charged scalar] (f1),
			(b) -- [edge label'=\(\overline{\psi_{4}}\)] (c),
			(c) -- [edge label'=\(\psi_{4}\), inner sep=6pt, insertion=0] (d) -- [charged scalar] (f2),
			(d) -- [fermion] (e),
	};
	\end{feynman}
\end{tikzpicture}}
\par\end{centering}
}\subfloat[]{\begin{centering}
\resizebox{.48\textwidth}{!}{
\begin{tikzpicture}
	\begin{feynman}
		\vertex (a) {\(\psi_{i}\)};
		\vertex [right=18mm of a] (b);
		\vertex [right=of b] (c) [label={ [xshift=0.1cm, yshift=0.1cm] \small $M^{\psi^{c}}_{4}$}];
		\vertex [right=of c] (d);
		\vertex [right=of d] (e) {\(\psi^{c}_{j}\)};
		\vertex [above=of b] (f1) {\(\overline{H}\)};
		\vertex [above=of d] (f2) {\(\overline{\phi}\)};
		\diagram* {
			(a) -- [fermion] (b) -- [charged scalar] (f1),
			(b) -- [edge label'=\(\overline{\psi^{c}_{4}}\)] (c),
			(c) -- [edge label'=\(\psi^{c}_{4}\), inner sep=6pt, insertion=0] (d) -- [charged scalar] (f2),
			(d) -- [fermion] (e),
	};
	\end{feynman}
\end{tikzpicture}}
\par\end{centering}
}\caption[Diagrams in the twin PS model which lead to the effective
Yukawa couplings in the mass insertion approximation]{ Diagrams in the model which lead to the effective Yukawa couplings
in the mass insertion approximation, $i,j=2,3$. \label{fig: mass_insertion_4thVL_TwinPS}}
\end{figure}

We may write the mass terms and couplings in Eq.~(\ref{eq:Lren_4thVL-1})
as a $5\times5$ matrix in flavour space (we also define 5-dimensional
vectors as $\psi_{\alpha}^{\mathrm{T}}$ and $\psi_{\beta}^{c}$),
\begin{equation}
\mathcal{L}_{\mathrm{mass}}^{ren}=\psi_{\alpha}^{\mathrm{T}}M^{\psi}\psi_{\beta}^{c}+\mathrm{h.c.}\,,
\end{equation}
\begin{equation}
\psi_{\alpha}^{\mathrm{T}}\equiv\left(\begin{array}{ccccc}
\psi_{1} & \psi_{2} & \psi_{3} & \psi_{4} & \overline{\psi_{4}^{c}}\end{array}\right)\,,\qquad\psi_{\beta}^{c}\equiv\left(\begin{array}{ccccc}
\psi_{1}^{c} & \psi_{2}^{c} & \psi_{3}^{c} & \psi_{4}^{c} & \overline{\psi_{4}}\end{array}\right)^{\mathrm{T}}\,,
\end{equation}
\begin{equation}
M^{\psi}=\left(
\global\long\def\arraystretch{1.3}%
\begin{array}{@{}llcccc@{}}
 & \multicolumn{1}{c@{}}{\psi_{1}^{c}} & \psi_{2}^{c} & \psi_{3}^{c} & \psi_{4}^{c} & \overline{\psi_{4}}\\
\cmidrule(l){2-6}\left.\psi_{1}\right| & 0 & 0 & 0 & 0 & 0\\
\left.\psi_{2}\right| & 0 & 0 & 0 & y_{24}^{\psi}\overline{H} & 0\\
\left.\psi_{3}\right| & 0 & 0 & 0 & y_{34}^{\psi}\overline{H} & x_{34}^{\psi}\phi\\
\left.\psi_{4}\right| & 0 & 0 & y_{43}^{\psi}H & 0 & M_{4}^{\psi}\\
\left.\overline{\psi_{4}^{c}}\right| & 0 & x_{42}^{\psi^{c}}\overline{\phi} & x_{43}^{\psi^{c}}\overline{\phi} & M_{4}^{\psi^{c}} & 0
\end{array}\right)\,.\label{eq:MassMatrix_4thVL}
\end{equation}
where the extra zeros have been achieved via suitable rotations of
$\psi_{i}$, $\psi_{j}^{c}$ ($i,j=1,2,3$), that leave unchanged
the upper $3\times3$ block (for further details see Section~\ref{subsec:Messenger_Dominance} and the discussion therein). There are several distinct mass scales
in this matrix: the Higgs VEVs $\left\langle H\right\rangle $ and
$\langle\overline{H}\rangle$, the Yukon VEVs $\left\langle \phi\right\rangle $
and $\langle\overline{\phi}\rangle$ and the vector-like fourth family masses
$M_{4}^{\psi}$, $M_{4}^{\psi^{c}}$. Assuming the latter are heavier
than all the VEVs, we may integrate out the fourth family to generate
effective Yukawa couplings for chiral quarks and leptons, as in the
diagrams of Fig.~\ref{fig: mass_insertion_4thVL_TwinPS}. This is denoted
as the mass insertion approximation (see Appendix~\ref{app:mixing_angle_formalism}).

However, from the diagrams in Fig.~\ref{fig: mass_insertion_4thVL_TwinPS}
one can anticipate that the heavy top mass requires $\left\langle \phi\right\rangle /M_{4}^{\psi}\sim1$,
and thus it is necessary to go beyond the mass insertion approximation
and work in the large mixing angle formalism (see Appendix~\ref{app:mixing_angle_formalism}). We shall block-diagonalise
the mass matrix in Eq.~(\ref{eq:MassMatrix_4thVL}) in order to obtain
the effective Yukawa couplings for the chiral families,
\begin{equation}
M^{\psi'}=\left(
\global\long\def\arraystretch{1.3}%
\begin{array}{@{}llcccc@{}}
 & \multicolumn{1}{c@{}}{\psi'{}_{1}^{c}} & \psi'{}_{2}^{c} & \psi'{}_{3}^{c} & \psi'{}_{4}^{c} & \overline{\psi'_{4}}\\
\cmidrule(l){2-6}\left.\psi'_{1}\right| &  &  &  &  & 0\\
\left.\psi'_{2}\right| &  &  &  &  & 0\\
\left.\psi'_{3}\right| &  &  & \widetilde{y}{}_{\alpha\beta}^{\psi'} &  & 0\\
\left.\psi'_{4}\right| &  &  &  &  & \hat{M}_{4}^{\psi}\\
\left.\overline{\psi'{}_{4}^{c}}\right| & 0 & 0 & 0 & \hat{M}_{4}^{\psi^{c}} & 0
\end{array}\right)\,,\label{MassMatrix_4thVL_decoupling}
\end{equation}
where $\widetilde{y}{}_{\alpha\beta}^{\psi'}$ is the upper $4\times4$
block of the mass matrix in this basis. The key feature of Eq.~(\ref{MassMatrix_4thVL_decoupling})
are the zeros in the fifth row and column which are achieved by rotating
the four families by the unitary $4\times4$ transformations,
\begin{flalign}
 & V_{\psi}=V_{34}^{\psi}=\left(\begin{array}{cccc}
1 & 0 & 0 & 0\\
0 & 1 & 0 & 0\\
0 & 0 & c_{34}^{\psi} & s_{34}^{\psi}\\
0 & 0 & -s_{34}^{\psi} & c_{34}^{\psi}
\end{array}\right)\,,\label{eq:34mixing}\\
 & V_{\psi^{c}}=V_{34}^{\psi^{c}}V_{24}^{\psi^{c}}=\left(\begin{array}{cccc}
1 & 0 & 0 & 0\\
0 & 1 & 0 & 0\\
0 & 0 & c_{34}^{\psi^{c}} & s_{34}^{\psi^{c}}\\
0 & 0 & -s_{34}^{\psi^{c}} & c_{34}^{\psi^{c}}
\end{array}\right)\left(\begin{array}{cccc}
1 & 0 & 0 & 0\\
0 & c_{24}^{\psi^{c}} & 0 & s_{24}^{\psi^{c}}\\
0 & 0 & 1 & 0\\
0 & -s_{24}^{\psi^{c}} & 0 & c_{24}^{\psi^{c}}
\end{array}\right)\,,\nonumber 
\end{flalign}
where we have defined $s_{i4}^{\psi^{(c)}}\equiv\sin\theta_{i4}^{\psi^{(c)}}$,~$c_{i4}^{\psi^{(c)}}\equiv\cos\theta_{i4}^{\psi^{(c)}}$,
with the mixing angles and the physical masses of the vector-like fermions
given by
\begin{flalign}
 & s_{34}^{\psi}=\frac{x_{34}^{\psi}\left\langle \phi\right\rangle }{\sqrt{\left(x_{34}^{\psi}\left\langle \phi\right\rangle \right)^{2}+\left(M_{4}^{\psi}\right)^{2}}}\,, &  & s_{24}^{\psi^{c}}=\frac{x_{42}^{\psi^{c}}\langle\overline{\phi}\rangle}{\sqrt{\left(x_{42}^{\psi^{c}}\langle\overline{\phi}\rangle\right)^{2}+\left(M_{4}^{\psi^{c}}\right)^{2}}}\,,\label{eq:34_mixing_extended-1}\\
 & s_{34}^{\psi^{c}}=\frac{x_{43}^{\psi^{c}}\langle\overline{\phi}\rangle}{\sqrt{\left(x_{42}^{\psi^{c}}\langle\overline{\phi}\rangle\right)^{2}+\left(x_{43}^{\psi^{c}}\langle\overline{\phi}\rangle\right)^{2}+\left(M_{4}^{\psi^{c}}\right)^{2}}}\,, &  & \hat{M}_{4}^{\psi}=\sqrt{\left(x_{34}^{\psi}\left\langle \phi\right\rangle \right)^{2}+\left(M_{4}^{\psi}\right)^{2}}\,,\label{eq:sqc24_mixing-1}\\
 & \hat{M}_{4}^{\psi^{c}}=\sqrt{\left(x_{42}^{\psi^{c}}\langle\overline{\phi}\rangle\right)^{2}+\left(x_{43}^{\psi^{c}}\langle\overline{\phi}\rangle\right)^{2}+\left(M_{4}^{\psi^{c}}\right)^{2}}\,,\label{eq:sqc34_mixing-1}
\end{flalign}
\\
Notice that we recover the expressions in the mass insertion approximation when $\left\langle \phi\right\rangle\ll M$, and in this case the mixing angles above are given by the usual NP scales ratios of a theory of flavour,
$\left\langle \phi\right\rangle/M$ (see Section~\ref{subsec:FromPlanckToEW}). As long as these ratios are held fixed to explain the SM Yukawa couplings, the independent scales $\left\langle \phi\right\rangle$ and $M$ may be anywhere \textit{from the Planck scale to the electroweak scale}. In our model, however, we shall see that some of these scales will be fixed to the TeV via the connection with the $B$-anomalies, leading to a testable theory of flavour with significant low-energy implications.

Now we apply the transformations in Eq.~(\ref{eq:34mixing}) to the
upper $4\times4$ block of (\ref{eq:MassMatrix_4thVL}), obtaining
effective Yukawa couplings for the chiral fermions as the upper $3\times3$
block of the mass matrix in the new basis,
\begin{equation}
\mathcal{L}_{eff}^{Yuk3\times3}=\psi'{}_{i}^{\mathrm{T}}V_{\psi}y_{\alpha\beta}^{\psi}V_{\psi^{c}}^{\dagger}\psi'{}_{j}^{c}+\mathrm{h.c.}\,,\label{eq:Primed_basis}
\end{equation}
\begin{equation}
\psi'{}_{\alpha}^{\mathrm{T}}=\psi_{\alpha}^{\mathrm{T}}V_{\psi}^{\dagger}\,,\qquad\psi'{}_{\alpha}^{c}=V_{\psi^{c}}\psi_{\alpha}^{c}\,,
\end{equation}
where $i,j=1,2,3$. We obtain
\begin{equation}
\mathcal{L}_{eff}^{Yuk,3\times3}=\left(
\global\long\def\arraystretch{0.7}%
\begin{array}{@{}llcc@{}}
 & \multicolumn{1}{c@{}}{\psi'{}_{1}^{c}} & \psi'{}_{2}^{c} & \psi'{}_{3}^{c}\\
\cmidrule(l){2-4}\left.\psi'_{1}\right| & 0 & 0 & 0\\
\left.\psi'_{2}\right| & 0 & 0 & 0\\
\left.\psi'_{3}\right| & 0 & 0 & c_{34}^{\psi^{c}}s_{34}^{\psi}y_{43}^{\psi}
\end{array}\right)H+\left(
\global\long\def\arraystretch{0.7}%
\begin{array}{@{}llcc@{}}
 & \multicolumn{1}{c@{}}{\psi'{}_{1}^{c}} & \psi'{}_{2}^{c} & \psi'{}_{3}^{c}\\
\cmidrule(l){2-4}\left.\psi'_{1}\right| & 0 & 0 & 0\\
\left.\psi'_{2}\right| & 0 & s_{24}^{\psi^{c}}y_{24}^{\psi} & c_{24}^{\psi^{c}}s_{34}^{\psi^{c}}y_{24}^{\psi}\\
\left.\psi'_{3}\right| & 0 & c_{34}^{\psi}s_{24}^{\psi^{c}}y_{34}^{\psi} & c_{34}^{\psi}c_{24}^{\psi^{c}}s_{34}^{\psi^{c}}y_{34}^{\psi}
\end{array}\right)\overline{H}\label{eq:MassMatrix_4thVL_effective-1-1}
\end{equation}
plus hermitian conjugate. Until the breaking of the twin PS symmetry, the matrix above is Pati-Salam
universal, so all fermions of the same family share the same effective
Yukawa couplings. If we impose the dominance of the doublet messenger fermions over the singlet messenger fermions, as we did in Chapter~\ref{Chapter:Fermiophobic},
\begin{equation}
M_{4}^{\psi}\ll M_{4}^{\psi^{c}}\,,\label{eq:hierarchy_scalesVL}
\end{equation}
then the first matrix in Eq.~(\ref{eq:MassMatrix_4thVL_effective-1-1})
generates larger effective third family Yukawa couplings, while the
second matrix generates suppressed second family Yukawa couplings
and mixings. This way, the hierarchy of quark and lepton masses in
the SM is re-expressed as the hierarchy of scales
in Eq.~(\ref{eq:hierarchy_scalesVL}). Remarkably, the hierarchical
relation in Eq.~(\ref{eq:hierarchy_scalesVL}) will lead to small
couplings of $\psi^{c}$ chiral fermions (eventually SM $SU(2)_{L}$
singlets) to $SU(4)^{I}_{PS}$ gauge bosons, hence obtaining dominantly
left-handed $U_{1}$ couplings. The couplings of the $\psi^{c}$ fermions
will be suppressed, connected to the origin of second family fermion
masses, and this way the tight high-$p_{T}$ constraints that afflict
other $U_{1}$ models can be relaxed (see more in Section~\ref{subsec:Colliders}).

On the other hand, since the sum of the two matrices in Eq.~(\ref{eq:MassMatrix_4thVL_effective-1-1})
has rank 1, the first family will be massless, which is a good first
order approximation. Indeed, the masses of first family fermions are protected by
an accidental $U(1)$ symmetry that will be broken at much higher energies
via new dynamics connected the second PS group, providing the small masses
of first family fermions as discussed in the next subsection.

After the symmetry breaking of the twin PS group to $G_{4321}$, the
Yukawa couplings $x_{34}^{\psi}$, $x_{42,43}^{\psi^{c}}$ and vector-like
masses $M_{4}^{\psi},$ $M_{4}^{\psi^{c}}$ remain universal up to
small RGE effects, however the Yukons decompose in a different way
for lepton and quarks as per Eq.~(\ref{eq:phi_decomposition}). Due
to this decomposition, the mixing angles in Eq.~(\ref{eq:MassMatrix_4thVL_effective-1-1})
are now different for quarks and leptons. The VEVs of the Yukons break
the remaining $SU(4)$ symmetry relating quarks and leptons, but the
Yukawa couplings still exhibit an accidental $SU(2)_{R}$ symmetry
relating $u^{c}$ and $d^{c}$ quarks. Hence, the mixing angles $s_{i4}^{u^{c}}$
and $s_{i4}^{d^{c}}$ are the same for up and down quarks, and we define
$q^{c}=u^{c},\,d^{c}$. On the other hand, the Higgs fields $H$ and
$\overline{H}$ decompose as personal Higgs doublets for the second
and third fermions as per Eqs.~(\ref{eq:H_Higgs}) and (\ref{eq:Hbar_Higgs}).
The personal Higgs are introduced in order to break the accidental
symmetry $SU(2)_{R}$ in the Yukawa couplings, otherwise the mass
matrices in the up and down sector would remain identical at tree-level.
A similar discussion applies to charged leptons and neutrinos, and
personal Higgses apply in the same way. Mass terms for second and
third family fermions will be obtained after the personal Higgs
develop VEVs $\left\langle H_{\alpha}\right\rangle \ll\left\langle \phi\right\rangle $,
which also play the role of breaking EW symmetry in the SM, see Section~\ref{subsec:High-scale-symmetry}.
This way, Eq.~(\ref{eq:MassMatrix_4thVL_effective-1-1}) decomposes
for each charged sector as the following effective mass matrices,
\begin{equation}
M_{\mathrm{eff}}^{u}=\left(
\global\long\def\arraystretch{0.7}%
\begin{array}{@{}llcc@{}}
 & \multicolumn{1}{c@{}}{\phantom{\!\,}u'{}_{1}^{c}} & \phantom{\!\,}u'{}_{2}^{c} & \phantom{\!\,}u'{}_{3}^{c}\\
\cmidrule(l){2-4}\left.Q'_{1}\right| & 0 & 0 & 0\\
\left.Q'_{2}\right| & 0 & 0 & 0\\
\left.Q'_{3}\right| & 0 & 0 & s_{34}^{Q}y_{43}^{\psi}
\end{array}\right)\left\langle H_{t}\right\rangle +\left(
\global\long\def\arraystretch{0.7}%
\begin{array}{@{}llcc@{}}
 & \multicolumn{1}{c@{}}{\phantom{\!\,}u'{}_{1}^{c}} & \phantom{\!\,}u'{}_{2}^{c} & \phantom{\!\,}u'{}_{3}^{c}\\
\cmidrule(l){2-4}\left.Q'_{1}\right| & 0 & 0 & 0\\
\left.Q'_{2}\right| & 0 & s_{24}^{q^{c}}y_{24}^{\psi} & s_{34}^{q^{c}}y_{24}^{\psi}\\
\left.Q'_{3}\right| & 0 & c_{34}^{Q}s_{24}^{q^{c}}y_{34}^{\psi} & c_{34}^{Q}s_{34}^{q^{c}}y_{34}^{\psi}
\end{array}\right)\left\langle H_{c}\right\rangle +\mathrm{h.c.}\,,\label{eq:MassMatrix_4thVL_effective_up}
\end{equation}
\begin{equation}
M_{\mathrm{eff}}^{d}=\left(
\global\long\def\arraystretch{0.7}%
\begin{array}{@{}llcc@{}}
 & \multicolumn{1}{c@{}}{\phantom{\!\,}d'{}_{1}^{c}} & \phantom{\!\,}d'{}_{2}^{c} & \phantom{\!\,}d'{}_{3}^{c}\\
\cmidrule(l){2-4}\left.Q'_{1}\right| & 0 & 0 & 0\\
\left.Q'_{2}\right| & 0 & 0 & 0\\
\left.Q'_{3}\right| & 0 & 0 & s_{34}^{Q}y_{43}^{\psi}
\end{array}\right)\left\langle H_{b}\right\rangle +\left(
\global\long\def\arraystretch{0.7}%
\begin{array}{@{}llcc@{}}
 & \multicolumn{1}{c@{}}{\phantom{\!\,}d'{}_{1}^{c}} & \phantom{\!\,}d'{}_{2}^{c} & \phantom{\!\,}d'{}_{3}^{c}\\
\cmidrule(l){2-4}\left.Q'_{1}\right| & 0 & 0 & 0\\
\left.Q'_{2}\right| & 0 & s_{24}^{q^{c}}y_{24}^{\psi} & s_{34}^{q^{c}}y_{24}^{\psi}\\
\left.Q'_{3}\right| & 0 & c_{34}^{Q}s_{24}^{q^{c}}y_{34}^{\psi} & c_{34}^{Q}s_{34}^{q^{c}}y_{34}^{\psi}
\end{array}\right)\left\langle H_{s}\right\rangle +\mathrm{h.c.}\,,\label{eq:MassMatrix_4thVL_effective_down}
\end{equation}
\begin{equation}
M_{\mathrm{eff}}^{e}=\left(
\global\long\def\arraystretch{0.7}%
\begin{array}{@{}llcc@{}}
 & \multicolumn{1}{c@{}}{\phantom{\!\,}e'{}_{1}^{c}} & \phantom{\!\,}e'{}_{2}^{c} & \phantom{\!\,}e'{}_{3}^{c}\\
\cmidrule(l){2-4}\left.L'_{1}\right| & 0 & 0 & 0\\
\left.L'_{2}\right| & 0 & 0 & 0\\
\left.L'_{3}\right| & 0 & 0 & s_{34}^{L}y_{43}^{\psi}
\end{array}\right)\left\langle H_{\tau}\right\rangle +\left(
\global\long\def\arraystretch{0.7}%
\begin{array}{@{}llcc@{}}
 & \multicolumn{1}{c@{}}{\phantom{\!\,}e'{}_{1}^{c}} & \phantom{\!\,}e'{}_{2}^{c} & \phantom{\!\,}e'{}_{3}^{c}\\
\cmidrule(l){2-4}\left.L'_{1}\right| & 0 & 0 & 0\\
\left.L'_{2}\right| & 0 & s_{24}^{e^{c}}y_{24}^{\psi} & s_{34}^{e^{c}}y_{24}^{\psi}\\
\left.L'_{3}\right| & 0 & c_{34}^{L}s_{24}^{e^{c}}y_{34}^{\psi} & c_{34}^{L}s_{34}^{e^{c}}y_{34}^{\psi}
\end{array}\right)\left\langle H_{\mu}\right\rangle +\mathrm{h.c.}\,,\label{eq:MassMatrix_4thVL_effective_leptons}
\end{equation}
where the Yukawas $y_{43}^{\psi}$ and $y_{24,34}^{\psi}$ are Pati-Salam
universal, and we have approximated all cosines related to $\psi^{c}$
fields to be 1 due to the hierarchy of vector-like masses in Eq.~(\ref{eq:hierarchy_scalesVL}).
We obtain a similar Dirac mass matrix for neutrinos, to be discussed
in the next subsection.

Due to the fact that vector-like fermions are much heavier than SM fermions,
the fourth row and column, that we have intentionally ignored when
writing Eqs.~(\ref{eq:MassMatrix_4thVL_effective_up}), (\ref{eq:MassMatrix_4thVL_effective_down})
and (\ref{eq:MassMatrix_4thVL_effective_leptons}), can be decoupled
from the $3\times3$ upper blocks, which we can diagonalise via independent
2-3 transformations for each charged sector $V_{23}^{u}$, $V_{23}^{d}$
and $V_{23}^{e}$. Similar transformations apply for $SU(2)_{L}$ singlet fermions
$u^{c}$, $d^{c}$, $e^{c}$, in such a way that the mass matrices
in Eqs.~(\ref{eq:MassMatrix_4thVL_effective_up}),~(\ref{eq:MassMatrix_4thVL_effective_down}),~(\ref{eq:MassMatrix_4thVL_effective_leptons})
are diagonalised as
\begin{flalign}
 & V_{23}^{u}M_{\mathrm{eff}}^{u}V_{23}^{u^{c}\dagger}=\mathrm{diag}(0,m_{c},m_{t})\,,\\
 & V_{23}^{d}M_{\mathrm{eff}}^{d}V_{23}^{d^{c}\dagger}=\mathrm{diag}(0,m_{s},m_{b})\,,\\
 & V_{23}^{e}M_{\mathrm{eff}}^{e}V_{23}^{e^{c}\dagger}=\mathrm{diag}(0,m_{\mu},m_{\tau})\,.
\end{flalign}
The CKM matrix is then given by
\begin{equation}
V_{\mathrm{CKM}}=V_{23}^{u}V_{23}^{d\dagger}=\left(\begin{array}{ccc}
1 & 0 & 0\\
0 & c_{23}^{u}c_{23}^{d}+s_{23}^{u}s_{23}^{d} & c_{23}^{d}s_{23}^{u}-c_{23}^{u}s_{23}^{d}\\
0 & -\left(c_{23}^{d}s_{23}^{u}-c_{23}^{u}s_{23}^{d}\right) & c_{23}^{u}c_{23}^{d}+s_{23}^{u}s_{23}^{d}
\end{array}\right)\approx\left(\begin{array}{ccc}
1 & 0 & 0\\
0 & V_{cs} & V_{cb}\\
0 & V_{ts} & V_{tb}
\end{array}\right).\label{eq:CKM_matrixTwinPS}
\end{equation}
We do not address the mixing involving the first family nor the $CP$-violating
phase since so far first family fermions remain massless,
as previously discussed. We are however required to preserve $V_{cb}$
as \cite{PDG:2022ynf}
\begin{equation}
V_{cb}=\left(40.8\pm1.4\right)\times10^{-3}\approx s_{23}^{u}-s_{23}^{d}\,,\label{eq:Vcb_model}
\end{equation}
 where in the last step we have approximated the cosines to be 1.
We will not fit $V_{ts}$, $V_{tb}$ and $V_{cs}$ up to the experimental
precision, as corrections related to the first family mixing (and
$CP$-violating phase) are required.

In the following we explore the parameters in the mass matrices of
Eqs.~(\ref{eq:MassMatrix_4thVL_effective_up}), (\ref{eq:MassMatrix_4thVL_effective_down}), (\ref{eq:MassMatrix_4thVL_effective_leptons}),
and their impact over the diagonalisation of the mass matrices:
\begin{itemize}
\item In very good approximation, the mass of the top quark is given by
the (3,3) entry in the first matrix of Eq.~(\ref{eq:MassMatrix_4thVL_effective_up}),
i.e.
\begin{equation}
m_{t}\approx s_{34}^{Q}y_{43}^{\psi}\left\langle H_{t}\right\rangle =s_{34}^{Q}y_{43}^{\psi}\alpha_{u}\frac{1}{\sqrt{1+\tan^{-2}\beta}}\frac{v_{\mathrm{SM}}}{\sqrt{2}}\,,
\end{equation}
where we have applied $\left\langle H_{t}\right\rangle =\alpha_{u}v_{u}$
as per Eq.~(\ref{eq:Higgs_matching}), with
\begin{equation}
v_{u}=\sin\beta\frac{v_{\mathrm{SM}}}{\sqrt{2}}=\frac{1}{\sqrt{1+\tan^{-2}\beta}}\frac{v_{\mathrm{SM}}}{\sqrt{2}}\,,
\end{equation}
as in usual 2HDM. If we consider $\tan\beta\approx10$ and $\alpha_{u}\sim\mathcal{O}(1)$,
then we obtain
\begin{equation}
m_{t}\approx s_{34}^{Q}y_{43}^{\psi}\frac{v_{\mathrm{SM}}}{\sqrt{2}}\equiv y_{t}\frac{v_{\mathrm{SM}}}{\sqrt{2}}\,.\label{eq:Top_Effective Yukawa}
\end{equation}
From the expression above, it is clear that very large mixing $s_{34}^{Q}\approx1$
is required in order to preserve a natural $y_{43}^{\psi}$, and to
avoid perturbativity issues in the top Yukawa. Moreover, we will see
that maximal values for $s_{34}^{Q}$ are also well motivated by the
$R_{D^{(*)}}$ anomaly, leading to a clear connection between $B$-physics
and the flavour puzzle only present in this model.
\item In the bullet point above, the effective top Yukawa coupling in the
Higgs basis has been estimated as $y_{t}\approx1$. By following the
same procedure, we can see that all fermion masses can be accommodated
with natural parameters. Remarkably, we obtain that all the effective
Yukawa couplings are SM-like in the Higgs basis, explaining the observed
pattern of SM Yukawa couplings at low energies.
\item The mixing between left-handed quark fields arise mainly from the
off-diagonal (2,3) entry in the quark mass matrices, which is controlled
by $s_{34}^{q^{c}}$. This mixing can be estimated for each sector
by the ratio of the (2,3) entry over the (3,3) entry, i.e.
\begin{equation}
\theta_{23}^{u}\approx\frac{s_{34}^{q^{c}}y_{24}^{\psi}\left\langle H_{c}\right\rangle }{s_{34}^{Q}y_{43}^{\psi}\left\langle H_{t}\right\rangle }\approx\frac{m_{c}}{m_{t}}\simeq\mathcal{O}(0.1V_{cb})\,,\qquad\theta_{23}^{d}\approx\frac{s_{34}^{q^{c}}y_{24}^{\psi}\left\langle H_{s}\right\rangle }{s_{34}^{Q}y_{43}^{\psi}\left\langle H_{b}\right\rangle }\approx\frac{m_{s}}{m_{b}}\simeq\mathcal{O}(V_{cb})\,,\label{eq:up_mixing_4thVL}
\end{equation}
obtained under the assumption $s_{34}^{q^{c}}\approx s_{24}^{q^{c}}$,
which is reasonable given that both are suppressed by the same scale
$M_{4}^{\psi^{c}}$. Therefore, the model predicts that $V_{cb}$
originates mainly from the down sector, while the mixing in
the up sector is small, suppressed by the larger mass hierarchy of the up sector. The specific
values of the mixing angles can be different if we relax $s_{24}^{q^{c}}\approx s_{34}^{q^{c}}$,
but the CKM remains down-dominated in any case.
\item The charged lepton sector follows a similar discussion as that of
the quark sector. If $s_{34}^{Q}\approx1$, then $s_{34}^{L}$ is
expected to be large as well and we obtain $\left\langle H_{\tau}\right\rangle \approx m_{\tau}$.
Under the assumption $s_{24}^{e^{c}}\approx s_{34}^{e^{c}}$, the
charged lepton mixing is predicted as
\begin{equation}
\theta_{23}^{e}\approx\frac{s_{34}^{e^{c}}y_{24}^{\psi}\left\langle H_{\mu}\right\rangle }{s_{34}^{L}y_{43}^{\psi}\left\langle H_{\tau}\right\rangle }\approx\frac{m_{\mu}}{m_{\tau}}\simeq\mathcal{O}(V_{cb})\,.
\end{equation}
A particularly interesting situation arises when $s_{34}^{e^{c}}>s_{24}^{e^{c}}$,
where a larger $\theta_{23}^{e}$ contributing to large atmospheric
neutrino mixing is obtained. In this scenario, interesting signals
in CLFV processes such as $\tau\rightarrow3\mu$ or $\tau\rightarrow\mu\gamma$
arise, mediated at tree-level by $SU(4)^{I}_{PS}$ gauge bosons. This is
obtained if $x_{43}^{\psi^{c}}>x_{42}^{\psi^{c}}$, without the need
of any tuning.
\item Unlike private Higgs models \cite{Porto:2007ed,Porto:2008hb,BenTov:2012cx,Rodejohann:2019izm}, the personal Higgs VEVs are not hierarchical,
all of order 1-10 GeV, with the exception of the top one whose VEV
is approximately that of the SM Higgs doublet, as discussed above.
The reason is that the fermion mass hierarchies arise from the hierarchies
$s_{34}^{\psi}\gg s_{24}^{\psi^{c}},s_{34}^{\psi^{c}}$, which find
their natural origin in the messenger dominance $M_{4}^{\psi}\ll M_{4}^{\psi^{c}}$
of Eq.~(\ref{eq:hierarchy_scalesVL}). The latter simultaneously
leads to dominantly left-handed leptoquark currents, which is an interesting
connection only present in our model.
\end{itemize}

\subsection{First family fermion masses and comments about neutrino masses} \label{subsec:firstFamily}

So far we have shown how the masses and mixing of second and third
family fermions arise in the twin PS model, via dynamics connected to the $SU(4)^{I}_{PS}$
group broken at the TeV scale. The first family remains massless so
far, protected by an accidental $U(1)$ symmetry. In this section we
introduce small breaking of such symmetry via dynamics
connected to the second PS group $SU(4)^{II}_{PS}$, broken
at a much higher scale $M_{\mathrm{High}}\gtrsim1\;\mathrm{PeV}$.
In this sense, our model is a multi-scale theory of flavour where the
different flavour hierarchies are explained by hierarchical NP scales,
in the spirit of other theories of flavour such as \cite{Panico:2016ull,Bordone:2017bld,Allwicher:2020esa,Fuentes-Martin:2020pww,Barbieri:2021wrc,Fuentes-Martin:2022xnb,Davighi:2022bqf,Davighi:2022fer,FernandezNavarro:2023rhv},
which are commonly connected with the $B$-anomalies as well. However,
all those alternative theories require the family decomposition of the
SM group to obtain an approximate $U(2)^{5}$ flavour symmetry (see
Section~\ref{subsec:U(2)5}). In contrast, our model achieves the
same goals without the need of family decomposing the SM nor claiming
$U(2)^{5}$, but rather via mixing between SM and vector-like fermions controlled by the mechanism of
messenger dominance \cite{Ferretti:2006df}, this
way providing an alternative and novel approach to connect the $B$-anomalies
with the origin of the flavour structure of the SM. 
\begin{table}[t]
\begin{centering}
\begin{tabular}{lccccccc}
\toprule 
Field & $SU(4)_{PS}^{I}$ & $SU(2)_{L}^{I}$ & $SU(2)_{R}^{I}$ & $SU(4)_{PS}^{II}$ & $SU(2)_{L}^{II}$ & $SU(2)_{R}^{II}$ & $\mathbb{Z}_{2}$\tabularnewline
\midrule
\midrule 
$\psi_{1,2,3}$ & $\mathbf{1}$ & $\mathbf{1}$ & $\mathbf{1}$ & $\mathbf{4}$ & $\mathbf{2}$ & $\mathbf{1}$ & (-), (+), (+)\tabularnewline
$\psi_{1,2,3}^{c}$ & $\mathbf{1}$ & $\mathbf{1}$ & $\mathbf{1}$ & $\mathbf{\overline{4}}$ & $\mathbf{1}$ & $\mathbf{\overline{2}}$ & (-), (+), (+)\tabularnewline
\midrule 
$\psi_{5}$ & $\mathbf{1}$ & $\mathbf{2}$ & $\mathbf{1}$ & $\mathbf{4}$ & $\mathbf{1}$ & $\mathbf{1}$ & (+)\tabularnewline
$\overline{\psi_{5}}$ & $\mathbf{1}$ & $\mathbf{\overline{2}}$ & $\mathbf{1}$ & $\mathbf{\overline{4}}$ & $\mathbf{1}$ & $\mathbf{1}$ & (+)\tabularnewline
$\psi_{5}^{c}$ & $\mathbf{1}$ & $\mathbf{2}$ & $\mathbf{1}$ & $\mathbf{\overline{4}}$ & $\mathbf{\overline{2}}$ & $\mathbf{\overline{2}}$ & (+)\tabularnewline
$\overline{\psi_{5}^{c}}$ & $\mathbf{1}$ & $\mathbf{\overline{2}}$ & $\mathbf{1}$ & $\mathbf{4}$ & $\mathbf{2}$ & $\mathbf{2}$ & (+)\tabularnewline
\midrule 
$h$ & $\mathbf{1}$ & $\mathbf{\overline{2}}$ & $\mathbf{1}$ & $\mathbf{1}$ & $\mathbf{1}$ & $\mathbf{2}$ & (-)\tabularnewline
\midrule 
$\Phi$ & $\mathbf{1}$ & $\mathbf{2}$ & $\mathbf{1}$ & $\mathbf{1}$ & $\mathbf{\overline{2}}$ & $\mathbf{1}$ & (+)\tabularnewline
$\overline{\Phi}$ & $\mathbf{1}$ & $\mathbf{1}$ & $\mathbf{\overline{2}}$ & $\mathbf{1}$ & $\mathbf{1}$ & $\mathbf{2}$ & (+)\tabularnewline
\midrule
$\xi_{15}$ & $\mathbf{1}$ & $\mathbf{1}$ & $\mathbf{1}$ & $\mathbf{15}$ & $\mathbf{1}$ & $\mathbf{1}$ & (+)\tabularnewline
\bottomrule
\end{tabular}
\par\end{centering}
\caption[Fields in the twin PS model participating in the origin of first family
fermion masses and mixing]{Fields participating in the origin of first family fermion masses
and mixing. \label{tab:Field_content_TwinPS-1}}
\end{table}

In order to explain first family masses, we add an extra family of
vector-like fermions that we call the ``fifth'' family, transforming
in the fundamental of $SU(4)^{II}_{PS}$ but being a singlet under $SU(4)^{I}_{PS}$.
We also introduce a non-standard Higgs field, as shown in Table~\ref{tab:Field_content_TwinPS-1}.
This choice ensures that the new vector-like family will mix with chiral fermions,
including the first family, without providing couplings of the first
family to the TeV scale $SU(4)^{I}_{PS}$ gauge bosons that are relevant
for $B$-physics.

We also introduce a discrete symmetry $\mathbb{Z}_{2}$ to
distinguish the first family and the new Higgs $h$, in order to achieve
a texture zero in the (1,1) entry of the effective Yukawa matrices.
Such a texture has been suggested to explain the empirical
relation $V_{us}\sim\sqrt{m_{d}/m_{s}}$ \cite{Gatto:1968ss}, delivering a Cabibbo angle that mostly
originates from mixing in the down sector. The
new terms in the renormalisable Lagrangian are
\begin{equation}
\mathcal{L}_{\mathrm{5}}^{ren}=y_{15}^{\psi}h\psi_{1}\psi_{5}^{c}+y_{51}^{\psi}h\psi_{5}\psi_{1}^{c}+x_{i5}^{\psi}\Phi\psi_{i}\overline{\psi_{5}}+x_{5i}^{\psi^{c}}\overline{\psi_{5}^{c}}\Phi\psi_{i}^{c}+M_{5}^{\psi}\psi_{5}\overline{\psi_{5}}+M_{5}^{\psi^{c}}\psi_{5}^{c}\overline{\psi_{5}^{c}}\,,\label{eq:Lren_4thVL-1-1}
\end{equation}
plus h.c., where $i=2,3$. Note that $\Phi$ gets a VEV spontaneously breaking
$SU(2)_{L}^{I}$ and $SU(2)^{II}_{L}$ down to their diagonal subgroup
at very high energies. We can arrange these couplings in matrix form
along with those of Eq.~(\ref{eq:MassMatrix_4thVL}), obtaining
\begin{equation}
M^{\psi}=\left(
\global\long\def\arraystretch{1.3}%
\begin{array}{@{}llcccccc@{}}
 & \multicolumn{1}{c@{}}{\psi_{1}^{c}} & \psi_{2}^{c} & \psi_{3}^{c} & \psi_{4}^{c} & \overline{\psi_{4}} & \psi_{5}^{c} & \overline{\psi_{5}}\\
\cmidrule(l){2-8}\left.\psi_{1}\right| & 0 & 0 & 0 & 0 & 0 & y_{15}^{\psi}h & 0\\
\left.\psi_{2}\right| & 0 & 0 & 0 & y_{24}^{\psi}\overline{H} & 0 & 0 & x_{25}^{\psi}\Phi\\
\left.\psi_{3}\right| & 0 & 0 & 0 & y_{34}^{\psi}\overline{H} & x_{34}^{\psi}\phi & 0 & x_{35}^{\psi}\Phi\\
\left.\psi_{4}\right| & 0 & 0 & y_{43}^{\psi}H & 0 & M_{4}^{\psi} & 0 & 0\\
\left.\overline{\psi_{4}^{c}}\right| & 0 & x_{42}^{\psi^{c}}\overline{\phi} & x_{43}^{\psi^{c}}\overline{\phi} & M_{4}^{\psi^{c}} & 0 & 0 & 0\\
\left.\psi_{5}\right| & y_{51}^{\psi}h & 0 & 0 & 0 & 0 & 0 & M_{5}^{\psi}\\
\left.\overline{\psi_{5}^{c}}\right| & 0 & x_{52}^{\psi^{c}}\Phi & x_{53}^{\psi^{c}}\Phi & 0 & 0 & M_{5}^{\psi^{c}} & 0
\end{array}\right)\,.
\end{equation}
Assuming $\left\langle h\right\rangle ,\left\langle \Phi\right\rangle \ll M_{5}^{\psi,\psi^{c}}$,
effective Yukawa couplings for the first family can be extracted in the mass insertion approximation as
\begin{equation}
\mathcal{L}_{eff}^{Yuk,3\times3}=\left(
\global\long\def\arraystretch{0.7}%
\begin{array}{@{}llcc@{}}
 & \multicolumn{1}{c@{}}{\psi_{1}^{c}} & \psi_{2}^{c} & \psi_{3}^{c}\\
\cmidrule(l){2-4}\left.\psi_{1}\right| & 0 & 0 & 0\\
\left.\psi_{2}\right| & x_{25}^{\psi}y_{51}^{\psi} & 0 & 0\\
\left.\psi_{3}\right| & x_{35}^{\psi}y_{51}^{\psi} & 0 & 0
\end{array}\right)\frac{\left\langle \Phi\right\rangle }{M_{5}^{\psi}}h+\left(
\global\long\def\arraystretch{0.7}%
\begin{array}{@{}llcc@{}}
 & \multicolumn{1}{c@{}}{\psi_{1}^{c}} & \psi_{2}^{c} & \psi_{3}^{c}\\
\cmidrule(l){2-4}\left.\psi_{1}\right| & 0 & y_{15}^{\psi}x_{52}^{\psi^{c}} & y_{15}^{\psi}x_{53}^{\psi^{c}}\\
\left.\psi_{2}\right| & 0 & 0 & 0\\
\left.\psi_{3}\right| & 0 & 0 & 0
\end{array}\right)\frac{\left\langle \Phi\right\rangle }{M_{5}^{\psi^{c}}}h+\mathrm{h.c.}\,,
\end{equation}
which needs to be added to the matrix in Eq.~(\ref{eq:MassMatrix_4thVL_effective-1-1})
in order to obtain the full set of effective Yukawa couplings.
One can see that such effective Yukawa couplings provide masses for the first family,
and their hierarchical smallness with respect to the second and third family fermions can be explained if we extend the messenger dominance in Eq.~(\ref{eq:hierarchy_scalesVL})
to include the fifth family,
\begin{equation}
\frac{\left\langle \Phi\right\rangle }{M_{5}^{\psi}},\frac{\left\langle \Phi\right\rangle }{M_{5}^{\psi^{c}}}\ll\frac{\langle\overline{\phi}\rangle }{M_{4}^{\psi^{c}}}\ll\frac{\left\langle \phi\right\rangle }{M_{4}^{\psi}}\lesssim1\,.\label{eq:hierarchy2}
\end{equation}
In this manner, the hierarchies of quark and charged lepton masses
in the SM Yukawa couplings are re-expressed as the hierarchy of scales
in Eq.~(\ref{eq:hierarchy2}). This is not just a reparameterisation
of the hierarchies, since it involves extra dynamics and testable
experimental predictions, such as the vector-like fermion spectrum with $M_{4}^{\psi}\sim1\;\mathrm{TeV}$
as motivated to explain the $R_{D^{(*)}}$ anomalies.

The Higgs $h$ decomposes at low energies as a type II 2HDM, where
$h_{u}\sim(\mathbf{1,\overline{2}},-1/2)$ and $h_{d}\sim(\mathbf{1,\overline{2}},1/2)$
get effective VEVs like the personal Higgs as $\langle h_{u} \rangle=\epsilon_{u}\langle H_{u} \rangle$ and
$\langle h_{d} \rangle=\epsilon_{d}\langle H_{d} \rangle$. These VEVs split the up-quark and down-quark masses, which would otherwise be degenerate
due to the twin PS symmetry. However, the down-quark mass and the
electron mass remain degenerate so far. They can be split simply
by introducing a Higgs transforming in the adjoint representation
of $SU(4)^{II}_{PS}$, $\xi_{15}\sim(\mathbf{15,1,1})_{II}$, that gets
a VEV to split the masses of the fifth family fermions. This allows
to write the couplings
\begin{equation}
\mathcal{L}_{\mathrm{5}}^{ren}\supset\lambda^{\psi}\psi_{5}\overline{\psi_{5}}\xi_{15}+\lambda^{\psi^{c}}\psi_{5}^{c}\overline{\psi_{5}^{c}}\xi_{15}+\mathrm{h.c.}
\end{equation}
These couplings result in quark-lepton mass splittings proportional
to the generator $T_{15}^{II}=\mathrm{diag}(1,1,1,-3)/(2\sqrt{6})$
leading to different contributions to the fifth family quark and
lepton masses,\textbf{
\begin{equation}
M_{5}^{Q}\equiv M_{5}^{\psi}+\frac{\lambda^{\psi}\left\langle \xi_{15}\right\rangle }{2\sqrt{6}}\,,\quad M_{5}^{L}\equiv M_{5}^{\psi}-3\frac{\lambda^{\psi}\left\langle \xi_{15}\right\rangle }{2\sqrt{6}}\,,
\end{equation}
\begin{equation}
M_{5}^{q^{c}}\equiv M_{5}^{\psi^{c}}+\frac{\lambda^{\psi^{c}}\left\langle \xi_{15}\right\rangle }{2\sqrt{6}}\,,\quad M_{5}^{e^{c}}\equiv M_{5}^{\psi^{c}}-3\frac{\lambda^{\psi^{c}}\left\langle \xi_{15}\right\rangle }{2\sqrt{6}}\,.
\end{equation}
}If the mass terms proportional to $\left\langle \xi_{15}\right\rangle $
dominate over the original mass terms, then they can be responsible
for the smallness of the electron mass compared to the down-quark
mass. Notice that the VEV $\left\langle \xi_{15}\right\rangle $ breaks
$SU(4)^{II}_{PS}$ at very high scales, providing a natural suppression
for the first family masses. We also mention that an alternative mechanism
was presented in \cite{King:2021jeo}, where the mass of the fifth
family is split via a non-renormalisable operator containing the
Higgs $H'$ and $\overline{H}'$, which then combine into an adjoint
of $SU(4)^{II}_{PS}$ as discussed in \cite{King:1994he} to give the desired
splitting proportional to $T_{15}^{II}$.

Finally we comment on the origin of neutrino masses in our model.
In principle, neutrinos get a Dirac mass matrix similar to that of
up-type quarks, hence predicting $m_{\nu_{\tau}}=m_{t}$ as usual
in Pati-Salam models, along with a hierarchical pattern of neutrino masses
with small mixing. However, the singlet neutrinos $\nu^{c}$ can get
a further Majorana mass matrix via non-renormalisable operators containing
the Higgs $H'$ and $\overline{H}'$ that break $SU(4)^{II}_{PS}$. By
means of adding a $\mathbb{Z}_{6}$ family symmetry broken by a Majoron scalar,
it was shown in \cite{King:2021jeo} that the tiny masses of active
neutrinos can be obtained via a type I seesaw mechanism featuring
single right-handed neutrino dominance \cite{King:1998jw,King:1999mb}.
One right-handed neutrino is much heavier than the others, getting
a mass at the heavy scale $\left\langle H'\right\rangle $ which has
to be close to the GUT scale in order to explain the tiny neutrino masses.
The hierarchies of Majorana masses for right-handed neutrinos then
cancel the hierarchies of the Dirac neutrino matrix, allowing to reproduce
the PMNS mixing matrix with large mixing angles. Note that this type of mechanism
are common in Pati-Salam models. However, we will see later that the explanation
of the $B$-anomalies pushes the twin PS model close to the boundary
of the perturbative domain, requiring that $SU(4)^{II}_{PS}$ is broken not
far above $1\;\mathrm{PeV}$. This seems in tension with $\left\langle H'\right\rangle \approx10^{16}\;\mathrm{GeV}$
as in the seesaw mechanism of \cite{King:2021jeo}, suggesting the
implementation of a low scale seesaw mechanism. Given that in our
model the origin of neutrino masses is independent of the low-energy
phenomenology explaining the $B$-anomalies, we leave a further discussion
about the origin of neutrino masses for the future. In a similar manner,
given that the origin of first family masses is related to dynamics
at the very heavy scale of $SU(4)^{II}_{PS}$ breaking, with negligible
impact for low-energy phenomenology, in the rest of this chapter we
will neglect first family fermion masses and focus on $B$-physics
phenomenology.

\subsection{\texorpdfstring{The low energy theory $G_{4321}$}{The low-energy theory G4321}}

In this section we shall discuss the $G_{4321}$ theory that breaks
down to the SM symmetry group at low energies $G_{4321}\rightarrow G_{\mathrm{SM}}$,
achieved via the scalars $\phi_{3}(\mathbf{4,\overline{3},1\oplus3},-1/6)$ and $\phi_{1}(\mathbf{4,1,1\oplus3},1/2)$
developing the VEVs
\begin{equation}
\left\langle \phi_{3}\right\rangle =\left(\begin{array}{ccc}
\frac{v_{3}}{\sqrt{2}} & 0 & 0\\
0 & \frac{v_{3}}{\sqrt{2}} & 0\\
0 & 0 & \frac{v_{3}}{\sqrt{2}}\\
0 & 0 & 0
\end{array}\right)\,,\quad\left\langle \phi_{1}\right\rangle =\left(\begin{array}{c}
0\\
0\\
0\\
\frac{v_{1}}{\sqrt{2}}
\end{array}\right)\,,
\end{equation}
where $v_{1},\,v_{3}\lesssim1\,\mathrm{TeV}$, and analogously for
$\overline{\phi_{3}}$ and $\overline{\phi_{1}}$ developing VEVs
$\overline{v}_{3}$ and $\overline{v}_{1}$, leading to the symmetry
breaking of $G_{4321}$ down to the SM gauge group,
\begin{equation}
SU(4)_{PS}^{I}\times SU(3)_{c}^{II}\times SU(2)_{L}^{I+II}\times U(1)_{Y'}\rightarrow SU(3)_{c}\times SU(2)_{L}\times U(1)_{Y}\,.\label{eq:4321_breaking}
\end{equation}
Here $SU(4)_{PS}^{I}$ is broken to $SU(3)_{c}^{I}\times U(1)_{B-L}^{I}$(at
the level of fermion representations, vector-like quarks and leptons are split
$\mathbf{4}^{I}\rightarrow(\mathbf{3},1/6)^{I}\oplus(\mathbf{1},-1/2)^{I}$),
with $SU(3)_{c}^{I}\times SU(3)_{c}^{II}$ further broken to the diagonal
subgroup $SU(3)_{c}^{I+II}$, identified as SM QCD. On the other hand,
$SU(2)_{L}^{I+II}$ remains as the SM $SU(2)_{L}$. The abelian generators
are broken to SM hypercharge $U(1)_{Y}$ given by $Y=T_{B-L}^{I}+Y'=T_{B-L}^{I}+T_{B-L}^{II}+T^{I+II}_{3R}$.
The physical massive scalar spectrum includes a real colour octet,
three SM singlets and a complex scalar transforming as $(\mathbf{3,1},2/3)$.
The heavy gauge boson spectrum includes a vector leptoquark $U_{1}^{\mu}\sim(\mathbf{3,1},2/3)$,
a colour octet $g'_{\mu}\sim(\mathbf{8,1},0)$ also identified as
coloron, and a $Z'_{\mu}\sim(\mathbf{1,1},0)$. The heavy gauge bosons
arise from the different steps of the symmetry breaking,
\begin{alignat}{2}
 & SU(4)_{PS}^{I}\rightarrow SU(3)_{c}^{I}\times U(1)_{B-L}^{I} & \quad & \Rightarrow U_{1}^{\mu}(\mathbf{3,1},2/3)\,,\\
 & SU(3)_{c}^{I}\times SU(3)_{c}^{II}\rightarrow SU(3)_{c}^{I+II} & \quad & \Rightarrow g'_{\mu}(\mathbf{8,1},0)\,,\\
 & U(1)_{B-L}^{I}\times U(1)_{Y'}\rightarrow U(1)_{Y} & \quad & \Rightarrow Z'_{\mu}(\mathbf{1,1},0)\,.
\end{alignat}
The gauge boson masses resulting from the symmetry breaking in Eq.~(\ref{eq:hierarchy_scalesVL})
are a generalisation of the results in \cite{Diaz:2017lit,DiLuzio:2017vat},
\begin{flalign}
 & M_{U_{1}}=\frac{1}{\sqrt{2}}g_{4}\sqrt{v_{1}^{2}+v_{3}^{2}}\,,\nonumber \\
 & M_{g'}=\sqrt{g_{4}^{2}+g_{3}^{2}}v_{3}\,,\label{eq:MU1}\\
 & M_{Z'}=\frac{\sqrt{3}}{2}\sqrt{g_{4}^{2}+\frac{2}{3}g_{1}^{2}}\sqrt{v_{1}^{2}+\frac{1}{3}v_{3}^{2}}\,,\nonumber 
\end{flalign}
where we have assumed $\overline{v}_{3}\approx v_{3}$ and $\overline{v}_{1}\approx v_{1}$
for simplicity. The mass of the coloron depends only on $v_{3}$,
and the scenario $v_{3}\gg v_{1}$ leads to the approximate relation
$M_{g'}\approx\sqrt{2}M_{U_{1}}$. This way the coloron, which suffers from
stronger high-$p_{T}$ constraints, can be slightly
heavier than the vector leptoquark, as the latter is the only one required to be light in order
to explain the $R_{D^{*}}$ anomalies.

In the original interaction basis, the heavy gauge bosons couple to left-handed
fermions (including VL left-handed fermions) via the interactions\footnote{Notice that we continue working in the 2-component notation introduced
in Appendix~\ref{app:2-component_notation}}
\begin{equation}
\frac{g_{4}}{\sqrt{2}}\left(Q_{4}^{\dagger}\bar{\sigma}^{\mu}L_{4}+\mathrm{h.c.}\right)U_{1\mu}+\mathrm{h.c.}\,,\label{eq:U1couplings_4th}
\end{equation}
\begin{equation}
\frac{g_{4}g_{s}}{g_{3}}\left(Q_{4}^{\dagger}\bar{\sigma}^{\mu}T^{a}Q_{4}-\frac{g_{3}^{2}}{g_{4}^{2}}Q_{i}^{\dagger}\bar{\sigma}^{\mu}T^{a}Q_{i}\right)g'{}_{\mu}^{a}\,,\label{eq:ColoronCouplings_4th}
\end{equation}
\begin{equation}
\frac{\sqrt{3}}{\sqrt{2}}\frac{g_{4}g_{Y}}{g_{1}}\left(\frac{1}{6}Q_{4}^{\dagger}\bar{\sigma}^{\mu}Q_{4}-\frac{1}{2}L_{4}^{\dagger}\bar{\sigma}^{\mu}L_{4}-\frac{g_{1}^{2}}{9g_{4}^{2}}Q_{i}^{\dagger}\bar{\sigma}^{\mu}Q_{i}+\frac{g_{1}^{2}}{3g_{4}^{2}}L_{i}^{\dagger}\bar{\sigma}^{\mu}L_{i}\right)Z'_{\mu}\,.\label{eq:ZprimeCouplings_4th}
\end{equation}
Similar couplings are obtained for right-handed fermions, although
the small mixing between right-handed fermions and VL fermions heavily
suppresses the couplings of $U_{1}$ to right-handed fermions. This
is not a trivial result, as these mixing angles are small because
they are connected to the origin of second family fermion masses (see
Section~\ref{subsec:Effective-Yukawa-couplings}). Therefore, the
couplings of $U_{1}$ to right-handed fermions can be safely neglected.
This way, the $U_{1}$ couplings will be dominantly left-handed, which
can alleviate the stringent bounds from high-$p_{T}$ searches, in contrast
to the alternative models in the literature based on the family decomposition
of the SM \cite{Bordone:2017bld,Allwicher:2020esa,Fuentes-Martin:2020pww,Fuentes-Martin:2022xnb,Davighi:2022bqf}.
Similar couplings are obtained for the VL partners in the conjugate
representations, however those couplings are irrelevant for the phenomenology
since the conjugate partners do not mix with the SM fermions. The
couplings of the coloron and $Z'$ to chiral fermions remain flavour
universal, in contrast to the left-handed sector, hence providing
no meaningful phenomenology for flavour processes\footnote{Although these couplings can be relevant for direct production at
high-$p_{T}$ of the coloron and $Z'$.}.

The gauge couplings of $SU(3)_{c}$ and $U(1)_{Y}$ are given by
\begin{equation}
g_{s}=\frac{g_{4}g_{3}}{\sqrt{g_{4}^{2}+g_{3}^{2}}}\,,\qquad\qquad g_{Y}=\frac{g_{4}g_{1}}{\sqrt{g_{4}^{2}+\frac{2}{3}g_{1}^{2}}}\,,\label{eq:SM_gauge_couplings}
\end{equation}
where $g_{4,3,2,1}$ are the gauge couplings of $G_{4321}$. The scenario
$g_{4}\gg g_{3,2,1}$ is well motivated from the phenomenological point
of view, since here the flavour universal couplings of light fermions
to the heavy $Z'$ and $g'$ are suppressed by the ratios $g_{1}/g_{4}$
and $g_{3}/g_{4}$, which will inhibit the direct production of these
states at the LHC. In this scenario, the relations above yield the
simple expressions $g_{s}\approx g_{3}$ and $g_{Y}\approx g_{1}$
for the SM gauge couplings.
\begin{figure}[t]
\begin{centering}
\begin{tikzpicture}
	\begin{feynman}
		\vertex (a) {\(Q_{3}\)};
		\vertex [right=13mm of a] (b);
		\vertex [right=11mm of b] (c) [label={ [xshift=0.1cm, yshift=0.1cm] \small $M^{Q}_{4}$}];
		\vertex [right=11mm of c] (d);
		\vertex [right=11mm of d] (e) [label={ [xshift=0.1cm, yshift=0.1cm] \small $M^{L}_{4}$}];
		\vertex [right=11mm of e] (f);
		\vertex [right=11mm of f] (g) {\(L_{3}\)};
		\vertex [above=14mm of b] (f1) {\(\phi_{3}\)};
		\vertex [above=14mm of d] (f2) {\(U_{1}\)};
		\vertex [above=14mm of f] (f3) {\(\phi_{1}\)};
		\diagram* {
			(a) -- [fermion] (b) -- [charged scalar] (f1),
			(b) -- [edge label'=\(\overline{Q}_{4}\)] (c),
			(c) -- [edge label'=\(Q_{4}\), inner sep=6pt, insertion=0] (d) -- [boson, blue] (f2),
			(d) -- [edge label'=\(L_{4}\), inner sep=6pt] (e),
			(e) -- [edge label'=\(\overline{L}_{4}\), insertion=0] (f) -- [charged scalar] (f3),
			(f) -- [fermion] (g),
	};
	\end{feynman}
\end{tikzpicture}
\par\end{centering}
\caption[Diagram in the twin PS model which leads to the effective $U_{1}$
couplings in the mass insertion approximation]{Diagram in the model which leads to the effective $U_{1}$ couplings
in the mass insertion approximation. \label{fig:U1_couplings_mass_insertion}}
\end{figure}

A key feature of the gauge boson couplings in Eqs.~(\ref{eq:U1couplings_4th}-\ref{eq:ZprimeCouplings_4th})
is that, while the coloron $g'_{\mu}$ and the $Z'_{\mu}$ couple
to all chiral and vector-like quarks and leptons, the vector leptoquark $U_{1}^{\mu}$
only couples to the fourth family vector-like fermions. However, the couplings
in Eqs.~(\ref{eq:U1couplings_4th}-\ref{eq:ZprimeCouplings_4th})
are written in the original interaction basis. We shall perform the transformation
to the decoupling basis (primed) as per Eq.~(\ref{eq:34mixing}),
\begin{equation}
\mathcal{L}_{U_{1}}^{\mathrm{gauge}}=\frac{g_{4}}{\sqrt{2}}Q'{}_{\alpha}^{\dagger}V_{34}^{Q}\bar{\sigma}^{\mu}\mathrm{diag}\left(0,0,0,1\right)V_{34}^{L\dagger}L'_{\beta}U_{1\mu}+\mathrm{h.c.}\,,
\end{equation}
where $\alpha,\beta=1,..,4$ and the indices of the matrices are implicit.
We obtain an effective coupling for the third family due to mixing
with the fourth family,
\begin{equation}
\mathcal{L}_{U_{1}}^{\mathrm{gauge}}=\frac{g_{4}}{\sqrt{2}}Q'{}_{i}^{\dagger}\bar{\sigma}^{\mu}\mathrm{diag}(0,0,s_{34}^{Q}s_{34}^{L})L'_{j}U_{1\mu}+\mathrm{h.c.}\,,\label{eq:U1_couplings_decoupling_4th}
\end{equation}
where we have omitted the fourth column and row for simplicity. The
diagrams in Fig.~\ref{fig:U1_couplings_mass_insertion} are illustrative,
however it must be remembered that the mass insertion approximation
is not accurate here due to the heavy top mass, instead we have to
work in the large mixing angle formalism as presented in Section~\ref{subsec:Effective-Yukawa-couplings}.
In principle, the couplings in Eq.~(\ref{eq:U1_couplings_decoupling_4th})
can simultaneously contribute to both LFU ratios $R_{K^{(*)}}$ and
$R_{D^{(*)}}$ once the further 2-3 transformations required to diagonalise
the quark and lepton mass matrices are taken into account. At low
energies, these transformations provide different effective $U_{1}$
couplings for the different components of the $SU(2)_{L}$ doublets\footnote{Notice that first family fermions get couplings to $U_{1}$ as well
via powers of small CKM-like mixing angles, such that they are suppressed enough to be neglected for the phenomenology. },
\begin{flalign}
\mathcal{L}_{U_{1}}^{\mathrm{gauge}} & =\frac{g_{4}}{\sqrt{2}}\hat{u}_{i}^{\dagger}\bar{\sigma}^{\mu}\left(\begin{array}{ccc}
0 & 0 & 0\\
0 & 0 & s_{34}^{Q}s_{34}^{L}s_{23}^{u}\\
0 & 0 & s_{34}^{Q}s_{34}^{L}c_{23}^{u}
\end{array}\right)\hat{\nu}{}_{j}U_{1\mu}\label{eq:U1couplings_u_4th}\\
 & +\frac{g_{4}}{\sqrt{2}}\hat{d}_{i}^{\dagger}\bar{\sigma}^{\mu}\left(\begin{array}{ccc}
0 & 0 & 0\\
0 & s_{34}^{Q}s_{34}^{L}s_{23}^{d}s_{23}^{e} & s_{34}^{Q}s_{34}^{L}s_{23}^{d}c_{23}^{e}\\
0 & s_{34}^{Q}s_{34}^{L}c_{23}^{d}s_{23}^{e} & s_{34}^{Q}s_{34}^{L}c_{23}^{d}c_{23}^{e}
\end{array}\right)\hat{e}_{j}U_{1\mu}+\mathrm{h.c.}\nonumber 
\end{flalign}

It is clear now that the effective leptoquark
couplings that contribute to LFU ratios arise due to the same mixing
effects which diagonalise the mass matrices of the model, yielding
mass terms for the SM fermions. Therefore, the flavour puzzle and
the $B$-physics anomalies are dynamically and parametrically connected
in this model, leading to a predictive framework.

Following the same methodology, we obtain the coloron and $Z'$ couplings
in the basis of mass eigenstates,
\begin{flalign}
 & \mathcal{L}_{g'}^{\mathrm{gauge}}=\frac{g_{4}g_{s}}{g_{3}}\hat{d}_{i}^{\dagger}\bar{\sigma}^{\mu}T^{a}\label{eq:coloron_d_couplings_4thVL}\\
 & \times\left(\begin{array}{ccc}
-\frac{g_{3}^{2}}{g_{4}^{2}} & 0 & 0\\
0 & -\left(c_{23}^{d}\right)^{2}\frac{g_{3}^{2}}{g_{4}^{2}}+\left(s_{34}^{Q}s_{23}^{d}\right)^{2} & \left(s_{34}^{Q}\right)^{2}s_{23}^{d}c_{23}^{d}\\
0 & \left(s_{34}^{Q}\right)^{2}s_{23}^{d}c_{23}^{d} & \left(s_{34}^{Q}c_{23}^{d}\right)^{2}-\left(c_{34}^{Q}c_{23}^{d}\right)^{2}\frac{g_{3}^{2}}{g_{4}^{2}}
\end{array}\right)\hat{d}_{j}g_{\mu}^{a'}+\left(d\rightarrow u\right)\,,\nonumber 
\end{flalign}
\begin{flalign}
 & \mathcal{L}_{Z',q}^{\mathrm{gauge}}=\frac{\sqrt{3}}{\sqrt{2}}\frac{g_{4}g_{Y}}{g_{1}}\hat{d}_{i}^{\dagger}\bar{\sigma}^{\mu}\label{eq:Z'_quark_couplings_4thVL}\\
 & \times\left(\begin{array}{ccc}
-\frac{g_{1}^{2}}{9g_{4}^{2}} & 0 & 0\\
0 & -\left(c_{23}^{d}\right)^{2}\frac{g_{1}^{2}}{9g_{4}^{2}}+\frac{1}{6}\left(s_{34}^{Q}s_{23}^{d}\right)^{2} & \frac{1}{6}\left(s_{34}^{Q}\right)^{2}s_{23}^{d}c_{23}^{d}\\
0 & \frac{1}{6}\left(s_{34}^{Q}\right)^{2}s_{23}^{d}c_{23}^{d} & \frac{1}{6}\left(s_{34}^{Q}c_{23}^{d}\right)^{2}-\left(c_{34}^{Q}c_{23}^{d}\right)^{2}\frac{g_{1}^{2}}{9g_{4}^{2}}
\end{array}\right)\hat{d}_{j}Z'_{\mu}+\left(d\rightarrow u\right)\,,\nonumber 
\end{flalign}
\begin{flalign}
 & \mathcal{L}_{Z',e}^{\mathrm{gauge}}=\frac{\sqrt{3}}{\sqrt{2}}\frac{g_{4}g_{Y}}{g_{1}}\hat{e}_{i}^{\dagger}\bar{\sigma}^{\mu}\label{eq:Z'_leptons_couplings_4thVL}\\
 & \times\left(\begin{array}{ccc}
\frac{g_{1}^{2}}{3g_{4}^{2}} & 0 & 0\\
0 & \left(c_{23}^{e}\right)^{2}\frac{g_{1}^{2}}{3g_{4}^{2}}-\frac{1}{2}\left(s_{34}^{L}s_{23}^{e}\right)^{2} & -\frac{1}{2}\left(s_{34}^{L}\right)^{2}s_{23}^{e}c_{23}^{e}\\
0 & -\frac{1}{2}\left(s_{34}^{L}\right)^{2}s_{23}^{e}c_{23}^{e} & -\frac{1}{2}\left(s_{34}^{L}c_{23}^{e}\right)^{2}+\left(c_{34}^{L}c_{23}^{e}\right)^{2}\frac{g_{1}^{2}}{3g_{4}^{2}}
\end{array}\right)\hat{e}_{j}Z'_{\mu}+\left(e\rightarrow\nu\right)\,.\nonumber 
\end{flalign}

The flavour-violating couplings of $U_{1}$ in Eq.~(\ref{eq:U1couplings_u_4th})
are all proportional to mixing between chiral fermions. In principle,
such mixing is of order $V_{cb}$ in the down sector, and of order $0.1V_{cb}$
in the up sector (see the discussion in Section~\ref{subsec:Effective-Yukawa-couplings}).
The small mixing in the up sector leads to a small $U_{1}$ 2-3 coupling,
possibly too small for $R_{D^{(*)}}$, however a deeper analysis is
required as we shall see in the next few sections. Moreover, flavour-violating
couplings involving the coloron and $Z'$ could be sizable in the
2-3 down sector, since the CKM mostly originates from
the down sector in this model. We shall study whether this is compatible
or not with the stringent constraints coming from $B_{s}-\bar{B}_{s}$
meson mixing observables.

\section{Matching the twin Pati-Salam model to the SMEFT\label{subsec:EFT_model_Appendix}}

In order to systematically study the low-energy phenomenology of the
twin Pati-Salam model, here we include the set of 4-fermion operators
obtained at tree-level after integrating out the heavy $U_{1}$, $Z'$
and $g'$,
\begin{align}
\mathcal{L}_{\text{4-\ensuremath{\mathrm{fermion}}}}= & -\frac{2}{v_{\mathrm{SM}}^{2}}\left[\left[C_{lq}^{(1)}\right]^{\alpha\beta ij}\left[Q_{lq}^{(1)}\right]^{\alpha\beta ij}+\left[C_{lq}^{(3)}\right]^{\alpha\beta ij}\left[Q_{lq}^{(3)}\right]^{\alpha\beta ij}\right.\nonumber \\
{} & \left.+\left[C_{qq}^{(1)}\right]^{ijkl}\left[Q_{qq}^{(1)}\right]^{ijkl}+\left[C_{qq}^{(3)}\right]^{ijkl}\left[Q_{qq}^{(3)}\right]^{ijkl}+\left[C_{ll}\right]^{\alpha\beta\delta\lambda}\left[Q_{ll}\right]^{\alpha\beta\delta\lambda}\right]\,,\label{eq:SMEFT_4fermion}
\end{align}
where we have chosen latin indices for quark flavours and greek indices
for lepton flavours. The SMEFT operators above can be matched to the
LEFT via the formalism introduced in Appendix~\ref{app:EFT_Matching},
and then they enter directly in the expressions for low-energy observables
introduced in Chapter~\ref{chap:2}. In our model, the
Wilson coefficients are given by 
\begin{align}
{} & \left[C_{lq}^{(1)}\right]^{\alpha\beta ij}=\frac{1}{2}C_{U}\beta_{i\alpha}\beta_{j\beta}^{*}-2C_{Z'}\xi_{ij}\xi_{\alpha\beta}\,, & {} & \left[C_{lq}^{(3)}\right]^{\alpha\beta ij}=\frac{1}{2}C_{U}\beta_{i\alpha}\beta_{j\beta}^{*}\,,\\
{} & \left[C_{qq}^{(1)}\right]^{ijkl}=\frac{1}{4}C_{g'}\kappa_{il}\kappa_{jk}-\frac{1}{6}C_{g'}\kappa_{ij}\kappa_{kl}+C_{Z'}\xi_{ij}\xi_{kl}\,, & {} & \left[C_{qq}^{(3)}\right]^{ijkl}=\frac{1}{4}C_{g'}\kappa_{il}\kappa_{jk}\,,\\
{} & \left[C_{ll}\right]^{\alpha\beta\delta\lambda}=C_{Z'}\xi_{\alpha\beta}\xi_{\delta\lambda}\,, & {} & {}
\end{align}
where we have defined
\begin{equation}
C_{U}=\frac{g_{4}^{2}v_{\mathrm{SM}}^{2}}{4M_{U_{1}}^{2}}\,,\qquad C_{g'}=\frac{g_{4}^{2}g_{s}^{2}}{2g_{3}^{2}}\frac{v_{\mathrm{SM}}^{2}}{M_{g'}^{2}}\,,\qquad C_{Z'}=\frac{3g_{4}^{2}g_{Y}^{2}}{8g_{1}^{2}}\frac{v_{\mathrm{SM}}^{2}}{M_{Z'}^{2}}\,.
\end{equation}

We consider all the fields in (\ref{eq:SMEFT_4fermion}) to be mass
eigenstates, as the effects of fermion mixing are encoded into the
$U_{1}$ ($\beta_{i\alpha}$), $g'$ ($\kappa_{ij}$) and $Z'$ ($\xi_{ij}$,\,$\xi_{\alpha\beta}$)
couplings given in Eqs.~(\ref{eq:U1couplings_u_4th}-\ref{eq:Z'_leptons_couplings_4thVL})
for the simplified model. For the extended model of Section~\ref{sec:Twin-Pati-Salam-Theory_3VL},
the same set of SMEFT operators applies, just changing the expressions
for the couplings by those of Eqs.~\eqref{eq:LQ_couplings}, \eqref{eq:Coloron_couplings_3VL}, (\ref{eq:Z'_couplings_3VL_q}-\ref{eq:Z'_couplings_3VL_e}).

\section{Phenomenology of the simplified model} \label{subsec:pheno_simplified}

\subsection{\texorpdfstring{$R_{K^{(*)}}$ and $R_{D^{(*)}}$}{RD(*) and RK(*)}\label{subsec:RK_RD}}

New contributions to the $R_{D^{(*)}}$ and $R_{K^{(*)}}$ ratios
arise in our model via tree-level contributions mediated by the $U_{1}$
vector leptoquark as in Fig.~\ref{fig:Leptoquark_RK_RD}, see also the SMEFT matching of the model in the previous section,
along with the expressions for $R_{D^{(*)}}$ and $R_{K^{(*)}}$ given in
Chapter~\ref{chap:2}. After integrating out $U_{1}$,
we observe the following scaling in terms of mixing angles of the model,
\begin{figure}[t]
\noindent \begin{centering}
\subfloat[\label{fig:Leptoquark_RK}]{\noindent \begin{centering}
\begin{tikzpicture}
	\begin{feynman}
		\vertex (a) {\(b_{L}\)};
		\vertex [below right=18mm of a] (b)  [label={ [yshift=0.1cm] \small $\beta_{b\mu}^{*}$}];
		\vertex [above right=of b] (c) {\(\mu_{L}\)};
		\vertex [below=of b] (f1) [label={ [xshift=0.02cm,yshift=-0.7cm] \small $\beta_{s\mu}$}];
		\vertex [below left=of f1] (f2) {\(s_{L}\)};
		\vertex [below right=of f1] (f3) {\(\mu_{L}\)};
		\diagram* {
			(a) -- [fermion] (b) -- [fermion] (c),
			(b) -- [boson, blue, edge label'=\(U_{1}\)] (f1),
			(f1) -- [fermion] (f2),
			(f1) -- [anti fermion] (f3),
	};
	\end{feynman}
\end{tikzpicture}
\par\end{centering}
}$\qquad$$\qquad$$\qquad$$\qquad$\subfloat[\label{fig:Leptoquark_RD}]{\noindent \begin{centering}
\begin{tikzpicture}
	\begin{feynman}
		\vertex (a) {\(b_{L}\)};
		\vertex [below right=18mm of a] (b)  [label={ [yshift=0.1cm] \small $\beta_{b\tau}^{*}$}];
		\vertex [above right=of b] (c) {\(\tau_{L}\)};
		\vertex [below=of b] (f1) [label={ [xshift=0.08cm,yshift=-0.7cm] \small $\beta_{c\nu_{\tau}}$}];
		\vertex [below left=of f1] (f2) {\(c_{L}\)};
		\vertex [below right=of f1] (f3) {\(\nu_{\tau L}\)};
		\diagram* {
			(a) -- [fermion] (b) -- [fermion] (c),
			(b) -- [boson, blue, edge label'=\(U_{1}\)] (f1),
			(f1) -- [fermion] (f2),
			(f1) -- [anti fermion] (f3),
	};
	\end{feynman}
\end{tikzpicture}
\par\end{centering}
}
\par\end{centering}
\caption[$U_{1}$-mediated tree-level diagrams contributing to $R_{D^{(*)}}$ and $R_{K^{(*)}}$]{$U_{1}$-mediated tree-level diagrams contributing to $b\rightarrow s\mu\mu$
(left panel) and $b\rightarrow c\tau\nu$ (right
panel). \label{fig:Leptoquark_RK_RD}}
\end{figure}
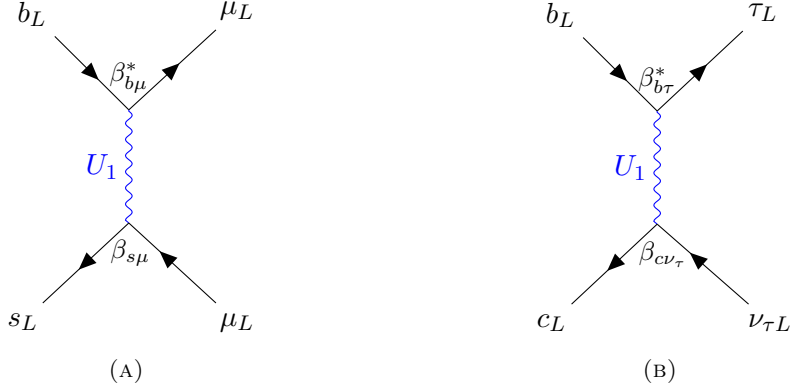
\begin{equation}
\left|\Delta R_{D^{(*)}}\right|\propto\left(s_{34}^{L}s_{34}^{Q}\right)^{2}s_{23}^{u}c_{23}^{u}\,,\label{eq:RD_U1}
\end{equation}
\begin{equation}
\left|\Delta R_{K^{(*)}}\right|\propto\left(s_{34}^{L}s_{34}^{Q}\right)^{2}\left(s_{23}^{e}\right)^{2}s_{23}^{d}c_{23}^{d}\,.\label{eq:RK_U1}
\end{equation}
From Eq.~(\ref{eq:RD_U1}) it can be seen that our contribution to
$R_{D^{(*)}}$ is proportional to the mixing angle $\theta_{23}^{u}$.
Such angle is naturally small in our model, roughly $\mathcal{O}(0.1V_{cb})$
as per Eq.~(\ref{eq:up_mixing_4thVL}), due to the fact that the
CKM originates mostly from the down sector. As a consequence, the
contribution to $R_{D^{(*)}}$ is suppressed. On the other hand, the
contribution of $U_{1}$ to $R_{K^{(*)}}$ is further suppressed by
the $\mathcal{O}(V_{cb})$ mixing angles $\theta_{23}^{d}$ and $\theta_{23}^{e}$,
for a total expected suppression of $\mathcal{O}(V_{cb}^{3})$.

\subsection{\texorpdfstring{$B_{s}-\bar{B}_{s}$ mixing}{Bs-Bsbar mixing}}\label{subsec:Bs_Mixing}

Flavour-violating couplings involving the coloron and $Z'$ could
be sizable in the 2-3 down sector, since the CKM originates mostly
from the down sector in this model. The general description
of $B_{s}-\bar{B}_{s}$ mixing in terms of effective operators is
included in Section~\ref{subsec:BsMixing}. Given that the flavour-violating
couplings in our model are dominantly left-handed, after integrating out the
heavy gauge bosons we obtain effective contributions to $C_{1}^{bs}$ only.

The bounds are highly constraining over this model because both the
coloron and $Z'$ mediate tree-level contributions to $\Delta M_{s}$,
which interfere positively with the SM prediction, while the latter
is already larger than the experimental result. We estimate that,
in order to satisfy the bound $\Delta M_{s}^{\mathrm{NP}}/\Delta M_{s}^{\mathrm{SM}}<0.11$ (95\% CL),
the 2-3 down-quark mixing needs to satisfy $\left|s_{23}^{d}\right|\apprle0.1V_{cb}$
if the 3-4 mixing is maximal $s_{34}^{Q}\approx1$.

\subsection{Parameter space of the simplified model}

\begin{figure}[t]
\subfloat[\label{fig:sQ34_sd_plane}]{\includegraphics[scale=0.44]{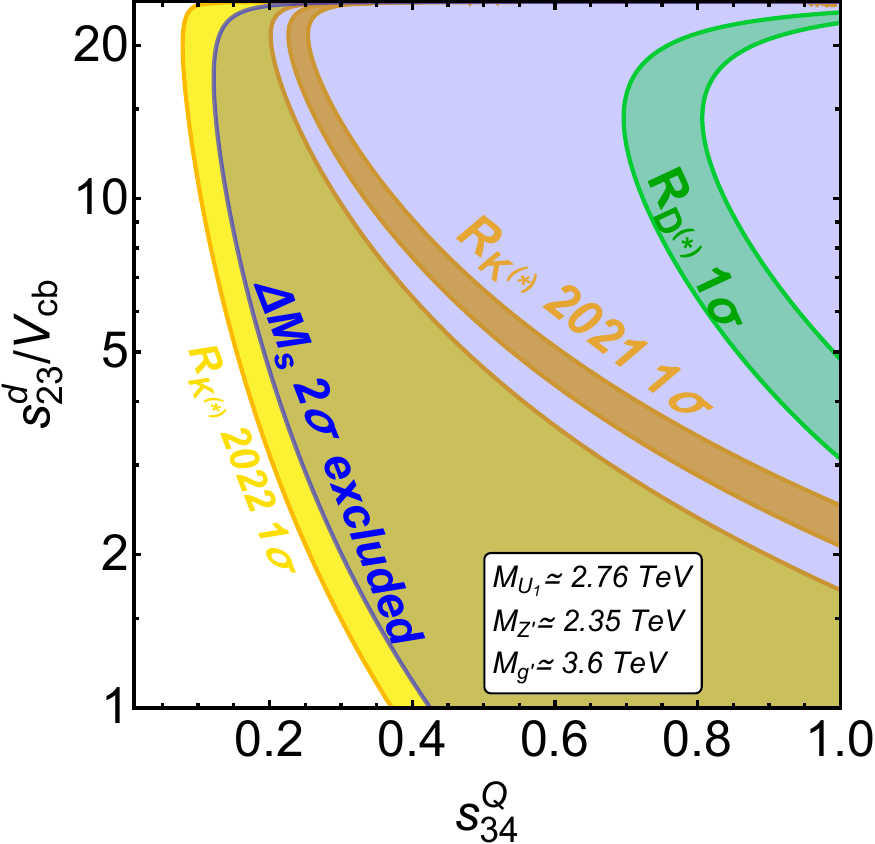}

}$\qquad$\subfloat[\label{fig:v3_v1}]{\includegraphics[scale=0.45]{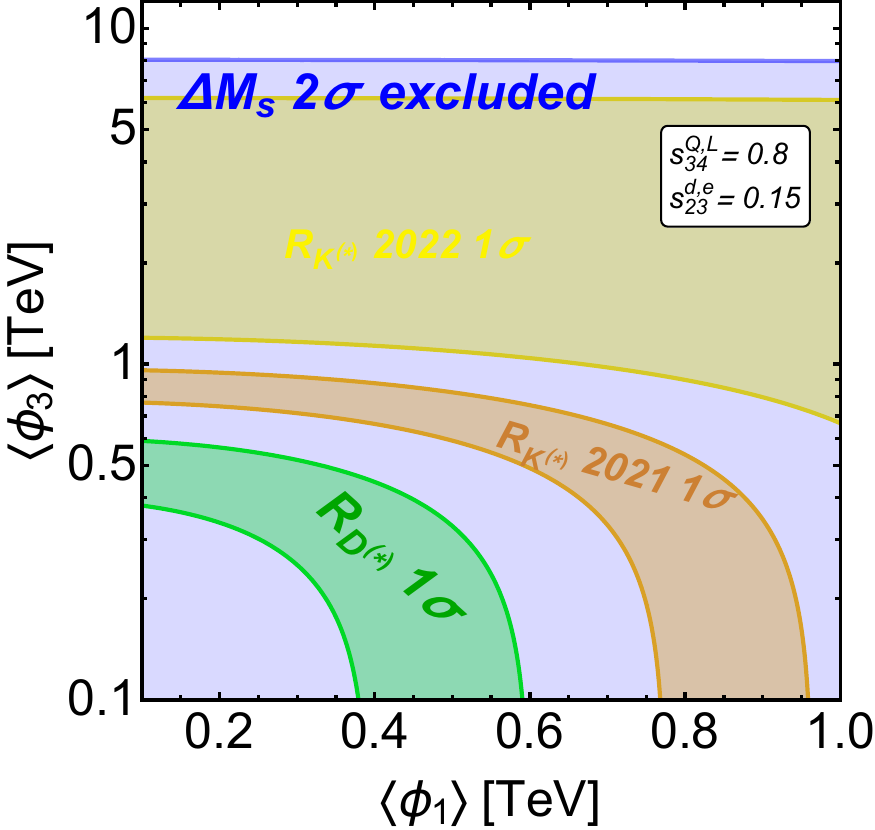}

}

\caption[Parameter space of the simplified twin PS model]{\textbf{\textit{Left:}} Regions compatible with $R_{D^{(*)}}$ and
$R_{K^{(*)}}$ (2022 and 2021 data) in the plane ($s_{34}^{Q}$, $s_{23}^{d}$),
the heavy gauge boson masses are fixed as depicted in the panel. \textbf{\textit{Right:}}
Regions compatible with $R_{D^{(*)}}$ and $R_{K^{(*)}}$ (2022
and 2021 data) in the plane ($\left\langle \phi_{1}\right\rangle $,$\left\langle \phi_{3}\right\rangle $),
which allows to explore the spectrum of heavy gauge boson masses.
The mixing angles are fixed as depicted in the plot. In both panels
the blue regions are excluded by the $\Delta M_{s}$ bound, see Eq.~(\ref{eq:bound_Bsmixing}).}
\end{figure}

As anticipated in the previous sections, the contribution of the vector leptoquark
to the $R_{D^{(*)}}$ anomalies is strongly suppressed by a naturally
small mixing angle $\theta_{23}^{u}\approx m_{c}/m_{t}$, leading
to a suppression of $\mathcal{O}(0.1V_{cb})$. In Fig.~\ref{fig:sQ34_sd_plane}
it can be seen that for a typical benchmark mass $M_{U_{1}}=3\,\mathrm{TeV}$,
a large $s_{23}^{u}\apprge4V_{cb}$ is needed in order to give a significant
contribution to the $R_{D^{(*)}}$ anomaly, provided that the 3-4
mixing is maximal.

The contribution to $R_{K^{(*)}}$ also suffers from an overall suppression
of $\mathcal{O}(V_{cb}^{3})$. We can go beyond the natural value
of $\theta_{23}^{u}$ by increasing the mixing angle $s_{34}^{q_{c}}$
(i.e.~increasing the fundamental Yukawa $x_{34}^{\psi^{c}}$, or
reducing the VL mass $M_{4}^{\psi^{c}}$), which controls the overall
size of the off-diagonal (2,3) entry in the effective mass matrices
of Eqs.~(\ref{eq:MassMatrix_4thVL_effective_up}) and ~(\ref{eq:MassMatrix_4thVL_effective_down}).
This way, we can explore the parameter space of larger 2-3 mixing
angles, provided that the experimental value of $V_{cb}$ is preserved
through Eq.~(\ref{eq:Vcb_model}), which relates both quark mixings
$\theta_{23}^{u}$ and $\theta_{23}^{d}$. We further assume $s_{34}^{Q}=s_{34}^{L}$
and $s_{23}^{d}=s_{23}^{e}$ for simplicity, both assumptions
are well motivated due to the underlying twin Pati-Salam symmetry.

Our results are depicted in Fig.~\ref{fig:sQ34_sd_plane} for a spectrum
of heavy gauge boson masses compatible with high-$p_{T}$ searches
(see Section~\ref{subsec:Colliders}). We find that for the given
benchmark, a small region of the parameter space is compatible with
the 2022 data of $R_{K^{(*)}}$, however the 1$\sigma$ region of
$R_{D^{(*)}}$ is not compatible with $\Delta M_{s}$. This version
of the model was already unable to explain the 2021 data of both LFU
anomalies due to the large constraints from tree-level $Z'$ and coloron
contributions to $\Delta M_{s}$.

In Fig.~\ref{fig:v3_v1} we have varied the VEVs of $\phi_{3}$ and
$\phi_{1}$, effectively exploring the parameter space of gauge boson masses,
in line with Eq.~(\ref{eq:MU1}). However, we find that the stringent
constraints from $\Delta M_{s}$ are only alleviated when $\left\langle \phi_{3}\right\rangle \apprge8\,\mathrm{TeV}$,
which corresponds to a coloron with mass $M_{g'}\gtrsim50\,\mathrm{TeV}$
and a vector leptoquark with mass $M_{U_{1}}\apprge34\,\mathrm{TeV}$,
too heavy to address $R_{D^{(*)}}$.

We conclude that the model in this simplified version is over-constrained
by large tree-level contributions to $\Delta M_{s}$ mediated by the
coloron and $Z'$. Such FCNCs arise due to the 2-3 CKM mixing having
its origin in the down sector. Moreover, the same small 2-3 mixing
angles suppress the contribution of the model to $R_{D^{(*)}}$. However,
we shall show that the proper flavour structure to be compatible with
all data is achieved in the extended version of the model presented
in Section~\ref{sec:Twin-Pati-Salam-Theory_3VL}, which is successful to address
the $B$-anomalies.

\section{Extending the simplified twin Pati-Salam theory of flavour\label{sec:Twin-Pati-Salam-Theory_3VL}}

In this section we present an extended version of the simplified twin
Pati-Salam model, featuring extra matter content and a discrete flavour
symmetry. This new version can achieve the proper flavour structure
required to be compatible with all data, solving the problems of the
simplified twin Pati-Salam model discussed in Section~\ref{subsec:Twin-Pati-Salam-Theory_1VL}.
Firstly, we will introduce the extended version of the model. Secondly,
we will revisit the diagonalisation of the mass matrix, leading to
the fermion masses and to the new couplings with the heavy gauge bosons.
Finally, we will study the phenomenology, showing that the model is
compatible with all data while predicting promising signals in flavour-violating
observables, rare $B$-decays and high-$p_{T}$ searches.

\subsection{New matter content and discrete flavour symmetry}

As firstly identified in \cite{DiLuzio:2018zxy}, when one considers
a 4321 model with all chiral fermions transforming as $SU(4)$ singlets
(fermiophobic framework), three vector-like fermion families can achieve
the proper flavour structure to explain the $B$-anomalies. Such flavour
structure can provide a GIM-like suppression of FCNCs, along with
large leptoquark couplings that can contribute to the LFU ratios.
Hence, as depicted in Table~\ref{tab:The-field-content_ExtendedModel},
we extend now the simplified model by two extra vector-like families,
to a total of three vector-like families charged under $SU(4)^{I}_{PS}$,
\begin{equation}
\begin{array}{c}
\psi_{4,5,6}(\mathbf{4,2,1;1,1,1})_{(1,1,\alpha)}\,,\overline{\psi}_{4,5,6}(\mathbf{\overline{4},\overline{2},1;1,1,1})_{(1,1,\alpha^{3})}\,,\\
\,\\
\psi_{4,5,6}^{c}(\mathbf{\overline{4},1,\overline{2};1,1,1})_{(1,1,\alpha)}\,,\overline{\psi^{c}}_{4,5,6}(\mathbf{4,1,2;1,1,1})_{(1,1,\alpha^{3})}\,,
\end{array}
\end{equation}
where it can be seen that the so-called fourth, fifth\footnote{Note that we have relabeled the fifth VL family
in this section with respect to the fifth family of the simplified model that provides
first family fermion masses and mixing, which should then be relabeled
to a different notation in the extended model. However, since the presence
of this family does not induce any couplings to $SU(4)^{I}_{PS}$ gauge
bosons, it has no low-energy phenomenological implications and we shall omit
it throughout this section.} and sixth VL families originate from the first Pati-Salam group,
being singlets under the second. They are indistinguishable under
the twin Pati-Salam symmetry in Eq.~(\ref{eq:TwinPS_symmetry}),
however a newly introduced $\mathbb{Z}_{4}$ flavour symmetry discriminates
the sixth family from the fourth and fifth, via different powers of
the $\mathbb{Z}_{4}$ charge $\alpha=e^{i\pi/2}$. This way, the total symmetry
group of the high energy model is extended to
\begin{table}[t]
\begin{centering}
\begin{tabular}{lccccccc}
\toprule 
Field & $SU(4)_{PS}^{I}$ & $SU(2)_{L}^{I}$ & $SU(2)_{R}^{I}$ & $SU(4)_{PS}^{II}$ & $SU(2)_{L}^{II}$ & $SU(2)_{R}^{II}$ & $\mathbb{Z}_{4}$\tabularnewline
\midrule
\midrule 
$\psi_{1,2,3}$ & $\mathbf{1}$ & $\mathbf{1}$ & $\mathbf{1}$ & $\mathbf{4}$ & $\mathbf{2}$ & $\mathbf{1}$ & $\alpha$, 1, 1\tabularnewline
$\psi_{1,2,3}^{c}$ & $\mathbf{1}$ & $\mathbf{1}$ & $\mathbf{1}$ & $\mathbf{\overline{4}}$ & $\mathbf{1}$ & $\mathbf{\overline{2}}$ & $\alpha$, $\alpha^{2}$, 1\tabularnewline
\midrule 
$\psi_{4,5,6}$ & $\mathbf{4}$ & $\mathbf{2}$ & $\mathbf{1}$ & $\mathbf{1}$ & $\mathbf{1}$ & $\mathbf{1}$ & 1, 1, $\alpha$\tabularnewline
$\overline{\psi}_{4,5,6}$ & $\mathbf{\overline{4}}$ & $\mathbf{\overline{2}}$ & $\mathbf{1}$ & $\mathbf{1}$ & $\mathbf{1}$ & $\mathbf{1}$ & 1, 1, $\alpha^{3}$\tabularnewline
$\psi_{4,5,6}^{c}$ & $\mathbf{\overline{4}}$ & $\mathbf{1}$ & $\mathbf{\overline{2}}$ & $\mathbf{1}$ & $\mathbf{1}$ & $\mathbf{1}$ & 1, 1, $\alpha$\tabularnewline
$\overline{\psi^{c}}_{4,5,6}$ & $\mathbf{4}$ & $\mathbf{1}$ & $\mathbf{2}$ & $\mathbf{1}$ & $\mathbf{1}$ & $\mathbf{1}$ & 1, 1, $\alpha^{3}$\tabularnewline
\midrule 
$\phi$ & $\mathbf{4}$ & $\mathbf{2}$ & $\mathbf{1}$ & $\mathbf{\overline{4}}$ & $\mathbf{\overline{2}}$ & $\mathbf{1}$ & 1\tabularnewline
$\overline{\phi}$,$\,\overline{\phi'}$ & $\mathbf{\overline{4}}$ & $\mathbf{1}$ & $\mathbf{\overline{2}}$ & $\mathbf{4}$ & $\mathbf{1}$ & $\mathbf{2}$ & 1, $\alpha^{2}$\tabularnewline
\midrule
$H$ & $\mathbf{\overline{4}}$ & $\mathbf{\overline{2}}$ & $\mathbf{1}$ & $\mathbf{4}$ & $\mathbf{1}$ & $\mathbf{2}$ & 1\tabularnewline
$\overline{H}$ & $\mathbf{4}$ & $\mathbf{1}$ & $\mathbf{2}$ & $\mathbf{\overline{4}}$ & $\mathbf{\overline{2}}$ & $\mathbf{1}$ & 1\tabularnewline
\midrule
$H'$ & $\mathbf{1}$ & $\mathbf{1}$ & $\mathbf{1}$ & $\mathbf{4}$ & $\mathbf{1}$ & $\mathbf{2}$ & 1\tabularnewline
$\overline{H}'$ & $\mathbf{1}$ & $\mathbf{1}$ & $\mathbf{1}$ & $\mathbf{\overline{4}}$ & $\mathbf{1}$ & $\mathbf{\overline{2}}$ & 1\tabularnewline
$\Phi$ & $\mathbf{1}$ & $\mathbf{2}$ & $\mathbf{1}$ & $\mathbf{1}$ & $\mathbf{\overline{2}}$ & $\mathbf{1}$ & 1\tabularnewline
$\overline{\Phi}$ & $\mathbf{1}$ & $\mathbf{1}$ & $\mathbf{\overline{2}}$ & $\mathbf{1}$ & $\mathbf{1}$ & $\mathbf{2}$ & 1\tabularnewline
\midrule
$\Omega_{15}$ & $\mathbf{15}$ & $\mathbf{1}$ & $\mathbf{1}$ & $\mathbf{1}$ & $\mathbf{1}$ & $\mathbf{1}$ & 1\tabularnewline
\bottomrule
\end{tabular}
\par\end{centering}
\caption[Field content in the extended twin PS model]{The field content under $G_{422}^{I}\times G_{422}^{II}\times \mathbb{Z}_{4}$,
see the main text for details. Fields transform under $\mathbb{Z}_{4}$ via powers of $\alpha=e^{i\pi/2}$. We do not include here extra content related to the origin of first family
fermion masses and mixing, which is discussed in Section~\ref{subsec:firstFamily}.\label{tab:The-field-content_ExtendedModel}}
\end{table}
\begin{equation}
G_{422}^{I}\times G_{422}^{II}\times \mathbb{Z}_{4}\,.
\end{equation}
The new $\mathbb{Z}_{4}$ discrete symmetry is introduced for phenomenological
purposes, as it will prevent fine-tuning, reduce the total number
of parameters of the model and protect from FCNCs involving the first
family of SM-like chiral fermions. Moreover, $\mathbb{Z}_{4}$ will simplify
the diagonalisation of the full mass matrices and preserve the success
of the effective Yukawa couplings for SM fermions given in Section~\ref{subsec:Effective-Yukawa-couplings},
with specific modifications. The origin of the chiral fermion families
is still the second Pati-Salam group, however now they transform in
a non-trivial way under $\mathbb{Z}_{4}$,
\begin{equation}
\psi_{1,2,3}(\mathbf{1,1,1;4,2,1})_{(\alpha,1,1)}\,,\quad\psi_{1,2,3}^{c}(\mathbf{1,1,1;\overline{4},1,\overline{2}})_{(\alpha,\alpha^{2},1)}\,.
\end{equation}
Finally, the scalar content is extended by an additional scalar $\Omega_{15}$
which transforms in the adjoint representation of $SU(4)^{I}_{PS}$, whose
VEV $\left\langle \Omega_{15}\right\rangle =T_{15}^{I}v_{15}$ splits
the vector-like masses of quarks and leptons, where $T^{I}_{15}=\mathrm{diag}(1,1,1,-3)/(2\sqrt{6})\,.$

We also include an additional copy of the Yukon $\overline{\phi}$,
denoted as $\overline{\phi'}$, featuring $\alpha^{2}$ charge under
$\mathbb{Z}_{4}$. The simplified Lagrangian in Eq.~(\ref{eq:Lren_4thVL-1})
is extended by the new matter content to
\begin{align}
\begin{aligned}\mathcal{L}_{\mathrm{mass}}^{ren} & =y_{ia}^{\psi}\overline{H}\psi_{i}\psi_{a}^{c}+y_{a3}^{\psi}H\psi_{a}\psi_{3}^{c}+x_{ia}^{\psi}\phi\psi_{i}\overline{\psi}_{a}+x_{a2}^{\psi^{c}}\overline{\psi_{a}^{c}\phi'}\psi_{2}^{c}+x_{a3}^{\psi^{c}}\overline{\psi_{a}^{c}\phi}\psi_{3}^{c} & {}\\
{} & +x_{16}^{\psi}\phi\psi_{1}\overline{\psi}_{6}+x_{61}^{\psi^{c}}\overline{\psi_{6}^{c}\phi}\psi_{1}^{c}+M_{ab}^{\psi}\psi_{a}\overline{\psi_{b}}+M_{ab}^{\psi^{c}}\psi_{a}^{c}\overline{\psi_{b}^{c}}+M_{66}^{\psi}\psi_{6}\overline{\psi_{6}}+M_{66}^{\psi^{c}}\psi_{6}^{c}\overline{\psi_{6}^{c}} & {}\\
{} & +\lambda_{15}^{aa}\Omega_{15}\psi_{a}\overline{\psi_{a}}+\lambda_{15}^{66}\Omega_{15}\psi_{6}\overline{\psi_{6}}+\bar{\lambda}_{15}^{aa}\Omega_{15}\psi_{a}^{c}\overline{\psi_{a}^{c}}+\bar{\lambda}_{15}^{66}\Omega_{15}\psi_{6}^{c}\overline{\psi_{6}^{c}}+\mathrm{h.c.}\,, & {}
\end{aligned}
\label{eq: full_lagrangian}
\end{align}
where $i=2,3$ and $a,b=4,5$ (terms $i=1$ and $a,b=6$ forbidden
by $\mathbb{Z}_{4}$). The symmetry breaking and the decomposition of the various
fields proceeds just like in the simplified model, see Section~\ref{subsec:High-scale-symmetry},
however the VEVs of the additional scalars $\overline{\phi'}$ and
$\Omega_{15}$ play a role in the spontaneous breaking of the 4321
symmetry, and the corresponding gauge boson masses become (assuming
$v_{1,3}\approx\overline{v}_{1,3}\approx\overline{v'}_{1,3}$ for
simplicity)
\begin{flalign}
 & M_{U_{1}}=\frac{1}{2}g_{4}\sqrt{3v_{1}^{2}+3v_{3}^{2}+\frac{4}{3}v_{15}^{2}}\,,\nonumber \\
 & M_{g'}=\frac{\sqrt{3}}{\sqrt{2}}\sqrt{g_{4}^{2}+g_{3}^{2}}v_{3}\,,\\
 & M_{Z'}=\frac{1}{2}\sqrt{\frac{3}{2}}\sqrt{g_{4}^{2}+\frac{2}{3}g_{1}^{2}}\sqrt{3v_{1}^{2}+v_{3}^{2}}\,.\nonumber 
\end{flalign}

\subsection{Effective Yukawa couplings in the extended model\label{subsec:Effective_Yukawa_3VL}}

In this section, we diagonalise the full mass matrix of the extended
model, following the same procedure as in Section~\ref{subsec:Effective-Yukawa-couplings},
but including the extra matter content of the extended model. We may
write the mass terms and couplings in Eq.~(\ref{eq: full_lagrangian})
as a $9\times9$ matrix in flavour space (we also define 9-dimensional
vectors as $\psi_{\alpha}$ and $\psi_{\beta}^{c}$ below),
\begin{equation}
\mathcal{L}_{4,5,6}^{ren}=\psi_{\alpha}^{\mathrm{T}}M^{\psi}\psi_{\beta}^{c}+\mathrm{h.c.}\,,
\end{equation}
\begin{equation}
\psi_{\alpha}=\left(\begin{array}{ccccccccc}
\psi_{1} & \psi_{2} & \psi_{3} & \psi_{4} & \psi_{5} & \psi_{6} & \overline{\psi_{4}^{c}} & \overline{\psi_{5}^{c}} & \overline{\psi_{6}^{c}}\end{array}\right)^{\mathrm{T}}\,,
\end{equation}
\begin{equation}
\psi_{\beta}^{c}=\left(\begin{array}{ccccccccc}
\psi_{1}^{c} & \psi_{2}^{c} & \psi_{3}^{c} & \psi_{4}^{c} & \psi_{5}^{c} & \psi_{6}^{c} & \overline{\psi_{4}} & \overline{\psi_{5}} & \overline{\psi_{6}}\end{array}\right)^{\mathrm{T}}\,,
\end{equation}
\begin{equation}
M^{\psi}=\left(
\global\long\def\arraystretch{1.3}%
\begin{array}{@{}llcccccccc@{}}
 & \multicolumn{1}{c@{}}{\psi_{1}^{c}} & \psi_{2}^{c} & \psi_{3}^{c} & \psi_{4}^{c} & \psi_{5}^{c} & \psi_{6}^{c} & \overline{\psi_{4}} & \overline{\psi_{5}} & \overline{\psi_{6}}\\
\cmidrule(l){2-10}\left.\psi_{1}\right| & 0 & 0 & 0 & 0 & 0 & 0 & 0 & 0 & x_{16}^{\psi}\phi\\
\left.\psi_{2}\right| & 0 & 0 & 0 & y_{24}^{\psi}\overline{H} & y_{25}^{\psi}\overline{H} & 0 & 0 & x_{25}^{\psi}\phi & 0\\
\left.\psi_{3}\right| & 0 & 0 & 0 & y_{34}^{\psi}\overline{H} & y_{35}^{\psi}\overline{H} & 0 & x_{34}^{\psi}\phi & x_{35}^{\psi}\phi & 0\\
\left.\psi_{4}\right| & 0 & 0 & y_{43}^{\psi}H & 0 & 0 & 0 & M_{44}^{Q,L} & M_{45}^{\psi} & 0\\
\left.\psi_{5}\right| & 0 & 0 & y_{53}^{\psi}H & 0 & 0 & 0 & M_{54}^{\psi} & M_{55}^{Q,L} & 0\\
\left.\psi_{6}\right| & 0 & 0 & 0 & 0 & 0 & 0 & 0 & 0 & M_{66}^{Q,L}\\
\left.\overline{\psi_{4}^{c}}\right| & 0 & x_{42}^{\psi^{c}}\overline{\phi'} & x_{43}^{\psi^{c}}\overline{\phi} & M_{44}^{\psi^{c}} & M_{45}^{\psi^{c}} & 0 & 0 & 0 & 0\\
\left.\overline{\psi_{5}^{c}}\right| & 0 & x_{52}^{\psi^{c}}\overline{\phi'} & x_{53}^{\psi^{c}}\overline{\phi} & M_{54}^{\psi^{c}} & M_{55}^{\psi^{c}} & 0 & 0 & 0 & 0\\
\left.\overline{\psi_{6}^{c}}\right| & x_{61}^{\psi^{c}}\overline{\phi} & 0 & 0 & 0 & 0 & M_{66}^{\psi^{c}} & 0 & 0 & 0
\end{array}\right)\,,\label{eq:9x9_mass_matrix}
\end{equation}
where the diagonal mass parameters $M_{44,55,66}^{Q,L}$ are split
for quarks and leptons due to the VEV of $\Omega_{15}$,
\begin{equation}
M_{aa}^{Q}\equiv M_{aa}^{\psi}+\frac{\lambda_{15}^{aa}\left\langle \Omega_{15}\right\rangle }{2\sqrt{6}}\,,\qquad M_{aa}^{L}\equiv M_{aa}^{\psi}-3\frac{\lambda_{15}^{aa}\left\langle \Omega_{15}\right\rangle }{2\sqrt{6}}\,,
\end{equation}
where $a=4,5,6$. Similar equations are obtained for the $\psi^{c}$
sector, however in the $\psi^{c}$ sector the mass splitting is minimal
due to $\left\langle \Omega_{15}\right\rangle $ being of order a
few hundreds GeV while $M_{aa}^{\psi^{c}}$ are much heavier due to
a generalisation of the messenger dominance in Eq.~(\ref{eq:hierarchy_scalesVL-1}).
The zeros in Eq.~(\ref{eq:9x9_mass_matrix}) are enforced by the
$\mathbb{Z}_{4}$ symmetry, except for the zero in the (2,7) entry which is
obtained by rotating $\psi_{2}$ and $\psi_{3}$, without loss of
generality thanks to the zeros in the upper $3\times3$ block (see
Section~\ref{subsec:Messenger_Dominance} and the discussion therein).

The matrix in Eq.~(\ref{eq:9x9_mass_matrix}) features three different
mass scales, the Higgs VEVs $\left\langle H\right\rangle $ and $\langle\overline{H}\rangle$,
the Yukon VEVs $\left\langle \phi\right\rangle $, $\langle\overline{\phi}\rangle$,
$\langle\overline{\phi'}\rangle$ and the vector-like mass terms $M_{ab}^{\psi}$
and $M_{ab}^{\psi^{c}}$. We can block diagonalise the matrix above
by taking advantage of the different mass scales. Firstly, we diagonalise
the $2\times2$ sub-blocks containing the heavy masses $M_{ab}^{\psi}$
and $M_{ab}^{\psi^{c}}$, 
\begin{flalign}
 & \left(\begin{array}{cc}
M_{4}^{Q} & 0\\
0 & M_{5}^{Q}
\end{array}\right)=V_{45}^{Q}\left(\begin{array}{cc}
M_{44}^{Q} & M_{45}^{\psi}\\
M_{54}^{\psi} & M_{55}^{Q}
\end{array}\right)V_{45}^{\overline{Q}\dagger}\,,\label{eq:VL_splitting_quark}\\
 & \left(\begin{array}{cc}
M_{4}^{L} & 0\\
0 & M_{5}^{L}
\end{array}\right)=V_{45}^{L}\left(\begin{array}{cc}
M_{44}^{L} & M_{45}^{\psi}\\
M_{54}^{\psi} & M_{55}^{L}
\end{array}\right)V_{45}^{\overline{L}\dagger}\,,\nonumber 
\end{flalign}
and similarly in the $\psi^{c}$ sector. The 4-5 rotations above just
redefine the elements in the 4th, 5th, 7th and 8th rows and columns
of the full mass matrix, leaving the upper $3\times3$ blocks unchanged
(plus we are allowed to introduce now the zero in the (2,7) entry
by performing the rotation of $\psi_{2}$ and $\psi_{3}$). Then we
perform a further sequence of rotations to go to the decoupling basis,
where no large elements appear apart from the diagonal heavy masses
(i.e.~those terms in the seventh, eighth and ninth rows and columns
involving the fields $\phi$ and $\overline{\phi}$ are all absorbed
into a redefinition of the heavy masses), and we obtain a block-diagonal
matrix similar to that of Eq.~(\ref{MassMatrix_4thVL_decoupling})
but enlarged with the fifth and sixth vector-like families. The total set of
unitary transformations is given by
\begin{equation}
\begin{array}{c}
V_{\psi}=V_{16}^{\psi}V_{35}^{\psi}V_{25}^{\psi}V_{34}^{\psi}V_{45}^{\psi}V_{45}^{\overline{\psi^{c}}}\,,\end{array}\label{eq:mixing_matrix_psi}
\end{equation}
\begin{equation}
V_{\psi^{c}}=V_{16}^{\psi^{c}}V_{35}^{\psi^{c}}V_{25}^{\psi^{c}}V_{34}^{\psi^{c}}V_{24}^{\psi^{c}}V_{45}^{\psi^{c}}V_{45}^{\overline{\psi}}\approx V_{34}^{\psi^{c}}V_{24}^{\psi^{c}}\,.\label{eq:mixing_matrix_psic}
\end{equation}
The mixing angles controlling the unitary transformations can be obtained
in the large mixing angle formalism (see Section~\ref{subsec:Effective-Yukawa-couplings})
as {\footnotesize{}
\begin{flalign}
 & s_{34}^{Q}=\frac{x_{34}^{\psi}\left\langle \phi_{3}\right\rangle }{\sqrt{\left(x_{34}^{\psi}\left\langle \phi_{3}\right\rangle \right)^{2}+\left(M_{4}^{Q}\right)^{2}}}\,, &  & s_{34}^{L}=\frac{x_{34}^{\psi}\left\langle \phi_{1}\right\rangle }{\sqrt{\left(x_{34}^{\psi}\left\langle \phi_{1}\right\rangle \right)^{2}+\left(M_{4}^{L}\right)^{2}}}\,,\label{eq:34_mixing_extended}\\
 & s_{25}^{Q}=\frac{x_{25}^{\psi}\left\langle \phi_{3}\right\rangle }{\sqrt{\left(x_{25}^{\psi}\left\langle \phi_{3}\right\rangle \right)^{2}+\left(M_{5}^{Q}\right)^{2}}}\,, &  & s_{25}^{L}=\frac{x_{25}^{\psi}\left\langle \phi_{1}\right\rangle }{\sqrt{\left(x_{25}^{\psi}\left\langle \phi_{1}\right\rangle \right)^{2}+\left(M_{5}^{L}\right)^{2}}}\,,\label{eq:25_mixing_extended}\\
 & s_{35}^{Q}=\frac{c_{34}^{Q}x_{35}^{\psi}\left\langle \phi_{3}\right\rangle }{\sqrt{\left(c_{34}^{Q}x_{35}^{\psi}\left\langle \phi_{3}\right\rangle \right)^{2}+\left(x_{25}^{\psi}\left\langle \phi_{3}\right\rangle \right)^{2}+\left(M_{5}^{Q}\right)^{2}}}, &  & s_{35}^{L}=\frac{c_{34}^{L}x_{35}^{\psi}\left\langle \phi_{1}\right\rangle }{\sqrt{\left(c_{34}^{L}x_{35}^{\psi}\left\langle \phi_{1}\right\rangle \right)^{2}+\left(x_{25}^{\psi}\left\langle \phi_{1}\right\rangle \right)^{2}+\left(M_{5}^{L}\right)^{2}}}\,,\label{eq:35_mixing}\\
 & s_{16}^{Q}=\frac{x_{16}^{\psi}\left\langle \phi_{3}\right\rangle }{\sqrt{\left(x_{16}^{\psi}\left\langle \phi_{3}\right\rangle \right)^{2}+\left(M_{6}^{Q}\right)^{2}}}\,, &  & s_{16}^{L}=\frac{x_{16}^{\psi}\left\langle \phi_{1}\right\rangle }{\sqrt{\left(x_{16}^{\psi}\left\langle \phi_{1}\right\rangle \right)^{2}+\left(M_{6}^{L}\right)^{2}}}\,,\\
 & s_{24}^{q^{c}}=\frac{x_{42}^{\psi^{c}}\langle\overline{\phi_{3}}\rangle}{\sqrt{\left(x_{42}^{\psi^{c}}\langle\overline{\phi_{3}}\rangle\right)^{2}+\left(M_{4}^{\psi^{c}}\right)^{2}}}\,, &  & s_{24}^{e^{c}}=\frac{x_{42}^{\psi^{c}}\langle\overline{\phi_{1}}\rangle}{\sqrt{\left(x_{42}^{\psi^{c}}\langle\overline{\phi_{1}}\rangle\right)^{2}+\left(M_{4}^{\psi^{c}}\right)^{2}}}\,,\label{eq:sqc24_mixing}\\
 & s_{34}^{q^{c}}=\frac{x_{43}^{\psi^{c}}\langle\overline{\phi_{3}}\rangle}{\sqrt{\left(x_{42}^{\psi^{c}}\langle\overline{\phi_{3}}\rangle\right)^{2}+\left(x_{43}^{\psi^{c}}\langle\overline{\phi_{3}}\rangle\right)^{2}+\left(M_{4}^{\psi^{c}}\right)^{2}}}\,, &  & s_{34}^{e^{c}}=\frac{x_{43}^{\psi^{c}}\langle\overline{\phi}_{1}\rangle}{\sqrt{\left(x_{42}^{\psi^{c}}\langle\overline{\phi_{1}}\rangle\right)^{2}+\left(x_{43}^{\psi^{c}}\langle\overline{\phi_{1}}\rangle\right)^{2}+\left(M_{4}^{\psi^{c}}\right)^{2}}}\,,\label{eq:sqc34_mixing}\\
 & \hat{M}_{4}^{Q}=\sqrt{\left(x_{34}^{\psi}\left\langle \phi_{3}\right\rangle \right)^{2}+\left(M_{4}^{Q}\right)^{2}}\,, &  & \hat{M}_{4}^{L}=\sqrt{\left(x_{34}^{\psi}\left\langle \phi_{1}\right\rangle \right)^{2}+\left(M_{4}^{L}\right)^{2}}\,,\label{eq:Mass-_4th}\\
 & \hat{M}_{5}^{Q}=\sqrt{\left(x_{25}^{\psi}\left\langle \phi_{3}\right\rangle \right)^{2}+\left(x_{35}^{\psi}\left\langle \phi_{3}\right\rangle \right)^{2}+\left(M_{5}^{Q}\right)^{2}}\,, &  & \hat{M}_{5}^{L}=\sqrt{\left(x_{25}^{\psi}\left\langle \phi_{1}\right\rangle \right)^{2}+\left(x_{35}^{\psi}\left\langle \phi_{1}\right\rangle \right)^{2}+\left(M_{5}^{L}\right)^{2}}\,,\label{eq:Mass_5th}\\
 & \hat{M}_{6}^{Q}=\sqrt{\left(x_{16}^{\psi}\left\langle \phi_{3}\right\rangle \right)^{2}+\left(M_{6}^{Q}\right)^{2}}\,, &  & \hat{M}_{6}^{L}=\sqrt{\left(x_{16}^{\psi}\left\langle \phi_{1}\right\rangle \right)^{2}+\left(M_{6}^{L}\right)^{2}}\,,\label{eq:Mass_6th}\\
 & \hat{M}_{4}^{q^{c}}=\sqrt{\left(x_{42}^{\psi^{c}}\langle\overline{\phi_{3}}\rangle\right)^{2}+\left(x_{43}^{\psi^{c}}\langle\overline{\phi_{3}}\rangle\right)^{2}+\left(M_{4}^{\psi^{c}}\right)^{2}}\,, &  & \hat{M}_{4}^{e^{c}}=\sqrt{\left(x_{42}^{\psi^{c}}\langle\overline{\phi_{1}}\rangle\right)^{2}+\left(x_{43}^{\psi^{c}}\langle\overline{\phi_{1}}\rangle\right)^{2}+\left(M_{4}^{\psi^{c}}\right)^{2}}\,.
\end{flalign}
}The transformations in the $\psi^{c}$ sector shown in Eq.~(\ref{eq:mixing_matrix_psic})
can be described by $V_{34}^{\psi^{c}}V_{24}^{\psi^{c}}$ in good
approximation, whose mixing angles are given above. This approximation
is accurate as far as the mixing involving the 5th and 6th $\psi^{c}$
fields is further suppressed by a generalisation of the messenger dominance
in Eq.~(\ref{eq:hierarchy_scalesVL}) to three vector-like families,
namely
\begin{equation}
M_{44}^{Q,L}\ll M_{55}^{Q,L}\sim M_{66}^{Q,L}\ll M_{44}^{\psi^{c}}\ll M_{55}^{\psi^{c}},M_{66}^{\psi^{c}}\,.\label{eq:hierarchy_scalesVL-1}
\end{equation}
The hierarchy above will preserve most features of the simplified
model, including large third family Yukawa couplings arising from mixing
with $\psi_{4}$ fermions, and small second family Yukawa couplings
arising from mixing with $\psi_{4}^{c}$ fermions. The couplings of
$U_{1}$ to chiral fermions will remain dominantly left-handed, since
the couplings to $\psi^{c}$ chiral fermions (or equivalently right-handed
fermions) will remain suppressed by small mixing angles. On the other
hand, the hierarchy $M_{44}^{Q,L}\ll M_{55}^{Q,L}$ will provide hierarchical
couplings of $U_{1}$ to third family and second family fermions,
so we anticipate a small contribution to $R_{K^{(*)}}$ and a large
contribution to $R_{D^{(*)}}$.

We obtain the effective Yukawa couplings for SM fermions by applying
the set of unitary transformations in Eqs.~(\ref{eq:mixing_matrix_psi})
and (\ref{eq:mixing_matrix_psic}) to the upper $6\times6$ block
of Eq.~(\ref{eq:9x9_mass_matrix}), in the same way as in Eq.~(\ref{eq:Primed_basis}).
In this basis (primed), the mass matrix for each charged sector reads
(assuming a small $x_{35}^{\psi}$, see Section~\ref{subsec:BsMixing_revisited}
for the motivation, and approximating cosines in the $\psi^{c}$ sector
to be 1),
\begin{equation}
M_{\mathrm{eff}}^{u}=\left(
\global\long\def\arraystretch{0.7}%
\begin{array}{@{}llcc@{}}
 & \multicolumn{1}{c@{}}{\phantom{\!\,}u'{}_{1}^{c}} & \phantom{\!\,}u'{}_{2}^{c} & \phantom{\!\,}u'{}_{3}^{c}\\
\cmidrule(l){2-4}\left.Q'_{1}\right| & 0 & 0 & 0\\
\left.Q'_{2}\right| & 0 & 0 & s_{25}^{Q}y_{53}^{\psi}\\
\left.Q'_{3}\right| & 0 & 0 & s_{34}^{Q}y_{43}^{\psi}
\end{array}\right)\left\langle H_{t}\right\rangle +\left(
\global\long\def\arraystretch{0.7}%
\begin{array}{@{}llcc@{}}
 & \multicolumn{1}{c@{}}{\phantom{\!\,}u'{}_{1}^{c}} & \phantom{\!\,}u'{}_{2}^{c} & \phantom{\!\,}u'{}_{3}^{c}\\
\cmidrule(l){2-4}\left.Q'_{1}\right| & 0 & 0 & 0\\
\left.Q'_{2}\right| & 0 & c_{25}^{Q}s_{24}^{q^{c}}y_{24}^{\psi} & c_{25}^{Q}s_{34}^{q^{c}}y_{24}^{\psi}\\
\left.Q'_{3}\right| & 0 & c_{34}^{Q}s_{24}^{q^{c}}y_{34}^{\psi} & c_{34}^{Q}s_{34}^{q^{c}}y_{34}^{\psi}
\end{array}\right)\left\langle H_{c}\right\rangle +\mathrm{h.c.}\,,\label{eq:MassMatrix_4thVL_effective_up-1}
\end{equation}
\begin{equation}
M_{\mathrm{eff}}^{d}=\left(
\global\long\def\arraystretch{0.7}%
\begin{array}{@{}llcc@{}}
 & \multicolumn{1}{c@{}}{\phantom{\!\,}d'{}_{1}^{c}} & \phantom{\!\,}d'{}_{2}^{c} & \phantom{\!\,}d'{}_{3}^{c}\\
\cmidrule(l){2-4}\left.Q'_{1}\right| & 0 & 0 & 0\\
\left.Q'_{2}\right| & 0 & 0 & s_{25}^{Q}y_{53}^{\psi}\\
\left.Q'_{3}\right| & 0 & 0 & s_{34}^{Q}y_{43}^{\psi}
\end{array}\right)\left\langle H_{b}\right\rangle +\left(
\global\long\def\arraystretch{0.7}%
\begin{array}{@{}llcc@{}}
 & \multicolumn{1}{c@{}}{\phantom{\!\,}d'{}_{1}^{c}} & \phantom{\!\,}d'{}_{2}^{c} & \phantom{\!\,}d'{}_{3}^{c}\\
\cmidrule(l){2-4}\left.Q'_{1}\right| & 0 & 0 & 0\\
\left.Q'_{2}\right| & 0 & c_{25}^{Q}s_{24}^{q^{c}}y_{24}^{\psi} & c_{25}^{Q}s_{34}^{q^{c}}y_{24}^{\psi}\\
\left.Q'_{3}\right| & 0 & c_{34}^{Q}s_{24}^{q^{c}}y_{34}^{\psi} & c_{34}^{Q}s_{34}^{q^{c}}y_{34}^{\psi}
\end{array}\right)\left\langle H_{s}\right\rangle +\mathrm{h.c.}\,,\label{eq:MassMatrix_4thVL_effective_down-1}
\end{equation}
\begin{equation}
M_{\mathrm{eff}}^{e}=\left(
\global\long\def\arraystretch{0.7}%
\begin{array}{@{}llcc@{}}
 & \multicolumn{1}{c@{}}{\phantom{\!\,}e'{}_{1}^{c}} & \phantom{\!\,}e'{}_{2}^{c} & \phantom{\!\,}e'{}_{3}^{c}\\
\cmidrule(l){2-4}\left.L'_{1}\right| & 0 & 0 & 0\\
\left.L'_{2}\right| & 0 & 0 & s_{25}^{L}y_{53}^{\psi}\\
\left.L'_{3}\right| & 0 & 0 & s_{34}^{L}y_{43}^{\psi}
\end{array}\right)\left\langle H_{\tau}\right\rangle +\left(
\global\long\def\arraystretch{0.7}%
\begin{array}{@{}llcc@{}}
 & \multicolumn{1}{c@{}}{\phantom{\!\,}e'{}_{1}^{c}} & \phantom{\!\,}e'{}_{2}^{c} & \phantom{\!\,}e'{}_{3}^{c}\\
\cmidrule(l){2-4}\left.L'_{1}\right| & 0 & 0 & 0\\
\left.L'_{2}\right| & 0 & c_{25}^{L}s_{24}^{e^{c}}y_{24}^{\psi} & c_{25}^{L}s_{34}^{e^{c}}y_{24}^{\psi}\\
\left.L'_{3}\right| & 0 & c_{34}^{L}s_{24}^{e^{c}}y_{34}^{\psi} & c_{34}^{L}s_{34}^{e^{c}}y_{34}^{\psi}
\end{array}\right)\left\langle H_{\mu}\right\rangle +\mathrm{h.c.}\,,\label{eq:MassMatrix_4thVL_effective_leptons-1}
\end{equation}
which are diagonalised by 2-3 rotations, and the CKM matrix is obtained
via Eq.~(\ref{eq:CKM_matrixTwinPS}). The first family masses and mixings can be obtained by adding a heavier family of vector-like fermions that originates from the second PS group, as shown in Section~\ref{subsec:firstFamily}, however this has no implications for low-energy phenomenology. The mass matrices above are of similar
form to Eqs.~(\ref{eq:MassMatrix_4thVL_effective_up}), (\ref{eq:MassMatrix_4thVL_effective_down}), (\ref{eq:MassMatrix_4thVL_effective_leptons}),
just featuring an extra off-diagonal component in the (2,3) entry
of the first matrix in each sector, arising from mixing with the 5th
family. This new term can be used to partially cancel the down-quark 2-3
mixing while simultaneously enhancing the up-quark mixing to preserve the CKM,
involving a mild tuning:
\begin{itemize}
\item Let us impose that the total (2,3) entry in the down-quark mass matrix
is small, i.e.
\begin{equation}
-s_{25}^{Q}\left|y_{53}^{\psi}\right|\left\langle H_{b}\right\rangle +c_{25}^{Q}s_{34}^{q^{c}}y_{24}^{\psi}\left\langle H_{s}\right\rangle \approx0\,.
\end{equation}
Following the discussion of Section~\ref{subsec:Effective-Yukawa-couplings},
a natural benchmark is $\left\langle H_{b}\right\rangle \approx m_{b}$
and $s_{34}^{q^{c}}y_{24}^{\psi}\left\langle H_{s}\right\rangle \approx m_{s}$,
hence
\begin{equation}
-s_{25}^{Q}\left|y_{53}^{\psi}\right|m_{b}+m_{s}\approx0\Rightarrow\left|y_{53}^{\psi}\right|=\frac{m_{s}}{s_{25}^{Q}m_{b}}\,.
\end{equation}
On the other hand, the mixing angle $s_{25}^{Q}$ is very relevant
for the $B$-decays and related phenomenology, and we obtain the typical
value $s_{25}^{Q}\approx0.2$ in Section~\ref{subsec:Low-energy-phenomenology},
featuring another connection between the flavour puzzle and $B$-physics
in our model. With this input, we obtain 
\begin{equation}
\left|y_{53}^{\psi}\right|\approx\mathcal{O}(0.1).
\end{equation}
In particular, the benchmark in Table \ref{tab:BP} suppresses the
down mixing with the choice $y_{53}^{\psi}=-0.3$, obtaining $s_{23}^{d}\approx\mathcal{O}(10^{-3})$
which is enough to control the stringent constraints from $B_{s}-\bar{B}_{s}$
meson mixing (see Section~\ref{subsec:Bs_Mixing}).
\item At the same time that $y_{53}^{\psi}$ partially cancels the down
mixing, it leads to large up mixing which preserves the CKM. Let us
now estimate the 2-3 mixing in the up sector as the ratio of the (2,3)
entry over the (3,3) entry in the up-quark mass matrix,
\begin{equation}
\frac{-s_{25}^{Q}\left|y_{53}^{\psi}\right|\left\langle H_{t}\right\rangle +c_{25}^{Q}s_{34}^{q^{c}}y_{24}^{\psi}\left\langle H_{c}\right\rangle }{s_{34}^{Q}y_{43}^{\psi}\left\langle H_{t}\right\rangle }\approx\frac{s_{25}^{Q}\left|y_{53}^{\psi}\right|\left\langle H_{t}\right\rangle +m_{c}}{m_{t}}\approx s_{25}^{Q}\left|y_{53}^{\psi}\right|\approx\mathcal{O}(V_{cb})\,,
\end{equation}
where we have considered $y_{43}^{\psi}=1$, $s_{34}^{Q}\approx1$,
as required to explain the top mass (see the discussion in the first
bullet point in Section~\ref{subsec:Effective-Yukawa-couplings})
and we have neglected the $m_{c}/m_{t}$ small factor. This way, we have taken advantage
of the new contribution via the 5th family (and of the different hierarchies
$m_{c}/m_{t}$ and $m_{s}/m_{b}$) to cancel the dangerous down-quark mixing
while preserving the CKM via up-quark mixing.
\item The situation in the lepton sector is similar due to Pati-Salam universality
of the parameters, i.e.
\begin{equation}
\frac{-s_{25}^{L}\left|y_{53}^{\psi}\right|\left\langle H_{\tau}\right\rangle +c_{25}^{L}s_{34}^{e^{c}}y_{24}^{\psi}\left\langle H_{\mu}\right\rangle }{s_{34}^{L}y_{43}^{\psi}\left\langle H_{\tau}\right\rangle }\approx\frac{-s_{25}^{L}\left|y_{53}^{\psi}\right|\left\langle H_{\tau}\right\rangle +m_{\mu}}{m_{\tau}}\approx s_{25}^{L}\left|y_{53}^{\psi}\right|\approx\mathcal{O}(V_{cb})\,.
\end{equation}
However, the leptonic mixing angles $s_{24}^{e^{c}}$ and $s_{34}^{e^{c}}$
are smaller than the quark ones if we assume the phenomenological
relation $\left\langle \phi_{3}\right\rangle \gg\left\langle \phi_{1}\right\rangle $.
This leads to $\left\langle H_{\mu}\right\rangle $ being above the
scale of the muon mass, which predicts a quick growth of lepton mixing
in the scenario $s_{34}^{e^{c}}>s_{24}^{e^{c}}$. This can be easily
achieved in natural benchmarks. In this scenario, interesting signals
arise in LFV processes such as $\tau\rightarrow3\mu$ or $\tau\rightarrow\mu\gamma$,
mediated at tree-level by the $Z'$ boson, see Section~\ref{subsec:LFV_processes}.
\end{itemize}
Other than the bullet points above, the mass matrices in Eqs.~(\ref{eq:MassMatrix_4thVL_effective_up-1}),
(\ref{eq:MassMatrix_4thVL_effective_down-1}), (\ref{eq:MassMatrix_4thVL_effective_leptons-1})
lead to similar predictions as those of the simplified model in Section~\ref{subsec:Effective-Yukawa-couplings}.

\subsection{Vector-fermion interactions in the extended model\label{subsec:Couplings_3VL}}

In this section we shall compute the vector-fermion couplings involving
the heavy gauge bosons $U_{1}$, $g'$, $Z'$ for the extended twin
Pati-Salam model. We omit the couplings of the vector-like partners in
the conjugate representations $\overline{\psi}_{\alpha}$ and $\overline{\psi_{\alpha}^{c}}$,
since they do not mix with SM fermions. 

\subsubsection*{$U_{1}$ couplings}

In the original interaction basis, the vector leptoquark couples to the
heavy $SU(2)_{L}$ doublets via the left-handed interactions\footnote{Notice that we continue working in the 2-component notation introduced
in Appendix~\ref{app:2-component_notation}.},
\begin{equation}
\mathcal{L}_{U_{1}}^{\mathrm{gauge}}=\frac{g_{4}}{\sqrt{2}}\left(Q_{4}^{\dagger}\bar{\sigma}_{\mu}L_{4}+Q_{5}^{\dagger}\bar{\sigma}_{\mu}L_{5}+Q_{6}^{\dagger}\bar{\sigma}_{\mu}L_{6}+\mathrm{h.c.}\right)U_{1}^{\mu}\,,
\end{equation}
where similar couplings to the heavy $SU(2)_{L}$ singlets $\psi^{c}$
are also present, however they lead to suppressed couplings to SM
fermions due to the messenger dominance in Eq.~(\ref{eq:hierarchy_scalesVL-1}).
This way, we obtain dominantly left-handed $U_{1}$ couplings in good
approximation. Now we shall apply the unitary transformations in Eq.~(\ref{eq:mixing_matrix_psi})
to rotate the fields from the original interaction basis to the decoupling
basis (primed),
\begin{equation}
\mathcal{L}_{U_{1}}^{\mathrm{gauge}}=\frac{g_{4}}{\sqrt{2}}Q'{}_{\alpha}^{\dagger}V_{Q}\bar{\sigma}_{\mu}\mathrm{diag}(0,0,0,1,1,1)V_{L}^{\dagger}L'_{\beta}U_{1}^{\mu}+\mathrm{h.c.}\,,\label{eq:rotation_decoupling_U1_6VL}
\end{equation}
where
\begin{equation}
V_{Q}=V_{16}^{Q}V_{35}^{Q}V_{25}^{Q}V_{34}^{Q}V_{45}^{Q}\,,\qquad V_{L}=V_{16}^{L}V_{35}^{L}V_{25}^{L}V_{34}^{L}V_{45}^{L}\,.\label{eq:mixing_matrix_Q}
\end{equation}
The 4-5 rotations are different for quarks and leptons due to $\left\langle \Omega_{15}\right\rangle $
splitting the mass terms of the vector-like fermions. They lead to a non-trivial
CKM-like matrix for the $U_{1}$ couplings,
\begin{equation}
W_{LQ}=V_{45}^{Q}V_{45}^{L\dagger}=\left(\begin{array}{ccc}
c_{\theta_{LQ}} & -s_{\theta_{LQ}} & 0\\
s_{\theta_{LQ}} & c_{\theta_{LQ}} & 0\\
0 & 0 & 1
\end{array}\right)\,,
\end{equation}
where $s_{\theta_{LQ}}$ depends on the angles $s_{45}^{Q}$ and $s_{45}^{L}$,
obtained from the diagonalisation in Eq.~(\ref{eq:VL_splitting_quark}).
The unitary matrix $W_{LQ}$ can be regarded as a generalisation of
the CKM matrix to $SU(4)^{I}_{PS}$ or quark-lepton flavour space. Similarly
to the CKM case, the $W_{LQ}$ matrix is the only source of flavour-changing
transitions among $SU(4)^{I}_{PS}$ states, and it appears only in interactions
mediated by $U_{1}$. In this sense, the vector leptoquark $U_{1}$
is analogous to the SM $W^{\pm}$ bosons. Similarly, the $Z'$ and $g'$
are analogous to the SM $Z$ boson, and we will show that their interactions
are $SU(4)^{I}_{PS}$ flavour-conserving at tree-level. In analogy to the
SM, we will denote $U_{1}$ transitions as charged currents and $Z'$,
$g'$ transitions as neutral currents. As in the SM, FCNCs proportional
to the $W_{LQ}$ matrix are generated at loop level. This mechanism
was firstly identified in \cite{DiLuzio:2018zxy} for a similar 4321
framework.

The same mixing that leads to the SM fermion masses and mixings, see
Eq.~(\ref{eq:mixing_matrix_Q}), also leads to effective $U_{1}$
couplings to SM fermions which can contribute to the LFU ratios,
\begin{equation}
\mathcal{L}_{U_{1}}^{\mathrm{gauge}}=\frac{g_{4}}{\sqrt{2}}Q'{}_{i}^{\dagger}\bar{\sigma}_{\mu}\left(\begin{array}{ccc}
s_{16}^{Q}s_{16}^{L}\epsilon & 0 & 0\\
0 & c_{\theta_{LQ}}s_{25}^{Q}s_{25}^{L} & s_{\theta_{LQ}}s_{25}^{Q}s_{34}^{L}\\
0 & -s_{\theta_{LQ}}s_{34}^{Q}s_{25}^{L} & c_{\theta_{LQ}}s_{34}^{Q}s_{34}^{L}
\end{array}\right)L'_{j}U_{1}^{\mu}+\mathrm{h.c.}\,,\label{eq:LQ_couplings}
\end{equation}
where we have considered that $s_{35}^{Q,L}$ are small, see Section\textbf{s}~\ref{subsec:Effective_Yukawa_3VL}
and \ref{subsec:BsMixing_revisited}. The first family coupling can
be diluted via mixing with vector-like fermions, which is parameterised
via the effective parameter $\epsilon$ (see Section~\ref{subsec:LFV_processes}
for more details\textbf{)}. The couplings above receive small corrections
due to 2-3 fermion mixing arising after diagonalising the effective
mass matrices in Eqs.~(\ref{eq:MassMatrix_4thVL_effective_up-1}),~(\ref{eq:MassMatrix_4thVL_effective_down-1}),~(\ref{eq:MassMatrix_4thVL_effective_leptons-1}).
It can be seen from Eq.~(\ref{eq:LQ_couplings}) that a large (2,3)
coupling $\beta_{c\nu_{\tau}}$ arises now, proportional to the large
sines $s_{\theta_{LQ}}$, $s_{34}^{L}$ and $s_{25}^{Q}$. This solves
one important issue of the simplified model, where the flavour-violating
couplings $\beta_{c\nu_{\tau}}$ and $\beta_{b\mu}$ where connected
to small 2-3 mixing angles, suppressing the contributions of $U_{1}$
to the LFU ratios. In any case, the leptoquark couplings that contribute
to $B$-decays arise due to the same mixing effects which diagonalise
the mass matrices of the model, yielding mass terms for the SM fermions.
This way, the flavour puzzle and the $B$-anomalies are dynamically
and parametrically connected in this model, leading to a predictive
framework. Notice that thanks to the connection with the flavour puzzle, our
$U_{1}$ couplings are dominantly left-handed, in contrast with the alternative
theories in the market\cite{Bordone:2017bld,Allwicher:2020esa,Fuentes-Martin:2020pww,Fuentes-Martin:2022xnb,Davighi:2022bqf}.

\subsubsection*{Coloron couplings and GIM-like mechanism}

In the original interaction basis, the coloron couplings are flavour-diagonal,
featuring the following couplings to $SU(2)_{L}$ doublets,
\begin{equation}
\mathcal{L}_{g'}^{\mathrm{gauge}}=\frac{g_{4}g_{s}}{g_{3}}\left(Q_{4}^{\dagger}\bar{\sigma}^{\mu}T^{a}Q_{4}+Q_{5}^{\dagger}\bar{\sigma}^{\mu}T^{a}Q_{5}+Q_{6}^{\dagger}\bar{\sigma}^{\mu}T^{a}Q_{6}-\frac{g_{3}^{2}}{g_{4}^{2}}Q_{i}^{\dagger}\bar{\sigma}^{\mu}T^{a}Q_{i}\right)g'{}_{\mu}^{a}\,,\label{eq:interaction_basis_g'}
\end{equation}
where $i=1,2,3$. Now we rotate to the decoupling basis by applying
the transformations in Eq.~(\ref{eq:mixing_matrix_Q}), (assuming
small $x_{35}^{\psi}$ as discussed in Section \ref{subsec:BsMixing_revisited})
obtaining
\begin{flalign}
\mathcal{L}_{g'}^{\mathrm{gauge}} & =\frac{g_{4}g_{s}}{g_{3}}Q'{}_{i}^{\dagger}\bar{\sigma}^{\mu}T^{a}\label{eq:Coloron_couplings_3VL}\\
 & \times\left(\begin{array}{ccc}
\left(s_{16}^{Q}\right)^{2}-\left(c_{16}^{Q}\right)^{2}\frac{g_{3}^{2}}{g_{4}^{2}} & 0 & 0\\
0 & \left(s_{25}^{Q}\right)^{2}-\left(c_{25}^{Q}\right)^{2}\frac{g_{3}^{2}}{g_{4}^{2}} & 0\\
0 & 0 & \left(s_{34}^{Q}\right)^{2}-\left(c_{34}^{Q}\right)^{2}\frac{g_{3}^{2}}{g_{4}^{2}}
\end{array}\right)Q'_{j}g'{}_{\mu}^{a}\,,\nonumber 
\end{flalign}
Here $V_{45}^{Q}$ cancels due to unitarity and due to the coloron couplings
between vector-like quarks being flavour universal in the original basis of
Eq.~(\ref{eq:interaction_basis_g'}). Therefore, as anticipated, the
CKM-like matrix $W_{LQ}$ does not affect the neutral currents mediated
by $g'$ (and similarly for the $Z'$ boson). The coloron couplings in Eq.~(\ref{eq:Coloron_couplings_3VL})
receive small corrections due to 2-3 mixing arising after diagonalising
the effective mass matrices in Eqs.~(\ref{eq:MassMatrix_4thVL_effective_up-1}),~(\ref{eq:MassMatrix_4thVL_effective_down-1}),~(\ref{eq:MassMatrix_4thVL_effective_leptons-1}),
predominantly in the up sector due to the down-aligned flavour structure
achieved in Section~\ref{subsec:Effective_Yukawa_3VL}. We obtain
similar couplings for $SU(2)_{L}$ singlets, however their mixing angles
are suppressed by the messenger dominance in Eq.~(\ref{eq:hierarchy_scalesVL-1}),
and so they remain as in the original interaction basis to good approximation.

The coloron couplings of Eq.~(\ref{eq:Coloron_couplings_3VL}) are
family universal if
\begin{equation}
s_{34}^{Q}=s_{25}^{Q}=s_{16}^{Q}\,,\label{eq:GIM-like_condition}
\end{equation}
leading to a GIM-like protection from tree-level FCNCs mediated by
the coloron. The condition above was already identified in \cite{DiLuzio:2018zxy},
denoted as \textit{full alignment limit}. However, we have seen that
maximal $s_{34}^{Q}\approx1$ is well motivated in our model to protect
the perturbativity of the top Yukawa, by the fit of the $R_{D^{(*)}}$
anomaly, and furthermore it naturally suppresses $s_{35}^{Q}$ via
a small $c_{34}^{Q}$. The caveat is that if the condition in Eq.~(\ref{eq:GIM-like_condition})
is implemented, then $s_{16}^{Q}$ and $s_{25}^{Q}$ would also be
maximal, leading to large couplings to valence quarks which would
blow up the production of the coloron at the LHC. This fact was already
identified in \cite{DiLuzio:2018zxy}, where large $s_{34}^{Q}$ was
also suggested by the $B$-anomalies, and this motivated a partial
alignment limit, 
\begin{equation}
s_{25}^{Q}=s_{16}^{Q}\,,\label{eq:GIM-like_condition-1}
\end{equation}
which suppresses FCNCs between the first and second quark families,
proportional to the largest off-diagonal elements of the CKM matrix. FCNCs between
the second and third families still arise, however we are protected
from the stringent constraints of $B_{s}-\bar{B}_{s}$ meson mixing
due to the down-aligned flavour structure achieved in Section~\ref{subsec:Effective_Yukawa_3VL}.
Finally, FCNCs between the first and third families are also under control,
as they are proportional to the smaller elements of the CKM matrix.

The GIM-like condition of Eq.~(\ref{eq:GIM-like_condition-1}) translates,
in terms of fundamental parameters of our model, into
\begin{equation}
\frac{x_{25}^{\psi}\left\langle \phi_{3}\right\rangle }{\sqrt{\left(x_{25}^{\psi}\left\langle \phi_{3}\right\rangle \right)^{2}+\left(M_{5}^{Q}\right)^{2}}}=\frac{x_{16}^{\psi}\left\langle \phi_{3}\right\rangle }{\sqrt{\left(x_{16}^{\psi}\left\langle \phi_{3}\right\rangle \right)^{2}+\left(M_{6}^{Q}\right)^{2}}}\,,\label{eq:GIM-like_2}
\end{equation}
which could be naively achieved with natural couplings and $M_{5}^{Q}$,
$M_{6}^{Q}$ being of the same order, as allowed by the messenger
dominance in Eq.~(\ref{eq:hierarchy_scalesVL-1}).
The couplings and vector-like mass terms can also be chosen differently, as far
as Eq.~(\ref{eq:GIM-like_2}) is preserved. At the moment, the GIM-like
mechanism is accidental. However, Eq.~(\ref{eq:GIM-like_2}) suggests
that the sixth and fifth family, and also the first and second families,
might transform as doublets under a global $SU(2)$ symmetry, enforcing
the parametric relations of Eq.~(\ref{eq:GIM-like_2}).

\subsubsection*{$Z'$ couplings}

We can follow the same procedure to extract the couplings of the $Z'$
boson to chiral fermions,
\begin{flalign}
\mathcal{L}_{Z',q}^{\mathrm{gauge}} & =\frac{\sqrt{3}}{\sqrt{2}}\frac{g_{4}g_{Y}}{g_{1}}Q'{}_{i}^{\dagger}\bar{\sigma}^{\mu}\label{eq:Z'_couplings_3VL_q}\\
\times & \left(\begin{array}{ccc}
\frac{1}{6}\left(s_{16}^{Q}\right)^{2}-\left(c_{16}^{Q}\right)^{2}\frac{g_{1}^{2}}{9g_{4}^{2}} & 0 & 0\\
0 & \frac{1}{6}\left(s_{25}^{Q}\right)^{2}-\left(c_{25}^{Q}\right)^{2}\frac{g_{1}^{2}}{9g_{4}^{2}} & 0\\
0 & 0 & \frac{1}{6}\left(s_{34}^{Q}\right)^{2}-\left(c_{34}^{Q}\right)^{2}\frac{g_{1}^{2}}{9g_{4}^{2}}
\end{array}\right)Q'_{j}Z'_{\mu}\,,\nonumber 
\end{flalign}
\begin{flalign}
\mathcal{L}_{Z',\ell}^{\mathrm{gauge}} & =-\frac{\sqrt{3}}{\sqrt{2}}\frac{g_{4}g_{Y}}{g_{1}}L'{}_{i}^{\dagger}\bar{\sigma}^{\mu}\label{eq:Z'_couplings_3VL_l}\\
 & \times\left(\begin{array}{ccc}
\frac{1}{2}\left(s_{16}^{L}\right)^{2}-\left(c_{16}^{L}\right)^{2}\frac{g_{1}^{2}}{3g_{4}^{2}} & 0 & 0\\
0 & \frac{1}{2}\left(s_{25}^{L}\right)^{2}-\left(c_{25}^{L}\right)^{2}\frac{g_{1}^{2}}{3g_{4}^{2}} & 0\\
0 & 0 & \frac{1}{2}\left(s_{34}^{L}\right)^{2}-\left(c_{34}^{L}\right)^{2}\frac{g_{1}^{2}}{3g_{4}^{2}}
\end{array}\right)L'_{j}Z'_{\mu}\,,\nonumber 
\end{flalign}
where the up-quark couplings above receive small corrections due
to 2-3 mixing arising after diagonalising the effective mass matrices
in Eqs.~(\ref{eq:MassMatrix_4thVL_effective_up-1}),~(\ref{eq:MassMatrix_4thVL_effective_down-1}).
However, larger 2-3 charged lepton mixing is possible (see Section~\ref{subsec:Effective_Yukawa_3VL}),
obtaining for charged leptons:
{\small{}
\begin{flalign}
 & \mathcal{L}_{Z',e}^{\mathrm{gauge}}\approx-\frac{\sqrt{3}}{\sqrt{2}}\frac{g_{4}g_{Y}}{g_{1}}\hat{e}{}_{i}^{\dagger}\bar{\sigma}^{\mu}\label{eq:Z'_couplings_3VL_e}\\
 & \times\left(\begin{array}{ccc}
\frac{1}{2}\left(s_{16}^{L}\right)^{2}-\left(c_{16}^{L}\right)^{2}\frac{g_{1}^{2}}{3g_{4}^{2}} & 0 & 0\\
0 & \frac{1}{2}\left(s_{25}^{L}\right)^{2}-\left(c_{25}^{L}\right)^{2}\frac{g_{1}^{2}}{3g_{4}^{2}} & \frac{1}{2}\left[\left(s_{34}^{L}\right)^{2}-\left(s_{25}^{L}\right)^{2}\right]s_{23}^{e}\\
0 & \frac{1}{2}\left[\left(s_{34}^{L}\right)^{2}-\left(s_{25}^{L}\right)^{2}\right]s_{23}^{e} & \frac{1}{2}\left(s_{34}^{L}\right)^{2}-\left(c_{34}^{L}\right)^{2}\frac{g_{1}^{2}}{3g_{4}^{2}}
\end{array}\right)\hat{e}_{j}Z'_{\mu}\,,\nonumber 
\end{flalign}
}at first order in $s^{e}_{23}$ and taking $c^{e}_{23}\approx 1$. The flavour-violating couplings above can lead to interesting signals in LFV processes such as $\tau\rightarrow3\mu$
and $\tau\rightarrow\mu\gamma$, see more in Section~\ref{subsec:LFV_processes}. 

In order to suppress LFV between the first and second lepton families,
a similar condition similar to Eq.~(\ref{eq:GIM-like_condition-1})
but for leptons can be implemented,
\begin{equation}
s_{25}^{L}=s_{16}^{L}\,.\label{eq:GIM-like_leptons-1}
\end{equation}
Remarkably, if the condition of Eq.~(\ref{eq:GIM-like_condition-1})
is fulfilled, then Eq.~(\ref{eq:GIM-like_leptons-1}) would also
be fulfilled in good approximation thanks to the underlying twin Pati-Salam
symmetry, the small breaking effects given by the splitting of vector-like
masses via $\left\langle \Omega_{15}\right\rangle $.

\section{Phenomenology of the extended model\label{subsec:Low-energy-phenomenology}}

The twin Pati-Salam model features a fermiophobic low-energy 4321 theory with
a rich phenomenology. Although extensive analyses of general 4321
models have been performed during the last few years, the vast majority
of them have been performed in the framework of non-fermiophobic 4321
models \cite{Cornella:2019hct,Cornella:2021sby,Barbieri:2022ikw,Bordone:2017bld,Bordone:2018nbg}.
Instead, the twin Pati-Salam model offers a fermiophobic scenario with a different
phenomenology. Being a theory of flavour, extra constraints and correlations
arise via the generation of the SM Yukawa couplings and the prediction
of fermion masses and mixing, including striking signals in LFV processes. Moreover,
the underlying twin Pati-Salam symmetry introduces universality (and
perturbativity) constraints over several parameters, which are not
present in alternative models. These features motivate a dedicated analysis.
We will highlight key observables for which the intrinsic nature of
the model can be disentangled from all alternative proposals. All low-energy observables considered are listed in Table~\ref{tab:Observables},
with references to current experimental bounds and links to theory
expressions.
\begin{table}[t]
\noindent \begin{centering}
\begin{tabular}{ccc}
\toprule 
\multicolumn{1}{c}{Observable} & Experiment/constraint  & Th. expr.\tabularnewline
\midrule
\midrule 
$\left[C_{\nu edu}^{*}\right]^{\tau\tau32}$ ($R_{D^{(*)}}$)  & $0.08\pm0.02$ (68\% CL)\cite{Angelescu:2021lln}  & (\ref{eq:R_D_WilsonCoefficient})\tabularnewline
\midrule 
$C_{9}^{\mu\mu}=-C_{10}^{\mu\mu}$ ($R_{K^{(*)}}^{2021}$)  & $[-0.31,-0.48]$ (68\% CL)\cite{Geng:2021nhg}  & (\ref{eq:R_K_WilsonCoefficients})\tabularnewline
\midrule 
$C_{9}^{\mu\mu}=-C_{10}^{\mu\mu}$ ($R_{K^{(*)}}^{2022}$)  & $[-0.01,-0.14]$ (68\% CL)(\ref{eq:C9_-C10})  & (\ref{eq:R_K_WilsonCoefficients})\tabularnewline
\midrule 
$\delta(\Delta M_{s})$  & $\apprle0.11$ (95\% CL) \cite{DiLuzio:2019jyq}  & (\ref{eq:delta_DeltaMs})\tabularnewline
\midrule 
$\mathcal{B}\left(\tau\rightarrow3\mu\right)$  & $<2.1\times10^{-8}$ (90\% CL)\cite{Hayasaka:2010np}  & (\ref{eq:tau_3mu})\tabularnewline
\midrule 
$\mathcal{B}\left(\tau\rightarrow\mu\gamma\right)$  & $<5.0\times10^{-8}$ (90\% CL)\cite{HFLAV:2019otj}  & (\ref{eq:tau_mu_photon})\tabularnewline
\midrule 
$\mathcal{B}\left(B_{s}\rightarrow\tau^{\pm}\mu^{\mp}\right)$  & $<3.4\times10^{-5}$ (90\% CL)\cite{LHCb:2019ujz}  & (\ref{eq:Bs_tau_mu})\tabularnewline
\midrule 
$\mathcal{B}\left(B^{+}\rightarrow K^{+}\tau^{\pm}\mu^{\mp}\right)$  & $<2.8\times10^{-5}$ (90\% CL)\cite{BaBar:2012azg}  & (\ref{eq:BKLFV1})\tabularnewline
\midrule 
$\mathcal{B}\left(\tau\rightarrow\mu\phi\right)$  & $<8.4\times10^{-8}$ (90\% CL)\cite{Belle:2011ogy}  & (\ref{eq:tau_muphi})\tabularnewline
\midrule 
$\mathcal{B}\left(K_{L}\rightarrow\mu^{\pm}e^{\mp}\right)$  & $<4.7\times10^{-12}$ (90\% CL) \cite{BNL:1998apv}  & (\ref{eq:KL_mue})\tabularnewline
\midrule 
$(g_{\tau}/g_{e,\mu})_{\ell+\pi+K}$  & $1.0003\pm0.0014$ \cite{HFLAV:2022wzx}  & (\ref{eq:LFUratios_tau})\tabularnewline
\midrule 
$\mathcal{B}\left(B_{s}\rightarrow\tau^{+}\tau^{-}\right)$  & $<5.2\times10^{-3}$ (90\% CL)\cite{LHCb:2017myy}  & (\ref{eq:Bs_tautau})\tabularnewline
\midrule 
$\mathcal{B}\left(B\rightarrow K\tau^{+}\tau^{-}\right)$  & $<2.25\times10^{-3}$ (90\% CL)\cite{BaBar:2016wgb}  & (\ref{eq:B_Ktautau})\tabularnewline
\midrule 
$\mathcal{B}\left(B^{+}\rightarrow K^{+}\nu\bar{\nu}\right)/\mathcal{B}\left(B^{+}\rightarrow K^{+}\nu\bar{\nu}\right)_{\mathrm{SM}}$  & $2.8\pm0.8$ (68\% CL) \cite{BelleIIEPS:2023} & (\ref{eq:BtoK_nunu})\tabularnewline
\midrule 
$\mathcal{B}\left(B\rightarrow K^{*}\nu\bar{\nu}\right)/\mathcal{B}\left(B\rightarrow K^{*}\nu\bar{\nu}\right)_{\mathrm{SM}}$  & $<2.7$ (90\% CL)\cite{BaBar:2013npw,Belle:2017oht}  & (\ref{eq:BtoK_nunu})\tabularnewline
\bottomrule
\end{tabular}
\par\end{centering}
\caption[Set of observables explored in the phenomenological analysis of the
extended twin PS model]{Set of observables explored in the phenomenological analysis, including
current experimental constraints.\label{tab:Observables}}
\end{table}

The benchmark points BP1 and BP2 in Table~\ref{tab:BP} address
the $R_{D^{(*)}}$ anomalies and are compatible with the 2021 and
2022 data on $R_{K^{(*)}}$, respectively, plus all the considered low-energy observables
and high-$p_{T}$ searches. They provide a good starting point to
study the relevant phenomenology, featuring typical configurations
of the model, and allow us to confront the 2021 picture of the model
versus the new situation with LFU preserved in $\mu/e$ ratios. Moreover,
they fit second and third family charged fermion masses and mixings,
featuring a down-aligned flavour structure with $\mathcal{O}(0.1)$
$\mu-\tau$ lepton mixing. The latter is more benchmark dependent,
with the common range being $s_{23}^{e}=[V_{cb},5V_{cb}]$. The case
$s_{23}^{e}\approx0.1$ is interesting because it leads to intriguing
signals in LFV processes, as we shall see. BP1 and BP2 also feature
$x_{25}^{\psi}\approx x_{16}^{\psi}$ and $M_{5}^{Q,L}\approx M_{6}^{Q,L}$,
providing a GIM-like suppression of 1-2 FCNCs.
\begin{table}[t]
\noindent \begin{centering}
\resizebox{\textwidth}{!}{
\begin{tabular}{lllllllll}
\toprule 
\multicolumn{4}{c}{Benchmark} &  & \multicolumn{4}{c}{Output}\tabularnewline
\midrule
\midrule 
$g_{4}$ & 3.5 & $\lambda_{15}^{44}$ & -0.5 &  & $s_{34}^{Q}$ & 0.978 & $M_{g'}$ & 3782.9 GeV\tabularnewline
\midrule 
$g_{3,2,1}$ & 1, 0.65, 0.36 & $\lambda_{15}^{55},\,\lambda_{15}^{66}$ & 2.5, 1.1 &  & $s_{34}^{L}$ & 0.977 & $M_{Z'}$ & 2414.3 GeV\tabularnewline
\midrule 
$x_{34}^{\psi}$ & 2 & $x_{42}^{\psi^{c}}$ & 0.4 &  & $s_{25}^{Q}=s_{16}^{Q}$ & $0.20^{*}$, $0.17^{**}$ & $s_{23}^{u}$ & 0.042556\tabularnewline
\midrule 
$x_{25}^{\psi}=x_{16}^{\psi}$ & $0.41^{*}$ , $0.35^{**}$ & $x_{43}^{\psi^{c}}$ & 1 &  & $s_{25}^{L}=s_{16}^{L}$ & 0.1455 & $s_{23}^{d}$ & 0.001497\tabularnewline
\midrule 
$M_{44}^{\psi}$ & 320 GeV & $M_{44}^{\psi^{c}}$ & 5 TeV &  & $s_{\theta_{LQ}}$ & 0.7097 & $s_{23}^{e}$ & -0.111\tabularnewline
\midrule 
$M_{55}^{\psi}$ & 780 GeV & $y_{53,43,34,24}^{\psi}$ & -0.3, 1, 1, 1 &  & $\hat{M}_{4}^{Q}$ & 1226.8 GeV & $V_{cb}$ & 0.04106\tabularnewline
\midrule 
$M_{66}^{\psi}$ & 1120 GeV & $\left\langle H_{t}\right\rangle $ & 177.2 GeV &  & $\hat{M}_{5}^{Q}$ & 1238.7 GeV & $m_{t}$ & 172.91 GeV\tabularnewline
\midrule 
$M_{45}^{\psi}$ & -700 GeV & $\left\langle H_{c}\right\rangle $ & 26.8 GeV &  & $\hat{M}_{4}^{L}$ & 614.04 GeV & $m_{c}$ & 1.270 GeV\tabularnewline
\midrule
$M_{54}^{\psi}$ & 50 GeV & $\left\langle H_{b}\right\rangle $ & 4.25 GeV &  & $\hat{M}_{5}^{L}$ & 845.26 GeV & $m_{b}$ & 4.180 GeV\tabularnewline
\midrule
$\left\langle \phi_{3}\right\rangle $ & 0.6 TeV & $\left\langle H_{s}\right\rangle $ & 2.1 GeV &  & $\hat{M}_{6}^{Q}$ & 1234.6 GeV & $m_{s}$ & 0.0987 GeV\tabularnewline
\midrule
$\left\langle \phi_{1}\right\rangle $ & 0.3 TeV & $\left\langle H_{\tau}\right\rangle $ & 1.75 GeV &  & $\hat{M}_{6}^{L}$ & 859.4 GeV & $m_{\tau}$ & 1.7765 GeV\tabularnewline
\midrule
$\left\langle \Omega_{15}\right\rangle $ & 0.4 TeV & $\left\langle H_{\mu}\right\rangle $ & 4.58 GeV &  & $M_{U_{1}}$ & 2987.1 GeV & $m_{\mu}$ & 105.65 MeV\tabularnewline
\bottomrule
\end{tabular}}
\par\end{centering}
\caption[Benchmarks for the extended twin PS model]{Input and output parameters for the benchmark points BP1 and BP2,
{*} indicates BP1 while {*}{*} indicates BP2, otherwise both benchmarks
share the same parameters. BP1 is compatible with 2021 data on $R_{K^{(*)}}$,
while BP2 is compatible with the 2022 updates by LHCb. \label{tab:BP}}
\end{table}

In the forthcoming sections we will assume the couplings of the fundamental
Lagrangian to be universal, such as $x_{34}^{\psi}$ and $x_{25}^{\psi}$,
however their universality is broken by small RGE effects which we
estimate in Section~\ref{subsec:Pertubativity} to be below 8\%.
We neglect the small RGE effects and preserve universal parameters
for the phenomenological analysis, in order to simplify the exploration
of the parameter space and highlight the underlying twin Pati-Salam
symmetry.

\subsection{\texorpdfstring{$R_{D^{(*)}}$ and $R_{K^{(*)}}$}{RD(*) and RK(*)}}

In our model, the left-handed WCs $\left[C_{\nu edu}^{*}\right]^{\tau\tau32}$
and $C_{9}^{\mu\mu}=-C_{10}^{\mu\mu}$ are obtained at tree-level
after integrating out the heavy gauge bosons, with the overall contribution
being dominated by $U_{1}$ tree-level exchange as in Fig.~\ref{fig:Leptoquark_RK_RD},
\begin{equation}
\left[C_{\nu edu}^{*}\right]^{\tau\tau32}(m_{b})=\frac{2\eta_{V}^{\nu\tau}}{V_{cb}}\left[C_{lq}^{(3)}\right]^{\tau\tau23}(\Lambda)\,,\label{eq:R_D_WilsonCoefficient}
\end{equation}
\begin{equation}
C_{9}^{\mu\mu}(m_{b})=-C_{10}^{\mu\mu}(m_{b})=-\frac{2\pi}{\alpha_{\mathrm{EM}}V_{tb}V_{ts}^{*}}\eta_{V}^{\ell\ell}\left(\left[C_{lq}^{(3)}\right]^{\mu\mu23}(\Lambda)+\left[C_{lq}^{(1)}\right]^{\mu\mu23}(\Lambda)\right)\,,\label{eq:R_K_WilsonCoefficients}
\end{equation}
where the negligible RGE effect is encoded as $\eta_{V}^{\nu\tau}\approx1.00144$,
$\eta_{V}^{\ell\ell}\approx0.974$ and has been computed with \texttt{DsixTools 2.1}
\cite{Fuentes-Martin:2020zaz} for $\Lambda=1\,\mathrm{TeV}$.
Notice also the tree-level matching of the twin Pati-Salam model to the SMEFT
in Section~\ref{subsec:EFT_model_Appendix}. In order to be compatible
with the SM-like $R_{K^{(*)}}$ ratios, constraints over $C_{9}^{\mu\mu}=-C_{10}^{\mu\mu}$
are shown in Table~\ref{tab:Observables}. Similarly, in order to
explain the $R_{D^{(*)}}$ anomalies there is a preferred region for
the $\left[C_{\nu edu}^{*}\right]^{\tau\tau32}$ coefficient, as shown in Table~\ref{tab:Observables}.
The full EFT description of these observables (including the definition
of the Wilson coefficients) plus a discussion of current experimental
data can be found in Sections~\ref{subsec:bsll} and \ref{subsec:bctaunu}.

In terms of fundamental parameters of the model, the deviations from
the SM in the LFU ratios scale as follows,
\begin{equation}
\left|\Delta R_{D^{(*)}}\right|\propto\left|\left(x_{34}^{\psi}\right)^{3}x_{25}^{\psi}\right|\,,\label{eq:RD_U1-1}
\end{equation}
\begin{equation}
\left|\Delta R_{K^{(*)}}\right|\propto\left|x_{34}^{\psi}\left(x_{25}^{\psi}\right)^{3}\right|\,,\label{eq:RK_U1-1}
\end{equation}
where we have fixed the vector-like masses and the 4321-breaking VEVs to the
values of our benchmark in Table~\ref{tab:BP}. This way, the Yukawa
couplings above control the contributions to most of the relevant
phenomenology, including the LFU ratios. The Pati-Salam universality
of $x_{34}^{\psi}$ and $x_{25}^{\psi}$ provides here a welcome constraint,
not present in other 4321 models. In particular, one can see that
both $R_{D^{(*)}}$ and $R_{K^{(*)}}$ are connected via the same
parameters and deviations in both are expected, while in general 4321 models
(such as the fermiophobic model of Refs.~\cite{DiLuzio:2017vat,DiLuzio:2018zxy}) the analog of a single $x_{i\alpha}^{\psi}$ decomposes into several
parameters for quarks and leptons, which decouple $R_{K^{(*)}}$ from
$R_{D^{(*)}}$.

Following from Eqs.~(\ref{eq:RD_U1-1}) and (\ref{eq:RK_U1-1}),
the cubic dependence of $R_{K^{(*)}}$ on $x_{25}^{\psi}$ anticipates
that we can suppress the contribution to $R_{K^{(*)}}$, while preserving
a large contribution to $R_{D^{(*)}}$ thanks to its linear dependence
on $x_{25}^{\psi}$. As a consequence, the yellow band of parameter
space preferred by 2022 $R_{K^{(*)}}$ is just shifted below the orange
band of 2021 $R_{K^{(*)}}$ in Fig.~\ref{fig:BsMixing_parameter_space}.
The 2022 $R_{K^{(*)}}$ band is compatible with $R_{D^{(*)}}$ at
$1\sigma$ only in a narrow region of the parameter space. This is
encouraging, given the fact that the model was built to address the
2021 tensions in both LFU ratios. However, in order to explain $R_{D^{(*)}}$,
small deviations from the SM in the $R_{K^{(*)}}$ ratios are unavoidable,
to be tested in the future via more precise measurements of LFU by
the LHCb collaboration. Moreover, lower central values for $R_{D^{(*)}}$
are also expected.

Remarkably, the fact that the twin Pati-Salam model only generates the effective
operator $(\bar{c}_{L}\gamma_{\mu}b_{L})$ $(\bar{\text{\ensuremath{\tau}}}_{L}\gamma^{\mu}\nu_{\tau L})$
implies that both $R_{D}$ and $R_{D^{*}}$ are corrected in the same
direction and with the same size, i.e.~$\Delta R_{D}=\Delta R_{D^{*}}$.
Instead, non-fermiophobic 4321 models also predict the scalar operator
$(\bar{c}_{L}b_{R})(\bar{\text{\ensuremath{\tau}}}_{R}\nu_{\tau L})$,
which leads to a larger correction for $R_{D}$ than that of $R_{D^{*}}$
(about 5/2 larger for the $\mathrm{PS}^{3}$ model, see Eq.~(27)
in \cite{Bordone:2017bld}). Current data is equally compatible with both
according to recent global fits \cite{Aebischer:2022oqe}.

\subsection{Off-shell photon penguin with tau leptons\label{subsec:Off-shell-photon-penguin}}

\begin{figure}[t]
\subfloat[\label{fig:Off-shell penguin}]{\noindent \begin{centering}
\resizebox{.44\textwidth}{!}{
\begin{tikzpicture}
	\begin{feynman}
		\vertex (a) {\(b_{L}\)};
		\vertex [right=20mm of a] (b);
		\vertex [below right=16mm of b] (c);
		\vertex [below=16mm of c] (f);
		\vertex [below left=16mm of f] (f1) {\(\ell^{+}\)};
		\vertex [below right=16mm of f] (f2) {\(\ell^{-}\)};
		\vertex [above right=16mm of c] (d);
		\vertex [right=16mm of d] (e) {\(s_{L}\)};
		\diagram* {
			(a) -- [fermion] (b) -- [boson, half left, blue, edge label'=\(U_{1}\)] (d) -- [fermion] (e),
			(b) -- [fermion, quarter right, edge label'={\(\tau_{L},\,E_{L4},\,E_{L5}\)}, inner sep=2pt] (c) -- [boson, blue, edge label'=\(\gamma\)] (f),
			(f1) -- [fermion] (f) -- [fermion] (f2),
			(c) -- [fermion, quarter right, edge label'={\(\tau_{L},\,E_{L4},\,E_{L5}\)}, inner sep=2pt] (d),
	};
	\end{feynman}
\end{tikzpicture}}
\par\end{centering}
}\subfloat[\label{fig:C9U}]{\includegraphics[scale=0.30]{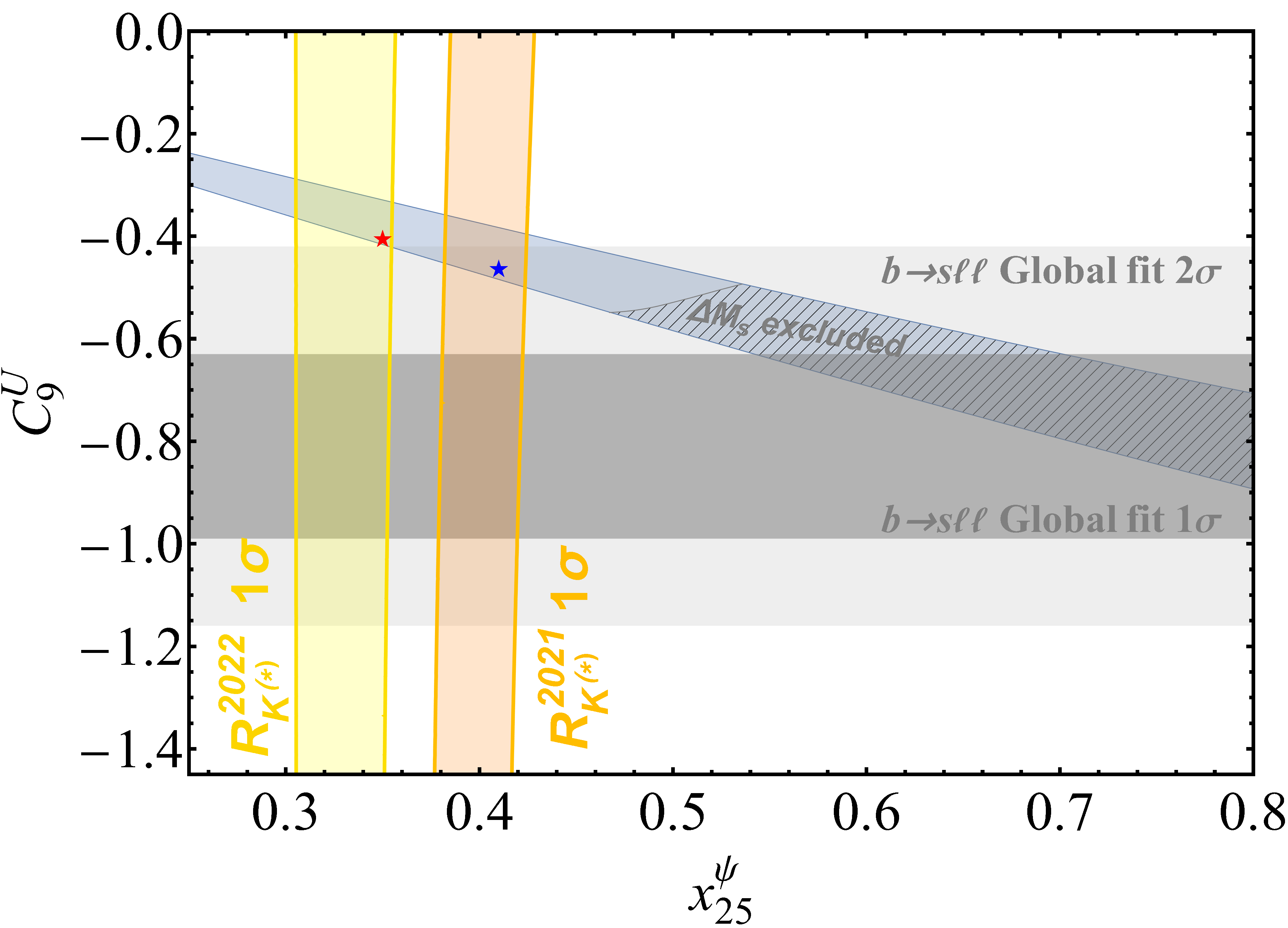}

}

\caption[$C_{9}^{U}$ in the extended twin PS model]{\textbf{\textit{Left:}}\textit{ }Off-shell photon penguin with tau
leptons in the loop that generates $C_{9}^{U}$, a contribution to
the lepton universal operator $\mathcal{O}_{9}^{23\ell\ell}$ that
participates in $b\rightarrow s\ell\ell$ transitions. \textbf{\textit{Right:}}
$C_{9}^{U}$ as a function of $x_{25}^{\psi}$ via Eq.~(\ref{eq:C9U}),
with $x_{34}^{\psi}$ varied in the range {[}1, 3.5{]} as preferred
by $R_{D^{(*)}}$ (blue region), with the rest of parameters fixed
as in Table~\ref{tab:BP}. The grey (light grey) region denotes the
1$\sigma$ ($2\sigma$) contour of $C_{9}^{U}$ as preferred by a
global fit to $b\rightarrow s\ell\ell$ data taken from \cite{Alguero:2021anc}.
The yellow (orange) band denotes the $1\sigma$ region preferred by
$R_{K^{(*)}}^{2022}$ ($R_{K^{(*)}}^{2021}$). The blue and red stars
denote BP1 and BP2 respectively.}
\end{figure}

As discussed in Section~\ref{subsec:ConnectionRDbsll}, the SMEFT
scenario $\left[C_{lq}^{(1)}\right]^{\tau\tau23}=\left[C_{lq}^{(3)}\right]^{\tau\tau23}$
correlates $R_{D^{(*)}}$ with a large contribution to $C_{9}^{\tau\tau}=-C_{10}^{\tau\tau}$,
which then mixes into a lepton universal contribution to the operator $C_{9}$
via RGE effects. The $U_{1}$ model is an specific example of this
scenario, where the same leptoquark couplings that explain $R_{D^{(*)}}$
are correlated to those in $b\rightarrow s\tau\tau$ due to $SU(2)_{L}$
invariance. The specific contributions to $b\rightarrow s\tau\tau$
will be explored in Section~\ref{subsec:Signals-in-rare-processes},
but in this section we explore the impact of the universal contribution
to $\mathcal{O}_{9}^{\ell\ell}$, denoted as $C_{9}^{U}$, over the
anomalous $b\rightarrow s\mu\mu$ transition. The contribution of $U_{1}$ to
$C_{9}^{U}$ originates from a 1-loop off-shell photon penguin diagram
with tau leptons running in the loop, shown in Fig.~\ref{fig:Off-shell penguin}.
Notice that this diagram is similar to that of Fig.~\ref{fig:Off-shell penguin_EFT} but exchanging
the insertion of the 4-fermion operator by the insertion of the $U_{1}$
leptoquark in the loop, plus adding new contributions provided by
the vector-like leptons of our model running in the loop. We finally obtain
\begin{flalign}
C_{9}^{U}=-\frac{v_{\mathrm{SM}}^{2}g_{4}^{2}}{6V_{tb}V_{ts}^{*}M_{U_{\text{1}}}^{2}} & \left(\log\left[\frac{2m_{b}^{2}}{g_{4}^{2}M_{U_{1}}^{2}}\right]\beta_{s\tau}\beta_{b\tau}^{*}+\log\left[\frac{2m_{E_{5}}^{2}}{g_{4}^{2}M_{U_{1}}^{2}}\right]\beta_{sE_{5}}\beta_{bE_{5}}^{*}\right.\label{eq:C9U}\\
 & \left.+\log\left[\frac{2m_{E_{4}}^{2}}{g_{4}^{2}M_{U_{1}}^{2}}\right]\beta_{sE_{4}}\beta_{bE_{4}}^{*}\right)\,,\nonumber 
\end{flalign}
which is explicitly correlated to $b\rightarrow s\tau\tau$, as well
as to $R_{D^{(*)}}$ since $SU(2)_{L}$ invariance implies $\beta_{s\tau}\approx\beta_{c\nu_{\tau}}$
for the $U_{1}$ couplings. Therefore, the scaling is $|C_{9}^{U}|\propto(x_{34}^{\psi})^{3}x_{25}^{\psi}$,
just like $R_{D^{(*)}}$. Notice that the contribution of our model
to $C_{9}^{U}$ differs from those of alternative models in the literature
\cite{Crivellin:2018yvo}
due to the vector-like leptons $E_{4,5}$ running in the loop. Unfortunately,
due to the flavour structure of our model, the contributions via vector-like
leptons interfere negatively with the leading contribution via the
tau loop, and hence our overall contribution to $C_{9}^{U}$ is generally
smaller than in the alternative models. The contribution from $E_{4}$
is negligible, but the contribution from $E_{5}$ reduces $C_{9}^{U}$
by a 20\% factor of the tau loop contribution.

In our model, $R_{D^{(*)}}$ and $R_{K^{(*)}}$ are correlated, as
can be seen from Eqs.~(\ref{eq:RD_U1-1}) and (\ref{eq:RK_U1-1}).
Therefore, $C_{9}^{U}$ is not only correlated with $R_{D^{(*)}}$
but also with $R_{K^{(*)}}$. Given that deviations from 1 in $R_{K^{(*)}}$
are now constrained by the new LHCb measurements, our final contribution
to $C_{9}^{U}$ is constrained to be $C_{9}^{U}\approx-0.4$, as can
be seen in Fig.~\ref{fig:C9U}. However, global fits of $b\rightarrow s\ell\ell$
data (see e.g.~\cite{Alguero:2021anc,Greljo:2022jac,Alguero:2023jeh}),
mostly driven by anomalies in $\mathrm{Br}(B\rightarrow K\mu\mu)$,
$\mathrm{Br}(B_{s}\rightarrow\phi\mu\mu)$ and $P'_{5}(B\rightarrow K^{*}\mu\mu)$
(see the discussion in Section~\ref{subsec:bsll}), prefer a larger
value $C_{9}^{U}\approx-0.8$. Therefore, we conclude that our model
is not able to fully address the anomalies in $b\rightarrow s\ell\ell$
via the off-shell photon penguin, although our contribution
to $C_{9}^{U}$ ameliorates the tensions. Performing a more
ambitious analysis would require to make assumptions about the hadronic
uncertainties afflicting $\mathrm{Br}(B\rightarrow K\mu\mu)$, $\mathrm{Br}(B_{s}\rightarrow\phi\mu\mu)$
and $P'_{5}(B\rightarrow K^{*}\mu\mu)$, which is beyond the scope
of this work.

However, we notice that in principle our model has the ideal structure
to the address all the anomalies in $B$-physics, i.e.~we reproduce
the preferred scenario presented in \cite{Alguero:2023jeh} (see also
Section~\ref{subsec:ConnectionRDbsll}) where a large
contribution to $\left[C_{\nu edu}^{*}\right]^{\tau\tau32}$ addresses $R_{D^{(*)}}$
and ameliorates tensions in $b\rightarrow s\mu\mu$ data, while a
small contribution to $C_{9}^{\mu\mu}=-C_{10}^{\mu\mu}$ further
improves the overall fit. Even though our contribution to $C_{9}^{U}$
does not reach the preferred values by the global fits, it improves
the overall description of $b\rightarrow s\mu\mu$ data with respect
to the SM.

\subsection{\texorpdfstring{$B_{s}-\bar{B}_{s}$ mixing}{Bs-Bsbar mixing} \label{subsec:BsMixing_revisited}}

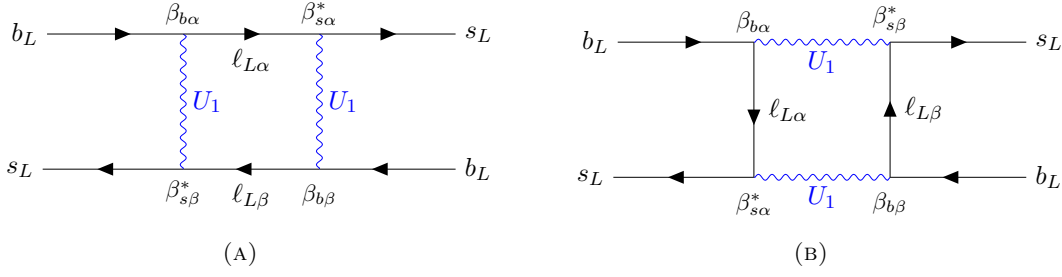
\begin{figure}[t]
\subfloat[]{\noindent \begin{centering}
\resizebox{.46\textwidth}{!}{
\begin{tikzpicture}
	\begin{feynman}
		\vertex (a) {\(s_{L}\)};
		\vertex [right=24mm of a] (b) [label={ [yshift=-0.7cm] \small $\beta_{s\beta}^{*}$}];
		\vertex [right=20mm of b] (c) [label={ [yshift=-0.7cm] \small $\beta_{b\beta}$}];
		\vertex [right=20mm of c] (d) {\(b_{L}\)};
		\vertex [above=20mm of b] (f1) [label={ \small $\beta_{b\alpha}$}];
		\vertex [above=20mm of c] (f2) [label={ \small $\beta_{s\alpha}^{*}$}];
		\vertex [left=20mm of f1] (f3) {\(b_{L}\)};
		\vertex [right=20mm of f2] (f4) {\(s_{L}\)};
		\diagram* {
			(a) -- [anti fermion] (b) -- [boson, blue, edge label'=\(U_{1}\)] (f1) -- [anti fermion] (f3),
			(b) -- [anti fermion, edge label'=\(\ell_{L\beta}\), inner sep=6pt] (c) -- [boson, blue, edge label'=\(U_{1}\)] (f2) -- [fermion] (f4),
			(c) -- [anti fermion] (d),
			(f1) -- [fermion, edge label'=\(\ell_{L\alpha}\), inner sep=6pt] (f2),
	};
	\end{feynman}
\end{tikzpicture}}
\par\end{centering}
}$\quad$\subfloat[]{\noindent \begin{centering}
\resizebox{.46\textwidth}{!}{
\begin{tikzpicture}
	\begin{feynman}
		\vertex (a) {\(s_{L}\)};
		\vertex [right=24mm of a] (b) [label={ [yshift=-0.7cm] \small $\beta_{s\alpha}^{*}$}];
		\vertex [right=20mm of b] (c) [label={ [yshift=-0.7cm] \small $\beta_{b\beta}$}];
		\vertex [right=20mm of c] (d) {\(b_{L}\)};
		\vertex [above=20mm of b] (f1) [label={ \small $\beta_{b\alpha}$}];
		\vertex [above=20mm of c] (f2) [label={ \small $\beta_{s\beta}^{*}$}];
		\vertex [left=20mm of f1] (f3) {\(b_{L}\)};
		\vertex [right=20mm of f2] (f4) {\(s_{L}\)};
		\diagram* {
			(a) -- [anti fermion] (b) -- [anti fermion, edge label'=\(\ell_{L\alpha}\), inner sep=6pt)] (f1) -- [anti fermion] (f3),
			(b) -- [boson, blue, edge label'=\(U_{1}\)] (c) -- [fermion, edge label'=\(\ell_{L\beta}\), inner sep=6pt)] (f2) -- [fermion] (f4),
			(c) -- [anti fermion] (d),
			(f1) -- [boson, blue, edge label'=\(U_{1}\)] (f2),
	};
	\end{feynman}
\end{tikzpicture}}
\par\end{centering}
}

\caption[$U_{1}$-mediated 1-loop diagrams contributing to $B_{s}-\bar{B}_{s}$
mixing]{$U_{1}$-mediated 1-loop diagrams contributing to $B_{s}-\bar{B}_{s}$
mixing. The indices $\alpha,\beta$ run for all charged leptons including
vector-like, i.e.~$\ell_{L\alpha}=\left(\mu_{L},\tau_{L},E_{L4},E_{L5}\right)$.
\label{fig:Box_BsMixing}}
\end{figure}

In the extended twin Pati-Salam model, tree-level contributions to $B_{s}-\bar{B}_{s}$
mixing via 2-3 quark mixing are suppressed due to the down-aligned
flavour structure achieved in Section~\ref{subsec:Effective_Yukawa_3VL}.
A further 1-loop contribution mediated by $U_{1}$ has been studied
in the literature \cite{DiLuzio:2018zxy,Cornella:2021sby,Fuentes-Martin:2020hvc}
for other 4321 models, and vector-like charged leptons are known to
play a crucial role. In Ref.~\cite{DiLuzio:2018zxy}, a framework
with three vector-like charged leptons was considered, however the loop function
was generalised from the SM $W^{\pm}$ box diagram, so the bounds were expected
to be slightly overestimated. Instead, in \cite{Fuentes-Martin:2020hvc}
the proper loop function was derived, but a framework with only one
vector-like charged lepton was considered. For this work, we have generalised
the loop function of \cite{Fuentes-Martin:2020hvc} to the case of
three vector-like leptons. The 1-loop contribution to the effective Wilson
coefficient $C_{1}^{bs}$ (see Section~\ref{subsec:BsMixing}) mediated
by $U_{1}$ reads, 
\begin{equation}
\left.C_{1}^{bs}\right|_{\mathrm{NP}}^{\mathrm{loop}}=\frac{g_{4}^{4}}{\left(8\pi M_{U_{1}}\right)^{2}}\sum_{\alpha,\beta}\left(\beta_{s\alpha}^{*}\beta_{b\alpha}\right)\left(\beta_{s\beta}^{*}\beta_{b\beta}\right)F(x_{\alpha},x_{\beta})\,,\label{eq:Cbs_U1}
\end{equation}
where $\alpha,\beta=\mu,\tau,E_{4},E_{5}$ run for all charged leptons,
including the vector-like partners (except for electrons and the sixth
charged lepton which do not couple to the second or third generation),
and $x_{\alpha}=(m_{\alpha}/M_{U_{1}})^{2}$. The contribution corresponds
to the box diagrams in Fig.~\ref{fig:Box_BsMixing}. The product
of couplings $\beta_{s\alpha}^{*}\beta_{b\alpha}$ has the fundamental
property
\begin{equation}
\sum_{\alpha}\beta_{s\alpha}^{*}\beta_{b\alpha}=0\,,\label{eq:unitarity_loop}
\end{equation}
which arises from unitarity of the transformations in Eq.~(\ref{eq:mixing_matrix_Q}).
This property, similarly to the GIM mechanism in the SM, is essential
to render the loop finite. However, the property holds as long as
the 2-3 down mixing and $s_{35}^{Q}$ are small. In particular, $s_{35}^{Q}$
is naturally small in the scenario $s_{34}^{Q}\approx1$, as it is
suppressed by the small cosine $c_{34}^{Q}$, see the definition of
$s_{35}^{Q}$ in Eq.~(\ref{eq:35_mixing}). Ultimately, the mixing
angle $s_{35}^{Q}$ is controlled by the fundamental parameter $x_{35}^{\psi}$,
and we obtained that $x_{35}^{\psi}\apprle0.09$ is required to pass
the $\Delta M_{s}$ bound.

The loop function reads
\begin{equation}
F(x_{\alpha},x_{\beta})=\left(1+\frac{x_{\alpha}x_{\beta}}{4}\right)B(x_{\alpha},x_{\beta})\,,
\end{equation}
where
\begin{equation}
B(x_{\alpha},x_{\beta})=\frac{1}{\left(1-x_{\alpha}\right)\left(1-x_{\beta}\right)}+\frac{x_{\alpha}^{2}\log x_{\alpha}}{\left(x_{\beta}-x_{\alpha}\right)\left(1-x_{\alpha}^{2}\right)}+\frac{x_{\beta}^{2}\log x_{\beta}}{\left(x_{\alpha}-x_{\beta}\right)\left(1-x_{\beta}^{2}\right)}\,.\label{eq:loop_function}
\end{equation}
In this manner, the loop function is dominated by the vector-like
partners because they are much heavier than chiral charged leptons.
In particular, in the motivated scenario with maximal $s_{34}^{L}$,
the couplings with the fourth family $\beta_{sE_{4}}^{*}\beta_{bE_{4}}$
are suppressed by the small cosine $c_{34}^{L}$. This way, the loop
is dominated by $E_{5}$ in good approximation. We obtain the effective
loop function in this scenario by removing all constants in $x_{\alpha,\beta}$,
which vanish due to the property (\ref{eq:unitarity_loop}), 
\begin{equation}
\tilde{F}(x)\approx F(x,x)-2F(x,0)+F(0,0)=\frac{x\left(x+4\right)\left(-1+x^{2}-2x\log x\right)}{4\left(x-1\right)^{3}}\,,\label{eq:F_tilde}
\end{equation}
and in this approximation the contribution to $C_{1}^{bs}$ reads
\begin{figure}[t]
\subfloat[\label{fig:ML5_deltaMs}]{\includegraphics[scale=0.36]{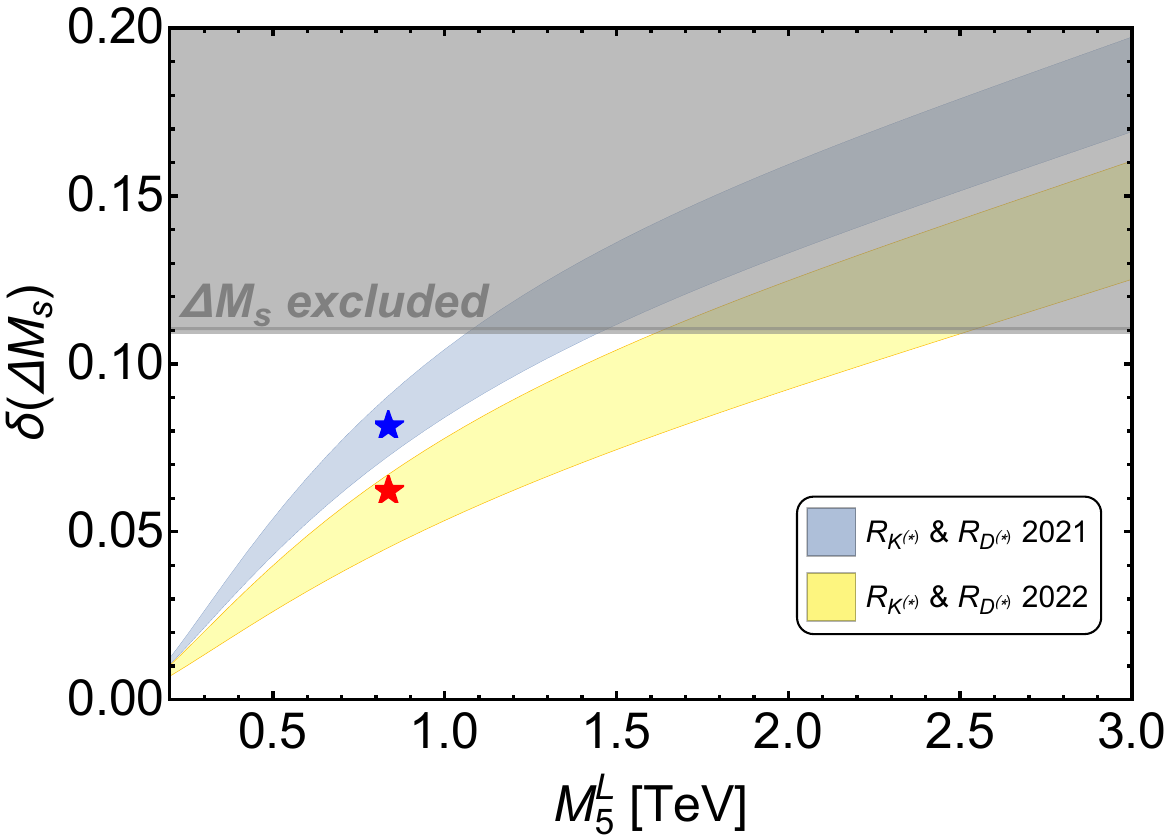}

}$\quad$\subfloat[\label{fig:BsMixing_parameter_space}]{\includegraphics[scale=0.36]{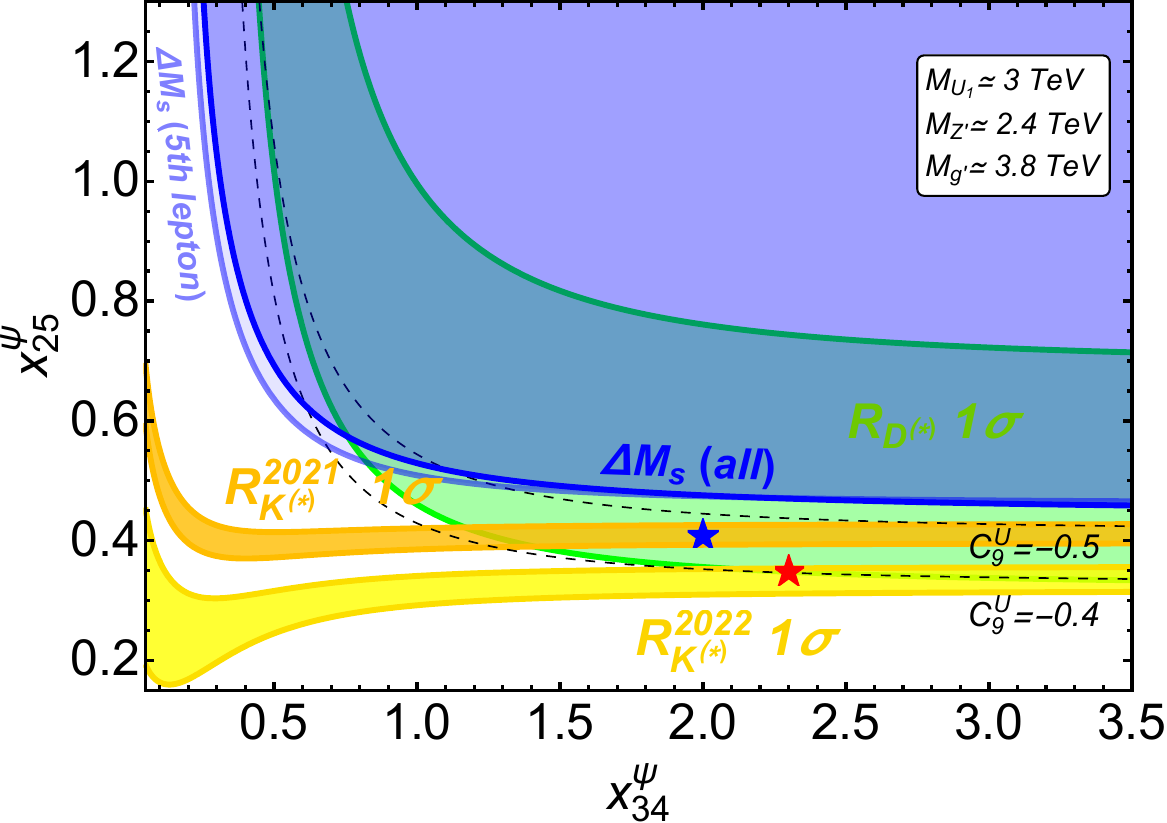}

}

\caption[Parameter space in the extended twin PS model relevant for $R_{D^{(*)}}$,
$R_{K^{(*)}}$ and $\Delta M_{s}$]{\textbf{\textit{Left:}}\textit{ }$\delta(\Delta M_{s})$ (Eq.~(\ref{eq:delta_DeltaMs}))
as a function of the 5th vector-like mass term. $x_{25}^{\psi}$ is
varied in the range $x_{25}^{\psi}=[0.3,\,0.35]$ ($[0.4,\,0.45]$)
preferred by $R_{K^{(*)}}^{2022}$ ($R_{K^{(*)}}^{2021}$), obtaining
the yellow (blue) band. The grey region is excluded by the $\Delta M_{s}$
bound, see Eq.~(\ref{eq:DeltaMs_bound}). \textbf{\textit{Right:}}
Parameter space in the plane ($x_{34}^{\psi}$, $x_{25}^{\psi}$)
compatible with $R_{D^{(*)}}$ and $R_{K^{(*)}}$ at 1$\sigma$. The
remaining parameters are fixed as in Table~\ref{tab:BP} for both
panels. The dashed lines show contours of constant $C_{9}^{U}$. The
blue region is excluded by the $\Delta M_{s}$ bound, the region excluded
only due to the contribution via the 5th lepton is also shown in lighter
blue for comparison. The blue and red stars denote BP1 and BP2 respectively.}
\end{figure}
\begin{equation}
\left.C_{1}^{bs}\right|_{\mathrm{NP}}^{\mathrm{loop}}=\frac{g_{4}^{4}}{\left(8\pi M_{U_{1}}\right)^{2}}\left(\beta_{sE_{5}}^{*}\beta_{bE_{5}}\right)^{2}\tilde{F}(x_{E_{5}})\,.\label{eq:Cbs_U1_5th}
\end{equation}
The loop function grows with $x_{E_{5}}$. However, in the limit of
large bare mass term $M_{5}^{L}$ the effective coupling $\beta_{sE_{5}}^{*}\propto s_{25}^{Q}$
vanishes (since large $M_{5}^{L}$ also implies large $M_{5}^{Q}$
due to the Pati-Salam symmetry), hence both the contribution to $C_{1}^{bs}$
and $R_{D^{(*)}}$ go away. In Fig.~\ref{fig:ML5_deltaMs} we plot
$\delta(\Delta M_{s})$ defined in Eq.~(\ref{eq:delta_DeltaMs})
in terms of $M_{5}^{L}$, and we vary $x_{25}^{\psi}$ in the ranges
compatible with $R_{D^{(*)}}$ and $R_{K^{(*)}}^{2022}$ ($R_{K^{(*)}}^{2021}$).
We can see that the $\Delta M_{s}$ bound requires a vector-like lepton
around 1.5-2 TeV in the 2022 case, while 2021 data was pointing to
a vector-like lepton with a mass around 1 TeV.

In Fig.~\ref{fig:BsMixing_parameter_space} we show that Eq.~(\ref{eq:Cbs_U1_5th})
is indeed a good approximation, up to small interference effects between
the 4th and 5th family contributions in the small $x_{34}^{\psi}$
region, where the fourth lepton is lighter. We also show the parameter
space compatible with $\Delta M_{s}$ and the LFU ratios in our benchmark
scenario. In particular, $\Delta M_{s}$ turns out to be the strongest
constraint over the parameter space other than $R_{K^{(*)}}^{2022}$.

\subsection{LFV processes\label{subsec:LFV_processes}}

\subsubsection*{$\boldsymbol{\tau\rightarrow3\mu}$}

The partial alignment condition of Eq.~(\ref{eq:GIM-like_leptons-1})
allows for $Z'$-mediated FCNCs in $\tau\mu$ processes, due to the
fact that the model predicts significant mixing between the muon and
tau charged leptons. This is a crucial prediction of the twin Pati-Salam theory
of flavour, not present in general 4321 models. Of particular interest
is the process $\tau\rightarrow3\mu$, which receives a tree-level
$Z'$-mediated contribution that grows with the $\tau\mu$ mixing angle $s_{23}^{e}$.
Beyond the latter, $\tau\rightarrow3\mu$ also receives a $U_{1}$-mediated
1-loop contribution
\begin{equation}
\left.\left[C_{ee}^{V,LL}\right]^{\mu\tau\mu\mu}\right|_{U_{1}}^{\mathrm{loop}}=\frac{3g_{4}^{4}}{128\pi^{2}M_{U_{1}}^{2}}\beta_{D_{5}\mu}^{*}\beta_{D_{5}\tau}\left(\beta_{D_{5}\mu}\right)^{2}\tilde{F}(x_{D_{5}})\,.\label{eq:Cbs_U1_5th-1}
\end{equation}
The effective coupling $\beta_{D_{5}\mu}$ is proportional to $s_{25}^{L}\approx0.1$,
which provides a further suppression of $\mathcal{O}((s_{25}^{L})^{3})$
that renders the loop negligible against the much larger tree-level
$Z'$-mediated contribution $\left.\left[C_{ee}^{V,LL}\right]^{\mu\tau\mu\mu}\right|_{Z'}$,
that is obtained from the tree-level matching to the SMEFT in Section~\ref{subsec:EFT_model_Appendix}.
The typical benchmark $s_{25}^{L}\approx0.1$ naturally suppresses
the $\mu\mu Z'$ coupling, keeping the $Z'$ contribution to $\tau\rightarrow3\mu$
under control, and simultaneously protects from $Z'\rightarrow\mu\mu$
dilepton searches at the LHC (see Section~\ref{subsec:Colliders}). 
\begin{figure}[t]
\subfloat[\label{fig:tau_3mu}]{\includegraphics[scale=0.36]{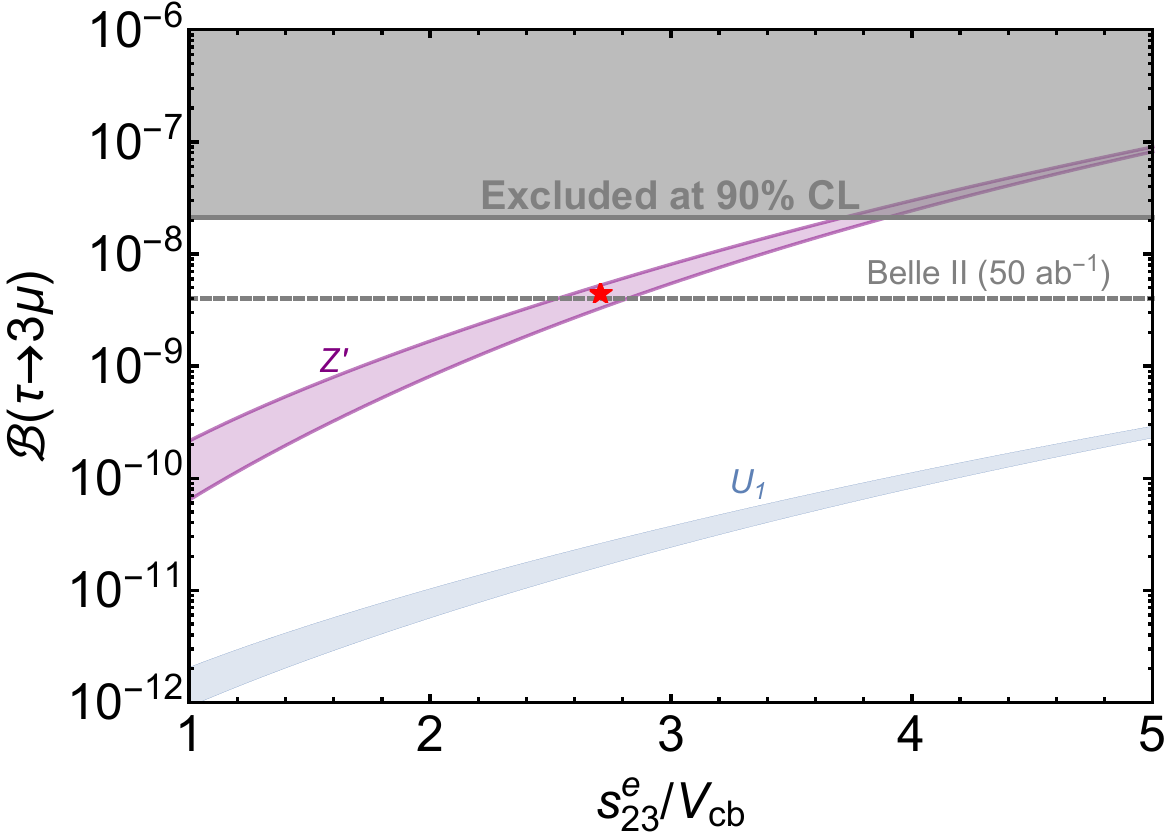}

}$\quad$\subfloat[\label{fig:tau_muphoton}]{\centering{}\includegraphics[scale=0.36]{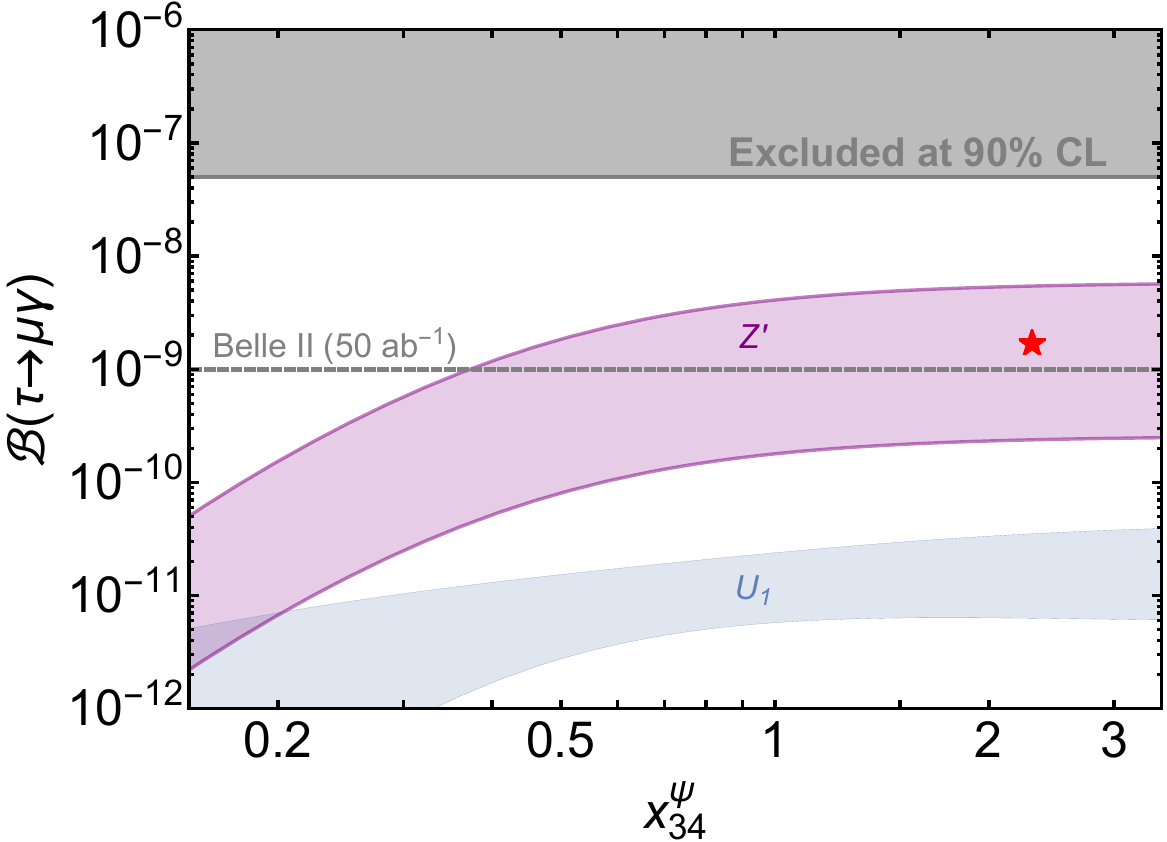}}

\caption[$\mathcal{B}\left(\tau\rightarrow3\mu\right)$ and $\mathcal{B}\left(\tau\rightarrow\mu\gamma\right)$
in the extended twin PS model]{\textbf{\textit{Left:}} $\mathcal{B}\left(\tau\rightarrow3\mu\right)$
as a function of the 2-3 charged lepton mixing sine $s_{23}^{e}$.
The purple region denotes the $Z'$ contribution while the blue region
denotes the $U_{1}$ contribution, for both we have varied $x_{25}^{\psi}=[0.3,0.35]$
which is compatible with $R_{K^{(*)}}^{2022}$. \textbf{\textit{Right:}}
$\mathcal{B}\left(\tau\rightarrow\mu\gamma\right)$ as a function
of $x_{34}^{\psi}$. The purple region denotes the $Z'$ contribution
for which we have varied $s_{23}^{e}=[V_{cb},5V_{cb}]$. The blue
region denotes the $U_{1}$ contribution, for which we have varied
$x_{25}^{\psi}=[0.1,1]$. The grey regions are excluded by the experiment,
the dashed lines show the projected future bound. The red star shows
BP2.}
\end{figure}

As depicted in Fig.~\ref{fig:tau_3mu}, the $Z'$ contribution dominates
over the $U_{1}$ contribution, and the regions of the parameter space
with very large $s_{23}^{e}$ are already excluded by the experiment.
We have chosen to plot the results of the 2022 case only, since this
observable depends mostly on $s_{23}^{e}$ and there is little variation
with 2021 data. The Belle~II collaboration will test a further region
of the parameter space \cite{Belle-II:2018jsg}, setting the bound
$s_{23}^{e}<2.8V_{cb}$ if no signal is detected. In general 4321
models (such as \cite{DiLuzio:2017vat,DiLuzio:2018zxy,Cornella:2019hct,Cornella:2021sby})
the $\mu-\tau$ mixing is unspecified, so only the small $U_{1}$
signal is predicted. Therefore, the large $Z'$ signal offers the
opportunity to disentangle the twin Pati-Salam model from alternative
4321 proposals.

As depicted in Fig.~\ref{fig:LFV_processes}, $\tau\rightarrow3\mu$
is the most constraining signal over the parameter space out of all
the LFV processes, provided that the 2-3 charged lepton mixing is
$\mathcal{O}(0.1)$.

\subsubsection*{$\boldsymbol{\tau\rightarrow\mu\gamma}$}

The dipole operator $[\mathcal{O}_{e\gamma}]^{\mu\tau}$ receives
1-loop contributions in our model via both $U_{1}$ and $Z'$,
\begin{equation}
[C_{e\gamma}]^{\mu\tau}=\left.[C_{e\gamma}]^{\mu\tau}\right|_{U_{1}}+\left.[C_{e\gamma}]^{\mu\tau}\right|_{Z'}\,,
\end{equation}
where
\begin{equation}
\left.[C_{e\gamma}]^{\mu\tau}\right|_{U_{1}}(\Lambda)=-\frac{C_{U}}{32\pi^{2}}\sum_{i}\beta_{i\mu}^{*}\beta_{i\tau}\left[G_{1}(x_{i})-2G_{2}(x_{i})\right]\,,
\end{equation}
\begin{equation}
\left.[C_{e\gamma}]^{\mu\tau}\right|_{Z'}(\Lambda)=-\frac{C_{Z'}}{32\pi^{2}}\sum_{\alpha}\xi_{\tau\alpha}\xi_{\alpha\mu}\widetilde{G}(x_{\alpha})\,,
\end{equation}
where $i=s,b,D_{4},D_{5}$ and $\alpha=\mu,\tau,E_{4},E_{5}$. The
loop functions are given by \cite{Cornella:2021sby,Fuentes-Martin:2020hvc,CarcamoHernandez:2019ydc}
\begin{equation}
G_{1}(x)=x\left[\frac{2-5x}{2\left(x-1\right)^{4}}\log x-\frac{4-13x+3x^{2}}{4\left(x-1\right)^{3}}\right]\,,
\end{equation}
\begin{equation}
G_{2}(x)=x\left[\frac{4x-1}{2\left(x-1\right)^{4}}x\log x-\frac{2-5x-3x^{2}}{4\left(x-1\right)^{3}}\right]\,,
\end{equation}
\begin{equation}
\widetilde{G}(x)=\frac{5x^{4}-14x^{3}+39x^{2}-38x-18x^{2}\log x+8}{12(1-x)^{4}}\,.
\end{equation}
The running of the dipole operator from $\Lambda=2\;\mathrm{TeV}$
to the scale $\mu\sim m_{\tau}$ is given by $[C_{e\gamma}]^{\mu\tau}(m_{\tau})\approx0.92[C_{e\gamma}]^{\mu\tau}(\Lambda)$,
as estimated with \texttt{DsixTools 2.1} \cite{Fuentes-Martin:2020zaz}. Neglecting
the muon mass, the branching ratio $\mathcal{B}(\tau\rightarrow\mu\gamma)$
is given by Eq.~(\ref{eq:tau_mu_photon}).

Provided that the 3-4 mixing is maximal, the $U_{1}$ loop is dominated
by the 5th vector-like quark, and in this situation the couplings
$\beta_{D_{5}\mu}^{*}\beta_{D_{5}\tau}$ are controlled by $x_{25}^{\psi}$.
The $Z'$ loop is dominated by chiral leptons, in particular by the
$\tau$ lepton, since the coupling $\xi_{\tau\tau}$ is maximal while
$\xi_{\mu\mu}$ is suppressed. In this scenario, the overall $Z'$
contribution is controlled by $\xi_{\tau\mu}$ which grows with the
$\mu-\tau$ mixing angle $s_{23}^{e}$, and the variation via $x_{25}^{\psi}$
is minimal.
\begin{figure}[t]
\noindent \begin{centering}
\subfloat[]{\begin{centering}
\resizebox{.46\textwidth}{!}{
\begin{tikzpicture}	
	\begin{feynman}
		\vertex (a) {\(\tau_{R}\)};
		\vertex [right=28mm of a] (b) [label={ [xshift=0.1cm, yshift=-0.55cm] \small $\beta_{i\tau}$}];
		\vertex [right=20mm of b] (c) [label={ [xshift=0.1cm, yshift=-0.6cm] \small $\beta_{i\mu}^{*}$}];
		\vertex [right=20mm of c] (d){\(\mu_{L}\)};
		\diagram* {
			(a) -- [edge label'=\(\tau_{L}\), near end, inner sep=6pt, insertion=0.5] (b) -- [boson, half left, edge label=$U_{1}$] (c),
			(b) --  [fermion, edge label'=\(d_{Li}\), inner sep=6pt] (c),
			(c) -- [fermion] (d),
	};
	\end{feynman}
\end{tikzpicture}}
\par\end{centering}
}$\;$\subfloat[]{\begin{centering}
\resizebox{.46\textwidth}{!}{
\begin{tikzpicture}	
	\begin{feynman}
		\vertex (a) {\(\tau_{R}\)};
		\vertex [right=28mm of a] (b) [label={ [xshift=0.1cm, yshift=-0.65cm] \small $\xi_{\tau\alpha}$}];
		\vertex [right=20mm of b] (c) [label={ [xshift=0.1cm, yshift=-0.7cm] \small $\xi_{\alpha\mu}$}];
		\vertex [right=20mm of c] (d){\(\mu_{L}\)};
		\diagram* {
			(a) -- [edge label'=\(\tau_{L}\), near end, inner sep=6pt, insertion=0.5] (b) -- [boson, half left, edge label=$Z'$] (c),
			(b) --  [fermion, edge label'=\(\ell_{L\alpha}\), inner sep=4pt] (c),
			(c) -- [fermion] (d),
	};
	\end{feynman}
\end{tikzpicture}}
\par\end{centering}
}
\par\end{centering}
\caption[Diagrams contributing to $\mathcal{B}\left(\tau\rightarrow\mu\gamma\right)$
in the extended twin PS model]{$U_{1}$ (left panel) and $Z'$ (right panel) 1-loop contributions
to $\tau\rightarrow\mu\gamma$. Photon lines are implicit. The index
$i$ runs for all down-quarks including vector-like, i.e~ $d_{Li}=\left(s_{L},b_{L},D_{L4},D_{L5}\right)$,
while $\alpha$ runs for all charged leptons including vector-like,
i.e.~$\ell_{L\alpha}=\left(\mu_{L},\tau_{L},E_{L4},E_{L5}\right)$.
\label{fig:diagram_tau_mu_photon}}
\end{figure}
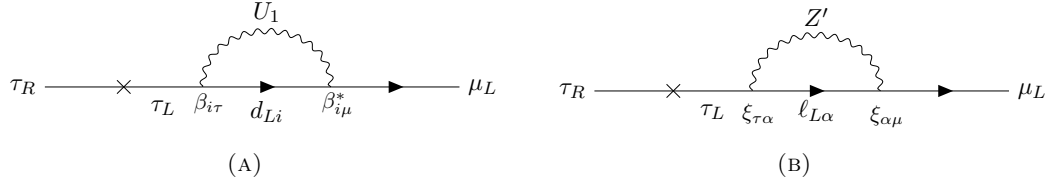

In Fig.~\ref{fig:tau_muphoton} we can see that the $Z'$ contribution
dominates the branching fraction in the range of large $x_{34}^{\psi}$
motivated by $R_{D^{(*)}}$, leading to the predictions for $\mathcal{B}\left(\tau\rightarrow\mu\gamma\right)$
being one/three orders of magnitude below the current experimental
limit depending on the value of $s_{23}^{e}$. We have also included
the bound projected by Belle~II \cite{Belle-II:2018jsg},
which will partially test the parameter space. In the 4321 models
of \cite{DiLuzio:2017vat,DiLuzio:2018zxy} the $\mu-\tau$ mixing
is unspecified, so only the blue $U_{1}$ signal is predicted. For
non-fermiophobic models, this signal is largely enhanced via a chirality
flip involving the bottom quark in the loop \cite{Cornella:2019hct,Cornella:2021sby,Barbieri:2022ikw,Fuentes-Martin:2020bnh,Fuentes-Martin:2020hvc},
predicting a larger signal $\mathcal{B}\left(\tau\rightarrow\mu\gamma\right)\approx10^{-8}$.
Instead, our $Z'$ signal lies below, offering the opportunity to
disentangle the twin Pati-Salam model from all alternative proposals.

\subsubsection*{$\boldsymbol{B_{s}\rightarrow\tau\mu}$, $\boldsymbol{B\rightarrow K\tau\mu}$
and $\boldsymbol{\tau\rightarrow\mu\phi}$ }

\begin{figure}[t]
\subfloat[\label{fig:Bs_to_taumu}]{\includegraphics[scale=0.36]{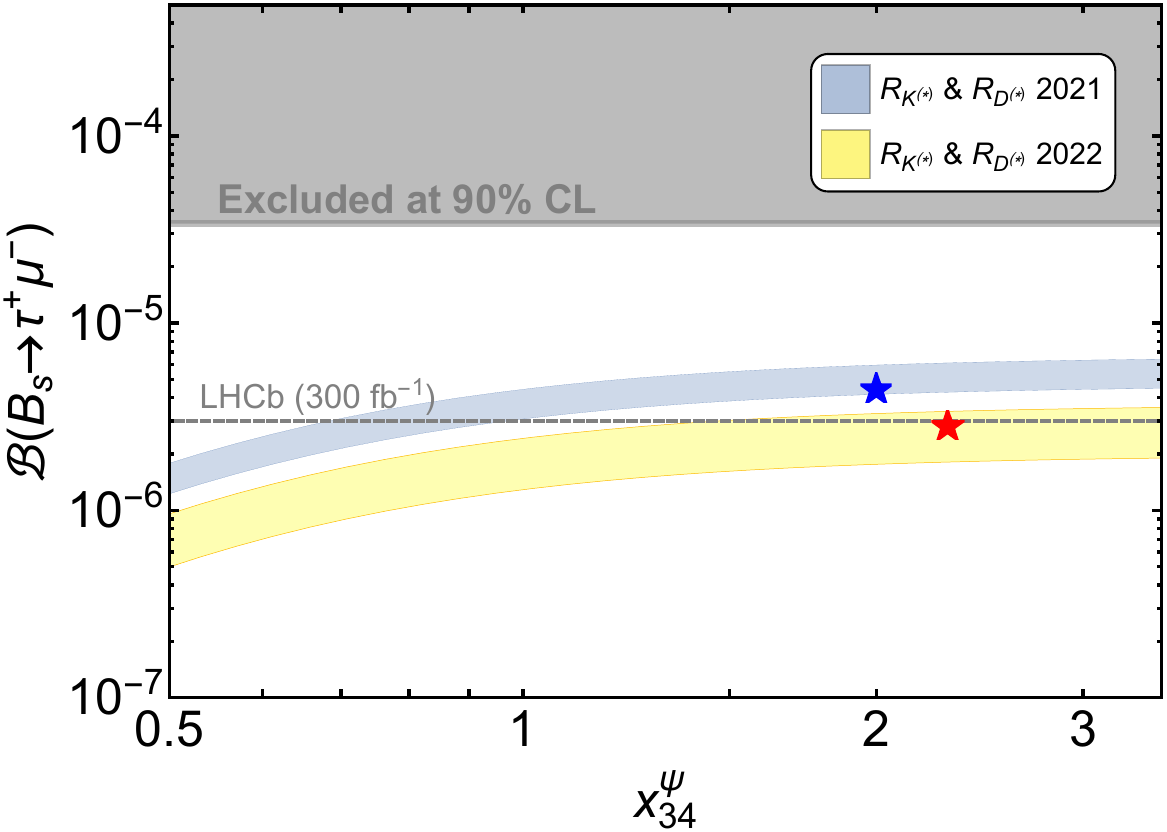}

}$\quad$\subfloat[\label{fig:LFV_processes}]{\centering{}\includegraphics[scale=0.36]{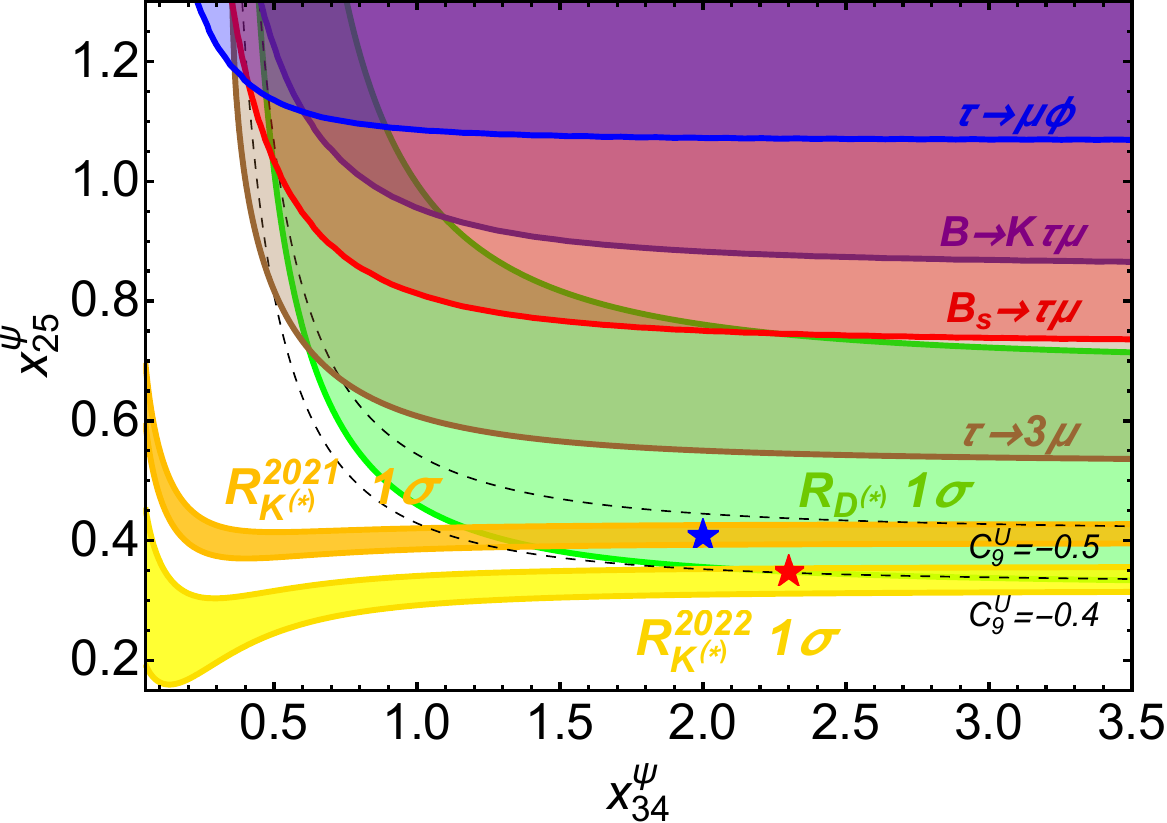}}

\caption[Constraints over the parameter space of the extended twin PS model
via LFV observables]{\textbf{\textit{Left:}} $\mathcal{B}\left(B_{s}\rightarrow\tau^{+}\mu^{-}\right)$
as a function of $x_{34}^{\psi}$. The yellow (blue) band is obtained
by varying $x_{25}^{\psi}$ in the range $x_{25}^{\psi}=[0.3,\,0.35]$($[0.4,\,0.45]$)
preferred by $R_{K^{(*)}}^{2022}$ ($R_{K^{(*)}}^{2021}$). The grey
region is excluded by the experiment, the dashed line shows the projected
future bound. \textbf{\textit{Right: }}Parameter space in the plane
($x_{34}^{\psi}$, $x_{25}^{\psi}$) compatible with $R_{D^{(*)}}$
and $R_{K^{(*)}}$ at 1$\sigma$. The remaining parameters are fixed
as in Table \ref{tab:BP}. The dashed lines show contours of constant
$C_{9}^{U}$. The regions excluded by LFV violating processes are
displayed. The blue (red) star shows BP1 (BP2). \label{fig:LFV_Btaumu}}
\end{figure}
The vector leptoquark $U_{1}$ mediates tree-level contributions to
flavour-violating (semi) leptonic $B$-decays to (kaons), taus and
muons. The experimental bound for $B_{s}\rightarrow\tau\mu$ was obtained
by LHCb \cite{LHCb:2019ujz}, while for $B\rightarrow K\tau\mu$ experimental
bounds are only available for the decays $B^{+}\rightarrow K^{+}\tau\mu$
\cite{BaBar:2012azg}. The process $\tau\rightarrow\mu\phi$ receives
tree-level contributions from both $U_{1}$ and $Z'$. However, $\tau\rightarrow\mu\phi$
turns out to be suppressed by the small effective couplings $\beta_{s\mu}\propto s_{25}^{Q}s_{25}^{L}$
and $\xi_{ss}\propto\left(s_{25}^{Q}\right)^{2}$, so that we find
$\mathcal{B}\left(\tau\rightarrow\mu\phi\right)\approx10^{-9}$, roughly
two orders of magnitude below the current experimental bounds, and
just below the future sensitivity of Belle~II \cite{Belle-II:2018jsg}.

As can be seen in Fig.~\ref{fig:LFV_processes}, $B_{s}\rightarrow\tau^{+}\mu^{-}$
implies the strongest constraint over the parameter space out of all
semileptonic LFV processes involving $\tau$ leptons, followed by
$B^{+}\rightarrow K^{+}\tau^{+}\mu^{-}$ and $\tau\rightarrow\mu\phi$.
The present experimental bounds lead to mild constraints over the
parameter space compatible with $R_{D^{(*)}}$. As depicted in Fig.~\ref{fig:Bs_to_taumu},
the 2021 region for $B_{s}\rightarrow\tau^{+}\mu^{-}$ was partially
within LHCb projected sensitivity, but the 2022 region will mostly
remain untested.

\subsubsection*{$\boldsymbol{K_{L}\rightarrow\mu e}$}
\begin{table}[t]
\begin{centering}
\begin{tabular}{cc}
\toprule 
field & $\mathbb{Z}_{2}$\tabularnewline
\midrule
\midrule 
$\overline{\psi}_{6}$, $\psi_{6}$ & 1\tabularnewline
\midrule 
$\overline{\psi'}_{6}$, $\psi'_{6}$ & -1\tabularnewline
\midrule 
$\chi$ & -1\tabularnewline
\bottomrule
\end{tabular}$\qquad\qquad$%
\begin{tabular}{cccc}
\toprule 
\multicolumn{2}{c}{Input} & \multicolumn{2}{c}{Output}\tabularnewline
\midrule
\midrule 
$M_{66}^{\psi}$ & 900 GeV & $\hat{M}_{66}^{Q}$ & 1211 GeV\tabularnewline
\midrule 
$M_{66'}^{\psi}$ & 1100 GeV & $\hat{M}_{66}^{L}$ & 834 GeV\tabularnewline
\midrule 
$x_{66}\left\langle \chi\right\rangle $ & -700 & $s_{66}^{Q}$ & 0.298\tabularnewline
\midrule 
$x'_{66}\left\langle \chi\right\rangle $ & 680 & $s_{66}^{L}$ & 0.967\tabularnewline
\midrule 
$\lambda_{15}^{66}$, $\lambda_{15}^{66'}$ & 1.5, 2.5 & $\cos\theta_{6}$ & 0.045\tabularnewline
\bottomrule
\end{tabular}
\par\end{centering}
\caption[New fields and benchmark parameters to dillute $\beta_{de}$]{\textbf{\textit{Left:}} Charge assignments under $\mathbb{Z}_{2}$ that allow
the desired mixing. \textbf{\textit{Right:}} Benchmark parameters
which lead to a dilution $\epsilon<0.1$. \label{tab:benchmark_suppression}}
\end{table}
The LFV process $K_{L}\rightarrow\mu e$ sets a strong bound
over all models featuring a vector leptoquark $U_{1}$ coupled
to the first and second families. In our model, the contribution is
proportional to the couplings $\left|\beta_{de}\beta_{s\mu}^{*}\right|^{2}$
(see the tree-level matching to the SMEFT in Section~\ref{subsec:EFT_model_Appendix}
and the EFT description of $K_{L}\rightarrow\mu e$ in Section~\ref{subsec:Semileptonic-CLFV-processes}).
The first family coupling $\beta_{de}$ can be diluted via mixing
with vector-like fermions, which we parameterised via the effective
parameter $\epsilon$ in Eq.~(\ref{eq:LQ_couplings}), so that $\beta_{se}\approx s_{16}^{Q}s_{16}^{L}\epsilon$. 

\begin{figure}[t]
\subfloat[\label{fig:KL_mue}]{\includegraphics[scale=0.35]{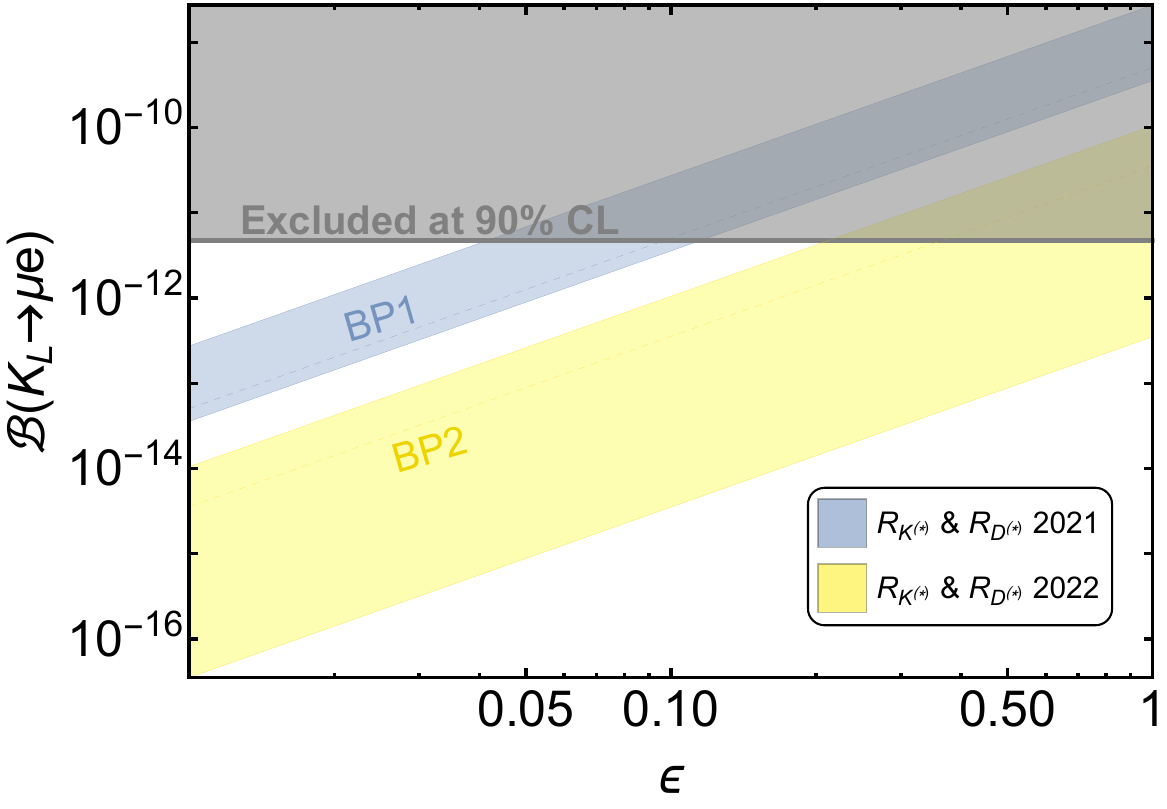}

}$\quad$\subfloat[\label{fig:LFU_ratios_tau}]{\includegraphics[scale=0.36]{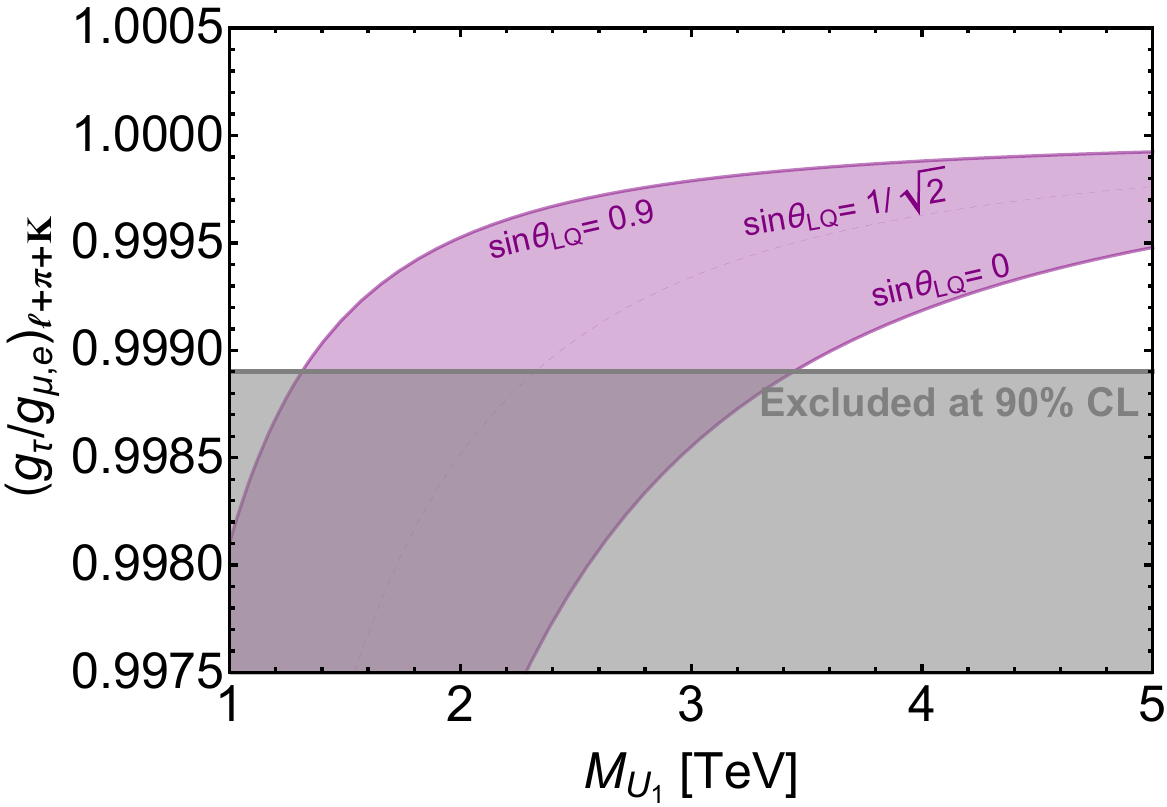}

}

\caption[$\mathcal{B}\left(K_{L}\rightarrow\mu^{\pm}e^{\mp}\right)$ and LFU ratios originated
from $\tau$ decays in the extended twin PS model]{\textbf{\textit{Left:}}\textit{ }$\mathcal{B}\left(K_{L}\rightarrow\mu^{\pm}e^{\mp}\right)$
(Eq.~(\ref{eq:KL_mue})) as a function of $\epsilon$ (see main text
for details). $x_{25}^{\psi}$ is varied in the range $x_{25}^{\psi}=[0.3,\,0.35]$
($[0.4,\,0.45]$) preferred by $R_{K^{(*)}}^{2022}$ ($R_{K^{(*)}}^{2021}$),
obtaining the yellow (blue) band. \textbf{\textit{Right:}} LFU
ratios originated from $\tau$ decays (Eq.~(\ref{eq:LFUratios_tau}))
as a function of the mass of the vector leptoquark $M_{U_{1}}$, $\sin\theta_{LQ}$
is varied in the range $\sin\theta_{LQ}=[0,0.9]$ and $g_{4}=3.5$.
The remaining parameters are fixed as in Table~\ref{tab:BP} for
both panels, and current exclusion limits are shown.}
\end{figure}

This can be done by adding an extra sixth-primed vector-like family transforming
in the same way as the sixth family under the twin Pati-Salam symmetry,
but discriminated by a flavour symmetry which we assume as $\mathbb{Z}_{2}$
for simplicity (we could use the $\mathbb{Z}_{4}$ symmetry of the model as
well), which forbids mixing between the sixth-primed family and any
chiral family. Instead, mixing between the sixth and sixth-primed
fermion families is allowed via a twin Pati-Salam singlet charged
under the new $\mathbb{Z}_{2}$, i.e.
\begin{equation}
\mathcal{L}_{\mathrm{mix}}=x_{66}\chi\psi'_{6}\overline{\psi}_{6}+x'_{66}\chi^{*}\psi{}_{6}\overline{\psi}'_{6}\,+\mathrm{h.c.}
\end{equation}
The mass terms of the sixth and sixth-primed fields are split via
$\Omega_{15}$ in the usual way,
\begin{equation}
\mathcal{L}_{\mathrm{mass}}=(M_{66}^{\psi}+\lambda_{15}^{66}T_{15}\Omega_{15})\psi_{6}\overline{\psi}_{6}+(M_{66'}^{\psi}+\lambda_{15}^{66'}T_{15}\Omega_{15})\psi'_{6}\overline{\psi}'_{6}+\mathrm{h.c.}
\end{equation}
After $\Omega_{15}$ and the singlet $\chi$ develop VEVs, we obtain
the following mass matrices for quarks and leptons
\begin{equation}
\mathcal{L}_{\mathrm{mass}}+\mathcal{L}_{\mathrm{mix}}=\left(\begin{array}{@{}lcc@{}}
 & Q_{6} & Q'_{6}\\
\cmidrule(l){2-3}\left.\overline{Q}{}_{6}\right| & M_{66}^{Q} & x_{66}\left\langle \chi\right\rangle \\
\left.\overline{Q}'_{6}\right| & x'_{66}\left\langle \chi\right\rangle  & M_{66'}^{Q}
\end{array}\right)+\left(\begin{array}{@{}lcc@{}}
 & L_{6} & L'_{6}\\
\cmidrule(l){2-3}\left.\overline{L}{}_{6}\right| & M_{66}^{L} & x_{66}\left\langle \chi\right\rangle \\
\left.\overline{L}'_{6}\right| & x'_{66}\left\langle \chi\right\rangle  & M_{66'}^{L}
\end{array}\right)+\mathrm{h.c.}\,,\label{eq:MassMatrix_4thVL_effective_up-1-1}
\end{equation}
where we have defined
\begin{equation}
M_{66}^{Q}=M_{66}^{\psi}+\frac{\lambda_{15}^{66}}{2\sqrt{6}}\left\langle \Omega_{15}\right\rangle \,,\quad M_{66}^{L}=M_{66}^{\psi}-3\frac{\lambda_{15}^{66}}{2\sqrt{6}}\left\langle \Omega_{15}\right\rangle \,,
\end{equation}
\begin{equation}
M_{66'}^{Q}=M_{66'}^{\psi}+\frac{\lambda_{15}^{66'}}{2\sqrt{6}}\left\langle \Omega_{15}\right\rangle \,,\quad M_{66'}^{L}=M_{66'}^{\psi}-3\frac{\lambda_{15}^{66'}}{2\sqrt{6}}\left\langle \Omega_{15}\right\rangle \,.
\end{equation}
The mass matrices in Eq.~(\ref{eq:MassMatrix_4thVL_effective_up-1-1})
are diagonalised by different unitary transformations in the quark
and lepton sector, $V_{66'}^{Q}$ and $V_{66'}^{L}$, in such a way
that the $U_{1}$ couplings are given by
\begin{equation}
\mathcal{L}_{U_{1}}=\frac{g_{4}}{\sqrt{2}}\left(\begin{array}{cc}
Q_{6}^{\dagger} & Q_{6}^{\dagger'}\end{array}\right)\gamma_{\mu}V_{66'}^{Q}\mathrm{diag}(1,1)V_{66'}^{L\dagger}\left(\begin{array}{c}
L_{6}\\
L'_{6}
\end{array}\right)U_{1}^{\mu}+\mathrm{h.c.}
\end{equation}
If we define 
\begin{equation}
V_{66'}^{Q}V_{66'}^{L\dagger}\equiv\left(\begin{array}{cc}
\cos\theta_{6} & \sin\theta_{6}\\
-\sin\theta_{6} & \cos\theta_{6}
\end{array}\right)\,,
\end{equation}
then the first family $U_{1}$ coupling receives a suppression via
$\cos\theta_{6}$ as
\begin{equation}
\beta_{de}=s_{16}^{Q}s_{16}^{L}\cos\theta_{6}\,.
\end{equation}
which is identified with the suppression parameter $\epsilon$ in
Eq.~(\ref{eq:LQ_couplings}),
\begin{equation}
\epsilon\equiv\cos\theta_{6}.
\end{equation}
We can achieve values of $\cos\theta_{6}$ smaller than 0.1 without
any aggressive tuning of the parameters, obtaining the mild suppression
desired for $K_{L}\rightarrow\mu e$ as per Fig.~\ref{fig:KL_mue}.
A suitable benchmark can be found in Table~\ref{tab:benchmark_suppression}.
Interestingly, this mechanism does not affect the $Z'$ and $g'$
interactions, as the unitary matrices $V_{66'}^{Q}$ and $V_{66'}^{L}$
cancel in neutral currents. This allows the GIM-like protection from
1-2 FCNCs to remain in place for both the quark and lepton sectors
via $s_{16}^{Q}=s_{25}^{Q}$ and $s_{16}^{L}=s_{25}^{L}$, without
entering in conflict with $K_{L}\rightarrow\mu e$ nor with $B$-physics.

In Fig.~\ref{fig:KL_mue} we can see that for the 2022 case, some
region of the parameter space is compatible with $K_{L}\rightarrow\mu e$
without the need of diluting the coupling. Instead, for the benchmark
values BP1 and BP2, a mild suppression is required. This signal is a direct consequence
of the underlying twin Pati-Salam symmetry and the GIM-like mechanism, which
lead to quasi-degenerate mixing angles $s_{16}^{Q}\approx s_{16}^{L}$
that are equal to their 25 counterparts, and as a consequence $\beta_{de}\neq0$.
Therefore, it is not present in other 4321 models \cite{DiLuzio:2017vat,DiLuzio:2018zxy,Cornella:2019hct,Cornella:2021sby,Barbieri:2022ikw}.

\subsection{Tests of universality in leptonic tau decays}

NP contributions to $R_{D^{(*)}}$ commonly involve large couplings
to tau leptons, which can have an important effect over LFU ratios
originated from tau decays. Such tests are constructed by performing
ratios of the partial widths of the tau lepton decaying to lighter leptons
and/or hadrons. We find all ratios in our model to be well approximated
by (see the tree-level matching to the SMEFT in Section~\ref{subsec:EFT_model_Appendix}
and the EFT description of $\tau$ LFU ratios in Section~\ref{subsec:Universality-in-tau}),
\begin{equation}
\left(\frac{g_{\tau}}{g_{\mu,e}}\right)_{\ell+\pi+K}\approx1-0.079C_{U}\left|\beta_{b\tau}\right|^{2}\,,\label{eq:LFUratios_tau}
\end{equation}
where $\beta_{b\tau}\approx\cos\theta_{LQ}$ assuming maximal 3-4
mixing. Therefore, it can be seen as a constraint over the $\beta_{b\tau}$
coupling, and hence is not directly related to $R_{K^{(*)}}$ so we
do not plot two bands here. The high-precision measurements of these
effective ratios only allow for per mil modifications, see the HFLAV
average \cite{HFLAV:2022wzx} in Table~\ref{tab:Observables}. As
depicted in Fig.~\ref{fig:LFU_ratios_tau}, this constraint sets
the lower bound $M_{U_{1}}\gtrsim2.2\,\mathrm{TeV}$ for $\sin\theta_{LQ}=1/\sqrt{2}$
and $g_{4}=3.5$. This bound becomes more restrictive for $\cos\theta_{LQ}\approx1$,
or equivalently $\beta_{b\tau}\approx1$, for which we find $M_{U_{1}}\gtrsim3.3\,\mathrm{TeV}$
if $g_{4}=3.5$ and $M_{U_{1}}\gtrsim2.9\,\mathrm{TeV}$ if $g_{4}=3$.
The latter case, more constrained by data, is predicted by non-fermiophobic
4321 models such as $\mathrm{PS^{3}}$ and variants \cite{Cornella:2019hct,Cornella:2021sby,Barbieri:2022ikw,Bordone:2017bld,Bordone:2018nbg}.

\subsection{\texorpdfstring{Signals in rare $B$-decays}{Signals in rare B-decays} \label{subsec:Signals-in-rare-processes}}

\subsubsection*{$\boldsymbol{B_{s}\rightarrow\tau\tau}$ and $\boldsymbol{B\rightarrow K\tau\tau}$}

\begin{figure}[t]
\subfloat[\label{fig:Bs_tautau}]{\includegraphics[scale=0.355]{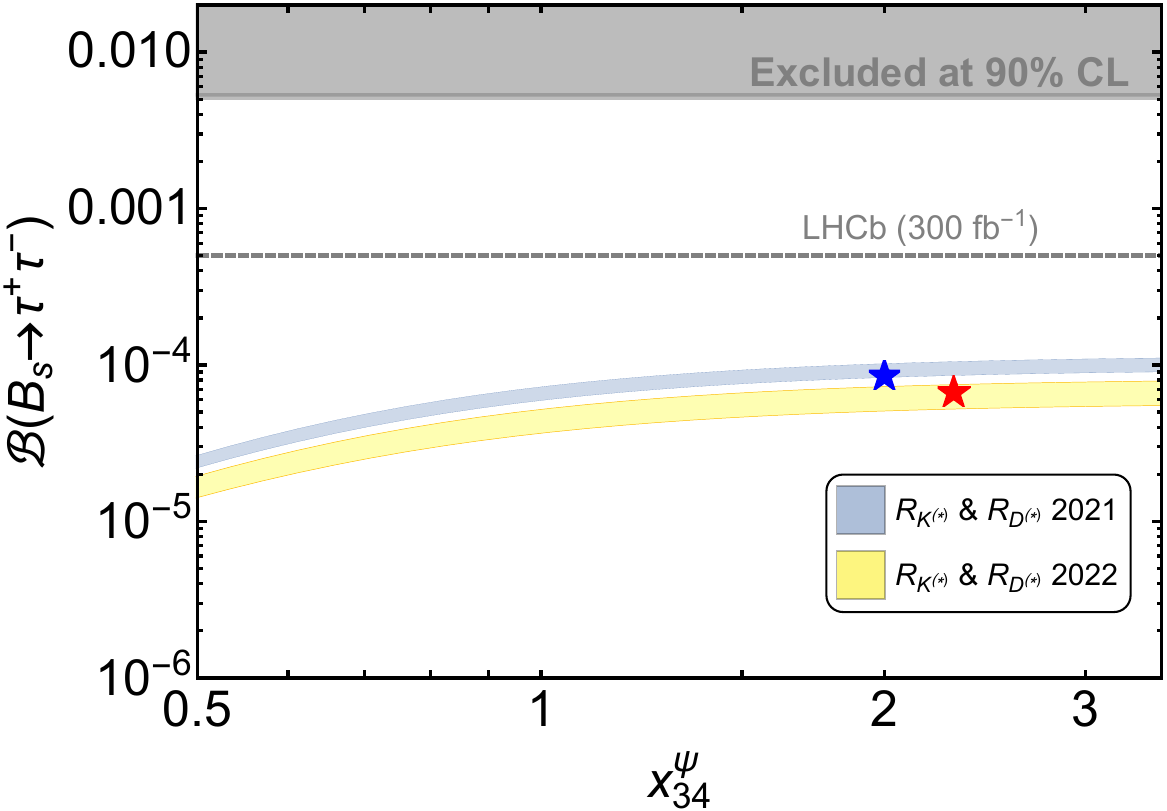}

}$\quad$\subfloat[\label{fig:B_K_tautau}]{\includegraphics[scale=0.355]{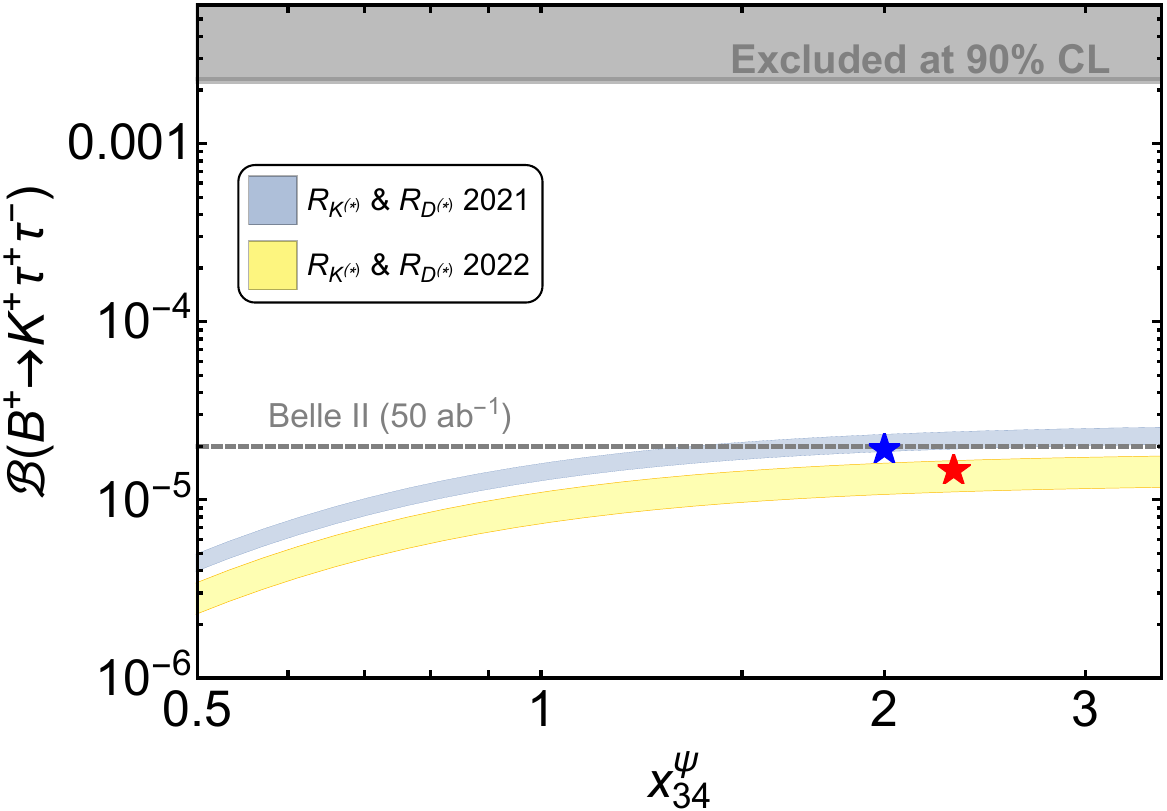}

}

\caption[$\mathcal{B}\left(B_{s}\rightarrow\tau^{+}\tau^{-}\right)$ and $\mathcal{B}\left(B^{+}\rightarrow K^{+}\tau^{+}\tau^{-}\right)$
in the extended twin PS model]{The branching fractions $\mathcal{B}\left(B_{s}\rightarrow\tau^{+}\tau^{-}\right)$
(left) and $\mathcal{B}\left(B^{+}\rightarrow K^{+}\tau^{+}\tau^{-}\right)$
(right) as a function of $x_{34}^{\psi}$, with $x_{25}^{\psi}$ varied
in the range $x_{25}^{\psi}=[0.3,\,0.35]$ ($[0.4,\,0.45]$) preferred
by $R_{K^{(*)}}^{2022}$ ($R_{K^{(*)}}^{2021}$), obtaining the yellow
(blue) band. The rest of the parameters are fixed as in Table~\ref{tab:BP}.
Current exclusion limits are displayed, along with their future projections.
The blue (red) star shows BP1 (BP2).\label{fig:B_tautau}}
\end{figure}
As anticipated in Section~\ref{subsec:Off-shell-photon-penguin},
the enhancement of $R_{D^{(*)}}$ via the $U_{1}$ leptoquark is correlated
to an enhancement of $b\rightarrow s\tau\tau$ via $SU(2)_{L}$ invariance
of the $U_{1}$ couplings to fermions. The respective branching fractions
are of order $10^{-7}$ in the SM and mild upper bounds have been
obtained by LHCb \cite{LHCb:2017myy} and BaBar \cite{BaBar:2016wgb},
respectively. See Section~\ref{subsec:bstautau} for a further
discussion of these observables. 

In Fig.~\ref{fig:B_tautau}, we plot the branching fractions as a
function of $x_{34}^{\psi}$, while $x_{25}^{\psi}$ is varied in
the ranges compatible with 2021 and 2022 $R_{K^{(*)}}$, respectively.
We find that the predictions are far below the current bounds, however
they lie closer to the expected future bounds from LHCb and Belle
II data \cite{LHCb:2018roe,Belle-II:2018jsg}. This prediction is
different in non-fermiophobic 4321 models \cite{Cornella:2019hct,Cornella:2021sby,Barbieri:2022ikw},
where these contributions are chirally enhanced due to the presence of scalar operators, and all the parameter space
is expected to be tested in the $B^{+}\rightarrow K^{+}\tau^{+}\tau^{-}$
process by Belle~II, see the full discussion in Section~\ref{subsec:LifetimeRationU1}. 

The lifetime ratio $\tau_{B_{s}}/\tau_{B_{d}}$ introduced in Section~\ref{subsec:LifetimeRationU1}
can potentially provide significant direct bounds over $\mathcal{B}\left(B_{s}\rightarrow\tau^{+}\tau^{-}\right)$
in the future \cite{Bordone:2023ybl}, which can discriminate as well between the twin Pati-Salam
model and non-fermiophobic 4321 models, see Fig.~\ref{Fig:U1_LifetimeRatios}.

\subsubsection*{$\boldsymbol{B\rightarrow K\nu\nu}$}

The $U_{1}$ leptoquark does not contribute at tree-level to $b\rightarrow s\nu\nu$
transitions, and the tree-level exchange of the $Z'$ is suppressed
due to the down-aligned flavour structure of the model. However, loop-level
corrections can lead to an important enhancement of the channel $B\rightarrow K\nu_{\tau}\bar{\nu}_{\tau}$
\cite{Cornella:2021sby}. We parameterise corrections to the SM branching
fraction as in Eq.~(\ref{eq:BtoK_nunu}), where the EFT and the Wilson
coefficients are defined in Section~\ref{subsec:bsnunu}.

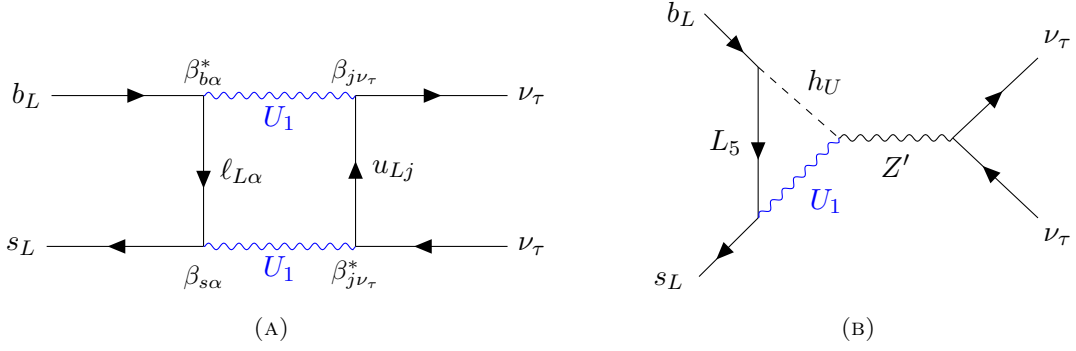
\begin{figure}[t]
\begin{centering}
\subfloat[\label{fig:Box_Diagram_BKnunu}]{\noindent \begin{centering}
\begin{tikzpicture}
	\begin{feynman}
		\vertex (a) {\(s_{L}\)};
		\vertex [right=24mm of a] (b) [label={ [yshift=-0.7cm] \small $\beta_{s\alpha}$}];
		\vertex [right=20mm of b] (c) [label={ [yshift=-0.7cm] \small $\beta_{j\nu_{\tau}}^{*}$}];
		\vertex [right=20mm of c] (d) {\(\nu_{\tau}\)};
		\vertex [above=20mm of b] (f1) [label={ \small $\beta_{b\alpha}^{*}$}];
		\vertex [above=20mm of c] (f2) [label={ \small $\beta_{j\nu_{\tau}}$}];
		\vertex [left=20mm of f1] (f3) {\(b_{L}\)};
		\vertex [right=20mm of f2] (f4) {\(\nu_{\tau}\)};
		\diagram* {
			(a) -- [anti fermion] (b) -- [anti fermion, edge label'=\(\ell_{L\alpha}\), inner sep=6pt)] (f1) -- [anti fermion] (f3),
			(b) -- [boson, blue, edge label'=\(U_{1}\)] (c) -- [fermion, edge label'=\(u_{Lj}\), inner sep=6pt)] (f2) -- [fermion] (f4),
			(c) -- [anti fermion] (d),
			(f1) -- [boson, blue, edge label'=\(U_{1}\)] (f2),
	};
	\end{feynman}
\end{tikzpicture}
\par\end{centering}
}$\qquad$\subfloat[\label{fig:Radial_Diagram_BKnunu}]{\noindent \begin{centering}
\begin{tikzpicture}
	\begin{feynman}
		\vertex (a) [label={ [xshift=-0.4cm,yshift=-0.3cm] $s_{L}$}];
		\vertex [above right=11mm of a] (b);
		\vertex [above right=of b] (c);
		\vertex [above=20mm of b] (d);
		\vertex [above left=10mm of d] (e)  [label={ [xshift=-0.3cm,yshift=-0.3cm] $b_{L}$}];
		\vertex [right=of c] (f);
		\vertex [above right=of f] (f2) {\(\nu_{\tau}\)};
		\vertex [below right=of f] (f3) {\(\nu_{\tau}\)};
		\diagram* {
			(a) -- [anti fermion] (b) -- [anti fermion,  edge label=\(L_{5}\),  inner sep=6pt] (d) -- [anti fermion] (e),
			(b) -- [boson, blue, edge label'=\(U_{1}\), inner sep=4pt] (c),
			(d) -- [scalar,  edge label=\(h_{U}\), inner sep=3pt] (c),
			(c) -- [boson, edge label'=\(Z'\), inner sep=6pt] (f),
			(f) -- [fermion] (f2),
			(f) -- [anti fermion] (f3),
	};
	\end{feynman}
\end{tikzpicture}
\par\end{centering}
}
\par\end{centering}
\caption[Box and penguin diagrams contributing to $B\rightarrow K\nu\nu$ in
the extended twin PS model]{Box and penguin diagrams contributing to $B\rightarrow K\nu\nu$.
The index $\alpha$ runs for all charged leptons including vector-like,
i.e.~$\ell_{L\alpha}=\left(\mu_{L},\tau_{L},E_{L4},E_{L5}\right)$,
and the index $j$ runs for all up-type quarks, including vector-like
$u_{Lj}=\left(c_{L},t_{L},U_{L4},U_{L5}\right)$. See more details
in the main text.\label{fig:Box-and-penguin_BtoKnunu}}
\end{figure}

We only obtain sizable contributions in the $\tau\tau$ channel due
to the flavour structure of the model, and we split the NP effects
into $Z'$ and $U_{1}$ contributions as follows,
\begin{equation}
C_{\nu,\mathrm{NP}}^{\tau\tau}=-\frac{1}{V_{tb}V_{ts}^{*}}\frac{\sqrt{2}}{4G_{F}}\left(C_{\nu,Z'}^{\tau\tau}+C_{\nu,U_{1}}^{\tau\tau}\right)\,.
\end{equation}
The $U_{1}$ contribution at NLO accuracy reads \cite{Fuentes-Martin:2020hvc}
\begin{equation}
C_{\nu,U_{1}}^{\tau\tau}\approx C_{\nu,U_{1}}^{\mathrm{RGE}}+\frac{g_{4}^{4}}{32\pi^{2}M^{2}_{U_{1}}}\sum_{\alpha,j}\left(\beta_{s\alpha}^{*}\beta_{b\alpha}\right)\left(\beta_{j\nu_{\tau}}\right)^{2}F(x_{\alpha},x_{j})\,,
\end{equation}
where the second term arises from the semileptonic box diagram in
Fig.~\ref{fig:Box_Diagram_BKnunu}, and the first term encodes the
RGE-induced contribution from the tree-level leptoquark-mediated operator
$[C_{ed}^{V,LL}]^{\tau\tau23}$, computed with  \texttt{DsixTools 2.1} \cite{Fuentes-Martin:2020zaz}
as
\begin{equation}
C_{\nu,U_{1}}^{\mathrm{RGE}}=0.047\frac{g_{4}^{2}}{2M_{U_{1}}^{2}}\beta_{b\tau}\beta_{s\tau}\,.
\end{equation}
The $Z'$ contribution to NLO accuracy reads
\begin{flalign}
C_{\nu,Z'}^{\tau\tau}\approx\frac{3g_{4}^{2}}{2M_{Z'}^{2}} & \left[\xi_{bs}\xi_{\nu_{\tau}\nu_{\tau}}\left(1+\frac{3}{2}\frac{g_{4}^{2}}{16\pi^{2}}\xi_{\nu_{\tau}\nu_{\tau}}^{2}\right)\right.\label{eq:Z'_BKnunu}\\
 & \left.\frac{g_{4}^{2}}{16\pi^{2}}\beta_{sE_{5}}^{*}\beta_{bE_{5}}\xi_{\nu_{\tau}\nu_{\tau}}G_{\Delta Q=1}(x_{E_{5}},x_{Z'},x_{R})\right]\,,\nonumber 
\end{flalign}
where $x_{E_{5}}\equiv(M_{5}^{L}/M_{U_{1}})^{2}$, $x_{Z'}\equiv M_{Z'}^{2}/M_{U_{1}}^{2}$
and $x_{R}\equiv M_{R}^{2}/M_{U_{1}}^{2}$ with $M_{R}$ being a scale
associated to the radial mode $h_{U}(\mathbf{3,1},2/3)$ arising from
$\phi_{3,1}$. The first term in Eq.~(\ref{eq:Z'_BKnunu}) corresponds
to the tree-level contribution plus a 1-loop $Z'$ correction to the
leptonic vertex. The coupling $\xi_{bs}$ is suppressed by the small
down mixing angle $\theta_{23}^{d}\approx0.001$, leading to per cent
corrections to $\mathcal{B}(B\rightarrow K^{(*)}\nu\bar{\nu})$. The
second term in Eq.~(\ref{eq:Z'_BKnunu}) corresponds to a 1-loop
correction to the flavour-violating $Z'$ vertex, with $U_{1}$, the
fifth vector-like lepton $E_{5}$ and $h_{U}$ running in the loop,
see Fig.~\ref{fig:Radial_Diagram_BKnunu}. The loop function is given
by \cite{Cornella:2021sby,Fuentes-Martin:2020hvc}
\begin{equation}
G_{\Delta Q=1}(x_{1},x_{2},x_{3})\approx\frac{5}{4}x_{1}+\frac{x_{1}}{2}\left(x_{2}-\frac{3}{2}\right)\left(\ln x_{3}-\frac{5}{2}\right)\,.
\end{equation}
In the twin Pati-Salam framework, we expect extra radial modes associated
to $\overline{\phi}_{3,1}$ and $\overline{\phi'}_{3,1}$, however they only
couple to right-handed chiral fermions and hence they cannot contribute to $C_{\nu,\mathrm{NP}}^{\tau\tau}$
which involves only left-handed chiral fermions.

\begin{figure}[t]
\begin{centering}
\includegraphics[scale=0.4]{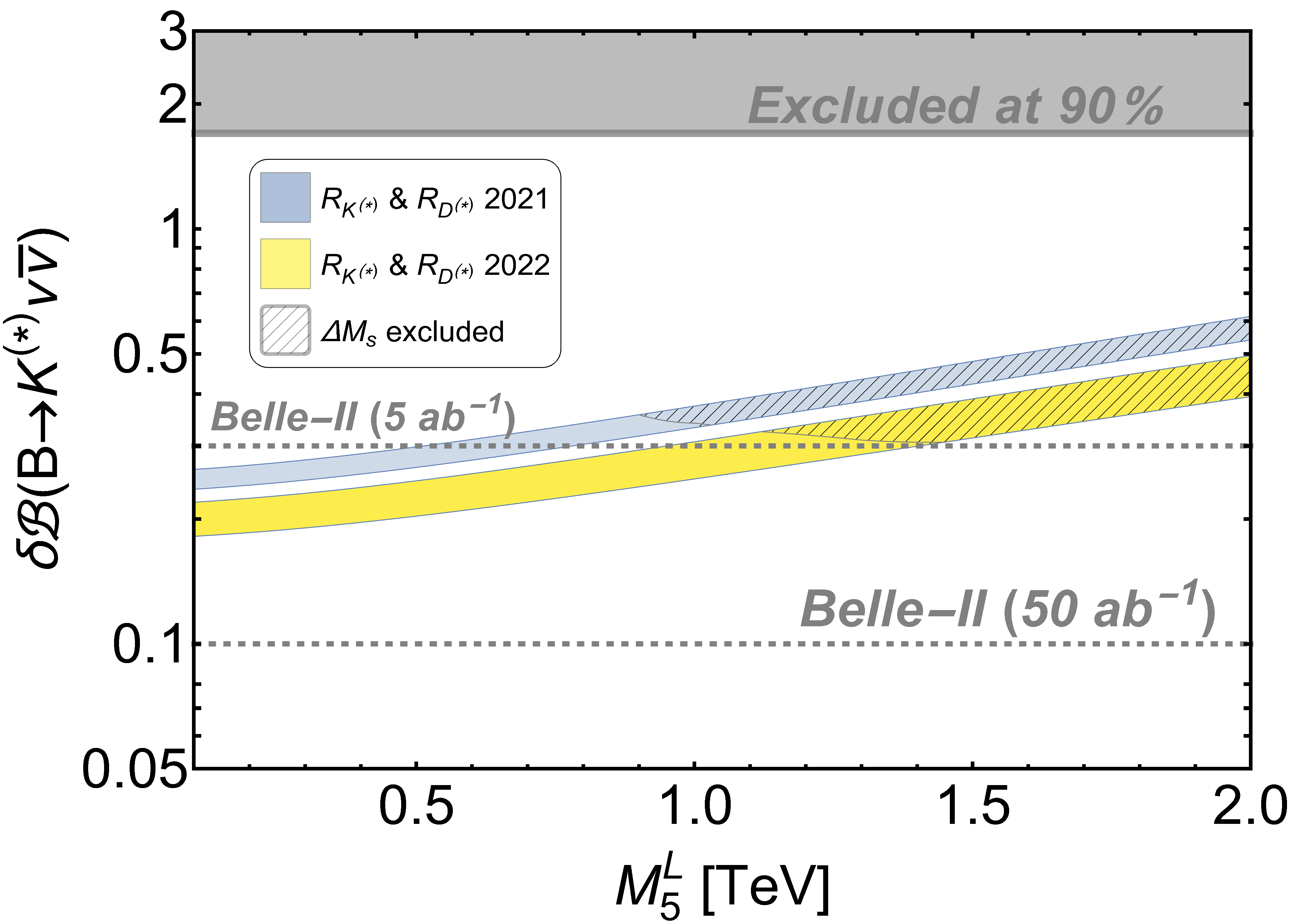}
\par\end{centering}
\caption[$B\rightarrow K^{(*)}\nu\bar{\nu}$ in the extended twin PS model]{$\delta\mathcal{B}(B\rightarrow K^{(*)}\nu\bar{\nu})$ (Eq.~(\ref{eq:BtoK_nunu}))
as a function of the 5th family vector-like mass term. $x_{25}^{\psi}$
is varied in the range $x_{25}^{\psi}=[0.3,\,0.35]$ ($[0.4,\,0.45]$)
preferred by $R_{K^{(*)}}^{2022}$ ($R_{K^{(*)}}^{2021}$), obtaining
the yellow (blue) band. The hatched region is excluded by the $\Delta M_{s}$
bound, see Eq.~(\ref{eq:DeltaMs_bound}). The grey region is excluded
by current experimental measurements, the dashed line indicates the
projected future bound. \label{fig:BtoK_nunu}}
\end{figure}

Both 1-loop contributions are dominated by the fifth vector-like charged lepton
and grow with its bare mass, $M_{5}^{L}$. This way, the overall contribution
to $B\rightarrow K\nu\nu$ can be sizable, yielding up to $\mathcal{O}(1)$
corrections with respect to the SM value, as depicted in Fig.~\ref{fig:BtoK_nunu}.

For low $M_{5}^{L}$, the enhancement of $\delta\mathcal{B}(B\rightarrow K^{(*)}\nu\bar{\nu})$
corresponds mostly to $C_{\nu,U}^{\mathrm{RGE}}$. For large $M_{5}^{L}$,
however, we have seen that stringent constraints from $B_{s}-\bar{B}_{s}$
meson mixing play an important role, see Section~\ref{subsec:BsMixing_revisited}.
This constraint is depicted as the hatched region in Fig.~\ref{fig:BtoK_nunu},
correlating $B\rightarrow K\nu\nu$ and $\Delta M_{s}$, a feature
which has not been highlighted in previous analyses. In particular,
$\Delta M_{s}$ rules out the region where $\delta\mathcal{B}(B\rightarrow K^{(*)}\nu\bar{\nu})$
can reach values close to current experimental limits, making impossible to address the $2.8\sigma$ anomaly in $\mathcal{B}(B^{+}\rightarrow K^{+}\nu \bar{\nu})$ suggested by the recent Belle~II data \cite{BelleIIEPS:2023}. Nevertheless,
the Belle~II collaboration is expected to measure $\mathcal{B}(B\rightarrow K^{(*)}\nu\bar{\nu})$
up to 10\% of the SM value \cite{Belle-II:2018jsg}, hence testing
all the parameter space of our model. 

Our signal of $B\rightarrow K^{(*)}\nu\bar{\nu}$ also offers a great
opportunity to disentangle our twin Pati-Salam framework from non-fermiophobic
4321 models and from the $\mathrm{PS}^{3}$ model \cite{Cornella:2019hct,Cornella:2021sby,Barbieri:2022ikw,Bordone:2017bld},
as they predict a much smaller signal (see Fig.~4.4 of \cite{Cornella:2021sby}
and compare their purple region with our Fig.~\ref{fig:BtoK_nunu}).

\subsection{Perturbativity\label{subsec:Pertubativity}}

\begin{figure}
\begin{centering}
\includegraphics[scale=0.4]{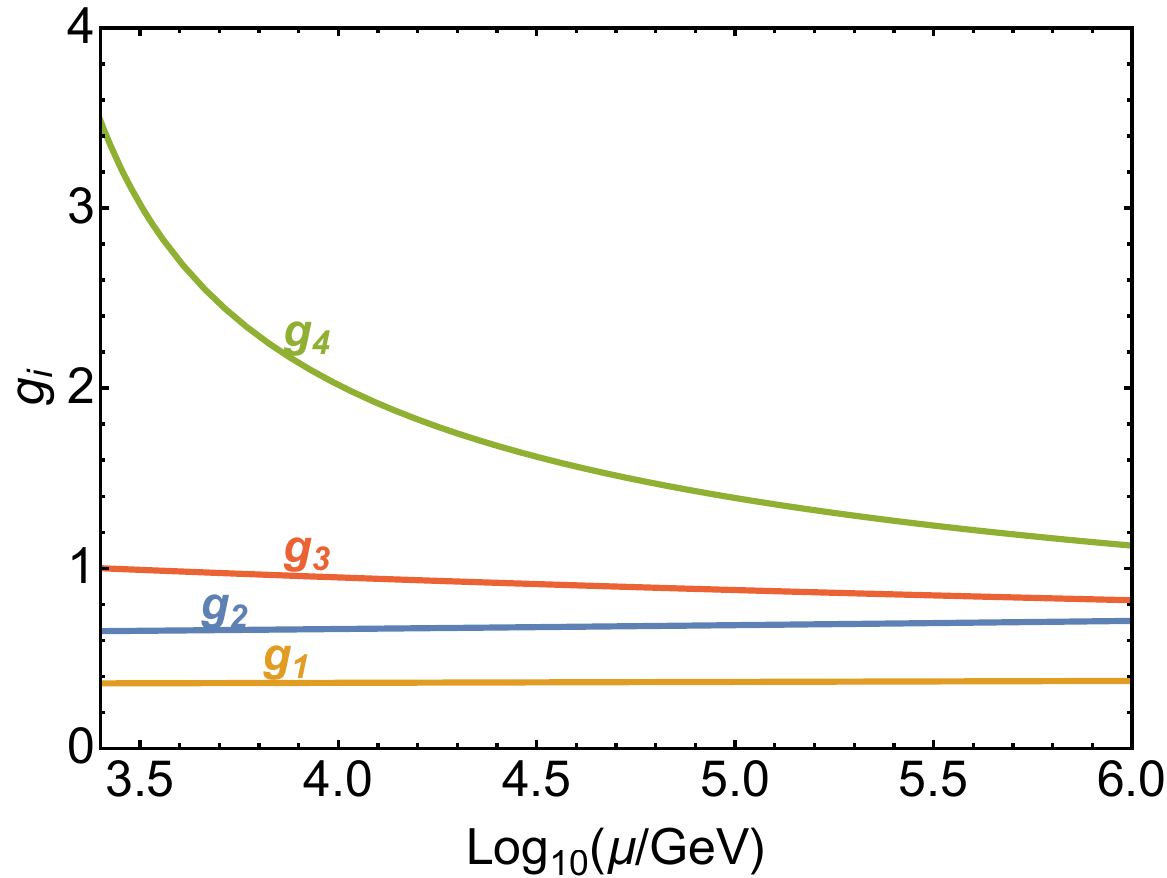}
\par\end{centering}
\caption[RGE of the gauge couplings in the extended twin PS model]{RGE of the gauge couplings in our benchmark scenario from the TeV
scale to the scale of the twin Pati-Salam symmetry $\mu\sim1\,\mathrm{PeV}$.\label{fig:Perturbativity_gauge}}
\end{figure}

The explanation of the $R_{D^{(*)}}$ anomaly requires large mixing
angles $s_{34}^{Q}$ and $s_{34}^{L}$, which translate into a sizeable
Yukawa coupling $x_{34}^{\psi}$, thus pushing the model close to
the boundary of the perturbative domain. Perturbativity is a serious
constraint over our model, since we need the low-energy 4321 theory
to remain perturbative until the high scale of the twin Pati-Salam
symmetry. When assessing the issue of perturbativity, two conditions
must be satisfied:
\begin{itemize}
\item Firstly, the low-energy observables must be calculable in perturbation
theory. For Yukawa couplings, we consider the bound $x_{34}^{\psi}<\sqrt{4\pi}$.
Regarding the gauge coupling $g_{4}$, standard perturbativity criteria
imposes the beta function criterion \cite{Goertz:2015nkp} $\left|\beta_{g_{4}}/g_{4}\right|<1$,
which yields $g_{4}<4\pi\sqrt{3}/\sqrt{28}\approx4.11$.
\item Secondly, the couplings must remain perturbative up to the energy
scale of the UV completion, i.e.~we have to check that the couplings
of the model do not face a Landau pole below the energy scale of the
second PS breaking, namely $\mu\approx1\,\mathrm{PeV}$.
\end{itemize}
The phenomenologically convenient choice of large $g_{4}$ is not
a problem for the extrapolation in the UV, thanks to the asymptotic
freedom of the $SU(4)$ gauge factor (see Fig.~\ref{fig:Perturbativity_gauge}).
To investigate the running of the most problematic Yukawa $x_{34}^{\psi}$,
we use the 1-loop renormalisation group equations of the
4321 model. For the gauge coupling beta functions $\beta_{g_{i}}=(dg_{i}/d\mu)/\mu$
we have \cite{DiLuzio:2018zxy}
\begin{equation}
\left(4\pi\right)^{2}\beta_{g_{1}}=\frac{131}{18}g_{1}^{3}\,,\quad\left(4\pi\right)^{2}\beta_{g_{2}}=\left(-\frac{19}{6}+\frac{8n_{\Psi}}{3}\right)g_{2}^{3}\,,
\end{equation}
\begin{equation}
\left(4\pi\right)^{2}\beta_{g_{3}}=-\frac{19}{3}g_{3}^{3}\,,\quad\left(4\pi\right)^{2}\beta_{g_{4}}=\left(-\frac{40}{3}+\frac{4n_{\Psi}}{3}\right)g_{4}^{3}\,,
\end{equation}
where $n_{\Psi}=3$ is the number of vector-like fermion families.
The Pati-Salam universality of the Yukawas $x_{i\alpha}^{\psi}$ is
broken by RGE effects which we quantify through the equations
\begin{align}
\begin{aligned}\left(4\pi\right)^{2}\beta_{x_{Q}} & =\frac{7}{2}x_{Q}x_{Q}^{\dagger}x_{Q}+\frac{1}{2}x_{Q}x_{L}^{\dagger}x_{L}+\frac{15}{8}x_{Q}\lambda_{15}\lambda_{15}^{\dagger}+2\mathrm{Tr}\left(x_{Q}x_{Q}^{\dagger}\right)x_{Q}\\
{} & -\frac{1}{12}g_{1}^{2}x_{Q}-\frac{9}{2}g_{2}^{2}x_{Q}-4g_{3}^{2}x_{Q}-\frac{45}{8}g_{4}^{2}x_{Q}\,,
\end{aligned}
\label{eq: full_lagrangian-1-2}
\end{align}
\begin{align}
\begin{aligned}\left(4\pi\right)^{2}\beta_{x_{L}} & =\frac{5}{2}x_{L}x_{L}^{\dagger}x_{L}+\frac{3}{2}x_{L}x_{Q}^{\dagger}x_{Q}+\frac{15}{8}x_{L}\lambda_{15}\lambda_{15}^{\dagger}+2\mathrm{Tr}\left(x_{L}x_{L}^{\dagger}\right)x_{L}\\
{} & -\frac{3}{4}g_{1}^{2}x_{L}-\frac{9}{2}g_{2}^{2}x_{L}-\frac{45}{8}g_{4}^{2}x_{L}\,,
\end{aligned}
\label{eq: full_lagrangian-1-1-2}
\end{align}
\begin{align}
\begin{aligned}\left(4\pi\right)^{2}\beta_{\lambda_{15}} & =\frac{21}{4}\lambda_{15}\lambda_{15}\lambda_{15}^{\dagger}+\frac{3}{2}\lambda_{15}x_{Q}^{\dagger}x_{Q}+\frac{1}{2}\lambda_{15}x_{L}^{\dagger}x_{L}+4\mathrm{Tr}\left(\lambda_{15}\lambda_{15}^{\dagger}\right)\lambda_{15}\\
{} & -\frac{9}{2}g_{2}^{2}\lambda_{15}-\frac{45}{4}g_{4}^{2}\lambda_{15}\,,
\end{aligned}
\label{eq: full_lagrangian-1-1-1-1}
\end{align}
where any contributions from the Yukawas of the personal Higgs, $y_{i\alpha}^{\psi}$,
are negligible as they are all 1 or smaller. 

The running of the effective Yukawa couplings is protected, as the
top Yukawa is order 1 and all of the others are smaller, SM-like (see
the discussion in Section~\ref{subsec:Effective_Yukawa_3VL}). This
feature is different from \cite{DiLuzio:2018zxy} which was not a
theory of flavour, causing the top mass to be accidentally suppressed
by the equivalent of $c_{34}^{Q}$ in our model, hence requiring a
large, non-perturbative top Yukawa to preserve the top mass. Instead,
in our model the effective top Yukawa arises proportional to the maximal
angle $s_{34}^{Q}$, rendering the top Yukawa natural and perturbative.
The matrices of couplings $x_{Q,L}$ and $\lambda_{15}$ are defined
as (assuming small $x_{35}^{\psi}$ as discussed in Section~\ref{subsec:BsMixing_revisited})
\begin{equation}
x_{\psi}=\left(\begin{array}{ccc}
x_{16}^{\psi} & 0 & 0\\
0 & x_{25}^{\psi} & 0\\
0 & 0 & x_{34}^{\psi}
\end{array}\right)\,,\quad\lambda_{15}=\left(\begin{array}{ccc}
\lambda_{15}^{6} & 0 & 0\\
0 & \lambda_{15}^{5} & 0\\
0 & 0 & \lambda_{15}^{4}
\end{array}\right)\,,\quad\psi=Q,\,L\,.
\end{equation}
The Yukawas $x_{25}^{\psi}$ and $x_{16}^{\psi}$ are not dangerous
as they are order 1 of smaller. The problematic Yukawa is $x_{34}^{\psi}$,
which is required to be large in order to address $R_{D^{(*)}}$,
and it is also connected with the physical mass of the fourth family lepton
as per Eq.~(\ref{eq:Mass-_4th}). Large $\lambda_{15}^{5}$ is also
required to obtain a large splitting of vector-like masses, which provides
a large $\theta_{LQ}$ as required by $R_{D^{(*)}}$.
\begin{figure}[t]
\begin{centering}
\subfloat[\label{fig:x_Yukawas}]{\includegraphics[scale=0.36]{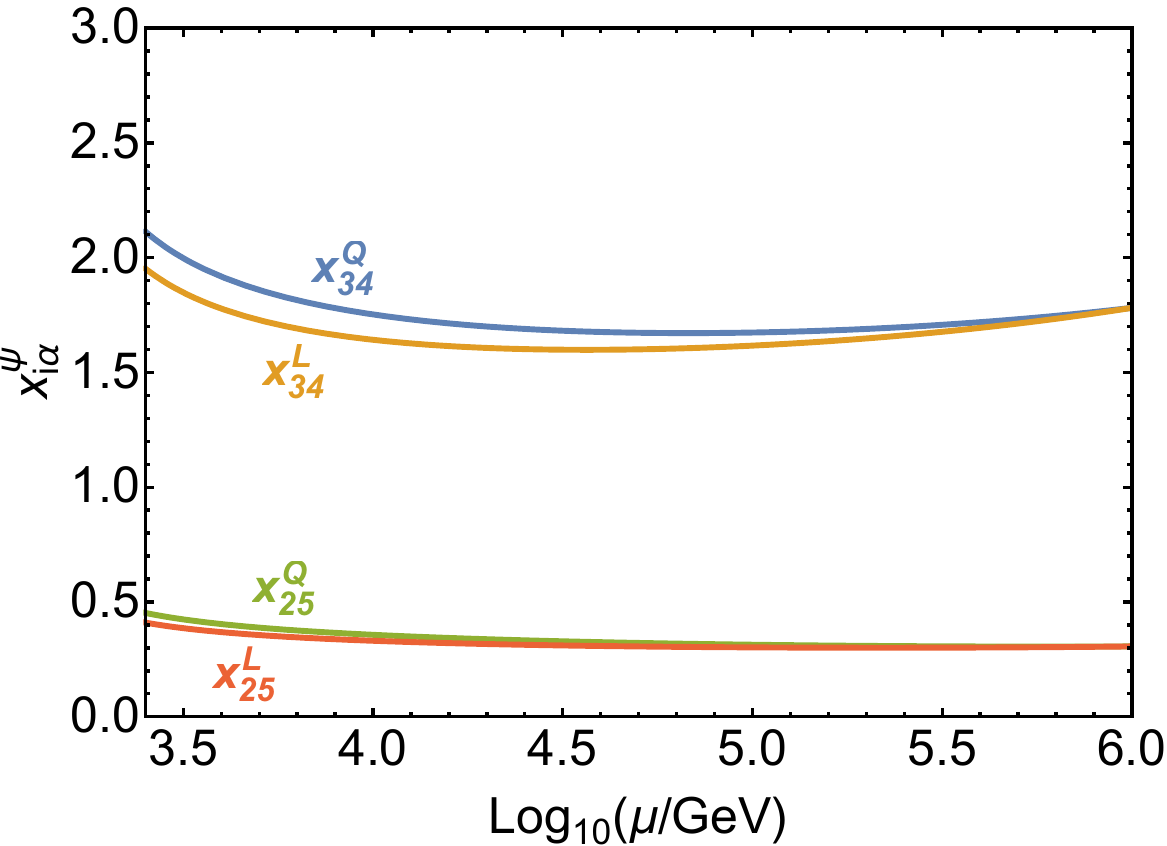}

}$\quad$\subfloat[\label{fig:Lambda_15}]{\includegraphics[scale=0.36]{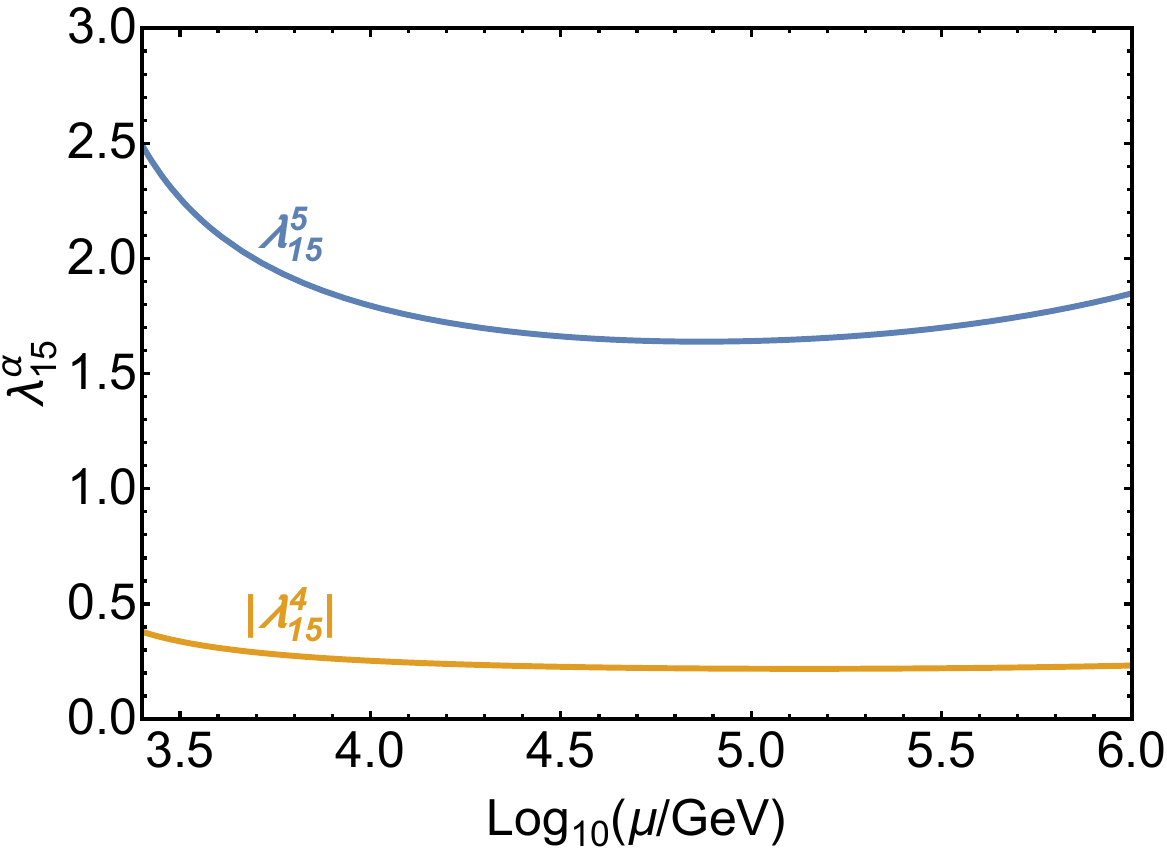}

}
\par\end{centering}
\caption[RGE of the Lagrangian couplings in the extended twin PS model]{RGE of the fundamental Yukawa couplings in our benchmark scenario
(Table~\ref{tab:BP}) from the TeV scale to the scale of the twin
Pati-Salam symmetry $\mu\sim1\,\mathrm{PeV}$. The left panel shows
the $x_{i\alpha}^{\psi}$ Yukawas which lead to the mixing between
SM fermions and vector-like partners. The right panel shows the $\lambda_{15}$
Yukawas which split the vector-like masses of quarks and leptons.\label{fig:Perturbativity_Yukawas}}
\end{figure}

Fig.~\ref{fig:Perturbativity_Yukawas} shows that the Yukawas of
our benchmark scenario remain perturbative up to the high energy scale
$\mu\approx1\,\mathrm{PeV}$, thanks to the choice of a large $g_{4}=3.5$.
However, we have checked that the Landau pole is hit when $x_{34}^{\psi}>2.5$,
hence this region should be considered as disfavoured by the perturbativity
criteria.

The small RGE effects that break the Pati-Salam universality of the Yukawa
couplings are below 8\% in any case, hence the universality of the
couplings is preserved at the TeV scale to good approximation.

\subsection{\texorpdfstring{High-$p_{T}$ signatures}{High-pT signatures}\label{subsec:Colliders}}

General 4321 models predict a plethora of high-$p_{T}$ signatures
involving the heavy gauge bosons and at least one family of vector-like
fermions, requiring dedicated analyses such as those in \cite{DiLuzio:2018zxy,Baker:2019sli,Cornella:2021sby}.
In particular, our model predicts a similar high-$p_{T}$ phenomenology
as that of \cite{DiLuzio:2018zxy}, which also considers effective
$U_{1}$ couplings via mixing with three families of vector-like fermions.
However, the bounds obtained in the high-$p_{T}$ analysis of \cite{DiLuzio:2018zxy}
are mostly outdated. Moreover, certain differences arise due to the
underlying twin Pati-Salam symmetry in our model, plus the different
implementation of the scalar sector and VEV structure. In contrast,
the most recent high-$p_{T}$ studies of the 4321 model assume a non-fermiophobic
framework \cite{Cornella:2021sby,Baker:2019sli}, where the heavy
gauge bosons have large couplings to right-handed third family fermions.
In this manner, some of the bounds in \cite{Cornella:2021sby,Baker:2019sli}
are overestimated for our model. This motivates a dedicated high-$p_{T}$
analysis of the twin Pati-Salam model.
\begin{figure}[t]
\noindent \begin{centering}
\subfloat[\label{fig:Spectrum}]{\begin{raggedright}
\begin{tabular}{c}
\includegraphics[scale=0.305]{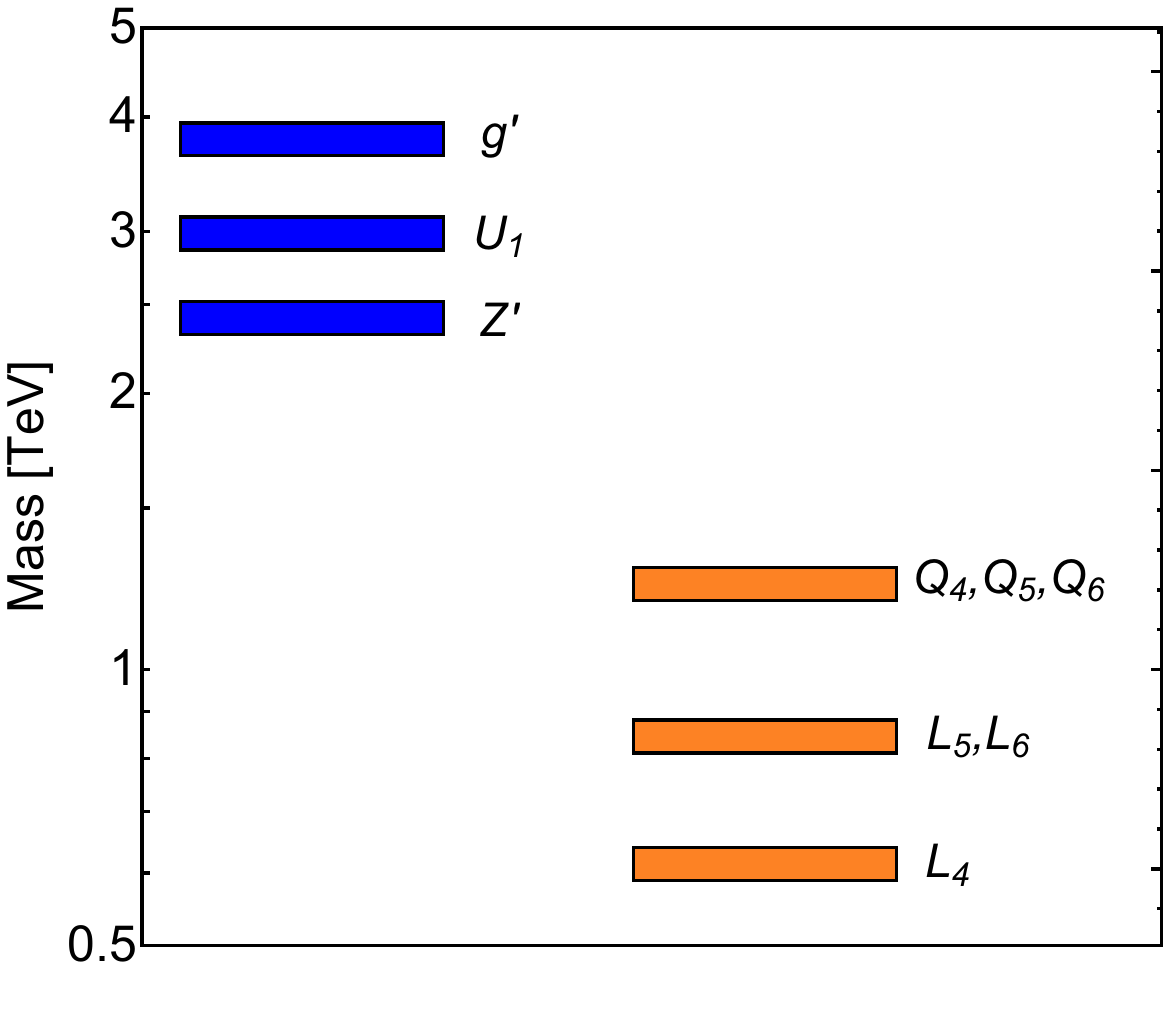}\tabularnewline
\end{tabular}
\par\end{raggedright}
}\subfloat[\label{fig:Decay_Channels_Vectors}]{\begin{centering}
\resizebox{.516\textwidth}{!}{
\begin{tabular}{cccc}
\toprule 
Particle & Decay mode & $\mathcal{B}(\mathrm{BP})$ & $\Gamma/M$\tabularnewline
\midrule
\midrule 
\multirow{4}{*}{$U_{1}$} & $Q_{3}L_{5}+Q_{5}L_{3}$ & $\sim0.47$ & \multirow{4}{*}{0.32}\tabularnewline
 & $Q_{3}L_{3}$ & $\sim0.22$ & \tabularnewline
 & $Q_{5}L_{5}$ & $\sim0.24$ & \tabularnewline
 & $Q_{i}L_{a}+Q_{a}L_{i}$ & $\sim0.07$ & \tabularnewline
\midrule 
\multirow{4}{*}{$g'$} & $Q_{3}Q_{3}$ & $\sim0.3$ & \multirow{4}{*}{0.5}\tabularnewline
 & $Q_{5}Q_{5}$ & $\sim0.3$ & \tabularnewline
 & $Q_{6}Q_{6}$ & $\sim0.3$ & \tabularnewline
 & $Q_{1}Q_{6}$+$Q_{2}Q_{5}+Q_{3}Q_{4}$ & $\sim0.1$ & \tabularnewline
\midrule 
\multirow{5}{*}{$Z'$} & $L_{5}L_{5}$ & $\sim0.29$ & \multirow{5}{*}{0.24}\tabularnewline
 & $L_{6}L_{6}$ & $\sim0.29$ & \tabularnewline
 & $L_{3}L_{3}$ & $\sim0.27$ & \tabularnewline
 & $Q_{3}Q_{3}$+$Q_{5}Q_{5}+Q_{6}Q_{6}$ & $\sim0.09$ & \tabularnewline
 & $L_{1}L_{6}+L_{2}L_{5}+L_{3}L_{4}$ & $\sim0.06$ & \tabularnewline
\end{tabular}}
\par\end{centering}
}
\par\end{centering}
\caption[TeV-scale spectrum of particles and main decay channels in the extended
twin PS model]{\textbf{\textit{Left: }}Spectrum of new vector bosons and fermions in our
benchmark scenario (BP, Table~\ref{tab:BP}) around the TeV scale.
\textbf{\textit{Right:}} Main decay channels of the new vectors $U_{1}$,
$g'$ and $Z'$ in BP. Addition (+) implies that the depicted channels
have been summed when computing the branching fraction $\mathcal{B}(\mathrm{BP})$.
$i=1,2$ and $a=5,6$.}
\end{figure}

We have included the particle spectrum of our benchmark scenario in
Fig.~\ref{fig:Spectrum}, as a typical configuration for the masses of
the new vectors and fermions. Table~\ref{fig:Decay_Channels_Vectors}
shows the main decay channels of the new vector bosons, which feature
large decay widths $\Gamma/M$ due to all the available decay channels to
vector-like fermions, plus the choice of large $g_{4}=3.5$ close
to the boundary of the perturbative regime.

In this section, we revisit some of the most simple collider
signals, including coloron dijet searches and $Z'$ dilepton searches.
We will also comment on $U_{1}$ searches, coloron ditop searches
and vector-like fermion searches. We will point out the differences
between our framework and general 4321 models, motivating a further
dedicated high-$p_{T}$ analysis of the twin Pati-Salam model.

\subsubsection*{Coloron signals}

The heavy colour octet has a large impact over collider searches for
4321 models, and its production usually sets the lower bound on the
scale of the model. In our case, the heavy coloron has a gauge origin,
hence the coloron couplings to two gluons are absent at tree-level,
reducing the coloron production at the LHC. Moreover, in the motivated
scenario $\left\langle \phi_{3}\right\rangle \gg\left\langle \phi_{1}\right\rangle $,
the coloron is slightly heavier than the vector leptoquark at roughly
$M_{g'}\approx\sqrt{2}M_{U_{1}}$, helping to suppress the impact
of the coloron over collider searches while preserving a slightly
lighter $U_{1}$ to explain the $B$-anomalies. In the scenario $g_{4}\gg g_{3,1}$,
the coupling strength of the coloron is roughly $g_{4}$, which receives
NLO corrections via the $K$-factor \cite{Fuentes-Martin:2019ign,Fuentes-Martin:2020luw}
\begin{equation}
K_{\mathrm{NLO}}\approx\left(1+2.65\frac{g_{4}^{2}}{16\pi^{2}}+8.92\frac{g_{s}^{2}}{16\pi^{2}}\right)^{-1/2}\,,\quad g_{g'}\approx K_{\mathrm{NLO}}g_{4}\,.
\end{equation}

\begin{figure}[t]
\subfloat[\label{fig:Coloron_Production}]{\includegraphics[scale=0.357]{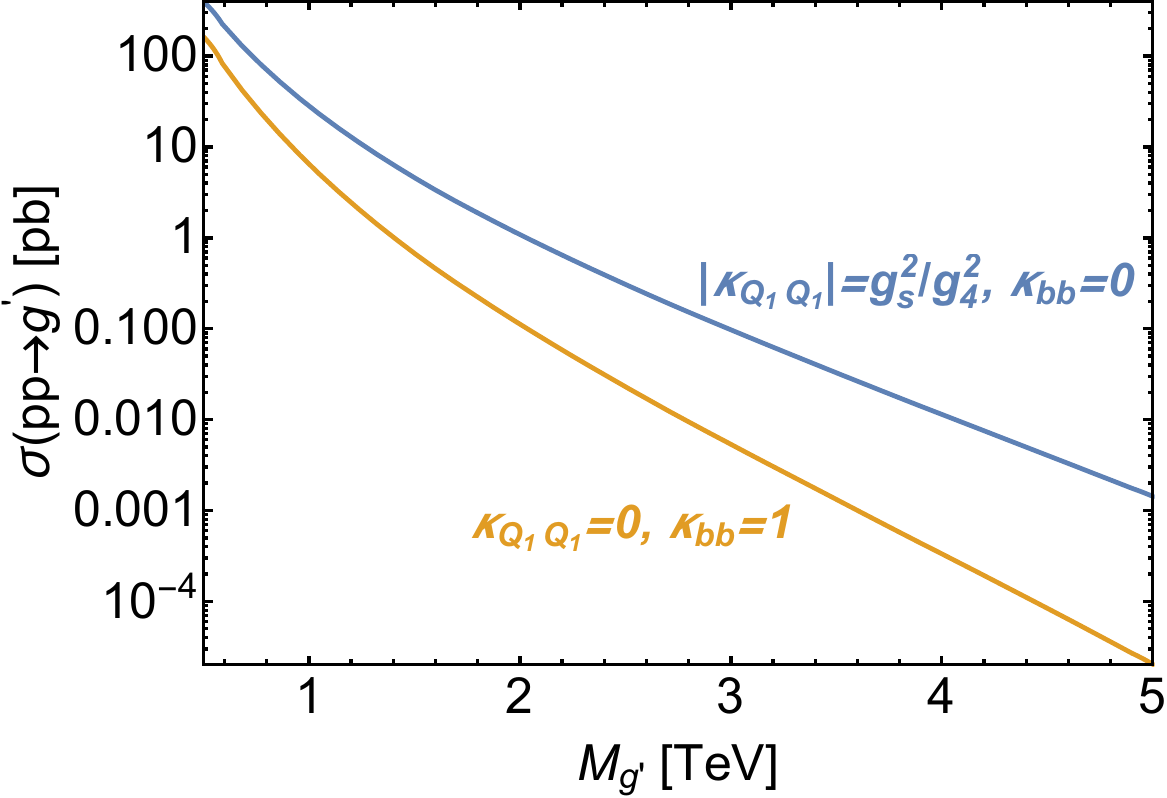}

}$\quad$\subfloat[\label{fig:Zprime_Production}]{\includegraphics[scale=0.357]{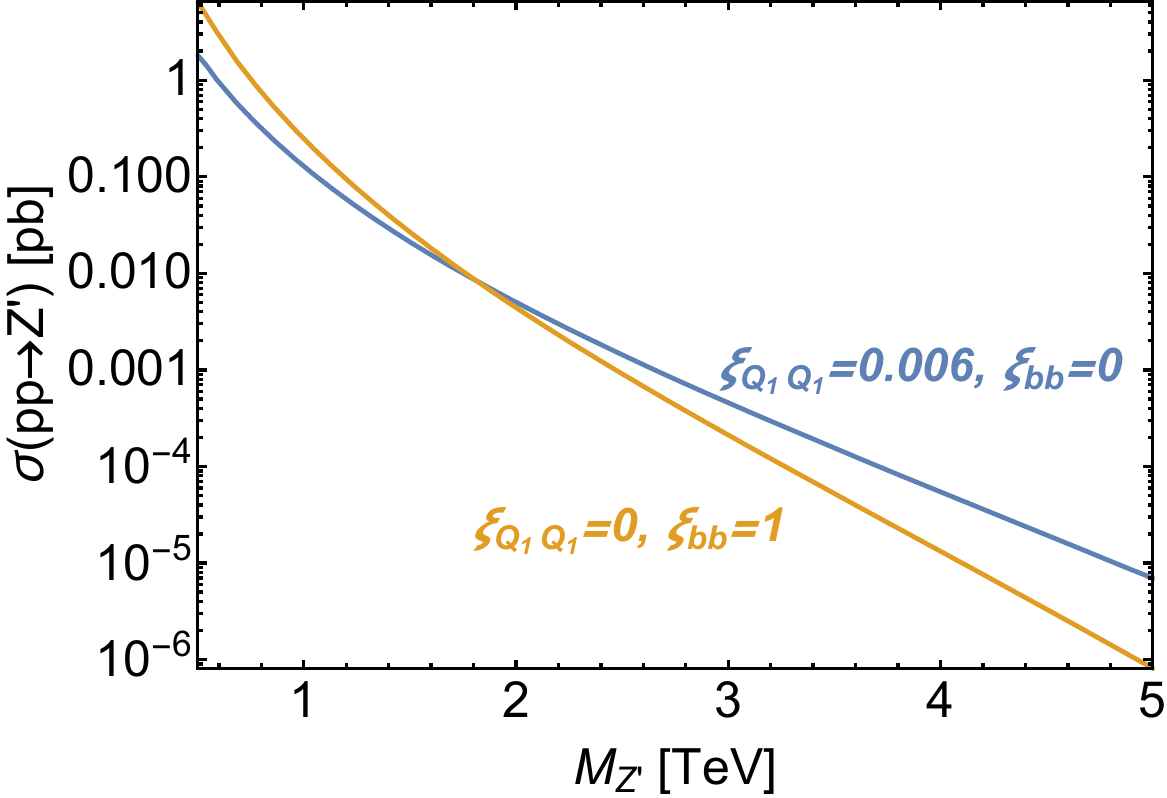}

}

\caption[Production cross sections via 13 TeV $pp$ collisions for the coloron and
$Z'$ in the extended twin PS model]{Production cross sections via 13 TeV $pp$ collisions for the coloron (left)
and $Z'$ (right), via their typical couplings to valence quarks (blue)
and bottoms (orange). The choice of $\xi_{Q_{1}Q_{1}}=0.006$ corresponds
to a mixing angle $s_{16}^{Q}\approx0.2$.\label{fig:B_tautau-1}}
\end{figure}

We have computed the coloron production cross section from 13 TeV
$pp$ collisions with \texttt{Madgraph5} \cite{Madgraph:2014hca}
using the default \texttt{NNPDF23LO} PDF set and the coloron UFO model,
publicly available in the \texttt{FeynRules} \cite{Feynrules:2013bka}
model database\footnote{\url{https://feynrules.irmp.ucl.ac.be/wiki/LeptoQuark}}.
We verify in Fig.~\ref{fig:Coloron_Production} that coloron production
is dominated by valence quarks, even though the coupling to left-handed
bottoms is maximal while the coupling to valence quarks is suppressed.
The coloron couples to light left-handed quarks (see Eq.~(\ref{eq:Coloron_couplings_3VL}))
via the mixing $s_{25}^{Q}\approx s_{16}^{Q}$ of $\mathcal{O}(0.1)$,
which interferes destructively with the flavour universal term, allowing
for a certain cancellation of the left-handed couplings to light quarks.
However, this partial cancellation is not possible for the flavour--universal
couplings to right-handed quarks.

\begin{figure}[t]
\subfloat[\label{fig:Coloron_dijet}]{\includegraphics[scale=0.365]{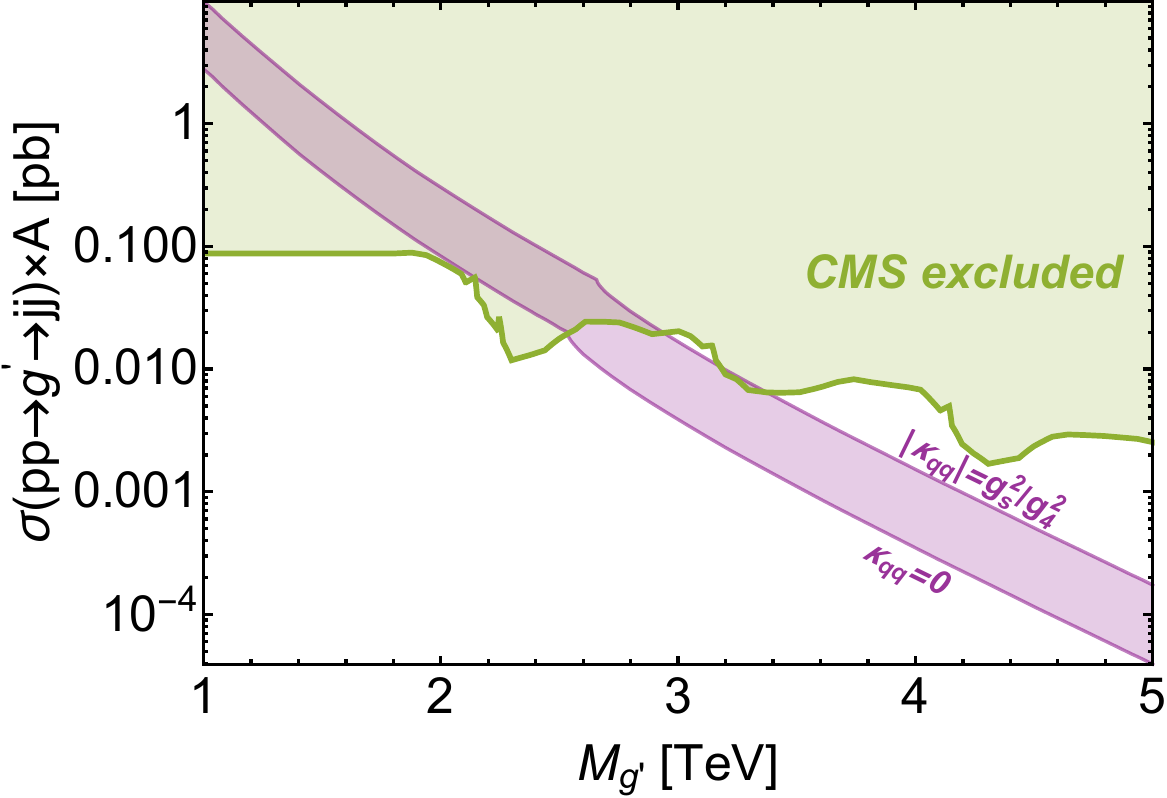}

}$\quad$\subfloat[\label{fig:Colliders_All}]{\includegraphics[scale=0.355]{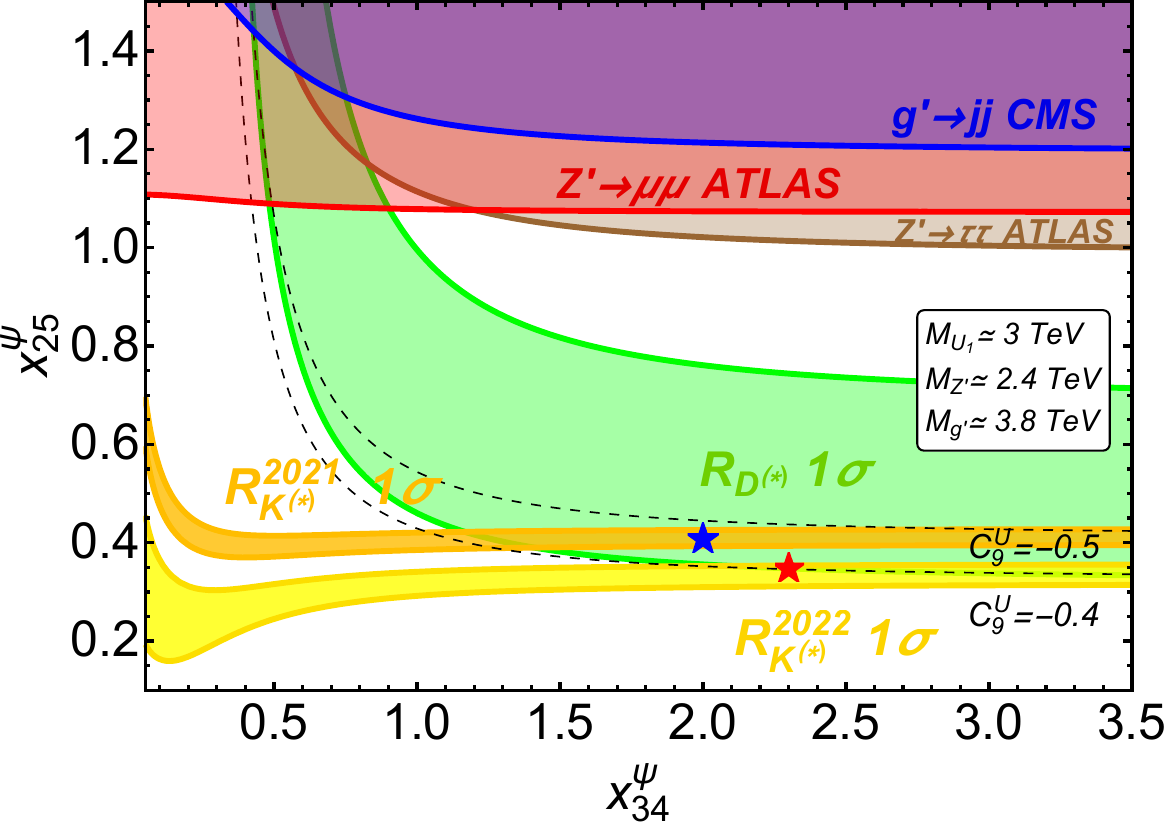}

}

\caption[Coloron dijet cross section and parameter space of the extended twin
PS model including high-$p_{T}$ bounds.]{\textbf{\textit{Left:}} Total cross section for the coloron dijet
channel in the narrow width approximation, with $\left|\kappa_{qq}\right|$
varied in the range $\left|\kappa_{qq}\right|=[0,g_{s}^{2}/g_{4}^{2}]$,
where $q=Q_{1},Q_{2}$. The remaining parameters are fixed as in Table~\ref{tab:BP}
for both panels. The exclusion bound from CMS is shown in green. \textbf{\textit{Right:}}
Parameter space in the plane ($x_{34}^{\psi}$, $x_{25}^{\psi}$)
compatible with the LFU ratios. The dashed lines show contours of
constant $C_{9}^{U}$. The regions excluded by the collider searches
considered are included. The blue (red) star shows BP1 (BP2).}
\end{figure}

We estimate analytically the branching fraction to all SM quarks excluding
tops, and then we compute the total cross section via the narrow width
approximation. Finally, we confront our results with the limits for
a $q\bar{q}$-initiated spin-1 resonance provided by CMS in Fig.~10
of \cite{CMS:2019gwf} (with acceptance $A\approx0.5$). The results are displayed in Fig.~\ref{fig:Coloron_dijet},
where we have varied the coupling to light left-handed quarks $\kappa_{qq}$
and fixed the rest of parameters as in Table~\ref{tab:BP}. We find
bounds ranging from $M_{g'}\apprge2.5\,\mathrm{TeV}$ when $\kappa_{qq}\approx0$
and $M_{g'}\apprge3\,\mathrm{TeV}$ when $\kappa_{qq}\approx g_{s}^{2}/g_{4}^{2}$.
These bounds are slightly milder than those obtained in \cite{Cornella:2021sby},
the reason being that in \cite{Cornella:2021sby} right-handed bottom
quarks are assumed to couple maximally to the coloron, while in our
model this coupling is suppressed.

We expect to find more stringent bounds in resonant coloron production
with $t\bar{t}$ final states, due to the maximal couplings of the
coloron to the third generation $SU(2)_{L}$ quark doublet. According
to the recent analysis in \cite{Cornella:2021sby}, our benchmark
scenario would lie below current bounds, due to the large decay width
$\Gamma_{g'}/M_{g'}\approx0.5$ provided by extra decay channels to
TeV scale vector-like quarks. The limit over the coloron mass is roughly
3.5 TeV, however this bound might be overestimated again for our model
due to the different couplings of the coloron to right-handed third family quarks.
Reconstructing the $t\bar{t}$ channel requires a dedicated analysis
and a different methodology, which is beyond the scope of this work.

\subsubsection*{$Z'$ signals}

For the $Z'$ boson, the flavour universal couplings to valence quarks
are more heavily suppressed than those of the coloron, via the small
ratio $g_{Y}^{2}/g_{4}^{2}$. Therefore, cancellation between the
term proportional to $s_{25}^{Q}\approx s_{16}^{Q}$
and the flavour universal one is not possible here. In contrast with
the coloron, the large left-handed couplings to bottoms can play a significant role in
$Z'$ production. The production cross section is estimated via the
same methodology as for the coloron above. We do not consider any
NLO corrections in this case, following the methodology of \cite{Baker:2019sli}.
In Fig.~\ref{fig:Zprime_Production} we show that the production
via bottoms is larger than the production via valence quarks for a
light $Z'$, however the production via valence quarks is larger for
$M_{Z'}\apprge2\,\mathrm{TeV}$, and shall not be neglected as it
commonly happens in the literature (see e.g.~\cite{Baker:2019sli}).

\begin{figure}[t]
\subfloat[\label{fig:Zprime_muons}]{\includegraphics[scale=0.365]{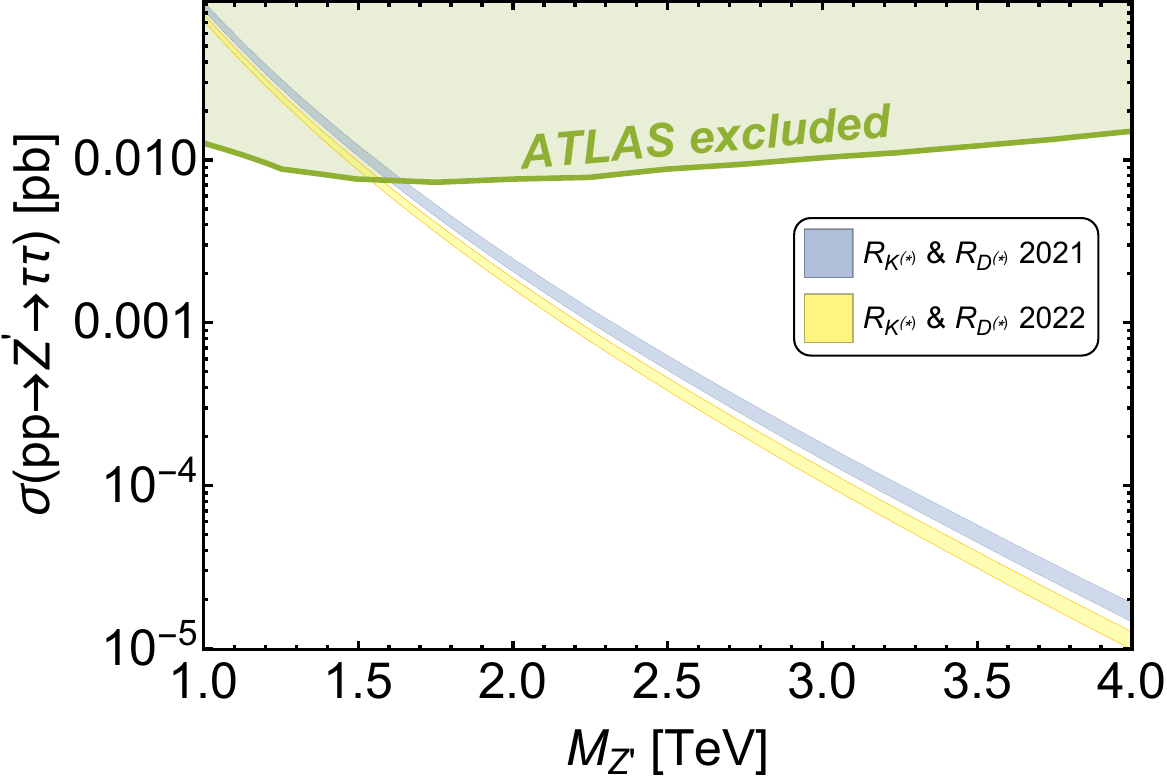}

}$\quad$\subfloat[\label{fig:Zprime_taus}]{\includegraphics[scale=0.35]{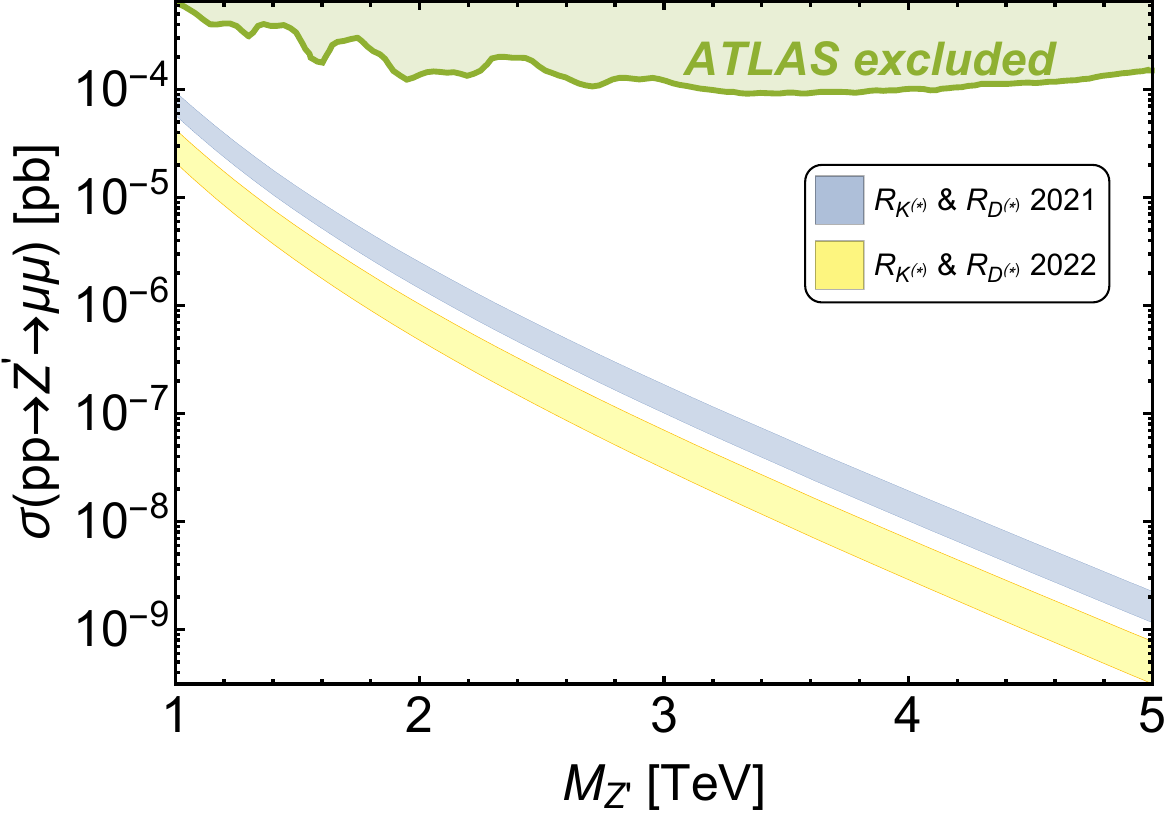}

}

\caption[$Z'$ dilepton production in the extended twin PS model]{Total cross section for ditau (left) and dimuon (right) production
via a heavy $Z'$ in the narrow width approximation, with $x_{25}^{\psi}$
varied in the range $x_{25}^{\psi}=[0.3,\,0.35]$ ($[0.4,\,0.45]$)
preferred by $R_{K^{(*)}}^{2022}$ ($R_{K^{(*)}}^{2021}$), obtaining
the yellow (blue) band. The exclusion bounds from ATLAS are shown
in green.}
\end{figure}

We estimate the branching fraction to muons and taus, and we compute
the total decay width via the narrow width approximation. We confront
our results with the limits from the dilepton resonance searches by
ATLAS \cite{ATLAS:2019erb} for muons and Fig.~7 (c) of \cite{ATLAS:2017eiz}
for taus. We display the results in Figs.~\ref{fig:Zprime_muons}
and \ref{fig:Zprime_taus}. In Fig.~\ref{fig:Colliders_All} we see
that these processes, along with coloron dijet searches, mildly constrain
the region of large $x_{25}^{\psi}$. Ditau searches are more competitive
than dimuon searches or coloron dijet searches despite the lesser integrated luminosity, due to the branching
fractions to muons and light quarks being suppressed by smaller mixing angles
$s_{25}^{Q,L}\sim\mathcal{O}(0.1)$. In contrast, the ditau channel is
enhanced by maximal 3-4 mixing, and sets bounds of roughly $M_{Z'}>1.5\,\mathrm{TeV}$,
see Fig.~\ref{fig:Zprime_taus}.

\subsubsection*{$U_{1}$ signals}

Leptoquark pair-production cross sections at the LHC are dominated
by QCD dynamics, and thus are largely independent of the leptoquark
couplings to fermions. Therefore, we are able to safely compare with
the analyses of Refs.~\cite{Cornella:2021sby,Baker:2019sli}. A certain
model dependence is present in the form of non-minimal couplings to
gluons, however these couplings are absent in models where $U_{1}$
has a gauge origin. According to Fig.~3.3 of \cite{Cornella:2021sby},
current bounds over direct production exclude $M_{U_{1}}<1.7\,\mathrm{TeV}$,
and the future bound is expected to exclude $M_{U_{1}}<2.1\,\mathrm{TeV}$
if no NP signal is found during the high-luminosity phase of the LHC.

An important constraint over $U_{1}$ arises from modifications of
the high-$p_{T}$ tail in the dilepton invariance mass distribution
of the Drell-Yan process $pp\rightarrow\tau^{+}\tau^{-}+X$, induced
by $t$-channel $U_{1}$ exchange \cite{Diaz:2017lit,Baker:2019sli,Angelescu:2021lln,Cornella:2021sby}.
This channel is well motivated by the $U_{1}$ explanation of $R_{D^{(*)}}$,
which unavoidably predicts a large $b\tau U_{1}$ coupling. The scenario
$\beta_{b\tau}^{R}=0$ considered in the study of \cite{Baker:2019sli,Cornella:2021sby}
fits well the twin Pati-Salam framework, up to a re-scaling of the
$U_{1}$ coupling strength as $g_{U}\rightarrow g_{U}\beta_{b\tau}^{L}$,
in order to account for the fact that our $\beta_{b\tau}^{L}$ coupling is not
maximal but $\beta_{b\tau}^{L}\approx c_{\theta_{LQ}}\approx0.67$
in our benchmark scenario, obtaining $g_{U}\approx2.3$. According
to the left panel of Fig.~3.3 in \cite{Cornella:2021sby}, the 3
TeV leptoquark of our benchmark easily satisfies the current bounds,
but is within projected limits for the high luminosity phase of LHC.
Finding $U_{1}$ much below 3 TeV is in tension with $pp\rightarrow g'\rightarrow t\bar{t}$
as explained before, due to the approximate relation $M_{g'}\approx\sqrt{2}M_{U_{1}}$
that entangles the masses of $U_{1}$ and the coloron (although the
$pp\rightarrow g'\rightarrow t\bar{t}$ bound is probably overestimated
for our dominantly left-handed model). 

The twin Pati-Salam model could provide a good $U_{1}$ candidate
for the 3$\sigma$ excess at CMS \cite{CMS-PAS-EXO-19-016} pointing
to a 2 TeV $U_{1}$ leptoquark in the well motivated channel $pp\rightarrow U_{1}\rightarrow\tau\tau$,
once the extra decay channels to vector-like fermions are considered,
assuming that the bound from $pp\rightarrow g'\rightarrow t\bar{t}$
is indeed overestimated for our model.

\subsubsection*{Vector-like fermions}

The presence of vector-like fermions is of fundamental importance
to discriminate between the different implementations of the 4321
model addressing the $B$-anomalies. A common constraint arises from
$\Delta F=2$ transitions at low energies, which require that the
vector-like charged lepton that mixes with muons is light (see Fig.~\ref{fig:ML5_deltaMs}).
The natural mass of the quark partner of $L_{5}$ should not lie far
away due to the approximate Pati-Salam universality, the small breaking
effects given by the VEV $\left\langle \Omega_{15}\right\rangle $.
In particular, in our benchmark scenario we obtained $M_{5}^{L}\approx0.8\,\mathrm{TeV}$
and $M_{5}^{Q}\approx1.2\,\mathrm{TeV}$. The flavour structure of
the model naturally predicts that both $Q_{5}$ and $L_{5}$ have
sizable couplings to the third generation of SM fermions.

The twin Pati-Salam model features also $L_{4}$ and $Q_{4}$ as a
relevant pair of vector-like fermions, which mix maximally with the
third generation in order to obtain the large couplings required for
$R_{D^{(*)}}$, and also to explain the top mass without perturbativity
issues. This implies that their bare mass terms in the original Lagrangian
are small, therefore their physical masses are dominated by $x_{34}^{\psi}\left\langle \phi_{3,1}\right\rangle $,
see Eqs.~(\ref{eq:34_mixing_extended}) and (\ref{eq:Mass-_4th}).
In the motivated scenario $\left\langle \phi_{3}\right\rangle \gg\left\langle \phi_{1}\right\rangle $
which slightly suppresses the production of the coloron, we found
$L_{4}$ to be very light, roughly 600 GeV in our benchmark.
Instead, $Q_{4}$ can lie above 1 TeV, being roughly 1.2 TeV in
our benchmark. The couplings of $L_{4}$ to SM fermions are smaller
than those of $L_{5}$, but it is dominantly coupled to third family
fermions.
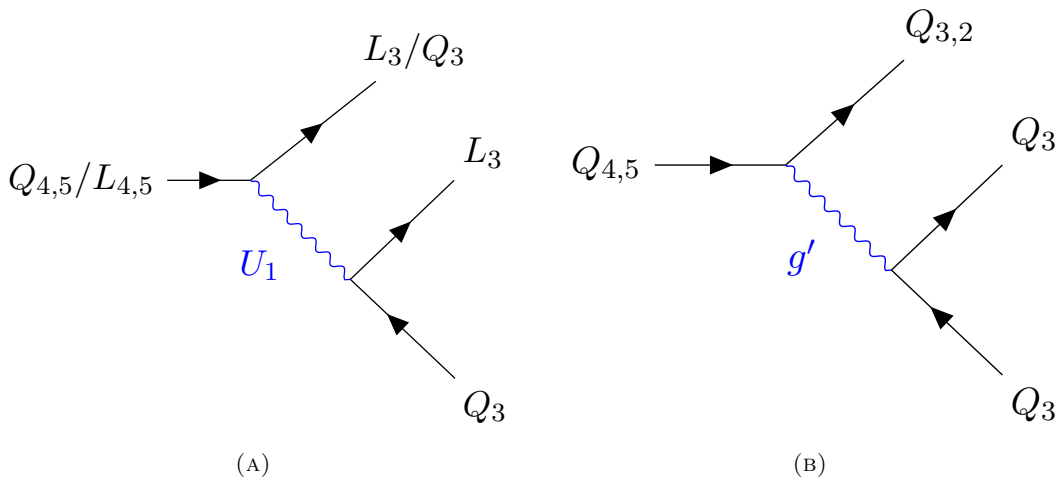
\begin{figure}
\subfloat[]{\resizebox{.47\textwidth}{!}{\begin{tikzpicture}	\begin{feynman}
		\vertex (a) {\(Q_{4,5}/L_{4,5}\)};
		\vertex [right=18mm of a] (b);
		\vertex [below right=of b] (c);
		\vertex [above right=of b] (f1) {\(L_{3}/Q_{3}\)};
		\vertex [above right=of c] (f2) {\(L_{3}\)};
		\vertex [below right=of c] (f3) {\(Q_{3}\)};
		\diagram* {
			(a) -- [fermion] (b) -- [fermion] (f1),
			(b) -- [boson, blue, edge label'=\(U_{1}\), inner sep=6pt] (c),
			(c) -- [fermion] (f2),
			(c) -- [anti fermion] (f3),
	};
	\end{feynman}
\end{tikzpicture}}

}$\quad$\subfloat[]{\resizebox{.46\textwidth}{!}{\begin{tikzpicture}	\begin{feynman}
		\vertex (a) {\(Q_{4,5}\)};
		\vertex [right=18mm of a] (b);
		\vertex [below right=of b] (c);
		\vertex [above right=of b] (f1) {\(Q_{3,2}\)};
		\vertex [above right=of c] (f2) {\(Q_{3}\)};
		\vertex [below right=of c] (f3) {\(Q_{3}\)};
		\diagram* {
			(a) -- [fermion] (b) -- [fermion] (f1),
			(b) -- [boson, blue, edge label'=\(g'\), inner sep=6pt] (c),
			(c) -- [fermion] (f2),
			(c) -- [anti fermion] (f3),
	};
	\end{feynman}
\end{tikzpicture}}

}

\caption[Examples of main decay channels for vector-like fermions in the extended twin PS model]{Examples of main decay channels for vector-like fermions. \label{fig:Vector-like_decays}}
\end{figure}

Interestingly, CMS recently performed a search for the vector-like
leptons of the 4321 model \cite{CMS:2022cpe}, finding a 2.8$\sigma$
preference for a vector-like lepton with a 600 GeV mass, however the analysis
assumes electroweak production only and maximal couplings to the third family.
If $Z'$-assisted production is included, $L_{5}$ with 800 GeV mass
could be a good candidate for the anomaly. Furthermore, $L_{4}$
at 600 GeV could also provide a good fit once non-maximal couplings
are considered, however this requires verification in a dedicated
analysis. Non-fermiophobic 4321 models, such as \cite{Cornella:2019hct,Cornella:2021sby},
predict a heavier vector-like lepton, while \cite{DiLuzio:2018zxy}
also predicts $L_{5}$ at around 800 GeV but a heavier $L_{4}$. Regarding
the sixth vector-like fermions $L_{6}$ and $Q_{6}$, we expect them
to have similar masses as $L_{5}$ and $Q_{5}$ in order to preserve
the GIM-like protection from 1-2 FCNCs, they are feebly coupled to
first family fermions but not to the second nor third families.

Current bounds on vector-like quark masses lie around 1 TeV, however the
strongest bounds are usually model dependent. Our vector-like quarks
are pair produced through gluon fusion and through the decay of the
coloron, which is very likely to be kinematically allowed. Their decays
leave a large number of third generation fermions in the final state,
following a similar pattern as the one discussed in \cite{DiLuzio:2018zxy},
see the example diagrams in Fig.~\ref{fig:Vector-like_decays}.
The twin Pati-Salam model naturally predicts light vector-like quarks
with masses around 1 TeV, which is a feature not present in all other
4321 models and may motivate specific searches.

\section{Comparison with alternative models\label{sec:Comparison_models}}

Table~\ref{tab:Comparison_Models} includes a simplified set of observables
that allows to disentangle the twin Pati-Salam model from the ones that are
already in the market. A further discussion can be found in the two
following subsections. 

\subsection{\texorpdfstring{Non-fermiophobic 4321 models}{Non-fermiophobic 4321 models}}

The twin Pati-Salam model is built as a fermiophobic framework, where all
chiral fermions are singlets under the TeV scale $SU(4)$. This is
a crucial difference between our model and the non-fermiophobic 4321
models \cite{Cornella:2019hct,Cornella:2021sby,Barbieri:2022ikw,Fuentes-Martin:2020bnh,Fuentes-Martin:2020hvc}
and their UV completions (including the $\mathrm{PS}^{3}$ model \cite{Bordone:2017bld}),
where the third family of chiral fermions transforms under the TeV
scale $SU(4)$. This implies large left- and right-handed third family
couplings to $SU(4)$ gauge bosons. By contrast, in our theory of
flavour, the right-handed couplings of SM fermions to $U_{1}$ (and
also to $Z'$ and $g'$) arise via small mixing angles connected to
the origin of second family fermion masses, hence the twin Pati-Salam model
predicts dominantly left-handed $U_{1}$ couplings. The low-energy
phenomenology between both approaches is radically different. 

In terms of the charged current anomalies $R_{D^{(*)}}$, the twin
Pati-Salam model only predicts the effective operator $(\bar{c}_{L}\gamma_{\mu}b_{L})(\bar{\text{\ensuremath{\tau}}}_{L}\gamma^{\mu}\nu_{\tau L})$,
and hence both $R_{D}$ and $R_{D^{*}}$ are corrected in the same
direction and with the same size. In contrast, non-fermiophobic 4321 models
also predict the scalar operator $(\bar{c}_{L}b_{R})(\bar{\text{\ensuremath{\tau}}}_{R}\nu_{\tau L})$.
Due to the presence of this operator, the NP effect on $\Delta R_{D}$
is larger than on $\Delta R_{D^{*}}$ (about 5/2 larger for the $\mathrm{PS}^{3}$
model, see Eq.~(27) in \cite{Bordone:2017bld}).

Another key observable is $B\rightarrow K\nu\nu$, for which the twin
Pati-Salam model predicts a larger branching fraction that will be fully tested
by Belle~II. Instead, non-fermiophobic 4321 models predict a smaller
branching fraction, see Fig.~4.4 of \cite{Cornella:2021sby} and
compare their purple region with our Fig.~\ref{fig:BtoK_nunu}. Moreover,
in our analysis of $B\rightarrow K\nu\nu$ we have highlighted correlations
with $B_{s}-\bar{B}_{s}$ mixing due to the loops being dominated
by the same vector-like charged lepton, a feature which is missing in the analysis
of \cite{Cornella:2021sby}.
\begin{table}[t]
\resizebox{\textwidth}{!}{
\begin{tabular}{ccccc}
\toprule 
 & twin PS & fermiophobic 4321 & $\mathrm{PS}^{3}$ & non-fermiophobic 4321\tabularnewline
\midrule
\midrule 
Refs. & this thesis & \cite{DiLuzio:2017vat,DiLuzio:2018zxy} & \cite{Bordone:2017bld,Bordone:2018nbg} & \cite{Cornella:2019hct,Cornella:2021sby,Barbieri:2022ikw,Greljo:2018tuh}\tabularnewline
\midrule 
Theory of flavour & Yes & No & Yes & No\tabularnewline
\midrule 
$R_{D^{(*)}}$ & $\Delta R_{D}=\Delta R_{D^{*}}$ & $\Delta R_{D}=\Delta R_{D^{*}}$ & $\Delta R_{D}>\Delta R_{D^{*}}$ & $\Delta R_{D}>\Delta R_{D^{*}}$\tabularnewline
\midrule 
$\mathcal{B}\left(\tau\rightarrow3\mu\right)$ & $10^{-8}$ & $\lesssim10^{-11}$ & $10^{-9}$ & -\tabularnewline
\midrule 
$\mathcal{B}\left(\tau\rightarrow\mu\gamma\right)$ & $10^{-9}$ & $\lesssim10^{-11}$ & $10^{-8}$ & $10^{-8}$\tabularnewline
\midrule 
$\mathcal{B}\left(\tau\rightarrow\mu\phi\right)$ & $10^{-9}$ & $10^{-11}$ & $10^{-10}$ & $10^{-10}$\tabularnewline
\midrule 
$\mathcal{B}\left(B_{s}\rightarrow\tau\mu\right)$ & $10^{-6}$ & \multicolumn{1}{c}{$10^{-7}$} & $10^{-5}$ & $10^{-5}$\tabularnewline
\midrule 
$\mathcal{B}\left(B_{s}\rightarrow\tau\tau\right)$ & $10^{-5}$ & - & $10^{-3}$ & $10^{-3}$\tabularnewline
\midrule 
$\mathcal{B}\left(B\rightarrow K\tau\tau\right)$ & $10^{-5}$ & - & $10^{-4}$ & $10^{-4}$\tabularnewline
\midrule 
$\delta\mathcal{B}(B\rightarrow K^{(*)}\nu\bar{\nu})$ (\ref{eq:BtoK_nunu}) & $0.3$ & - & $0.2$ & $0.2$\tabularnewline
\midrule 
vector-like fermion families & 3 & 3 & 1 & 1\tabularnewline
\midrule 
High-$p_{T}$ constraints & Mild & Mild & Tight & Tight\tabularnewline
\bottomrule
\end{tabular}}\caption[Main observables to distinguish the twin PS model from other proposals.]{Main observables to distinguish the twin Pati-Salam model from other proposals.
The numbers are only indicative, as these predictions may vary along
the parameter space of the different models. The dash (-) indicates
that the observable was not considered or numbers were not given in
the corresponding references. In the high-$p_{T}$ row we broadly
refer to how constrained is the model by high-$p_{T}$ searches~\cite{Aebischer:2022oqe}.\label{tab:Comparison_Models}}
\end{table}

Regarding the rest of the observables, broadly speaking the twin PS
model predicts larger branching fractions for LFV processes. The exception
is $\tau\rightarrow\mu\gamma$, which is enhanced in non-fermiophobic
models via a chirality flip with the bottom quark running in the loop.
The rare decays $B_{s}\rightarrow\tau\tau$ and $B\rightarrow K\tau\tau$
are chirally enhanced in non-fermiophobic models
due to the presence of scalar operators connected to the third family
right-handed couplings. The LHCb and Belle~II collaborations will test significant regions
of the parameter space to disentangle the different
4321 approaches. \\ The lifetime ratio $\tau_{B_{s}}/\tau_{B_{d}}$ introduced
in Section~\ref{subsec:LifetimeRationU1} can potentially discriminate
as well between the twin PS model and non-fermiophobic 4321 models \cite{Bordone:2023ybl},
see Fig.~\ref{Fig:U1_LifetimeRatios}.

High-$p_{T}$ searches also offer a window to disentangle both approaches,
since most of the constraints afflicting non-fermiophobic 4321 scenarios
are relaxed in the dominantly left-handed scenario of the twin PS
model. Particularly relevant are also the different implementations
of vector-like fermions.

\subsection{Fermiophobic 4321 models}

To our knowledge, the only fermiophobic 4321 model proposed in
the literature is that of Ref.~\cite{DiLuzio:2017vat}, whose phenomenology
was studied in detail in \cite{DiLuzio:2018zxy}. This model presents
a simplified fermiophobic scenario with a rather ad-hoc flavour structure
motivated by the phenomenology, including an ad-hoc alignment of SM-like
Yukawas and VL-chiral fermion mixing. Furthermore, \cite{DiLuzio:2017vat}
does not address the question of quark and lepton masses (is not a
theory of flavour), unlike the model proposed here. It lacks
quark-lepton unification of SM fermions and leads to a less predictive
framework than the twin Pati-Salam model.

The twin Pati-Salam model leads to an effective fermiophobic 4321
model at the TeV scale. However, the underlying twin PS symmetry implies
correlations of key parameters, leading to extra constraints and correlations
between observables, which are not present in the analyses of \cite{DiLuzio:2017vat,DiLuzio:2018zxy}.
For example, $R_{D^{(*)}}$ and $R_{K^{(*)}}$ are correlated here
due to the universality of $x_{25}^{\psi}$ and $x_{34}^{\psi}$,
leading to quasi-universal mixing angles $s_{25}^{Q,L}$ and $s_{34}^{Q,L}$.
By contrast, such mixing angles are free parameters in \cite{DiLuzio:2017vat,DiLuzio:2018zxy}
and one can explain $R_{D^{(*)}}$ without giving any contribution
to $R_{K^{(*)}}$. As a consequence, a dedicated analysis was required
to show that $R_{D^{(*)}}$ can be explained in the twin PS model
while being compatible with the recent data on $R_{K^{(*)}}$, as
we did in this chapter.

Since the twin Pati-Salam model is a theory of flavour, while \cite{DiLuzio:2017vat,DiLuzio:2018zxy}
is not, new signals are predicted in LFV processes connected to the
origin of fermion masses and mixings. The twin PS model predicts non-vanishing
$\text{\ensuremath{\mu\tau}}$ mixing, leading to striking signals
in $\tau\rightarrow3\mu$ and $\tau\rightarrow\mu\gamma$ close to
current experimental bounds, as summarised in Figs.~\ref{fig:tau_3mu}
and \ref{fig:tau_muphoton}. The large contributions to the branching
fractions of $\tau\rightarrow3\mu$ and $\tau\rightarrow\mu\gamma$
are mediated by the $Z'$ boson, see the purple region in Figs.~\ref{fig:tau_3mu}
and \ref{fig:tau_muphoton}. In contrast, in simplified fermiophobic
4321 models \cite{DiLuzio:2017vat,DiLuzio:2018zxy} only a much smaller
1-loop $U_{1}$-mediated signal is predicted. This signal was not
computed in Refs.~\cite{DiLuzio:2017vat,DiLuzio:2018zxy} as it is
very small compared to the experimental bounds, but we have computed
it here for the sake of comparison, and it is depicted as the blue
region in Figs.~\ref{fig:tau_3mu} and \ref{fig:tau_muphoton}. In
a similar way, we obtain $\mathcal{B}(\tau\rightarrow\mu\phi)$ two
orders of magnitude larger than in \cite{DiLuzio:2017vat,DiLuzio:2018zxy}
due to the $\tau\mu$ mixing predicted by the twin Pati-Salam model.

Finally, the fermion mixing predicted by the twin Pati-Salam model avoids
current constraints coming from CKM unitarity, $\Delta F=2$ and electroweak
precision observables as presented in \cite{Fajfer:2013wca}. The reasons
are the absence of SM-like Yukawa couplings for chiral fermions in
the original basis (as they will be generated indeed via this mixing),
along with the fact that vector-like quark $SU(2)_{L}$ doublets and SM quark
$SU(2)_{L}$ singlets do not mix, hence the vector-like quark doublet remains
unsplit. Remarkably, this is different from \cite{DiLuzio:2017vat,DiLuzio:2018zxy},
where mixing between the chiral quark singlets and the vector-like (right-handed)
quark doublet was induced due to the presence of the SM-like Yukawa
couplings for chiral fermions, leading to possible splitting of the
vector-like quark doublet, which constrains the mixing angles for third family
quarks according to the analysis in \cite{Fajfer:2013wca}.

\section{Conclusions\label{sec:Conclusions}}

We have proposed the twin Pati-Salam model as a multi-scale theory
of flavour able to explain the origin of the flavour structure of
the SM, connecting the origin of the SM Yukawa couplings with the
origin of the couplings to fermions of a TeV scale vector leptoquark
$U_{1}$ that explains the anomalies in $B$-physics. The basic idea
of this model is that all three families of SM chiral fermions transform
under one PS group, while families of vector-like fermions transform
under the other one. Vector leptoquark couplings and SM Yukawa couplings
emerge together after mixing of the chiral fermions with the vector-like
fermions, thereby providing a direct link between $B$-physics and
fermion masses and mixings. In this manner, the twin Pati-Salam model features a fermiophobic framework in complete analogy with the fermiophobic $Z'$ models of Chapter~\ref{Chapter:Fermiophobic}, such that both the $B$-anomalies and the flavour hierarchies are explained via the mechanism of messenger dominance \cite{Ferretti:2006df}. Remarkably, the need to explain the low-energy
$B$-anomalies fixes the NP scales of flavour $\left\langle \phi\right\rangle $
and $M$ (which \textit{a priori} could be anywhere \textit{from the
Planck scale to the electroweak scale})
to live close to the TeV scale, at least those involving the origin
of second and third family fermion masses.

Firstly, we presented a simplified version of the model where second
and third family fermion masses originate from mixing with a ``fourth''
vector-like family of fermions charged under $SU(4)^{I}_{PS}$, broken
at the TeV scale. $SU(2)_{L}$ doublet vector-like fermions with masses around
the TeV scale are assumed to have a large mixing with left-handed
third family fermions, in order to provide their effective Yukawa
couplings. In the spirit of messenger dominance \cite{Ferretti:2006df}, hierarchically heavier $SU(2)_{L}$
vector-like fermion singlets mix with second and third family right-handed
fermions, in order to provide second family fermion masses and the
small $V_{cb}$ CKM element. The origin of first family fermion masses
and their mixing is connected to vector-like fermions charged under the second
PS group, broken around the much heavier PeV scale, completing the
multi-scale picture for the origin of flavour.

However, with only a single vector-like family charged under $SU(4)^{I}_{PS}$,
the model is unable to explain the $B$-anomalies in a natural way,
as it does not achieve the required flavour structure to avoid constraints
by $B_{s}-\bar{B}_{s}$ meson mixing mediated by the heavy neutral
vectors of the model. We then extended the simplified model to include
three vector-like families, together with a $\mathbb{Z}_{4}$ discrete symmetry
to control the flavour structure. This version of the model allows
for larger flavour-violating and dominantly left-handed $U_{1}$ couplings
as required to address $R_{D^{(*)}}$, thanks to mixing between a
fourth and fifth vector-like families which also mix with the second
and third generations of SM fermions. A sixth vector-like family is
included to mix with the first SM family, for the sake of suppressing
any FCNCs involving first and second family fermions. The mechanism
resembles the GIM suppression of FCNCs in the SM, featuring a similar
Cabbibo-like matrix which is present in leptoquark currents, but not
in neutral currents mediated by the coloron and $Z'$.

This version of the model can explain the $R_{D^{(*)}}$ anomalies
at 1$\sigma$ while being compatible with all data, however we expect
small deviations from the SM on the $R_{K^{(*)}}$ ratios, to be tested
in the future via more precise tests of LFU by the LHCb collaboration.
In contrast to the alternative models, our model predicts dominantly
left-handed $U_{1}$ couplings to fermions, leading to $\Delta R_{D}=\Delta R_{D^{*}}$.
The contribution to $R_{D^{(*)}}$ is correlated as well with an universal
contribution to the operator $\mathcal{O}_{9}$, that further improves
the overall fit of the model to $b\rightarrow s\mu\mu$ data.

Non-negligible $\mu\tau$ mixing is predicted by the theory of flavour,
leading to interesting signals in $\tau\rightarrow3\mu$ and $\tau\rightarrow\mu\gamma$,
mostly due to $Z'$ exchange, which are close to present experimental
bounds in some region of the parameter space. Signals in LFV semileptonic
processes mediated by $U_{1}$ at tree-level are found to lie well
below current experimental limits, with the exception of $K_{L}\rightarrow e\mu$
which constrains a small region of the parameter space. However, this
tension can be alleviated if the first family $U_{1}$ coupling is
diluted via mixing with vector-like fermions. Tests of LFU in tau
decays set important bounds over the mass of $U_{1}$ depending on
its coupling to third family fermions. Contributions of $U_{1}$
to the rare decays $B_{s}\rightarrow\tau\tau$ and $B\rightarrow K\tau\tau$
are broadly below current and projected experimental sensitivity.
Instead, the rare decay $B\rightarrow K^{(*)}\nu\bar{\nu}$ offers
the opportunity to fully test the model in the near future, since
Belle~II is expected to cover all the parameter space compatible with
the $B$-anomalies. Remarkably, the model can be easily disentangled from
all other proposals via the previous set of observables, as discussed
in Section~\ref{sec:Comparison_models}.

Apart from the aforementioned low-energy predictions at LHCb and Belle~II,
the model is also testable via high-$p_{T}$ searches at the LHC.
The study of the 1-loop contribution of vector leptoquark $U_{1}$
exchange to $B_{s}-\bar{B}_{s}$ mixing revealed that the fifth vector-like
lepton has to be light, around 1-2 TeV, to be compatible with the
stringent bound from $\Delta M_{s}$. This is easily achieved in the
twin Pati-Salam model, where light vector-like fermions are well motivated
in order to naturally obtain the large mixing needed to fit the $R_{D^{(*)}}$
anomaly, and also to fit the heavy top mass without perturbativity
issues. In particular, the fourth and fifth charged leptons are suggested
as good candidates to explain the CMS excess \cite{CMS:2022cpe},
but further study is required in this direction. Vector-like quarks
are found to lie not far above 1 TeV in the suggested benchmark, hence
motivating specific searches at LHC to be performed. Regarding the
heavy vectors, dijet searches and dilepton searches set mild bounds
over the mass of the coloron and $Z'$, respectively. The more stringent
bound over the scale of the model arises from the ditop searches in
\cite{Baker:2019sli,Cornella:2021sby}, which push the mass of the
coloron to lie above 3.5 TeV, however those bounds could be slightly
overestimated for our model as they are obtained for different 4321
scenarios. Finally, the mass range for $U_{1}$ is compatible with
current bounds, and mostly lie within the projected sensitivity of
the high luminosity phase of LHC. A good fit for the $3\sigma$ CMS
excess in $U_{1}$ searches \cite{CMS-PAS-EXO-19-016} could be provided
if the extra decay channels to vector-like fermions are considered,
assuming that the bound from ditop searches is indeed overestimated.

As we have shown, the twin Pati-Salam model predicts the NP scenario preferred
by the global fits \cite{Alguero:2023jeh} in order to explain $B$-physics
data, and connects explicitly the effective $U_{1}$ leptoquark couplings
with the origin of Yukawa couplings in the SM. The model exhibits
a rich flavour phenomenology with key observables that will allow
to disentangle it from all other proposals, along with a TeV-scale
phenomenology that can be probed in the near future by high-$p_{T}$
experiments at the LHC.
%% ----------------------------------------------------------------
%% Trihypercharge.tex
%% ---------------------------------------------------------------- 
\chapter{Tri-hypercharge: a path to the origin of flavour} \label{Chapter:Tri-hypercharge}

\begin{quote}
    ``You’re telling me that people at CERN dug
    out millions of tons of earth just to smash tiny particles?''
    Kohler shrugged. ``Sometimes to find truth, one must move mountains.''
  \begin{flushright}
  \hfill \hfill $-$ Dan Brown, \textit{Angels \& Demons}
  \par\end{flushright}
\end{quote}

\noindent In this chapter, based on Ref.~\cite{FernandezNavarro:2023rhv}, we introduce a theory of flavour based on assigning
a separate gauged weak hypercharge to each family of chiral fermions. Assuming that the Higgs doublet only
carries third family hypercharge, then only third family fermions
get renormalisable Yukawa couplings, explaining their heaviness. Light
charged fermion masses and CKM mixing may arise from non-renormalisable operators, connected to the new
scalar fields that break the three hypercharge groups down to SM hypercharge.
We shall conclude that neutrino masses and mixing may be explained via the addition of vector-like
singlet neutrinos that carry cancelling family hypercharges. Finally, we will see that this model
has a rich phenomenology if the new physics scales are low, including
flavour-violating observables, LHC physics and electroweak precision
observables.

\section{Introduction}

Theories of flavour may involve new symmetries (global, local, continuous,
discrete, abelian, non-abelian...) beyond the SM group, possibly broken at
some high scale down to the SM. Traditionally, new gauge structures
beyond the SM have been considered to be flavour universal, as grand
unified theories (GUTs) which usually embed all three families in
an identical way, or even extended GUTs which embed all three families
in a single representation (usually along with extra exotic fermions).
Alternatively, there exists the well-motivated case of family symmetries
which commute with the SM gauge group, and are then spontaneously
broken, leading to family structure. However, there are other less
explored ways in which the SM gauge group could be embedded into a
larger gauge structure in a flavour non-universal way. In particular,
the \textit{family decomposition} of the SM gauge group (including a hierarchical
symmetry breaking pattern down to the SM) was first proposed during
the 80s and 90s, with the purpose of motivating lepton non-universality
\cite{Li:1981nk,Ma:1987ds,Ma:1988dn,Li:1992fi} or assisting technicolor
model building \cite{Hill:1994hp,Muller:1996dj,Malkawi:1996fs}. However,
the natural origin of flavour hierarchies in such a framework was
not explored until more recently in \cite{Craig:2011yk,Panico:2016ull,Barbieri:2021wrc}.
Here it was proposed that the flavour non-universality of Yukawa couplings
in the SM might well find its origin in a flavour non-universal gauge
sector, broken in a hierarchical way down to the SM. Interestingly,
model building in this direction has received particular attention
in recent years \cite{Bordone:2017bld,Allwicher:2020esa,Fuentes-Martin:2020pww,Fuentes-Martin:2022xnb,Davighi:2022bqf,Davighi:2022fer}.
With the exception of Ref.~\cite{Davighi:2022fer}, the remaining
recent attempts have been motivated by the need to obtain a TeV scale
vector leptoquark from Pati-Salam unification in order to address
the $B$-anomalies\footnote{Remarkably, flavour non-universality is not the only way to connect
the TeV-scale Pati-Salam vector leptoquark addressing the $B$-anomalies
with the origin of flavour hierarchies, see the twin Pati-Salam theory
of flavour discussed in Chapter~\ref{Chapter:TwinPS} of this thesis,
which considers the mechanism of messenger dominance \cite{Ferretti:2006df}}. Therefore, all these setups share a similar feature: a low scale
$SU(4)$ gauge group under which only the third family of SM fermions
transforms in a non-trivial way. In contrast, in this work we want
to explore the capabilities of flavour non-universality to address
the flavour puzzle in a more minimal, simple and bottom-up approach.

We propose that the SM symmetry originates from
a larger gauge group in the UV that contains three separate weak hypercharge
gauge factors, 
\begin{equation}
SU(3)_{c}\times SU(2)_{L}\times U(1)_{Y_{1}}\times U(1)_{Y_{2}}\times U(1)_{Y_{3}}\,,
\end{equation}
which we will denote as the \textit{tri-hypercharge} (TH) $U(1)_{Y}^{3}$
gauge group. We will associate each of the three hypercharge gauge
groups with a separate SM family, such that each fermion family $i$
only carries hypercharge under the corresponding $U(1)_{Y_{i}}$ factor.
This ensures that each family transforms differently under the gauge
group $U(1)_{Y}^{3}$, which avoids the family repetition of the SM,
and provides the starting point for a theory of flavour. For example,
assuming that a single Higgs doublet only carries third family hypercharge,
then only the third family Yukawa couplings are allowed at renormalisable
level. With two Higgs doublets carrying third family hypercharge,
we show that the naturalness of the scheme increases. This simple
and economical framework naturally explains the heaviness of the third
family, the smallness of $V_{cb}$ and $V_{ub}$, and delivers Yukawa
couplings that preserve an accidental and global $U(2)^{5}$ flavour
symmetry acting on the light families, which is known to provide a
good first order description of the SM spectrum plus an efficient
suppression of flavour-violating effects for new physics \cite{Barbieri:2011ci}.
Remarkably, this appears to be the simplest way to provide the $U(2)^{5}$
flavour symmetry\footnote{An alternative way to deliver $U(2)^{5}$ consists in decomposing
$SU(2)_{L}$ only and taking advantage of the fact that right-handed rotations
remain unphysical in order to remove the remaining $U(2)^{5}$-breaking
entries of the Yukawa matrices (see the complete review of Ref.~\cite{Davighi:2023iks}).
Another example \cite{Allanach:2018lvl} considered an extension of
the SM by a $U(1)_{Y_{3}}$ gauge group under which only third family
fermions (and the Higgs) are hypercharge-like charged, where $U(1)_{Y_{3}}$
commutes with SM hypercharge, leading to an accidental $U(2)^{5}$.}. The masses of first and second family fermions, along with the CKM
mixing, then appear as small breaking sources of $U(2)^{5}$ that
arise after the cascade spontaneous symmetry breaking of $U(1)_{Y}^{3}$
down to SM hypercharge, which can be parameterised in a model-independent
way in terms of spurions. In a realistic model, the spurions will
be realised by a choice of ``hyperon'' scalars which transform under
the different family hypercharge groups, breaking the tri-hypercharge
symmetry. We will motivate a specific symmetry breaking chain where
dynamics at a low scale, which could be as low as the TeV, explain
the flavour hierarchies $m_{2}/m_{3}$, while dynamics at a heavier
scale explain $m_{1}/m_{2}$. This symmetry breaking pattern will
sequentially recover the approximate flavour symmetry of the SM, and
provide a natural suppression of FCNCs for TeV new physics, while
the rest of flavour-violating effects are suppressed by a naturally
heavier scale. In this manner, the tri-hypercharge gauge group is
an example of a multi-scale origin of flavour as introduced in Section~\ref{subsec:FromPlanckToEW}.
Moreover, later in Chapter~\ref{Chapter:Tri-unification} we shall see that the tri-hypercharge gauge group,
among other gauge non-universal theories, may arise from a gauge unified framework.

The chapter is organised as follows. In Section~\ref{sec:The-Tri-Hypercharge-model}
we introduce the TH gauge group, along with the fermion and Higgs
doublet content of the model. We discuss the implications for third
family fermion masses along with the mass hierarchy between the top
and bottom/tau fermions. In Section~\ref{sec:Charged-fermion-masses-mixing}
we study the origin of charged fermion masses and mixing in the TH
model, firstly via a spurion formalism which reveals model-independent
considerations, and secondly by introducing example models with hyperons.
In Section~\ref{sec:Neutrino-masses-and-Mixing} we study the origin
of neutrino masses and mixing in the TH model. In particular, we discuss
the impact of the $U(2)^{5}$ flavour symmetry over the dimension-5
Weinberg operator, and afterwards we provide an example type I seesaw
model where neutrino masses and mixing can be accommodated. In Section~\ref{sec:Phenomenology}
we perform a preliminary exploration of the phenomenological implications
and discovery prospects of the $U(1)_{Y}^{3}$ theory of flavour.
Finally, Section~\ref{sec:ConclusionsTri-hypercharge} outlines our main conclusions.

\section{Tri-hypercharge gauge theory\label{sec:The-Tri-Hypercharge-model}}

The tri-hypercharge gauge group is based on assigning a separate gauged weak hypercharge
to each fermion family, 
\begin{table}[t]
\begin{centering}
\begin{tabular}{lccccc}
\toprule 
Field & $SU(3)_{c}$ & $SU(2)_{L}$ & $U(1)_{Y_{1}}$ & $U(1)_{Y_{2}}$ & $U(1)_{Y_{3}}$\tabularnewline
\midrule
\midrule 
$Q_{1}$ & $\mathrm{\mathbf{3}}$ & $\mathbf{2}$ & $1/6$ & 0 & 0\tabularnewline
$u_{1}^{c}$ & $\overline{\mathbf{3}}$ & $\mathbf{1}$ & $-2/3$ & 0 & 0\tabularnewline
$d_{1}^{c}$ & $\overline{\mathbf{3}}$ & $\mathbf{1}$ & $1/3$ & 0 & 0\tabularnewline
$L_{1}$ & $\mathbf{1}$ & $\mathbf{2}$ & $-1/2$ & 0 & 0\tabularnewline
$e_{1}^{c}$ & $\mathbf{1}$ & $\mathbf{1}$ & $1$ & 0 & 0\tabularnewline
\midrule 
$Q_{2}$ & $\mathrm{\mathbf{3}}$ & $\mathbf{2}$ & 0 & $1/6$ & 0\tabularnewline
$u_{2}^{c}$ & $\overline{\mathbf{3}}$ & $\mathbf{1}$ & 0 & $-2/3$ & 0\tabularnewline
$d_{2}^{c}$ & $\overline{\mathbf{3}}$ & $\mathbf{1}$ & 0 & $1/3$ & 0\tabularnewline
$L_{2}$ & $\mathbf{1}$ & $\mathbf{2}$ & 0 & $-1/2$ & 0\tabularnewline
$e_{2}^{c}$ & $\mathbf{1}$ & $\mathbf{1}$ & 0 & $1$ & 0\tabularnewline
\midrule 
$Q_{3}$ & $\mathrm{\mathbf{3}}$ & $\mathbf{2}$ & 0 & 0 & $1/6$\tabularnewline
$u_{3}^{c}$ & $\overline{\mathbf{3}}$ & $\mathbf{1}$ & 0 & 0 & $-2/3$\tabularnewline
$d_{3}^{c}$ & $\overline{\mathbf{3}}$ & $\mathbf{1}$ & 0 & 0 & $1/3$\tabularnewline
$L_{3}$ & $\mathbf{1}$ & $\mathbf{2}$ & 0 & 0 & $-1/2$\tabularnewline
$e_{3}^{c}$ & $\mathbf{1}$ & $\mathbf{1}$ & 0 & 0 & $1$\tabularnewline
\bottomrule
\end{tabular}
\par\end{centering}
\caption[Charge assignments of the SM fermions under the tri-hypercharge gauge
group]{ Charge assignments of the SM fermions under the TH gauge group.
$Q_{i}$ and $L_{i}$ (where $i=1,2,3$) are left-handed $SU(2)_{L}$
doublets of chiral quarks and leptons, while $u_{i}^{c}$, $d_{i}^{c}$
and $e_{i}^{c}$ are the $CP$-conjugate right-handed quarks and
leptons (so that they become left-handed\footnotemark). \label{tab:Field_content}}
\end{table}
\begin{equation}
SU(3)_{c}\times SU(2)_{L}\times U(1)_{Y_{1}}\times U(1)_{Y_{2}}\times U(1)_{Y_{3}}\,,
\end{equation}
in such a way that the $i$th fermion family only carries $Y_{i}$
hypercharge, with the other hypercharges set equal to zero (see Table~\ref{tab:Field_content}),
where $Y=Y_{1}+Y_{2}+Y_{3}$ is equal to SM weak hypercharge. Anomalies
cancel separately for each family, as in the SM, but without family
replication. The TH gauge group is broken down to the SM via appropriate
SM singlet scalars, which however carry family hypercharges. We denote
these fields linking the family hypercharges as \textit{hyperons}.
The TH group could be broken down to the SM in different ways, however
we motivate the following symmetry breaking pattern, \allowdisplaybreaks[0]
\begin{alignat}{1}
 & SU(3)_{c}\times SU(2)_{L}\times{\displaystyle U(1)_{Y_{1}}\times U(1)_{Y_{2}}\times U(1)_{Y_{3}}}\nonumber \\
 & {\displaystyle \overset{v_{12}}{\rightarrow}SU(3)_{c}\times SU(2)_{L}\times U(1)_{Y_{1}+Y_{2}}\times U(1)_{Y_{3}}}\label{eq:Symmetry_Breaking}\\
 & {\displaystyle \overset{v_{23}}{\rightarrow}SU(3)_{c}\times SU(2)_{L}\times U(1)_{Y_{1}+Y_{2}+Y_{3}}\,.}\nonumber 
\end{alignat}\allowdisplaybreaks
\footnotetext{The reader who is not familiar with this notation based on left-handed
2-component Weyl spinors can find the connection with the traditional
4-component, left-right notation in Appendix~\ref{app:2-component_notation}.}This choice is well supported by symmetry arguments that will have
phenomenological consequences: At high energies, the TH group discriminates
between the three SM fermion families, explicitly breaking the approximate
$U(3)^{5}$ flavour symmetry of the SM. At a heavy scale $v_{12}$,
the first and second hypercharges are broken down to their diagonal
subgroup, and the associated $Z'$ boson potentially mediates dangerous
1-2 FCNCs. Nevertheless, the gauge group below the scale $v_{12}$
preserves an accidental $U(2)^{5}$ flavour symmetry. The groups $U(1)_{Y_{1}+Y_{2}}\times U(1)_{Y_{3}}$
are broken down to their diagonal subgroup (SM hypercharge) at a scale
$v_{23}$, and the associated $Z'$ boson is protected from mediating
the most dangerous FCNCs thanks to the $U(2)^{5}$ symmetry. In this
manner, the most dangerous FCNCs are suppressed by the heavier scale
$v_{12}$, while the scale $v_{23}$ can be very low with interesting
phenomenological implications. We will see that dynamics connected
to the scale $v_{12}$ will play a role in the origin of the family
hierarchy $m_{1}/m_{2}$, while dynamics connected to the scale $v_{23}$
will play a role in the origin of $m_{2}/m_{3}$. The distribution
of the various scales in the model reproduces what we would expect in a multi-scale theory of flavour based on $U(2)^{5}$, as anticipated in Section~\ref{subsec:U(2)5} and in Fig.~\ref{fig:Multiscale-picture-flavour}. Despite the apparently complex gauge sector of the tri-hypercharge setup,
consisting of five arbitrary gauge couplings, in Chapter~\ref{Chapter:Tri-unification} we shall see that such a theory may arise from a gauge unified framework.

Provided that the SM Higgs only carries third family hypercharge,
$H(\mathbf{1},\mathbf{2})_{(0,0,-\frac{1}{2})}$, then only third
family Yukawa couplings are allowed at renormalisable level and an
accidental $U(2)^{5}$ flavour symmetry acting on the light families
emerges in the Yukawa sector, 
\begin{equation}
\mathcal{L}=y_{t}Q_{3}\tilde{H}u_{3}^{c}+y_{b}Q_{3}Hd_{3}^{c}+y_{\tau}L_{3}He_{3}^{c}+\mathrm{h.c.}
\end{equation}
where $\tilde{H}$ is the $CP$-conjugate of $H$. This setup already
provides an explanation for the smallness of light fermion masses
with respect to the third family, along with the smallness of quark
mixing, as they all must arise from non-renormalisable operators which
minimally break the $U(2)^{5}$ symmetry. Although this is a good
first order description of the SM spectrum, the question of why the
bottom and tau fermions are much lighter than the top remains unanswered,
and assuming only a single Higgs doublet, a tuning of order 2\% for
the bottom coupling and of 1\% for the tau coupling would be required.
Given that $m_{s,\mu}\propto\lambda^{5}m_{t}$ while $m_{c}\propto\lambda^{3}m_{t}$,
this setup also requires to generate a stronger fermion hierarchy
in the down and charged lepton sectors with respect to the up sector,
unless the tuning in the bottom and tau couplings is extended to the
second family. As we shall see shortly, the $U(1)_{Y}^{3}$ model
(and very likely a more general set of theories of flavour based on
the family decomposition of the SM group) predicts a similar mass
hierarchy for all charged sectors.

Due to the above considerations, it seems natural to consider a type
II two Higgs doublet model (2HDM), where both Higgs doublets only
carry third family hypercharge,
\begin{equation}
H_{u}(\mathbf{1},\mathbf{2})_{(0,0,\frac{1}{2})},\ \ \ \ H_{d}(\mathbf{1},\mathbf{2})_{(0,0,-\frac{1}{2})}\,,
\end{equation}
where as usual for a type II 2HDM, FCNCs can be forbidden by e.g.~a softly broken $\mathbb{Z}_{2}$ discrete symmetry or by Supersymmetry (not
necessarily low scale), which we however do not specify in order to
preserve the bottom-up spirit of this work. In any case, $\tan\beta=v_{u}/v_{d}\sim\lambda^{-2}\approx20$,
which is compatible with current data (see e.g.~\cite{Beniwal:2022kyv,deGiorgi:2023wjh}),
will provide the hierarchy between the top and bottom/tau masses with
all dimensionless couplings being $\mathcal{O}(1)$. Such an overall hierarchy
between the down and charged lepton sectors with respect to the up
sector is extended to all families, providing a better description
of second family charged fermion masses as we shall see.

\section{Charged fermion masses and mixing\label{sec:Charged-fermion-masses-mixing}}

\subsection{Lessons from the spurion formalism\label{subsec:Lessons-from-the}}

In all generality, we introduce $U(2)^{5}$-breaking \textit{spurions}
$\Phi$ in the Yukawa matrices of charged fermions
\begin{flalign}
{\cal L} & =\begin{pmatrix}Q_{1} & Q_{2} & Q_{3}\end{pmatrix}\begin{pmatrix}{\Phi}(\frac{1}{2},0,-\frac{1}{2}) & {\Phi}(-\frac{1}{6},\frac{2}{3},-\frac{1}{2}) & {\Phi}(-\frac{1}{6},0,\frac{1}{6})\\
{\Phi}(\frac{2}{3},-\frac{1}{6},-\frac{1}{2}) & {\Phi}(0,\frac{1}{2},-\frac{1}{2}) & {\Phi}(0,-\frac{1}{6},\frac{1}{6})\\
{\Phi}(\frac{2}{3},0,-\frac{2}{3}) & {\Phi}(0,\frac{2}{3},-\frac{2}{3}) & 1
\end{pmatrix}\begin{pmatrix}u_{1}^{c}\\
u_{2}^{c}\\
u_{3}^{c}
\end{pmatrix}H_{u}\nonumber \\
 & +\begin{pmatrix}Q_{1} & Q_{2} & Q_{3}\end{pmatrix}\begin{pmatrix}{\Phi}(-\frac{1}{2},0,\frac{1}{2}) & {\Phi}(-\frac{1}{6},-\frac{1}{3},\frac{1}{2}) & {\Phi}(-\frac{1}{6},0,\frac{1}{6})\\
{\Phi}(-\frac{1}{3},-\frac{1}{6},\frac{1}{2}) & {\Phi}(0,-\frac{1}{2},\frac{1}{2}) & {\Phi}(0,-\frac{1}{6},\frac{1}{6})\\
{\Phi}(-\frac{1}{3},0,\frac{1}{3}) & {\Phi}(0,-\frac{1}{3},\frac{1}{3}) & 1
\end{pmatrix}\begin{pmatrix}d_{1}^{c}\\
d_{2}^{c}\\
d_{3}^{c}
\end{pmatrix}H_{d}\label{eq:Spurions}\\
 & +\begin{pmatrix}L_{1} & L_{2} & L_{3}\end{pmatrix}\begin{pmatrix}{\Phi}(-\frac{1}{2},0,\frac{1}{2}) & {\Phi}(\frac{1}{2},-1,\frac{1}{2}) & {\Phi}(\frac{1}{2},0,-\frac{1}{2})\\
{\Phi}(-1,\frac{1}{2},\frac{1}{2}) & {\Phi}(0,-\frac{1}{2},\frac{1}{2}) & {\Phi}(0,\frac{1}{2},-\frac{1}{2})\\
{\Phi}(-1,0,1) & {\Phi}(0,-1,1) & 1
\end{pmatrix}\begin{pmatrix}e_{1}^{c}\\
e_{2}^{c}\\
e_{3}^{c}
\end{pmatrix}H_{d}\,+\mathrm{h.c.}\,,\nonumber 
\end{flalign}
where each spurion carries non-trivial charge assignments under $U(1)_{Y}^{3}$.
In an EFT approach, each spurion above can be matched to specific
ratios of hyperons $\phi_{i}$ over EFT cut-off scales $\Lambda_{i}$,
i.e. 
\begin{equation}
\Phi=\frac{\phi_{1}...\phi_{n}}{\Lambda_{1}...\Lambda_{n}}\,,
\end{equation}
where we have suppressed dimensionless couplings. The choice of hyperons
and $\Lambda_{i}$ above carries all the model dependence.

Assuming that the cut-off scales $\Lambda_{i}$ of the EFT are universal,
i.e.~that all $\Lambda_{i}$ are common to all charged sectors,
then the spurion formalism reveals some general considerations about
the origin of charged fermion masses and mixing: 
\begin{itemize}
\item The same spurions (up to conjugation) appear in the diagonal entries
of all matrices. Therefore, unless texture zeros are introduced in
specific models, this means that the masses of second family fermions
are likely to be degenerate up to dimensionless couplings, and the
same discussion applies to first family fermions. This motivates again
the addition of the second Higgs doublet or an alternative mechanism
in order to generate the hierarchy between the charm mass and the
lighter strange and muon masses. 
\item The same spurions appear in the (2,3) entries of the up and down Yukawa
matrices. Therefore, the 2-3 mixing in both the up and down sectors
is expected to be of a similar size, giving no predictions about the
alignment of the CKM element $V_{cb}$. The similar argument applies
to 1-3 mixing and $V_{ub}$. 
\item The spurions in the (1,2) entry of the up and down matrices are different.
Therefore, specific models have the potential to give predictions
about the alignment of the CKM element $V_{us}$. 
\item The same spurion (up to conjugation) that enters in all (2,2) entries
also populates the (2,3) entry of the charged lepton Yukawa matrix.
Similarly, the same spurion (up to conjugation) that enters in all
(1,1) entries also populates the (1,3) entry of the charged lepton
Yukawa matrix. In general, this predicts left-handed $\mu-\tau$ ($e-\tau$)
mixing of $\mathcal{O}(m_{2}/m_{3})$ ($\mathcal{O}(m_{1}/m_{3})$),
unless texture zeros are introduced in specific models (see Section~\ref{subsec:From-Spurions}).
This leads to a sizable enhancement of LFV $\tau\rightarrow\mu$ and
$\tau\rightarrow e$ transitions above the SM predictions, mediated
by heavy $Z'$ bosons in the model (see Section~\ref{sec:Phenomenology}). 
\item The spurions in the lower off-diagonal entries of the Yukawa matrices
all carry independent charge assignments, so right-handed fermion
mixing is model-dependent and can be different in all charged sectors. 
\end{itemize}
In the following, we go beyond the spurion formalism and introduce
different sets of hyperons. As we shall wee, the hyperons will provide
small $U(2)^{5}$-breaking effects via non-renormalisable operators,
leading to the masses of first and second family charged fermions,
along with CKM mixing. In the next few subsections, we will describe
example scenarios which provide a good description of charged fermion
masses and mixing.

\subsection{From spurions to hyperons \label{subsec:From-Spurions}}

The physical origin of the spurions of the previous subsection will
correspond to new Higgs scalar fields that break the $U(1)_{Y}^{3}$
symmetry, which we call \textit{hyperons}. The hyperons induce small
$U(2)^{5}$-breaking effects at the non-renormalisable level that
will lead to the masses and mixings of charged fermions. As the most
straightforward scenario, we could promote the spurions in the diagonal
entries of the matrices in Eq.~(\ref{eq:Spurions}) to hyperons,
along with the off-diagonal spurions in the upper half of the down
matrix\footnote{Notice that the same spurions enter in both the (1,3) and (2,3) entries
of the up and down matrices in Eq.~(\ref{eq:Spurions}).}. In an EFT approach, the set of hyperons that we have assumed generates
the following Yukawa matrices, 
\begin{flalign}
{\cal L}^{d\leq5} & =\begin{pmatrix}Q_{1} & Q_{2} & Q_{3}\end{pmatrix}\begin{pmatrix}{\phi}_{\ell13}^{(\frac{1}{2},0,-\frac{1}{2})}/\Lambda & 0 & {\phi}_{q13}^{(-\frac{1}{6},0,\frac{1}{6})}/\Lambda\\
0 & {\phi}_{\ell23}^{(0,\frac{1}{2},-\frac{1}{2})}/\Lambda & {\phi}_{q23}^{(0,-\frac{1}{6},\frac{1}{6})}/\Lambda\\
0 & 0 & 1
\end{pmatrix}\begin{pmatrix}u_{1}^{c}\\
u_{2}^{c}\\
u_{3}^{c}
\end{pmatrix}H_{u}\\
 & +\begin{pmatrix}Q_{1} & Q_{2} & Q_{3}\end{pmatrix}\begin{pmatrix}{\tilde{\phi}}_{\ell13}^{(-\frac{1}{2},0,\frac{1}{2})}/\Lambda & {\phi}_{d12}^{(-\frac{1}{6},-\frac{1}{3},\frac{1}{2})}/\Lambda & {\phi}_{q13}^{(-\frac{1}{6},0,\frac{1}{6})}/\Lambda\\
0 & {\tilde{\phi}}_{\ell23}^{(0,-\frac{1}{2},\frac{1}{2})}/\Lambda & {\phi}_{q23}^{(0,-\frac{1}{6},\frac{1}{6})}/\Lambda\\
0 & 0 & 1
\end{pmatrix}\begin{pmatrix}d_{1}^{c}\\
d_{2}^{c}\\
d_{3}^{c}
\end{pmatrix}H_{d}\\
 & +\begin{pmatrix}L_{1} & L_{2} & L_{3}\end{pmatrix}\begin{pmatrix}{\tilde{\phi}}_{\ell13}^{(-\frac{1}{2},0,\frac{1}{2})}/\Lambda & 0 & {\phi}_{\ell13}^{(\frac{1}{2},0,-\frac{1}{2})}/\Lambda\\
0 & {\tilde{\phi}}_{\ell23}^{(0,-\frac{1}{2},\frac{1}{2})}/\Lambda & {\phi}_{\ell23}^{(0,\frac{1}{2},-\frac{1}{2})}/\Lambda\\
0 & 0 & 1
\end{pmatrix}\begin{pmatrix}e_{1}^{c}\\
e_{2}^{c}\\
e_{3}^{c}
\end{pmatrix}H_{d}\,+\mathrm{h.c.}\,,
\end{flalign}
where the universal scale $\Lambda$ is the high cut-off scale of
the EFT, and we ignore the $\mathcal{O}(1)$ dimensionless couplings
of each entry. Although we have chosen only the specific set of hyperons
shown, leaving some zeros in the matrices, these zeros may be filled
in by higher order operators with dimension larger than 5, which so
far we are ignoring.

When the hyperons develop VEVs, assumed to be smaller than the cut-off
scale $\Lambda$, then each entry of the matrix will receive a suppressed
numerical effective coupling given by ratios of the form $\left\langle {\phi}\right\rangle /\Lambda$,
whose values can be assumed arbitrarily. Having the freedom to choose
arbitrary VEVs for each hyperon, the Yukawa matrices above could provide
a good first order description of charged fermion masses and CKM mixing.
We choose to fix various $\left\langle {\phi}\right\rangle /\Lambda$ ratios in terms of powers
of the Wolfenstein parameter $\lambda\simeq0.225$, obtaining 
\begin{flalign}
{\cal L} & =\begin{pmatrix}u_{1} & u_{2} & u_{3}\end{pmatrix}\begin{pmatrix}\lambda^{6} & 0 & \lambda^{3}\\
0 & \lambda^{3} & \lambda^{2}\\
0 & 0 & 1
\end{pmatrix}\begin{pmatrix}u_{1}^{c}\\
u_{2}^{c}\\
u_{3}^{c}
\end{pmatrix}\frac{v_{\mathrm{SM}}}{\sqrt{2}}\label{eq:Texture_a}\\
 & +\begin{pmatrix}d_{1} & d_{2} & d_{3}\end{pmatrix}\begin{pmatrix}\lambda^{6} & \lambda^{4} & \lambda^{3}\\
0 & \lambda^{3} & \lambda^{2}\\
0 & 0 & 1
\end{pmatrix}\begin{pmatrix}d_{1}^{c}\\
d_{2}^{c}\\
d_{3}^{c}
\end{pmatrix}\lambda^{2}\frac{v_{\mathrm{SM}}}{\sqrt{2}}\label{eq:Texture_b}\\
 & +\begin{pmatrix}e_{1} & e_{2} & e_{3}\end{pmatrix}\begin{pmatrix}\lambda^{6} & 0 & \lambda^{6}\\
0 & \lambda^{3} & \lambda^{3}\\
0 & 0 & 1
\end{pmatrix}\begin{pmatrix}e_{1}^{c}\\
e_{2}^{c}\\
e_{3}^{c}
\end{pmatrix}\lambda^{2}\frac{v_{\mathrm{SM}}}{\sqrt{2}}\,+\mathrm{h.c.}\label{eq:Texture_c}
\end{flalign}
As anticipated from the spurion formalism, the alignment of $V_{cb}$
and $V_{ub}$ is not predicted by the model. In contrast, the model
predicts a relevant left-handed $\mu-\tau$ ($e-\tau$) mixing connected to
the same hyperon that provides the second family (first family) effective
Yukawa couplings. Thanks to the addition of the second Higgs doublet,
the model successfully explains third and second family fermion masses
with $\mathcal{O}(1)$ dimensionless couplings. The down-quark and
electron masses are also reasonably explained, although the up-quark
mass is naively a factor $\mathcal{O}(\lambda^{-1.5})$ larger than current data.
Notice that so far we are only assuming one universal cut-off scale
$\Lambda$, while in realistic models several cut-off scales $\Lambda$
may be associated to different messengers in the UV theory, which
could provide a larger suppression for the up-quark effective coupling.
Therefore, within the limitations of our EFT approach, the description
of charged fermion masses given by the set of Eqs.~(\ref{eq:Texture_a}-\ref{eq:Texture_c})
is very successful.

As an alternative example, one could also consider a model where the (1,1)
spurion in Eq.~(\ref{eq:Spurions}) is not promoted to hyperon, but
instead all the spurions in the (1,2) and (2,1) entries are promoted,
so that the Yukawa matrices show an exact texture zero in the (1,1) entry,
\begin{flalign}
{\cal L}^{d\leq5} & =\begin{pmatrix}Q_{1} & Q_{2} & Q_{3}\end{pmatrix}\begin{pmatrix}0 & {\phi}_{u12}^{(-\frac{1}{6},\frac{2}{3},-\frac{1}{2})} & {\phi}_{q13}^{(-\frac{1}{6},0,\frac{1}{6})}\\
{\phi}_{u21}^{(\frac{2}{3},-\frac{1}{6},-\frac{1}{2})} & {\phi}_{\ell23}^{(0,\frac{1}{2},-\frac{1}{2})} & {\phi}_{q23}^{(0,-\frac{1}{6},\frac{1}{6})}\\
0 & 0 & 1
\end{pmatrix}\begin{pmatrix}u_{1}^{c}\\
u_{2}^{c}\\
u_{3}^{c}
\end{pmatrix}H_{u}\\
 & +\begin{pmatrix}Q_{1} & Q_{2} & Q_{3}\end{pmatrix}\begin{pmatrix}0 & {\phi}_{d12}^{(-\frac{1}{6},-\frac{1}{3},\frac{1}{2})} & {\phi}_{q13}^{(-\frac{1}{6},0,\frac{1}{6})}\\
{\phi}_{d21}^{(-\frac{1}{3},-\frac{1}{6},\frac{1}{2})} & {\tilde{\phi}}_{\ell23}^{(0,-\frac{1}{2},\frac{1}{2})} & {\phi}_{q23}^{(0,-\frac{1}{6},\frac{1}{6})}\\
0 & 0 & 1
\end{pmatrix}\begin{pmatrix}d_{1}^{c}\\
d_{2}^{c}\\
d_{3}^{c}
\end{pmatrix}H_{d}\\
 & +\begin{pmatrix}L_{1} & L_{2} & L_{3}\end{pmatrix}\begin{pmatrix}0 & {\phi}_{e12}^{(\frac{1}{2},-1,\frac{1}{2})} & 0\\
{\phi}_{e21}^{(-1,\frac{1}{2},\frac{1}{2})} & {\tilde{\phi}}_{\ell23}^{(0,-\frac{1}{2},\frac{1}{2})} & {\phi}_{\ell23}^{(0,\frac{1}{2},-\frac{1}{2})}\\
0 & 0 & 1
\end{pmatrix}\begin{pmatrix}e_{1}^{c}\\
e_{2}^{c}\\
e_{3}^{c}
\end{pmatrix}H_{d}\,+\mathrm{h.c.}\,,
\end{flalign}
where we have omitted the high cut-off $\Lambda$ suppressing each
dimension-5 operator above. The VEV over $\Lambda$ ratios of
the new hyperons can be fixed by the requirement of addressing first
family fermion masses, obtaining Yukawa matrices with texture zeros
in the (1,1) entry as
\begin{flalign}
{\cal L} & =\begin{pmatrix}u_{1} & u_{2} & u_{3}\end{pmatrix}\begin{pmatrix}0 & \lambda^{5} & \lambda^{3}\\
\lambda^{5.5} & \lambda^{3} & \lambda^{2}\\
0 & 0 & 1
\end{pmatrix}\begin{pmatrix}u_{1}^{c}\\
u_{2}^{c}\\
u_{3}^{c}
\end{pmatrix}\frac{v_{\mathrm{SM}}}{\sqrt{2}}\\
 & +\begin{pmatrix}d_{1} & d_{2} & d_{3}\end{pmatrix}\begin{pmatrix}0 & \lambda^{4} & \lambda^{3}\\
\lambda^{4} & \lambda^{3} & \lambda^{2}\\
0 & 0 & 1
\end{pmatrix}\begin{pmatrix}d_{1}^{c}\\
d_{2}^{c}\\
d_{3}^{c}
\end{pmatrix}\lambda^{2}\frac{v_{\mathrm{SM}}}{\sqrt{2}}\\
 & +\begin{pmatrix}e_{1} & e_{2} & e_{3}\end{pmatrix}\begin{pmatrix}0 & \lambda^{5} & 0\\
\lambda^{4.4} & \lambda^{3} & \lambda^{3}\\
0 & 0 & 1
\end{pmatrix}\begin{pmatrix}e_{1}^{c}\\
e_{2}^{c}\\
e_{3}^{c}
\end{pmatrix}\lambda^{2}\frac{v_{\mathrm{SM}}}{\sqrt{2}}\,+\mathrm{h.c.}\,,
\end{flalign}
which provide an even better description of first family fermion masses
than the original simplified model. Notice that in this scenario,
a sizable left-handed $e-\tau$ mixing is no longer predicted.

We conclude that the most straightforward choices of hyperons, motivated
by the spurion formalism, already provide a good description of charged
fermion masses and mixings. However, these simplified models leave
some questions unanswered. Given that we are assuming the symmetry
breaking highlighted in Eq.~(\ref{eq:Symmetry_Breaking}), we notice
that there are no hyperons breaking the first and second hypercharges
down to their diagonal subgroup, and we would expect those to play
a role in the origin of fermion hierarchies and mixing. Moreover,
in the simplified models introduced so far, several hyperons display
unexplained large hierarchies of VEVs whose values are assumed \textit{a
posteriori} to fit the fermion masses. Given that all these hyperons
participate in the 23-breaking step of Eq.~(\ref{eq:Symmetry_Breaking}),
we would expect all of them to develop VEVs at a similar scale, rather
than the hierarchical scales assumed. This motivates further model
building. In the following subsections we discuss a couple of example
models which address these issues.

\subsection{Model 1: Minimal case with three hyperons\label{subsec:Model-1:-Minimal}}

We introduce here the following set of three hyperons,
\begin{equation}
{\phi}_{\ell23}^{(0,\frac{1}{2},-\frac{1}{2})}\,,\qquad{\phi}_{q23}^{(0,-\frac{1}{6},\frac{1}{6})}\,,\qquad{\phi}_{q12}^{(-\frac{1}{6},\frac{1}{6},0)}\,.
\end{equation}
Following the EFT approach of the previous subsection, we now analyse
the effective Yukawa matrices obtained by combining the SM charged
fermions, the Higgs doublets and the hyperons, in a tower of non-renormalisable
operators preserving the $U(1)_{Y}^{3}$ gauge symmetry, 
\begin{flalign}
{\cal L} & =\begin{pmatrix}Q_{1} & Q_{2} & Q_{3}\end{pmatrix}\begin{pmatrix}\tilde{\phi}_{q12}^{3}{\phi}_{\ell23} & {\phi}_{q12}{\phi}_{\ell23} & {\phi}_{q12}{\phi}_{q23}\\
\tilde{\phi}_{q12}^{4}{\phi}_{\ell23} & {\phi}_{\ell23} & {\phi}_{q23}\\
\tilde{\phi}_{q12}^{4}{\phi}_{\ell23}\tilde{\phi}_{q23} & {\phi}_{\ell23}\tilde{\phi}_{q23} & 1
\end{pmatrix}\begin{pmatrix}u_{1}^{c}\\
u_{2}^{c}\\
u_{3}^{c}
\end{pmatrix}H_{u}\\
 & +\begin{pmatrix}Q_{1} & Q_{2} & Q_{3}\end{pmatrix}\begin{pmatrix}\phi_{q12}^{3}\tilde{\phi}_{\ell23} & {\phi}_{q12}\tilde{\phi}_{\ell23} & {\phi}_{q12}{\phi}_{q23}\\
{\phi}_{q12}^{2}\tilde{\phi}_{\ell23} & \tilde{\phi}_{\ell23} & {\phi}_{q23}\\
{\phi}_{q12}^{2}{\phi}_{q23}^{2} & {\phi}_{q23}^{2} & 1
\end{pmatrix}\begin{pmatrix}d_{1}^{c}\\
d_{2}^{c}\\
d_{3}^{c}
\end{pmatrix}H_{d}\\
 & +\begin{pmatrix}L_{1} & L_{2} & L_{3}\end{pmatrix}\begin{pmatrix}\phi_{q12}^{3}\tilde{\phi}_{\ell23} & \tilde{\phi}_{q12}^{3}\tilde{\phi}_{\ell23} & \tilde{\phi}_{q12}^{3}{\phi}_{\ell23}\\
\phi_{q12}^{6}\tilde{\phi}_{\ell23} & \tilde{\phi}_{\ell23} & {\phi}_{\ell23}\\
\phi_{q12}^{6}\tilde{\phi}_{\ell23}^{2} & \tilde{\phi}_{\ell23}^{2} & 1
\end{pmatrix}\begin{pmatrix}e_{1}^{c}\\
e_{2}^{c}\\
e_{3}^{c}
\end{pmatrix}H_{d}\,+\mathrm{h.c.}\,,
\end{flalign}
where the powers of $\Lambda$ in the denominator and the dimensionless
couplings of each entry are not shown. Once the hyperons above develop
VEVs, we obtain very economical and efficient Yukawa textures for
modeling the observed pattern of SM Yukawa couplings. In particular,
the masses of second family fermions arise at dimension 5 in the EFT,
while first family masses have an extra suppression as they arise
from dimension-8 operators. Regarding CKM mixing, 2-3 quark mixing
leading to $V_{cb}$ arises from dimension-5 operators, while $V_{ub}$
has an extra mild suppression as it arises from dimension-6 operators.
In all cases, right-handed fermion mixing is suppressed with respect
to left-handed mixing. This is a highly desirable feature, given the
strong phenomenological constraints on right-handed flavour-changing
currents \cite{UTfit:2007eik,Isidori:2014rba} (see Section~\ref{subsec:BsMixing}),
which may be mediated by heavy $Z'$ bosons arising from the symmetry
breaking of $U(1)_{Y}^{3}$.

In good approximation, quark mixing leading to $V_{us}$ arises as
the ratio of the (1,2) and (2,2) entries of the quark matrices above,
therefore we expect
\begin{equation}
\frac{\left\langle {\phi}_{q12}\right\rangle }{\Lambda}\sim V_{us}\simeq\lambda\,,
\end{equation}
where $\lambda=\sin\theta_{C}\simeq0.225$. In a similar manner, we
can fix the ratio $\left\langle {\phi}_{q23}\right\rangle /\Lambda$
by reproducing the observed $V_{cb}$
\begin{equation}
\frac{\left\langle {\phi}_{q23}\right\rangle }{\Lambda}\sim V_{cb}\simeq\lambda^{2}\,.
\end{equation}
Given that both $\left\langle {\phi}_{q23}\right\rangle $ and $\left\langle {\phi}_{\ell23}\right\rangle $
play a role in the last step of the symmetry breaking cascade (see
Eq.~(\ref{eq:Symmetry_Breaking})), it is expected that both VEVs
live at a similar scale, although they are not expected to be degenerate
but differ by an $\mathcal{O}(1)$ factor. This way, we are free to
choose
\begin{equation}
\frac{\left\langle {\phi}_{\ell23}\right\rangle }{\Lambda}\sim\frac{m_{c}}{m_{t}}\sim\lambda^{3}\,,
\end{equation}
which, given that $\left\langle H_{d}\right\rangle $ provides an
extra suppression of $\mathcal{O}(\lambda^{2})$ for down-quarks and charged
lepton Yukawas, allows to predict all second family masses with $\mathcal{O}(1)$
dimensionless couplings. In contrast with the simplified models of Section~\ref{subsec:From-Spurions},
this model provides all the 23-breaking VEVs at the same scale, plus
a larger 12-breaking VEV, following a mild hierarchy given by $v_{23}/v_{12}\sim\lambda$.
This way, the symmetry breaking of the $U(1)_{Y}^{3}$ gauge group
proceeds just like in Eq.~(\ref{eq:Symmetry_Breaking}), as desired.

Having fixed all the hyperon VEVs with respect to $\Lambda$, now
we are able to write the full mass matrices for each sector in terms
of the Wolfenstein parameter $\lambda$, 
\begin{flalign}
{\cal L} & =\begin{pmatrix}u_{1} & u_{2} & u_{3}\end{pmatrix}\begin{pmatrix}\lambda^{6} & \lambda^{4} & \lambda^{3}\\
\lambda^{7} & \lambda^{3} & \lambda^{2}\\
\lambda^{9} & \lambda^{5} & 1
\end{pmatrix}\begin{pmatrix}u_{1}^{c}\\
u_{2}^{c}\\
u_{3}^{c}
\end{pmatrix}\frac{v_{\mathrm{SM}}}{\sqrt{2}}\\
 & +\begin{pmatrix}d_{1} & d_{2} & d_{3}\end{pmatrix}\begin{pmatrix}\lambda^{6} & \lambda^{4} & \lambda^{3}\\
\lambda^{5} & \lambda^{3} & \lambda^{2}\\
\lambda^{6} & \lambda^{4} & 1
\end{pmatrix}\begin{pmatrix}d_{1}^{c}\\
d_{2}^{c}\\
d_{3}^{c}
\end{pmatrix}\lambda^{2}\frac{v_{\mathrm{SM}}}{\sqrt{2}}\\
 & +\begin{pmatrix}e_{1} & e_{2} & e_{3}\end{pmatrix}\begin{pmatrix}\lambda^{6} & \lambda^{6} & \lambda^{6}\\
\lambda^{9} & \lambda^{3} & \lambda^{3}\\
\lambda^{12} & \lambda^{6} & 1
\end{pmatrix}\begin{pmatrix}e_{1}^{c}\\
e_{2}^{c}\\
e_{3}^{c}
\end{pmatrix}\lambda^{2}\frac{v_{\mathrm{SM}}}{\sqrt{2}}\,+\mathrm{h.c.}
\end{flalign}
We can see that this setup provides a reasonable description of charged
fermion masses and mixing. Although the up and down-quark masses are
slightly off by $\mathcal{O}(\lambda)$ factors, we remember that
we are only assuming one universal cut-off scale $\Lambda$, while
in realistic models several scales $\Lambda$ may be associated to
different messengers in the UV theory, further improving the fit of
first family quark masses, as discussed in Section~\ref{subsec:From-Spurions}.
All things considered, the description of fermion masses seems very
efficient, considering the limitations of our EFT framework.

However, the model does not predict the alignment of $V_{us}$. Moreover,
we also notice that right-handed $s-d$ mixing is just mildly suppressed
as $s_{12}^{d_{R}}\simeq\mathcal{O}(\lambda^{2})$ in this model.
Given the stringent bounds over left-right scalar operators contributing
to $K-\bar{K}$ meson mixing \cite{UTfit:2007eik,Isidori:2014rba}
(which might arise in this kind of models as we shall see in Section~\ref{sec:Phenomenology}),
the scale $v_{12}$ can be pushed far above the TeV if $V_{us}$ originates
from the down sector. From a phenomenological point of view, it would
be interesting to find models which give clear predictions about the
alignment of $V_{us}$, and ideally provide a more efficient suppression
of right-handed quark mixing. We shall see in the next subsection
that this can be achieved by minimally extending the set of hyperons
of this model.

\subsection{Model 2: Five hyperons for a more predictive setup\label{subsec:Model-2:-Five}}

Model 1 proposed in the previous subsection, despite its simplicity
and minimality, does not give clear predictions about the alignment
of the CKM matrix. Heavy $Z'$ bosons arising from the symmetry breaking
of $U(1)_{Y}^{3}$ have the potential to mediate contributions to
$K-\bar{K}$ meson mixing, which could set a lower bound over
the scale of $U(1)_{Y}^{3}$-breaking, but such contributions depend on the
alignment of $V_{us}$. Moreover, the largest contributions to $K-\bar{K}$
mixing depend both on the alignment of $V_{us}$ and on right-handed
$s-d$ mixing, which is just mildly suppressed in Model 1. Therefore,
we propose here a similar model with slightly extended hyperon content
that can account for a clear prediction about the alignment of $V_{us}$,
plus a more efficient suppression of right-handed fermion mixing. We consider
here the hyperons 
\begin{equation}
{\phi}_{\ell23}^{(0,\frac{1}{2},-\frac{1}{2})},\qquad{\phi}_{q23}^{(0,-\frac{1}{6},\frac{1}{6})},\qquad{\phi}_{q13}^{(-\frac{1}{6},0,\frac{1}{6})},\qquad{\phi}_{d12}^{(-\frac{1}{6},-\frac{1}{3},\frac{1}{2})},\qquad{\phi}_{e12}^{(\frac{1}{4},-\frac{1}{4},0)}.
\end{equation}
With this set of hyperons, the effective Yukawa couplings in the EFT
are (suppressing as usual powers of $\Lambda$ and dimensionless couplings)
%\allowdisplaybreaks
\begin{flalign}
{\cal L} & =\begin{pmatrix}Q_{1} & Q_{2} & Q_{3}\end{pmatrix}\begin{pmatrix}\phi_{e12}^{2}\tilde{\phi}_{\ell23} & {\phi}_{q13}{\tilde{\phi}}_{q23}{\phi}_{\ell23} & {\phi}_{q13}\\
{\phi}_{e12}^{2}{\tilde{\phi}}_{q13}{\phi}_{q23}{\phi}_{\ell23} & {\phi}_{\ell23} & {\phi}_{q23}\\
{\phi}_{e12}^{2}{\tilde{\phi}}_{q13}{\phi}_{\ell23} & {\phi}_{\ell23}\tilde{\phi}_{q23} & 1
\end{pmatrix}\begin{pmatrix}u_{1}^{c}\\
u_{2}^{c}\\
u_{3}^{c}
\end{pmatrix}H_{u}\\
 & +\begin{pmatrix}Q_{1} & Q_{2} & Q_{3}\end{pmatrix}\begin{pmatrix}\phi_{e12}^{2}\tilde{\phi}_{\ell23} & {\phi}_{d12} & {\phi}_{q13}\\
{\phi}_{q13}^{2}{\phi}_{q23} & \tilde{\phi}_{\ell23} & {\phi}_{q23}\\
{\phi}_{q13}^{2} & {\phi}_{q23}^{2} & 1
\end{pmatrix}\begin{pmatrix}d_{1}^{c}\\
d_{2}^{c}\\
d_{3}^{c}
\end{pmatrix}H_{d}\\
 & +\begin{pmatrix}L_{1} & L_{2} & L_{3}\end{pmatrix}\begin{pmatrix}\tilde{\phi}_{e12}^{2}\tilde{\phi}_{\ell23} & \phi_{e12}^{2}\tilde{\phi}_{\ell23} & \phi_{e12}^{2}{\phi}_{\ell23}\\
\tilde{\phi}_{e12}^{4}\tilde{\phi}_{\ell23} & \tilde{\phi}_{\ell23} & {\phi}_{\ell23}\\
\tilde{\phi}_{e12}^{4}\tilde{\phi}_{\ell23}^{2} & {\tilde{\phi}}_{\ell23}^{2} & 1
\end{pmatrix}\begin{pmatrix}e_{1}^{c}\\
e_{2}^{c}\\
e_{3}^{c}
\end{pmatrix}H_{d}\,+\mathrm{h.c.}
\end{flalign}
Following the same approach as with Model 1, we assign the following
powers of $\lambda$ to the VEV over $\Lambda$ ratios in order to
reproduce fermion masses and CKM mixing, 
\begin{equation}
\frac{\left\langle {\phi}_{\ell23}\right\rangle }{\Lambda}=\frac{\left\langle {\phi}_{q13}\right\rangle }{\Lambda}\simeq\lambda^{3}\,,\quad\frac{\left\langle {\phi}_{q23}\right\rangle }{\Lambda}\simeq\lambda^{2}\,,\quad\frac{\left\langle {\phi}_{d12}\right\rangle }{\Lambda}\simeq\lambda^{4}\,,\quad\frac{\left\langle {\phi}_{e12}\right\rangle }{\Lambda}\simeq\lambda\,.
\end{equation}
Although it would seem that in this scenario there exists a mild hierarchy
between 23-breaking VEVs of $\mathcal{O}(\lambda^{2})$, since ${\phi}_{d12}$
only appears in the 12 entry of the down matrix, it would be very
reasonable that the dimensionless coupling in that entry provides
a factor $\lambda$ suppression, such that all 23-breaking VEVs live
at the same scale. The largest VEV is still the 12-breaking one, which
is now associated to the hyperon ${\phi}_{e12}$, and the mild hierarchy
between scales remains as $v_{12}/v_{23}\simeq\lambda$. With these
assignments of VEVs over $\Lambda$ ratios, the Yukawa textures are given
by 
\begin{flalign}
{\cal L} & =\begin{pmatrix}u_{1} & u_{2} & u_{3}\end{pmatrix}\begin{pmatrix}\lambda^{5} & \lambda^{8} & \lambda^{3}\\
\lambda^{10} & \lambda^{3} & \lambda^{2}\\
\lambda^{7} & \lambda^{5} & 1
\end{pmatrix}\begin{pmatrix}u_{1}^{c}\\
u_{2}^{c}\\
u_{3}^{c}
\end{pmatrix}\frac{v_{\mathrm{SM}}}{\sqrt{2}}\\
 & +\begin{pmatrix}d_{1} & d_{2} & d_{3}\end{pmatrix}\begin{pmatrix}\lambda^{5} & \lambda^{4} & \lambda^{3}\\
\lambda^{8} & \lambda^{3} & \lambda^{2}\\
\lambda^{6} & \lambda^{4} & 1
\end{pmatrix}\begin{pmatrix}d_{1}^{c}\\
d_{2}^{c}\\
d_{3}^{c}
\end{pmatrix}\lambda^{2}\frac{v_{\mathrm{SM}}}{\sqrt{2}}\\
 & +\begin{pmatrix}e_{1} & e_{2} & e_{3}\end{pmatrix}\begin{pmatrix}\lambda^{5} & \lambda^{5} & \lambda^{5}\\
\lambda^{7} & \lambda^{3} & \lambda^{3}\\
\lambda^{10} & \lambda^{6} & 1
\end{pmatrix}\begin{pmatrix}e_{1}^{c}\\
e_{2}^{c}\\
e_{3}^{c}
\end{pmatrix}\lambda^{2}\frac{v_{\mathrm{SM}}}{\sqrt{2}}\,+\mathrm{h.c.}
\end{flalign}
Just like in Model 1, this model provides a compelling description
of all charged fermion masses and mixing. Notice that this scenario
provides a very efficient suppression of right-handed fermion mixing.
Moreover, it is clear that here $V_{us}$ mixing originates from the
down sector, providing a more predictive setup than Model 1, which
will be useful for phenomenological purposes as discussed in Section~\ref{sec:Phenomenology}.

Finally we comment that the Higgs doublets and hyperons can mediate
FCNCs such as $K-\bar{K}$ mixing. In the present model, this
may arise from tree-level exchange of $\phi_{d12}$ hyperons which
can mediate down-strange transitions. Such a coupling originates from
$Q_{1}\phi_{d12}d_{2}^{c}H_{d}/\Lambda$ which will lead to a suppressed
down-strange coupling of order $\lambda^{2}v_{\mathrm{SM}}/\Lambda\approx10^{-5}$ (taking $\Lambda\approx100\;\mathrm{TeV}$ as expected if $v_{23}\approx\mathcal{O}(\mathrm{TeV})$).
Assuming the mass of the hyperons to be at the scale $v_{23}$, then we expect these FCNCs to be under
control. Since we assume a type II 2HDM, tree-level Higgs doublets
exchange contributions are forbidden, and FCNCs mediated by the Higgs
doublets can only proceed via their mixing with hyperons, carrying
therefore an extra suppression via the mixing angle, along with the
suppression of hyperon couplings already discussed. In more general
models of this kind, such contributions to $K-\bar{K}$ mixing
could be even further suppressed depending on the order of the operator
and the alignment of $V_{us}$, and in all generality we expect all
hyperon couplings to be suppressed by at least a factor $v_{\mathrm{SM}}/\Lambda\approx10^{-3}$.
A more detailed study of the phenomenology of the scalar sector in
this general class of models is beyond the scope of this chapter.

\section{Neutrino masses and mixing\label{sec:Neutrino-masses-and-Mixing}}

\subsection{General considerations and spurion formalism}

The origin of neutrino masses and mixing requires a dedicated analysis
due to their particular properties. We start by introducing $U(2)^{5}$-breaking
spurions (carrying inverse of mass dimension) for the Weinberg operator
\begin{equation}
\mathcal{L}_{\mathrm{Weinberg}}=\begin{pmatrix}L_{1} & L_{2} & L_{3}\end{pmatrix}\begin{pmatrix}{\Phi}(1,0,-1) & {\Phi}(\frac{1}{2},\frac{1}{2},-1) & {\Phi}(\frac{1}{2},0,-\frac{1}{2})\\
{\Phi}(\frac{1}{2},\frac{1}{2},-1) & {\Phi}(0,1,-1) & {\Phi}(0,\frac{1}{2},-\frac{1}{2})\\
{\Phi}(\frac{1}{2},0,-\frac{1}{2}) & {\Phi}(0,\frac{1}{2},-\frac{1}{2}) & 1
\end{pmatrix}\begin{pmatrix}L_{1}\\
L_{2}\\
L_{3}
\end{pmatrix}H_{u}H_{u}\,,
\end{equation}
which reveals that, as expected, the $U(2)^{5}$ approximate symmetry
is naively present in the neutrino sector as well. As a consequence, one generally expects one neutrino to be much heavier than the others, displaying
tiny mixing with the other neutrino flavours. In the spirit of the type I
seesaw mechanism, one could think of adding a $U(1)_{Y}^{3}$ singlet
neutrino as $N(0,0,0)$. Such a singlet neutrino can only couple to
the third family active neutrino at renormalisable level, i.e.~$\mathcal{L}_{N}\supset L_{3}H_{u}N+m_{N}NN$,
where all fermion fields are written in a left-handed 2-component convention.
This way, the coupling $L_{2}H_{u}N$, which is required for large
atmospheric neutrino mixing, can only arise at the non-renormalisable
level. Therefore, it is expected to be suppressed with respect to
$L_{3}H_{u}N$. This seems to be inconsistent with large atmospheric
neutrino mixing, at least within the validity of our EFT framework.
As anticipated before, this is a consequence of the accidental $U(2)^{5}$
flavour symmetry delivered by the TH model.

Given such general considerations, we conclude that in order to obtain
neutrino masses and mixing from the type I seesaw mechanism, it is
required to add SM singlet neutrinos that carry tri-hypercharges (but
whose hypercharges add up to zero). These neutrinos will allow to
introduce $U(2)^{5}$-breaking operators similar for all neutrino
flavours, providing a mechanism to obtain the adequate neutrino mixing
in a natural way. In order to cancel gauge anomalies,
the most simple option is that these singlet neutrinos are vector-like.
In the following subsection we include an example of successful type
I seesaw mechanism based on this idea.

\subsection{Example of successful neutrino mixing from the seesaw mechanism\label{subsec:Example-of-seesaw}}

In the following, we provide an example scenario which reproduces
the observed pattern of neutrino mixing, as a proof of principle.
According to the discussion in the previous subsection, in order to
implement a type I seesaw mechanism that delivers large neutrino mixing,
we need to add vector-like neutrinos that carry tri-hypercharges (but
whose hypercharges add up to zero). We also need to introduce hyperons
that will provide small Dirac mass terms for the active neutrinos
in the form of non-renormalisable operators. Under these considerations,
we start by adding the following vector-like neutrino\footnote{We remind the reader that in our convention, all fermion fields including
$N_{\mathrm{atm}}$ and $\overline{N}_{\mathrm{atm}}$ are left-handed,
see Appendix~\ref{app:2-component_notation}.} and hyperon 
\begin{equation}
N_{\mathrm{atm}}^{(0,\frac{1}{4},-\frac{1}{4})}\,,\qquad\overline{N}_{\mathrm{atm}}^{(0,-\frac{1}{4},\frac{1}{4})}\,,\qquad\phi_{\mathrm{atm}}^{(0,\frac{1}{4},-\frac{1}{4})}\,,
\end{equation}
where the charge assignments are chosen to provide large \textit{atmospheric
neutrino mixing}. This way, we can write the following non-renormalisable
operators along with the Majorana and vector-like masses of $N_{\mathrm{atm}}$,
\begin{align}
\mathcal{L}_{N_{\mathrm{atm}}}\supset & \frac{1}{\Lambda_{\mathrm{atm}}}(\phi_{\mathrm{atm}}L_{2}+\tilde{\phi}_{\mathrm{atm}}L_{3})H_{u}N_{\mathrm{atm}}+\frac{\phi_{\mathrm{atm}}}{\Lambda_{\mathrm{atm}}}L_{3}H_{u}\overline{N}_{\mathrm{atm}}\label{eq:Latm}\\
 & +\phi_{\ell23}N_{\mathrm{atm}}N_{\mathrm{atm}}+\tilde{\phi}_{\ell23}\overline{N}_{\mathrm{atm}}\overline{N}_{\mathrm{atm}}+M_{N_{\mathrm{atm}}}\overline{N}_{\mathrm{atm}}N_{\mathrm{atm}}\,,\nonumber 
\end{align}
where we have ignored the $\mathcal{O}(1)$ dimensionless couplings,
and the hyperon $\phi_{\ell23}^{(0,\frac{1}{2},-\frac{1}{2})}$ is
already present in both Model 1 and Model 2 for the charged fermion
sector. Notice that $N_{\mathrm{atm}}$ provides the couplings $\mathcal{L}_{N_{\mathrm{atm}}}\supset L_{3}H_{u}N_{\mathrm{atm}}+L_{2}H_{u}N_{\mathrm{atm}}$,
as required to explain atmospheric mixing. Notice also that the conjugate
neutrino $\overline{N}_{\mathrm{atm}}$ unavoidably couples to $L_{3}$
as $\phi_{\mathrm{atm}}L_{3}H_{u}\overline{N}_{\mathrm{atm}}$, but
not to $L_{2}$ because we have not included any hyperon providing
this dimension-5 operator.

In a similar spirit, we introduce another vector-like neutrino and
other hyperons in order to obtain large \textit{solar neutrino mixing}
\begin{equation}
N_{\mathrm{sol}}^{(\frac{1}{4},\frac{1}{4},-\frac{1}{2})}\,,\qquad\overline{N}_{\mathrm{sol}}^{(-\frac{1}{4},-\frac{1}{4},\frac{1}{2})}\,,\qquad\phi_{\mathrm{sol}}^{(-\frac{1}{2},-\frac{1}{2},1)},\qquad\phi_{\nu13}^{(-\frac{1}{4},-\frac{1}{4},\frac{1}{2})}\,,
\end{equation}
which provide the following non-renormalisable operators and mass
terms, 
\begin{align}
\mathcal{L}_{N_{\mathrm{sol}}}\supset & \frac{1}{\Lambda_{\mathrm{sol}}}(\phi_{e12}L_{1}+\tilde{\phi}_{e12}L_{2}+\phi_{\nu13}L_{3})H_{u}N_{\mathrm{sol}}+\frac{\phi_{\nu13}}{\Lambda_{\mathrm{sol}}}L_{3}H_{u}\overline{N}_{\mathrm{sol}}\label{eq:Lsol}\\
 & +\phi_{\mathrm{sol}}N_{\mathrm{sol}}N_{\mathrm{sol}}+\tilde{\phi}_{\mathrm{sol}}\overline{N}_{\mathrm{sol}}\overline{N}_{\mathrm{sol}}+M_{N_{\mathrm{sol}}}\overline{N}_{\mathrm{sol}}N_{\mathrm{sol}}\,,\nonumber 
\end{align}
where we have ignored again the $\mathcal{O}(1)$ dimensionless couplings,
and the hyperon $\phi_{e12}^{(\frac{1}{2},-\frac{1}{2},0)}$ is already
present in Model 2 for the charged fermion sector. The hyperon $\phi_{\nu13}$
(which will eventually populate the (1,3) entry of the effective neutrino
mass matrix) is not required to obtain non-zero reactor mixing, which
would already arise from the other operators, but it is required in
order to have enough free parameters to fit all observed neutrino
mixing angles and mass splittings. We could have included the $U(1)_{Y}^{3}$
singlet neutrino $N(0,0,0)$, but as discussed in the previous subsection,
its contributions to the Weinberg operator may be suppressed by its
large Majorana mass $m_{N}$, resulting in possibly negligible contributions
to the seesaw mechanism. We are therefore free to assume that such
a neutrino $N(0,0,0)$, if exists, is in any case decoupled from the
seesaw, while the atmospheric and solar SM singlet neutrinos $N_{\mathrm{atm}}$
and $N_{\mathrm{sol}}$ could yield dominant and subdominant contributions,
resulting in a natural normal neutrino mass hierarchy as in sequential
dominance \cite{King:1998jw,King:1999mb,King:2002nf}.

Notice that the vector-like neutrinos $N_{\mathrm{atm}}$ and $N_{\mathrm{sol}}$
get contributions to their masses from the VEVs of the hyperons $\phi_{\ell23}$
and $\phi_{\mathrm{sol}}$, respectively, which we denote generically
as $v_{23}$ since they both take part in the 23-breaking step of
Eq.~(\ref{eq:Symmetry_Breaking}). In addition, $N_{\mathrm{atm}}$
and $N_{\mathrm{sol}}$ get contributions to their masses from the
unspecified vector-like mass terms, that we generically denote as
$M_{\mathrm{VL}}$. Now we arrange all the couplings of the neutrino
sector into Dirac-type mass matrices and Majorana-type mass matrices,
i.e.
\begin{flalign}
 &  &  & m_{D_{L}}=\left(
\global\long\def\arraystretch{0.7}%
\begin{array}{@{}llc@{}}
 & \multicolumn{1}{c@{}}{\phantom{\!\,}\overline{N}_{\mathrm{sol}}} & \phantom{\!\,}\overline{N}_{\mathrm{atm}}\\
\cmidrule(l){2-3}\left.L_{1}\right| & 0 & 0\\
\left.L_{2}\right| & 0 & 0\\
\left.L_{3}\right| & \frac{\tilde{\phi}_{\nu13}}{\Lambda_{\mathrm{sol}}} & \frac{\phi_{\mathrm{atm}}}{\Lambda_{\mathrm{atm}}}
\end{array}\right)H_{u}\,, &  & m_{D_{R}}=\left(
\global\long\def\arraystretch{0.7}%
\begin{array}{@{}llc@{}}
 & \multicolumn{1}{c@{}}{\phantom{\!\,}N_{\mathrm{sol}}} & \phantom{\!\,}N_{\mathrm{atm}}\\
\cmidrule(l){2-3}\left.L_{1}\right| & \frac{\phi_{e12}}{\Lambda_{\mathrm{sol}}} & 0\\
\left.L_{2}\right| & \frac{\tilde{\phi}_{e12}}{\Lambda_{\mathrm{sol}}} & \frac{\phi_{\mathrm{atm}}}{\Lambda_{\mathrm{atm}}}\\
\left.L_{3}\right| & \frac{\phi_{\nu13}}{\Lambda_{\mathrm{sol}}} & \frac{\tilde{\phi}_{\mathrm{atm}}}{\Lambda_{\mathrm{atm}}}
\end{array}\right)H_{u}\,,\label{eq:mDL_mDR}\\
\nonumber \\
 &  &  & M_{L}=\left(
\global\long\def\arraystretch{0.7}%
\begin{array}{@{}llc@{}}
 & \multicolumn{1}{c@{}}{\phantom{\!\,}\overline{N}_{\mathrm{sol}}} & \phantom{\!\,}\overline{N}_{\mathrm{atm}}\\
\cmidrule(l){2-3}\left.\:\:\overline{N}_{\mathrm{sol}}\right| & \tilde{\phi}_{\mathrm{sol}} & 0\\
\left.\overline{N}_{\mathrm{atm}}\right| & 0 & \tilde{\phi}_{\ell23}
\end{array}\right)\approx v_{23}\mathbb{I}_{2\times2}\,, &  & M_{R}\approx\left(
\global\long\def\arraystretch{0.7}%
\begin{array}{@{}llc@{}}
 & \multicolumn{1}{c@{}}{\phantom{\!\,}N_{\mathrm{sol}}} & \phantom{\!\,}N_{\mathrm{atm}}\\
\cmidrule(l){2-3}\left.\:\:N_{\mathrm{sol}}\right| & \phi_{\mathrm{sol}} & 0\\
\left.N_{\mathrm{atm}}\right| & 0 & \phi_{\ell23}
\end{array}\right)\approx v_{23}\mathbb{I}_{2\times2}\,,\label{eq:ML_MR}
\end{flalign}
\begin{equation}
M_{LR}=\left(
\global\long\def\arraystretch{0.7}%
\begin{array}{@{}llc@{}}
 & \multicolumn{1}{c@{}}{\phantom{\!\,}N_{\mathrm{sol}}} & \phantom{\!\,}N_{\mathrm{atm}}\\
\cmidrule(l){2-3}\left.\:\:\overline{N}_{\mathrm{sol}}\right| & M_{N_{\mathrm{sol}}} & 0\\
\left.\overline{N}_{\mathrm{atm}}\right| & 0 & M_{N_{\mathrm{atm}}}
\end{array}\right)\approx M_{\mathrm{VL}}\mathbb{I}_{2\times2}\,,\label{eq:MLR}
\end{equation}
where $\mathbb{I}_{2\times2}$ is the $2\times2$ identity matrix
and we have ignored $\mathcal{O}(1)$ dimensionless couplings. In
Eqs.~(\ref{eq:ML_MR}) and (\ref{eq:MLR}) above, we have considered
the following approximations and assumptions: 
\begin{itemize}
\item We have neglected $\mathcal{O}(1)$ dimensionless couplings generally
present for each non-zero entry of Eq.~(\ref{eq:ML_MR}). With this
consideration, we find $M_{L}=M_{R}$ after the hyperons develop their
VEVs. Furthermore, since the two hyperons appearing in $M_{L}$ and
$M_{R}$ participate in the 23-breaking step of Eq.~(\ref{eq:Symmetry_Breaking}),
we have assumed that they both develop a similar VEV $\left\langle \phi_{\ell23}\right\rangle \approx\left\langle \phi_{\mathrm{sol}}\right\rangle \approx\mathcal{O}(v_{23})$.
For simplicity we take them to be equal, although the same conclusions
hold as long as they just differ by $\mathcal{O}(1)$ factors, as
naturally expected.
\item For simplicity, we have assumed a similar vector-like mass for both
neutrinos in Eq.~(\ref{eq:MLR}), i.e.~$M_{N_{\mathrm{sol}}}\approx M_{N_{\mathrm{atm}}}\equiv M_{\mathrm{VL}}$. 
\end{itemize}
With these definitions, the full mass matrix of the neutrino sector
$M_{\nu}$ can be written in a compact form as
\begin{equation}
M_{\nu}=\left(
\global\long\def\arraystretch{0.7}%
\begin{array}{@{}llcc@{}}
 & \multicolumn{1}{c@{}}{\phantom{\!\,}\nu} & \phantom{\!\,}\overline{N} & \phantom{\!\,}N\\
\cmidrule(l){2-4}\left.\:\,\nu\right| & 0 & m_{D_{L}} & m_{D_{R}}\\
\left.\overline{N}\right| & m_{D_{L}}^{\mathrm{T}} & M_{L} & M_{LR}\\
\left.N\right| & m_{D_{R}}^{\mathrm{T}} & M_{LR}^{\mathrm{T}} & M_{R}
\end{array}\right)\equiv\left(\begin{array}{cc}
0 & m_{D}\\
m_{D}^{\mathrm{T}} & M_{N}
\end{array}\right)\,,\label{eq:Full_Mnu}
\end{equation}
where we have defined $\nu$ as a 3-component vector containing the
weak eigenstates of active neutrinos, while $N$ and $\overline{N}$
are 2-component vectors containing the SM singlet neutrinos $N_{\mathrm{atm}}$,
$N_{\mathrm{sol}}$ and conjugate neutrinos $\overline{N}_{\mathrm{atm}}$
,$\overline{N}_{\mathrm{sol}}$, respectively. Dirac-type masses in
$m_{D_{L,R}}$ may be orders of magnitude samller than the electroweak scale,
because they arise from non-renormalisable operators proportional
to the SM VEV. In contrast, the eigenvalues of $M_{N}$ are not smaller
than $\mathcal{O}(v_{23})$, which is at least TeV. Therefore, the
condition $m_{D}\ll M_{N}$ is fulfilled in Eq.~(\ref{eq:Full_Mnu})
and we can safely apply the seesaw formula as 
\begin{align}
m_{\nu} & =m_{D}M_{N}^{-1}m_{D}^{\mathrm{T}}\label{eq:Seesaw_General}\\
 & =\left(\begin{array}{cc}
m_{D_{L}} & m_{D_{R}}\end{array}\right)\left(\begin{array}{cc}
v_{23} & -M_{\mathrm{VL}}\\
-M_{\mathrm{VL}} & v_{23}
\end{array}\right)\left(\begin{array}{c}
m_{D_{L}}^{\mathrm{T}}\\
m_{D_{R}}^{\mathrm{T}}
\end{array}\right)\frac{1}{v_{23}^{2}-M_{\mathrm{VL}}^{2}}\nonumber \\
 & =\left[m_{D_{L}}m_{D_{L}}^{\mathrm{T}}v_{23}-m_{D_{L}}m_{D_{R}}^{\mathrm{T}}M_{\mathrm{VL}}-m_{D_{R}}m_{D_{L}}^{\mathrm{T}}M_{\mathrm{VL}}+m_{D_{R}}m_{D_{R}}^{\mathrm{T}}v_{23}\right]\frac{1}{v_{23}^{2}-M_{\mathrm{VL}}^{2}}\,.\nonumber 
\end{align}
Given the structure of $m_{D_{L}}$ and $m_{D_{R}}$ in Eq.~(\ref{eq:mDL_mDR}),
the products above involving $m_{D_{L}}$ lead to a hierarchical $m_{\nu}$
matrix where only the entries in the third row and column are populated
and the others are zero. In contrast, the product $m_{D_{R}}m_{D_{R}}^{\mathrm{T}}$
provides a matrix $m_{\nu}$ where all entries are populated. Therefore,
if $M_{\mathrm{VL}}\gg v_{23}$, then the effective neutrino matrix
becomes hierarchical, rendering impossible to explain the observed
pattern of neutrino mixing and mass splittings with $\mathcal{O}(1)$
parameters. Instead, if $M_{\mathrm{VL}}$ is of the same order or
smaller than $v_{23}$, i.e.~$M_{\mathrm{VL}}\apprle v_{23}$, then
the resulting matrix is in any case a matrix where all entries are
populated, which has the potential to explain the observed patterns
of neutrino mixing.

This argument holds as long as $m_{D_{L}}$ is populated by zeros
in at least some of the entries involving $L_{1}$ and $L_{2}$, like
in our example model. Instead, in the very particular case where both
$m_{D_{L}}$ and $m_{D_{R}}$ are similarly populated matrices and
of the same order, then the terms proportional to $M_{\mathrm{VL}}$
in Eq.~(\ref{eq:Seesaw_General}) can provide an effective neutrino
mass matrix where all entries are populated. In this case, $M_{\mathrm{VL}}>v_{23}$
is possible. Nevertheless, even in this scenario we expect $M_{\mathrm{VL}}$
not to be very large, since the smallness of $m_{D_{L}}$ and $m_{D_{R}}$
(that arise from non-renormalisable operators) may potentially provides most
of the suppression for the small neutrino masses. Furthermore, this scenario
involves the addition of several extra hyperons with very particular
charges, making the model more complicated, so we do not consider
it.

Given the above considerations, we proceed the calculation by expanding
the term $m_{D_{R}}m_{D_{R}}^{\mathrm{T}}v_{23}$ in Eq.~(\ref{eq:Seesaw_General})
\begin{flalign}
m_{\nu} & \simeq m_{D_{R}}m_{D_{R}}^{\mathrm{T}}v_{23}\\
 & =\left(\begin{array}{ccc}
\frac{\phi_{e12}^{2}}{\Lambda_{\mathrm{sol}}^{2}} & \frac{\phi_{e12}\tilde{\phi}_{e12}}{\Lambda_{\mathrm{sol}}^{2}} & \frac{\phi_{e12}\phi_{\nu13}}{\Lambda_{\mathrm{sol}}^{2}}\\
\frac{\phi_{e12}\tilde{\phi}_{e12}}{\Lambda_{\mathrm{sol}}^{2}} & \frac{\phi_{\mathrm{atm}}^{2}}{\Lambda_{\mathrm{atm}}^{2}}+\frac{\tilde{\phi}_{e12}^{2}}{\Lambda_{\mathrm{sol}}^{2}} & \frac{\phi_{\mathrm{atm}}\tilde{\phi}_{\mathrm{atm}}}{\Lambda_{\mathrm{atm}}^{2}}+\frac{\tilde{\phi}_{e12}\phi_{\nu13}}{\Lambda_{\mathrm{sol}}^{2}}\\
\frac{\phi_{e12}\phi_{\nu13}}{\Lambda_{\mathrm{sol}}^{2}} & \frac{\phi_{\mathrm{atm}}\tilde{\phi}_{\mathrm{atm}}}{\Lambda_{\mathrm{atm}}^{2}}+\frac{\tilde{\phi}_{e12}\phi_{\nu13}}{\Lambda_{\mathrm{sol}}^{2}} & \frac{\phi_{\mathrm{atm}}^{2}}{\Lambda_{\mathrm{atm}}^{2}}+\frac{\phi_{\nu13}^{2}}{\Lambda_{\mathrm{sol}}^{2}}
\end{array}\right)\frac{H_{u}H_{u}}{v_{23}}\,.\nonumber 
\end{flalign}
Given the symmetry breaking pattern of the model shown in Eq.~(\ref{eq:Symmetry_Breaking}),
we take $\left\langle {\phi}_{e12}\right\rangle \simeq\mathcal{O}(v_{12})$
and $\left\langle \phi_{\nu13}\right\rangle \approx\left\langle \phi_{\ell23}\right\rangle \approx\left\langle \phi_{\mathrm{atm}}\right\rangle \approx\left\langle \phi_{\mathrm{sol}}\right\rangle \approx\mathcal{O}(v_{23})$\footnote{In the calculations that follow, we assume for simplicity that these
VEVs are equal. However, the same conclusions hold as long as the
VEVs vary by $\mathcal{O}(1)$ factors, which is the natural expectation.}. Motivated by our discussion of the charged fermion sector (see Section~\ref{sec:Charged-fermion-masses-mixing}),
we consider the relation $v_{23}/v_{12}\simeq\lambda$. By inserting
such VEVs, we obtain
\begin{equation}
m_{\nu}\simeq\left(\begin{array}{ccc}
0 & 0 & 0\\
0 & 1 & 1\\
0 & 1 & 1
\end{array}\right)v_{23}\frac{H_{u}H_{u}}{\Lambda_{\mathrm{atm}}^{2}}\,+\left(\begin{array}{ccc}
1 & 1 & \lambda\\
1 & 1 & \lambda\\
\lambda & \lambda & \lambda^{2}
\end{array}\right)v_{23}\frac{H_{u}H_{u}}{\lambda^{2}\Lambda_{\mathrm{sol}}^{2}}\,.
\end{equation}
If $\Lambda_{\mathrm{sol}}=\Lambda_{\mathrm{atm}}$, we observe that
there exists a mild hierarchy of order $\lambda^{2}$ between the
12 and 23 sectors in the matrix above. Considering the dimensionless
coefficients that we have ignored so far, the numerical diagonalisation
of $m_{\nu}$ would require some parameters of $\mathcal{O}(0.01)$
in order to explain the observed neutrino mixing angles and mass splittings
\cite{deSalas:2020pgw,Gonzalez-Garcia:2021dve}. The situation can
be improved if we assume a mild hierarchy between cut-off scales $\Lambda_{\mathrm{atm}}/\Lambda_{\mathrm{sol}}\simeq\lambda$,
obtaining to leading order for each entry (ignoring dimensionless
coefficients), 
\begin{equation}
m_{\nu}\simeq\left(\begin{array}{ccc}
1 & 1 & \lambda\\
1 & 1 & 1\\
\lambda & 1 & 1
\end{array}\right)v_{23}\frac{v_{\mathrm{SM}}^{2}}{\Lambda_{\mathrm{atm}}^{2}}\,,
\end{equation}
where we have introduced the SM VEV as $\left\langle H_{u}\right\rangle =v_{\mathrm{SM}}$
(ignoring the factor $1/\sqrt{2}$). Considering now the dimensionless
coefficients in the matrix above, we find that numerical diagonalisation
can accommodate all the observed neutrino mixing angles and mass splittings
\cite{deSalas:2020pgw,Gonzalez-Garcia:2021dve} with $\mathcal{O}(1)$
parameters, and we are able to reproduce both normal and inverted
ordered scenarios.

Notice that we have been driven to a scenario where the vector-like
neutrinos get Majorana masses from the VEVs of hyperons in the model.
Furthermore, the vector-like masses necessarily have to be of the
same or smaller order than the VEVs of the hyperons in order to explain
the observed pattern of neutrino mixing. Therefore, in the particular
example included in this section, the vector-like neutrinos get a
mass at the scale $v_{23}$ of the 23-breaking step in Eq.~(\ref{eq:Symmetry_Breaking}),
which could happen at a relatively low scale as we shall see in Section~\ref{sec:Phenomenology}.
As a consequence, the vector-like neutrinos involved in the seesaw
mechanism are expected to be relatively light, and the high energy
cut-offs of the EFT $\Lambda_{\mathrm{atm}}$ and $\Lambda_{\mathrm{sol}}$
are expected to provide most the suppression of light neutrino masses.
We conclude that, due to the $U(2)^{5}$ flavour symmetry provided
by the TH model, we have been driven to a low scale seesaw in order
to predict the observed pattern of neutrino mixing.

\section{Symmetry breaking and gauge mixing}

We are assuming that the symmetry breaking of the tri-hypercharge
gauge group down to the SM follows the following pattern
\allowdisplaybreaks[0]
\begin{alignat}{1}
 & SU(3)_{c}\times SU(2)_{L}\times{\displaystyle U(1)_{Y_{1}}\times U(1)_{Y_{2}}\times U(1)_{Y_{3}}}\nonumber \\
 & {\displaystyle \overset{v_{12}}{\rightarrow}SU(3)_{c}\times SU(2)_{L}\times U(1)_{Y_{1}+Y_{2}}\times U(1)_{Y_{3}}}\label{eq:Symmetry_Breaking-1}\\
 & {\displaystyle \overset{v_{23}}{\rightarrow}SU(3)_{c}\times SU(2)_{L}\times U(1)_{Y_{1}+Y_{2}+Y_{3}}}\nonumber \\
 & {\displaystyle \overset{v_{\mathrm{SM}}}{\rightarrow}SU(3)_{c}\times U(1)_{Q}\,.}
\end{alignat}
\allowdisplaybreaks
Therefore, first and second hypercharges are broken at a high scale
down to the diagonal subgroup, and then the remaining factors $U(1)_{Y_{1}+Y_{2}}\times U(1)_{Y_{3}}$
are broken down to SM hypercharge. In this process, heavy $Z'$ bosons
arise, with their masses connected to the different scales of symmetry
breaking. Given that the tri-hypercharge gauge group is based on a
family decomposition of SM hypercharge, the heavy $Z'$ bosons can
potentially mix with the SM $Z$ boson, even if kinetic mixing is
absent. This mixing breaks custodial symmetry, having significant
phenomenological implications if the NP scales are not very heavy.
Therefore, in the next subsections we study the several steps of symmetry
breaking, including electroweak symmetry breaking, and extract the
masses of the massive neutral gauge bosons that arise in this process.

\subsection{High scale symmetry breaking\label{sec:High-scale-symmetry}}

Assuming that the 12-breaking scale is far above the electroweak scale,
at very high energies we consider only the factors $U(1)_{Y_{1}}\times U(1)_{Y_{2}}$
with renormalisable Lagrangian (neglecting fermion content and any
kinetic mixing\footnote{Considering kinetic mixing in the Lagrangian of Eq.~(\ref{eq:Lagrangian_12})
only leads to a redefinition of either the $g_{1}$ or $g_{2}$ couplings
in the canonical basis (where the kinetic terms are diagonal).} for simplicity),
\begin{align}
\mathcal{L} & =-\frac{1}{4}F_{\mu\nu}^{(1)}F^{\mu\nu(1)}-\frac{1}{4}F_{\mu\nu}^{(2)}F^{\mu\nu(2)}+(D_{\mu}\phi_{12})^{*}D^{\mu}\phi_{12}-V(\phi_{12})\,,\label{eq:Lagrangian_12}
\end{align}
where for simplicity we assume only one hyperon $\phi_{12}(q,-q)$,
which develops a VEV $\left\langle \phi_{12}\right\rangle =v_{12}/\sqrt{2}$
spontaneously breaking $U(1)_{Y_{1}}\times U(1)_{Y_{2}}$ down to
its diagonal subgroup. The covariant derivative reads
\begin{equation}
D_{\mu}=\partial_{\mu}-ig_{1}Y_{1}B_{1\mu}-ig_{2}Y_{2}B_{2\mu}\,.
\end{equation}
Expanding the kinetic term of $\phi_{12}$, we obtain mass terms for
the gauge bosons as
\begin{equation}
{\displaystyle \mathcal{M}^{2}=\frac{q^{2}v_{12}^{2}}{2}\left(
\global\long\def\arraystretch{0.7}%
\begin{array}{@{}lcc@{}}
 & \phantom{\!\,}B{}_{1}^{\mu} & \phantom{\!\,}B{}_{2}^{\mu}\\
\cmidrule(l){2-3}\left.B_{1\mu}\right| & g_{1}^{2} & -g_{1}g_{2}\\
\left.B_{2\mu}\right| & -g_{1}g_{2} & g_{2}^{2}
\end{array}\right)\,.}
\end{equation}
The diagonalisation of the matrix above reveals
\begin{equation}
\hat{\mathcal{M}}^{2}=\frac{q^{2}v_{12}^{2}}{2}\left(
\global\long\def\arraystretch{0.7}%
\begin{array}{@{}lcc@{}}
 & \phantom{\!\,}Y{}_{12}^{\mu} & \phantom{\!\,}Z'{}_{12}^{\mu}\\
\cmidrule(l){2-3}\left.\;\,Y_{12\mu}\right| & 0 & 0\\
\left.\,\,Z'_{12\mu}\right| & 0 & g_{1}^{2}+g_{2}^{2}
\end{array}\right)\,,\label{eq:MassMatrixGauge_Simple}
\end{equation}
in the basis of mass eigenstates given by
\begin{equation}
\left(\begin{array}{c}
Y_{12\mu}\\
Z'_{12\mu}
\end{array}\right)=\left(\begin{array}{cc}
\cos\theta_{12} & \sin\theta_{12}\\
-\sin\theta_{12} & \cos\theta_{12}
\end{array}\right)\left(\begin{array}{c}
B_{1\mu}\\
B_{2\mu}
\end{array}\right)\,,\qquad\sin\theta_{12}=\frac{g_{1}}{\sqrt{g_{1}^{2}+g_{2}^{2}}}\,.
\end{equation}
Therefore, we obtain a massive gauge boson $Z'_{12\mu}$ at the scale
$v_{12}$, while $Y_{1}+Y_{2}$ associated to the gauge boson $Y_{12\mu}$
remains unbroken. These results are trivially generalised for the
case of more hyperons. The covariant derivative in the new basis is
given by
\begin{align}
D_{\mu} & =\partial_{\mu}-i\frac{g_{1}g_{2}}{\sqrt{g_{1}^{2}+g_{2}^{2}}}(Y_{1}+Y_{2})Y_{12\mu}-i\left(-\frac{g_{1}^{2}}{\sqrt{g_{1}^{2}+g_{2}^{2}}}Y_{1}+\frac{g_{2}^{2}}{\sqrt{g_{1}^{2}+g^{2}}}Y_{2}\right)Z'_{12\mu}\label{eq:Covariant_Derivative_12}\\
 & =\partial_{\mu}-ig_{12}(Y_{1}+Y_{2})Y_{12\mu}-i\left(-g_{1}\sin\theta_{12}Y_{1}+g_{2}\cos\theta_{12}Y_{2}\right)Z'_{12\mu}\,.\nonumber 
\end{align}
The fermion couplings in Eq.~(\ref{eq:Z12_couplings}) are readily
extracted by expanding the fermion kinetic terms applying Eq.~(\ref{eq:Covariant_Derivative_12}).

\subsection{Low scale symmetry breaking\label{sec:Low-scale-symmetry}}

The renormalisable Lagrangian of a theory $SU(2)_{L}\times U(1)_{Y_{1}+Y_{2}}\times U(1)_{Y_{3}}$
with $H(\mathbf{2})_{(0,\frac{1}{2})}$ and $\phi_{23}(\mathbf{1})_{(q,-q)}$
reads (neglecting fermion content and kinetic mixing, although we
will consider the effect of kinetic mixing at the end of the section),
\begin{align}
\mathcal{L}_{\mathrm{ren}} & =-\frac{1}{4}F_{\mu\nu}^{(12)}F^{\mu\nu(12)}-\frac{1}{4}F_{\mu\nu}^{(3)}F^{\mu\nu(3)}-\frac{1}{4}W_{\mu\nu}^{a}W_{a}^{\mu\nu}\label{eq:23_breaking_Lagrangian}\\
 & +(D_{\mu}H)^{\dagger}D^{\mu}H+(D_{\mu}\phi_{23})^{*}D^{\mu}\phi_{23}\nonumber \\
 & -V(H,\phi_{23})\,,\nonumber 
\end{align}
where the covariant derivatives read
\begin{equation}
D_{\mu}H=(\partial_{\mu}-ig_{L}\frac{\sigma^{a}}{2}W_{\mu}^{a}-i\frac{g_{3}}{2}B_{3\mu})H\,,\label{eq:cov1}
\end{equation}
\begin{equation}
D_{\mu}\phi_{23}=(\partial_{\mu}-ig_{12}qB_{12\mu}+ig_{3}qB_{3\mu})\phi_{23}\,,\label{eq:cov2}
\end{equation}
and $\sigma^{a}$ with $a=1,2,3$ are the Pauli matrices. The Higgs
doublet develops the usual electroweak symmetry breaking VEV as
\begin{equation}
\left\langle H\right\rangle =\frac{1}{\sqrt{2}}\left(\begin{array}{c}
0\\
v_{\mathrm{SM}}
\end{array}\right)\,,
\end{equation}
while the hyperon develops a higher scale VEV as 
\begin{equation}
\left\langle \phi_{23}\right\rangle =\frac{v_{23}}{\sqrt{2}}\,,
\end{equation}
which spontaneously breaks the group $U(1)_{Y_{1}+Y_{2}}\times U(1)_{Y_{3}}$
down to its diagonal subgroup.

Expanding the following kinetic terms with the expressions of the
covariant derivatives of Eqs.~(\ref{eq:cov1}) and (\ref{eq:cov2}),
we obtain
\begin{flalign}
 & (D_{\mu}H)^{\dagger}D^{\mu}H+(D_{\mu}\phi_{23})^{*}D^{\mu}\phi_{23}\\
 & =\frac{v_{\mathrm{SM}}^{2}g_{L}^{2}}{4}W_{\mu}W^{\mu\dagger}+\frac{q^{2}v_{23}^{2}}{2}\left(
\global\long\def\arraystretch{0.7}%
\begin{array}{@{}llcc@{}}
 & \multicolumn{1}{c@{}}{\phantom{\!\,}W_{3}^{\mu}} & \phantom{\!\,}B_{12}^{\mu} & \phantom{\!\,}B_{3}^{\mu}\\
\cmidrule(l){2-4}\left.\,W_{3\mu}\right| & g_{L}^{2}r^{2} & 0 & -g_{L}g_{3}r^{2}\\
\left.B_{12\mu}\right| & 0 & g_{12}^{2} & -g_{12}g_{3}\\
\left.\;\,B_{3\mu}\right| & -g_{L}g_{3}r^{2} & -g_{12}g_{3} & g_{3}^{2}+g_{3}^{2}r^{2}
\end{array}\right),\nonumber 
\end{flalign}
where $r=\frac{v_{\mathrm{SM}}}{2qv_{23}}\ll1$, we have defined $W_{\mu}=(W_{\mu}^{1}+iW_{\mu}^{2})/\sqrt{2}$,
and we denote $M_{\mathrm{gauge}}^{2}$ as the off-diagonal matrix
above. Given the two different scales in the mass matrix above, we first apply the following transformation 
\begin{equation}
\left(\begin{array}{c}
W_{3}^{\mu}\\
Y^{\mu}\\
X^{\mu}
\end{array}\right)=\left(\begin{array}{ccc}
1 & 0 & 0\\
0 & \cos\theta_{23} & \sin\theta_{23}\\
0 & -\sin\theta_{23} & \cos\theta_{23}
\end{array}\right)\left(\begin{array}{c}
W_{3}^{\mu}\\
B_{12}^{\mu}\\
B_{3}^{\mu}
\end{array}\right)=\left(\begin{array}{c}
W_{3}^{\mu}\\
\cos\theta_{23}B_{12}^{\mu}+\sin\theta_{23}B_{3}^{\mu}\\
-\sin\theta_{23}B_{12}^{\mu}+\cos\theta_{23}B_{3}^{\mu}
\end{array}\right)\,,\label{eq:C.7}
\end{equation}
where 
\begin{equation}
\sin\theta_{23}=\frac{g_{12}}{\sqrt{g_{12}^{2}+g_{3}^{2}}}\,,
\end{equation}
and we denote the rotation in Eq.~(\ref{eq:C.7}) as $V_{\theta_{23}}$,
obtaining 
\begin{equation}
V_{\theta_{23}}M_{\mathrm{gauge}}^{2}V_{\theta_{23}}^{\dagger}=\frac{q^{2}v_{23}^{2}}{2}\left(
\global\long\def\arraystretch{0.7}%
\begin{array}{@{}llcc@{}}
 & \multicolumn{1}{c@{}}{\phantom{\!\,}W_{3}^{\mu}} & \phantom{\!\,}Y^{\mu} & \phantom{\!\,}X^{\mu}\\
\cmidrule(l){2-4}\left.W_{3\mu}\right| & g_{L}^{2}r^{2} & -g_{L}g_{Y}r^{2} & -g_{L}g_{X}r^{2}\\
\left.\;\,\,\,Y_{\mu}\right| & -g_{L}g_{Y}r^{2} & g_{Y}^{2}r^{2} & g_{Y}g_{X}r^{2}\\
\left.\;\;X_{\mu}\right| & -g_{L}g_{X}r^{2} & g_{Y}g_{X}r^{2} & g_{F}^{2}+g_{X}^{2}r^{2}
\end{array}\right)\,,\label{eq:MassMatrix_MatchLiterature}
\end{equation}
where $Y^{\mu}$ is the SM hypercharge gauge boson with gauge coupling
\begin{equation}
g_{Y}=\frac{g_{12}g_{3}}{\sqrt{g_{12}^{2}+g_{3}^{2}}}\simeq0.36\,,
\end{equation}
where the numeric value depicted is evaluated at the electroweak scale, and
$X^{\mu}$ can be interpreted as an effective gauge boson with effective
couplings 
\begin{equation}
g_{X}=\frac{g_{3}^{2}}{\sqrt{g_{12}^{2}+g_{3}^{2}}}\,,\qquad\qquad g_{F}=\sqrt{g_{12}^{2}+g_{3}^{2}}\,,
\end{equation}
to the Higgs boson and to $\phi_{23}$, respectively. In this basis,
the covariant derivatives read 
\begin{equation}
D_{\mu}H=(\partial_{\mu}-ig_{L}\frac{\sigma^{a}}{2}W_{\mu}^{a}-i\frac{g_{Y}}{2}Y_{\mu}-i\frac{g_{X}}{2}X_{\mu})H\,,
\end{equation}
\begin{equation}
D_{\mu}\phi_{23}=(\partial_{\mu}-iqg_{F}X_{\mu})\phi_{23}\,.
\end{equation}
The mass matrix in this basis can be block-diagonalised by applying
the following transformation 
\begin{equation}
\left(\begin{array}{c}
A^{\mu}\\
(Z^{0})^{\mu}\\
X^{\mu}
\end{array}\right)=\left(\begin{array}{ccc}
\sin\theta_{W} & \cos\theta_{W} & 0\\
\cos\theta_{W} & -\sin\theta_{W} & 0\\
0 & 0 & 1
\end{array}\right)\left(\begin{array}{c}
W_{3}^{\mu}\\
Y^{\mu}\\
X^{\mu}
\end{array}\right)=\left(\begin{array}{c}
\cos\theta_{W}Y^{\mu}+\sin\theta_{W}W_{3}^{\mu}\\
-\sin\theta_{W}Y^{\mu}+\cos\theta_{W}W_{3}^{\mu}\\
X^{\mu}
\end{array}\right)\,,\label{eq:Basis_BeforeZ-ZprimeMixing}
\end{equation}
where the mixing angle is identified with the usual weak mixing angle
as 
\begin{equation}
\sin\theta_{W}=\frac{g_{Y}}{\sqrt{g_{Y}^{2}+g_{L}^{2}}}\,,
\end{equation}
and we denote the rotation in Eq.~(\ref{eq:Basis_BeforeZ-ZprimeMixing})
as $V_{\theta_{W}}$\footnote{Notice that this is not the usual SM convention, because we have ordered
$W_{\mu}^{3}$ and $Y_{\mu}$ differently.}, obtaining 
\begin{flalign}
 & V_{\theta_{W}}V_{\theta_{23}}M_{\mathrm{gauge}}^{2}(V_{\theta_{W}}V_{\theta_{23}})^{\dagger}\\
 & =\frac{q^{2}v_{23}^{2}}{2}\left(
\global\long\def\arraystretch{0.7}%
\begin{array}{@{}llcc@{}}
 & \multicolumn{1}{c@{}}{\phantom{\!\,}A^{\mu}} & \phantom{\!\,}(Z^{0})^{\mu} & \phantom{\!\,}X^{\mu}\\
\cmidrule(l){2-4}\left.\quad\;A_{\mu}\right| & 0 & 0 & 0\\
\left.(Z^{0})_{\mu}\right| & 0 & (g_{L}^{2}+g_{Y}^{2})r^{2} & -g_{X}\sqrt{g_{Y}^{2}+g_{L}^{2}}r^{2}\\
\left.\quad\,X_{\mu}\right| & 0 & -g_{X}\sqrt{g_{Y}^{2}+g_{L}^{2}}r^{2} & g_{F}^{2}+g_{X}^{2}r^{2}
\end{array}\right)\,,\nonumber 
\end{flalign}
where we have already identified the massless photon. Now we diagonalise
the remaining $2\times2$ sub-block in the limit of small $r^{2}$.
We obtain 
\begin{equation}
Z_{\mu}=\cos\theta_{Z-Z'_{23}}\left(-\sin\theta_{W}Y_{\mu}+\cos\theta_{W}W_{3\mu}\right)+\sin\theta_{Z-Z'_{23}}X_{\mu}\,,
\end{equation}
\begin{equation}
Z'_{23\mu}=-\sin\theta_{Z-Z'_{23}}\left(-\sin\theta_{W}Y_{\mu}+\cos\theta_{W}W_{3\mu}\right)+\cos\theta_{Z-Z'_{23}}X_{\mu}\,,
\end{equation}
where to leading order in $r^{2}$ 
\begin{equation}
\sin\theta_{Z-Z'_{23}}\approx\frac{\sqrt{g_{Y}^{2}+g_{L}^{2}}g_{X}}{g_{F}^{2}}r^{2}=\frac{g_{3}\cos\theta_{23}}{\sqrt{g_{Y}^{2}+g_{L}^{2}}}\left(\frac{M_{Z}^{0}}{M_{Z'_{23}}^{0}}\right)^{2}=\frac{\sqrt{g_{3}^{2}-g_{Y}^{2}}}{\sqrt{g_{Y}^{2}+g_{L}^{2}}}\left(\frac{M_{Z}^{0}}{M_{Z'_{23}}^{0}}\right)^{2}\,,\label{eq:Z_Zp_mixing}
\end{equation}
where we have used the matching condition with SM hypercharge to write
everything in terms of $g_{3}$ and $g_{Y}$. We can see that the
SM $Z$ boson carries a small admixture of the $X_{\mu}$ boson, which
provides a small shift to its mass as 
\begin{equation}
M_{Z}^{2}\approx q^{2}v_{23}^{2}\left(g_{Y}^{2}+g_{L}^{2}\right)\left(r^{2}-\frac{g_{X}^{2}}{g_{F}^{2}}r^{4}\right)=(M_{Z}^{0})^{2}\left[1-\frac{g_{3}^{2}-g_{Y}^{2}}{g_{Y}^{2}+g_{L}^{2}}\left(\frac{M_{Z}^{0}}{M_{Z'_{23}}^{0}}\right)^{2}\right]\,,
\end{equation}
\begin{equation}
M_{Z'_{23}}^{2}\approx q^{2}v_{23}^{2}g_{F}^{2}\left(1+\frac{g_{X}^{2}}{g_{F}^{2}}r^{2}\right)=(M_{Z'_{23}}^{0})^{2}\left[1+\frac{g_{3}^{2}-g_{Y}^{2}}{g_{Y}^{2}+g_{L}^{2}}\left(\frac{M_{Z}^{0}}{M_{Z'_{23}}^{0}}\right)^{2}\right]\,,
\end{equation}
where 
\begin{equation}
M_{Z}^{0}=\frac{v_{\mathrm{SM}}}{2}\sqrt{g_{Y}^{2}+g_{L}^{2}}\,,\qquad\qquad M_{Z'_{23}}^{0}=qv_{\mathrm{23}}\sqrt{g_{12}^{2}+g_{3}^{2}}=qv_{\mathrm{23}}\frac{g_{3}^{2}}{\sqrt{g_{3}^{2}-g_{Y}^{2}}},
\end{equation}
are the masses of the $Z$ boson in the SM and the mass of the $Z'_{23}$
boson in absence of $Z-Z'_{23}$ mixing, respectively. All these results
can be easily generalised for the case of more hyperons or more Higgs
doublets.

As expected, the SM $Z$ boson mass arises at order $r^{2}$, with
a leading correction from $Z-Z'_{23}$ mixing arising at order $r^{4}$.
Instead, the $Z'_{23}$ boson arises at leading order in the power
expansion, with the leading correction from $Z-Z'_{23}$ mixing arising
at order $r^{2}$. Remarkably, the presence of $Z-Z'_{23}$ mixing
always shifts the mass of the $Z$ boson to smaller values with respect
to the SM prediction.

The equations obtained match general results in the literature \cite{Allanach:2018lvl,Langacker:2008yv,Bandyopadhyay:2018cwu},
which consider scenarios where the starting point is a matrix such
as Eq.~\eqref{eq:MassMatrix_MatchLiterature} with $g_{F}=g_{X}$.
Our equations match those of these papers when $g_{F}=g_{X}$ (and
taking into account that we need to perform an extra rotation $\theta_{23}$
to arrive to Eq.~\eqref{eq:MassMatrix_MatchLiterature}). 

In the case that a kinetic mixing term $\sin\chi F_{\mu\nu}^{(12)}F_{\mu\nu}^{(3)}/2$
is included in Eq.~\eqref{eq:23_breaking_Lagrangian}, then one can
repeat the calculations of this section to finally obtain
\begin{equation}
\sin\theta_{Z-Z'_{23}}=\frac{g_{3}\cos\theta_{23}}{\sqrt{g_{Y}^{2}+g_{L}^{2}}}(1-\sin\chi)\left(\frac{M_{Z}^{0}}{M_{Z'_{23}}^{0}}\right)^{2}\,,\label{eq:Z_Zp_mixing_kinetic}
\end{equation}
where now
\begin{equation}
\cos\theta_{23}=\frac{g_{3}\sec^{2}\chi}{\sqrt{g_{12}^{2}+g_{3}^{2}\left(\sec\chi-\tan\chi\right)^{2}}}\,,\qquad g_{Y}=\frac{g_{12}g_{3}\sec^{2}\chi}{\sqrt{g_{12}^{2}+g_{3}^{2}\left(\sec\chi-\tan\chi\right)^{2}}}\,,
\end{equation}
\begin{equation}
M_{Z'_{23}}^{0}=qv_{\mathrm{23}}\sqrt{g_{12}^{2}+g_{3}^{2}\left(\frac{1-\sin\chi}{1+\sin\chi}\right)}\,.
\end{equation}
Assuming that the kinetic mixing parameter $\sin\chi$ is small compared
to unity (e.g.~absent at tree-level and generated by loop diagrams),
then the dominant effect is always the original gauge mixing and kinetic
mixing can be neglected. As an example, the hyperon $\phi_{23}(\mathbf{1})_{(q,-q)}$
charged under both $U(1)$ groups generates kinetic mixing at 1-loop
as $d\sin\chi/d\log\mu=-g_{12}g_{3}q^{2}/(16\pi^{2})$, which leads
to $\sin\chi(\mu)=g_{12}g_{3}q^{2}\log(m_{\phi_{23}}^{2}/\mu^{2})/(16\pi^{2})$.
For the natural benchmark $g_{12}\approx\sqrt{3/2}g_{Y}$ and $g_{3}\approx\sqrt{3}g_{Y}$
motivated in Section~\ref{subsec:Couplings_Zprime}, along with typical
values $q=1/2$ and $m_{\phi_{23}}=1\,\mathrm{TeV}$, we obtain $\sin\chi(M_{Z})\simeq0.002$.

Neglecting the small $Z-Z'_{23}$ mixing, the fermion couplings of
the $Z'_{23}$ gauge boson given in Eq.~\eqref{eq:Z23_couplings}
are obtained by expanding the fermion kinetic terms in the usual way,
using the covariant derivative (where $T_{3}$ is the third-component
$SU(2)_{L}$ isospin, and we do not include the terms associated to
charge currents nor QCD interactions) 
\begin{flalign}
D_{\mu} & =\partial_{\mu}-i\left[eQA_{\mu}+\left(T_{3}g_{L}\cos\theta_{W}-g_{Y}\sin\theta_{W}(Y_{1}+Y_{2}+Y_{3})\right)Z_{\mu}^{0}\right.\label{eq:cov_23}\\
 & \left.+\left(-g_{12}\sin\theta_{23}(Y_{1}+Y_{2})+g_{3}\cos\theta_{23}Y_{3}\right)Z'_{23\mu}\right]\nonumber \\
 & =D_{\mu}^{\mathrm{SM}}-i\left(-\frac{g_{Y}^{2}}{\sqrt{g_{3}^{2}-g_{Y}^{2}}}(Y_{1}+Y_{2})+\sqrt{g_{3}^{2}-g_{Y}^{2}}Y_{3}\right)Z'_{23\mu}\,,
\end{flalign}
which is an excellent approximation for all practical purposes other
than precision $Z$ boson phenomenology. In that case, one has to
consider that the couplings of the $Z$ boson to fermions are shifted
due to $Z-Z'_{23}$ mixing as
\begin{equation}
g_{Z}^{f_{L}f_{L}}=\left(g_{Z}^{f_{L}f_{L}}\right)^{0}+\sin\theta_{Z-Z'_{23}}g_{Z'_{23}}^{f_{L}f_{L}}\,,
\end{equation}
where $g_{Z'_{23}}^{f_{L}f_{L}}$ are the fermion couplings of $Z'_{23}$
in the absence of $Z-Z'_{23}$ mixing, as given in Eq.~\eqref{eq:Z23_couplings},
and similarly for right-handed fermions by just replacing $L$ by
$R$ everywhere. We can see that in any case, the shift in the $Z$
boson couplings is suppressed by the small ratio $(M_{Z}^{0}/M_{Z'_{23}}^{0})^{2}$.

\section{Phenomenology\label{sec:Phenomenology}}

\subsection{\texorpdfstring{Couplings of the heavy $Z'$ bosons to fermions}{Couplings of the heavy Z' bosons to fermions} \label{subsec:Couplings_Zprime}}

In Sections~\ref{sec:Charged-fermion-masses-mixing} and \ref{sec:Neutrino-masses-and-Mixing}
we have discussed examples of $U(1)_{Y}^{3}$ models which provide
a compelling description of all fermion masses and mixings, and we have highlighted
model-independent features which are intrinsic to the $U(1)^{3}_{Y}$ framework. We
have assumed that the symmetry breaking pattern of the $U(1)_{Y}^{3}$ group down
to the SM is described by Eq.~(\ref{eq:Symmetry_Breaking}), in such
a way that at a high scale $v_{12}$, the group $U(1)_{Y_{1}}\times U(1)_{Y_{2}}$
is broken down to its diagonal subgroup. The remaining group $U(1)_{Y_{1}+Y_{2}}\times U(1)_{Y_{3}}$
is broken down to SM hypercharge at a lower scale $v_{23}$. The hierarchy between the scales $v_{12}$ and $v_{23}$ generally plays
a role on the origin of flavour hierarchies in the SM, although in specific models we have found that a mild hierarchy $v_{23}/v_{12}\simeq\lambda$ is enough.

A massive gauge boson $Z'_{12}$ is predicted to live at the higher
scale $v_{12}$, displaying \textit{intrinsically} flavour non-universal
couplings to the first two families of SM fermions. Similarly,
another massive boson $Z'_{23}$ lives at the lower scale $v_{23}$, displaying
family universal couplings to first and second family fermions, while the
couplings to the third family are intrinsically different. In the
following, we show the coupling matrices in family space (obtained
from Eqs.~\eqref{eq:Covariant_Derivative_12}
and~\eqref{eq:cov_23}), ignoring fermion mass mixing,
\begin{flalign}
 & \mathcal{L}_{Z'_{12}}\supset Y_{\psi_{L,R}}\overline{\psi}_{L,R}\gamma^{\mu}\left(\begin{array}{ccc}
-g_{1}\sin\theta_{12} & 0 & 0\\
0 & g_{2}\cos\theta_{12} & 0\\
0 & 0 & 0
\end{array}\right)\psi_{L,R}Z'_{12\mu}\,,\label{eq:Z12_couplings}\\
 & \sin\theta_{12}=\frac{g_{1}}{\sqrt{g_{1}^{2}+g_{2}^{2}}}\,,\qquad\cos\theta_{12}=\frac{g_{2}}{\sqrt{g_{1}^{2}+g_{2}^{2}}}\,,\nonumber \\
 & \mathcal{L}_{Z'_{23}}\supset Y_{\psi_{L,R}}\overline{\psi}_{L,R}\gamma^{\mu}\left(\begin{array}{ccc}
-g_{12}\sin\theta_{23} & 0 & 0\\
0 & -g_{12}\sin\theta_{23} & 0\\
0 & 0 & g_{3}\cos\theta_{23}
\end{array}\right)\psi_{L,R}Z'_{23\mu}\,,\label{eq:Z23_couplings}\\
 & \sin\theta_{23}=\frac{g_{12}}{\sqrt{g_{12}^{2}+g_{3}^{2}}}\,,\qquad\cos\theta_{23}=\frac{g_{3}}{\sqrt{g_{12}^{2}+g_{3}^{2}}}\,,\nonumber 
\end{flalign}
where $Y_{\psi_{L,R}}$ is the SM hypercharge of $\psi_{L,R}$\footnote{Note that we have departed from our 2-component and purely left-handed notation, used in the rest of the chapter, to use instead a 4-component left-right notation, which is more familiar in phenomenological studies. The connection
between both conventions can be found in Appendix~\ref{app:2-component_notation}.}, where $\psi$ is a 3-component column vector containing the three
families. Explicitly, $\psi_{L}=Q_{L}^{i},\,L_{L}^{i},$ with $Y_{\psi_{L}}=1/6,\,-1/2,$
and $\psi_{R}=u_{R}^{i},d_{R}^{i},e_{R}^{i}$ with $Y_{\psi_{R}}=2/3,-1/3,-1$,
respectively, ignoring couplings to the SM singlet neutrinos\footnote{Note that such low scale SM singlet neutrinos may be observable via their gauge couplings to $Z'_{23}$, which can be obtained
from the covariant derivative in Eq.~(\ref{eq:cov_23}).} discussed
in the Section~\ref{sec:Neutrino-masses-and-Mixing}. In order to include fermion mass mixing we need to define the $SU(2)_{L}$
doublets and singlets accordingly, see Eqs.~(\ref{eq:fermion_mixing1}-\ref{eq:fermion_mixing3}).

Notice that the couplings to\textit{ right-handed} fermions are larger
since their hypercharges are larger in magnitude than those
of left-handed fermions. The SM hypercharge gauge coupling $g_{Y}(M_{Z})\simeq0.36$
is entangled to the $g_{i}$ couplings via the relations 
\begin{equation}
g_{Y}=\frac{g_{12}g_{3}}{\sqrt{g_{12}^{2}+g_{3}^{2}}}\,,\qquad\qquad\qquad\qquad g_{12}=\frac{g_{1}g_{2}}{\sqrt{g_{1}^{2}+g_{2}^{2}}}\,.\label{eq:gauge_couplingsTH}
\end{equation}
The expressions above reveal a lower bound on the gauge couplings
$g_{i}\apprge g_{Y}$. Moreover, we may use the matching condition
with SM hypercharge to exchange $g_{12}$ in favour of $g_{3}$ and
$g_{Y}$ for the couplings of $Z'_{23}$,
\begin{equation}
\mathcal{L}_{Z'_{23}}\supset Y_{\psi_{L,R}}\overline{\psi}_{L,R}\gamma^{\mu}\left(\begin{array}{ccc}
-{\displaystyle \frac{g_{Y}^{2}}{\sqrt{g_{3}^{2}-g_{Y}^{2}}}} & 0 & 0\\
0 & -{\displaystyle \frac{g_{Y}^{2}}{\sqrt{g_{3}^{2}-g_{Y}^{2}}}} & 0\\
0 & 0 & \sqrt{g_{3}^{2}-g_{Y}^{2}}
\end{array}\right)\psi_{L,R}Z'_{23\mu}\,,
\end{equation}
Therefore, the phenomenology of $Z'_{23}$ can be completely described
in terms of its mass and the $g_{3}$ coupling. When $g_{3}$ is large,
$Z'_{23}$ is mostly coupled to the third family, while for $g_{3}$
small $Z'_{23}$ is mostly coupled to the first and second families.
In contrast, the couplings of $Z'_{12}$ need to be described in general
by two free gauge couplings. Here we choose $g_{1}$ and $g_{2}$,
but one may exchange one of these by e.g.~$g_{3}$ and $g_{Y}$ through
the matching conditions.

Throughout this work we have considered a bottom-up
approach where the $U(1)_{Y}^{3}$ model is just the next step in
our understanding of Nature, which reveals information about the origin
of flavour, but nevertheless is an EFT remnant of a more fundamental
UV-complete theory. In this spirit, we have studied the RGE evolution
of the gauge couplings $g_{i}$, obtaining that for $g_{i}(\mathrm{TeV})\simeq1$
the model can be extrapolated to the Planck scale (and beyond). Instead,
for $g_{i}(\mathrm{TeV})\simeq2$, a Landau pole is found at a scale
$\mathcal{O}(10^{4}\:\mathrm{TeV})$, which anyway seems like a reasonable
scale for an UV embedding, given that we expect the cut-off scale of
the effective Yukawa operators (see Section~\ref{sec:Charged-fermion-masses-mixing})
to be around $\mathcal{O}(10^{2}\:\mathrm{TeV})$ in order to provide the required suppression for charged fermion masses. Therefore, in order to protect the perturbativity
of the model, we avoid considering $g_{i}>2$ in the phenomenological
analysis. Nevertheless, we highlight a natural scenario where the
three gauge couplings have a similar size $g_{1}\simeq g_{2}\simeq g_{3}\simeq\sqrt{3}g_{Y}$,
which could be connected to a possible \textit{gauge unification}, as we shall see in Chapter~\ref{Chapter:Tri-unification}.
This benchmark is depicted as a dashed horizontal line in Figs.~\ref{fig:Zp12_plot}
and \ref{fig:Zp23_plot}.

\subsection{Tree-level SMEFT matching for 4-fermion operators \label{subsec:Tree-level-SMEFT-matching-Tri-hypercharge}}

In order to write the tree-level SMEFT matching for 4-fermion operators
in our model, we write the $Z'$ couplings of Eqs.~(\ref{eq:Z12_couplings})
and (\ref{eq:Z23_couplings}) in a more compact form as
\begin{equation}
\mathcal{L}_{\mathrm{int}}=\sum_{f=q,\ell,u,d,e}\sum_{i,j}(\kappa_{ij}^{f}\overline{f}_{i}\gamma^{\mu}f_{j}Z'_{12\mu}+\xi_{ij}^{f}\overline{f}_{i}\gamma^{\mu}f_{j}Z'_{23\mu})\,.
\end{equation}
The Wilson coefficients of SMEFT operators $Q_{\alpha}$ (see Appendix~\ref{app:SMEFT_Operators})
are then given in Table~\ref{tab:SMEFT_matching_tri-hypercharge}
as a function of the couplings $\kappa_{ij}^{f}$, $\xi_{ij}^{f}$ and
the auxiliary variables
\begin{equation}
C_{12}=-\frac{1}{2M_{Z'_{12}}^{2}}\,,\qquad C_{23}=-\frac{1}{2M_{Z'_{23}}^{2}}\,.
\end{equation}
In general, the TH model predicts non-trivial fermion mixing in all
charged sectors, and we shall take this into account by defining the
$SU(2)_{L}$ doublets and singlets in the SMEFT accordingly, i.e.
\begin{flalign}
 & \hat{q}_{L}^{i}=(\begin{array}{cc}
V_{u_{L}}^{ij}u_{L}^{i} & V_{d_{L}}^{ij}d_{L}^{j}\end{array})^{\mathrm{T}}\,, &  & \hat{\ell}_{L}^{i}=(\begin{array}{cc}
V_{\nu_{L}}^{ij}\nu_{L}^{j} & V_{e_{L}}^{ij}e_{L}^{i}\end{array})^{\mathrm{T}}\,, & \text{\,}\label{eq:fermion_mixing1}\\
 & \hat{u}_{R}^{i}=V_{u_{R}}^{ij}u_{R}^{i}\,, &  & \hat{d}_{R}^{i}=V_{d_{R}}^{ij}d_{R}^{i}\,, & \,\\
 & \hat{e}_{R}^{i}=V_{e_{R}}^{ij}e_{R}^{i}\,. &  &  & \,\label{eq:fermion_mixing3}
\end{flalign}
Notice as well that different implementations of the TH model may
lead to different fermion mixing (see Section~\ref{sec:Charged-fermion-masses-mixing}),
therefore only the different mixing matrices above need to be exchanged
when studying different model variations.

\begin{table}
\begin{centering}
\begin{tabular}{cccc}
\toprule 
\multicolumn{2}{c}{$(\overline{R}R)(\overline{R}R)$} & \multicolumn{2}{c}{$(\overline{L}L)(\overline{R}R)$}\tabularnewline
\midrule 
$Q_{\alpha}$ & $C_{\alpha}$ & $Q_{\alpha}$ & $C_{\alpha}$\tabularnewline
\midrule
\midrule 
$\left[Q_{uu}\right]_{ijkl}$ & $C_{12}\kappa_{ij}^{u}\kappa_{kl}^{u}$+$C_{23}\xi_{ij}^{u}\xi_{kl}^{u}$ & $\left[Q_{qu}^{(1)}\right]_{ijkl}$ & $2C_{12}\kappa_{ij}^{q}\kappa_{kl}^{u}$+$2C_{23}\xi_{ij}^{q}\xi_{kl}^{u}$\tabularnewline
\midrule 
$\left[Q_{dd}\right]_{ijkl}$ & $C_{12}\kappa_{ij}^{d}\kappa_{kl}^{d}$+$C_{23}\xi_{ij}^{d}\xi_{kl}^{d}$ & $\left[Q_{qd}^{(1)}\right]_{ijkl}$ & $2C_{12}\kappa_{ij}^{q}\kappa_{kl}^{d}$+$2C_{23}\xi_{ij}^{q}\xi_{kl}^{d}$\tabularnewline
\midrule 
$\left[Q_{ud}^{(1)}\right]_{ijkl}$ & $2C_{12}\kappa_{ij}^{u}\kappa_{kl}^{d}$+$2C_{23}\xi_{ij}^{u}\xi_{kl}^{d}$ & $\left[Q_{qe}\right]_{ijkl}$ & $2C_{12}\kappa_{ij}^{q}\kappa_{kl}^{e}$+$2C_{23}\xi_{ij}^{q}\xi_{kl}^{e}$\tabularnewline
\midrule 
$\left[Q_{eu}\right]_{ijkl}$ & $2C_{12}\kappa_{ij}^{e}\kappa_{kl}^{u}$+$2C_{23}\xi_{ij}^{e}\xi_{kl}^{u}$ & $\left[Q_{\ell u}\right]_{ijkl}$ & $2C_{12}\kappa_{ij}^{\ell}\kappa_{kl}^{u}$+$2C_{23}\xi_{ij}^{\ell}\xi_{kl}^{u}$\tabularnewline
\midrule 
$\left[Q_{ed}\right]_{ijkl}$ & $2C_{12}\kappa_{ij}^{e}\kappa_{kl}^{d}$+$2C_{23}\xi_{ij}^{e}\xi_{kl}^{d}$ & $\left[Q_{\ell d}\right]_{ijkl}$ & $2C_{12}\kappa_{ij}^{\ell}\kappa_{kl}^{d}$+$2C_{23}\xi_{ij}^{\ell}\xi_{kl}^{d}$\tabularnewline
\midrule 
$\left[Q_{ee}\right]_{ijkl}$ & $C_{12}\kappa_{ij}^{e}\kappa_{kl}^{e}$+$C_{23}\xi_{ij}^{e}\xi_{kl}^{e}$ & $\left[Q_{\ell e}\right]_{ijkl}$ & $2C_{12}\kappa_{ij}^{\ell}\kappa_{kl}^{e}$+$2C_{23}\xi_{ij}^{\ell}\xi_{kl}^{e}$\tabularnewline
\midrule
\midrule 
\multicolumn{2}{c}{$(\overline{L}L)(\overline{L}L)$} & \multirow{5}{*}{} & \multirow{5}{*}{}\tabularnewline
\cmidrule{1-2} \cmidrule{2-2} 
$Q_{\alpha}$ & $C_{\alpha}$ &  & \tabularnewline
\cmidrule{1-2} \cmidrule{2-2} 
$\left[Q_{qq}^{(1)}\right]_{ijkl}$ & $C_{12}\kappa_{ij}^{q}\kappa_{kl}^{q}$+$C_{23}\xi_{ij}^{q}\xi_{kl}^{q}$ &  & \tabularnewline
\cmidrule{1-2} \cmidrule{2-2} 
$\left[Q_{\ell q}^{(1)}\right]_{ijkl}$ & $2C_{12}\kappa_{ij}^{\ell}\kappa_{kl}^{q}$+$2C_{23}\xi_{ij}^{\ell}\xi_{kl}^{q}$ &  & \tabularnewline
\cmidrule{1-2} \cmidrule{2-2} 
$\left[Q_{\ell\ell}\right]_{ijkl}$ & $C_{12}\kappa_{ij}^{\ell}\kappa_{kl}^{\ell}$+$C_{23}\xi_{ij}^{\ell}\xi_{kl}^{\ell}$ &  & \tabularnewline
\bottomrule
\end{tabular}
\par\end{centering}
\caption[Tree-level SMEFT matching for 4-fermion operators in the $U(1)_{Y}^{3}$
model]{Tree-level SMEFT matching for 4-fermion operators in the $U(1)_{Y}^{3}$
model.\label{tab:SMEFT_matching_tri-hypercharge}}
\end{table}

With the above considerations, the model can then be matched to the
LEFT as shown in Appendix~\ref{app:Matching}, and the contributions
to low-energy observables can be computed via the EFT formalism introduced
in Chapter~\ref{chap:2}.

Finally, we note that beyond 4-fermion operators, Higgs-bifermion
and purely bosonic SMEFT operators are also induced via $Z-Z'_{23}$
mixing. We do not include them in this section because we discuss
electroweak symmetry breaking and $Z-Z'_{23}$ mixing separately in
Section~\ref{sec:Low-scale-symmetry}.

\subsection{\texorpdfstring{The high scale boson $Z'_{12}$}{The high scale boson Z'12}}

In any implementation of the $U(1)_{Y}^{3}$ model, $Z'_{12}$ is
expected to mediate sizable tree-level transitions between first and
second generation left-handed quarks, either in the up or down sectors
depending on the alignment of the CKM matrix predicted by the specific
model. Furthermore, our analysis in Section~\ref{sec:Charged-fermion-masses-mixing}
reveals that $U(1)_{Y}^{3}$ models generally predict non-vanishing
charged lepton mixing and mixing among right-handed quarks. This way,
contributions to $K-\bar{K}$ and $D-\bar{D}$ meson
mixing (Section~\ref{subsec:BsMixing}), along with CLFV processes
such as $\mu\rightarrow e\gamma$ (Section~\ref{subsec:leptonic_CLFV}), have the potential
to push the scale $v_{12}$ far above the $\mathrm{TeV}$. 
\begin{figure}[t]
\begin{centering}
\includegraphics[scale=0.5]{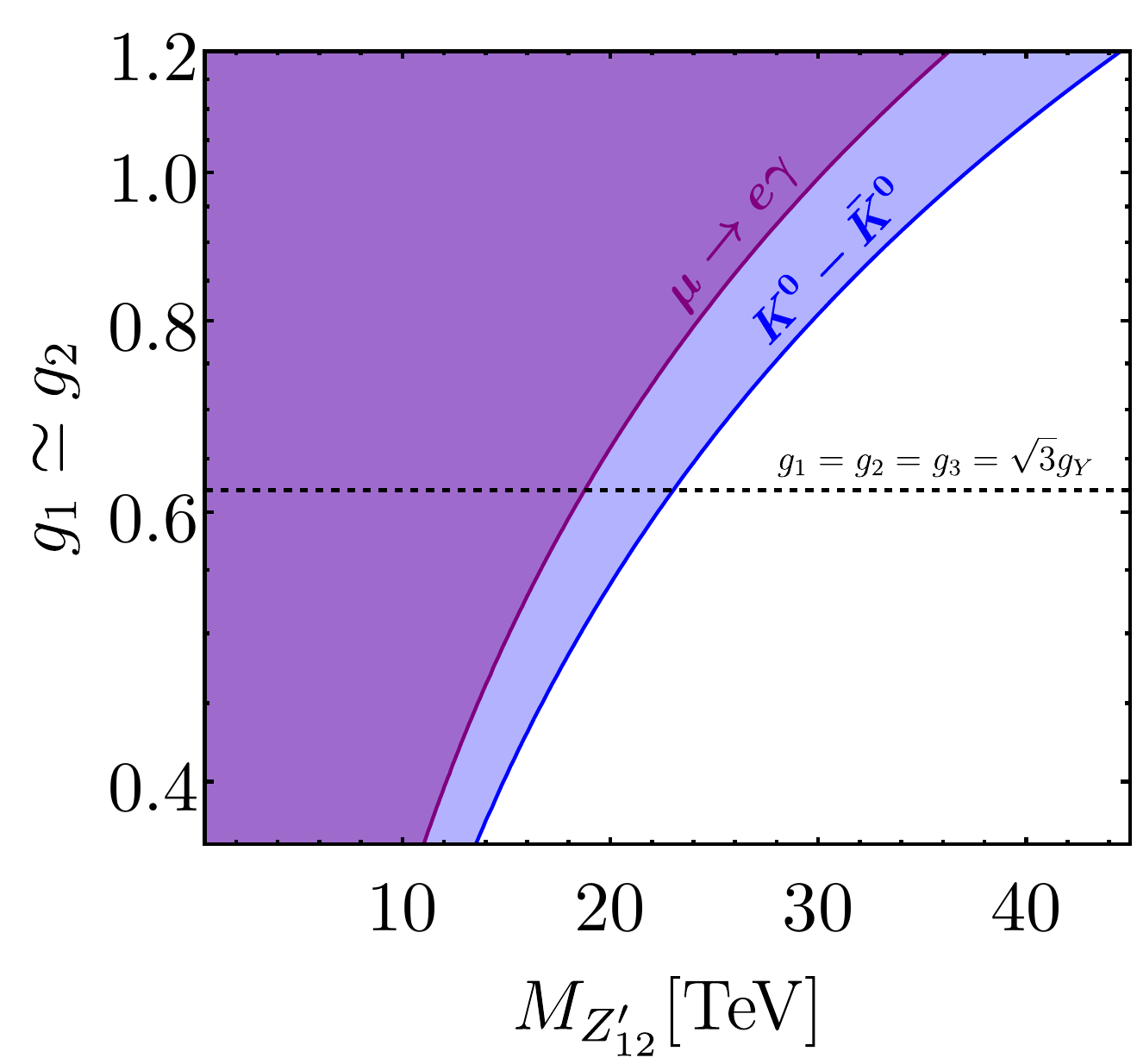}
\par\end{centering}
\caption[Parameter space of the high scale breaking in the $U(1)_{Y}^{3}$
model]{Parameter space of the high scale breaking, where $M_{Z'_{12}}$
is the mass of the heavy $Z'_{12}$ gauge boson and $g_{1}$, $g_{2}$
are the gauge couplings of the $U(1)_{Y_{1}}$ and $U(1)_{Y_{2}}$
groups, respectively. For simplicity, we assume $g_{1}$ and $g_{2}$
to be similar, and the non-generic fermion mixing predicted by Model
2 in Section~\ref{subsec:Model-2:-Five}. Shaded regions in the plot
depict 95\% CL exclusions over the parameter space. The dashed line
represents the natural benchmark $g_{1}\simeq g_{2}\simeq g_{3}\simeq\sqrt{3}g_{Y}$
motivated in the main text. \label{fig:Zp12_plot}}
\end{figure}

Being more specific, for Model 1 described in Section~\ref{subsec:Model-1:-Minimal}
we find the stringent bounds over $v_{12}$ to come from the scalar
and coloured operator $(\bar{s}_{L}^{\alpha}d_{R}^{\beta})(\bar{s}_{R}^{\beta}d_{L}^{\alpha})$
obtained after integrating out $Z'_{12}$ at tree-level (and applying
a Fierz rearrangement), which contributes to $K-\bar{K}$
mixing. Model 1 predicts the up and down left-handed mixings to be
similar up to dimensionless couplings, which must therefore play some
role in the alignment of the CKM matrix. In either case, mildly suppressed
right-handed $s-d$ mixing $s_{12}^{d_{R}}\simeq\mathcal{O}(\lambda^{2})$
is predicted. If $V_{us}$ originates mostly from the down sector,
then $K-\bar{K}$ mixing imposes the stringent bound $M_{Z'_{12}}>170\:\mathrm{TeV}$
for gauge couplings of $\mathcal{O}(0.5)$. Instead, if the dimensionless
coupling provides a mild suppression of $\mathcal{O}(0.1)$ in left-handed
$s-d$ mixing, such that $V_{us}$ originates mostly from the up sector,
then the bound is relaxed to $M_{Z'_{12}}>55\:\mathrm{TeV}$. We find
bounds from $D-\bar{D}$ mixing to be always weaker, even
if $V_{us}$ originates from the up sector, since right-handed up
mixing is strongly suppressed in Model 1.

In contrast with Model 1, Model 2 described in Section~\ref{subsec:Model-2:-Five}
provides a more predictive scenario where $V_{us}$ originates unambiguously
from the down sector. Here right-handed quark mixing is more suppressed,
obtaining $s_{12}^{d_{R}}\simeq\mathcal{O}(\lambda^{5})$. Nevertheless,
$K-\bar{K}$ mixing still imposes the strongest bounds over
the parameter case, as can be seen in Fig.~\ref{fig:Zp12_plot}.
In this case, the lower bound over the mass of $Z'_{12}$ can be as
low as 10-50 TeV, depending on the values of the gauge couplings.
We find the CLFV process $\mu\rightarrow e\gamma$ to provide a slightly
weaker bound over the parameter space, because charged lepton mixing
is generally suppressed with respect to quark mixing in Model 2. We
find the bound from $\mu\rightarrow3e$ to be very similar to the
bound from $\mu\rightarrow e\gamma$.

\subsection{\texorpdfstring{The low scale boson $Z'_{23}$}{The low scale boson Z'23}}

Given that the high scale symmetry breaking can be as low as 10-20
TeV for specific models, and considering the hierarchy of scales $v_{23}/v_{12}\simeq\lambda$
suggested by these specific models in Section~\ref{sec:Charged-fermion-masses-mixing},
it is possible to find the low scale breaking $v_{23}$ near the TeV. Since $Z'_{23}$ features flavour universal couplings to the
first and second families, the stringent bounds from $K-\bar{K}$
mixing and $\mu\rightarrow e\gamma$ are avoided, in the spirit of
the \textit{GIM mechanism}. This way, $Z'_{23}$ can live at the TeV
scale, within the reach of the LHC and future colliders.

Any implementation of the $U(1)_{Y}^{3}$ model predicts small \textit{mixing}
between $Z'_{23}$ and the SM $Z$ boson given by the mixing angle
(see Appendix~\ref{sec:Low-scale-symmetry})\footnote{In this section we only discuss the impact of $Z-Z'_{23}$ gauge mixing,
while kinetic mixing is found to be negligible as long as the kinetic
mixing parameter is smaller than $\mathcal{O}(1)$ (see Appendix~\ref{sec:Low-scale-symmetry}).}
\begin{equation}
\mathrm{sin}\theta_{Z-Z'_{23}}=\frac{\sqrt{g_{3}^{2}-g_{Y}^{2}}}{\sqrt{g_{Y}^{2}+g_{L}^{2}}}\left(\frac{M_{Z}^{0}}{M_{Z'_{23}}^{0}}\right)^{2}\,,
\end{equation}
where $M_{Z}^{0}$ and $M_{Z'_{23}}^{0}$ are the masses of the $Z$
and $Z'_{23}$ bosons in the absence of mixing, respectively, and
$g_{L}$ is the gauge coupling of $SU(2)_{L}$. This mixing leads
to a \textit{small shift} on the mass of the $Z$ boson, which has
an impact on the so-called $\rho$ parameter 
\begin{equation}
\rho=\frac{M_{W}^{2}}{M_{Z}^{2}\cos^{2}\theta_{W}}=\frac{1}{1-(g_{3}^{2}-g_{Y}^{2})\left(\frac{v_{\mathrm{SM}}}{2M_{Z'_{23}}^{0}}\right)^{2}}\,,
\end{equation}
which is predicted as $\rho=1$ at tree-level in the SM. This is a
consequence of custodial symmetry in the Higgs potential, introduced
as well in Section~\ref{sec:ScalarSectorandEWSSB}, which is explicitly
broken in our model via $Z-Z'_{23}$ mixing leading to deviations
in $\rho$. The fact that in our model $M_{Z}$ is always shifted
to smaller values leads to $\rho>1$ at tree-level. Given that $M_{Z}$
is commonly an input experimental parameter of the SM used in the
determination of $g_{Y}$ and $g_{L}$, the downward shift of $M_{Z}$
with respect to the SM prediction would be seen from the experimental
point of view as an upward shift of $M_{W}$ with respect to the SM
prediction. Nevertheless, the experimental picture of $M_{W}$ is
puzzling after the recent measurement by CDF \cite{CDF:2022hxs}.
This measurement points towards $M_{W}$ being larger than the SM
prediction with high significance, but it is in tension with the combination
of measurements by LHC, LEP and Tevatron D0 \cite{PDG:2022ynf}. Neglecting
the recent CDF measurement for the moment, current data\footnote{The current world average (without the latest CDF measurement) of
$M_{W}$ does not consider the very recent $M_{W}$ update by ATLAS~\cite{ATLAS:2023fsi}.
Given that the central value and the uncertainty of this measurement
are just slightly reduced with respect to the 2017 measurement \cite{ATLAS:2017rzl},
we do not expect a big impact over the world average.} provides $\rho=1.0003\pm0.0005$ \cite{PDG:2022ynf} (assuming that
both the oblique parameters $T$ and $S$ are non-zero, as we expect
in our model). We obtain the approximate bound $g_{3}/M_{Z'}<3.1\,\mathrm{TeV}$
at 95\% CL, which translates to an approximate bound over the mixing
angle of $\mathrm{sin}\theta_{Z-Z'_{23}}<0.001$. 
\begin{figure}[t]
\begin{centering}
\includegraphics[scale=0.53]{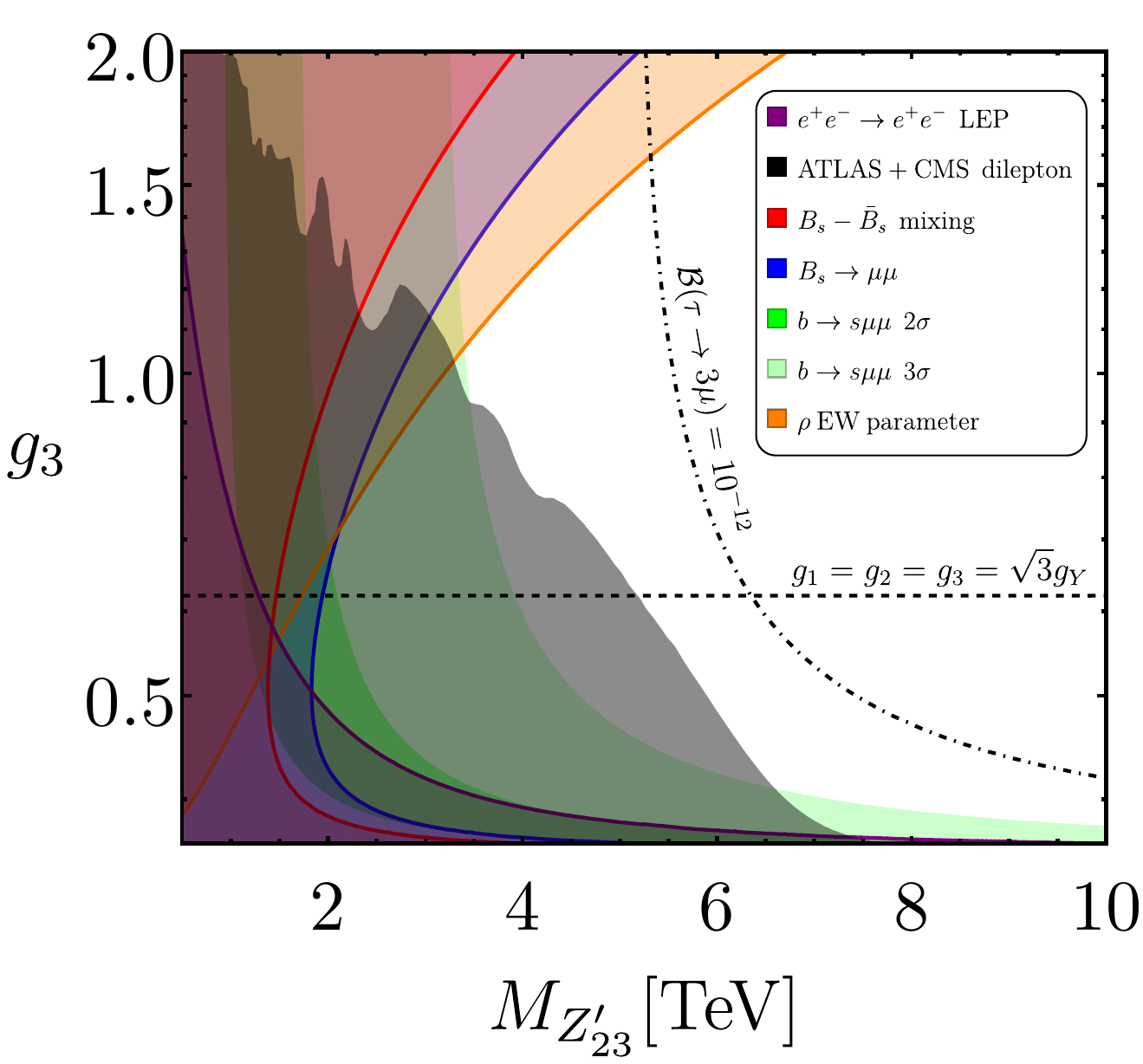}
\par\end{centering}
\caption[Parameter space of the low scale breaking in the $U(1)_{Y}^{3}$ model]{Parameter space of the low scale breaking, where $M_{Z'_{23}}$ is
the mass of the heavy $Z'_{23}$ gauge boson and $g_{3}$ is the gauge
coupling of the $U(1)_{Y_{3}}$ group. The gauge coupling $g_{12}$
is fixed in terms of $g_{3}$ and $g_{Y}$ via Eq.~(\ref{eq:gauge_couplingsTH}),
and we consider the non-generic fermion mixing predicted by Model
2 in Section~\ref{subsec:Model-2:-Five}. Shaded regions in the plot
depict 95\% CL exclusions over the parameter space, with the exception
of the green (light green) region which is preferred by a global fit
to $b\rightarrow s\mu\mu$ data at 2$\sigma$ (3$\sigma$) \cite{Alguero:2023jeh}.
The dashed line represents the natural benchmark $g_{1}\simeq g_{2}\simeq g_{3}\simeq\sqrt{3}g_{Y}$
motivated in the main text. The dashed-dotted line represents the
contour where $\mathcal{B}(\tau\rightarrow3\mu)=10^{-12}$.\label{fig:Zp23_plot}}
\end{figure}

$Z-Z'_{23}$ mixing also shifts the couplings of the $Z$ boson to
fermions, leading to an important impact over $Z$-pole EWPOs if $Z'_{23}$
lives at the TeV scale. We find bounds coming from tests of $Z$ boson
lepton universality and flavour-violating $Z$ decays not to be competitive
with the bound from $\rho$. The electron asymmetry parameter $A_{e}$,
which already deviates from the SM by almost $2\sigma$ \cite{ALEPH:2005ab},
is expected to deviate further in our model. Nevertheless, we expect
our model to improve the fit of $A_{b}^{\mathrm{FB}}$, which is in
tension with the SM prediction by more than $2\sigma$ \cite{ALEPH:2005ab}.
In conclusion, the global effect of our model over EWPOs can only
by captured by performing a global fit, which we leave for future
work. Global fits of EWPOs in the context of other $Z'$ models have
been performed in the literature, see e.g.~\cite{Erler:2009ut,Allanach:2021kzj,Allanach:2022bik},
which obtain 95\% CL maximum values of $\mathrm{sin}\theta_{Z-Z'}$
ranging from $0.002$ to $0.0006$ depending on the model. We expect
our model to lie on the more restrictive side of that range. We do not 
expect our model to explain the anomalous CDF $M_{W}$ measurement, because predicting
such a large $M_{W}$ via large $Z-Z'_{23}$ mixing would lead to intolerably large contributions to other EWPOs,
worsening the global fit.

The massive $Z'_{23}$ boson has sizable couplings to light quarks
and light charged leptons unless $g_{3}$ is very large, which we
do not expect based on naturalness arguments and also to protect the
extrapolation of the model in the UV, as discussed in Section~\ref{subsec:Couplings_Zprime}.
Consequently, on general grounds we expect a significant production of a
TeV-scale $Z'_{23}$ at the LHC, plus a sizable branching fraction
to electrons and muons. We have prepared the UFO model of $Z'_{23}$
using \texttt{FeynRules} \cite{Feynrules:2013bka}, and then we have
computed the $Z'_{23}$ production cross section for 13 TeV $pp$
collisions using \texttt{Madgraph5} \cite{Madgraph:2014hca} with
the default PDF \texttt{NNPDF23LO}. We estimated analytically the
branching fraction to electrons and muons, and we computed the total
decay width via the narrow width approximation. We confront our results
with the limits from the most recent dilepton resonance searches by
ATLAS \cite{ATLAS:2019erb} and CMS \cite{CMS:2021ctt} in order to
obtain 95\% CL exclusion bounds. The bounds from ditau \cite{ATLAS:2017eiz}
and ditop \cite{ATLAS:2020lks} searches turn out not to be competitive
even for the region of large $g_{3}$, where the bound from the $\rho$
EW parameter is stronger. Our results are depicted as the black-shaded
region in Fig.~\ref{fig:Zp23_plot}. As expected, the bounds become
weaker in the region $g_{3}>1$ where the couplings to light fermions
become mildly suppressed. In the region of small $g_{3}$ we find
the opposite behavior, such that LHC limits can exclude $Z'_{23}$
as heavy as 6-7 TeV. After combining the LHC exclusion with the bounds
coming from the $\rho$ EW parameter, we conclude that we can find
$Z'_{23}$ as light as 3.5 TeV for $g_{3}=1$, while for the benchmark
$g_{i}\simeq\sqrt{3}g_{Y}$ we obtain $M_{Z'_{23}}\apprge5\;\mathrm{TeV}$.
\begin{table}
\centering{}%
\begin{tabular}{cc}
\toprule 
$Z'_{23}$ decay mode & $\mathcal{B}$\tabularnewline
\midrule
\midrule 
$\bar{t}t$ & $\sim0.28$\tabularnewline
$\bar{u}u+\bar{c}c$ & $\sim0.14$\tabularnewline
$\bar{t}c+\bar{c}t$ & $\sim10^{-4}$\tabularnewline
\midrule 
$\bar{b}b$ & $\sim0.08$\tabularnewline
$\bar{d}d+\bar{s}s$ & $\sim0.04$\tabularnewline
$\bar{b}s+\bar{s}b$ & $\sim10^{-4}$\tabularnewline
\midrule 
$\tau^{+}\tau^{-}$ & $\sim0.25$\tabularnewline
$e^{+}e^{-}+\mu^{+}\mu^{-}$ & $\sim0.12$\tabularnewline
$\tau^{+}\mu^{-}+\tau^{-}\mu^{+}$ & $\sim10^{-5}$\tabularnewline
\midrule 
$\bar{\nu}\nu$ & $\sim0.08$\tabularnewline
\bottomrule
\end{tabular}\caption[Main decay modes of TeV-scale $Z'_{23}$ for the natural benchmark]{Main decay modes of $Z'_{23}$ for the natural benchmark $g_{1}\simeq g_{2}\simeq g_{3}\simeq\sqrt{3}g_{Y}$.
We assume that decays into SM singlet neutrinos are kinematically
forbidden or suppressed.\textcolor{orange}{{} \label{tab:Main-decay-modes}}}
\end{table}

Given that $Z'_{23}$ has sizable couplings to electrons, we have
studied the bounds over contact interactions obtained at LEP \cite{Electroweak:2003ram}.
For our model, the most competitive bounds arise from contact interactions
involving only electrons. Assuming vector-like interactions, the bounds
by LEP are only sensitive to regions with very small $g_{3}$, where the couplings of $Z'_{23}$
to electrons (and muons) are larger, but
can exclude $Z'_{23}$ masses beyond 10 TeV. However, we expect this
bound to be slightly overestimated for our model, since the interactions
of $Z'_{23}$ are not exactly vector-like due to the different hypercharge
of $e_{L}$ and $e_{R}$, as depicted in Eq.~(\ref{eq:Z23_couplings}).
Nevertheless, the bounds over chiral operators are much weaker than
the bound over the vector-like operator, and a dedicated reanalysis
of the data would be required in order to obtain the proper bound
for our model, which is beyond the scope of this work. Therefore,
we prefer to be conservative and depict the largest bound of the vector-like
operator as the purple region in Fig.~\ref{fig:Zp23_plot}.

We have also considered implications for $B$-physics. The heavy boson
$Z'_{23}$ has a sizable left-handed $b_{L}s_{L}$ coupling and an
approximately vector-like and universal coupling to electron and muon
pairs. Given these features, a $Z'_{23}$ with a mass of 2 TeV mediates
a meaningful contribution to the effective operator $\mathcal{O}_{9}^{\ell\ell}$
(with $\ell=e,\mu$), where sizable NP contributions are preferred
according to the most recent global fits \cite{Alguero:2023jeh},
without contributing to the SM-like $R_{K^{(*)}}$ ratios \cite{LHCb:2022qnv}.
However, as depicted in Fig.~\ref{fig:Zp23_plot}, the region where
the model could address the anomalies in $b\rightarrow s\mu\mu$ data
are in tension with the bounds obtained by dilepton searches, as expected
for a $Z'$ which has sizable couplings to light quarks. Nevertheless,
we can see that a relevant $C_{9}^{\ell\ell}\sim0.1$ can be obtained
for a heavier $Z'_{23}$ in the region where $g_{3}<0.5$, as in this
region couplings to light fermions are enhanced. 

$\mathcal{B}(B_{s}\rightarrow\mu\mu)$ is also enhanced above the
SM prediction due to both $Z'_{23}$ and $Z$ exchange diagrams. In
the region of small $g_{3}$ the $Z'_{23}$ exchange dominates, while
for large $g_{3}$ the $Z$ exchange dominates. As anticipated before,
the couplings of the $Z'_{23}$ boson to muons are approximately,
but not completely, vector-like. Therefore, a small contribution to
the operator $\mathcal{O}_{10}^{\mu\mu}$ is generated in the region
of small $g_{3}$, where $Z'_{23}$ couplings to muons are larger,
leading to $Z'_{23}$ exchange being dominant. However, in the region
of large $g_{3}$, the flavour-violating coupling $\bar{s}_{L}b_{L}Z'_{23}$
mixes into $\bar{s}_{L}b_{L}Z$ via $Z-Z'_{23}$ mixing. This provides an
effective contribution to $\mathcal{O}_{10}^{\mu\mu}$ mediated by
the $Z$ boson that enhances $\mathcal{B}(B_{s}\rightarrow\mu\mu)$
above the SM prediction. Overall, the region excluded at 95\% CL by
the current HFLAV average \cite{HFLAV:2022wzx} (including the latest
measurement by CMS \cite{CMS:2022mgd}) is depicted as blue-shaded
in Fig.~\ref{fig:Zp23_plot}. It is clear that the resulting bound
is not currently competitive with that from the $\rho$ EW parameter which constrains
the size of $Z-Z'_{23}$ mixing.

The flavour-violating structure of $Z'_{23}$ fermion couplings leads to sizable
contributions to $B_{s}-\bar{B}_{s}$ meson mixing \cite{DiLuzio:2019jyq}
and CLFV processes involving $\tau\rightarrow\mu$ transitions and
$\tau\rightarrow e$ transitions, although well below existing experimental
limits. $\tau\rightarrow\mu(e)$ transitions arise
from mixing angles connected to the flavour hierarchy $\mathcal{O}(m_{2}/m_{3})$
($\mathcal{O}(m_{1}/m_{3})$), see Section~\ref{subsec:Lessons-from-the}.
As an example, in Fig.~\ref{fig:Zp23_plot} we depict the contour
for $\mathcal{B}(\tau\rightarrow3\mu)=10^{-12}$. We find $\tau\rightarrow\mu\gamma$
to be more competitive than $\tau\rightarrow3\mu$ only in the region
$g_{3}>1$.

Beyond indirect detection, in the near future $Z'_{23}$ could be
directly produced at the LHC, HL-LHC and future colliders such as FCC
or a high energy muon collider. The particular pattern of $Z'_{23}$
fermion couplings will allow to disentangle our model from all other proposals.
For the natural benchmark $g_{1}\simeq g_{2}\simeq g_{3}\simeq\sqrt{3}g_{Y}$,
$Z'_{23}$ preferentially decays to top pairs and ditaus, as can be
seen in Table~\ref{tab:Main-decay-modes}. Furthermore, $Z'_{23}$
preferentially couples and decays to \textit{right-handed} charged
fermions, given their larger hypercharge with respect to left-handed
charged fermions, and this prediction may be tested via suitable asymmetry observables.
Similarly, decays to down-type quarks are generally
suppressed with respect to (right-handed) up-type quarks and charged
leptons, given the smaller hypercharge of the former. Due to the modification
of EWPOs, our model can also be tested in an electroweak precision
machine such as FCC-$ee$. An alternative way of discovery would be the
detection of the hyperon scalars breaking the $U(1)_{Y}^{3}$ group
down to SM hypercharge, since the hyperons participating in the 23-breaking may be
as light as the TeV, however we leave a study about the related
phenomenology for future work. In the same spirit, the model naturally
predicts SM singlet neutrinos which could be as light as the TeV scale
(see Section~\ref{subsec:Example-of-seesaw}), with phenomenological
implications yet to be explored in a future work.

\section{Conclusions\label{sec:ConclusionsTri-hypercharge}}

In this chapter we have proposed a tri-hypercharge (TH) embedding of the SM, based on assigning a separate gauged weak hypercharge to each
family. The idea is that each fermion family $i$ only carries hypercharge
under a corresponding $U(1)_{Y_{i}}$ factor. This ensures that each
family transforms differently under the TH gauge group $U(1)_{Y}^{3}$,
which avoids the family repetition of the SM and provides the starting
point for a theory of flavour.

The three family specific hypercharge groups are spontaneously broken
in a cascade symmetry breaking down to SM hypercharge. We have motivated
a particular symmetry breaking pattern, where in a first step $U(1)_{Y_{1}}\times U(1)_{Y_{2}}$
is broken down to its diagonal subgroup at a high scale $v_{12}$.
The remaining group $U(1)_{Y_{1}+Y_{2}}\times U(1)_{Y_{3}}$ is broken
down to SM hypercharge at a scale $v_{23}$. This symmetry breaking
pattern sequentially recovers the accidental flavour symmetry of the
SM, providing protection versus FCNCs that allows the NP scales to be relatively low. Dynamics connected
to the scale $v_{12}$ play a role in the origin of the family hierarchy
$m_{1}/m_{2}$, while dynamics connected to the scale $v_{23}$ play
a role in the origin of $m_{2}/m_{3}$. The hierarchy of scales $v_{23}/v_{12}$
generally plays a role on the origin of flavour hierarchies as well,
although we have found that a mild hierarchy $v_{23}/v_{12}\simeq\lambda$
is enough for specific implementations of the model, where $\lambda\simeq0.225$
is the Wolfenstein parameter.

Assuming that the SM Higgs only carries third family hypercharge,
then only the third family Yukawa couplings are allowed at renormalisable
level. This explains the heaviness of the third family, the smallness
of $V_{cb}$ and $V_{ub}$ quark mixing, and delivers an accidental
$U(2)^{5}$ flavour symmetry in the Yukawa sector acting on the light families, which provides
a reasonable first order description of the SM spectrum. However,
$U(2)^{5}$ does not explain the hierarchical heaviness of the top
quark with respect to the bottom and tau fermions. Furthermore, we have proven that the model generates a similar mass hierarchy for all charged sectors, being unable to explain the heaviness of the charm quark with respect to the strange and muon without small couplings. We have motivated the addition of a
second Higgs doublet as a natural and elegant solution, which allows
a more natural description of the hierarchies between the different
charged fermion sectors.

We have explored the capabilities of the $U(1)_{Y}^{3}$ model to
explain the observed hierarchies in the charged fermion
sector, via the addition of non-renormalisable operators containing
$U(1)_{Y}^{3}$-breaking scalars which act as small breaking effects
of $U(2)^{5}$. After extracting model-independent considerations
from the spurion formalism, we have presented example models where
all charged fermion masses and mixings are addressed. Following a
similar methodology, we have studied the origin of neutrino masses
and mixing in the TH model. We have shown that due to the $U(1)_{Y}^{3}$
gauge symmetry, the implementation of a type I seesaw mechanism naturally
leads to a low scale seesaw, where the SM singlet neutrinos in the
model may be as light as the TeV scale. We have provided an example
model compatible with the observed pattern of neutrino mixing.

As usual for theories of flavour, the NP scales $\left\langle \phi\right\rangle $
and $\Lambda$ that explain the origin of flavour hierarchies in the
SM may be anywhere \textit{from the Planck scale to the electroweak
scale}, provided that the ratios $\left\langle \phi\right\rangle /\Lambda$
are held fixed (see Section~\ref{subsec:FromPlanckToEW}). Intriguingly,
a preliminary phenomenological analysis shows that current data allows
relatively low NP scales. The heavy gauge boson $Z'_{12}$ arising
from the 12-breaking displays completely flavour non-universal couplings
to fermions, and generally contributes to $\Delta F=2$ and CLFV processes.
The size of the most dangerous contributions are however model-dependent.
In selected specific models provided in this chapter, we have found
that the most dangerous contributions to $K-\bar{K}$ mixing
and $\mu\rightarrow e\gamma$ are strongly suppressed, allowing for
$Z'_{12}$ to be as light as 10-50 TeV. Therefore, the lightest gauge
boson $Z'_{23}$ arising from the 23-breaking can live at the TeV
scale, within the reach of LHC and future colliders, since $Z'_{23}$
avoids bounds from $K-\bar{K}$ mixing and $\mu\rightarrow e\gamma$
thanks to an accidental GIM mechanism for light fermions.

We find the gauge boson $Z'_{23}$ to have a rich low-energy phenomenology:
mixing with the SM $Z$ boson leads to implications for the $W$ boson
mass and EWPOs, plus we expect contributions to flavour-violating
processes involving the third family, such as $\tau\rightarrow3\mu(e)$
and $B_{s}-\bar{B}_{s}$ meson mixing. After our preliminary analysis,
we find that current data allows $Z'_{23}$ to be as light as 3-4
TeV in some regions of the parameter space, the strongest bounds coming
from dilepton searches at LHC along with the contribution to the $\rho$
EW parameter. In the case of discovery, the particular pattern of
$Z'_{23}$ couplings and decays to fermions will allow to disentangle
our model from all other proposals. However, most of the phenomenological
consequences are yet to be explored in detail: a global fit to EWPOs
and flavour observables will allow to properly confront our model
versus current data. An alternative way of discovery would be the
detection of the Higgs scalars (hyperons) breaking the $U(1)_{Y}^{3}$
down to SM hypercharge, however we leave a discussion about the related
phenomenology for future work. In the same spirit, the model naturally
predicts SM singlet neutrinos which could be as light as the TeV,
with phenomenological implications yet to be explored. The tri-hypercharge
gauge group may be the first step towards understanding the origin
of three fermion families in Nature, the hierarchical charged fermion masses
and CKM mixing, revealing the existence of a flavour non-universal
gauge structure encoded in Nature at energies above the electroweak
scale. 

Despite the apparent complexity of the gauge sector of the tri-hypercharge gauge group,
consisting of five arbitrary gauge couplings, in Chapter~\ref{Chapter:Tri-unification} we 
shall see that such a theory may arise from a gauge unified framework.
%% ----------------------------------------------------------------
%% Triunification.tex
%% ---------------------------------------------------------------- 
\chapter{Tri-unification: the origin of gauge non-universal theories of flavour} \label{Chapter:Tri-unification}

\begin{quote}
    ``As a child, I considered such unknowns sinister.\\
    Now, though, I understand they bear no ill will.\\ 
    The Universe is, and we are.''
  \begin{flushright}
  \hfill \hfill $-$ Solanum in \textit{Outer Wilds}
  \par\end{flushright}
\end{quote}

\noindent In this chapter, based on Ref.~\cite{FernandezNavarro:2023hrf},
we propose a grand unified framework consisting on assigning a separate
$SU(5)$ to each fermion family. The three $SU(5)$ groups are related
by a cyclic permutation symmetry $\mathbb{Z}_{3}$, such that the model is described by
a single gauge coupling in the UV, despite $SU(5)^{3}$ being a non-simple
gauge group. We motivate that such a tri-unification framework may
embed gauge non-universal theories, such as the tri-hypercharge model
proposed in the previous chapter to address the origin of flavour.
In this manner, we discuss a minimal tri-hypercharge example which
can account for all the quark and lepton (including neutrino) masses
and mixing parameters, with the many gauge couplings of the tri-hypercharge
group unifying at the GUT scale into a single gauge coupling associated
to the cyclic $SU(5)^{3}$ group.

\section{Introduction}

Flavour non-universal gauge embeddings of the SM gauge group were
first proposed during the 80s and 90s. Back then, it was already highlighted
that these frameworks may have many applications, such as motivating
lepton non-universality \cite{Li:1981nk,Ma:1987ds,Ma:1988dn,Li:1992fi}
or assisting technicolor model building \cite{Hill:1994hp,Muller:1996dj}.
Other applications include addressing tensions with the SM in electroweak
precision data \cite{Malkawi:1996fs}, or explaining the baryon asymmetry
of the Universe due to the presence of non-universal $SU(2)$ factors
embedding the usual $SU(2)_{L}$ \cite{Shu:2006mm}. More recently,
such non-universal gauge structures had been shown to explain the
flavour structure of the SM if spontaneously broken in a desired way.
Most notable examples include the $SU(2)_{L}^{3}$ model \cite{Chiang:2009kb,Davighi:2023xqn,Capdevila:2024gki},
the $SU(3)_{c}^{3}$ model (which is only able to explain the smallness
of quark mixing) \cite{Carone:1995ge} and the tri-hypercharge model
$U(1)_{Y}^{3}$ \cite{FernandezNavarro:2023rhv} discussed in Chapter~\ref{Chapter:Tri-hypercharge}.

However, these theories explain the flavour structure of the SM at
the price of complicating the gauge sector with the introduction of
extra, arbitrary gauge couplings. The existing UV completions are
all based on variations of the Pati-Salam gauge group \cite{Bordone:2017bld,Allwicher:2020esa,Fuentes-Martin:2020pww,Fuentes-Martin:2022xnb,Davighi:2022bqf,Davighi:2022fer,Davighi:2023iks},
leaving up to nine arbitrary gauge couplings in the UV. In contrast,
here we shall attempt to construct a gauge unified framework from
which gauge non-universal theories may emerge at relatively low-energies,
opening the possibility to build consistent non-universal descriptions
of Nature that are valid all the way up to the scale of grand unification.
For this we propose a non-supersymmetric $SU(5)^{3}$ framework,
\begin{equation}
SU(5)_{1}\times SU(5)_{2}\times SU(5)_{3}\,,
\end{equation}
together with a cyclic symmetry $\mathbb{Z}_{3}$ that relates the three $SU(5)$
factors. This is a generalisation of $SU(5)$ grand unification \cite{Georgi:1974sy}
in which we assign a separate $SU(5)$ group to each fermion family.
The cyclic symmetry that relates the three $SU(5)$ factors ensures
that at the GUT scale the three gauge couplings are equal, such that
the gauge sector is fundamentally described by one gauge coupling.
Therefore, although $SU(5)^{3}$ is not a simple group, it may be
regarded as a unified gauge theory. Gauge unified theories including
different gauge factors for each fermion family were first sketched
in the early days of the GUT program \cite{Salam:1979p,Rajpoot:1980ib,Georgi:1981gj},
and were later considered in supersymmetric scenarios\footnote{However it is worth mentioning that supersymmetric $SU(5)^{3}$ scenarios
suffer from rapid proton decay via dimension-5 operators generated
by coloured triplet Higgs exchange \cite{Barbieri:1994cx,Babu:2007mb},
which have large couplings (of order unity) to first generation fermions
as a consequence of the discrete symmetry (e.g.~cyclic symmetry)
that enforces a single gauge coupling at the GUT scale.} during the 90s and 2000s, where the main motivations were achieving
GUT symmetry breaking without adjoint fields \cite{Barbieri:1994cx,Chou:1998pra},
solving the doublet-triplet splitting problem \cite{Asaka:2004ry},
or unifying all chiral fermions into the same representation \cite{Babu:2007mb}.

In contrast with the previous work, here we propose our non-supersymmetric
$SU(5)^{3}$ tri-unification framework as a realistic origin for gauge
non-universal physics at lower scales, that had been shown to have
many applications for model building purposes as described before.
As a proof of concept, in this chapter we will discuss $SU(5)^{3}$
as an embedding of the tri-hypercharge model \cite{FernandezNavarro:2023rhv},
showing that it is possible to unify the various gauge couplings of
tri-hypercharge into a single gauge coupling associated with the cyclic
$SU(5)^{3}$ gauge group. We also study proton decay in this example,
and present the predictions of the proton lifetime in the dominant
$e^{+}\pi^{0}$ channel.

The layout of the remainder of the chapter is as follows. In Section~\ref{sec:General-framework}
we discuss a general $SU(5)^{3}$ tri-unification framework for model building. In
rather lengthy Section~\ref{sec:Tri-Hypercharge-example} we analyse
an example $SU(5)^{3}$ unification model breaking to tri-hypercharge,
including the charged fermion mass hierarchies and quark mixing, neutrino
masses and mixing, gauge coupling unification and proton decay. Section
\ref{sec:conclusions} concludes the chapter.

\section{General \texorpdfstring{$SU(5)^{3}$}{tri-unification}
framework for model building\label{sec:General-framework}}
\begin{table}[t]
\centering{}%
\begin{tabular}{cccc}
\toprule 
\textbf{Field}  & $\boldsymbol{SU(5)_{1}}$  & $\boldsymbol{SU(5)_{2}}$  & $\boldsymbol{SU(5)_{3}}$\tabularnewline
\midrule 
$F_{1}$  & \fiveS  & \one  & \one\tabularnewline
$F_{2}$  & \one  & \fiveS  & \one\tabularnewline
$F_{3}$  & \one  & \one  & \fiveS\tabularnewline
$T_{1}$  & \ten  & \one  & \one\tabularnewline
$T_{2}$  & \one  & \ten  & \one\tabularnewline
$T_{3}$  & \one  & \one  & \ten\tabularnewline
\midrule 
$\Omega_{12}$ & \24 & \24 & \one\tabularnewline
$\Omega_{13}$ & \24 & \one & \24\tabularnewline
$\Omega_{23}$ & \one & \24 & \24\tabularnewline
\midrule 
$H_{1}$  & \five  & \one  & \one\tabularnewline
$H_{2}$  & \one  & \five  & \one\tabularnewline
$H_{3}$  & \one  & \one  & \five\tabularnewline
\bottomrule
\end{tabular}\caption[Minimal content for the general $SU(5)^{3}$ tri-unification framework]{Minimal content for the general $SU(5)^{3}$ tri-unification framework.
Due to the cyclic symmetry, there are only four independent representations,
one for each of the fermions $F_{i},T_{i}$ and one for each of the
scalars $\Omega_{ij},H_{i}$. \label{tab:ParticleContent_General}}
\end{table}
The basic idea of tri-unification is to embed the SM gauge group into
a semi-simple gauge group containing three $SU(5)$ factors, 
\begin{equation}
SU(5)_{1}\times SU(5)_{2}\times SU(5)_{3}\,,\label{eq:SU(5)^3-1}
\end{equation}
where each $SU(5)$ factor is associated to one family of chiral fermions
$i=1,2,3$. Moreover, we incorporate a cyclic permutation symmetry
$\mathbb{Z}_{3}$ that relates the three $SU(5)$ factors, in the
spirit of the trinification model \cite{Glashow:1984gc}. This implies
that at the high energy GUT scale where $SU(5)^{3}$ is broken (typically
in excess of $10^{16}$ GeV) the gauge couplings of the three $SU(5)$
factors are equal by cyclic symmetry, such that the gauge sector is
fundamentally described by one gauge coupling. Therefore, although
$SU(5)^{3}$ is not a simple group, it may be regarded as a unified
gauge theory. Moreover, the cyclic symmetry also ensures that all SM fermions
belong to a single irreducible representation of the complete symmetry group.

The motivation for considering such an $SU(5)^{3}$ with cyclic symmetry
is that it allows gauge non-universal theories of flavour to emerge
at low energies\footnote{$SU(5)^{3}$ tri-unification may provide a unified origin for many
gauge non-universal theories proposed in the literature to address
different questions beyond the flavour puzzle, see e.g.~\cite{Malkawi:1996fs,Shu:2006mm}.} from a gauge universal theory, depending on the symmetry breaking
chain. In the first step, $SU(5)^{3}$ may be\footnote{This first step of symmetry breaking is optional, but may be convenient
to control the scale of gauge unification as discussed in Section~\ref{sec:gcu}.} broken to three copies of the SM gauge group SM$^{3}$. Then at lower
energies, SM$^{3}$ is broken to some universal piece $G_{\mathrm{universal}}$
consisting of some diagonal subgroups, together with some remaining
family groups $G_{1}\times G_{2}\times G_{3}$. If the light Higgs
doublet(s) transform non-trivially under the third family group $G_{3}$,
but not under the first nor second, then third family fermions get
natural masses at the electroweak scale, while first and second family
fermions are massless in first approximation. Their small masses naturally
arise from the breaking of the non-universal gauge group down to the
SM, which is the diagonal subgroup, and an approximate $U(2)^{5}$
flavour symmetry emerges, which is known to provide an efficient suppression
of the most dangerous flavour-violating effects for new physics \cite{Barbieri:2011ci,Allwicher:2023shc}.

At still lower energies, the non-diagonal group factors $G_{1}\times G_{2}\times G_{3}$
are broken down to their diagonal subgroup, eventually leading to
a flavour universal SM gauge group factor. This may happen in stages.
In particular, the symmetry breaking pattern 
\begin{equation}
G_{1}\times G_{2}\times G_{3}\rightarrow G_{1+2}\times G_{3}\rightarrow\mathrm{G_{1+2+3}}
\end{equation}
may naturally explain the origin of fermion mass hierarchies and the
smallness of quark mixing, while anarchic neutrino mixing may be incorporated
via variations of the type I seesaw mechanism \cite{FernandezNavarro:2023rhv,Fuentes-Martin:2020pww}.

Minimal examples of this class of theories include the tri-hypercharge
model \cite{FernandezNavarro:2023rhv}, already discussed in Chapter~\ref{Chapter:Tri-hypercharge},
where the universal (diagonal) group consists of the non-Abelian SM
gauge group factors $G_{\mathrm{universal}}=SU(3)_{c}\times SU(2)_{L}$
while the remaining groups are the three gauge weak hypercharge factors
$G_{1}\times G_{2}\times G_{3}=U(1)_{Y_{1}}\times U(1)_{Y_{2}}\times U(1)_{Y_{3}}$.
Another example is the $SU(2)_{L}^{3}$ model \cite{Li:1981nk,Ma:1987ds,Ma:1988dn,Li:1992fi,Muller:1996dj,Chiang:2009kb,Allwicher:2020esa,Davighi:2023xqn,Capdevila:2024gki},
where $G_{\mathrm{universal}}=SU(3)_{c}\times U(1)_{Y}$ and $G_{1}\times G_{2}\times G_{3}=SU(2)_{L1}\times SU(2)_{L2}\times SU(2)_{L3}$.
There also exists the $SU(3)_{c}^{3}$ model \cite{Carone:1995ge}
(which is only able to explain the smallness of quark mixing), where
$G_{\mathrm{universal}}=SU(2)_{L}\times U(1)_{Y}$ and $G_{1}\times G_{2}\times G_{3}=SU(3)_{c1}\times SU(3)_{c2}\times SU(3)_{c3}$.
Variations of these theories have been proposed in recent years, several
of them assuming a possible embedding into (variations of) a Pati-Salam
setup \cite{Bordone:2017bld,Fuentes-Martin:2020pww,Fuentes-Martin:2022xnb,Davighi:2022bqf,Davighi:2022fer,Davighi:2023iks}.

All these theories share a common feature: they explain the origin
of the flavour structure of the SM at the price of complicating the
gauge sector, which may now contain up to nine arbitrary gauge couplings.
We will motivate that $SU(5)^{3}$ as the embedding of general theories
$G_{\mathrm{universal}}\times G_{1}\times G_{2}\times G_{3}$ resolves
this issue, by unifying the complicated gauge sector of these theories
into a single gauge coupling. The main ingredients of our general
setup are as follows: 
\begin{itemize}
\item The presence of the $\mathbb{Z}_{3}$ symmetry, which is of fundamental
importance to achieve gauge unification, imposes that the matter content
of the model shall be invariant under cyclic permutations of the three
$SU(5)$ factors. This enforces that each $SU(5)$ factor contains
the same representations of fermions and scalars, i.e.~if the representation
$(\mathbf{A,B,C})$ is included, then $(\mathbf{C,A,B})$ and $(\mathbf{B,C,A})$
must be included too. 
\item Each family of chiral fermions $i$ is embedded in the usual way into
$\mathbf{\overline{5}}$ and $\mathbf{10}$ representations of their
corresponding $SU(5)_{i}$ factor, that we denote as $F_{i}=(d_{i}^{c},\ell_{i})\sim\mathbf{\overline{5}}_{i}$
and $T_{i}=(q_{i},u_{i}^{c},e_{i}^{c})\sim\mathbf{10}_{i}$ as shown
in Table~\ref{tab:ParticleContent_General}. This choice is naturally
consistent with the $\mathbb{Z}_{3}$ symmetry. 
\item In a similar manner, three Higgs doublets $H_{1}$, $H_{2}$ and $H_{3}$
are embedded into $\mathbf{5}$ representations, one for
each $SU(5)_{i}$ factor. Notice that in non-universal theories of
flavour it is commonly assumed the existence of only one Higgs doublet
$H_{3}$, which transforms only under the third site in order to explain
the heaviness of the third family. This way, the $SU(5)^{3}$ framework
involves the restriction of having three Higgses rather than only
one, but we will argue that if the $\mathbb{Z}_{3}$ symmetry is broken
below the GUT scale, then only the third family Higgs $H_{3}$ may
be light and perform electroweak symmetry breaking, while $H_{1}$
and $H_{2}$ are heavier and may play the role of heavy messengers
for the effective Yukawa couplings of the light families. 
\item Higgs scalars in bi-representations connecting the different sites
may be needed to generate the SM flavour structure at the level of
the $G_{\mathrm{universal}}\times G_{1}\times G_{2}\times G_{3}$
theory, e.g.~$(\mathbf{2,\overline{2}})$ scalars in $SU(2)_{L}^{3}$
or $(Y,-Y)$ scalars in tri-hypercharge (the so-called hyperons).
These can be embedded in the associated bi-representations of $SU(5)^{3}$,
e.g.~$(\mathbf{5,\overline{5}})$ scalars, $(\mathbf{10,\overline{10}})$
scalars and so on. In Appendix~\ref{app:Hyperons} we tabulate all
such scalars from $SU(5)^{3}$ representations with dimension up to
$\mathbf{45}$, along with the hyperons that they generate at low
energies.
\item Finally, three scalar fields in bi-adjoint representations of each $SU(5)$,
$\Omega_{ij}$, spontaneously break the tri-unification symmetry. The three 
$\Omega_{ij}$ are enough to perform \textit{both} horizontal and vertical breaking of the three $SU(5)$ groups at the GUT scale, down to 
the non-universal gauge group $G_{\mathrm{universal}}\times G_{1}\times G_{2}\times G_{3}$ of choice
that later explains the flavour structure of the SM (e.g.~tri-hypercharge
or $SU(2)_{L}^{3}$). Another possibility that we will explore is
breaking $SU(5)^{3}$ first to three copies of the SM (one for each
family) and then to $G_{\mathrm{universal}}\times G_{1}\times G_{2}\times G_{3}$
in a second step. 
\end{itemize}
To summarise, the general pattern of symmetry breaking we assume is
as follows\footnote{One should note that none of the individual groups, $SU(3)$, $SU(2)$
or $U(1)$, in each SM$_{i}$ group correspond to the SM's $SU(3)_{c}$,
$SU(2)_{L}$ or $U(1)_{Y}$. The latter emerge after symmetry breaking
from the diagonal sub-groups of the former. Nevertheless, we will
denote each $\left(SU(3)\times SU(2)\times U(1)\right)_{i}$ as SM$_{i}$
and the total $\left(SU(3)\times SU(2)\times U(1)\right)^{3}$ group
as SM$^{3}$ for the sake of brevity.}, 
\begin{flalign}
SU(5)^{3} & \rightarrow\mathrm{SM_{1}}\times\mathrm{SM_{2}}\times\mathrm{SM_{3}}\\
 & \rightarrow G_{\mathrm{universal}}\times G_{1}\times G_{2}\times G_{3}\\
 & \rightarrow G_{\mathrm{universal}}\times G_{1+2}\times G_{3}\\
 & \rightarrow\mathrm{SM_{1+2+3}}\,,
\end{flalign}
where the $\mathrm{SM}^{3}$ step is optional but may be convenient
to achieve unification. In particular, the first step of symmetry
breaking makes use of three SM singlets contained in $\Omega_{ij}$,
while the second step may be performed via the remaining degrees of
freedom in the $\Omega_{ij}$, depending on the details of the low energy gauge
theory that survives. The two final breaking steps are performed by
Higgs scalars connecting the different sites that need to be specified
for each particular model.

Beyond the general considerations listed in this section, when building
a specific model one needs to choose the symmetry group $G_{\mathrm{universal}}\times G_{1}\times G_{2}\times G_{3}$,
and add explicit scalars and/or fermion messengers that mediate the
effective Yukawa couplings of light fermions.

Finally, one needs to study the Renormalization Group Equations (RGEs)
of the various gauge couplings at the different steps all the way
up to the $SU(5)^{3}$ scale where all gauge couplings need to unify.
This is not a simple task, but we shall see that the relatively light
messengers required to generate the effective Yukawa couplings, along
with the presence of the approximate $\mathbb{Z}_{3}$ symmetry at
low energies, may naturally help to achieve unification. In the following,
we shall illustrate this by describing a working example of the $SU(5)^{3}$
framework based on tri-hypercharge (see Chapter~\ref{Chapter:Tri-hypercharge}), where the various
gauge couplings of the tri-hypercharge model unify at the GUT scale
into a single gauge coupling.

\section{An example \texorpdfstring{$SU(5)^{3}$}{SU(5){3}}
unification model breaking to tri-hypercharge\label{sec:Tri-Hypercharge-example}}

In the following we discuss an example of the tri-hypercharge model, i.e.~$G_{\mathrm{universal}}=SU(3)_{c}\times SU(2)_{L}$
and $G_{1}\times G_{2}\times G_{3}=U(1)_{Y_{1}}\times U(1)_{Y_{2}}\times U(1)_{Y_{3}}$ (see Chapter~\ref{Chapter:Tri-hypercharge}), originating from the 
$SU(5)^{3}$ tri-unification framework.
In this example, the basic idea is that $SU(5)^{3}$ breaks, via a
sequence of scales, to the low energy (well below the GUT scale) tri-hypercharge
gauge group with a separate gauged weak hypercharge for each fermion
family, 
\begin{table}[htp]
\centering{}%
\begin{tabular}{cccc}
\toprule 
\textbf{Field}  & $\boldsymbol{SU(5)_{1}}$  & $\boldsymbol{SU(5)_{2}}$  & $\boldsymbol{SU(5)_{3}}$\tabularnewline
\midrule 
$F_{1}$  & \fiveS  & \one  & \one\tabularnewline
$F_{2}$  & \one  & \fiveS  & \one\tabularnewline
$F_{3}$  & \one  & \one  & \fiveS\tabularnewline
$T_{1}$  & \ten  & \one  & \one\tabularnewline
$T_{2}$  & \one  & \ten  & \one\tabularnewline
$T_{3}$  & \one  & \one  & \ten\tabularnewline
\midrule 
\rowcolor{yellow!10} $\chi_{1}$  & \ten  & \one  & \one\tabularnewline
\rowcolor{yellow!10} $\chi_{2}$  & \one  & \ten  & \one\tabularnewline
\rowcolor{yellow!10} $\chi_{3}$  & \one  & \one  & \ten\tabularnewline
\midrule 
\rowcolor{yellow!10} $\Xi_{0}$  & \one  & \one  & \one\tabularnewline
\rowcolor{yellow!10} $\Xi_{12}$  & \five  & \fiveS  & \one\tabularnewline
\rowcolor{yellow!10} $\Xi_{13}$  & \fiveS  & \one  & \five\tabularnewline
\rowcolor{yellow!10} $\Xi_{23}$  & \one  & \five  & \fiveS\tabularnewline
\midrule 
\rowcolor{yellow!10} $\Sigma_{\mathrm{atm}}$  & \one  & \ten  & \tenS \tabularnewline
\rowcolor{yellow!10} $\Sigma_{\mathrm{sol}}$  & \ten  & \one  & \tenS \tabularnewline
\rowcolor{yellow!10} $\Sigma_{\mathrm{cyclic}}$  & \ten  & \tenS  & \one \tabularnewline
\midrule 
$\Omega_{12}$ & \24 & \24 & \one\tabularnewline
$\Omega_{13}$ & \24 & \one & \24\tabularnewline
$\Omega_{23}$ & \one & \24 & \24\tabularnewline
\midrule 
$H_{1}^{u}$  & \five  & \one  & \one\tabularnewline
$H_{2}^{u}$  & \one  & \five  & \one\tabularnewline
$H_{3}^{u}$  & \one  & \one  & \five\tabularnewline
$H_{1}^{\mathbf{\overline{5}}}$  & \fiveS  & \one  & \one\tabularnewline
$H_{2}^{\mathbf{\overline{5}}}$  & \one  & \fiveS  & \one\tabularnewline
$H_{3}^{\mathbf{\overline{5}}}$  & \one  & \one  & \fiveS\tabularnewline
$H_{1}^{\mathbf{45}}$  & $\mathbf{45}$  & \one  & \one\tabularnewline
$H_{2}^{\mathbf{45}}$  & \one  & $\mathbf{45}$  & \one\tabularnewline
$H_{3}^{\mathbf{45}}$  & \one  & \one  & $\mathbf{45}$\tabularnewline
\midrule 
$\Phi_{12}^{F}$  & \five  & \fiveS  & \one\tabularnewline
$\Phi_{13}^{F}$  & \fiveS  & \one  & \five\tabularnewline
$\Phi_{23}^{F}$  & \one  & \five  & \fiveS\tabularnewline
$\Phi_{12}^{T}$  & \tenS  & \ten  & \one\tabularnewline
$\Phi_{13}^{T}$  & \ten  & \one  & \tenS\tabularnewline
$\Phi_{23}^{T}$  & \one  & \tenS  & \ten\tabularnewline
\midrule 
$\Phi_{12}^{\mathbf{45}}$  & \one  & $\mathbf{\overline{45}}$  & $\mathbf{45}$ \tabularnewline
$\Phi_{13}^{\mathbf{45}}$  & $\mathbf{\overline{45}}$  & \one  & $\mathbf{45}$ \tabularnewline
$\Phi_{12}^{\mathbf{45}}$  & $\mathbf{\overline{45}}$  & $\mathbf{45}$  & \one \tabularnewline
$\Phi^{TFT}$  & \ten  & \five  & \ten \tabularnewline
$\Phi^{FTT}$  & \five  & \ten  & \ten \tabularnewline
$\Phi^{TTF}$  & \ten  & \ten  & \five \tabularnewline
\bottomrule
\end{tabular}\caption[Fermion and scalar particle content and representations for the example
$SU(5)^{3}$ tri-unification model]{Fermion and scalar particle content and representations under $SU(5)^{3}$.
$F_{i}$ and $T_{i}$ include the chiral fermions of the SM in the
usual way, while $\chi_{i}$, $\xi$'s and $\Xi$'s (highlighted in
yellow) are vector-like fermions, thus the conjugate partners must
be considered. $\Omega$'s, $H$'s and $\Phi$'s are scalars. \label{tab:ParticleContent_model}}
\end{table}
\begin{equation}
SU(5)^{3}\rightarrow\dots\rightarrow SU(3)_{c}\times SU(2)_{L}\times U(1)_{Y_{1}}\times U(1)_{Y_{2}}\times U(1)_{Y_{3}}\,.\label{GUTbreaking}
\end{equation}
In Chapter~\ref{Chapter:Tri-hypercharge} it was shown that the low energy tri-hypercharge model
can naturally generate the flavour structure of the SM if spontaneously
broken to SM hypercharge in a convenient way. The minimal setup involves
the vacuum expectation values (VEVs) of the new Higgs ``hyperons''
\begin{equation}
\phi_{q12}\sim(\mathbf{1,1})_{(-1/6,1/6,0)},\qquad\phi_{q23}\sim(\mathbf{1,1})_{(0,-1/6,1/6)},\qquad\phi_{\ell23}\sim(\mathbf{1,1})_{(0,1/2,-1/2)}\,.
\end{equation}
At the GUT scale, the hyperons are embedded into bi-$\mathbf{\overline{5}}$
and bi-$\mathbf{10}$ representations of $SU(5)^{3}$ expressed as
$\Phi_{ij}^{T,F}$, which must preserve the cyclic symmetry, as shown
in Table~\ref{tab:ParticleContent_model}. Although this involves
the appearance of many hyperons (and other scalars) beyond the minimal
set of hyperons that we need, we shall assume that only the desired
hyperons get a VEV (and the rest of scalars may remain very heavy).
Moreover, the $SU(5)^{3}$ framework also poses constraints on the
possible family hypercharges of the hyperons, as collected in Appendix~\ref{app:Hyperons}.
For the $SU(5)^{3}$ setup, it is convenient to add 
\begin{equation}
\phi_{q13}\sim(\mathbf{1,1})_{(-1/6,0,1/6)}\,,\qquad\phi_{\ell13}\sim(\mathbf{1,1})_{(1/2,0,-1/2)}\,,
\end{equation}
which are anyway required by the cyclic symmetry, to the set of hyperons
which get a VEV.

The hyperons allow to write a set of non-renormalisable operators
that provide effective Yukawa couplings for light fermions, as described
in Chapter~\ref{Chapter:Tri-hypercharge} by working in an EFT framework. However, in our unified
model, we need to introduce heavy messengers that mediate such effective
operators in order to obtain a UV complete setup. For this, we add
one set of vector-like fermions transforming in the $\mathbf{10}$
representation for each $SU(5)$ factor, i.e.~$\chi_{i}\sim\mathbf{10}_{i}$
and $\overline{\chi}_{i}\sim\overline{\mathbf{10}}_{i}$. We shall
assume that only the quark doublets $Q_{i}\sim(\mathbf{3,2})_{1/6_{i}}$
and $\overline{Q}_{i}\sim(\mathbf{\overline{3},2})_{-1/6_{i}}$ are
relatively light and play a role in the effective Yukawa couplings,
while the remaining degrees of freedom in $\chi_{i}$ and $\overline{\chi}_{i}$
remain very heavy, 
\begin{equation}
\chi_{i}\sim\mathbf{10}_{i}\rightarrow Q_{i}\sim(\mathbf{3,2})_{1/6_{i}}\,,\qquad\overline{\chi}_{i}\sim\overline{\mathbf{10}}_{i}\rightarrow\overline{Q}_{i}\sim(\mathbf{\overline{3},2})_{-1/6_{i}}\,.
\end{equation}
We shall see that $Q_{i}$ and $\overline{Q}_{i}$ also contribute
to the RGEs in the desired way to achieve gauge unification. The full
field content of this model also includes extra vector-like fermions
$\Sigma$ and $\Xi$ as shown in Table~\ref{tab:ParticleContent_model}.
These play a role in the origin of neutrino masses as discussed in
Section~\ref{subsec:Neutrinos}.

Finally, beyond the minimal set of Higgs doublets introduced in Section~\ref{sec:General-framework},
we shall introduce here three pairs of $\mathbf{5}$, $\overline{\mathbf{5}}$
and $\mathbf{45}$ Higgs representations preserving the cyclic symmetry.
The doublets in the $\overline{\mathbf{5}}$ and $\mathbf{45}$ mix,
leaving light linear combinations that couple differently to down-quarks
and charged leptons in the usual way \cite{Georgi:1979df}, which
we denote as $H_{i}^{d}$.

Therefore, below the GUT scale we effectively have three pairs of
Higgs doublets $H_{1}^{u,d}$, $H_{2}^{u,d}$ and $H_{3}^{u,d}$,
such that the $u$- and $d$- labeled Higgs only couple to up-quarks
(and neutrinos) and to down-quarks and charged leptons, respectively,
in the spirit of the type II 2HDM. This choice is motivated to explain
the mass hierarchies between the different charged sectors, as originally
identified in Chapter~\ref{Chapter:Tri-hypercharge}, and could be enforced e.g.~by a $\mathbb{Z}_{2}$
discrete symmetry. We assume that the third family Higgs $H_{3}^{u,d}$
are the lightest, they perform electroweak symmetry breaking and provide
Yukawa couplings for the third family with $\mathcal{O}(1)$ coefficients
if $\tan\beta\approx20$. In contrast, we assume that the Higgs $H_{1}^{u,d}$,
$H_{2}^{u,d}$ have masses above the TeV (but much below the GUT scale)
and act as messengers of the effective Yukawa couplings for the light
families.

\begin{figure}
\includegraphics[scale=0.80]{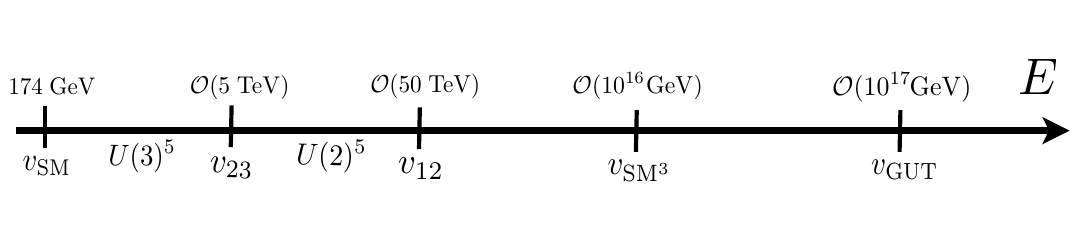}
\caption[Diagram showing the different scales of spontaneous symmetry breaking in the example $SU(3)^{5}$ tri-unification model]{Diagram showing the different scales of spontaneous symmetry breaking in our example model (see also Eqs.(\ref{eq:GUT_breaking}-\ref{eq:SM})),
along with the accidental, approximate flavour symmetries ($U(3)^{5}$ and $U(2)^{5}$) that arise at low energies.}
\label{fig:scales}

\end{figure}

In detail, we assume that the $SU(5)^{3}$ group is broken down to
the SM through the following symmetry breaking chain 
\begin{flalign}
SU(5)^{3} & \xrightarrow{v_{\mathrm{GUT}}}\mathrm{SM_{1}}\times\mathrm{SM_{2}}\times\mathrm{SM_{3}}\label{eq:GUT_breaking}\\
 & \xrightarrow{v_{{\rm SM^{3}}}}SU(3)_{1+2+3}\times SU(2)_{1+2+3}\times U(1)_{1}\times U(1)_{2}\times U(1)_{3}\\
 & \xrightarrow{v_{12}}SU(3)_{1+2+3}\times SU(2)_{1+2+3}\times U(1)_{1+2}\times U(1)_{3}\label{eq:23_breaking}\\
 & \xrightarrow{v_{23}}SU(3)_{1+2+3}\times SU(2)_{1+2+3}\times U(1)_{1+2+3}\,.\label{eq:SM}
\end{flalign}
The $SU(5)^{3}$ breaking happens at the GUT scale, while the tri-hypercharge
breaking may happen as low as the TeV scale, as allowed by current
data (see Section~\ref{sec:Phenomenology}), while the $\mathrm{SM}^{3}$ breaking step
is optional but may be convenient to achieve unification, and may
be regarded as free parameter. This second breaking step is performed
by the $SU(3)_{i}$ octets and $SU(2)_{i}$ triplets contained in
$\Omega_{ij}$. See also Fig.~\ref{fig:scales}
for an illustrative diagram.

We shall show that within this setup, achieving gauge unification
just requires further assuming that three colour octets that live
in $\Omega_{ij}$ are light, while the remaining
degrees of freedom of the bi-adjoints remain very heavy. Before that,
we shall study in detail how our model explains the origin of the
flavour structure of the SM. 

\subsection{Charged fermion mass hierarchies and quark mixing \label{subsec:Charged_fermions}}

In this section we shall discuss the origin of charged fermion mass
hierarchies and quark mixing at the tri-hypercharge layer. In contrast
with Chapter~3, here we shall specify the UV origin of the effective
Yukawa couplings as we intend to build a UV-complete GUT model. In
this manner, we will show a much more complete framework for the generation
of the SM flavour structure than in Chapter~\ref{Chapter:Tri-hypercharge}, where an EFT framework
with a common cut-off scale for all effective Yukawa couplings was
considered.

The Higgs doublets in the cyclic $\mathbf{\overline{5}}$ and $\mathbf{45}$
split the couplings of down-quarks and charged leptons in the usual
way \cite{Georgi:1979df}. We denote as $H_{i}^{d}$ the linear combinations
that remain light, with their effective couplings to down-quarks and
charged leptons given by 
\begin{equation}
\widetilde{y}_{i}^{d}H_{i}^{d}T_{i}F_{i}\rightarrow y_{i}^{d}H_{i}^{d}q_{i}d_{i}^{c}+y_{i}^{e}H_{i}^{d}\ell_{i}e_{i}^{c},
\end{equation}
where 
\begin{equation}
y_{i}^{d}=y_{i}^{\mathbf{\overline{5}}}+y_{i}^{\mathbf{\overline{45}}}\,,\qquad y_{i}^{e}=y_{i}^{\mathbf{\overline{5}}}-3y_{i}^{\mathbf{\overline{45}}}\,.
\end{equation}

We focus now on the following set of couplings involving the hyperons,
the vector-like fermions $\chi_{i}$ and the light linear combinations
of Higgs doublets, 
\begin{align}
\mathcal{L} & \supset x_{ij}\,\Phi_{ij}^{T}T_{i}\overline{\chi}_{j}+z_{i}^{u}\,H_{i}^{u}\chi_{i}T_{i}+z_{i}^{d}\,H_{i}^{d}\chi_{i}F_{i}\nonumber \\
 & +y_{i}^{u}\,H_{i}^{u}T_{i}T_{i}+\widetilde{y}_{i}^{d}\,H_{i}^{d}T_{i}F_{i}+f_{ij}^{u}\,H_{i}^{u}\widetilde{H}_{j}^{u}\widetilde{\Phi}_{ij}^{F}+f_{ij}^{d}\,H_{i}^{d}\widetilde{H}_{j}^{d}\Phi_{ij}^{F}\,,
\end{align}
where $i,j=1,2,3$, $f_{ij}^{u,d}$ have mass dimension and the rest
of the couplings are dimensionless. After integrating out the heavy
vector-like fermions $\chi_{i}$, $\overline{\chi}_{i}$ and Higgs
doublets $H_{1,2}^{u,d}$, we obtain the following set of effective
Yukawa couplings, 
\begin{flalign}
\mathcal{L} & =\begin{pmatrix}q_{1} & q_{2} & q_{3}\end{pmatrix}\begin{pmatrix}{\displaystyle c_{11}^{u}\frac{{\phi}_{\ell13}}{M_{H_{1}^{u}}}} & {\displaystyle c_{12}^{u}\frac{{\phi}_{\ell23}}{M_{H_{2}^{u}}}\frac{{\phi}_{q12}}{M_{Q_{2}}}} & {\displaystyle c_{13}^{u}\frac{{\phi}_{q13}}{M_{Q_{3}}}}\\
\vspace{-0.1cm} & \vspace{-0.1cm} & \vspace{-0.1cm}\\
{\displaystyle c_{21}^{u}\frac{{\phi}_{\ell13}}{M_{H_{1}^{u}}}\frac{{\phi}_{q12}}{M_{Q_{1}}}} & {\displaystyle c_{22}^{u}\frac{{\phi}_{\ell23}}{M_{H_{2}^{u}}}} & {\displaystyle c_{23}^{u}\frac{{\phi}_{q23}}{M_{Q_{3}}}}\\
\vspace{-0.1cm} & \vspace{-0.1cm} & \vspace{-0.1cm}\\
{\displaystyle c_{31}^{u}\frac{{\phi}_{\ell13}}{M_{H_{1}^{u}}}\frac{{\widetilde{\phi}}_{q13}}{M_{Q_{1}}}} & {\displaystyle c_{32}^{u}\frac{{\phi}_{\ell23}}{M_{H_{2}^{u}}}\frac{{\widetilde{\phi}}_{q23}}{M_{Q_{2}}}} & {\displaystyle c_{33}^{u}}
\end{pmatrix}\begin{pmatrix}u_{1}^{c}\\
u_{2}^{c}\\
u_{3}^{c}
\end{pmatrix}H_{3}^{u}\label{eq:eff_Yukawa_ups}\\
 & +\begin{pmatrix}q_{1} & q_{2} & q_{3}\end{pmatrix}\begin{pmatrix}{\displaystyle c_{11}^{d}\frac{{\widetilde{\phi}}_{\ell13}}{M_{H_{1}^{d}}}} & {\displaystyle c_{12}^{d}\frac{{\widetilde{\phi}}_{\ell23}}{M_{H_{2}^{d}}}\frac{{\phi}_{q12}}{M_{Q_{2}}}} & {\displaystyle c_{13}^{d}\frac{{\phi}_{q13}}{M_{Q_{3}}}}\\
\vspace{-0.1cm} & \vspace{-0.1cm} & \vspace{-0.1cm}\\
{\displaystyle c_{21}^{d}\frac{{\phi}_{\ell13}}{M_{H_{1}^{d}}}\frac{{\phi}_{q12}}{M_{Q_{1}}}} & {\displaystyle c_{22}^{d}\frac{{\widetilde{\phi}}_{\ell23}}{M_{H_{2}^{d}}}} & {\displaystyle c_{23}^{d}\frac{{\phi}_{q23}}{M_{Q_{3}}}}\\
\vspace{-0.1cm} & \vspace{-0.1cm} & \vspace{-0.1cm}\\
{\displaystyle c_{31}^{d}\frac{{\phi}_{\ell13}}{M_{H_{1}^{d}}}\frac{{\widetilde{\phi}}_{q13}}{M_{Q_{1}}}} & {\displaystyle c_{32}^{d}\frac{{\phi}_{\ell23}}{M_{H_{2}^{d}}}\frac{{\widetilde{\phi}}_{q23}}{M_{Q_{2}}}} & {\displaystyle c_{33}^{d}}
\end{pmatrix}\begin{pmatrix}d_{1}^{c}\\
d_{2}^{c}\\
d_{3}^{c}
\end{pmatrix}H_{3}^{d}\\
 & +\begin{pmatrix}\ell_{1} & \ell_{2} & \ell_{3}\end{pmatrix}\begin{pmatrix}{\displaystyle c_{11}^{e}\frac{{\widetilde{\phi}}_{\ell13}}{M_{H_{1}^{d}}}} & {\displaystyle 0} & {\displaystyle 0}\\
{\displaystyle 0} & {\displaystyle c_{22}^{e}\frac{{\widetilde{\phi}}_{\ell23}}{M_{H_{2}^{d}}}} & {\displaystyle 0}\\
{\displaystyle 0} & {\displaystyle 0} & {\displaystyle c_{33}^{e}}
\end{pmatrix}\begin{pmatrix}e_{1}^{c}\\
e_{2}^{c}\\
e_{3}^{c}
\end{pmatrix}H_{3}^{d}+\mathrm{h.c.}\,,\label{eq:Eff_Yukawa_charged_leptons}
\end{flalign}
where the dimensionless coefficients $c_{ij}^{u,d,e}$ are given by
\begin{equation}
c_{ij}^{u}={\displaystyle \begin{pmatrix}{\displaystyle y_{1}^{u}\frac{f_{13}^{u}}{M_{H_{1}}^{u}}} & {\displaystyle x_{12}y_{2}^{u}\frac{f_{23}^{u}}{M_{H_{2}}^{u}}} & {\displaystyle x_{13}z_{3}^{u}}\\
\vspace{-0.1cm} & \vspace{-0.1cm} & \vspace{-0.1cm}\\
{\displaystyle x_{21}y_{1}^{u}\frac{f_{13}^{u}}{M_{H_{1}}^{u}}} & {\displaystyle y_{2}^{u}\frac{f_{23}^{u}}{M_{H_{2}}^{u}}} & {\displaystyle x_{23}z_{3}^{u}}\\
\vspace{-0.1cm} & \vspace{-0.1cm} & \vspace{-0.1cm}\\
{\displaystyle x_{31}y_{1}^{u}\frac{f_{13}^{u}}{M_{H_{1}}^{u}}} & {\displaystyle x_{32}y_{2}^{u}\frac{f_{23}^{u}}{M_{H_{2}}^{u}}} & {\displaystyle y_{3}^{u}}
\end{pmatrix}}\,,
\end{equation}
\begin{equation}
c_{ij}^{d}=\begin{pmatrix}{\displaystyle y_{1}^{d}\frac{f_{13}^{d}}{M_{H_{1}}^{d}}} & {\displaystyle x_{12}y_{2}^{d}\frac{f_{23}^{d}}{M_{H_{2}}^{d}}} & {\displaystyle x_{13}z_{3}^{d}}\\
\vspace{-0.1cm} & \vspace{-0.1cm} & \vspace{-0.1cm}\\
{\displaystyle x_{21}y_{1}^{d}\frac{f_{13}^{d}}{M_{H_{1}}^{d}}} & {\displaystyle y_{2}^{d}\frac{f_{23}^{d}}{M_{H_{2}}^{d}}} & {\displaystyle x_{23}z_{3}^{d}}\\
\vspace{-0.1cm} & \vspace{-0.1cm} & \vspace{-0.1cm}\\
{\displaystyle x_{31}y_{1}^{d}\frac{f_{13}^{d}}{M_{H_{1}}^{d}}} & {\displaystyle x_{32}y_{2}^{d}\frac{f_{23}^{d}}{M_{H_{2}}^{d}}} & {\displaystyle y_{3}^{d}}
\end{pmatrix}\,,
\end{equation}
\begin{equation}
c_{ij}^{e}=\mathrm{diag}\left({\displaystyle y_{1}^{e}\frac{f_{13}^{d}}{M_{H_{1}}^{d}}},{\displaystyle y_{2}^{e}\frac{f_{23}^{d}}{M_{H_{2}}^{d}}},{\displaystyle y_{3}^{e}}\right)\,.
\end{equation}

\begin{figure}[t]
\subfloat[]{\resizebox{.34\textwidth}{!}{ \begin{tikzpicture}
	\begin{feynman}
		\vertex (a);
		\vertex [right=18mm of a] (b);
		\vertex [right=of b] (c);
		\vertex [right=of c] (d);
		\vertex [right=of d] (e);
		\vertex [above=of b] (f1) {\(\phi_{qi3}\)};
		\vertex [above=of d] (f2) {\(H_{3}^{u,d}\)};
		\diagram* {
			(a) -- [fermion, edge label'=\(q_{i}\), inner sep=6pt] (b) -- [scalar] (f1),
			(b) -- [edge label'=\(\overline{Q}_{3}\)] (c),
			(c) -- [edge label'=\(Q_{3}\),inner sep=6pt, insertion=0] (d) -- [scalar] (f2),
			(d) -- [fermion, edge label'=\(u_{3}^{c}{,}\,d_{3}^{c}\)] (e),
	};
	\end{feynman}
\end{tikzpicture} }

}$\;$\subfloat[]{\begin{centering}
\resizebox{.37\textwidth}{!}{ \begin{tikzpicture}
	\begin{feynman}
		\vertex (a);
		\vertex [right=20mm of a] (b);
        \vertex [above=of b] (h1) {\(\phi_{q12}\)};
        %\vertex [right=of b] (c) [label={ [xshift=0.1cm, yshift=0.1cm] \small $M_{Q_{2}}$}];
		\vertex [right=of b] (c);
        \vertex [above=of c] (h2);
		\vertex [right=of c] (d);
        \vertex [right=of d] (e);
		%\vertex [above=of d] (f1) [label={ [xshift=0.7cm, yshift=-0.4cm] \small $M_{H^{u,d}_{2}}$}];
		\vertex [above=of d] (f1);
		\vertex [above=of f1] (f2);
		\vertex [above left=of f2] (g1) {\(\phi_{\ell 23}\)};
		\vertex [above right=of f2] (g2) {\(H^{u,d}_{3}\)};
		\diagram* {
			(a) -- [fermion, edge label'=\(q_{1}\), inner sep=6pt] (b) -- [scalar] (h1),
			(b) -- [edge label'=\(\overline{Q}_{2}\)] (c) -- [edge label'=\(Q_{2}\), inner sep=6pt, insertion=0] (d),
            (d) -- [fermion, edge label'=\(u_{2}^{c}{,}\,d_{2}^{c}\)] (e),
			(d) -- [scalar, edge label'=\(H_{2}^{u,d}\)] (f1) -- [scalar, edge label'=\(H_{2}^{u,d}\), inner sep=6pt, insertion=0] (f2),
			(f2) -- [scalar] (g1),
			(f2) -- [scalar] (g2),
	};
	\end{feynman}
\end{tikzpicture}} 
\par\end{centering}
}$\;$\subfloat[]{\begin{centering}
\resizebox{.22\textwidth}{!}{ \begin{tikzpicture}
	\begin{feynman}
		\vertex (a);
		\vertex [right=16mm of a] (b);
		\vertex [right=of b] (c);
		%\vertex [above=of b] (f1) [label={ [xshift=0.8cm, yshift=-0.5cm] \small $M_{H_{i}^{u,d}}$}];
		\vertex [above=of b] (f1);
		\vertex [above=of f1] (f2);
		\vertex [above left=of f2] (g1) {\(\phi_{\ell i3}\)};
		\vertex [above right=of f2] (g2) {\(H^{u,d}_{3}\)};
		\diagram* {
			(a) -- [fermion, edge label'=\(q_{i}{,}\,\ell_{i}\), inner sep=6pt] (b) -- [fermion, edge label'=\(u^{c}_{i}{,}\,d^{c}_{i}{,}\,e^{c}_{i}\)] (c),
			(b) -- [scalar, edge label'=\(H_{i}^{u,d}\)] (f1),
			(f1) -- [scalar, edge label'=\(H_{i}^{u,d}\),inner sep=6pt, insertion=0] (f2),
			(f2) -- [scalar] (g1),
			(f2) -- [scalar] (g2),
	};
	\end{feynman}
\end{tikzpicture}} 
\par\end{centering}
}\caption[Diagrams in the example $SU(5)^{3}$ tri-unification model which lead to the origin of light charged fermion
masses and quark mixing]{Diagrams in the model which lead to the origin of light charged fermion
masses and quark mixing, where $i=1,2$. \label{fig:charged_fermions_diagrams}}
\end{figure}
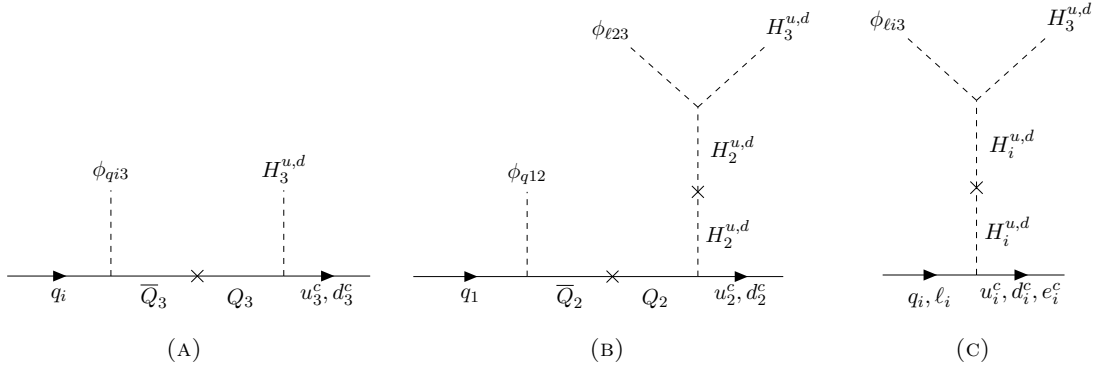

It is clear that third family charged fermions get their masses from
$\mathcal{O}(1)$ Yukawa couplings to the Higgs doublets $H_{3}^{u,d}$,
where the mass hierarchies $m_{b,\tau}/m_{t}$ are explained via $\tan\beta\approx\lambda^{-2}$,
where $\lambda\simeq0.224$ is the Wolfenstein parameter. In contrast,
quark mixing and the masses of first and second family charged fermions
arise from effective Yukawa couplings involving the heavy messengers
of the model, once the hyperons develop their VEVs. The heavy Higgs
doublets $H_{1}^{u,d}$ and $H_{2}^{u,d}$ play a role in the origin
of the family mass hierarchies, while the origin of quark mixing involves
both the heavy Higgs and the vector-like quarks $Q_{i}$ and $\overline{Q}_{i}$,
as shown in Fig.~\ref{fig:charged_fermions_diagrams}. We fix the
various $\left\langle \phi\right\rangle /M$ ratios in terms of the
Wolfenstein parameter $\lambda\simeq0.224$ 
\begin{equation}
\frac{\left\langle \phi_{q23}\right\rangle }{M_{Q_{i}}}\approx\lambda^{2}\,,\qquad\frac{\left\langle \phi_{q13}\right\rangle }{M_{Q_{i}}}\approx\lambda^{3},\qquad\frac{\left\langle \phi_{\ell23}\right\rangle }{M_{H_{2}^{u,d}}}\approx\lambda^{3}\,,\qquad\frac{\left\langle \phi_{q12}\right\rangle }{M_{Q_{i}}}\approx\lambda\,,\qquad\frac{\left\langle \phi_{\ell23}\right\rangle }{M_{H_{1}^{u,d}}}\approx\lambda^{6}\,.\label{eq:VEV_M_ratios}
\end{equation}
We notice that the tiny masses of the first family are explained via
the hierarchies of Higgs messengers 
\begin{equation}
M_{H_{3}^{u,d}}\ll M_{H_{2}^{u,d}}\ll M_{H_{1}^{u,d}}\,,
\end{equation}
in the spirit of messenger dominance \cite{Ferretti:2006df}. In other
words, the heavy Higgs doublets $H_{1}^{u,d}$ and $H_{2}^{u,d}$
can be thought of gaining small effective VEVs from mixing with $H_{3}^{u,d}$,
which are light and perform electroweak symmetry breaking, and these
effective VEVs provide naturally small masses for light charged fermions.
This is in contrast with the original spirit of tri-hypercharge, where
the $m_{1}/m_{2}$ mass hierarchies find their natural origin due
to the higher dimension of the effective Yukawa couplings involving
the first family (see e.g.~Section~\ref{subsec:Model-1:-Minimal}). However, we note that in
the $SU(5)^{3}$ framework, the three pairs of Higgs doublets $H_{i}^{u,d}$
are required by the $\mathbb{Z}_{3}$ symmetry, hence it seems natural
that they play a role on the origin of fermion masses. Moreover, the
introduction of these Higgs provides a very minimal framework to UV-complete
the effective Yukawa couplings of tri-hypercharge, which otherwise
would require a much larger set of heavy messengers that are not
desired, as they may enhance too much the RGE of the gauge couplings,
eventually leading to a non-perturbative gauge coupling at the GUT
scale. 

The numerical values for the ratios in Eq.~(\ref{eq:VEV_M_ratios})
provide the following Yukawa textures (ignoring dimensionless coefficients)
\begin{flalign}
\mathcal{L} & =\begin{pmatrix}q_{1} & q_{2} & q_{3}\end{pmatrix}\begin{pmatrix}\lambda^{6} & \lambda^{4} & \lambda^{3}\\
\lambda^{7} & \lambda^{3} & \lambda^{2}\\
\lambda^{9} & \lambda^{5} & 1
\end{pmatrix}\begin{pmatrix}u_{1}^{c}\\
u_{2}^{c}\\
u_{3}^{c}
\end{pmatrix}v_{\mathrm{SM}}\label{eq:eff_Yukawa_ups-1}\\
 & +\begin{pmatrix}q_{1} & q_{2} & q_{3}\end{pmatrix}\begin{pmatrix}\lambda^{6} & \lambda^{4} & \lambda^{3}\\
\lambda^{7} & \lambda^{3} & \lambda^{2}\\
\lambda^{9} & \lambda^{5} & 1
\end{pmatrix}\begin{pmatrix}d_{1}^{c}\\
d_{2}^{c}\\
d_{3}^{c}
\end{pmatrix}\lambda^{2}\,v_{\mathrm{SM}}\\
 & +\begin{pmatrix}\ell_{1} & \ell_{2} & \ell_{3}\end{pmatrix}\begin{pmatrix}\lambda^{6} & 0 & 0\\
0 & \lambda^{3} & 0\\
0 & 0 & 1
\end{pmatrix}\begin{pmatrix}e_{1}^{c}\\
e_{2}^{c}\\
e_{3}^{c}
\end{pmatrix}\lambda^{2}\,v_{\mathrm{SM}}+\mathrm{h.c.}\,,\label{eq:Eff_Yukawa_charged_leptons-1}
\end{flalign}
where $v_{\mathrm{SM}}$ is the usual SM electroweak VEV and the fit
of the up-quark mass may be improved by assuming a mild difference
between $M_{H_{1}^{u}}$ and $M_{H_{1}^{d}}$. In general, the alignment
of the CKM matrix is not predicted but depends on the choice of dimensionless
coefficients and on the difference between $M_{H_{2}^{u}}$ and $M_{H_{2}^{d}}$.
Any charged lepton mixing is suppressed by the very heavy masses of
the required messengers contained in $\chi_{i}$ and $\overline{\chi}_{i}$,
leading to the off-diagonal zeros in Eq.~(\ref{eq:Eff_Yukawa_charged_leptons-1}),
in such a way that the PMNS matrix must dominantly arise from the
neutrino sector, as we shall see. We notice that a mild hierarchy
of dimensionless couplings $y_{1}^{e}/y_{1}^{d}\approx\lambda^{1.4}$
may be needed to account for the mass hierarchy between the down-quark
and the electron.

The larger suppression of the (2,1), (3,1) and (3,2) entries in the
quark Yukawa textures ensures a significant suppression of right-handed
quark mixing. This is a very desirable feature, given the strong phenomenological
constraints on right-handed flavour-changing currents \cite{UTfit:2007eik,Isidori:2014rba}.
This way, we expect the model to reproduce the low energy phenomenology
of Model 2 in Chapter~\ref{Chapter:Tri-hypercharge}, where the VEVs of the 23 and 13 hyperons
may be as low as the TeV scale, while the VEVs of the 12 hyperons
may be as low as 50 TeV or so. In this manner, we provide the following
benchmark values for the mass scales involved in the flavour sector\footnote{We note that all VEVs and masses listed here may vary by $\mathcal{O}(1)$
factors, as naturally expected, without affecting our final conclusions.} 
\begin{equation}
\left\langle \phi_{q23}\right\rangle \approx\left\langle \phi_{q13}\right\rangle \approx\left\langle \phi_{\ell23}\right\rangle \approx\left\langle \phi_{\ell13}\right\rangle \sim\mathcal{O}(5\,\mathrm{TeV})\,,
\end{equation}
\begin{equation}
\left\langle \phi_{q12}\right\rangle \sim\mathcal{O}(50\,\mathrm{TeV})\,,
\end{equation}
\begin{equation}
M_{Q_{i}}\sim\mathcal{O}(100\,\mathrm{TeV})\,,
\end{equation}
\begin{equation}
M_{H_{2}^{u,d}}\sim\mathcal{O}(100\,\mathrm{TeV})\,,
\end{equation}
\begin{equation}
M_{H_{1}^{u,d}}\sim\mathcal{O}(10^{4}\,\mathrm{TeV})\,.
\end{equation}

\subsection{Neutrino masses and mixing\label{subsec:Neutrinos}}

\begin{figure}[t]
\subfloat[]{\resizebox{.32\textwidth}{!}{ \begin{tikzpicture}
	\begin{feynman}
		\vertex (a);
		\vertex [right=16mm of a] (b);
		\vertex [right=of b] (c);
		\vertex [right=of c] (d);
		\vertex [right=of d] (e);
		\vertex [above=of b] (f1) {\(H^{u}_{3}\)};
		\vertex [above=of d] (f2) {\(\phi_{u32}{,}\,\phi_{u31}\)};
		\diagram* {
			(a) -- [fermion, edge label'=\(\ell_{3}\)] (b) -- [charged scalar] (f1),
			(b) -- [edge label'=\(\xi_0\)] (c),
			(c) -- [edge label'=\(\xi_0\), insertion=0] (d) -- [charged scalar] (f2),
			(d) -- [fermion, edge label'=\(N_{\mathrm{atm}}{,}\,N_{\mathrm{sol}}\)] (e),
	};
	\end{feynman}
\end{tikzpicture} }

}\subfloat[]{\begin{centering}
\resizebox{.32\textwidth}{!}{ \begin{tikzpicture}
	\begin{feynman}
		\vertex (a);
		\vertex [right=16mm of a] (b);
		\vertex [right=of b] (c);
		\vertex [right=of c] (d);
		\vertex [right=of d] (e);
		\vertex [above=of b] (f1) {\(H^{u}_{3}\)};
		\vertex [above=of d] (f2) {\(\phi_{q23}{,}\,\phi_{u123}\)};
		\diagram* {
			(a) -- [fermion, edge label'=\(\ell_{2}\)] (b) -- [charged scalar] (f1),
			(b) -- [edge label'=\(\xi_{23}\)] (c),
			(c) -- [edge label'=\(\xi_{23}\), insertion=0] (d) -- [charged scalar] (f2),
			(d) -- [fermion, edge label'=\(N_{\mathrm{atm}}{,}\,N_{\mathrm{sol}}\)] (e),
	};
	\end{feynman}
\end{tikzpicture}} 
\par\end{centering}
}\subfloat[]{\begin{centering}
\resizebox{.32\textwidth}{!}{ \begin{tikzpicture}
	\begin{feynman}
		\vertex (a);
		\vertex [right=16mm of a] (b);
		\vertex [right=of b] (c);
		\vertex [right=of c] (d);
		\vertex [right=of d] (e);
		\vertex [above=of b] (f1) {\(H^{u}_{3}\)};
		\vertex [above=of d] (f2) {\(\phi_{q13}{,}\)};
		\diagram* {
			(a) -- [fermion, edge label'=\(\ell_{1}\)] (b) -- [charged scalar] (f1),
			(b) -- [edge label'=\(\xi_{13}\)] (c),
			(c) -- [edge label'=\(\xi_{13}\), insertion=0] (d) -- [charged scalar] (f2),
			(d) -- [fermion, edge label'=\(N_{\mathrm{sol}}\)] (e),
	};
	\end{feynman}
\end{tikzpicture}} 
\par\end{centering}
}\caption[Diagrams in the example $SU(5)^{3}$ tri-unification model leading to effective Yukawa couplings in the neutrino sector]{Diagrams leading to effective Yukawa couplings in the neutrino sector.
\label{fig:Diagrams_neutrinos}}
\end{figure}
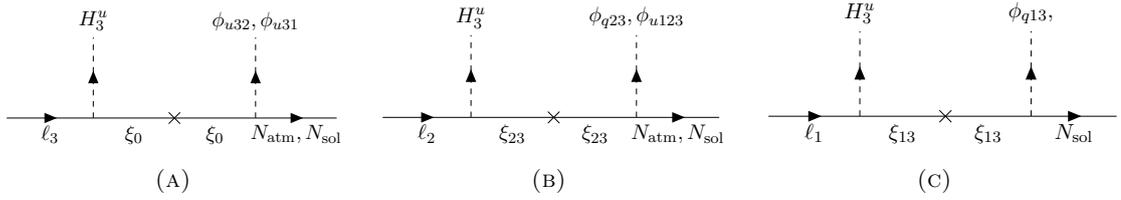

Explaining the observed pattern of neutrino mixing and mass splittings
in gauge non-universal theories of flavour is usually difficult, due
to the accidental $U(2)^{5}$ flavour symmetry predicted by these
models, which is naively present in the neutrino sector as well. However,
exotic variations of the type I seesaw mechanism have been shown to
be successful in accommodating neutrino observations within non-universal
theories of flavour, see Refs.~\cite{FernandezNavarro:2023rhv,Fuentes-Martin:2020pww}.
Here we will incorporate the mechanism of \cite{FernandezNavarro:2023rhv}
(also described in Section~\ref{sec:Neutrino-masses-and-Mixing} of this thesis), which consists of
adding SM singlet neutrinos which carry family hypercharges (although
their sum must of course vanish). These neutrinos can be seen as the
fermionic counterpart of hyperons, as they will connect the different
hypercharge sites, therefore breaking the $U(2)^{5}$ flavour symmetry
in the neutrino sector. In this manner, these neutrinos allow to write
effective operators which may provide a successful pattern for neutrino
mixing. However, the particular model presented in Section~\ref{sec:Neutrino-masses-and-Mixing} incorporates
SM singlet neutrinos with 1/4 family hypercharge factors, which cannot
be obtained from $SU(5)^{3}$, at least not from representations with
dimension smaller than $\mathbf{45}$\footnote{Since these singlet neutrinos can be seen as the fermionic counterpart
of hyperons, the search for $SU(5)^{3}$ hyperon embeddings shown
in Appendix~\ref{app:Hyperons} shows that no neutrinos with 1/4
family hypercharge factors are found from $SU(5)^{3}$ representations
with dimension up to $\mathbf{45}$.} according to a search with \texttt{GroupMath} \cite{Fonseca:2020vke}.

Following the recipe of Section~\ref{sec:Neutrino-masses-and-Mixing}, we start by introducing two right-handed
neutrinos: $N_{\mathrm{atm}}\sim(\mathbf{1,1})_{(0,2/3,-2/3)}$ and
$N_{\mathrm{sol}}\sim(\mathbf{1,1})_{(2/3,0,-2/3)}$, which will be
responsible for atmospheric and solar neutrino mixing, respectively.
These neutrinos are embedded in $\Sigma_{23}\sim(\mathbf{1,10,\overline{10}})$
and $\Sigma_{13}\sim(\mathbf{10,1,\overline{10}})$ representations
of $SU(5)^{3}$, respectively. We also need to add the cyclic permutation
$N_{\mathrm{cyclic}}$ embedded in $\Sigma_{12}\sim(\mathbf{10,\overline{10},1})$
to preserve the cyclic symmetry of $SU(5)^{3}$. However, we find
that if the ``cyclic'' neutrino contained in $\Sigma_{12}$ is much
heavier than the other neutrinos, then we can ignore it as it decouples
from the seesaw, and we recover the minimal framework of Section~\ref{sec:Neutrino-masses-and-Mixing}.
Finally, in order to cancel gauge anomalies, we choose to make these
neutrinos vector-like by introducing the three corresponding conjugate
neutrinos.

The next step is adding hyperons that provide effective Yukawa couplings
and Majorana masses for the singlet neutrinos. These are summarised
in the Dirac and Majorana mass matrices that follow (ignoring the
$\mathcal{O}(1)$ dimensionless couplings and the much heavier cyclic
neutrinos) 
\begin{flalign}
 &  &  & m_{D_{L}}=\left(\global\long
\global\long\def\arraystretch{0.7}%
\begin{array}{@{}llc@{}}
 & \multicolumn{1}{c@{}}{\phantom{\!\,}\overline{N}_{\mathrm{sol}}} & \phantom{\!\,}\overline{N}_{\mathrm{atm}}\\
\cmidrule(l){2-3}\left.L_{1}\right| & 0 & 0\\
\left.L_{2}\right| & 0 & 0\\
\left.L_{3}\right| & \widetilde{\phi}_{u31}^{(-\frac{2}{3},0,\frac{2}{3})} & \widetilde{\phi}_{u32}^{(0,-\frac{2}{3},\frac{2}{3})}
\end{array}\right)\frac{H_{u}}{M_{\xi}}\,, &  & m_{D_{R}}=\left(\global\long
\global\long\def\arraystretch{0.7}%
\begin{array}{@{}llc@{}}
 & \multicolumn{1}{c@{}}{\phantom{\!\,}N_{\mathrm{sol}}} & \phantom{\!\,}N_{\mathrm{atm}}\\
\cmidrule(l){2-3}\left.L_{1}\right| & \phi_{q13}^{(-\frac{1}{6},0,\frac{1}{6})} & 0\\
\left.L_{2}\right| & \phi_{u123}^{(-\frac{2}{3},\frac{1}{2},\frac{1}{6})} & \phi_{q23}^{(0,-\frac{1}{6},\frac{1}{6})}\\
\left.L_{3}\right| & \phi_{u31}^{(-\frac{2}{3},0,\frac{2}{3})} & \phi_{u32}^{(0,-\frac{2}{3},\frac{2}{3})}
\end{array}\right)\frac{H_{u}}{M_{\xi}}\,,\label{eq:mDL_mDR2}\\
\nonumber \\
 &  &  & M_{L}=\left(\global\long
\global\long\def\arraystretch{0.7}%
\begin{array}{@{}llc@{}}
 & \multicolumn{1}{c@{}}{\phantom{\!\,}\overline{N}_{\mathrm{sol}}} & \phantom{\!\,}\overline{N}_{\mathrm{atm}}\\
\cmidrule(l){2-3}\left.\:\:\overline{N}_{\mathrm{sol}}\right| & \tilde{\phi}_{\mathrm{sol}}^{(-\frac{4}{3},0,\frac{4}{3})} & 0\\
\left.\overline{N}_{\mathrm{atm}}\right| & 0 & \widetilde{\phi}_{\mathrm{atm}}^{(0,-\frac{4}{3},\frac{4}{3})}
\end{array}\right)\,, &  & M_{R}=\left(\global\long
\global\long\def\arraystretch{0.7}%
\begin{array}{@{}llc@{}}
 & \multicolumn{1}{c@{}}{\phantom{\!\,}N_{\mathrm{sol}}} & \phantom{\!\,}N_{\mathrm{atm}}\\
\cmidrule(l){2-3}\left.\:\:N_{\mathrm{sol}}\right| & \phi_{\mathrm{sol}}^{(-\frac{4}{3},0,\frac{4}{3})} & 0\\
\left.N_{\mathrm{atm}}\right| & 0 & \phi_{\mathrm{atm}}^{(0,-\frac{4}{3},\frac{4}{3})}
\end{array}\right)\,,\label{eq:ML_MR2}
\end{flalign}
\begin{equation}
M_{LR}=\left(\global\long
\global\long\def\arraystretch{0.7}%
\begin{array}{@{}llc@{}}
 & \multicolumn{1}{c@{}}{\phantom{\!\,}N_{\mathrm{sol}}} & \phantom{\!\,}N_{\mathrm{atm}}\\
\cmidrule(l){2-3}\left.\:\:\overline{N}_{\mathrm{sol}}\right| & M_{N_{\mathrm{sol}}} & 0\\
\left.\overline{N}_{\mathrm{atm}}\right| & 0 & M_{N_{\mathrm{atm}}}
\end{array}\right)\,,\label{eq:MLR2}
\end{equation}

where the heavy scale $M_{\xi}$ is associated to the mass of the
heavy vector-like fermions $\xi_{0}\sim(\mathbf{1,1})_{(0,0,0)}$,
$\xi_{23}\sim(\mathbf{1,1})_{(0,1/2,-1/2)}$ (plus cyclic permutations),
which are embedded in the representations $\Xi_{0}\sim(\mathbf{1,1,1})$
and $\Xi_{23}\sim(\mathbf{1,5,\overline{5}})$ (plus conjugate, plus
cyclic permutations) of $SU(5)^{3}$. Example diagrams are shown in
Fig~\ref{fig:Diagrams_neutrinos}. We now construct the full neutrino
mass matrix as 
\begin{equation}
M_{\nu}=\left(\global\long
\global\long\def\arraystretch{0.7}%
\begin{array}{@{}llcc@{}}
 & \multicolumn{1}{c@{}}{\phantom{\!\,}\nu} & \phantom{\!\,}\overline{N} & \phantom{\!\,}N\\
\cmidrule(l){2-4}\left.\:\,\nu\right| & 0 & m_{D_{L}} & m_{D_{R}}\\
\left.\overline{N}\right| & m_{D_{L}}^{\mathrm{T}} & M_{L} & M_{LR}\\
\left.N\right| & m_{D_{R}}^{\mathrm{T}} & M_{LR}^{\mathrm{T}} & M_{R}
\end{array}\right)\equiv\left(\begin{array}{cc}
0 & m_{D}\\
m_{D}^{\mathrm{T}} & M_{N}
\end{array}\right)\,,\label{eq:Full_Mnu2}
\end{equation}
where we have defined $\nu$ as a 3-component vector containing the
weak eigenstates of active neutrinos, while $N$ and $\overline{N}$
are 2-component vectors containing the SM singlets $N$ and conjugate
neutrinos $\overline{N}$ , respectively. Now we assume that all the
hyperons in Eqs.~(\ref{eq:ML_MR2}-\ref{eq:MLR2}) get VEVs at the
scale $v_{23}$ of 23 hypercharge breaking according to Eq.~(\ref{eq:23_breaking}),
and we have into account that $\left\langle \phi_{q13}\right\rangle /\left\langle \phi_{q23}\right\rangle \approx\lambda$
as obtained from the discussion of the charged fermion sector in Section~\ref{subsec:Charged_fermions}.
It is also required to assume $M_{N_{\mathrm{atm}}},\,M_{N_{\mathrm{atm}}}\apprle v_{23}$
in order to obtain the observed neutrino mixing with the textures
of Eqs.~(\ref{eq:ML_MR2}-\ref{eq:MLR2}).

Dirac-type masses in $m_{D_{L,R}}$ may be orders of magnitude smaller
than the electroweak scale, because they arise from non-renormalisable
operators proportional to the SM VEV. In contrast, the eigenvalues
of $M_{N}$ are not smaller than $\mathcal{O}(v_{23})$, which is
at least TeV. Therefore, the condition $m_{D}\ll M_{N}$ is fulfilled
in Eq.~(\ref{eq:Full_Mnu2}) and we can safely apply the seesaw formula
to obtain, up to $\mathcal{O}(1)$ factors, 
\begin{equation}
m_{\nu}\simeq m_{D}M_{N}^{-1}m_{D}^{\mathrm{T}}\approx\left(\begin{array}{ccc}
1 & 1 & \lambda\\
1 & 1 & 1\\
\lambda & 1 & 1
\end{array}\right)v_{23}\frac{v_{\mathrm{SM}}^{2}}{M_{\xi}^{2}}\,.\label{eq:Neutrino_texture}
\end{equation}
This is the same texture that was obtained in Section~\ref{sec:Neutrino-masses-and-Mixing}, which is
able to accommodate all the observed neutrino mixing angles and mass
splittings \cite{deSalas:2020pgw,Gonzalez-Garcia:2021dve} with $\mathcal{O}(1)$
parameters once the dimensionless coefficients implicit in Eq.~(\ref{eq:Neutrino_texture})
are considered. Remarkably, the singlet neutrinos $N_{\mathrm{atm}}$
and $N_{\mathrm{sol}}$ get masses around the TeV scale $(v_{23})$
and contribute to the RGE, while the cyclic neutrino is assumed to
get a very heavy vector-like mass and decouples, as mentioned before.

\subsection{Energy regimes, symmetries and particle content}

\label{sec:app}

Having discussed how the flavour structure of the SM is dynamically
generated by the tri-hypercharge layer, now we discuss in detail the
symmetries and particle content of our model at each energy regime
between the GUT and electroweak scales.

\subsection*{Regime 1: $\boldsymbol{SU(5)^{3}}$ breaking scale $\boldsymbol{\to\left(SU(3)\times SU(2)\times U(1)\right)^{3}}$
breaking scale}

{ %\renewcommand{\arraystretch}{1.1}
\begin{table}[t]
\centering %
\resizebox{14.5cm}{!}{
\begin{tabular}{cccccccccc}
\toprule 
\textbf{Field}  & $\boldsymbol{SU(3)_{1}}$  & $\boldsymbol{SU(2)_{1}}$  & $\boldsymbol{U(1)_{1}}$  & $\boldsymbol{SU(3)_{2}}$  & $\boldsymbol{SU(2)_{2}}$  & $\boldsymbol{U(1)_{2}}$  & $\boldsymbol{SU(3)_{3}}$  & $\boldsymbol{SU(2)_{3}}$  & $\boldsymbol{U(1)_{3}}$ \tabularnewline
\midrule 
$q_{1}$  & \three  & \two  & $\frac{1}{6}$  & \one  & \one  & 0  & \one  & \one  & 0 \tabularnewline
$u_{1}^{c}$  & \threeS  & \one  & $-\frac{2}{3}$  & \one  & \one  & 0  & \one  & \one  & 0 \tabularnewline
$d_{1}^{c}$  & \threeS  & \one  & $\frac{1}{3}$  & \one  & \one  & 0  & \one  & \one  & 0 \tabularnewline
$\ell_{1}$  & \one  & \two  & $-\frac{1}{2}$  & \one  & \one  & 0  & \one  & \one  & 0 \tabularnewline
$e_{1}^{c}$  & \one  & \one  & 1  & \one  & \one  & 0  & \one  & \one  & 0 \tabularnewline
$q_{2}$  & \one  & \one  & 0  & \three  & \two  & $\frac{1}{6}$  & \one  & \one  & 0 \tabularnewline
$u_{2}^{c}$  & \one  & \one  & 0  & \threeS  & \one  & $-\frac{2}{3}$  & \one  & \one  & 0 \tabularnewline
$d_{2}^{c}$  & \one  & \one  & 0  & \threeS  & \one  & $\frac{1}{3}$  & \one  & \one  & 0 \tabularnewline
$\ell_{2}$  & \one  & \one  & 0  & \one  & \two  & $-\frac{1}{2}$  & \one  & \one  & 0 \tabularnewline
$e_{2}^{c}$  & \one  & \one  & 0  & \one  & \one  & 1  & \one  & \one  & 0 \tabularnewline
$q_{3}$  & \one  & \one  & 0  & \one  & \one  & 0  & \three  & \two  & $\frac{1}{6}$ \tabularnewline
$u_{3}^{c}$  & \one  & \one  & 0  & \one  & \one  & 0  & \threeS  & \one  & $-\frac{2}{3}$ \tabularnewline
$d_{3}^{c}$  & \one  & \one  & 0  & \one  & \one  & 0  & \threeS  & \one  & $\frac{1}{3}$ \tabularnewline
$\ell_{3}$  & \one  & \one  & 0  & \one  & \one  & 0  & \one  & \two  & $-\frac{1}{2}$ \tabularnewline
$e_{3}^{c}$  & \one  & \one  & 0  & \one  & \one  & 0  & \one  & \one  & 1 \tabularnewline
\midrule 
$\xi_{0}$  & \one  & \one  & 0  & \one  & \one  & 0  & \one  & \one  & 0 \tabularnewline
\rowcolor{yellow!10} $\xi_{12}$  & \one  & \one  & $\frac{1}{2}$  & \one  & \one  & $-\frac{1}{2}$  & \one  & \one  & 0 \tabularnewline
\rowcolor{yellow!10} $\xi_{13}$  & \one  & \one  & $\frac{1}{2}$  & \one  & \one  & 0  & \one  & \one  & $-\frac{1}{2}$ \tabularnewline
\rowcolor{yellow!10} $\xi_{23}$  & \one  & \one  & 0  & \one  & \one  & $\frac{1}{2}$  & \one  & \one  & $-\frac{1}{2}$ \tabularnewline
\rowcolor{yellow!10} $Q_{1}$  & \three  & \two  & $\frac{1}{6}$  & \one  & \one  & 0  & \one  & \one  & 0 \tabularnewline
\rowcolor{yellow!10} $Q_{2}$  & \one  & \one  & 0  & \three  & \two  & $\frac{1}{6}$  & \one  & \one  & 0 \tabularnewline
\rowcolor{yellow!10} $Q_{3}$  & \one  & \one  & 0  & \one  & \one  & 0  & \three  & \two  & $\frac{1}{6}$ \tabularnewline
\rowcolor{yellow!10} $N_{{\rm atm}}$  & \one  & \one  & 0  & \one  & \one  & $\frac{2}{3}$  & \one  & \one  & $-\frac{2}{3}$ \tabularnewline
\rowcolor{yellow!10} $N_{{\rm sol}}$  & \one  & \one  & $\frac{2}{3}$  & \one  & \one  & 0  & \one  & \one  & $-\frac{2}{3}$ \tabularnewline
\rowcolor{yellow!10} $N_{{\rm cyclic}}$  & \one  & \one  & $\frac{2}{3}$  & \one  & \one  & $-\frac{2}{3}$  & \one  & \one  & 0 \tabularnewline
\midrule 
$\Theta_{1}$  & \eight  & \one  & 0  & \one  & \one  & 0  & \one  & \one  & 0 \tabularnewline
$\Theta_{2}$  & \one  & \one  & 0  & \eight  & \one  & 0  & \one  & \one  & 0 \tabularnewline
$\Theta_{3}$  & \one  & \one  & 0  & \one  & \one  & 0  & \eight  & \one  & 0 \tabularnewline
$\Delta_{1}$  & \one  & \three  & 0  & \one  & \one  & 0  & \one  & \one  & 0 \tabularnewline
$\Delta_{2}$  & \one  & \one  & 0  & \one  & \three  & 0  & \one  & \one  & 0 \tabularnewline
$\Delta_{3}$  & \one  & \one  & 0  & \one  & \one  & 0  & \one  & \three  & 0 \tabularnewline
$H_{1}^{u}$  & \one  & \two  & $\frac{1}{2}$  & \one  & \one  & 0  & \one  & \one  & 0 \tabularnewline
$H_{1}^{d}$  & \one  & \two  & $-\frac{1}{2}$  & \one  & \one  & 0  & \one  & \one  & 0 \tabularnewline
$H_{2}^{u}$  & \one  & \one  & 0  & \one  & \two  & $\frac{1}{2}$  & \one  & \one  & 0 \tabularnewline
$H_{2}^{d}$  & \one  & \one  & 0  & \one  & \two  & $-\frac{1}{2}$  & \one  & \one  & 0 \tabularnewline
$H_{3}^{u}$  & \one  & \one  & 0  & \one  & \one  & 0  & \one  & \two  & $\frac{1}{2}$ \tabularnewline
$H_{3}^{d}$  & \one  & \one  & 0  & \one  & \one  & 0  & \one  & \two  & $-\frac{1}{2}$ \tabularnewline
\midrule 
$\Phi_{\ell12}$  & \one  & \one  & $\frac{1}{2}$  & \one  & \one  & $-\frac{1}{2}$  & \one  & \one  & 0 \tabularnewline
$\Phi_{\ell13}$  & \one  & \one  & $\frac{1}{2}$  & \one  & \one  & 0  & \one  & \one  & $-\frac{1}{2}$ \tabularnewline
$\Phi_{\ell23}$  & \one  & \one  & 0  & \one  & \one  & $\frac{1}{2}$  & \one  & \one  & $-\frac{1}{2}$ \tabularnewline
$\Phi_{q12}$  & \one  & \one  & $-\frac{1}{6}$  & \one  & \one  & $\frac{1}{6}$  & \one  & \one  & 0 \tabularnewline
$\Phi_{q13}$  & \one  & \one  & $-\frac{1}{6}$  & \one  & \one  & 0  & \one  & \one  & $\frac{1}{6}$ \tabularnewline
$\Phi_{q23}$  & \one  & \one  & 0  & \one  & \one  & $-\frac{1}{6}$  & \one  & \one  & $\frac{1}{6}$ \tabularnewline
$\Phi_{u12}$  & \one  & \one  & $-\frac{2}{3}$  & \one  & \one  & $\frac{2}{3}$  & \one  & \one  & 0 \tabularnewline
$\Phi_{u13}$  & \one  & \one  & $-\frac{2}{3}$  & \one  & \one  & 0  & \one  & \one  & $\frac{2}{3}$ \tabularnewline
$\Phi_{u23}$  & \one  & \one  & 0  & \one  & \one  & $-\frac{2}{3}$  & \one  & \one  & $\frac{2}{3}$ \tabularnewline
\bottomrule
\end{tabular}}\caption[Example $SU(5)^{3}$ tri-unification model: energy regime 1]{Fermion and scalar representations under $\left(SU(3)\times SU(2)\times U(1)\right)^{3}$
in energy regime 1. Fermions highlighted in yellow belong to a vector-like
pair and thus have a conjugate representation not shown in this table.}
\label{tab:ParticleContent2}
\end{table}

}

As a result of $SU(5)^{3}$ breaking, each of the fermion representations
$F_{i}$ and $T_{i}$ becomes charged under an $SU(3)\times SU(2)\times U(1)$
factor. Regarding the rest of the fields, most get masses at the $M_{{\rm GUT}}\sim v_{\mathrm{GUT}}$
unification scale and decouple. We will assume that only those explicitly
required at low energies remain light. For instance, out of all the
components of the $\Omega_{ij}$ scalars, only the $\Theta_{i}$ and $\Delta_{i}$
states, belonging to the adjoint representations of $SU(3)_{i}$ and
$SU(2)_{i}$, respectively, remain in the particle spectrum. Similarly,
only some SM singlets in the $\Phi_{i}$ scalar fields are assumed
to be present at this energy scale. For instance, this is the case
of $\Phi_{\ell23}$, contained in $\Phi_{23}^{(5)}$, a (\one,\five,\fiveS)
representation of $SU(5)^{3}$, as shown in Table~\ref{tab:hyperons1}.
These representations eventually become the tri-hypercharge hyperons
at lower energies. Similarly, the $Q_{i}$ vector-like quarks in the
$\chi_{i}$ and $\overline{\chi}_{i}$ multiplets are also assumed
to be present at this energy scale. The full fermion and scalar particle
content of the model in this energy regime is shown in Table~\ref{tab:ParticleContent2}.

\subsection*{Regime 2: $\boldsymbol{\left(SU(3)\times SU(2)\times U(1)\right)^{3}}$
breaking scale $\boldsymbol{\to\xi}$ scale}

{ %\renewcommand{\arraystretch}{1.4}
\begin{table}[tb]
\centering %
\resizebox{9.1cm}{!}{
\begin{tabular}{cccccc}
\toprule 
\textbf{Field}  & $\boldsymbol{SU(3)_{c}}$  & $\boldsymbol{SU(2)_{L}}$  & $\boldsymbol{U(1)_{Y_{1}}}$  & $\boldsymbol{U(1)_{Y_{2}}}$  & $\boldsymbol{U(1)_{Y_{3}}}$ \tabularnewline
\midrule 
$q_{1}$  & \three  & \two  & $\frac{1}{6}$  & 0  & 0 \tabularnewline
$u_{1}^{c}$  & \threeS  & \one  & $-\frac{2}{3}$  & 0  & 0 \tabularnewline
$d_{1}^{c}$  & \threeS  & \one  & $\frac{1}{3}$  & 0  & 0 \tabularnewline
$\ell_{1}$  & \one  & \two  & $-\frac{1}{2}$  & 0  & 0 \tabularnewline
$e_{1}^{c}$  & \one  & \one  & 1  & 0  & 0 \tabularnewline
$q_{2}$  & \three  & \two  & 0  & $\frac{1}{6}$  & 0 \tabularnewline
$u_{2}^{c}$  & \threeS  & \one  & 0  & $-\frac{2}{3}$  & 0 \tabularnewline
$d_{2}^{c}$  & \threeS  & \one  & 0  & $\frac{1}{3}$  & 0 \tabularnewline
$\ell_{2}$  & \one  & \two  & 0  & $-\frac{1}{2}$  & 0 \tabularnewline
$e_{2}^{c}$  & \one  & \one  & 0  & 1  & 0 \tabularnewline
$q_{3}$  & \three  & \two  & 0  & 0  & $\frac{1}{6}$ \tabularnewline
$u_{3}^{c}$  & \threeS  & \one  & 0  & 0  & $-\frac{2}{3}$ \tabularnewline
$d_{3}^{c}$  & \threeS  & \one  & 0  & 0  & $\frac{1}{3}$ \tabularnewline
$\ell_{3}$  & \one  & \two  & 0  & 0  & $-\frac{1}{2}$ \tabularnewline
$e_{3}^{c}$  & \one  & \one  & 0  & 0  & 1 \tabularnewline
\midrule 
$\xi_{0}$  & \one  & \one  & 0  & 0  & 0 \tabularnewline
\rowcolor{yellow!10} $\xi_{12}$  & \one  & \one  & $\frac{1}{2}$  & $-\frac{1}{2}$  & 0 \tabularnewline
\rowcolor{yellow!10} $\xi_{13}$  & \one  & \one  & $\frac{1}{2}$  & 0  & $-\frac{1}{2}$ \tabularnewline
\rowcolor{yellow!10} $\xi_{23}$  & \one  & \one  & 0  & $\frac{1}{2}$  & $-\frac{1}{2}$ \tabularnewline
\rowcolor{yellow!10} $Q_{1}$  & \three  & \two  & $\frac{1}{6}$  & 0  & 0 \tabularnewline
\rowcolor{yellow!10} $Q_{2}$  & \three  & \two  & 0  & $\frac{1}{6}$  & 0 \tabularnewline
\rowcolor{yellow!10} $Q_{3}$  & \three  & \two  & 0  & 0  & $\frac{1}{6}$ \tabularnewline
\rowcolor{yellow!10} $N_{{\rm atm}}$  & \one  & \one  & 0  & $\frac{2}{3}$  & $-\frac{2}{3}$ \tabularnewline
\rowcolor{yellow!10} $N_{{\rm sol}}$  & \one  & \one  & $\frac{2}{3}$  & 0  & $-\frac{2}{3}$ \tabularnewline
\midrule 
$\Theta_{1}$  & \eight  & \one  & 0  & 0  & 0 \tabularnewline
$\Theta_{2}$  & \eight  & \one  & 0  & 0  & 0 \tabularnewline
$\Theta_{3}$  & \eight  & \one  & 0  & 0  & 0 \tabularnewline
$H_{1}^{u}$  & \one  & \two  & $\frac{1}{2}$  & 0  & 0 \tabularnewline
$H_{1}^{d}$  & \one  & \two  & $-\frac{1}{2}$  & 0  & 0 \tabularnewline
$H_{2}^{u}$  & \one  & \two  & 0  & $\frac{1}{2}$  & 0 \tabularnewline
$H_{2}^{d}$  & \one  & \two  & 0  & $-\frac{1}{2}$  & 0 \tabularnewline
$H_{3}^{u}$  & \one  & \two  & 0  & 0  & $\frac{1}{2}$ \tabularnewline
$H_{3}^{d}$  & \one  & \two  & 0  & 0  & $-\frac{1}{2}$ \tabularnewline
\midrule 
$\phi_{\ell12}$  & \one  & \one  & $\frac{1}{2}$  & $-\frac{1}{2}$  & 0 \tabularnewline
$\phi_{\ell13}$  & \one  & \one  & $\frac{1}{2}$  & 0  & $-\frac{1}{2}$ \tabularnewline
$\phi_{\ell23}$  & \one  & \one  & 0  & $\frac{1}{2}$  & $-\frac{1}{2}$ \tabularnewline
$\phi_{q12}$  & \one  & \one  & $-\frac{1}{6}$  & $\frac{1}{6}$  & 0 \tabularnewline
$\phi_{q13}$  & \one  & \one  & $-\frac{1}{6}$  & 0  & $\frac{1}{6}$ \tabularnewline
$\phi_{q23}$  & \one  & \one  & 0  & $-\frac{1}{6}$  & $\frac{1}{6}$ \tabularnewline
$\phi_{u12}$  & \one  & \one  & $-\frac{2}{3}$  & $\frac{2}{3}$  & 0 \tabularnewline
$\phi_{u13}$  & \one  & \one  & $-\frac{2}{3}$  & 0  & $\frac{2}{3}$ \tabularnewline
$\phi_{u23}$  & \one  & \one  & 0  & $-\frac{2}{3}$  & $\frac{2}{3}$ \tabularnewline
\bottomrule
\end{tabular}}\caption[Example $SU(5)^{3}$ tri-unification model: energy regime 2, 3, 4, 5 and 6]{Fermion and scalar representations under $SU(3)_{c}\times SU(2)_{L}\times U(1)_{Y_{1}}\times U(1)_{Y_{2}}\times U(1)_{Y_{3}}$
in energy regimes 2, 3, 4, 5 and 6. Some states in this table get
decoupled at intermediate scales and are not present at all energy
regimes, see text for details. Fermions highlighted in yellow belong
to a vector-like pair and thus have a conjugate representation not
shown in this table. \label{tab:ParticleContent3} }
\end{table}

}

The $\left(SU(3)\times SU(2)\times U(1)\right)^{3}$ gauge symmetry
gets broken by the non-zero VEVs of the $\Theta_{i}$ and $\Delta_{i}$
scalars. The $\Theta_{i}$ octets break $SU(3)_{1}\times SU(3)_{2}\times SU(3)_{3}\to SU(3)_{1+2+3}\equiv SU(3)_{c}$,
while the $\Delta_{i}$ triplets play an analogous role for the $SU(2)$
factors. We assume these two breakings to take place simultaneously
at $v_{{\rm {SM}^{3}}}=\langle\Theta_{i}\rangle=\langle\Delta_{i}\rangle$,
slightly below the GUT scale, where we expect $\mathbb{Z}_{3}$ to remain exact in order to ensure the degenerate VEVs of $\Theta_{i}$ and $\Delta_{i}$. As a result of this, the remnant symmetry
is the tri-hypercharge group (see Chapter~\ref{Chapter:Tri-hypercharge}), $SU(3)_{c}\times SU(2)_{L}\times U(1)_{Y_{1}}\times U(1)_{Y_{2}}\times U(1)_{Y_{3}}$:
\begin{equation}
\left(SU(3)\times SU(2)\times U(1)\right)^{3}\quad\xrightarrow{\langle\Theta_{i}\rangle,\langle\Delta_{i}\rangle}\quad SU(3)_{c}\times SU(2)_{L}\times U(1)_{Y_{1}}\times U(1)_{Y_{2}}\times U(1)_{Y_{3}}
\end{equation}
The gauge couplings above ($g_{s_{i}}$ and $g_{L_{i}}$, with $i=1,2,3$)
and below ($g_{s}$ and $g_{L}$) the breaking scale verify the matching
relations 
\begin{align}
\frac{g_{s_{1}}\,g_{s_{2}}\,g_{s_{3}}}{\sqrt{g_{s_{1}}^{2}g_{s_{2}}^{2}+g_{s_{1}}^{2}g_{s_{3}}^{2}+g_{s_{2}}^{2}g_{s_{3}}^{2}}} & =g_{s}\,,\\
\frac{g_{L_{1}}\,g_{L_{2}}\,g_{L_{3}}}{\sqrt{g_{L_{1}}^{2}g_{L_{2}}^{2}+g_{L_{1}}^{2}g_{L_{3}}^{2}+g_{L_{2}}^{2}g_{L_{3}}^{2}}} & =g_{L}\,,
\end{align}
which are equivalent to 
\begin{align}
\alpha_{s_{1}}^{-1}+\alpha_{s_{2}}^{-1}+\alpha_{s_{3}}^{-1} & =\alpha_{s}^{-1}\,,\\
\alpha_{L_{1}}^{-1}+\alpha_{L_{2}}^{-1}+\alpha_{L_{3}}^{-1} & =\alpha_{L}^{-1}\,,
\end{align}
with $\alpha_{i}^{-1}=4\pi/g_{i}^{2}$.

The main difference with respect to the original tri-hypercharge model
proposed in Chapter~\ref{Chapter:Tri-hypercharge} is that a complete ultraviolet completion
for the generation of the flavour structure is provided in our setup.
As already explained in the previous two Sections, we achieve this
with the hyperons and vector-like fermions present in the particle
spectrum, which originate from $SU(5)^{3}$ representations. We assume
$N_{{\rm cyclic}}$ as well as the conjugate representation $\overline{N}_{{\rm cyclic}}$
to be decoupled at this energy scale. Similarly, the $\Delta_{i}$
triplets are also assumed to get masses of the order of the SM$^{3}$
breaking scale and decouple. The resulting fermion and scalar particle
content of the model is shown in Table~\ref{tab:ParticleContent3}.

\subsection*{Regime 3: $\boldsymbol{\xi}$ scale $\boldsymbol{\to H_{1}}$ scale}

The next energy threshold is given by the $\xi$ singlets, responsible
for the flavour structure of the neutrino sector, with masses $M_{\xi}\sim10^{10}$
GeV. At this scale, the $\xi_{0}$ as well as the $\xi_{12}$, $\xi_{13}$,
$\xi_{23}$ and their conjugate representations are integrated out
and no longer contribute to the running of the gauge couplings. The
gauge symmetry does not change and stays the same as in the previous
energy regime. The resulting particle spectrum is that of Table~\ref{tab:ParticleContent3}
removing the $\xi$ singlet fermions.

\subsection*{Regime 4: $\boldsymbol{H_{1}}$ scale $\boldsymbol{\to H_{2}}$ scale}

At energies of the order of $M_{H_{1}^{u,d}}\sim10^{4}$ TeV, the
$H_{1}^{u,d}$ scalar doublets decouple from the particle spectrum
of the model. Again, the gauge symmetry does not change. The particle
spectrum at this stage is that shown on Table~\ref{tab:ParticleContent3}
removing the $\xi$ singlet fermions and the $H_{1}^{u,d}$ scalar
doublets.

\subsection*{Regime 5: $\boldsymbol{H_{2}}$ scale $\boldsymbol{\to Q,\,\Theta}$
scale}

At energies of the order of $M_{H_{2}^{u,d}}\sim100$ TeV, the $H_{2}^{u,d}$
scalar doublets decouple from the particle spectrum of the model.
As in the previous two energy thresholds, the gauge symmetry remains
the same. The particle spectrum at this stage is that shown on Table~\ref{tab:ParticleContent3}
removing the $\xi$ singlet fermions and the $H_{1,2}^{u,d}$ scalar
doublets.

\subsection*{Regime 6: $\boldsymbol{Q,\,\Theta}$ scale $\boldsymbol{\to SU(3)_{c}\times SU(2)_{L}\times U(1)_{Y_{1}}\times U(1)_{Y_{2}}\times U(1)_{Y_{3}}}$
breaking scale}

{ %\renewcommand{\arraystretch}{1.4}
\begin{table}[tb]
\centering %
\begin{tabular}{ccccc}
\toprule 
\textbf{Field}  & $\boldsymbol{SU(3)_{c}}$  & $\boldsymbol{SU(2)_{L}}$  & $\boldsymbol{U(1)_{Y_{12}}}$  & $\boldsymbol{U(1)_{Y_{3}}}$ \tabularnewline
\midrule 
$q_{1}$  & \three  & \two  & $\frac{1}{6}$  & 0 \tabularnewline
$u_{1}^{c}$  & \threeS  & \one  & $-\frac{2}{3}$  & 0 \tabularnewline
$d_{1}^{c}$  & \threeS  & \one  & $\frac{1}{3}$  & 0 \tabularnewline
$\ell_{1}$  & \one  & \two  & $-\frac{1}{2}$  & 0 \tabularnewline
$e_{1}^{c}$  & \one  & \one  & 1  & 0 \tabularnewline
$q_{2}$  & \three  & \two  & $\frac{1}{6}$  & 0 \tabularnewline
$u_{2}^{c}$  & \threeS  & \one  & $-\frac{2}{3}$  & 0 \tabularnewline
$d_{2}^{c}$  & \threeS  & \one  & $\frac{1}{3}$  & 0 \tabularnewline
$\ell_{2}$  & \one  & \two  & $-\frac{1}{2}$  & 0 \tabularnewline
$e_{2}^{c}$  & \one  & \one  & 1  & 0 \tabularnewline
$q_{3}$  & \three  & \two  & 0  & $\frac{1}{6}$ \tabularnewline
$u_{3}^{c}$  & \threeS  & \one  & 0  & $-\frac{2}{3}$ \tabularnewline
$d_{3}^{c}$  & \threeS  & \one  & 0  & $\frac{1}{3}$ \tabularnewline
$\ell_{3}$  & \one  & \two  & 0  & $-\frac{1}{2}$ \tabularnewline
$e_{3}^{c}$  & \one  & \one  & 0  & 1 \tabularnewline
\midrule 
\rowcolor{yellow!10} $N_{{\rm atm}}$  & \one  & \one  & $\frac{2}{3}$  & $-\frac{2}{3}$ \tabularnewline
\rowcolor{yellow!10} $N_{{\rm sol}}$  & \one  & \one  & $\frac{2}{3}$  & $-\frac{2}{3}$ \tabularnewline
\midrule 
$H_{3}^{u}$  & \one  & \two  & 0  & $\frac{1}{2}$ \tabularnewline
$H_{3}^{d}$  & \one  & \two  & 0  & $-\frac{1}{2}$ \tabularnewline
\midrule 
$\phi_{\ell13}$  & \one  & \one  & $\frac{1}{2}$  & $-\frac{1}{2}$ \tabularnewline
$\phi_{\ell23}$  & \one  & \one  & $\frac{1}{2}$  & $-\frac{1}{2}$ \tabularnewline
$\phi_{q13}$  & \one  & \one  & $-\frac{1}{6}$  & $\frac{1}{6}$ \tabularnewline
$\phi_{q23}$  & \one  & \one  & $-\frac{1}{6}$  & $\frac{1}{6}$ \tabularnewline
$\phi_{u13}$  & \one  & \one  & $-\frac{2}{3}$  & $\frac{2}{3}$ \tabularnewline
$\phi_{u23}$  & \one  & \one  & $-\frac{2}{3}$  & $\frac{2}{3}$ \tabularnewline
\bottomrule
\end{tabular}\caption[Example $SU(5)^{3}$ tri-unification model: energy regime 7]{Fermion and scalar representations under $SU(3)_{c}\times SU(2)_{L}\times U(1)_{Y_{12}}\times U(1)_{Y_{3}}$
in energy regime 7. Fermions highlighted in yellow belong to a vector-like
pair and thus have a conjugate representation not shown in this table.
\label{tab:ParticleContent4} }
\end{table}

}

At $M_{Q}\lesssim M_{H_{2}^{u,d}}$, the $Q_{i}$ vector-like quarks and the $\Theta_{i}$ colour octets
decouple from the particle spectrum of the model. As in the previous
two energy thresholds, the gauge symmetry is not altered. The particle
spectrum at this stage is that shown on Table~\ref{tab:ParticleContent3}
removing the $\xi$ singlet fermions, the $H_{1,2}^{u,d}$ scalar
doublets, the $Q_{i}$ vector-like quarks and the $\Theta_{i}$ colour octets.

Hyperons are responsible for the breaking of the tri-hypercharge symmetry.
In a first hypercharge breaking step, $U(1)_{Y_{1}}\times U(1)_{Y_{2}}\times U(1)_{Y_{3}}$
gets broken to $U(1)_{Y_{12}}\times U(1)_{Y_{3}}$, where $Y_{12}=Y_{1}+Y_{2}$,
by the non-zero VEV of the $\phi_{q12}$ hyperon, $v_{12}=\langle\phi_{q12}\rangle\sim50$
TeV: 
\begin{equation}
SU(3)_{c}\times SU(2)_{L}\times U(1)_{Y_{1}}\times U(1)_{Y_{2}}\times U(1)_{Y_{3}}\xrightarrow{\langle\phi_{q12}\rangle} SU(3)_{c}\times SU(2)_{L}\times U(1)_{Y_{12}}\times U(1)_{Y_{3}}
\end{equation}
The gauge couplings above ($g_{Y_{1}}$ and $g_{Y_{2}}$) and below
($g_{Y_{12}}$) the breaking scale verify the matching relation 
\begin{equation}
\frac{g_{Y_{1}}\,g_{Y_{2}}}{\sqrt{g_{Y_{1}}^{2}+g_{Y_{2}}^{2}}}=g_{Y_{12}}\,,
\end{equation}
which is equivalent to 
\begin{equation}
\alpha_{Y_{1}}^{-1}+\alpha_{Y_{2}}^{-1}=\alpha_{Y_{12}}^{-1}\,.
\end{equation}
The ``12 hyperons'' $\phi_{\ell12}$, $\phi_{q12}$ and $\phi_{u12}$
get masses of the order of $\langle\phi_{q12}\rangle$ and decouple
at this stage. We also assume the $\Theta_{i}$ colour octets get
a mass $M_{\Theta}\sim M_{Q}$ and are integrated out at this scale
scale as well. The resulting fermion and scalar particle content is
shown in Table~\ref{tab:ParticleContent4}.

\subsection*{Regime 7: $\boldsymbol{SU(3)_{c}\times SU(2)_{L}\times U(1)_{Y_{1}}\times U(1)_{Y_{2}}\times U(1)_{Y_{3}}}$
breaking scale\protect \\
 \hspace*{2cm} $\boldsymbol{\to SU(3)_{c}\times SU(2)_{L}\times U(1)_{Y_{12}}\times U(1)_{Y_{3}}}$
breaking scale}

The $SU(3)_{c}\times SU(2)_{L}\times U(1)_{Y_{12}}\times U(1)_{Y_{3}}$
gauge symmetry also gets broken by hyperon VEVs, leaving as a remnant
the conventional SM gauge symmetry with $Y=Y_{12}+Y_{3}=Y_{1}+Y_{2}+Y_{3}$.
In this case, the hyperons responsible for the breaking are $\phi_{\ell13}$,
$\phi_{\ell23}$, $\phi_{q13}$ and $\phi_{q23}$, which get VEVs
of the order of $v_{23}\sim5$ TeV: 
\begin{equation}
SU(3)_{c}\times SU(2)_{L}\times U(1)_{Y_{12}}\times U(1)_{Y_{3}}\xrightarrow{\langle\phi_{\ell13,23}\rangle,\langle\phi_{q13,23}\rangle} SU(3)_{c}\times SU(2)_{L}\times U(1)_{Y}
\end{equation}
The gauge couplings above ($g_{Y_{12}}$ and $g_{Y_{3}}$) and below
($g_{Y}$) the breaking scale verify the matching relation 
\begin{equation}
\frac{g_{Y_{12}}\,g_{Y_{3}}}{\sqrt{g_{Y_{12}}^{2}+g_{Y_{3}}^{2}}}=g_{Y}\,,
\end{equation}
which is equivalent to 
\begin{equation}
\alpha_{Y_{12}}^{-1}+\alpha_{Y_{3}}^{-1}=\alpha_{Y}^{-1}\,.
\end{equation}
All the remaining hyperons as well as the neutrino mass messengers
$N_{{\rm atm}}$ and $N_{{\rm sol}}$ (as well as their conjugate
representations) decouple at this stage. The resulting particle spectrum
is that of a two Higgs doublet model, with universal charges for all
fermions.

\subsection*{Regime 8: $\boldsymbol{SU(3)_{c}\times SU(2)_{L}\times U(1)_{Y_{12}}\times U(1)_{Y_{3}}}$
breaking scale\protect \\
 \hspace*{2cm} $\boldsymbol{\to SU(3)_{c}\times SU(2)_{L}\times U(1)_{Y}}$
breaking scale}

Finally, at the scale $v_{\mathrm{SM}}$, the electroweak symmetry
gets broken in the usual way, by the VEVs of the $H_{3}^{u,d}$ scalar
doublets: 
\begin{equation}
SU(3)_{c}\times SU(2)_{L}\times U(1)_{Y}\quad\xrightarrow{\langle H_{3}^{u,d}\rangle}\quad SU(3)_{c}\times U(1)_{{\rm em}}
\end{equation}

\subsection{Gauge coupling unification}

\label{sec:gcu}

In order to ensure that the gauge couplings of our model do indeed
unify into a single value at some high energy scale, we must solve
their one-loop RGEs, which take the generic form~\cite{Machacek:1983tz}
\begin{equation}
\frac{dg_{i}}{d\ln\mu}=\frac{g_{i}^{3}}{16\,\pi^{2}}\,b_{i}\,.
\end{equation}
The $b_{i}$ coefficients depend on the specific group $G_{i}$, with
gauge coupling $g_{i}$, and the representations in the model. They
are given by 
\begin{equation}
b_{i}=-\frac{11}{3}\,C_{2}(G_{i})+\frac{4}{3}\,\kappa\,S_{2}(F_{i})+\frac{1}{6}\,\eta\,S_{2}(S_{i})\,.
\end{equation}
Here $\mu$ is the renormalization scale, $C_{2}(G_{i})$ is the quadratic
Casimir of the adjoint representation of $G_{i}$ and $S_{2}(F_{i})$
and $S_{2}(S_{i})$ are the sums of the Dynkin indices of all fermion
and scalar non-trivial representations under $G_{i}$. Finally, $\kappa=1\,(1/2)$
for Dirac (Weyl) fermions and $\eta=2\,(1)$ for complex (real) scalars.

\begin{table}[tb]
\centering{}%
\resizebox{14.5cm}{!}{
\begin{tabular}{ccc}
\toprule 
\textbf{Regime}  & \textbf{Gauge group}  & \textbf{$\boldsymbol{b_{i}}$ coefficients} \tabularnewline
\midrule 
1  & SM$^{3}$  & $\left(-\frac{22}{3},-3,\frac{46}{15},-\frac{22}{3},-3,\frac{46}{15},-\frac{22}{3},-3,\frac{46}{15}\right)$ \tabularnewline
2  & $SU(3)_{c}\times SU(2)_{L}\times U(1)_{Y_{1}}\times U(1)_{Y_{2}}\times U(1)_{Y_{3}}$  & $\left(0,\frac{11}{3},\frac{122}{45},\frac{122}{45},\frac{46}{15}\right)$ \tabularnewline
3  & $SU(3)_{c}\times SU(2)_{L}\times U(1)_{Y_{1}}\times U(1)_{Y_{2}}\times U(1)_{Y_{3}}$  & $\left(0,\frac{11}{3},\frac{104}{45},\frac{104}{45},\frac{8}{3}\right)$ \tabularnewline
4  & $SU(3)_{c}\times SU(2)_{L}\times U(1)_{Y_{1}}\times U(1)_{Y_{2}}\times U(1)_{Y_{3}}$  & $\left(0,\frac{10}{3},\frac{19}{9},\frac{104}{45},\frac{8}{3}\right)$ \tabularnewline
5  & $SU(3)_{c}\times SU(2)_{L}\times U(1)_{Y_{1}}\times U(1)_{Y_{2}}\times U(1)_{Y_{3}}$  & $\left(0,3,\frac{19}{9},\frac{19}{9},\frac{8}{3}\right)$ \tabularnewline
6  & $SU(3)_{c}\times SU(2)_{L}\times U(1)_{Y_{1}}\times U(1)_{Y_{2}}\times U(1)_{Y_{3}}$  & $\left(-4,-3,\frac{89}{45},\frac{89}{45},\frac{38}{15}\right)$ \tabularnewline
7  & $SU(3)_{c}\times SU(2)_{L}\times U(1)_{Y_{12}}\times U(1)_{Y_{3}}$  & $\left(-7,-3,\frac{11}{3},\frac{38}{15}\right)$ \tabularnewline
8  & $SU(3)_{c}\times SU(2)_{L}\times U(1)_{Y}$  & $\left(-7,-3,\frac{21}{5}\right)$ \tabularnewline
\bottomrule
\end{tabular}}\caption[Beta function coefficients of the example $SU(5)^{3}$ tri-unification model]{$b_{i}$ coefficients of our model. See Section~\ref{sec:app} for
details on the gauge symmetries and particle content at each energy
regime. \label{tab:bcoeffs}}
\end{table}
\begin{figure}[t]
\centering \includegraphics[width=0.75\linewidth]{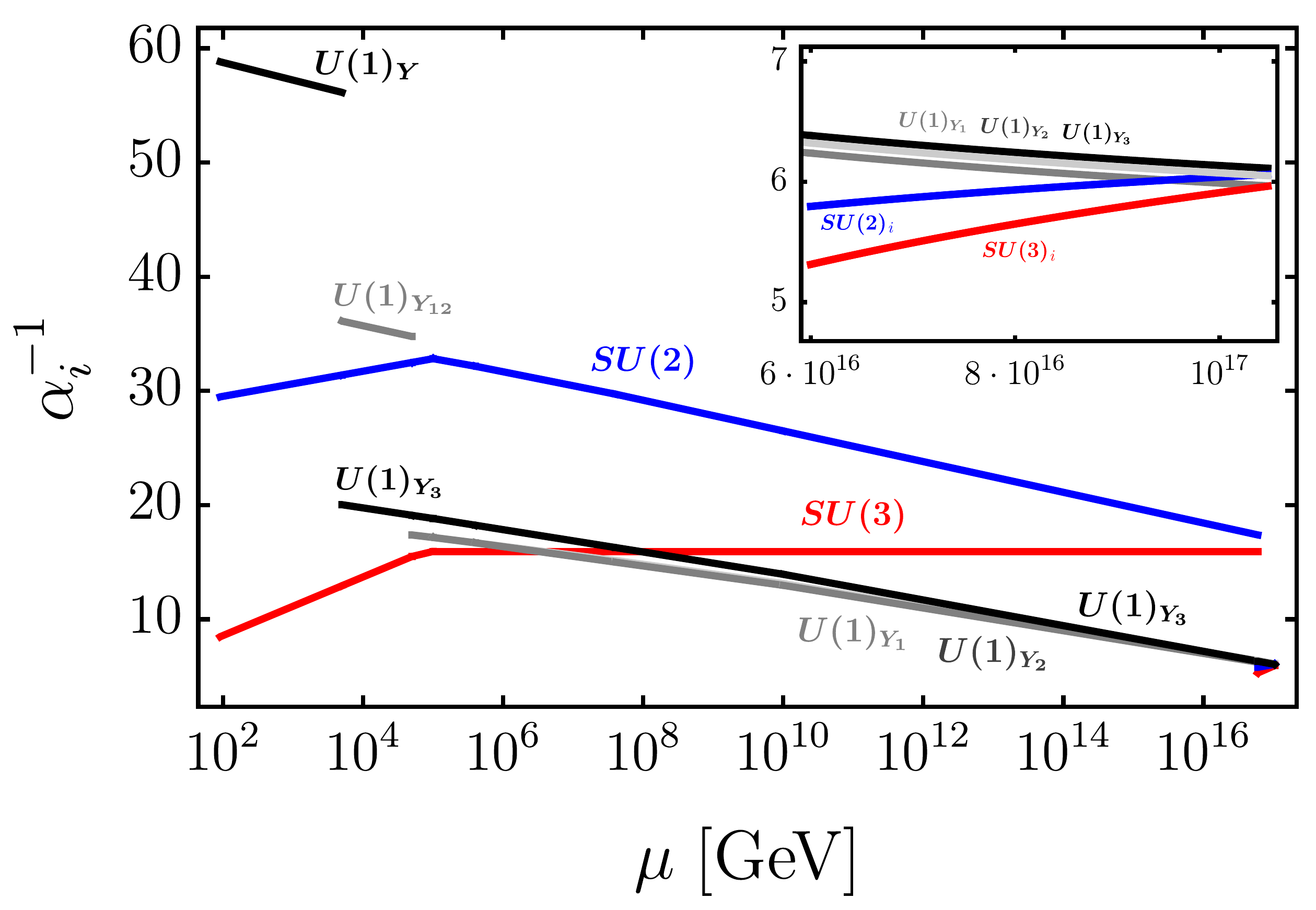}
\caption[Running of the gauge couplings in the example $SU(5)^{3}$ tri-unification model: first benchmark]{Running of the gauge couplings. The red lines correspond to the $SU(3)$
gauge couplings, the blue ones to the $SU(2)$ gauge couplings and
the black/grey ones to the $U(1)$ gauge couplings. A zoom-in with
the high-energy region close to the unification scale is also shown.
These results have been obtained with $v_{23}=5$ TeV, $v_{12}=50$
TeV, $M_{Q}=100$ TeV, $M_{H_{2}^{u,d}}=400$ TeV, $M_{H_{1}^{u,d}}=4\cdot10^{4}$
TeV, $M_{\xi}=10^{10}$ GeV and $v_{{\rm {SM}^{3}}}=6\cdot10^{16}$
GeV. \label{fig:running} }
\end{figure}
\begin{figure}[t]
\centering \includegraphics[width=0.49\linewidth]{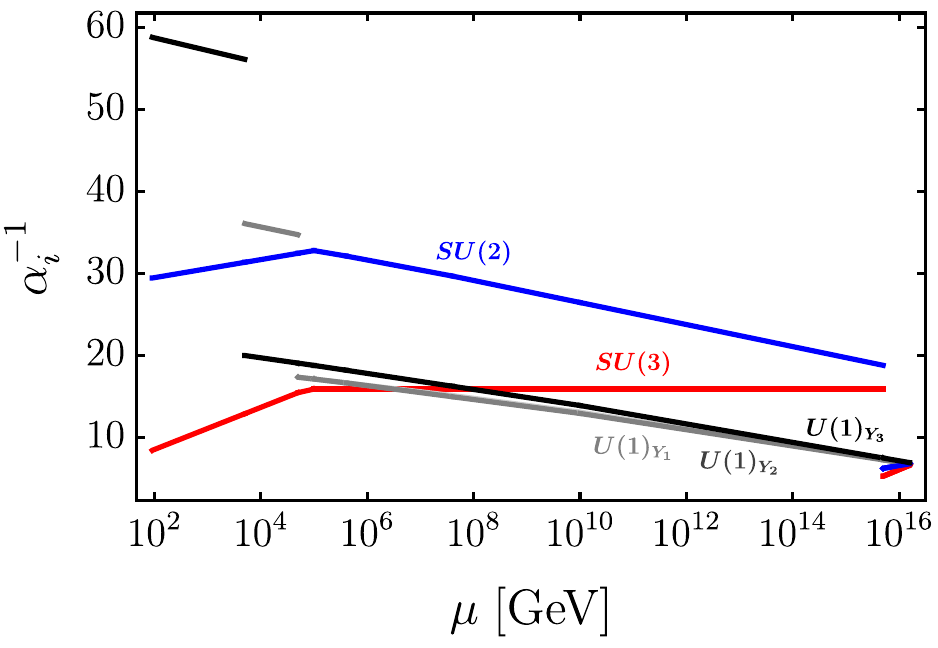}
\includegraphics[width=0.485\linewidth]{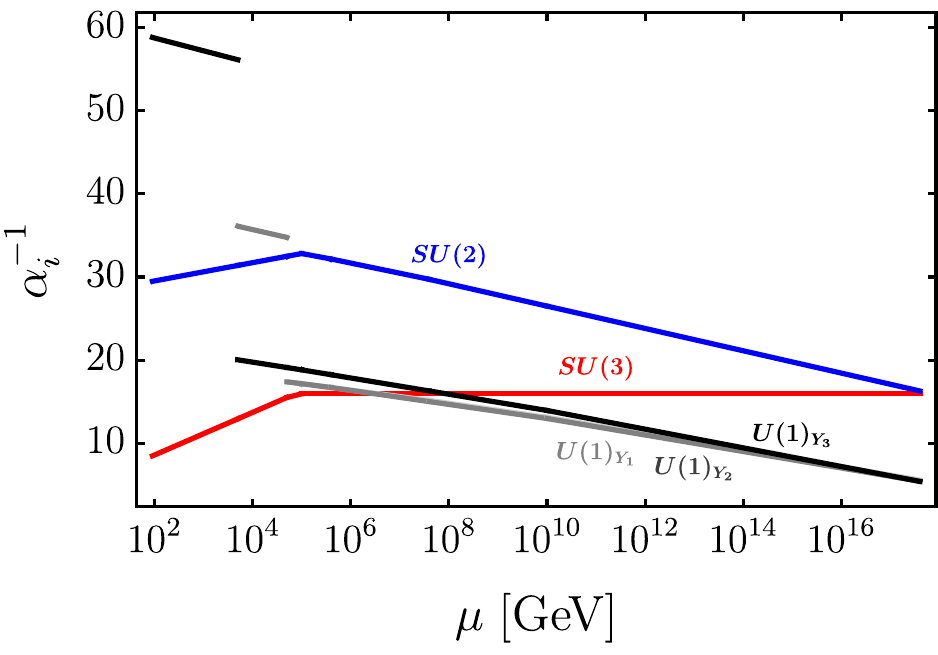}
\caption[Running of the gauge couplings in the example $SU(5)^{3}$ tri-unification model: alternative benchmarks]{Running of the gauge couplings. Colour code as in Fig.~\ref{fig:running}.
These results have been obtained with $v_{{\rm {SM}^{3}}}=5\cdot10^{15}$
GeV (left) and $v_{{\rm {SM}^{3}}}=4\cdot10^{17}$ GeV (right). The
rest of the intermediate scales have been chosen as in Eq.~(\ref{eq:scales}).
\label{fig:running2} }
\end{figure}
We computed the $b_{i}$ coefficients of our model, taking into account
not only the gauge group for each energy regime, but also the particle
content, since a particle decouples and does not contribute to the
running at energies below its mass. The gauge symmetries and particle
content at each energy regime are described in detail in the previous
section, whereas our results for the $b_{i}$ coefficients of the
model are given in Table~\ref{tab:bcoeffs}. Finally, we display
results for the running of the gauge couplings in Fig.~\ref{fig:running}.
This figure has been obtained by fixing the intermediate energy scales
to 
\begin{equation} \label{eq:scales}
  \begin{split}
    &v_{23} = 5 \; \text{TeV} \, , \\   
    &M_{\Theta} = 100 \; \text{TeV} \, , \\
    &M_\xi = 10^{10} \; \text{GeV} \, , 
  \end{split}
\quad
  \begin{split}
    &v_{12} = 50 \, \text{TeV} \; , \\
    &M_{H_2^{u,d}} = 400 \, \mathrm{TeV} , \\
    &v_{\rm{SM}^3} = 6 \cdot 10^{16} \; \text{GeV} \,.
  \end{split}
  \quad
  \begin{split}
   &M_Q = 100 \; \text{TeV} \; , \\
   &M_{H_1^{u,d}} = 4 \cdot 10^4 \; \text{TeV} \; , \\
   &
  \end{split}
\end{equation}
The nine gauge couplings of the SM$^{3}$ group unify at a very high
unification scale $M_{{\rm GUT}}\approx10^{17}$ GeV, slightly above
the SM$^{3}$ breaking scale, with a unified gauge coupling $g_{{\rm GUT}}\approx1.44$.
We note the important role played by three $\Theta_{i}$ colour
octets embedded into $\Omega_{ij}$, and by the $Q_{i}$ vector-like quarks which also act as heavy messengers of the flavour theory,
which are crucial to modify the running of the $SU(3)$ and $SU(2)$ gauge couplings
in order to achieve unification. We also highlight that the discontinuities in Figs.~\ref{fig:running} and \ref{fig:running2} are due to the gauge coupling matching conditions that apply at the steps in which the $U(1)_{Y}$ group is decomposed into two (first discontinuity) and three hypercharges (second discontinuity) and in which the $SU(3)$ and $SU(2)$ groups are decomposed into one for each family (third discontinuity).

Even though the $\mathbb{Z}_{3}$ symmetry
gets broken at the SM$^{3}$ breaking scale, it stays approximately
conserved at low energies, down to the tri-hypercharge breaking scale,
and only the running of $U(1)_{Y_{3}}$ is slightly different from
that of the other two hypercharge groups. In fact, the gauge couplings
of the $U(1)_{Y_{1}}$ and $U(1)_{Y_{2}}$ groups almost overlap and
cannot be distinguished in Fig.~\ref{fig:running}. This can be easily
understood by inspecting the $b_{i}$ coefficients on Table~\ref{tab:bcoeffs}.
Then, the matching conditions at $v_{12}=50$ TeV split the low energy
$g_{Y_{12}}$ and $g_{Y_{3}}$ couplings, which become clearly different:
$g_{Y_{12}}(v_{12})\approx0.59$ and $g_{Y_{3}}(v_{12})\approx0.79$.
Finally, at $v_{23}=5$ TeV one recovers the standard $SU(3)_{c}\times SU(2)_{L}\times U(1)_{Y}$
gauge group, which remains unbroken down to the electroweak scale.

\begin{figure}
\centering %
\begin{minipage}[c]{0.53\textwidth}%
 \centering \includegraphics[width=1\linewidth]{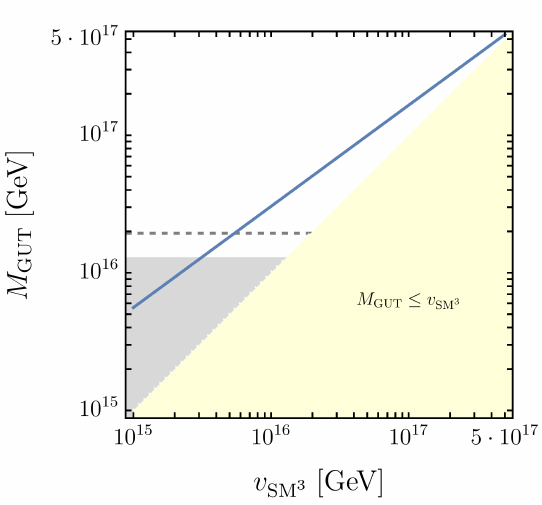} %
\end{minipage}%
\begin{minipage}[c]{0.445\textwidth}%
 \centering \vspace*{0.15cm}
 \includegraphics[width=1\linewidth]{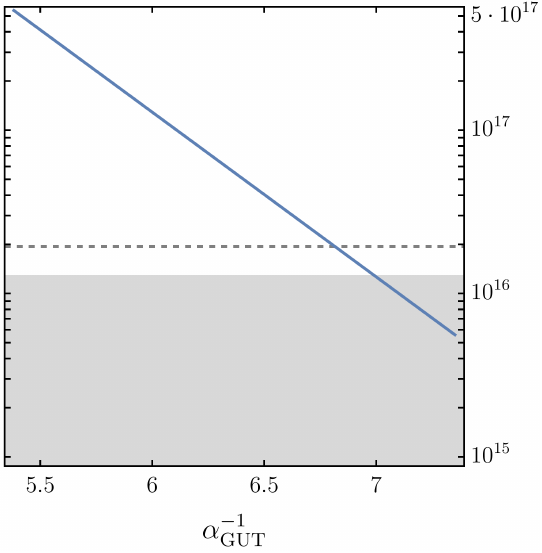} %
\end{minipage}\caption[$M_{{\rm GUT}}$ as a function of $v_{{\rm {SM}^{3}}}$ and
$\alpha_{{\rm GUT}}^{-1}$ in the example $SU(5)^{3}$ tri-unification model]{$M_{{\rm GUT}}$ as a function of $v_{{\rm {SM}^{3}}}$ (left) and
$\alpha_{{\rm GUT}}^{-1}$ (right). The SM$^{3}$ breaking scale $v_{{\rm {SM}^{3}}}$
varies in these plots, while the rest of the intermediate scales have
been fixed to the values in Eq.~(\ref{eq:scales}). The shaded grey
region is excluded by the existing Super-Kamiokande 90\% C.L. limit
on the $p\to e^{+}\pi^{0}$ lifetime, $\tau(p\to e^{+}\pi^{0})>2.4\cdot10^{34}$
years~\cite{Super-Kamiokande:2020wjk}, whereas the horizontal dashed
line corresponds to the projected Hyper-Kamiokande sensitivity at
90\% C.L. after 20 years of runtime, $\tau(p\to e^{+}\pi^{0})>1.2\cdot10^{35}$
years, obtained in~\cite{Bhattiprolu:2022xhm}. See Section~\ref{subsec:protondecay2}
for details on the proton decay calculation. Finally, the shaded yellow
region on the left-hand plot is excluded due to $M_{{\rm GUT}}\protect\leq v_{{\rm {SM}^{3}}}$.
\label{fig:running3} }
\end{figure}

In order to study how unification changes with the scale of SM$^{3}$
breaking, $v_{{\rm {SM}^{3}}}$, we consider the values $v_{{\rm {SM}^{3}}}=5\cdot10^{15}$
GeV and $v_{{\rm {SM}^{3}}}=4\cdot10^{17}$ GeV and fix the rest of
the intermediate scales as in Eq.~(\ref{eq:scales}). Results for
the running of the gauge couplings in these two scenarios are shown
in Fig.~\ref{fig:running2}. On the left-hand side we show the case
$v_{{\rm {SM}^{3}}}=5\cdot10^{15}$ GeV whereas on the right-hand
side we display our results for $v_{{\rm {SM}^{3}}}=4\cdot10^{17}$
GeV. In the first case, our choice of SM$^{3}$ breaking scale leads
to unification of the gauge couplings at a relatively low scale, $M_{{\rm GUT}}\approx1.8\cdot10^{16}$
GeV. This is potentially troublesome, as it may lead to too fast proton
decay, as explained below. In contrast, when the SM$^{3}$ breaking
scale is chosen to be very high, as in the second scenario, unification
also gets delayed to much higher energies. In fact, we note that with
our choice $v_{{\rm {SM}^{3}}}=4\cdot10^{17}$ GeV, gauge coupling
unification already takes place at the SM$^{3}$ breaking scale, $M_{{\rm GUT}}\approx v_{{\rm {SM}^{3}}}$.
In this case, $SU(5)^{3}$ breaks directly to the tri-hypercharge
group $SU(3)_{c}\times SU(2)_{L}\times U(1)_{Y_{1}}\times U(1)_{Y_{2}}\times U(1)_{Y_{3}}$
and there is no intermediate SM$^{3}$ scale. Finally, the impact
of $v_{{\rm {SM}^{3}}}$ is further illustrated in Fig.~\ref{fig:running3}.
Here we show the relation between $M_{{\rm GUT}}$, $v_{{\rm {SM}^{3}}}$
and $\alpha_{{\rm GUT}}^{-1}=4\pi/g_{{\rm GUT}}^{2}$. These two plots
have been made by varying $v_{{\rm {SM}^{3}}}$ and all the other
intermediate scales fixed as in Eq.~(\ref{eq:scales}). The left-hand
side of this figure confirms that larger $v_{{\rm {SM}^{3}}}$ values
lead to higher unification scales and smaller gaps between these two
energy scales. The right-hand side of the figure shows the relation
between the unified gauge coupling and the GUT scale. Again, the larger
$M_{{\rm GUT}}$ (or, equivalently, larger $v_{{\rm {SM}^{3}}}$)
is, the larger $g_{{\rm GUT}}$ (and smaller $\alpha_{{\rm GUT}}^{-1}$)
becomes. In particular, in this plot $g_{{\rm GUT}}$ ranges from
$\sim1.30$ to $\sim1.53$.

\subsection{Proton decay\label{subsec:protondecay2}}

As in any GUT, proton decay is a major prediction in our setup. In
standard, non-supersymmetric $SU(5)$ the most relevant proton decay
mode is usually $p\to e^{+}\pi^{0}$. This process is induced by the
tree-level exchange of the $X\sim(\mathbf{3},\mathbf{2})_{-5/6}$ (and
complex conjugate $X^{*}\sim(\mathbf{\overline{3}},\mathbf{\overline{2}})_{5/6}$)
gauge bosons contained in the $\mathbf{24}$ (adjoint) representation.
Integrating out these heavy vector leptoquarks leads to effective
dimension-6 operators that violate both baryon and lepton number,
as described in Section~\ref{subsec:ProtonDecay}. The resulting proton life time can be
roughly estimated as 
\begin{equation}
\tau_{p}\approx\frac{m_{X}^{4}}{\alpha_{{\rm GUT}}^{2}m_{p}^{5}}\,,\label{eq:proton_app}
\end{equation}
where $m_{X}$ is the mass of the heavy leptoquark, $m_{p}\approx0.938$
GeV is the proton mass and $\alpha_{{\rm GUT}}=g_{{\rm GUT}}^{2}/(4\pi)$
is the value of the fine structure constant at the unification scale.
One can easily estimate that for $m_{X}=10^{17}$ GeV and $g_{{\rm GUT}}\sim1.5$,
the proton life time is $\tau_{p}\sim10^{38}$ years, well above the
current experimental limit, $\tau(p\to e^{+}\pi^{0})>2.4\cdot10^{34}$
years at 90\% C.L. ~\cite{Super-Kamiokande:2020wjk}. Therefore,
a large unification scale suffices to guarantee that our model respects
the current limits on the proton lifetime. In fact, such a long life
time is beyond the reach of near future experiments, which will increase
the current limit by about one order of magnitude~\cite{Bhattiprolu:2022xhm}.
\begin{table}[tb]
\centering{}\resizebox{14.5cm}{!}{ %
\begin{tabular}{ccc}
\toprule 
\textbf{Gauge group}  & \textbf{$\gamma_{iL}$ coefficients}  & \textbf{$\gamma_{iR}$ coefficients} \tabularnewline
\midrule 
SM$^{3}$  & $\left(2,\frac{9}{4},\frac{23}{20},0,0,0,0,0,0\right)$  & $\left(2,\frac{9}{4},\frac{11}{20},0,0,0,0,0,0\right)$\tabularnewline
$SU(3)_{c}\times SU(2)_{L}\times U(1)_{Y_{1}}\times U(1)_{Y_{2}}\times U(1)_{Y_{3}}$  & $\left(2,\frac{9}{4},\frac{23}{20},0,0\right)$  & $\left(2,\frac{9}{4},\frac{11}{20},0,0\right)$\tabularnewline
$SU(3)_{c}\times SU(2)_{L}\times U(1)_{Y_{12}}\times U(1)_{Y_{3}}$  & $\left(2,\frac{9}{4},\frac{23}{20},0\right)$  & $\left(2,\frac{9}{4},\frac{11}{20},0\right)$\tabularnewline
$SU(3)_{c}\times SU(2)_{L}\times U(1)_{Y}$  & $\left(2,\frac{9}{4},\frac{23}{20}\right)$  & $\left(2,\frac{9}{4},\frac{11}{20}\right)$\tabularnewline
\bottomrule
\end{tabular}}\caption[Anomalous dimension coefficients for proton decay operators in the
example $SU(5)^{3}$ tri-unification model]{Anomalous dimension coefficients $\gamma_{iL,R}$ for proton decay
operators in our model. These are computed as 1-loop quantum corrections
(vertex corrections and self-energy corrections via the various gauge
bosons at each intermediate scale) to the proton decay operators $\mathcal{O}_{L}^{d=6}$
and $\mathcal{O}_{R}^{d=6}$ defined in Eqs.~\eqref{eq:operator_OL} and \eqref{eq:operator_OR}. We compute
them by following the algorithm of Appendix A in Ref.~\cite{Chakrabortty:2019fov}. \label{tab:gammacoeffs}}
\end{table}

A more precise determination of the $p\to e^{+}\pi^{0}$ decay width
is obtained by following the formalism of Section~\ref{subsec:ProtonDecay} and applying
Eq.~\eqref{eq:proton_decay_width}. One needs to know the $\beta$-function coeficients ($b_{i}$)
given for our model in Table~\ref{tab:bcoeffs}, along with the anomalous
dimensions ($\gamma_{iL(R)}$) of the proton decay operators, given
for each intermediate scale of our model in Table~\ref{tab:gammacoeffs}. 

Given that in our model we have three $SU(5)$ groups, we actually
have three generations of the usual $SU(5)$ leptoquarks, coupling
only to their corresponding family of chiral fermions. However, since
the three $SU(5)_{i}$ groups are all broken down to their $\mathrm{SM}_{i}$
subgroups at the same scale, in practice the model reproduces the
phenomenology of a flavour universal leptoquark as in conventional
$SU(5),$ albeit with the specific fermion mixing predicted by our
flavour model as shown in Section~\ref{subsec:Charged_fermions}.
The effect of fermion mixing is encoded via the coefficients $C_{L}$
and $C_{R}$ defined in Eqs.~ \eqref{eq:proton_CL} and \eqref{eq:proton_CR} in all generality. In our especific
model, their expression is obtained by just setting $\Lambda_{2}=0$\footnote{This is due to the fact that $\Lambda_{2}$ is associated to the $X'\sim(\mathbf{3},\mathbf{2})_{-1/6}$
gauge bosons which are not present in $SU(5)$ frameworks, as discussed
in Section~\ref{subsec:ProtonDecay}.}, i.e.
\begin{equation}
C_{L}=(V_{u^{c}}^{\dagger}V_{u})^{11}(V_{e^{c}}^{\dagger}V_{d})^{11}+(V_{u^{c}}^{\dagger}V_{u}V_{\mathrm{CKM}})^{11}(V_{e^{c}}^{\dagger}V_{d}V_{\mathrm{CKM}}^{\dagger})^{11}\,,
\end{equation}
\begin{equation}
C_{R}=(V_{u^{c}}^{\dagger}V_{u})^{11}(V_{d^{c}}^{\dagger}V_{e})^{11}\,,
\end{equation}
where $V_{\mathrm{CKM}}=V_{u}^{\dagger}V_{d}$. Notice that even though
our flavour model predicts non-generic fermion mixing, the alignment
of the CKM matrix is not univocally predicted but relies on the choice
of dimensionless coefficients. Assuming the CKM mixing to originate
mostly from the down sector we find $C_{L}\simeq1.946$ and $C_{R}\simeq0.999$,
while if the CKM mixing originates mostly from the up sector we find
$C_{L}\simeq1.946$ and $C_{R}\simeq0.974$. Both cases lead to very
similar low-energy phenomenology regarding proton decay.

\begin{figure}
\centering %
\begin{minipage}[c]{0.53\textwidth}%
\centering \includegraphics[width=1\linewidth]{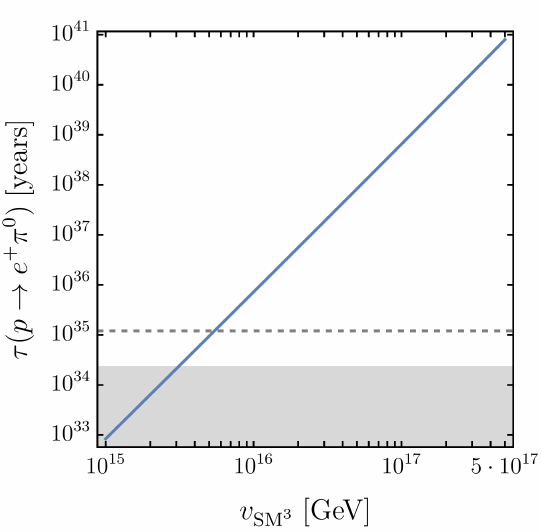} %
\end{minipage}%
\begin{minipage}[c]{0.445\textwidth}%
\centering \vspace*{-0.1cm}
 \includegraphics[width=1.01\linewidth]{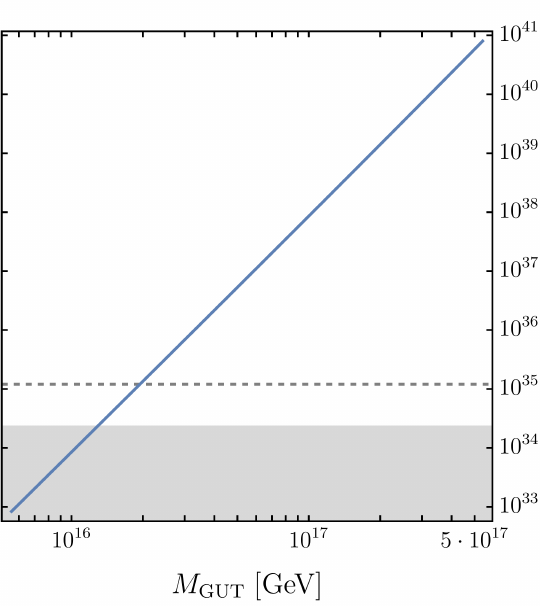} %
\end{minipage}\caption[$\tau(p\to e^{+}\pi^{0})$ as a function of $v_{{\rm {SM}^{3}}}$
and $M_{{\rm GUT}}$ in the example $SU(5)^{3}$ tri-unification model]{$\tau(p\to e^{+}\pi^{0})$ as a function of $v_{{\rm {SM}^{3}}}$
(left) and $M_{{\rm GUT}}$ (right). The SM$^{3}$ breaking scale
$v_{{\rm {SM}^{3}}}$ varies in these plots, while the rest of the
intermediate scales have been fixed to the values in Eq.~(\ref{eq:scales}).
The shaded grey region is excluded by the existing Super-Kamiokande
90\% C.L. limit on the $p\to e^{+}\pi^{0}$ lifetime, $\tau(p\to e^{+}\pi^{0})>2.4\cdot10^{34}$
years~\cite{Super-Kamiokande:2020wjk}, whereas the horizontal dashed
line corresponds to the projected Hyper-Kamiokande sensitivity at
90\% C.L. after 20 years of runtime, $\tau(p\to e^{+}\pi^{0})>1.2\cdot10^{35}$
years, obtained in~\cite{Bhattiprolu:2022xhm}. \label{fig:proton} }
\end{figure}

We show our numerical results for the $p\to e^{+}\pi^{0}$ lifetime
in Fig.~\ref{fig:proton}. Again, $v_{{\rm {SM}^{3}}}$ varies in
the left panel of this figure, while the rest of intermediate scales
have been chosen as in Eq.~(\ref{eq:scales}). The right panel shows
an equivalent plot with the $p\to e^{+}\pi^{0}$ lifetime as a function
of $M_{{\rm GUT}}$. This figure provides complementary information
to that already shown in Fig.~\ref{fig:running3}. In both cases
we have used the precise determination of the lifetime in Eq.~(\ref{eq:proton_decay_width}),
but we note that the estimate in Eq.~(\ref{eq:proton_app}) actually
provides a very good approximation, with $\tau_{p}/\tau_{p}^{{\rm app}}\in\left[0.5,1.2\right]$
in the parameter region covered in Fig.~\ref{fig:proton}. The current
Super-Kamiokande 90\% C.L. limit on the $p\to e^{+}\pi^{0}$ lifetime,
$\tau(p\to e^{+}\pi^{0})>2.4\cdot10^{34}$ years~\cite{Super-Kamiokande:2020wjk},
excludes values of the GUT scale below $M_{{\rm GUT}}\sim1.3\cdot10^{16}$
GeV, while the projected Hyper-Kamiokande sensitivity at 90\% C.L.
after 20 years of runtime, $\tau(p\to e^{+}\pi^{0})>1.2\cdot10^{35}$
years~\cite{Bhattiprolu:2022xhm}, would push this limit on the unification
scale in our model to $M_{{\rm GUT}}\sim2\cdot10^{16}$ GeV. Therefore,
our model will be probed in the next round of proton decay searches,
although large regions of the parameter space predict a long proton
lifetime, well beyond any foreseen experiment.

\section{Conclusions \label{sec:conclusions}}

In this chapter we have discussed $SU(5)^{3}$ with cyclic symmetry
(``tri-unification'') as the possible gauge unified origin of many
gauge non-universal theories at low energies. The $SU(5)^{3}$ with
cyclic symmetry setup consists of a single gauge coupling and unifies
all SM fermions into a single irreducible representation. When spontaneously
broken, such framework may lead to gauge non-universal theories and
family structure. These theories are known to have many applications
for model building purposes, and more recently have been proposed
as a possible explanation for the flavour structure of the SM.

As a proof of concept, we have developed an $SU(5)^{3}$ tri-unification
example that is spontaneously broken to the tri-hypercharge gauge
group discussed in Chapter~\ref{Chapter:Tri-hypercharge}, which dynamically generates the flavour
structure of the SM (including the neutrino sector). We have shown
how the five gauge couplings of tri-hypercharge unify into the
single gauge coupling of the cyclic $SU(5)^{3}$ group, by assuming
minimal multiplet splitting, together with a set of relatively light
colour octet scalars that also play a role in spontaneous symmetry
breaking. The approximate conservation of the cyclic symmetry at low
energies is also crucial to achieve gauge unification, plus the heavy
messengers required to generate the flavour structure also modify
the RGE in the desired way, highlighting the minimality of the framework.

We have also studied proton decay in this example, and presented the
predictions of the proton lifetime in the dominant $e^{+}\pi^{0}$
channel. If $SU(5)^{3}$ breaks to tri-hypercharge in one step, then
we have shown that gauge unification happens at a very high scale
$M_{\mathrm{GUT}}\sim10^{17}\,\mathrm{GeV}$, predicting a long proton
lifetime well beyond any foreseen experiment. In contrast, if we assume
that $SU(5)^{3}$ first breaks to three SM groups (which then break
to tri-hypercharge), then the scale of $\mathrm{SM}^{3}$ breaking
becomes essentially a free parameter that allows to lower $M_{\mathrm{GUT}}$,
predicting a proton lifetime to be possibly observable at Hyper-Kamiokande.
In this manner, the signals on proton decay may allow to test the
model at high scales, while low energy signals associated with tri-hypercharge
enable the model to be tested by collider and flavour experiments.

More generally, as a take home message, we conclude that $SU(5)^{3}$
tri-unification reconciles the idea of gauge non-universality with
the idea of gauge coupling unification, opening up the possibility
to build consistent non-universal descriptions of Nature that are
valid all the way up to the scale of grand unification. This is remarkable
given that gauge non-universal theories are known to predict a rather
complicated gauge sector consisting of many arbitrary gauge couplings.
Our work opens the possibility to take these theories seriously all
the way to the GUT scale, providing a consistent framework were new
phenomenology such as cosmological imprints of the spontaneous symmetry
breaking may be studied, being complementary to the low energy signals
predicted by the gauge non-universal layer of physics.
%% ----------------------------------------------------------------
%% Conclusions.tex
%% ---------------------------------------------------------------- 
\chapter{Final thoughts and the future ahead} \label{Chapter: Conclusions}
\begin{quote}
    -``After decades of flavour model building, we understood nothing about the origin of flavour.''\\
    -``Maybe it is you that understood nothing!''
  \begin{flushright}
  \hfill \hfill $-$ Graham G. Ross, to a comment from $\quad\;\;$\\Álvaro de Rújula in a physics conference\footnote{Michele Frigerio, private communication.}
  \par\end{flushright}
\end{quote}

\noindent If you are reading this, and you went through the whole thesis before
reaching these lines, thanks! Despite all the sacrifices, I had my
best time while writing this and I hope I could somehow transmit you
my personal view of physics throughout all those lengthy chapters.
You may be wondering now what are the final conclusions, maybe you
are wondering if I learnt something about Nature after writing all
those pages. The honest truth is that I do not have many definite conclusions,
though. Nature is not giving much hints right now about her unsolved
mysteries, and the flavour puzzle is no exception. As I tried to introduce
in Chapter~\ref{chap:Chapter1}, the new dynamics connected to the origin of flavour
might be hidden \textit{anywhere} from the Planck scale to the electroweak scale, giving its name to this thesis.

Unlike other solutions
to open puzzles in Nature, such as TeV-scale SUSY and the QCD axion,
the flavour puzzle does not point to any particularly accessible energy
scale or region in the parameter space, with most solutions being
decoupling theories where everything can always be very heavy, recovering
the low-energy phenomenology of the Standard Model. Some even think
that there is no puzzle to solve at all, claiming that all flavour
parameters are ``technically natural'' (in the sense that they are
stable under radiative corrections \cite{tHooft:1979rat}), or that we may never understand
what is Nature hiding behind the flavour puzzle, at least not in humanity's
lifetime. In this thesis, however, I hope I managed to convince you
that this is \textit{not} my view of particle physics. I believe that sooner or
later we will find that an existing (or yet to be written) well-motivated
theory is realised in Nature, just as it has happened several times
in the past when it was thought as well that (particle) physics was over.
Even if nothing is found, we will still have learnt more about our
Universe, just by failing once more. My work on this thesis was motivated
by this spirit. 

In Chapter~\ref{chap:Chapter1} I did a somewhat lengthy introduction to the SM, focusing on
its open questions and on the particular role of flavour. Surprisingly,
while doing this I learnt about fermion unification in GUTs, about
custodial symmetry, about the open problems in cosmology, about the
Froggatt-Nielsen mechanism and many more things. Then in Chapter~\ref{chap:2}
I discussed key flavour observables which show experimental anomalies
or which are correlated in well-motivated BSM scenarios to the anomalous
observables, hence being important for discovery prospects. I included
the EFT description of these observables in terms of the LEFT, which
can then be easily matched to the SMEFT, both being very useful EFTs
for phenomenological studies. During this writing, I learnt not only
about these observables but also about the aforementioned EFTs, and
most notably I learnt how difficult it can be to keep updated the
experimental data of several (flavour) observables during the few-months
period of writing a PhD thesis. I will always remember this when I
have the urge to ask experimentalists about timelines in their data
analyses.

In Chapter~\ref{Chapter:Fermiophobic} I studied a class of local $U(1)'$ extensions of the SM,
where chiral fermions are uncharged under the $U(1)'$ but an exotic
family of vector-like fermions is charged, providing effective $Z'$ couplings
for chiral fermions via mixing. This feature gives the name \textit{fermiophobic}
to this class of models. Despite the many theoretical puzzles
behind the $(g-2)_{\mu}$ anomaly, I decided to study its BSM interpretation
and I provided a very simple $U(1)'$ fermiophobic scenario where
the solution to $(g-2)_{\mu}$ is correlated to a suppression of the
decay of the Higgs boson to two photons, which can test the validity
of this scenario with the sufficient experimental precision. The basic
idea is the chiral enhancement provided by vector-like leptons that
couple to the SM Higgs doublet. Afterwards, I considered a theory
of flavour with fermiophobic $Z'$ that can explain the origin of
the SM flavour structure. In this model, both the effective Yukawa
couplings of second and third family chiral fermions along with their
effective $Z'$ couplings arise via mixing with the heavy vector-like
fermions, hence connecting the origin of Yukawa couplings in the SM
with the low-energy phenomenology of the model. First family masses
are then implemented via a heavy Higgs doublet that gets a small effective
VEV via mixing with the Higgs doublets that break the electroweak symmetry. The mechanism of messenger dominance
\cite{Ferretti:2006df} points to several NP scales in the UV, hinting
to a multi-scale origin of flavour. If the $U(1)'$ factor is broken
at a relatively low scale, then there could be some signals in low
energy observables that could test the model. A connection with the
$(g-2)_{\mu}$ anomaly requires a minimal extension of the model in
order to obtain an effective coupling of the vector-like leptons to
the SM Higgs, however we find that ultimately it is not possible to address
the anomaly in the flavour model due to a correlated chiral enhancement of $\mathcal{B}(\tau\rightarrow\mu\gamma)$.
Nevertheless, by dropping the extra content required for $(g-2)_{\mu}$,
we have found that current data allows for the $Z'$ boson to be as
light as 1 TeV, with the leading constraint being $B_{s}-\bar{B}_{s}$
mixing. By exploring this model I learnt about messenger dominance,
and I learnt how heavy Higgs doublets can be an alternative to vector-like
fermions in order to play the role of heavy messengers that UV-complete the effective Yukawa operators of
a theory of flavour. I also learnt about Higgs diphoton decay and
its future projections at HL-LHC, and of course I learnt about chiral
enhancement and its crucial relation with $(g-2)_{\mu}$. This work
was motivated by a previous work made in 2021 where we connected both the
$R_{K^{(*)}}$ and $(g-2)_{\mu}$ anomalies in a fermiophobic $Z'$ framework
\cite{FernandezNavarro:2021sfb}.

In Chapter~\ref{Chapter:TwinPS} I studied a twin Pati-Salam theory of flavour which
contains a TeV-scale vector leptoquark $U_{1}\sim(\mathbf{3,1},2/3)$
\cite{FernandezNavarro:2022gst}. This twin Pati-Salam symmetry is
broken down to the SM in a two-step process, such that in an intermediate
step at the TeV scale the model is described by a 4321 gauge group
\cite{DiLuzio:2017vat}. This two-step pattern of symmetry breaking
hints to a multi-scale origin of flavour. The model features a fermiophobic
framework as well, where both the effective Yukawa couplings for chiral
fermions and their effective $U_{1}$ couplings originate again from
mixing with heavy vector-like fermions. The mechanism of messenger
dominance plays a fundamental role here in order to simultaneously explain
the fermion mass hierarchies and deliver the flavour structure required
to explain the so-called $B$-anomalies. I discovered that one vector-like
fermion family is not enough to achieve such flavour structure, but
indeed three vector-like fermion families are required. In this case,
the model predicts a plethora of low-energy signals in flavour observables,
several of them fundamentally related to the origin of fermion mass
hierarchies and mixing. The model can also be tested via direct searches
of the new heavy degrees of freedom at the LHC, including the vector-like
fermions, the $U_{1}$ leptoquark, and a coloron $g'\sim(\mathbf{8,1},0)$
and $Z'$ gauge bosons. While doing this work I learnt much about
low-energy flavour observables and their description in an EFT framework,
via performing a lengthy phenomenological analysis involving each low-energy
process correlated to the explanation of the $B$-anomalies.
I also learnt more about very useful tools for phenomenological studies,
including \texttt{DsixTools} \cite{Fuentes-Martin:2020zaz}, \texttt{Madgraph5}
\cite{Madgraph:2014hca}, \texttt{FeynRules} \cite{Feynrules:2013bka}
and \texttt{package-X} \cite{Patel:2016fam}.

In Chapter~\ref{Chapter:Tri-hypercharge} I studied the possibility that the SM originates from
a non-universal gauge theory in the UV. We argue that one of the most simple ways to achieve
this is by assigning a separate gauge hypercharge to each fermion family
at high energies \cite{FernandezNavarro:2023rhv}, broken down to
the usual weak hypercharge which is the diagonal subgroup. This simple framework
avoids the family replication of the SM, and could be the first step
towards a deep non-universal gauge structure in the UV. If the Higgs
doublet(s) only carry third family hypercharge, then only third family
renormalisable Yukawa couplings are allowed by the gauge symmetry,
explaining their heaviness. The light fermions are massless in first
approximation, delivering an accidental $U(2)^{5}$ flavour symmetry
in the Yukawa couplings. The small masses and mixing of the light
charged fermions are then introduced via non-renormalisable operators,
as a minimal breaking of the $U(2)^{5}$ symmetry. These non-renormalisable
operators depend explicitly on the scalars linking the different hypercharge
groups (hyperons), which get VEVs in order to break the ``tri-hypercharge''
symmetry down to SM hypercharge. In minimal models with simple sets
of hyperons, we found that a mildly hierarchical breaking of tri-hypercharge
naturally explains the mass hierarchies between first and second family
charged fermions, which naturally arise via several mass insertions
of the hyperons. In this manner, the tri-hypercharge gauge model also
hints to a multi-scale origin of flavour. We also found that in order
to explain neutrino mixing, it is useful to introduce right-handed
neutrinos which carry non-zero hypercharges (although their sum must
vanish), which then turn out to get Majorana masses at the lowest scale of
symmetry breaking, that could be as low as a few TeV. Indeed, the
model has a rich phenomenology if the NP scales are low: from flavour
violating observables to LHC physics and electroweak precision observables.
We observe that the $Z'$ boson contributing to the most dangerous
FCNCs is naturally heavier, while a second $Z'$ boson may be
as light as a few TeV because it is protected by the $U(2)^{5}$ symmetry.
The most promising discovery channels for this $Z'$ boson are dilepton
searches at the LHC and electroweak precision observables, which are
altered via an unavoidable $Z-Z'$ gauge mixing. During this project,
I learnt much about model building, including the different implementations
of the type I seesaw mechanism and the different textures for neutrino mass
matrices. I also learnt about the very interesting physics of electroweak
precision observables and about gauge mixing and/or kinetic mixing.

Finally, in Chapter~\ref{Chapter:Tri-unification} I proposed a gauge unified origin for gauge
non-universal frameworks such as the aforementioned tri-hypercharge
theory. In this manner, such theories may be taken seriously all the
way up to the GUT scale. The model consists on assigning a separate
$SU(5)$ group to each fermion family. However, assuming that the
three $SU(5)$ groups are related by a cyclic permutation symmetry
$\mathbb{Z}_{3}$, then the model is described by a single gauge coupling in
the UV, despite $SU(5)^{3}$ being a non-simple group. Moreover, the
cyclic symmetry in such a framework also ensures that all SM fermions belong to a single representation of
the complete group. First, I discussed
a general $SU(5)^{3}$ ``tri-unification'' framework for model building,
where gauge non-universal theories of flavour may be embedded. Secondly,
I constructed an explicit, minimal tri-hypercharge example model originating
from $SU(5)^{3}$ tri-unification, which can account for all the quark
and lepton (including neutrino) masses and mixing parameters. I achieved
the successful unification of the five gauge couplings of the tri-hypercharge
group into a single gauge coupling associated to the cyclic $SU(5)^{3}$
group, and I also studied the implications for the stability of the
proton in such a setup. During this project, I learnt much about model
building, about the group theory of $SU(5)$ and about proton decay. I also learnt much
about UV-completing the effective Yukawa couplings of a theory of
flavour via different sets of various vector-like fermions and/or
scalar fields. I also increased my knowledge in the very interesting
physics of gauge unification and RGE, along with the model building
helpful to achieve the successful unification of gauge couplings.

So this is all, isn't it? During my PhD I have studied a bunch of
theories of flavour and explored the discovery prospects if (some of) the new
physics scales are low. A significant part of my thesis work was originally
motivated by the $R_{K^{(*)}}$ anomalies, that turned out to disappear
when I was just starting my final year of PhD. The flavour anomalies
that remain are mostly under question, either in the theory side such as
$(g-2)_{\mu}$ and $b\rightarrow s\mu\mu$, or in the experimental
side such as $R_{D^{(*)}}$, where an upcoming measurement by the BaBar
collaboration might lead to a decrease in the overall significance
\cite{Allanach:2023bgg}. Nevertheless, more data keeps coming from
the various particle physics experiments, and new anomalies in well-motivated
channels such as the recent measurement of $\mathcal{B}(B^{+}\rightarrow K^{+}\nu\bar{\nu})$
by Belle II \cite{BelleIIEPS:2023} may eventually establish as
solid evidence for physics beyond the Standard Model. The origin of flavour may still be
around the corner, waiting to be discovered in particle physics experiments
that test the origin of flavour \textit{from the bottom-up}.

Additionally, in recent years cosmological observations, especially
the detection of gravitational waves, have been suggested as a possible
way to test BSM scenarios that are inaccessible to current particle
physics experiments. For example, in the tri-hypercharge model discussed in Chapters \ref{Chapter:Tri-hypercharge} and \ref{Chapter:Tri-unification},
the hierarchical breaking of $U(1)$ gauge factors in the early Universe
may produce (metastable) cosmic strings if the symmetry breaking happens at very
high scales, with the associated emission of a characteristic
signal of gravitational waves. In the same spirit, if the breaking
of the $U(1)$ factors is associated to first order phase transitions
in the early Universe, a characteristic multi-peaked signal of gravitational
waves may be within the reach of current and upcoming gravitational
waves observatories. This opens the possibility of testing multi-scale
theories such as those proposed in this thesis, among others, which
may explain some of the most fundamental open questions of the Standard
Model. These cosmological probes may be complementary to the existing
searches at particle physics experiments, such that the origin of
flavour could be tested \textit{from the top-down and from the bottom-up}.
The future ahead of us is exciting, with the possibility of testing theories
that were thought inaccessible, such as the type I seesaw mechanism and Grand
Unified Theories. I am strongly convinced that only by learning and
studying the many faces and aspects of Nature we will obtain a better
understanding of our Universe, and I hope that in the near future
I can study the many open questions in fundamental physics, propose
new theories that address these problems and test such new theories
up to whatever new physics scales are possible, all the way \textit{from
the Planck scale to the electroweak scale}.

%\begin{lstlisting}[caption=Listing of what an example listing would be like]
%This is a test listing
%
%The test listing has several lines
%to show how the listings
%will be displayed
%\end{lstlisting}

\appendix
%% ----------------------------------------------------------------
%% AppendixA.tex
%% ---------------------------------------------------------------- 

\chapter{Four-component and two-component spinor notation}
\label{app:2-component_notation}
In this appendix, we introduce the two-component notation taking all fermions
as left-handed Weyl spinors, which is common in the literature for model building
studies. We show the connection with the four-component, left-right notation, highlighting explicit examples.

\section{Two-component spinors}
In the chiral spinor representation, a (4-component) Dirac spinor
$\Psi$ consists of two independent (2-component) Weyl spinors
$\psi_{L}$ and $\psi_{R}$ with well-defined chirality,
\begin{equation}
\Psi=\left(\begin{array}{c}
\psi_{L}\\
\psi_{R}
\end{array}\right)\,.
\end{equation}
The spinor fields $\psi_{L}$ and $\psi_{R}$ transform under the Lorentz group $SO^{+}(1,3)\cong SU(2)_{L}\times SU(2)_{R}$
as $(\mathbf{2},\mathbf{1})$ and $(\mathbf{1},\mathbf{2})$, respectively, and
by convention we denote $\psi_{L}$ as the \textit{left-handed} Weyl spinor
and $\psi_{R}$ as the \textit{right-handed} Weyl spinor. Notice that
if $\psi_{L}$ is a left-handed spinor, then the hermitian conjugate
$\psi_{L}^{\dagger}$ is a right-handed spinor. In a similar way,
if $\psi_{R}^{\dagger}$ is a right-handed spinor, then $\psi_{R}^{\dagger}$
is a left-handed spinor. Therefore, any particular fermionic degrees
of freedom can be described equally well using a left-handed Weyl
spinor or a right-handed one. Given this relation, it is tempting
to get rid of chiral subscripts and write our quantum field theory
in terms of Weyl spinors with the same chirality. Let us redefine
the Weyl spinors as
\begin{equation}
\left.\begin{array}{c}
\psi_{L}\equiv\psi\\
\psi_{R}\equiv\psi^{c\dagger}
\end{array}\right\} \Rightarrow\Psi=\left(\begin{array}{c}
\psi\\
\psi^{c\dagger}
\end{array}\right)\,.
\end{equation}
In this notation, $\psi$ is a left-handed spinor and $\psi^{c}$ is a left-handed spinor
as well. $\psi^{c}$ is usually referred as the $CP$-conjugate of
the original right-handed spinor $\psi_{R}$, because the $CP$ transformation
maps Weyl spinors to their \textit{own} hermitian conjugate, and this
is the reason behind the $c$ superscript in the notation. However,
notice that in general $\psi$ and $\psi^{c}$ are fundamentally independent
degrees of freedom, not related by any $c$ transformation despite
what the notation may suggest. Majorana fermions are a remarkable
exception, for which the two spinors are related through hermitian
conjugation, i.e.~$\psi^{c}=\psi$.

In our quest to describe our chiral fermions via left-handed Weyl spinors
(rather than with 4-component, left-right Dirac spinors), it is
convenient to introduce the adjoint Dirac spinor $\overline{\Psi}$
and the gamma matrices in the 2-component formalism,
\begin{equation}
\overline{\Psi}\equiv\Psi^{\dagger}\gamma^{0}=\left(\begin{array}{cc}
\psi^{c} & \psi^{\dagger}\end{array}\right)\,,\qquad\gamma^{\mu}=\left(\begin{array}{cc}
0 & \sigma^{\mu}\\
\bar{\sigma}^{\mu} & 0
\end{array}\right)\,,
\end{equation}
with
\begin{equation}
\sigma^{\mu}=\left(1,\vec{\sigma}\right)\,,\qquad\bar{\sigma}^{\mu}=\left(1,-\vec{\sigma}\right)\,,
\end{equation}
where $\vec{\sigma}=(\sigma_{1},\sigma_{2},\sigma_{3})$ contains the Pauli matrices. In the chiral representation,
left-handed and right-handed Dirac spinors in 4-component notation are given by
\begin{equation}
\Psi_{L}\equiv P_{L}\Psi=\left(\begin{array}{c}
\psi\\
0
\end{array}\right)\,,\qquad\Psi_{R}\equiv P_{R}\Psi=\left(\begin{array}{c}
0\\
\psi^{c\dagger}
\end{array}\right)\,,\label{eq:4comp_2comp}
\end{equation}
where $P_{L,R}=(1\mp\gamma_{5})/2$ are the usual chiral projectors.

In terms of SM fermions (and right-handed neutrinos), this formalism
allows to relate the 4-component, left-right spinors with the 2-component left-handed spinors as 
\begin{equation}
\left(Q_{i},L_{i}\right)\equiv\left(Q_{Li},L_{Li}\right)\,,\quad\left(u_{i}^{c},d_{i}^{c},\nu_{i}^{c},e_{i}^{c}\right)\overset{CP}{\rightarrow}\left(u_{Ri},d_{Ri},\nu_{Ri},e_{Ri}\right)\,,\label{eq:CP_LR_3}
\end{equation}
where $i=1,2,3$ is a flavour index. Notice that the SM quantum numbers of
$\psi^{c}$ are flipped with respect to those of $\psi_{R}$. In theories with
vector-like fermions, the relations above can be generalised to
\begin{flalign}
 & \left(Q_{a},L_{a}\right)\equiv\left(Q_{La},L_{La}\right)\,, &  & \left(\overline{Q}_{a},\overline{L}_{a}\right)\overset{CP}{\rightarrow}\left(\widetilde{Q}_{Ra},\widetilde{L}_{Ra}\right)\,,\label{eq:CP_LR_1}\\
 & \left(u_{a}^{c},d_{a}^{c},\nu_{a}^{c},e_{a}^{c}\right)\overset{CP}{\rightarrow}\left(u_{Ra},d_{Ra},\nu_{Ra},e_{Ra}\right)\,, &  & \left(\overline{u_{a}^{c}},\overline{d_{a}^{c}},\overline{\nu_{a}^{c}},\overline{e_{a}^{c}}\right)\equiv\left(\widetilde{u}_{La},\widetilde{d}_{La},\widetilde{\nu}_{La},\widetilde{e}_{La}\right)\,.\label{eq:CP_LR_2}
\end{flalign}
where $a$ is a flavour index that runs for the given vector-like
fermion generations of the specific model. We note that in 2-component notation,
the bar for vector-like fermions has nothing to do with any transformation,
but just denotes the conjugate fermion partner, which carries the opposite quantum numbers.

We highlight that the 4-component Dirac notation with explicit chiral
indices is common for phenomenological studies, while the 2-component
notation in which all spinors are left-handed is common for model building.
The reason is that the 2-component notation is useful to describe extended gauge groups or
grand unified theories like $SU(5)$, where left-handed fermions $\psi$
unify with conjugate right-handed fermions $\psi^{c}$ in the same
representations. It is also useful for supersymmetric theories 
\cite{Martin:1997ns}. For extended reviews of 2-component notation, we recommend
\cite{Dreiner:2008tw,Martin:2012us}.

As an example of these relations between 4-component and 2-component
notations, in the following we transform fermion kinetic terms, Yukawa
couplings and the Weinberg operator from 4-component to 2-component
notation.

\section{Kinetic terms and gauge interactions}

In 4-component notation, the fermion kinetic terms in the Lagrangian
are generically given by
\begin{equation}
\mathcal{L_{\mathrm{kin}}}\supset i\overline{\Psi}_{L}\gamma^{\mu}D_{\mu}\Psi_{L}+i\overline{\Psi}_{R}\gamma^{\mu}D_{\mu}\Psi_{R}\,.
\end{equation}
Let us now obtain the kinetic terms in 2-component notation by using
Eq.~(\ref{eq:4comp_2comp}),
\begin{flalign}
i\overline{\Psi}_{L}\gamma^{\mu}D_{\mu}\Psi_{L} & =i\left(\begin{array}{cc}
0 & \psi^{\dagger}\end{array}\right)\gamma^{\mu}D_{\mu}\left(\begin{array}{c}
\psi\\
0
\end{array}\right)\\
 & =i\left(\begin{array}{cc}
0 & \psi^{\dagger}\bar{\sigma}^{\mu}\end{array}\right)D_{\mu}\left(\begin{array}{c}
\psi\\
0
\end{array}\right)=i\psi^{\dagger}\bar{\sigma}^{\mu}D_{\mu}\psi\,,
\end{flalign}
\begin{flalign}
i\overline{\Psi}_{R}\gamma^{\mu}D_{\mu}\Psi_{R} & =i\left(\begin{array}{cc}
\psi^{c} & 0\end{array}\right)\gamma^{\mu}D_{\mu}\left(\begin{array}{c}
0\\
\psi^{c\dagger}
\end{array}\right)\\
 & =i\left(\begin{array}{cc}
\psi^{c}\sigma^{\mu} & 0\end{array}\right)D_{\mu}\left(\begin{array}{c}
0\\
\psi^{c\dagger}
\end{array}\right)=i\psi^{c}\sigma^{\mu}D_{\mu}\psi^{c\dagger}\,.
\end{flalign}
As an example, the couplings of the generic quarks $Q_{i}$ and $d_{i}^{c}$
to the $Z_{\mu}$ boson would be given as
\begin{equation}
\mathcal{L}_{Z_{\mu}}\supset\left(g_{Q}Q_{i}^{\dagger}\bar{\sigma}^{\mu}Q_{i}+g_{d^{c}}d_{i}^{c}\sigma^{\mu}d_{i}^{c\dagger}\right)Z_{\mu}\,.
\end{equation}
However, in the literature it is common to
heavily abuse the notation and write $\sigma^{\mu},\bar{\sigma}^{\mu}\rightarrow\gamma^{\mu}$, i.e.
\begin{equation}
\mathcal{L}_{Z_{\mu}}\supset\left(g_{Q}Q_{i}^{\dagger}\gamma^{\mu}Q_{i}+g_{d^{c}}d_{i}^{c}\gamma^{\mu}d_{i}^{c\dagger}\right)Z_{\mu}\,.
\end{equation}
We advise the reader to remember that in 2-component notation the gamma (also called Dirac) matrices
must be formally exchanged by $\sigma^{\mu}$ and $\bar{\sigma}^{\mu}$.

\section{Yukawa interactions}

In 4-component notation, the SM Yukawa interactions for a generic fermion
family are given by
\begin{equation}
\mathcal{L}_{\mathrm{Yukawa}}=y_{ij}\overline{\Psi}_{Li}H\Psi_{Rj}+\mathrm{h.c.}\,,
\end{equation}
where remember that $H\sim(\boldsymbol{1,2},1/2)$. In 2-component
notation
\begin{equation}
\mathcal{L}_{\mathrm{Yukawa}}=y_{ij}\left(\begin{array}{cc}
0 & \psi_{i}^{\dagger}\end{array}\right)H\left(\begin{array}{c}
0\\
\psi_{j}^{c\dagger}
\end{array}\right)+\mathrm{h.c.}=y_{ij}\psi_{i}^{\dagger}H\psi_{j}^{c\dagger}+y_{ji}^{*}\psi_{i}H^{\dagger}\psi_{j}^{c}\,.
\end{equation}
Then in 2-component notation it is convenient to redefine the Higgs
doublet as $H(\mathbf{1,2},1/2)\rightarrow H(\mathbf{1,2},-1/2)$\footnote{Notice that in doing this one has to be careful with the implicit
$SU(2)_{L}$ indices of the Yukawa operator.} (and exchange the Yukawa matrix by its hermitian conjugate), such
that the Yukawa couplings are
\begin{equation}
\mathcal{L}_{\mathrm{Yukawa}}=y_{ij}\psi_{i}H\psi_{j}^{c}+\mathrm{h.c.}
\end{equation}
In this manner, the SM Yukawa couplings for each fermion family are
given by
\begin{equation}
\mathcal{L}_{\mathrm{Yukawa}}=y_{ij}^{u}Q_{i}\widetilde{H}u_{j}^{c}+y_{ij}^{d}Q_{i}Hd_{j}^{c}+y_{ij}^{e}L_{i}He_{j}^{c}+\mathrm{h.c.}\,,
\end{equation}
where $\tilde{H}=i\sigma_{2}H^{\dagger}$ is the $CP$-conjugate of $H$.

\section{Weinberg operator}

In 4-component notation, the Weinberg operator (see Eq.~(\ref{eq:Weinberg})) is given by\footnote{Remember that we now define the Higgs doublet as $H\sim(\mathbf{1,2},-1/2)$ which is more convenient for the 2-component notation.}
\begin{equation}
\mathcal{L}^{d=5}_{\mathrm{Weinberg}}= c_{ij}\left(\bar{L}_{Li}^{C}\widetilde{H}\right)\left(L_{Lj}\widetilde{H}\right)+\mathrm{h.c.}\,,
\end{equation}
where we have absorbed the cut-off scale in the dimensionful
coefficients $c_{ij}$. We need to apply charge conjugation $C$ to
the 4-component Dirac spinor,
\begin{equation}
\Psi^{C}=C\overline{\Psi}^{\mathrm{T}}=\left(\begin{array}{c}
\psi^{c}\\
\psi^{\dagger}
\end{array}\right)\,,
\end{equation}
in order to obtain
\begin{flalign}
\mathcal{L}^{d=5}_{\mathrm{Weinberg}}= & c_{ij}\left(\bar{L}_{Li}^{C}\widetilde{H}\right)\left(L_{Lj}\widetilde{H}\right)+\mathrm{h.c.}=c_{ij}\left[\left(\begin{array}{cc}
L_{i} & 0\end{array}\right)\widetilde{H}\right]\left[\widetilde{H}\left(\begin{array}{c}
L_{j}\\
0
\end{array}\right)\right]+\mathrm{h.c.}\\
 & =c_{ij}\left(L_{i}\widetilde{H}\right)\left(L_{j}\widetilde{H}\right)+\mathrm{h.c.} \nonumber
\end{flalign}

\chapter{Large mixing angle formalism and mass insertion approximation} \label{app:mixing_angle_formalism}
In this thesis, we study several models where heavy vector-like fermions
mix with the chiral fermions of the SM. For illustration purposes, let
us assume the following generic terms in the mass Lagrangian,
\begin{equation}
\mathcal{L}_{\mathrm{mass}}\supset x_{34}^{\psi}\phi\psi_{3}\overline{\psi}_{4}+M_{4}^{\psi}\psi_{4}\overline{\psi}_{4}+\mathrm{h.c.}\,,\label{eq:Lagrangian_mixing_example}
\end{equation}
where we work in a 2-component notation with left-handed Weyl fermions,
as discussed in Appendix~\ref{app:2-component_notation}. The heavy vector-like ``fourth'' family
fermion obtains mass from the arbitrary vector-like mass term $M_{4}^{\psi}$.
Besides, in the example above, a scalar SM singlet couples the heavy
vector-like fermion to a third family chiral fermion. Once the scalar
singlet develops a VEV, $\left\langle \phi\right\rangle $, the first
term in Eq.~(\ref{eq:Lagrangian_mixing_example}) provides mixing
between the third and fourth family (left-handed) fermions. In all
generality, we obtain (see \cite{King:2020mau} for further details)
\begin{equation}
x_{34}^{\psi}\left\langle \phi\right\rangle \psi_{3}\overline{\psi}_{4}+M_{4}^{\psi}\psi_{4}\overline{\psi}_{4}=\left(x_{34}^{\psi}\left\langle \phi\right\rangle \psi_{3}+M_{4}^{\psi}\psi_{4}\right)\overline{\psi}_{4}=\hat{M}_{4}^{\psi}\frac{x_{34}^{\psi}\left\langle \phi\right\rangle \psi_{3}+M_{4}^{\psi}\psi_{4}}{\sqrt{\left(x_{34}^{\psi}\left\langle \phi\right\rangle \right)^{2}+\left(M_{4}^{\psi}\right)^{2}}}\overline{\psi}_{4}\,,
\end{equation}
where in the last step we have normalised the vector in order to obtain
\begin{equation}
\hat{M}_{4}^{\psi}=\sqrt{\left(x_{34}^{\psi}\left\langle \phi\right\rangle \right)^{2}+\left(M_{4}^{\psi}\right)^{2}}
\end{equation}
as the physical mass of the vector-like fermion. We can identify the
mixing angles as 
\begin{equation}
s_{34}^{\psi}=\frac{x_{34}^{\psi}\left\langle \phi\right\rangle }{\sqrt{\left(x_{34}^{\psi}\left\langle \phi\right\rangle \right)^{2}+\left(M_{4}^{\psi}\right)^{2}}}\,,\qquad c_{34}^{\psi}=\frac{M_{4}^{\psi}}{\sqrt{\left(x_{34}^{\psi}\left\langle \phi\right\rangle \right)^{2}+\left(M_{4}^{\psi}\right)^{2}}}\,.\label{eq:Mixing_angle_formalism}
\end{equation}
This way, the mass eigenstates are given by 
\begin{equation}
\hat{\psi}_{4}=c_{34}^{\psi}\psi_{4}+s_{34}^{\psi}\psi_{3}\,,\qquad\hat{\psi}_{3}=c_{34}^{\psi}\psi_{3}-s_{34}^{\psi}\psi_{4}\,.
\end{equation}
In this manner, we have obtained an accurate description of fermion
mixing valid in all regimes: this is called the \textit{large mixing
angle formalism}.

Notice that if the vector-like mass is much larger than the VEV of
the scalar singlet, i.e.~$\left\langle \phi\right\rangle \ll M_{4}^{\psi}$,
we may approximate the mass eigenstates of physical fermions as
\begin{equation}
\hat{\psi}_{4}\approx\psi_{4}+s_{34}^{\psi}\psi_{3}\,,\qquad\hat{\psi}_{3}\approx\psi_{3}-s_{34}^{\psi}\psi_{3}\,,
\end{equation}
where
\begin{equation}
s_{34}^{\psi}\approx x_{34}^{\psi}\frac{\left\langle \phi\right\rangle }{M_{4}^{\psi}}\ll1\,,
\end{equation}
such that in good approximation $\hat{\psi}_{4}\approx\psi_{4}$ and
$\hat{\psi}_{3}\approx\psi_{3}$, and the physical mass of the vector-like
fermion is given in good approximation by $M_{4}^{\psi}$. This is
denoted as the\textit{ mass insertion approximation}.

More generally, along with the mass terms of the Lagrangian in Eq.~(\ref{eq:Lagrangian_mixing_example}),
in the models studied in this thesis we shall also find couplings
that connect chiral and vector-like fermions via Higgs doublets,
\begin{equation}
\mathcal{L}_{\mathrm{mass}}\supset y_{43}^{\psi}H\psi_{4}\psi_{3}^{c}+x_{34}^{\psi}\phi\psi_{3}\overline{\psi}_{4}+M_{4}^{\psi}\psi_{4}\overline{\psi}_{4}+\mathrm{h.c.}\label{eq:Lagrangian_mixing_example-1}
\end{equation}
In the regime $\left\langle H\right\rangle \ll\left\langle \phi\right\rangle \,,M_{4}^{\psi}$,
which is accurate to describe models with heavy NP much above the electroweak
scale, the first term in Eq.~(\ref{eq:Lagrangian_mixing_example-1})
provides an effective Higgs Yukawa coupling for the third family fermion
via the mixing mediated by $\phi$, i.e.
\begin{figure}[t]
\begin{centering}
\begin{tikzpicture}
	\begin{feynman}
		\vertex (a) {\(\psi_{3}\)};
		\vertex [right=18mm of a] (b);
		\vertex [right=of b] (c) [label={ [xshift=0.1cm, yshift=0.1cm] \small $M^{\psi}_{4}$}];
		\vertex [right=of c] (d);
		\vertex [right=of d] (e) {\(\psi^{c}_{3}\)};
		\vertex [above=of b] (f1) {\(\phi\)};
		\vertex [above=of d] (f2) {\(H\)};
		\diagram* {
			(a) -- [fermion] (b) -- [charged scalar] (f1),
			(b) -- [edge label'=\(\overline{\psi}_{4}\)] (c),
			(c) -- [edge label'=\(\psi_{4}\), inner sep=6pt, insertion=0] (d) -- [charged scalar] (f2),
			(d) -- [fermion] (e),
	};
	\end{feynman}
\end{tikzpicture}
\par\end{centering}
\caption[Example of effective Yukawa coupling in the mass insertion approximation.]{Example of effective Yukawa coupling in the mass insertion approximation.
\label{fig:Example-of-effective_mass_insertion}}
\end{figure}
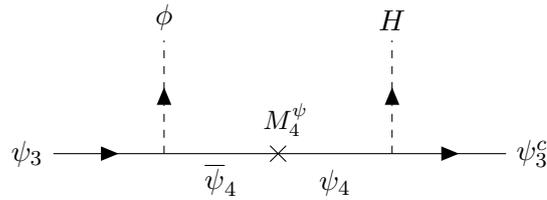
\begin{equation}
y_{43}^{\psi}H\psi_{4}\psi_{3}^{c}\approx y_{43}^{\psi}H(c_{34}^{\psi}\hat{\psi}_{4}+s_{34}^{\psi}\hat{\psi}_{3})\psi_{3}^{c}\rightarrow y_{43}^{\psi}s_{34}^{\psi}H\hat{\psi}_{3}\psi_{3}^{c}\,.
\end{equation}
Therefore, the effective Yukawa coupling is given by 
\begin{equation}
y_{3}=y_{43}^{\psi}s_{34}^{\psi}=y_{43}^{\psi}\frac{x_{34}^{\psi}\left\langle \phi\right\rangle }{\sqrt{\left(x_{34}^{\psi}\left\langle \phi\right\rangle \right)^{2}+\left(M_{4}^{\psi}\right)^{2}}}\,,
\end{equation}
where $s_{34}^{\psi}$ is given in the large mixing angle formalism
by Eq.~(\ref{eq:Mixing_angle_formalism}). By taking the limit
$\left\langle \phi\right\rangle \ll M_{4}^{\psi}$ above, one can
find the effective Yukawa coupling in the mass insertion approximation
as 
\begin{equation}
y_{3}\approx y_{43}^{\psi}x_{34}^{\psi}\frac{\left\langle \phi\right\rangle }{M_{4}^{\psi}}\,.
\end{equation}
This Yukawa coupling can also be extracted from the mass insertion
diagrams in Fig.~\ref{fig:Example-of-effective_mass_insertion}. This process can be easily generalised to vector-like fermions which mix with $CP$-conjugate right-handed fermions $\psi^{c}$. Notice that with the set of couplings introduced so far, the conjugate fermions $\overline{\psi}_{4}$ and $\overline{\psi^{c}_{4}}$ do not mix with chiral fermions (only the fermions $\psi_{4}$ and $\psi^{c}_{4}$ do mix with chiral fermions). They would mix if Yukawa couplings of the form $y^{\psi}_{4}H\psi_{4}\psi^{c}_{4}$ are introduced in the Lagrangian, as discussed in Chapter~\ref{Chapter:Fermiophobic}, although such mixing is usually negligible in the regime $\left\langle H\right\rangle \ll\left\langle \phi\right\rangle ,M_{4}^{\psi}$.

\chapter{Hyperons from \texorpdfstring{$SU(5)^{3}$}{SU(5)^{3}}} \label{app:Hyperons}
Gauge non-universal theories of flavour are usually spontaneously broken down to the SM via the VEVs of scalar fields in bi-representations of the different sites. In the case of the tri-hypercharge model (see Chapter~\ref{Chapter:Tri-hypercharge}) such fields carry family hypercharges that add up to zero, and we call them hyperons. The embedding of hyperons into the $SU(5)^{3}$ tri-unification framework proposed in Chapter~\ref{Chapter:Tri-unification} constrains their possible family hypercharge assignments. Tables~\ref{tab:hyperons1} and \ref{tab:hyperons2} list all possible hyperon embeddings in $SU(5)^{3}$ representations with dimension up to $\mathbf{45}$. These tables have been obtained with the help of~\texttt{GroupMath}~\cite{Fonseca:2020vke}.

{
\renewcommand{\arraystretch}{1.4}
\begin{table}[tb]
  \footnotesize
  \centering
  \begin{tabular}{|rrr|c|}
      \hline
      \multicolumn{3}{|c|}{\textbf{Hyperon}} & \textbf{$\boldsymbol{SU(5)^{3}}$ representations} \\
      \hline
\rowcolor{blue!10} 0 & $-\frac{1}{3}$ & $\frac{1}{3}$ & (\one,\five,\fiveS),(\one,\five,\ffiveS),(\one,\ffive,\fiveS),(\one,\ffive,\ffiveS),(\24,\five,\fiveS),(\24,\five,\ffiveS),(\24,\ffive,\fiveS),(\24,\ffive,\ffiveS) \\
  0 & $\frac{1}{2}$ & $-\frac{1}{2}$ & (\one,\five,\fiveS),(\one,\five,\ffiveS),(\one,\ffive,\fiveS),(\one,\ffive,\ffiveS),(\24,\five,\fiveS),(\24,\five,\ffiveS),(\24,\ffive,\fiveS),(\24,\ffive,\ffiveS) \\
  \rowcolor{blue!10} & & & (\one,\ten,\tenS),(\one,\ten,\fortyS),(\one,\fifteen,\fifteenS),(\one,\tfive,\tfiveS),(\one,\tfive,\fortyS),(\one,\forty,\tenS),(\one,\forty,\tfiveS),(\one,\forty,\fortyS), \\
  \rowcolor{blue!10} & & & (\24,\ten,\tenS),(\24,\ten,\fifteenS),(\24,\ten,\tfiveS),(\24,\ten,\fortyS),(\24,\fifteen,\tenS),(\24,\fifteen,\fifteenS),(\24,\fifteen,\fortyS), \\
  \rowcolor{blue!10} \multirow{-3}{*}{0} & \multirow{-3}{*}{$-\frac{2}{3}$} & \multirow{-3}{*}{$\frac{2}{3}$} & (\24,\tfive,\tenS),(\24,\tfive,\tfiveS),(\24,\tfive,\fortyS),(\24,\forty,\tenS),(\24,\forty,\fifteenS),(\24,\forty,\tfiveS),(\24,\forty,\fortyS) \\
  & & & (\one,\ten,\tenS),(\one,\ten,\fifteenS),(\one,\ten,\fortyS),(\one,\fifteen,\tenS),(\one,\fifteen,\fifteenS),(\one,\fifteen,\fortyS),(\one,\tfive,\tfiveS), \\
  & & & (\one,\tfive,\fortyS),(\one,\forty,\tenS),(\one,\forty,\fifteenS),(\one,\forty,\tfiveS),(\one,\forty,\fortyS),(\24,\ten,\tenS),(\24,\ten,\fifteenS), \\
  & & & (\24,\ten,\tfiveS),(\24,\ten,\fortyS),(\24,\fifteen,\tenS),(\24,\fifteen,\fifteenS),(\24,\fifteen,\tfiveS),(\24,\fifteen,\fortyS),(\24,\tfive,\tenS), \\
  \multirow{-4}{*}{0} & \multirow{-4}{*}{$\frac{1}{6}$} & \multirow{-4}{*}{$-\frac{1}{6}$} & (\24,\tfive,\fifteenS),(\24,\tfive,\tfiveS),(\24,\tfive,\fortyS),(\24,\forty,\tenS),(\24,\forty,\fifteenS),(\24,\forty,\tfiveS),(\24,\forty,\fortyS) \\
  \rowcolor{blue!10} & & & (\one,\ten,\tenS),(\one,\fifteen,\fifteenS),(\one,\tfive,\tfiveS),(\one,\forty,\fortyS),(\24,\ten,\tenS),(\24,\ten,\fifteenS),(\24,\ten,\fortyS), \\
  \rowcolor{blue!10} \multirow{-2}{*}{0} & \multirow{-2}{*}{$1$} & \multirow{-2}{*}{$-1$} & (\24,\fifteen,\tenS),(\24,\fifteen,\fifteenS),(\24,\tfive,\tfiveS),(\24,\tfive,\fortyS),(\24,\forty,\tenS),(\24,\forty,\tfiveS),(\24,\forty,\fortyS) \\
  0 & $\frac{5}{6}$ & $-\frac{5}{6}$ & (\one,\24,\24), (\24,\24,\24) \\
  \rowcolor{blue!10} 0 & $-\frac{3}{2}$ & $\frac{3}{2}$ & (\one,\tfive,\tfiveS),(\one,\forty,\fortyS),(\24,\tfive,\tfiveS),(\24,\tfive,\fortyS),(\24,\forty,\tfiveS),(\24,\forty,\fortyS) \\
  0 & $\frac{4}{3}$ & $-\frac{4}{3}$ & (\one,\ffive,\ffiveS), (\24,\ffive,\ffiveS) \\
  \rowcolor{blue!10} 0 & $-\frac{7}{6}$ & $\frac{7}{6}$ & (\one,\ffive,\ffiveS), (\24,\ffive,\ffiveS) \\
      \hline
  \end{tabular}
  \caption[Hyperons charged under two individual hypercharge groups and their \555 origin]{
  \small Hyperons charged under two individual hypercharge groups and their \555 origin. All \555 representations that involve up to $\ffive$ and $\ffiveS$ of $\GGM$ are included. Other hyperons can be obtained by reordering the hypercharge values or by conjugating the \555 representations.
  \label{tab:hyperons1}
  }
\end{table}
}

{
\renewcommand{\arraystretch}{1.4}
\begin{table}[tb]
  \footnotesize
  \centering
  \resizebox{14.5cm}{!}{
  \begin{tabular}{|rrr|c|}
      \hline
      \multicolumn{3}{|c|}{\textbf{Hyperon}} & \textbf{$\boldsymbol{SU(5)^{3}}$ representations} \\
      \hline
  & & & (\five,\five,\tenS),(\five,\five,\fifteenS),(\five,\five,\fortyS),(\five,\ffive,\tenS),(\five,\ffive,\fifteenS),(\five,\ffive,\tfiveS),(\five,\ffive,\fortyS),(\ffive,\five,\tenS), \\
  \multirow{-2}{*}{$-\frac{1}{3}$} & \multirow{-2}{*}{$-\frac{1}{3}$} & \multirow{-2}{*}{$\frac{2}{3}$} & (\ffive,\five,\fifteenS),(\ffive,\five,\tfiveS),(\ffive,\five,\fortyS),(\ffive,\ffive,\tenS),(\ffive,\ffive,\fifteenS),(\ffive,\ffive,\tfiveS),(\ffive,\ffive,\fortyS) \\
  \rowcolor{blue!10} & & & (\five,\five,\tenS),(\five,\five,\fifteenS),(\five,\five,\fortyS),(\five,\ffive,\tenS),(\five,\ffive,\fifteenS),(\five,\ffive,\tfiveS),(\five,\ffive,\fortyS),(\ffive,\five,\tenS), \\
  \rowcolor{blue!10} \multirow{-2}{*}{$-\frac{1}{3}$} & \multirow{-2}{*}{$\frac{1}{2}$} & \multirow{-2}{*}{$-\frac{1}{6}$} & (\ffive,\five,\fifteenS),(\ffive,\five,\tfiveS),(\ffive,\five,\fortyS),(\ffive,\ffive,\tenS),(\ffive,\ffive,\fifteenS),(\ffive,\ffive,\tfiveS),(\ffive,\ffive,\fortyS) \\
  & & & (\five,\five,\tenS),(\five,\five,\fifteenS),(\five,\ffive,\tenS),(\five,\ffive,\fifteenS),(\five,\ffive,\fortyS),(\ffive,\five,\tenS), \\
  \multirow{-2}{*}{$\frac{1}{2}$} & \multirow{-2}{*}{$\frac{1}{2}$} & \multirow{-2}{*}{$-1$} & (\ffive,\five,\fifteenS),(\ffive,\five,\fortyS),(\ffive,\ffive,\tenS),(\ffive,\ffive,\fifteenS),(\ffive,\ffive,\tfiveS),(\ffive,\ffive,\fortyS) \\
  \rowcolor{blue!10} $-\frac{1}{3}$ & $-\frac{1}{2}$ & $\frac{5}{6}$ & (\five,\fiveS,\24),(\five,\ffiveS,\24),(\ffive,\fiveS,\24),(\ffive,\ffiveS,\24) \\
  & & & (\five,\ten,\ten),(\five,\ten,\forty),(\five,\fifteen,\tfive),(\five,\fifteen,\forty),(\five,\tfive,\fifteen),(\five,\forty,\ten),(\five,\forty,\fifteen), \\
  & & & (\five,\forty,\forty),(\ffive,\ten,\ten),(\ffive,\ten,\fifteen),(\ffive,\ten,\forty),(\ffive,\fifteen,\ten),(\ffive,\fifteen,\tfive),(\ffive,\fifteen,\forty), \\
  \multirow{-3}{*}{$-\frac{1}{3}$} & \multirow{-3}{*}{$-\frac{2}{3}$} & \multirow{-3}{*}{$1$} & (\ffive,\tfive,\ten),(\ffive,\tfive,\fifteen),(\ffive,\tfive,\forty),(\ffive,\forty,\ten),(\ffive,\forty,\fifteen),(\ffive,\forty,\forty) \\
  \rowcolor{blue!10} & & & (\five,\ten,\ten),(\five,\ten,\fifteen),(\five,\ten,\tfive),(\five,\ten,\forty),(\five,\fifteen,\ten),(\five,\fifteen,\fifteen),(\five,\fifteen,\tfive), \\
  \rowcolor{blue!10} & & & (\five,\fifteen,\forty),(\five,\tfive,\ten),(\five,\tfive,\fifteen),(\five,\tfive,\forty),(\five,\forty,\ten),(\five,\forty,\fifteen),(\five,\forty,\tfive), \\
  \rowcolor{blue!10} & & & (\five,\forty,\forty),(\ffive,\ten,\ten),(\ffive,\ten,\fifteen),(\ffive,\ten,\tfive),(\ffive,\ten,\forty),(\ffive,\fifteen,\ten), \\
  \rowcolor{blue!10} & & & (\ffive,\fifteen,\fifteen),(\ffive,\fifteen,\tfive),(\ffive,\fifteen,\forty),(\ffive,\tfive,\ten),(\ffive,\tfive,\fifteen),(\ffive,\tfive,\tfive), \\
  \rowcolor{blue!10} \multirow{-5}{*}{$-\frac{1}{3}$} & \multirow{-5}{*}{$\frac{1}{6}$} & \multirow{-5}{*}{$\frac{1}{6}$} & (\ffive,\tfive,\forty),(\ffive,\forty,\ten),(\ffive,\forty,\fifteen),(\ffive,\forty,\tfive),(\ffive,\forty,\forty) \\
  & & & (\five,\ten,\ten),(\five,\ten,\fifteen),(\five,\ten,\forty),(\five,\fifteen,\tfive),(\five,\fifteen,\forty),(\five,\tfive,\ten),(\five,\tfive,\fifteen), \\
  & & & (\five,\tfive,\forty),(\five,\forty,\ten),(\five,\forty,\fifteen),(\five,\forty,\forty),(\ffive,\ten,\ten),(\ffive,\ten,\fifteen),(\ffive,\ten,\tfive), \\
  & & & (\ffive,\ten,\forty),(\ffive,\fifteen,\ten),(\ffive,\fifteen,\fifteen),(\ffive,\fifteen,\tfive),(\ffive,\fifteen,\forty),(\ffive,\tfive,\ten),(\ffive,\tfive,\fifteen), \\
  \multirow{-4}{*}{$\frac{1}{2}$} & \multirow{-4}{*}{$-\frac{2}{3}$} & \multirow{-4}{*}{$\frac{1}{6}$} & (\ffive,\tfive,\tfive),(\ffive,\tfive,\forty),(\ffive,\forty,\ten),(\ffive,\forty,\fifteen),(\ffive,\forty,\tfive),(\ffive,\forty,\forty) \\
  \rowcolor{blue!10} $\frac{1}{2}$ & $1$ & $-\frac{3}{2}$ & (\five,\ten,\forty),(\five,\fifteen,\tfive),(\five,\fifteen,\forty),(\ffive,\ten,\forty),(\ffive,\fifteen,\tfive),(\ffive,\fifteen,\forty),(\ffive,\forty,\forty) \\
  $-\frac{1}{3}$ & $-1$ & $\frac{4}{3}$ & (\five,\tenS,\ffive),(\five,\fortyS,\ffive),(\ffive,\tenS,\ffive),(\ffive,\fifteenS,\ffive),(\ffive,\tfiveS,\ffive),(\ffive,\fortyS,\ffive) \\
  \rowcolor{blue!10} $\frac{1}{2}$ & $\frac{2}{3}$ & $-\frac{7}{6}$ & (\five,\tenS,\ffive),(\five,\tfiveS,\ffive),(\five,\fortyS,\ffive),(\ffive,\tenS,\ffive),(\ffive,\fifteenS,\ffive),(\ffive,\tfiveS,\ffive),(\ffive,\fortyS,\ffive) \\
  $-\frac{1}{3}$ & $-\frac{5}{6}$ & $\frac{7}{6}$ & (\five,\24,\ffiveS),(\ffive,\24,\ffiveS) \\
  \rowcolor{blue!10} $\frac{1}{2}$ & $\frac{5}{6}$ & $-\frac{4}{3}$ & (\five,\24,\ffiveS),(\ffive,\24,\ffiveS) \\
  $-\frac{1}{3}$ & $\frac{3}{2}$ & $-\frac{7}{6}$ & (\five,\fortyS,\ffive),(\ffive,\tfiveS,\ffive),(\ffive,\fortyS,\ffive) \\
  \rowcolor{blue!10} & & & (\ten,\ten,\ffive),(\ten,\fifteen,\ffive),(\ten,\forty,\ffive),(\fifteen,\ten,\ffive),(\fifteen,\forty,\ffive),(\tfive,\tfive,\ffive), \\
  \rowcolor{blue!10} \multirow{-2}{*}{$-\frac{2}{3}$} & \multirow{-2}{*}{$-\frac{2}{3}$} & \multirow{-2}{*}{$\frac{4}{3}$} & (\tfive,\forty,\ffive),(\forty,\ten,\ffive),(\forty,\fifteen,\ffive),(\forty,\tfive,\ffive),(\forty,\forty,\ffive) \\
  & & & (\ten,\ten,\ffive),(\ten,\fifteen,\ffive),(\ten,\forty,\ffive),(\fifteen,\ten,\ffive),(\fifteen,\fifteen,\ffive), \\
  \multirow{-2}{*}{$\frac{1}{6}$} & \multirow{-2}{*}{$1$} & \multirow{-2}{*}{$-\frac{7}{6}$} & (\fifteen,\forty,\ffive),(\tfive,\forty,\ffive),(\forty,\ten,\ffive),(\forty,\fifteen,\ffive),(\forty,\forty,\ffive) \\
  \rowcolor{blue!10} & & & (\ten,\tenS,\24),(\ten,\fifteenS,\24),(\ten,\tfiveS,\24),(\ten,\fortyS,\24),(\fifteen,\tenS,\24),(\fifteen,\fifteenS,\24),(\fifteen,\fortyS,\24),(\tfive,\tenS,\24), \\
  \rowcolor{blue!10} \multirow{-2}{*}{$-\frac{2}{3}$} & \multirow{-2}{*}{$-\frac{1}{6}$} & \multirow{-2}{*}{$\frac{5}{6}$} & (\tfive,\fifteenS,\24),(\tfive,\tfiveS,\24),(\tfive,\fortyS,\24),(\forty,\tenS,\24),(\forty,\fifteenS,\24),(\forty,\tfiveS,\24),(\forty,\fortyS,\24) \\
  & & & (\ten,\tenS,\24),(\ten,\fifteenS,\24),(\ten,\fortyS,\24),(\fifteen,\tenS,\24),(\fifteen,\fifteenS,\24),(\fifteen,\fortyS,\24), \\
  \multirow{-2}{*}{$\frac{1}{6}$} & \multirow{-2}{*}{$-1$} & \multirow{-2}{*}{$\frac{5}{6}$} & (\tfive,\tfiveS,\24),(\tfive,\fortyS,\24),(\forty,\tenS,\24),(\forty,\fifteenS,\24),(\forty,\tfiveS,\24),(\forty,\fortyS,\24) \\
  \rowcolor{blue!10}  $-\frac{2}{3}$ & $-\frac{5}{6}$ & $\frac{3}{2}$ & (\ten,\24,\fortyS),(\tfive,\24,\tfiveS),(\tfive,\24,\fortyS),(\forty,\24,\tfiveS),(\forty,\24,\fortyS) \\
  $\frac{1}{6}$ & $-\frac{3}{2}$ & $\frac{4}{3}$ & (\ten,\forty,\ffive),(\fifteen,\forty,\ffive),(\forty,\forty,\ffive) \\
  \rowcolor{blue!10} $\frac{1}{6}$ & $-\frac{4}{3}$ & $\frac{7}{6}$ & (\ten,\ffiveS,\ffiveS),(\fifteen,\ffiveS,\ffiveS),(\tfive,\ffiveS,\ffiveS),(\forty,\ffiveS,\ffiveS) \\
      \hline
  \end{tabular}}
  \caption[Hyperons charged under the three individual hypercharge groups and their \555 origin]{
  \small Hyperons charged under the three individual hypercharge groups and their \555 origin. All \555 representations that involve up to $\ffive$ and $\ffiveS$ of $\GGM$ are included. Other hyperons can be obtained by reordering the hypercharge values or by conjugating the \555 representations.
  \label{tab:hyperons2}
  }
\end{table}
}

\chapter{EFT operators and tree-level matching}  \label{app:EFT_Matching}
In this appendix, we list the SMEFT and LEFT operators in the Warsaw and San Diego basis, respectively, including
the tree-level matching conditions between SMEFT and LEFT Wilson coefficients.

\section{The SMEFT operators} \label{app:SMEFT_Operators}
This appendix lists the SMEFT operators up to dimension six in the so-called Warsaw basis. The operators were
listed in Ref.~\cite{Grzadkowski:2010es}. They are reproduced in Tables~\ref{tab:smeft5ops}-\ref{tab:smeft6ops} since we make significant use of them in this thesis.

%%%%%%%%%%%%%%%%%%%%%%%%%%%%%%%%%%%%%%%%%%%%%%%%%%%%%%%%%%%%%%%%%%%%%%%
% SMEFT d=5 Operators
%%%%%%%%%%%%%%%%%%%%%%%%%%%%%%%%%%%%%%%%%%%%%%%%%%%%%%%%%%%%%%%%%%%%%%% 

\begin{table}[p]
\begin{center}
\small
%\centerline{SMEFT $d=5$}
%\vspace{0.25cm}
\begin{minipage}[t]{5.2cm}
\renewcommand{\arraystretch}{1.5}
\begin{tabular}[t]{c|c}
\multicolumn{2}{c}{\boldmath{$\Delta L = 2 \qquad (LL)HH+\hc$}} \\
\hline
$Q_{5}$      & $\epsilon^{ij} \epsilon^{k\ell} (l_{ip}^T C l_{kr} ) H_j H_\ell  $  \\
\end{tabular}
\end{minipage}
\end{center}
\caption[Dimension-five $\Delta L=2$ operator $Q_5$ in SMEFT (Weinberg operator)]{Dimension-five $\Delta L=2$ operator $Q_5$ in SMEFT (Weinberg operator).  There is also the Hermitian conjugate $\Delta L = -2$ operator $Q_5^\dagger$, as indicated by ${}+\hc$ in the table heading.
Subscripts $p$ and $r$ are weak-eigenstate indices. Table taken from \cite{Jenkins:2017jig}.}
\label{tab:smeft5ops}
\end{table}

\begin{table}[ht]
\vspace{-2cm}
\begin{center}
\small
%%\vspace{0.25cm}
\begin{minipage}[t]{5.2cm}
\renewcommand{\arraystretch}{1.5}
\begin{tabular}[t]{c|c}
\multicolumn{2}{c}{\boldmath{$\Delta B = \Delta L = 1 +\mathrm{h.c.}$}} \\
\hline
$Q_{duql}$      & $\epsilon^{\alpha \beta \gamma} \epsilon^{ij} (d^T_{\alpha p} C u_{\beta r} ) (q^T_{\gamma i s} C l_{jt})  $  \\
$Q_{qque}$      & $\epsilon^{\alpha \beta \gamma} \epsilon^{ij} (q^T_{\alpha i p} C q_{\beta j r} ) (u^T_{\gamma s} C e_{t})  $  \\
$Q_{qqql}$      & $\epsilon^{\alpha \beta \gamma} \epsilon^{i\ell} \epsilon^{jk} (q^T_{\alpha i p} C q_{\beta j r} ) (q^T_{\gamma k s} C l_{\ell t})  $  \\
$Q_{duue}$      & $\epsilon^{\alpha \beta \gamma} (d^T_{\alpha p} C u_{\beta r} ) (u^T_{\gamma s} C e_{t})  $  \\
\end{tabular}
\end{minipage}
\end{center}
\caption[Dimension-six $\Delta B= \Delta L=1$ operators in SMEFT]{Dimension-six $\Delta B= \Delta L=1$ operators in SMEFT.  There are also Hermitian conjugate $\Delta B = \Delta L = -1$ operators, as indicated by ${}+\mathrm{h.c.}$ in the table heading.
Subscripts $p$, $r$, $s$ and $t$ are flavour indices. Table taken from \cite{Jenkins:2017jig}.}
\label{tab:smeft6baryonops}
\end{table}

\begin{table}[t]
\hspace{-0.5cm}

\begin{center}
\begin{adjustbox}{width=0.89\textwidth,center}
\small
%\centerline{SMEFT $d=6$}
%\vspace{0.25cm}
\begin{minipage}[t]{4.45cm}
\renewcommand{\arraystretch}{1.5}
\begin{tabular}[t]{c|c}
\multicolumn{2}{c}{\boldmath$1:X^3$} \\
\hline
$Q_G$                & $f^{ABC} G_\mu^{A\nu} G_\nu^{B\rho} G_\rho^{C\mu} $ \\
$Q_{\widetilde G}$          & $f^{ABC} \widetilde G_\mu^{A\nu} G_\nu^{B\rho} G_\rho^{C\mu} $ \\
$Q_W$                & $\epsilon^{IJK} W_\mu^{I\nu} W_\nu^{J\rho} W_\rho^{K\mu}$ \\ 
$Q_{\widetilde W}$          & $\epsilon^{IJK} \widetilde W_\mu^{I\nu} W_\nu^{J\rho} W_\rho^{K\mu}$ \\
\end{tabular}
\end{minipage}
\begin{minipage}[t]{2.7cm}
\renewcommand{\arraystretch}{1.5}
\begin{tabular}[t]{c|c}
\multicolumn{2}{c}{\boldmath$2:H^6$} \\
\hline
$Q_H$       & $(H^\dag H)^3$ 
\end{tabular}
\end{minipage}
\begin{minipage}[t]{5.1cm}
\renewcommand{\arraystretch}{1.5}
\begin{tabular}[t]{c|c}
\multicolumn{2}{c}{\boldmath$3:H^4 D^2$} \\
\hline
$Q_{H\Box}$ & $(H^\dag H)\Box(H^\dag H)$ \\
$Q_{H D}$   & $\ \left(H^\dag D_\mu H\right)^* \left(H^\dag D_\mu H\right)$ 
\end{tabular}
\end{minipage}
\begin{minipage}[t]{2.7cm}
\renewcommand{\arraystretch}{1.5}
\begin{tabular}[t]{c|c}
\multicolumn{2}{c}{\boldmath$5: \psi^2H^3 + \mathrm{h.c.}$} \\
\hline
$Q_{eH}$           & $(H^\dag H)(\bar l_p e_r H)$ \\
$Q_{uH}$          & $(H^\dag H)(\bar q_p u_r \widetilde H )$ \\
$Q_{dH}$           & $(H^\dag H)(\bar q_p d_r H)$\\
\end{tabular}
\end{minipage}

\vspace{0.25cm}

\end{adjustbox}

\begin{adjustbox}{width=0.89\textwidth,center}

\begin{minipage}[t]{4.7cm}
\renewcommand{\arraystretch}{1.5}
\begin{tabular}[t]{c|c}
\multicolumn{2}{c}{\boldmath$4:X^2H^2$} \\
\hline
$Q_{H G}$     & $H^\dag H\, G^A_{\mu\nu} G^{A\mu\nu}$ \\
$Q_{H\widetilde G}$         & $H^\dag H\, \widetilde G^A_{\mu\nu} G^{A\mu\nu}$ \\
$Q_{H W}$     & $H^\dag H\, W^I_{\mu\nu} W^{I\mu\nu}$ \\
$Q_{H\widetilde W}$         & $H^\dag H\, \widetilde W^I_{\mu\nu} W^{I\mu\nu}$ \\
$Q_{H B}$     & $ H^\dag H\, B_{\mu\nu} B^{\mu\nu}$ \\
$Q_{H\widetilde B}$         & $H^\dag H\, \widetilde B_{\mu\nu} B^{\mu\nu}$ \\
$Q_{H WB}$     & $ H^\dag \tau^I H\, W^I_{\mu\nu} B^{\mu\nu}$ \\
$Q_{H\widetilde W B}$         & $H^\dag \tau^I H\, \widetilde W^I_{\mu\nu} B^{\mu\nu}$ 
\end{tabular}
\end{minipage}
\begin{minipage}[t]{5.2cm}
\renewcommand{\arraystretch}{1.5}
\begin{tabular}[t]{c|c}
\multicolumn{2}{c}{\boldmath$6:\psi^2 XH+\mathrm{h.c.}$} \\
\hline
$Q_{eW}$      & $(\bar l_p \sigma^{\mu\nu} e_r) \tau^I H W_{\mu\nu}^I$ \\
$Q_{eB}$        & $(\bar l_p \sigma^{\mu\nu} e_r) H B_{\mu\nu}$ \\
$Q_{uG}$        & $(\bar q_p \sigma^{\mu\nu} T^A u_r) \widetilde H \, G_{\mu\nu}^A$ \\
$Q_{uW}$        & $(\bar q_p \sigma^{\mu\nu} u_r) \tau^I \widetilde H \, W_{\mu\nu}^I$ \\
$Q_{uB}$        & $(\bar q_p \sigma^{\mu\nu} u_r) \widetilde H \, B_{\mu\nu}$ \\
$Q_{dG}$        & $(\bar q_p \sigma^{\mu\nu} T^A d_r) H\, G_{\mu\nu}^A$ \\
$Q_{dW}$         & $(\bar q_p \sigma^{\mu\nu} d_r) \tau^I H\, W_{\mu\nu}^I$ \\
$Q_{dB}$        & $(\bar q_p \sigma^{\mu\nu} d_r) H\, B_{\mu\nu}$ 
\end{tabular}
\end{minipage}
\begin{minipage}[t]{5.4cm}
\renewcommand{\arraystretch}{1.5}
\begin{tabular}[t]{c|c}
\multicolumn{2}{c}{\boldmath$7:\psi^2H^2 D$} \\
\hline
$Q_{H l}^{(1)}$      & $(H^\dag i\overleftrightarrow{D}_\mu H)(\bar l_p \gamma^\mu l_r)$\\
$Q_{H l}^{(3)}$      & $(H^\dag i\overleftrightarrow{D}^I_\mu H)(\bar l_p \tau^I \gamma^\mu l_r)$\\
$Q_{H e}$            & $(H^\dag i\overleftrightarrow{D}_\mu H)(\bar e_p \gamma^\mu e_r)$\\
$Q_{H q}^{(1)}$      & $(H^\dag i\overleftrightarrow{D}_\mu H)(\bar q_p \gamma^\mu q_r)$\\
$Q_{H q}^{(3)}$      & $(H^\dag i\overleftrightarrow{D}^I_\mu H)(\bar q_p \tau^I \gamma^\mu q_r)$\\
$Q_{H u}$            & $(H^\dag i\overleftrightarrow{D}_\mu H)(\bar u_p \gamma^\mu u_r)$\\
$Q_{H d}$            & $(H^\dag i\overleftrightarrow{D}_\mu H)(\bar d_p \gamma^\mu d_r)$\\
$Q_{H u d}$ + h.c.   & $i(\widetilde H ^\dag D_\mu H)(\bar u_p \gamma^\mu d_r)$\\
\end{tabular}
\end{minipage}
\end{adjustbox}
\end{center}

\begin{center}
\begin{adjustbox}{width=0.9\textwidth,center}
\begin{minipage}[t]{4.75cm}
\renewcommand{\arraystretch}{1.5}
\begin{tabular}[t]{c|c}
\multicolumn{2}{c}{\boldmath$8:(\bar L L)(\bar L L)$} \\
\hline
$Q_{ll}$        & $(\bar l_p \gamma^\mu l_r)(\bar l_s \gamma_\mu l_t)$ \\
$Q_{qq}^{(1)}$  & $(\bar q_p \gamma^\mu q_r)(\bar q_s \gamma_\mu q_t)$ \\
$Q_{qq}^{(3)}$  & $(\bar q_p \gamma^\mu \tau^I q_r)(\bar q_s \gamma_\mu \tau^I q_t)$ \\
$Q_{lq}^{(1)}$                & $(\bar l_p \gamma^\mu l_r)(\bar q_s \gamma_\mu q_t)$ \\
$Q_{lq}^{(3)}$                & $(\bar l_p \gamma^\mu \tau^I l_r)(\bar q_s \gamma_\mu \tau^I q_t)$ 
\end{tabular}
\end{minipage}
\hspace{0.2cm}
\begin{minipage}[t]{5.25cm}
\renewcommand{\arraystretch}{1.5}
\begin{tabular}[t]{c|c}
\multicolumn{2}{c}{\boldmath$8:(\bar R R)(\bar R R)$} \\
\hline
$Q_{ee}$               & $(\bar e_p \gamma^\mu e_r)(\bar e_s \gamma_\mu e_t)$ \\
$Q_{uu}$        & $(\bar u_p \gamma^\mu u_r)(\bar u_s \gamma_\mu u_t)$ \\
$Q_{dd}$        & $(\bar d_p \gamma^\mu d_r)(\bar d_s \gamma_\mu d_t)$ \\
$Q_{eu}$                      & $(\bar e_p \gamma^\mu e_r)(\bar u_s \gamma_\mu u_t)$ \\
$Q_{ed}$                      & $(\bar e_p \gamma^\mu e_r)(\bar d_s\gamma_\mu d_t)$ \\
$Q_{ud}^{(1)}$                & $(\bar u_p \gamma^\mu u_r)(\bar d_s \gamma_\mu d_t)$ \\
$Q_{ud}^{(8)}$                & $(\bar u_p \gamma^\mu T^A u_r)(\bar d_s \gamma_\mu T^A d_t)$ \\
\end{tabular}
\end{minipage}
\hspace{0.2cm}
\begin{minipage}[t]{4.75cm}
\renewcommand{\arraystretch}{1.5}
\begin{tabular}[t]{c|c}
\multicolumn{2}{c}{\boldmath$8:(\bar L L)(\bar R R)$} \\
\hline
$Q_{le}$               & $(\bar l_p \gamma^\mu l_r)(\bar e_s \gamma_\mu e_t)$ \\
$Q_{lu}$               & $(\bar l_p \gamma^\mu l_r)(\bar u_s \gamma_\mu u_t)$ \\
$Q_{ld}$               & $(\bar l_p \gamma^\mu l_r)(\bar d_s \gamma_\mu d_t)$ \\
$Q_{qe}$               & $(\bar q_p \gamma^\mu q_r)(\bar e_s \gamma_\mu e_t)$ \\
$Q_{qu}^{(1)}$         & $(\bar q_p \gamma^\mu q_r)(\bar u_s \gamma_\mu u_t)$ \\ 
$Q_{qu}^{(8)}$         & $(\bar q_p \gamma^\mu T^A q_r)(\bar u_s \gamma_\mu T^A u_t)$ \\ 
$Q_{qd}^{(1)}$ & $(\bar q_p \gamma^\mu q_r)(\bar d_s \gamma_\mu d_t)$ \\
$Q_{qd}^{(8)}$ & $(\bar q_p \gamma^\mu T^A q_r)(\bar d_s \gamma_\mu T^A d_t)$\\
\end{tabular}
\end{minipage}

\end{adjustbox}

\vspace{0.25cm}

\begin{adjustbox}{width=0.57\textwidth,center}

\begin{minipage}[t]{3.75cm}
\renewcommand{\arraystretch}{1.5}
\begin{tabular}[t]{c|c}
\multicolumn{2}{c}{\boldmath$8:(\bar LR)(\bar RL)+\mathrm{h.c.}$} \\
\hline
$Q_{ledq}$ & $(\bar l_p^j e_r)(\bar d_s q_{tj})$ 
\end{tabular}
\end{minipage}
\hspace{0.4cm}
\begin{minipage}[t]{5.5cm}
\renewcommand{\arraystretch}{1.5}
\begin{tabular}[t]{c|c}
\multicolumn{2}{c}{\boldmath$8:(\bar LR)(\bar L R)+\mathrm{h.c.}$} \\
\hline
$Q_{quqd}^{(1)}$ & $(\bar q_p^j u_r) \epsilon_{jk} (\bar q_s^k d_t)$ \\
$Q_{quqd}^{(8)}$ & $(\bar q_p^j T^A u_r) \epsilon_{jk} (\bar q_s^k T^A d_t)$ \\
$Q_{lequ}^{(1)}$ & $(\bar l_p^j e_r) \epsilon_{jk} (\bar q_s^k u_t)$ \\
$Q_{lequ}^{(3)}$ & $(\bar l_p^j \sigma_{\mu\nu} e_r) \epsilon_{jk} (\bar q_s^k \sigma^{\mu\nu} u_t)$
\end{tabular}
\end{minipage}
\end{adjustbox}
\end{center}
\caption[The 76 dimension-six operators that conserve baryon and lepton number in the SMEFT]{The 76 dimension-six operators that conserve baryon and lepton number in the SMEFT. The operators are divided into eight classes according to their field content. The class-8 four-fermion operators are further divided into subclasses according to their chiral properties. Operators with ${}+\mathrm{h.c.}$ have Hermitian conjugates. The subscripts $p, r, s, t$ are flavour indices indices. Table taken from \cite{Jenkins:2017jig}.}
\label{tab:smeft6ops}
\end{table}

\section{The LEFT operators} \label{app:LEFT_Operators}
This appendix lists the LEFT operators up to dimension six in the so-called San Diego basis. The operators
were listed in Ref.~\cite{Jenkins:2017jig}. They are reproduced in Tables~\ref{tab:oplist1} and \ref{tab:oplist2} since we make significant use of them in this thesis.
\begin{table}[ht]
\vspace{-2cm}
\begin{adjustbox}{width=0.85\textwidth,center}
\begin{minipage}[t]{3cm}
\renewcommand{\arraystretch}{1.51}
\small
\begin{align*}
\begin{array}[t]{c|c}
\multicolumn{2}{c}{\boldsymbol{\nu \nu+\mathrm{h.c.}}} \\
\hline
\O_{\nu} & (\nu_{Lp}^T C \nu_{Lr})  \\
\end{array}
\end{align*}
\end{minipage}
%%%
% --- END TABLE
%%
%
%%
% --- START TABLE
%%
\begin{minipage}[t]{3cm}
\renewcommand{\arraystretch}{1.51}
\small
\begin{align*}
\begin{array}[t]{c|c}
\multicolumn{2}{c}{\boldsymbol{(\nu \nu) X+\mathrm{h.c.}}} \\
\hline
\O_{\nu \gamma} & (\nu_{Lp}^T C   \sigma^{\mu \nu}  \nu_{Lr})  F_{\mu \nu}  \\
\end{array}
\end{align*}
\end{minipage}
%%%
% --- END TABLE
%%%
%%
% --- START TABLE
%%

\begin{minipage}[t]{3cm}
\renewcommand{\arraystretch}{1.51}
\small
\begin{align*}
\begin{array}[t]{c|c}
\multicolumn{2}{c}{\boldsymbol{(\overline L R ) X+\mathrm{h.c.}}} \\
\hline
\O_{e \gamma} & \bar e_{Lp}   \sigma^{\mu \nu} e_{Rr}\, F_{\mu \nu}  \\
\O_{u \gamma} & \bar u_{Lp}   \sigma^{\mu \nu}  u_{Rr}\, F_{\mu \nu}   \\
\O_{d \gamma} & \bar d_{Lp}  \sigma^{\mu \nu} d_{Rr}\, F_{\mu \nu}  \\
\O_{u G} & \bar u_{Lp}   \sigma^{\mu \nu}  T^A u_{Rr}\,  G_{\mu \nu}^A  \\
\O_{d G} & \bar d_{Lp}   \sigma^{\mu \nu} T^A d_{Rr}\,  G_{\mu \nu}^A \\
\end{array}
\end{align*}
\end{minipage}
\begin{minipage}[t]{3cm}
\renewcommand{\arraystretch}{1.51}
\small
\begin{align*}
\begin{array}[t]{c|c}
\multicolumn{2}{c}{\boldsymbol{X^3}} \\
\hline
\O_G     & f^{ABC} G_\mu^{A\nu} G_\nu^{B\rho} G_\rho^{C\mu}  \\
\O_{\widetilde G} & f^{ABC} \widetilde G_\mu^{A\nu} G_\nu^{B\rho} G_\rho^{C\mu}   \\
\end{array}
\end{align*}
\end{minipage}
\end{adjustbox}
%

%
%%
% --- START TABLE
%%
\mbox{}\\[-1.25cm]

\begin{adjustbox}{width=1\textwidth,center}
%%%
% --- END TABLE
%%%
\begin{minipage}[t]{3cm}
\renewcommand{\arraystretch}{1.51}
\small
\begin{align*}
\begin{array}[t]{c|c}
\multicolumn{2}{c}{\boldsymbol{(\overline L L)(\overline L  L)}} \\
\hline
\op{\nu\nu}{V}{LL} & (\bar \nu_{Lp}  \gamma^\mu \nu_{Lr} )(\bar \nu_{Ls} \gamma_\mu \nu_{Lt})   \\
\op{ee}{V}{LL}       & (\bar e_{Lp}  \gamma^\mu e_{Lr})(\bar e_{Ls} \gamma_\mu e_{Lt})   \\
\op{\nu e}{V}{LL}       & (\bar \nu_{Lp} \gamma^\mu \nu_{Lr})(\bar e_{Ls}  \gamma_\mu e_{Lt})  \\
\op{\nu u}{V}{LL}       & (\bar \nu_{Lp} \gamma^\mu \nu_{Lr}) (\bar u_{Ls}  \gamma_\mu u_{Lt})  \\
\op{\nu d}{V}{LL}       & (\bar \nu_{Lp} \gamma^\mu \nu_{Lr})(\bar d_{Ls} \gamma_\mu d_{Lt})     \\
\op{eu}{V}{LL}      & (\bar e_{Lp}  \gamma^\mu e_{Lr})(\bar u_{Ls} \gamma_\mu u_{Lt})   \\
\op{ed}{V}{LL}       & (\bar e_{Lp}  \gamma^\mu e_{Lr})(\bar d_{Ls} \gamma_\mu d_{Lt})  \\
\op{\nu edu}{V}{LL}      & (\bar \nu_{Lp} \gamma^\mu e_{Lr}) (\bar d_{Ls} \gamma_\mu u_{Lt})  + \mathrm{h.c.}   \\
\op{uu}{V}{LL}        & (\bar u_{Lp} \gamma^\mu u_{Lr})(\bar u_{Ls} \gamma_\mu u_{Lt})    \\
\op{dd}{V}{LL}   & (\bar d_{Lp} \gamma^\mu d_{Lr})(\bar d_{Ls} \gamma_\mu d_{Lt})    \\
\op{ud}{V1}{LL}     & (\bar u_{Lp} \gamma^\mu u_{Lr}) (\bar d_{Ls} \gamma_\mu d_{Lt})  \\
\op{ud}{V8}{LL}     & (\bar u_{Lp} \gamma^\mu T^A u_{Lr}) (\bar d_{Ls} \gamma_\mu T^A d_{Lt})   \\[-0.5cm]
\end{array}
\end{align*}
%\end{minipage}
%\begin{minipage}[t]{3cm}
\renewcommand{\arraystretch}{1.51}
\small
\begin{align*}
\begin{array}[t]{c|c}
\multicolumn{2}{c}{\boldsymbol{(\overline R  R)(\overline R R)}} \\
\hline
\op{ee}{V}{RR}     & (\bar e_{Rp} \gamma^\mu e_{Rr})(\bar e_{Rs} \gamma_\mu e_{Rt})  \\
\op{eu}{V}{RR}       & (\bar e_{Rp}  \gamma^\mu e_{Rr})(\bar u_{Rs} \gamma_\mu u_{Rt})   \\
\op{ed}{V}{RR}     & (\bar e_{Rp} \gamma^\mu e_{Rr})  (\bar d_{Rs} \gamma_\mu d_{Rt})   \\
\op{uu}{V}{RR}      & (\bar u_{Rp} \gamma^\mu u_{Rr})(\bar u_{Rs} \gamma_\mu u_{Rt})  \\
\op{dd}{V}{RR}      & (\bar d_{Rp} \gamma^\mu d_{Rr})(\bar d_{Rs} \gamma_\mu d_{Rt})    \\
\op{ud}{V1}{RR}       & (\bar u_{Rp} \gamma^\mu u_{Rr}) (\bar d_{Rs} \gamma_\mu d_{Rt})  \\
\op{ud}{V8}{RR}    & (\bar u_{Rp} \gamma^\mu T^A u_{Rr}) (\bar d_{Rs} \gamma_\mu T^A d_{Rt})  \\
\end{array}
\end{align*}
\end{minipage}
%
%\hspace{-1.5cm}
%
\begin{minipage}[t]{3cm}
\renewcommand{\arraystretch}{1.51}
\small
\begin{align*}
\begin{array}[t]{c|c}
\multicolumn{2}{c}{\boldsymbol{(\overline L  L)(\overline R  R)}} \\
\hline
\op{\nu e}{V}{LR}     & (\bar \nu_{Lp} \gamma^\mu \nu_{Lr})(\bar e_{Rs}  \gamma_\mu e_{Rt})  \\
\op{ee}{V}{LR}       & (\bar e_{Lp}  \gamma^\mu e_{Lr})(\bar e_{Rs} \gamma_\mu e_{Rt}) \\
\op{\nu u}{V}{LR}         & (\bar \nu_{Lp} \gamma^\mu \nu_{Lr})(\bar u_{Rs}  \gamma_\mu u_{Rt})    \\
\op{\nu d}{V}{LR}         & (\bar \nu_{Lp} \gamma^\mu \nu_{Lr})(\bar d_{Rs} \gamma_\mu d_{Rt})   \\
\op{eu}{V}{LR}        & (\bar e_{Lp}  \gamma^\mu e_{Lr})(\bar u_{Rs} \gamma_\mu u_{Rt})   \\
\op{ed}{V}{LR}        & (\bar e_{Lp}  \gamma^\mu e_{Lr})(\bar d_{Rs} \gamma_\mu d_{Rt})   \\
\op{ue}{V}{LR}        & (\bar u_{Lp} \gamma^\mu u_{Lr})(\bar e_{Rs}  \gamma_\mu e_{Rt})   \\
\op{de}{V}{LR}         & (\bar d_{Lp} \gamma^\mu d_{Lr}) (\bar e_{Rs} \gamma_\mu e_{Rt})   \\
\op{\nu edu}{V}{LR}        & (\bar \nu_{Lp} \gamma^\mu e_{Lr})(\bar d_{Rs} \gamma_\mu u_{Rt})  +\mathrm{h.c.} \\
\op{uu}{V1}{LR}        & (\bar u_{Lp} \gamma^\mu u_{Lr})(\bar u_{Rs} \gamma_\mu u_{Rt})   \\
\op{uu}{V8}{LR}       & (\bar u_{Lp} \gamma^\mu T^A u_{Lr})(\bar u_{Rs} \gamma_\mu T^A u_{Rt})    \\ 
\op{ud}{V1}{LR}       & (\bar u_{Lp} \gamma^\mu u_{Lr}) (\bar d_{Rs} \gamma_\mu d_{Rt})  \\
\op{ud}{V8}{LR}       & (\bar u_{Lp} \gamma^\mu T^A u_{Lr})  (\bar d_{Rs} \gamma_\mu T^A d_{Rt})  \\
\op{du}{V1}{LR}       & (\bar d_{Lp} \gamma^\mu d_{Lr})(\bar u_{Rs} \gamma_\mu u_{Rt})   \\
\op{du}{V8}{LR}       & (\bar d_{Lp} \gamma^\mu T^A d_{Lr})(\bar u_{Rs} \gamma_\mu T^A u_{Rt}) \\
\op{dd}{V1}{LR}      & (\bar d_{Lp} \gamma^\mu d_{Lr})(\bar d_{Rs} \gamma_\mu d_{Rt})  \\
\op{dd}{V8}{LR}   & (\bar d_{Lp} \gamma^\mu T^A d_{Lr})(\bar d_{Rs} \gamma_\mu T^A d_{Rt}) \\
\op{uddu}{V1}{LR}   & (\bar u_{Lp} \gamma^\mu d_{Lr})(\bar d_{Rs} \gamma_\mu u_{Rt})  + \mathrm{h.c.}  \\
\op{uddu}{V8}{LR}      & (\bar u_{Lp} \gamma^\mu T^A d_{Lr})(\bar d_{Rs} \gamma_\mu T^A  u_{Rt})  + \mathrm{h.c.} \\
\end{array}
\end{align*}
\end{minipage}

\begin{minipage}[t]{3cm}
\renewcommand{\arraystretch}{1.51}
\small
\begin{align*}
\begin{array}[t]{c|c}
\multicolumn{2}{c}{\boldsymbol{(\overline L R)(\overline L R)+\mathrm{h.c.}}} \\
\hline
\op{ee}{S}{RR} 		& (\bar e_{Lp}   e_{Rr}) (\bar e_{Ls} e_{Rt})   \\
\op{eu}{S}{RR}  & (\bar e_{Lp}   e_{Rr}) (\bar u_{Ls} u_{Rt})   \\
\op{eu}{T}{RR} & (\bar e_{Lp}   \sigma^{\mu \nu}   e_{Rr}) (\bar u_{Ls}  \sigma_{\mu \nu}  u_{Rt})  \\
\op{ed}{S}{RR}  & (\bar e_{Lp} e_{Rr})(\bar d_{Ls} d_{Rt})  \\
\op{ed}{T}{RR} & (\bar e_{Lp} \sigma^{\mu \nu} e_{Rr}) (\bar d_{Ls} \sigma_{\mu \nu} d_{Rt})   \\
\op{\nu edu}{S}{RR} & (\bar   \nu_{Lp} e_{Rr})  (\bar d_{Ls} u_{Rt} ) \\
\op{\nu edu}{T}{RR} &  (\bar  \nu_{Lp}  \sigma^{\mu \nu} e_{Rr} )  (\bar  d_{Ls}  \sigma_{\mu \nu} u_{Rt} )   \\
\op{uu}{S1}{RR}  & (\bar u_{Lp}   u_{Rr}) (\bar u_{Ls} u_{Rt})  \\
\op{uu}{S8}{RR}   & (\bar u_{Lp}   T^A u_{Rr}) (\bar u_{Ls} T^A u_{Rt})  \\
\op{ud}{S1}{RR}   & (\bar u_{Lp} u_{Rr})  (\bar d_{Ls} d_{Rt})   \\
\op{ud}{S8}{RR}  & (\bar u_{Lp} T^A u_{Rr})  (\bar d_{Ls} T^A d_{Rt})  \\
\op{dd}{S1}{RR}   & (\bar d_{Lp} d_{Rr}) (\bar d_{Ls} d_{Rt}) \\
\op{dd}{S8}{RR}  & (\bar d_{Lp} T^A d_{Rr}) (\bar d_{Ls} T^A d_{Rt})  \\
\op{uddu}{S1}{RR} &  (\bar u_{Lp} d_{Rr}) (\bar d_{Ls}  u_{Rt})   \\
\op{uddu}{S8}{RR}  &  (\bar u_{Lp} T^A d_{Rr}) (\bar d_{Ls}  T^A u_{Rt})  \\[-0.5cm]
\end{array}
\end{align*}
%\end{minipage}
%\hspace{2cm}
%\begin{minipage}[t]{3cm}
\renewcommand{\arraystretch}{1.51}
\small
\begin{align*}
\begin{array}[t]{c|c}
\multicolumn{2}{c}{\boldsymbol{(\overline L R)(\overline R L) +\mathrm{h.c.}}} \\
\hline
\op{eu}{S}{RL}  & (\bar e_{Lp} e_{Rr}) (\bar u_{Rs}  u_{Lt})  \\
\op{ed}{S}{RL} & (\bar e_{Lp} e_{Rr}) (\bar d_{Rs} d_{Lt}) \\
\op{\nu edu}{S}{RL}  & (\bar \nu_{Lp} e_{Rr}) (\bar d_{Rs}  u_{Lt})  \\
\end{array}
\end{align*}
\end{minipage}
\end{adjustbox}
%%%
% --- END TABLE
%%%
\setlength{\abovecaptionskip}{0.15cm}
\setlength{\belowcaptionskip}{-3cm}
\caption[The operators in LEFT of dimension three, five, and six that conserve baryon and lepton number, and the dimension-three and dimension-five $\Delta L=\pm 2$ operators]{The operators for LEFT of dimension three, five, and six that conserve baryon and lepton number, and the dimension-three and dimension-five $\Delta L=\pm 2$ operators.  The dimension-three $\Delta L=2$ operator $\mathcal{O}_\nu$ is the Majorana-neutrino mass operator, while the dimension-five $\Delta L=2$ operator $\mathcal{O}_{\nu\gamma}$ is the Majorana-neutrino dipole operator.  There are 5 additional dimension-five dipole operators $(\bar L R)X$.  The 80 dimension-six operators consist of 2 pure gauge operators $X^3$ and 78 four-fermion operators $\psi^4$, which are further divided by their chiral structure.  The $\psi^4$ operator superscripts $V$, $S$, $T$ refer to products of vector, scalar, and tensor fermion bilinears, and the additional two labels $L$ or $R$ refer to the chiral projectors in the bilinears.  Operators with ${}+\mathrm{h.c.}$ have Hermitian conjugates.  The subscripts $p, r, s, t$ are flavour indices. Table taken from \cite{Jenkins:2017jig}.}
\label{tab:oplist1}
\end{table}

\begin{table}[ht]
%
% --- START TABLE
%%
\centering
\begin{minipage}[t]{3cm}
\renewcommand{\arraystretch}{1.5}
\small
\begin{align*}
\begin{array}[t]{c|c}
\multicolumn{2}{c}{\boldsymbol{\Delta L = 4 + \mathrm{h.c.}}}  \\
\hline
\op{\nu\nu}{S}{LL} &  (\nu_{Lp}^T C \nu_{Lr}^{}) (\nu_{Ls}^T C \nu_{Lt}^{} )  \\
\end{array}
\end{align*}
\end{minipage}
%%%
% --- END TABLE
%%%
%%

% --- START TABLE
%%
%\hspace{-1cm}
\begin{adjustbox}{width=\textwidth,center}
\begin{minipage}[t]{3cm}
\renewcommand{\arraystretch}{1.5}
\small
\begin{align*}
\begin{array}[t]{c|c}
\multicolumn{2}{c}{\boldsymbol{\Delta L =2 + \mathrm{h.c.}}}  \\
\hline
\op{\nu e}{S}{LL}  &  (\nu_{Lp}^T C \nu_{Lr}) (\bar e_{Rs} e_{Lt})   \\
\op{\nu e}{T}{LL} &  (\nu_{Lp}^T C \sigma^{\mu \nu} \nu_{Lr}) (\bar e_{Rs}\sigma_{\mu \nu} e_{Lt} )  \\
\op{\nu e}{S}{LR} &  (\nu_{Lp}^T C \nu_{Lr}) (\bar e_{Ls} e_{Rt} )  \\
\op{\nu u}{S}{LL}  &  (\nu_{Lp}^T C \nu_{Lr}) (\bar u_{Rs} u_{Lt} )  \\
\op{\nu u}{T}{LL}  &  (\nu_{Lp}^T C \sigma^{\mu \nu} \nu_{Lr}) (\bar u_{Rs} \sigma_{\mu \nu} u_{Lt} ) \\
\op{\nu u}{S}{LR}  &  (\nu_{Lp}^T C \nu_{Lr}) (\bar u_{Ls} u_{Rt} )  \\
\op{\nu d}{S}{LL}   &  (\nu_{Lp}^T C \nu_{Lr}) (\bar d_{Rs} d_{Lt} ) \\
\op{\nu d}{T}{LL}   &  (\nu_{Lp}^T C \sigma^{\mu \nu}  \nu_{Lr}) (\bar d_{Rs} \sigma_{\mu \nu} d_{Lt} ) \\
\op{\nu d}{S}{LR}  &  (\nu_{Lp}^T C \nu_{Lr}) (\bar d_{Ls} d_{Rt} ) \\
\op{\nu edu}{S}{LL} &  (\nu_{Lp}^T C e_{Lr}) (\bar d_{Rs} u_{Lt} )  \\
\op{\nu edu}{T}{LL}  & (\nu_{Lp}^T C  \sigma^{\mu \nu} e_{Lr}) (\bar d_{Rs}  \sigma_{\mu \nu} u_{Lt} ) \\
\op{\nu edu}{S}{LR}   & (\nu_{Lp}^T C e_{Lr}) (\bar d_{Ls} u_{Rt} ) \\
\op{\nu edu}{V}{RL}   & (\nu_{Lp}^T C \gamma^\mu e_{Rr}) (\bar d_{Ls} \gamma_\mu u_{Lt} )  \\
\op{\nu edu}{V}{RR}   & (\nu_{Lp}^T C \gamma^\mu e_{Rr}) (\bar d_{Rs} \gamma_\mu u_{Rt} )  \\
\end{array}
\end{align*}
\end{minipage}
%%%
% --- END TABLE
%%%
%
%%
%%
% --- START TABLE
%%
\begin{minipage}[t]{3cm}
\renewcommand{\arraystretch}{1.5}
\small
\begin{align*}
\begin{array}[t]{c|c}
\multicolumn{2}{c}{\boldsymbol{\Delta B = \Delta L = 1 + \mathrm{h.c.}}} \\
\hline
\op{udd}{S}{LL} &  \epsilon_{\alpha\beta\gamma}  (u_{Lp}^{\alpha T} C d_{Lr}^{\beta}) (d_{Ls}^{\gamma T} C \nu_{Lt}^{})   \\
\op{duu}{S}{LL} & \epsilon_{\alpha\beta\gamma}  (d_{Lp}^{\alpha T} C u_{Lr}^{\beta}) (u_{Ls}^{\gamma T} C e_{Lt}^{})  \\
\op{uud}{S}{LR} & \epsilon_{\alpha\beta\gamma}  (u_{Lp}^{\alpha T} C u_{Lr}^{\beta}) (d_{Rs}^{\gamma T} C e_{Rt}^{})  \\
\op{duu}{S}{LR} & \epsilon_{\alpha\beta\gamma}  (d_{Lp}^{\alpha T} C u_{Lr}^{\beta}) (u_{Rs}^{\gamma T} C e_{Rt}^{})   \\
\op{uud}{S}{RL} & \epsilon_{\alpha\beta\gamma}  (u_{Rp}^{\alpha T} C u_{Rr}^{\beta}) (d_{Ls}^{\gamma T} C e_{Lt}^{})   \\
\op{duu}{S}{RL} & \epsilon_{\alpha\beta\gamma}  (d_{Rp}^{\alpha T} C u_{Rr}^{\beta}) (u_{Ls}^{\gamma T} C e_{Lt}^{})   \\
\op{dud}{S}{RL} & \epsilon_{\alpha\beta\gamma}  (d_{Rp}^{\alpha T} C u_{Rr}^{\beta}) (d_{Ls}^{\gamma T} C \nu_{Lt}^{})   \\
\op{ddu}{S}{RL} & \epsilon_{\alpha\beta\gamma}  (d_{Rp}^{\alpha T} C d_{Rr}^{\beta}) (u_{Ls}^{\gamma T} C \nu_{Lt}^{})   \\
\op{duu}{S}{RR}  & \epsilon_{\alpha\beta\gamma}  (d_{Rp}^{\alpha T} C u_{Rr}^{\beta}) (u_{Rs}^{\gamma T} C e_{Rt}^{})  \\
\end{array}
\end{align*}
\end{minipage}
%%%
% --- END TABLE
%%%
%
%%
% --- START TABLE
%%
\begin{minipage}[t]{3cm}
\renewcommand{\arraystretch}{1.5}
\small
\begin{align*}
\begin{array}[t]{c|c}
\multicolumn{2}{c}{\boldsymbol{\Delta B = - \Delta L = 1 + \mathrm{h.c.}}}  \\
\hline
\op{ddd}{S}{LL} & \epsilon_{\alpha\beta\gamma}  (d_{Lp}^{\alpha T} C d_{Lr}^{\beta}) (\bar e_{Rs}^{} d_{Lt}^\gamma )  \\
\op{udd}{S}{LR}  & \epsilon_{\alpha\beta\gamma}  (u_{Lp}^{\alpha T} C d_{Lr}^{\beta}) (\bar \nu_{Ls}^{} d_{Rt}^\gamma )  \\
\op{ddu}{S}{LR} & \epsilon_{\alpha\beta\gamma}  (d_{Lp}^{\alpha T} C d_{Lr}^{\beta})  (\bar \nu_{Ls}^{} u_{Rt}^\gamma )  \\
\op{ddd}{S}{LR} & \epsilon_{\alpha\beta\gamma}  (d_{Lp}^{\alpha T} C d_{Lr}^{\beta}) (\bar e_{Ls}^{} d_{Rt}^\gamma ) \\
\op{ddd}{S}{RL}  & \epsilon_{\alpha\beta\gamma}  (d_{Rp}^{\alpha T} C d_{Rr}^{\beta}) (\bar e_{Rs}^{} d_{Lt}^\gamma )  \\
\op{udd}{S}{RR}  & \epsilon_{\alpha\beta\gamma}  (u_{Rp}^{\alpha T} C d_{Rr}^{\beta}) (\bar \nu_{Ls}^{} d_{Rt}^\gamma )  \\
\op{ddd}{S}{RR}  & \epsilon_{\alpha\beta\gamma}  (d_{Rp}^{\alpha T} C d_{Rr}^{\beta}) (\bar e_{Ls}^{} d_{Rt}^\gamma )  \\
\end{array}
\end{align*}
\end{minipage}
\end{adjustbox}
%%%
% --- END TABLE
%%%
%
\caption[The LEFT dimension-six four-fermion operators that violate baryon and/or lepton number]{The LEFT dimension-six four-fermion operators that violate baryon and/or lepton number.  All operators have Hermitian conjugates.  The operator superscripts $V$, $S$, $T$ refer to products of vector, scalar, and tensor fermion bilinears, and the additional two labels $L$ or $R$ refer to the chiral projectors in the bilinears.  The subscripts $p, r, s, t$ are flavour indices. Table taken from \cite{Jenkins:2017jig}}
\label{tab:oplist2}
\end{table}

\section{Tree-level matching conditions} \label{app:Matching}
This appendix includes the tree-level matching
conditions between the Warsaw basis of SMEFT operators and the San
Diego basis of LEFT operators. This matching conditions were listed
in Ref.~\cite{Jenkins:2017jig}. They are reproduced in Tables~\ref{dim3}-\ref{dim6bml} since we
make significant use of them in this thesis. In all cases, we use
$C$ to denote SMEFT Wilson coefficients.

Note that electroweak symmetry breaking is modified in the SMEFT by
the presence of dimension-six operators. The scalar field can be written
in the unitary gauge as \cite{Jenkins:2017jig}
\begin{align}
H & =\frac{1}{\sqrt{2}}\left(\begin{array}{c}
0\\
\left[1+\ckin\right]h+v_{T}
\end{array}\right)\,,\label{eq:Hvev}
\end{align}
where 
\begin{align}
\ckin & \equiv\left(C_{H\Box}-\frac{1}{4}C_{HD}\right)v^{2}_{\mathrm{SM}}\,,\qquad v_{T}\equiv\left(1+\frac{3C_{H}v_{\mathrm{SM}}^{2}}{8\lambda}\right)v_{\mathrm{SM}}\,,\label{eq:chkindef}
\end{align}
The rescaling of $h$ in Eq.~(\ref{eq:Hvev}) is necessary for
$h$ to have a conventionally normalised kinetic energy term, and the
VEV $v_{T}$ in the SMEFT is not the same as $v_{\mathrm{SM}}$ in
the SM Lagrangian due to the dimension-six contributions to the Higgs
interactions $Q_{H}$, $Q_{H\Box}$ and $Q_{HD}$, which contribute
to the scalar potential and kinetic energy terms. 

One also needs to perform gauge field and gauge coupling redefinitions
to yield gauge kinetic terms and mass terms that are properly normalised
and diagonal
\begin{align}
G_{\mu}^{A} & =\mathcal{G}_{\mu}^{A}\left(1+C_{HG}v_{T}^{2}\right), & W_{\mu}^{I} & =\mathcal{W}_{\mu}^{I}\left(1+C_{HW}v_{T}^{2}\right), & B_{\mu} & =\mathcal{B}_{\mu}\left(1+C_{HB}v_{T}^{2}\right),\label{5.16a}
\end{align}
\begin{align}
\gcg & =g_{3}\left(1+C_{HG}\,v_{T}^{2}\right), & \overline{g}_{2} & =g_{2}\left(1+C_{HW}\,v_{T}^{2}\right), & \overline{g}_{1} & =g_{1}\left(1+C_{HB}\,v_{T}^{2}\right),\label{5.16b}
\end{align}
and 
\begin{align}
\begin{pmatrix}{\cal Z}_{\mu}\\
{\cal A}_{\mu}
\end{pmatrix} & =\begin{pmatrix}\bar{c}-\frac{\epsilon}{2}\bar{s} & \qquad-\bar{s}+\frac{\epsilon}{2}\bar{c}\\
\bar{s}+\frac{\epsilon}{2}\bar{c} & \qquad\bar{c}+\frac{\epsilon}{2}\bar{s}
\end{pmatrix}\begin{pmatrix}{\cal W}^{3}_{\mu}\\
{\cal B}_{\mu}
\end{pmatrix},\quad\epsilon\equiv C_{HWB}v_{T}^{2}\,.
\end{align}
The neutral gauge boson mass eigenstates ${\cal Z}^{\mu}$ and ${\cal A}^{\mu}$
in the above equation depend on the weak mixing angle $\tc$ through
\begin{align}
\cos\tc & \equiv\bar{c}=\frac{{\overline{g}_{2}}}{\sqrt{\overline{g}_{1}^{2}+\overline{g}_{2}^{2}}}\left[1-\frac{\epsilon}{2}\ \frac{{\overline{g}_{1}}}{{\overline{g}_{2}}}\left(\frac{{\overline{g}_{2}^{2}-\overline{g}_{1}^{2}}}{{\overline{g}_{1}^{2}+\overline{g}_{2}^{2}}}\right)\right],\nn\sin\tc & \equiv\bar{s}=\frac{{\overline{g}_{1}}}{\sqrt{\overline{g}_{1}^{2}+\overline{g}_{2}^{2}}}\left[1+\frac{\epsilon}{2}\ \frac{{\overline{g}_{2}}}{{\overline{g}_{1}}}\left(\frac{{\overline{g}_{2}^{2}-\overline{g}_{1}^{2}}}{{\overline{g}_{1}^{2}+\overline{g}_{2}^{2}}}\right)\right].\label{2.23}
\end{align}
The massive gauge bosons of the SM receive NP contributions to their
masses as 
\begin{align}
M_{{\cal W}}^{2} & =\frac{1}{4}\overline{g}_{2}^{2}v_{T}^{2},\nn M_{{\cal Z}}^{2} & =\frac{1}{4}\left(\overline{g}_{2}^{2}+\overline{g}_{1}^{2}\right)v_{T}^{2}\left(1+\frac{1}{2}C_{HD}v_{T}^{2}\right)+\frac{\epsilon}{2}\overline{g}_{1}\overline{g}_{2}v_{T}^{2}\,.
\end{align}
In the above equations, $G_{\mu}^{A}$, $W_{\mu}^{I}$, and $B_{\mu}$
and $g_{3}$, $g_{2}$, and $g_{1}$ are the gauge fields and coupling
constants in the SM (without NP contributions), while ${\cal G}_{\mu}^{A}$,
${\cal W}_{\mu}^{I}$, and ${\cal B}_{\mu}$ and $\gcg$, $\overline{g}_{2}$,
and $\overline{g}_{1}$ are the gauge fields and coupling constants modified by
NP contributions from dimension-six SMEFT operators. Note that products of gauge
couplings and gauge fields $g_{3}G_{\mu}^{A}=\gcg\mathcal{G}_{\mu}^{A}$,
$g_{2}W_{\mu}^{I}=\overline{g}_{2}\mathcal{W}_{\mu}^{I}$ and $g_{1}B_{\mu}=\overline{g}_{1}\mathcal{B}_{\mu}$
are unchanged by the above redefinitions. In a similar manner, one
can check that the effective couplings $\overline{e}$ and $\overline{g}_{Z}$ are given
by 
\begin{align}
\overline{e} & =\overline{g}_{2}\,\sin\tc-\frac{1}{2}\,\cos\tc\,\overline{g}_{2}\,v_{T}^{2}\,C_{HWB},\nn\overline{g}_{Z} & =\frac{\overline{e}}{\sin\tc\cos\tc}\left[1+\frac{\overline{g}_{1}^{2}+\overline{g}_{2}^{2}}{2\overline{g}_{1}\overline{g}_{2}}v_{T}^{2}C_{HWB}\right].
\end{align}
The fermion couplings to the massive gauge bosons $\mathcal{W}_{\mu}^{\pm}$
and $\mathcal{Z}_{\mu}$ in the SMEFT take the usual form 
\begin{align}
\mathcal{L} & =-\frac{\overline{g}_{2}}{\sqrt{2}}\left\{ \mathcal{W}_{\mu}^{+}j_{\mathcal{W}}^{\mu}+\mathrm{h.c.}\right\} -\overline{g}_{Z}\mathcal{Z}_{\mu}j_{\mathcal{Z}}^{\mu}\,.\label{14.9}
\end{align}
However, notice that the couplings of the massive gauge bosons to
fermions get modified via dimension-six SMEFT operators,
\begin{align}
j_{\mathcal{W}}^{\mu} & =[W_{l}]_{pr}\overline{\nu}_{Lp}\gamma^{\mu}e_{Lr}+[W_{q}]_{pr}\overline{u}_{Lp}\gamma^{\mu}d_{Lr}+[W_{R}]_{pr}\overline{u}_{Rp}\gamma^{\mu}d_{Rr},\nn j_{Z}^{\mu} & =[Z_{\nu_{L}}]_{pr}\overline{\nu}_{Lp}\gamma^{\mu}\nu_{Lr}+[Z_{e_{L}}]_{pr}\overline{e}_{Lp}\gamma^{\mu}e_{Lr}+[Z_{e_{R}}]_{pr}\overline{e}_{Rp}\gamma^{\mu}e_{Rr}\nn & +[Z_{u_{L}}]_{pr}\overline{u}_{Lp}\gamma^{\mu}u_{Lr}+[Z_{u_{R}}]_{pr}\overline{u}_{Rp}\gamma^{\mu}u_{Rr}+[Z_{d_{L}}]_{pr}\overline{d}_{Lp}\gamma^{\mu}d_{Lr}+[Z_{d_{R}}]_{pr}\overline{d}_{Rp}\gamma^{\mu}d_{Rr},
\end{align}
where {\small{}
\begin{align}
[W_{l}]_{pr} & =\mathrlap{\left[\delta_{pr}+v_{T}^{2}C_{\substack{Hl\\
pr
}
}^{(3)}\right],\qquad[W_{q}]_{pr}=\left[\delta_{pr}+v_{T}^{2}C_{\substack{Hq\\
pr
}
}^{(3)}\right],\qquad[W_{R}]_{pr}=\left[\frac{1}{2}v_{T}^{2}C_{\substack{Hud\\
pr
}
}\right],}\nn[Z_{\nu_{L}}]_{pr} & =\left[\delta_{pr}\left(\frac{1}{2}\right)-\frac{1}{2}v_{T}^{2}C_{\substack{Hl\\
pr
}
}^{(1)}+\frac{1}{2}v_{T}^{2}C_{\substack{Hl\\
pr
}
}^{(3)}\right],\nn[Z_{e_{L}}]_{pr} & =\left[\delta_{pr}\left(-\frac{1}{2}+\sc^{2}\right)-\frac{1}{2}v_{T}^{2}C_{\substack{Hl\\
pr
}
}^{(1)}-\frac{1}{2}v_{T}^{2}C_{\substack{Hl\\
pr
}
}^{(3)}\right], & [Z_{e_{R}}]_{pr} & =\left[\delta_{pr}\left(+\sc^{2}\right)-\frac{1}{2}v_{T}^{2}C_{\substack{He\\
pr
}
}\right],\nn[Z_{u_{L}}]_{pr} & =\left[\delta_{pr}\left(\frac{1}{2}-\frac{2}{3}\sc^{2}\right)-\frac{1}{2}v_{T}^{2}C_{\substack{Hq\\
pr
}
}^{(1)}+\frac{1}{2}v_{T}^{2}C_{\substack{Hq\\
pr
}
}^{(3)}\right], & [Z_{u_{R}}]_{pr} & =\left[\delta_{pr}\left(-\frac{2}{3}\sc^{2}\right)-\frac{1}{2}v_{T}^{2}C_{\substack{Hu\\
pr
}
}\right],\nn[Z_{d_{L}}]_{pr} & =\left[\delta_{pr}\left(-\frac{1}{2}+\frac{1}{3}\sc^{2}\right)-\frac{1}{2}v_{T}^{2}C_{\substack{Hq\\
pr
}
}^{(1)}-\frac{1}{2}v_{T}^{2}C_{\substack{Hq\\
pr
}
}^{(3)}\right], & [Z_{d_{R}}]_{pr} & =\left[\delta_{pr}\left(+\frac{1}{3}\sc^{2}\right)-\frac{1}{2}v_{T}^{2}C_{\substack{Hd\\
pr
}
}\right].\label{14.10b}
\end{align}
}Here $[W_{l}]_{pr}$, $[W_{q}]_{pr}$ and $[W_{R}]_{pr}$ are the
couplings of $\mathcal{W}_{\mu}^{+}$ to $(\overline{\nu}_{Lp}\gamma^{\mu}e_{Lr})$,
$(\overline{u}_{Lp}\gamma^{\mu}d_{Lr})$, and $(\overline{u}_{Rp}\gamma^{\mu}d_{Rr})$,
respectively, and $[Z_{\psi}]_{pr}$ are the couplings of $\mathcal{Z}_{\mu}$
to $(\overline{\psi}_{p}\gamma^{\mu}\psi_{r})$ for $\psi=\nu_{L},e_{L},e_{R},u_{L},u_{R},d_{L},d_{R}$.
Note that the couplings in Eq.~(\ref{14.10b}) are written
in the weak-eigenstate basis, so the SM contribution is proportional
to Kronecker-delta symbols. The dimension-six $\psi^{2}H^{2}D$ operators
in the SMEFT give the $1/\Lambda^{2}$ contributions proportional
to coefficients $C$ in Eq.~(\ref{14.10b}). Interestingly, in spontaneously
broken SMEFT, $\mathcal{W}_{\mu}^{+}$ couples to the right-handed
charged current $(\overline{u}_{Rp}\gamma^{\mu}d_{Rr})$ with coupling
$\left[W_{R}\right]_{pr}$ due to the dimension-six operator $Q_{Hud}$~.
In general, the modified couplings of the massive gauge bosons give
NP contributions to LEFT operators, as shown in Tables~\ref{dim3}-\ref{dim6bml} which contain
the tree-level matching conditions.

We also note that for the tree-level matching we assume the same normalisation for SMEFT and LEFT operators.

%%%
% --- START TABLE
%%%
\begin{table}[ht]
\vspace{-0.8cm}
\renewcommand{\arraystretch}{1.2}
\small
\begin{align*}
\begin{array}[t]{c|c|c|c}
\multicolumn{4}{c}{\boldsymbol{\Delta L=2 \qquad \nu \nu+{\rm h.c.}}} \\
\toprule
 & \text{Number} & \text{SM} &   \text{Matching} \\
\midrule\midrule
\O_{\nu} & \frac12 n_\nu (n_\nu+1) & 6  & \frac{1}{2} C_{\substack{5 \\ pr}} v_T^2 \\
\bottomrule
\end{array}
\end{align*}
\setlength{\belowcaptionskip}{-2cm}
\caption[Tree-level matching for dimension-three $\Delta L=2$ Majorana neutrino mass operators in LEFT-SMEFT]{Dimension-three $\Delta L=2$ Majorana neutrino mass operators in LEFT.  There are also Hermitian conjugate $\Delta L=-2$ operators $\mathcal{O}_{\nu}^\dagger$, as indicated in the table heading.  The second column is the number of operators for an arbitrary number of neutrino flavours $n_\nu$, and the third column is the number in the SM LEFT with $n_\nu =3$.  The last column is the tree-level matching coefficient in SMEFT. Table taken from \cite{Jenkins:2017jig}.}
\label{dim3}
\end{table}
%%%
% --- END TABLE
%%%

%%%
% --- START TABLE
%%%
\begin{table}[ht]
\vspace{1.5cm}
\renewcommand{\arraystretch}{1.2}
\small
\begin{align*}
\begin{array}[t]{c|c|c|c}
\multicolumn{4}{c}{\boldsymbol{\Delta L =2 \qquad (\nu \nu) X+{\rm h.c.}}} \\
\toprule
 & \text{Number} & \text{SM} &   \text{Matching} \\
\midrule\midrule
\O_{\nu \gamma} & \frac12 n_\nu (n_\nu-1) & 3  & 0 \\
\bottomrule
\end{array}
\end{align*}
\setlength{\abovecaptionskip}{0cm}
\caption[Tree-level matching for dimension-five $\Delta L=2$ Majorana neutrino dipole operators in LEFT-SMEFT]{Dimension-five $\Delta L=2$ Majorana neutrino dipole operators in LEFT.  There are also Hermitian conjugate $\Delta L=-2$ operators $\mathcal{O}_{\nu \gamma}^\dagger$, as indicated in the table heading.
The second column is the number of operators for an arbitrary number of neutrino flavours $n_\nu$, and the third column is number in the SM LEFT with $n_\nu=3$.  The last column is the tree-level matching coefficient in SMEFT, which vanishes. Table taken from \cite{Jenkins:2017jig}.}
\label{dim5n}
\end{table}
%%%
% --- END TABLE
%%%

%%%
% --- START TABLE
%%%
\begin{table}[ht]
\renewcommand{\arraystretch}{1.2}
\small
\begin{align*}
\begin{array}[t]{c|c|c|c}
\multicolumn{4}{c}{\boldsymbol{(\bar L R ) X+{\rm h.c.}}} \\
\toprule
 & \text{Number} & \text{SM} &   \text{Matching} \\
\midrule\midrule
\multicolumn{4}{c}{ \text{Leptonic} } \\
\hline
\O_{e \gamma} & n_e^2 & 9  & \frac{1}{\sqrt 2}\left( -C_{\substack{eW \\ pr}} \bar s + C_{\substack{eB \\ pr}}  \bar c  \right) v_T \\
\midrule
%\text{Total} & & n_e^2 & 9 & \\
%\hline
\multicolumn{4}{c}{ \text{Nonleptonic} } \\
\hline
\O_{u \gamma}   & n_u^2 & 4  &  \frac{1}{\sqrt 2} \left( C_{\substack{uW \\ pr}} \bar s + C_{\substack{uB \\ pr}} \bar c \right) v_T \\
\O_{d \gamma}   & n_d^2 & 9  &  \frac{1}{\sqrt 2}  \left( -C_{\substack{dW \\ pr}} \bar s + C_{\substack{dB \\ pr}} \bar c \right) v_T \\
\O_{u G}  & n_u^2 & 4  &  \frac{1}{\sqrt 2} C_{\substack{uG \\ pr}} v_T \\
\O_{d G}  & n_d^2 & 9  & \frac{1}{\sqrt 2} C_{\substack{dG \\ pr}} v_T \\
\hline
\text{Total}  & 2 n_u^2+ 2n_d^2 & 26 & \\
\bottomrule
%\hline \hline
%\text{Total} & & 2 n_u^2+ 2n_d^2+ n_e^2  & 35
%
\end{array}
\end{align*}
\setlength{\abovecaptionskip}{0cm}
\caption[Tree-level matching for dimension-five $(\bar L R)X$ dipole operators in LEFT-SMEFT]{Dimension-five $(\bar L R)X$ dipole operators in LEFT.  There are also Hermitian conjugate dipole operators $(\bar R L)X$, as indicated in the table heading.  The operators are divided into the leptonic and nonleptonic operators.  The second column is the number of operators for an arbitrary number of charged lepton flavours $n_e$, $u$-type quark flavours $n_u$, and $d$-type quark flavours $n_d$, and the third column is the number in the SM LEFT with $n_e=3$, $n_u=2$ and $n_d=3$.  The last column is the tree-level matching coefficient in SMEFT. $\bar s$ and $\bar c$ are defined in Eq.~(\ref{2.23}). Table taken from \cite{Jenkins:2017jig}.}
\label{dim5mag}
\end{table}
%%%
% --- END TABLE
%%%

%%%
% --- START TABLE
%%%
\begin{table}[ht]
\renewcommand{\arraystretch}{1.2}
\small
\begin{align*}
\begin{array}[t]{c|c|c|c}
\multicolumn{4}{c}{\boldsymbol{X^3}} \\
\toprule
 & \text{Number} & \text{SM} &   \text{Matching} \\
\midrule\midrule
\O_G      & 1 & 1 & C_G \\
\O_{\widetilde G}  & 1 & 1  & C_{\widetilde G} \\
\hline
\text{Total}  & 2 & 2 \\
\bottomrule
\end{array}
\end{align*}
\setlength{\abovecaptionskip}{0cm}
\caption[Tree-level matching for dimension-six triple-gauge-boson operators in LEFT-SMEFT]{Dimension-six triple-gauge-boson operators in LEFT.  The tree-level matching coefficient of each operator is equal to the coefficient of the corresponding operator in SMEFT. Table taken from \cite{Jenkins:2017jig}.}
\label{dim6X3}
\end{table}
%%%
% --- END TABLE
%%%

%%%
% --- START TABLE
%%%
\begin{table}[ht]
\renewcommand{\arraystretch}{1.2}
\small
\begin{align*}
\resizebox{0.92\textwidth}{!}{
\begin{adjustbox}{center}
\begin{array}[t]{c|c|c|c}
\multicolumn{4}{c}{\boldsymbol{(\bar L  L)(\bar L  L)}} \\
\toprule
& \text{Number} & \text{SM} &   \text{Matching} \\
\midrule\midrule
\multicolumn{4}{c}{ \text{Leptonic} } \\
\hline
\op{\nu\nu}{V}{LL}  & \frac14 n_\nu^2 (n_\nu+1)^2 & 36   & C_{\substack{ ll \\ prst}} -\frac{\overline{g}_{Z}^2}{4 M_Z^2} \left[Z_\nu \right]_{pr} \left[Z_\nu \right]_{st} -\frac{\overline{g}_{Z}^2}{4 M_Z^2} \left[Z_\nu \right]_{pt} \left[Z_\nu \right]_{sr}  \\
\op{ee}{V}{LL}      & \frac14 n_e^2 (n_e+1)^2 & 36   & C_{\substack{ ll \\ prst}} -\frac{\overline{g}_{Z}^2}{4 M_Z^2}   \left[Z_{e_L} \right]_{pr} \left[Z_{e_L} \right]_{st} -\frac{\overline{g}_{Z}^2}{4 M_Z^2}   \left[Z_{e_L} \right]_{pt} \left[Z_{e_L} \right]_{sr} \\
\op{\nu e}{V}{LL}    & n_e^2 n_\nu^2 & 81   &  C_{\substack{ ll \\ prst}} + C_{\substack{ ll \\ stpr}} -\frac{\overline{g}_{2}^2}{2 M_W^2} 
\left[W_l\right]_{pt} \left[W_l\right]_{rs}^*  -\frac{\overline{g}_{Z}^2}{M_Z^2} 
\left[Z_\nu \right]_{pr} \left[Z_{e_L} \right]_{st} \\
\hline
%\text{Total} & & \frac12 n_e^2(3 n_e^2 + 2n_e + 1)  & 153 \\
\text{Total}  & n_e^2 n_\nu^2 + \frac{1}{4} n_e^2(n_e+1)^2 & \\
	 & {} + \frac{1}{4} n_\nu^2(n_\nu+1)^2  & 153 & \\
\midrule
\multicolumn{4}{c}{ \text{Semileptonic} } \\
\hline
\op{\nu u}{V}{LL}       & n_\nu^2 n_u^2  & 36   &  C^{(1)}_{\substack{ lq \\ prst}} + C^{(3)}_{\substack{ lq \\ prst}} 
-\frac{\overline{g}_{Z}^2}{ M_Z^2}   \left[Z_\nu \right]_{pr} \left[Z_{u_L} \right]_{st} \\
\op{\nu d}{V}{LL}          & n_\nu^2 n_d^2  & 81    & C^{(1)}_{\substack{ lq \\ prst}} -  C^{(3)}_{\substack{ lq \\ prst}}
-\frac{\overline{g}_{Z}^2}{ M_Z^2} 
 \left[Z_\nu \right]_{pr} \left[Z_{d_L} \right]_{st}    \\
\op{eu}{V}{LL}     & n_e^2 n_u^2  & 36    & C^{(1)}_{\substack{ lq \\ prst}} -  C^{(3)}_{\substack{ lq \\ prst}}
-\frac{\overline{g}_{Z}^2}{ M_Z^2}   \left[Z_{e_L} \right]_{pr} \left[Z_{u_L} \right]_{st}  \\
\op{ed}{V}{LL}      & n_e^2 n_d^2  & 81   & C^{(1)}_{\substack{ lq \\ prst}} + C^{(3)}_{\substack{ lq \\ prst}} 
-\frac{\overline{g}_{Z}^2}{ M_Z^2}   \left[Z_{e_L} \right]_{pr} \left[Z_{d_L} \right]_{st}  \\
\op{\nu e d u}{V}{LL}   + {\rm h.c.}  & 2 \times n_e n_\nu n_u n_d   & 2\times 54    &  2 C^{(3)}_{\substack{ lq \\ prst}} -\frac{\overline{g}_{2}^2}{2 M_W^2} 
\left[W_l\right]_{pr} \left[W_q\right]_{ts}^* \\
\hline
%\text{Total} & & 2 n_e^2 (n_u^2+n_d^2+n_un_d) & 342 & \\
\text{Total}  & (n_e^2 + n_\nu^2)(n_u^2+n_d^2) & & \\
	 & {}+ 2 n_e n_\nu n_u n_d & 342 & \\
\midrule
\multicolumn{4}{c}{ \text{Nonleptonic} } \\
\hline
\op{uu}{V}{LL}       & \frac12 n_u^2 (n_u^2+1)   & 10   &  C^{(1)}_{\substack{ qq \\ prst}} + C^{(3)}_{\substack{ qq \\ prst}}  -\frac{\overline{g}_{Z}^2}{2 M_Z^2}   \left[Z_{u_L} \right]_{pr} \left[Z_{u_L} \right]_{st} \\
\op{dd}{V}{LL}     & \frac12 n_d^2 (n_d^2+1)  & 45   & C^{(1)}_{\substack{ qq \\ prst}} + C^{(3)}_{\substack{ qq \\ prst}}  -\frac{\overline{g}_{Z}^2}{2 M_Z^2}   \left[Z_{d_L} \right]_{pr} \left[Z_{d_L} \right]_{st} \\
\op{ud}{V1}{LL}    & n_u^2 n_d^2  & 36     &  C^{(1)}_{\substack{ qq \\ prst}} + C^{(1)}_{\substack{ qq \\ stpr}} - C^{(3)}_{\substack{ qq \\ prst}} - C^{(3)}_{\substack{ qq \\ stpr}}  +\frac{2}{N_c} C^{(3)}_{\substack{ qq \\ ptsr}} +\frac{2}{N_c} C^{(3)}_{\substack{ qq \\ srpt}} \\
& &   & -\frac{\overline{g}_{2}^2}{2 M_W^2} 
\left[W_q\right]_{pt} \left[W_q\right]_{rs}^* \frac{1}{N_c}   -\frac{\overline{g}_{Z}^2}{ M_Z^2}    \left[Z_{u_L} \right]_{pr} \left[Z_{d_L} \right]_{st} \\
\op{ud}{V8}{LL}      & n_u^2 n_d^2  & 36   &  4 C^{(3)}_{\substack{ qq \\ ptsr}} +4 C^{(3)}_{\substack{ qq \\ srpt}} -\frac{\overline{g}_{2}^2}{M_W^2} 
\left[W_q\right]_{pt} \left[W_q\right]_{rs}^*   \\
\hline 
\text{Total} &  2 n_u^2 n_d^2 + \frac12 n_u^2 (n_u^2+1)  & &  \\
	 &  {}+\frac12 n_d^2 (n_d^2+1)   & 127 &  \\
\bottomrule
\end{array}
\end{adjustbox}}
\end{align*}
\caption[Tree-level matching for dimension-six four-fermion operators: two left-handed currents in LEFT-SMEFT]{Dimension-six four-fermion operators: two left-handed currents in LEFT.  The $(\bar L  L)(\bar L  L)$ operators are divided into leptonic, semileptonic, and nonleptonic operators.  The semileptonic operator 
$\op{\nu e d u}{V}{LL}$ and its Hermitian conjugate ${\op{\nu e d u}{V}{LL}}^\dagger$ are both present.  All other operators are Hermitian.
The second column is the number of operators for an arbitrary number of neutrino flavours $n_\nu$, charged lepton flavours $n_e$, $u$-type quark flavours $n_u$, and $d$-type quark flavours $n_d$, and the third column is the number in the SM LEFT with $n_\nu=3$, $n_e=3$, $n_u=2$, and $n_d=3$.  The last column is the tree-level matching coefficient in SMEFT. Table taken from \cite{Jenkins:2017jig}.}
\label{dim6ll}
\end{table}
%%%
% --- END TABLE
%%%

%%%
% --- START TABLE
%%%
\begin{table}[ht]
\renewcommand{\arraystretch}{1.2}
\small
\begin{align*}
\begin{adjustbox}{center}
\begin{array}[t]{c|c|c|c}
\multicolumn{4}{c}{\boldsymbol{(\bar R  R)(\bar R  R)}} \\
\toprule
 & \text{Number} & \text{SM} &   \text{Matching} \\
\midrule\midrule
\multicolumn{4}{c}{ \text{Leptonic} } \\
\hline
\op{ee}{V}{RR}  & \frac14 n_e^2 (n_e+1)^2 & 36   &  C_{\substack{ ee \\ prst}} -\frac{\overline{g}_{Z}^2}{4 M_Z^2}   \left[Z_{e_R} \right]_{pr} \left[Z_{e_R} \right]_{st} -\frac{\overline{g}_{Z}^2}{4 M_Z^2}   \left[Z_{e_R} \right]_{pt} \left[Z_{e_R} \right]_{sr}  \\
\midrule
%
%\hline
%\text{Total} & & \frac14 n_e^2 (n_e+1)^2  & 36 \\
%
\multicolumn{4}{c}{ \text{Semileptonic} } \\
\hline
\op{eu}{V}{RR}   & n_e^2 n_u^2 & 36  & C_{\substack{ eu \\ prst}} -\frac{\overline{g}_{Z}^2}{ M_Z^2}   \left[Z_{e_R} \right]_{pr} \left[Z_{u_R} \right]_{st} \\
\op{ed}{V}{RR}   & n_e^2 n_d^2 & 81   &   C_{\substack{ ed \\ prst}} -\frac{\overline{g}_{Z}^2}{M_Z^2}   \left[Z_{e_R} \right]_{pr} \left[Z_{d_R} \right]_{st} \\
\hline
\text{Total}  & n_e^2 (n_u^2+n_d^2)  & 117 & \\
\midrule
\multicolumn{4}{c}{ \text{Nonleptonic} } \\
\hline
\op{uu}{V}{RR}   & \frac12 n_u^2 (n_u^2+1)   & 10   &  C_{\substack{ uu \\ prst}}
-\frac{\overline{g}_{Z}^2}{2 M_Z^2}   \left[Z_{u_R} \right]_{pr} \left[Z_{u_R} \right]_{st}
 \\
\op{dd}{V}{RR}   & \frac12 n_d^2 (n_d^2+1)  & 45   &  C_{\substack{ dd \\ prst}}
-\frac{\overline{g}_{Z}^2}{2 M_Z^2}   \left[Z_{d_R} \right]_{pr} \left[Z_{d_R} \right]_{st} \\
\op{ud}{V1}{RR}    & n_u^2 n_d^2  & 36 &  C^{(1)}_{\substack{ ud \\ prst}}-\frac{\overline{g}_{2}^2}{2 M_W^2} 
\left[W_R\right]_{pt} \left[W_R\right]_{rs}^* \frac{1}{N_c} -\frac{\overline{g}_{Z}^2}{ M_Z^2}   \left[Z_{u_R} \right]_{pr} \left[Z_{d_R} \right]_{st}  \\
\op{ud}{V8}{RR}    & n_u^2 n_d^2  & 36     & C^{(8)}_{\substack{ ud \\ prst}} -\frac{\overline{g}_{2}^2}{M_W^2} 
\left[W_R\right]_{pt} \left[W_R\right]_{rs}^*  \\
\hline
\text{Total} &  2 n_u^2 n_d^2 + \frac12 n_u^2 (n_u^2+1)    & & \\
	& {}  +\frac12 n_d^2 (n_d^2+1) & 127 & \\
\bottomrule
\end{array}
\end{adjustbox}
\end{align*}
\caption[Tree-level matching for dimension-six four-fermion operators: two right-handed currents in LEFT-SMEFT]{Dimension-six four-fermion operators: two right-handed currents in LEFT.  The $(\bar R  R)(\bar R  R)$ operators are divided into leptonic, semileptonic, and nonleptonic operators.  The second column is the number of operators for an arbitrary number of charged lepton flavours $n_e$, 
$u$-type quark flavours $n_u$, and $d$-type quark flavours $n_d$, and the third column is the number in the SM LEFT with $n_e=3$, $n_u=2$, and $n_d=3$.  The last column is the tree-level matching coefficient in SMEFT. Table taken from \cite{Jenkins:2017jig}.}
\label{dim6rr}
\end{table}
%%%
% --- END TABLE
%%%
%
%%%%
% --- START TABLE
%%%
\begin{table}[ht]
\renewcommand{\arraystretch}{1.2}
\small
\begin{align*}
\begin{adjustbox}{center}
\begin{array}[t]{c|c|c|c}
\multicolumn{4}{c}{} \\[-1cm]
\multicolumn{4}{c}{\boldsymbol{(\bar L L)(\bar R  R)}} \\
\toprule
& \text{Number} & \text{SM} &   \text{Matching} \\
\midrule\midrule
\multicolumn{4}{c}{ \text{Leptonic} } \\
\hline
\op{\nu e}{V}{LR}   & n_e^2 n_\nu^2 & 81   & C_{\substack{ le \\ prst}} 
-\frac{\overline{g}_{Z}^2}{ M_Z^2}  \left[Z_\nu \right]_{pr} \left[Z_{e_R} \right]_{st} 
%- \frac{\overline{g}_{2}^2}{2M_W^4} \frac12  \left[ \mathcal{U}_l \right]_{pt} \left[ \mathcal{U}_l \right]_{rs}^*
 \\
\op{ee}{V}{LR}  & n_e^4 & 81   & C_{\substack{ le \\ prst}}
-\frac{\overline{g}_{Z}^2}{ M_Z^2}   \left[Z_{e_L} \right]_{pr} \left[Z_{e_R} \right]_{st} 
%\nn
%& & & & - \frac{1}{2m_h^2} \left[ \mathcal{Y}_e \right]_{tp}^* \left[ \mathcal{Y}_e \right]_{sr} 
%- \frac{\overline{g}_{Z}^2}{2M_Z^2} \left[ \mathcal{Z}_e \right]_{tp}^* \left[ \mathcal{Z}_e \right]_{sr} 
\\
\hline
\text{Total}  & n_e^2( n_e^2 + n_\nu^2)  & 162 & \\
\midrule
\multicolumn{4}{c}{ \text{Semileptonic} } \\ 
\hline
\op{\nu u}{V}{LR}  &  n_\nu^2 n_u^2 & 36  &  C_{\substack{ lu \\ prst}}  
-\frac{\overline{g}_{Z}^2}{ M_Z^2}   \left[Z_\nu \right]_{pr} \left[Z_{u_R} \right]_{st}  \\
\op{\nu d}{V}{LR}    & n_\nu^2 n_d^2 &  81  & C_{\substack{ ld \\ prst}}  
-\frac{\overline{g}_{Z}^2}{ M_Z^2}   \left[Z_\nu \right]_{pr} \left[Z_{d_R} \right]_{st}  \\
\op{eu}{V}{LR}   & n_e^2 n_u^2 & 36   & C_{\substack{ lu \\ prst}} 
-\frac{\overline{g}_{Z}^2}{M_Z^2}   \left[Z_{e_L} \right]_{pr} \left[Z_{u_R} \right]_{st} \\
\op{ed}{V}{LR}   & n_e^2 n_d^2 & 81   & C_{\substack{ ld \\ prst}} 
-\frac{\overline{g}_{Z}^2}{M_Z^2}   \left[Z_{e_L} \right]_{pr} \left[Z_{d_R} \right]_{st} \\
\op{ue}{V}{LR}   & n_e^2 n_u^2 & 36  & C_{\substack{ qe \\ prst}}
 -\frac{\overline{g}_{Z}^2}{M_Z^2}   \left[Z_{u_L} \right]_{pr} \left[Z_{e_R} \right]_{st}  \\
\op{de}{V}{LR}     & n_e^2 n_d^2 & 81  &  C_{\substack{ qe \\ prst}} 
-\frac{\overline{g}_{Z}^2}{ M_Z^2}    \left[Z_{d_L} \right]_{pr} \left[Z_{e_R} \right]_{st}  \\
\op{\nu edu}{V}{LR}  +{ \rm h.c.}  & 2\times n_e n_\nu n_u n_d &  2\times 54   & -\frac{\overline{g}_{2}^2}{2 M_W^2} 
\left[W_l\right]_{pr} \left[W_R\right]_{ts}^* \\
\hline
%\text{Total} & & n_e^2(3n_u^2+3n_d^2+2n_un_d)  & 459 & \\
\text{Total} &  (2n_e^2 +n_\nu^2)(n_u^2 + n_d^2)  & & \\
	& {}+ 2 n_e n_\nu n_u n_d & 459 & \\
\midrule
\multicolumn{4}{c}{ \text{Nonleptonic} } \\
\hline
\op{uu}{V1}{LR}     & n_u^4 & 16   &  C^{(1)}_{\substack{ qu \\ prst}} 
 -\frac{\overline{g}_{Z}^2}{ M_Z^2}   \left[Z_{u_L} \right]_{pr} \left[Z_{u_R} \right]_{st}  \nn
\op{uu}{V8}{LR}   & n_u^4 & 16  &  C^{(8)}_{\substack{ qu \\ prst}}  \\
\op{ud}{V1}{LR}    & n_u^2 n_d^2  &  36   & C^{(1)}_{\substack{ qd \\ prst}} 
 -\frac{\overline{g}_{Z}^2}{ M_Z^2}   \left[Z_{u_L} \right]_{pr} \left[Z_{d_R} \right]_{st} \nn
\op{ud}{V8}{LR}   & n_u^2 n_d^2  &  36   & C^{(8)}_{\substack{ qd \\ prst}} \\
\op{du}{V1}{LR}     & n_u^2 n_d^2 & 36   & C^{(1)}_{\substack{ qu \\ prst}} 
-\frac{\overline{g}_{Z}^2}{M_Z^2}  \left[Z_{d_L} \right]_{pr} \left[Z_{u_R} \right]_{st}  \nn
\op{du}{V8}{LR}     & n_u^2 n_d^2  &  36   &  C^{(8)}_{\substack{ qu \\ prst}}
\\
\op{dd}{V1}{LR}   & n_d^4 & 81    & C^{(1)}_{\substack{ qd \\ prst}} 
-\frac{\overline{g}_{Z}^2}{ M_Z^2}   \left[Z_{d_L} \right]_{pr} \left[Z_{d_R} \right]_{st}  \nn
\op{dd}{V8}{LR}   & n_d^4 & 81    & C^{(8)}_{\substack{ qd \\ prst}} 
\\
\op{uddu}{V1}{LR}  + {\rm h.c.}  & 2\times n_u^2 n_d^2 & 2\times 36    &  -\frac{\overline{g}_{2}^2}{2 M_W^2} 
\left[W_q\right]_{pr} \left[W_R\right]_{ts}^* \nn
\op{uddu}{V8}{LR} + {\rm h.c.}  & 2\times n_u^2 n_d^2 & 2\times 36   & 0
\\
\hline
\text{Total}  & 2(n_u^4+n_d^4+4n_u^2n_d^2)  & 482 &  \\
\bottomrule
\end{array}
\end{adjustbox}
\end{align*}
\caption[Tree-level matching for dimension-six four-fermion operators: left-handed times right-handed currents in LEFT-SMEFT]{Dimension-six four-fermion operators: left-handed times right-handed currents in LEFT.  The $(\bar L  L)(\bar R  R)$ operators are divided into leptonic, semileptonic, and nonleptonic operators.  Semileptonic operators $\op{\nu edu}{V}{LR}$ and nonleptonic operators $\op{uddu}{V1}{LR}$
and $\op{uddu}{V8}{LR}$ all come with additional Hermitian conjugate operators.  All other operators are Hermitian.  The second column is the number of operators for an arbitrary number of neutrino flavours $n_\nu$, charged lepton flavours $n_e$, $u$-type quark flavours $n_u$, and $d$-type quark flavours $n_d$, and the third column is the number in the SM LEFT with $n_\nu=3$, $n_e=3$, $n_u=2$, and $n_d=3$.  The last column is the tree-level matching coefficient in SMEFT. Table taken from \cite{Jenkins:2017jig}.}
\label{dim6lr}
\end{table}
%%%
% --- END TABLE
%%%
%
%%%
% --- START TABLE
%%%
\begin{table}[ht]
\renewcommand{\arraystretch}{1.2}
\small
\begin{align*}
\begin{array}[t]{c|c|c|c}
\multicolumn{4}{c}{\boldsymbol{(\bar L R)(\bar R L) +{\rm h.c.}}} \\
\toprule
& \text{Number} & \text{SM} &   \text{Matching} \\
\midrule\midrule
\multicolumn{4}{c}{ \text{Semileptonic} } \\
\hline
\op{eu}{S}{RL}   & n_e^2 n_u^2  & 36 & 0
 \\
\op{ed}{S}{RL}  & n_e^2 n_d^2  & 81 & C_{\substack{ ledq \\ prst}} 
\\
\op{\nu edu}{S}{RL}  & n_e n_\nu n_u n_d  & 54 &  C_{\substack{ ledq \\ prst}} 
\\
\hline
\text{Total}  & n_e^2 (n_u^2+n_d^2) + n_e n_\nu n_u n_d & 171 \\
\bottomrule
\end{array}
\end{align*}
\caption[Tree-level matching for dimension-six four-fermion operators: $(\bar L R)(\bar R L)$ scalar bilinears in LEFT-SMEFT]{Dimension-six four-fermion operators: $(\bar L R)(\bar R L)$ scalar bilinears in LEFT.  There are also Hermitian conjugate operators, as indicated in the table heading.  All of the operators are semileptonic operators. The second column is the number of operators for an arbitrary number of neutrino flavours $n_\nu$, charged lepton flavours $n_e$, $u$-type quark flavours $n_u$, and $d$-type quark flavours $n_d$, and the third column is the number in the SM LEFT with $n_\nu=3$, $n_e=3$, $n_u=2$, and $n_d=3$.  The last column is the tree-level matching coefficient in SMEFT. Table taken from \cite{Jenkins:2017jig}.}
\label{dim6lrrl}
\end{table}
%%%
% --- END TABLE
%%%
%
%%%
% --- START TABLE
%%%
\begin{table}[ht]
\renewcommand{\arraystretch}{1.2}
\small
\begin{align*}
\begin{adjustbox}{center}
\begin{array}[t]{c|c|c|c}
\multicolumn{4}{c}{\boldsymbol{(\bar L R)(\bar L R)+{\rm h.c.}}} \\
\toprule
 & \text{Number} & \text{SM} &   \text{Matching} \\
\midrule\midrule
\multicolumn{4}{c}{ \text{Leptonic} } \\
\hline
\op{ee}{S}{RR}   & \frac12 n_e^2(n_e^2+1)  & 45 & 0 \nn
\midrule
\multicolumn{4}{c}{ \text{Semileptonic} } \\
\hline
\op{eu}{S}{RR}  & n_e^2 n_u^2 & 36 & -C^{(1)}_{\substack{ lequ \\ prst}}
 \\
\op{eu}{T}{RR} & n_e^2 n_u^2 & 36 & -C^{(3)}_{\substack{ lequ \\ prst}}  \\
\op{ed}{S}{RR} & n_e^2 n_d^2  & 81 &  0
 \\
\op{ed}{T}{RR} & n_e^2 n_d^2  & 81&  0 \\
\op{\nu edu}{S}{RR} & n_e n_\nu n_u n_d  & 54 &  C^{(1)}_{\substack{ lequ \\ prst}} 
\\
\op{\nu edu}{T}{RR}  & n_e n_\nu n_u n_d  & 54 &  C^{(3)}_{\substack{ lequ \\ prst}}  \\
\hline
\text{Total} &  2 n_e^2(n_u^2+n_d^2) + 2 n_e n_\nu n_u n_d & 342  \\
\midrule
\multicolumn{4}{c}{ \text{Nonleptonic} } \\
\hline
\op{uu}{S1}{RR}  & \frac12 n_u^2(n_u^2+1)  & 10 & 0 \nn
\op{uu}{S8}{RR}   & \frac12 n_u^2(n_u^2+1)  & 10 & 0 \\
\op{ud}{S1}{RR} &  n_u^2 n_d^2  & 36 &   C^{(1)}_{\substack{ quqd \\ prst}} \\
\op{ud}{S8}{RR}  &  n_u^2 n_d^2  & 36 &  C^{(8)}_{\substack{ quqd \\ prst}}  \\
\op{dd}{S1}{RR}  & \frac12 n_d^2(n_d^2+1)  & 45 &   0 \\
\op{dd}{S8}{RR}  & \frac12 n_d^2(n_d^2+1)  & 45 & 0 \\
\op{uddu}{S1}{RR} & n_u^2 n_d^2  & 36 &   -C^{(1)}_{\substack{ quqd \\ stpr}}
 \\
\op{uddu}{S8}{RR}   & n_u^2 n_d^2  & 36 &  -C^{(8)}_{\substack{ quqd \\ stpr}} \\
\hline
\text{Total}  & 4 n_u^2 n_d^2 + n_u^2(n_u^2+1) & & \\
	 & {}+n_d^2(n_d^2+1) & 254 & \\
\bottomrule
\end{array}
\end{adjustbox}
\end{align*}
\caption[Tree-level matching for dimension-six four-fermion operators: $(\bar L R)(\bar L R)$ scalar and tensor bilinears in LEFT-SMEFT]{Dimension-six four-fermion operators: $(\bar L R)(\bar L R)$ scalar and tensor bilinears in LEFT.  There are also Hermitian conjugate operators, as indicated in the table heading.  The operators are divided into leptonic, semileptonic, and nonleptonic operators.  The second column is the number of operators for an arbitrary number of neutrino flavours $n_\nu$, charged lepton flavours $n_e$, $u$-type quark flavours $n_u$, and $d$-type quark flavours $n_d$, and the third column is the number in the SM LEFT with $n_\nu=3$, $n_e=3$, $n_u=2$, and $n_d=3$.  The last column is the tree-level matching coefficient in SMEFT. Table taken from \cite{Jenkins:2017jig}.}
\label{dim6lrlr}
\end{table}
%%%
% --- END TABLE
%%%

%%%
% --- START TABLE
%%%
\begin{table}[ht]
\renewcommand{\arraystretch}{1.2}
\small
\begin{align*}
\begin{array}[t]{c|c|c|c}
\multicolumn{4}{c}{\boldsymbol{\Delta L = 4 + {\rm h.c.}}} \\
\toprule
 & \text{Number} &  \text{SM} &   \text{Matching} \\
\midrule\midrule
\op{\nu\nu}{S}{LL} & \frac{1}{12} n_\nu^2(n_\nu^2-1) & 6 & 0 
% \frac{1}{6m_h^2}  \left( [\mathcal{Y}_5]_{pr} [\mathcal{Y}_5]_{st} + [\mathcal{Y}_5]_{ps} [\mathcal{Y}_5]_{rt} +[\mathcal{Y}_5]_{pt} [\mathcal{Y}_5]_{sr} \right) 
\\
\bottomrule
\end{array}
\end{align*}
\caption[Tree-level matching for dimension-six $\Delta L=4$ operators in LEFT-SMEFT]{Dimension-six $\Delta L=4$ operators in LEFT.  There are also Hermitian conjugate operators, as indicated in the table heading.
The second column is the number of operators for an arbitrary number of neutrino flavours $n_\nu$, and the third column is the number in the SM LEFT with $n_\nu=3$.  The last column is the tree-level matching coefficient in SMEFT. Table taken from \cite{Jenkins:2017jig}.}
\label{dim6l4}
\end{table}
%%%
% --- END TABLE
%%%

%
%%%
% --- START TABLE
%%%
\begin{table}[ht]
\renewcommand{\arraystretch}{1.2}
\small
\begin{align*}
\begin{array}[t]{c|c|c|c}
\multicolumn{4}{c}{\boldsymbol{\Delta L =2 + {\rm h.c.}}} \\
\toprule
& \text{Number} & \text{SM} &   \text{Matching} \\
\midrule\midrule
\multicolumn{4}{c}{ \text{Leptonic} } \\
\hline
\op{\nu e}{S}{LL}  & \frac12 n_\nu(n_\nu+1) n_e^2 & 54 & % - \frac{1}{m_h^2}[\mathcal{Y}_5]_{pr}\left[ \mathcal{Y}_e\right]_{ts}^*  
0 \\
\op{\nu e}{T}{LL}  & \frac12 n_\nu(n_\nu-1) n_e^2  & 27 &  0 \\
\op{\nu e}{S}{LR}  & \frac12 n_\nu(n_\nu+1) n_e^2 & 54 & % - \frac{1}{m_h^2} [\mathcal{Y}_5]_{pr} \left[ \mathcal{Y}_e \right]_{st} 
0 \\
\hline
\text{Total} &  \frac{1}{2} n_\nu( 3n_\nu + 1 ) n_e^2 & 135 & \\
\midrule
\multicolumn{4}{c}{ \text{Semileptonic} } \\
\hline
\op{\nu u}{S}{LL}  & \frac12 n_\nu(n_\nu+1) n_u^2 & 24 &  % - \frac{1}{m_h^2} [\mathcal{Y}_5]_{pr} \left[ \mathcal{Y}_u\right]_{st} 
0 \\
\op{\nu u}{T}{LL}  &  \frac12 n_\nu(n_\nu-1) n_u^2  & 12 &  0 \\
\op{\nu u}{S}{LR}  & \frac12 n_\nu(n_\nu+1) n_u^2  & 24 &  % - \frac{1}{m_h^2}[\mathcal{Y}_5]_{pr} \left[ \mathcal{Y}_u\right]_{ts}^* 
0 \\
\op{\nu d}{S}{LL}  & \frac12 n_\nu(n_\nu+1) n_d^2 & 54 & %  - \frac{1}{m_h^2}[\mathcal{Y}_5]_{pr} \left[ \mathcal{Y}_d\right]_{ts}^*
0 \\
\op{\nu d}{T}{LL}  & \frac12 n_\nu(n_\nu-1) n_d^2   & 27 & 0 \\
\op{\nu d}{S}{LR} & \frac12 n_\nu(n_\nu+1) n_d^2  & 54 &  % - \frac{1}{m_h^2} [\mathcal{Y}_5]_{pr} \left[ \mathcal{Y}_d\right]_{st} 
0 \\
\op{\nu edu}{S}{LL} & n_e n_\nu n_u n_d & 54 & 0 \\
\op{\nu edu}{T}{LL} & n_e n_\nu n_u n_d & 54 & 0 \\
\op{\nu edu}{S}{LR} & n_e n_\nu n_u n_d & 54 &  0\\
\op{\nu edu}{V}{RL}   & n_e n_\nu n_u n_d & 54 &  0 \\
\op{\nu edu}{V}{RR}  & n_e n_\nu n_u n_d & 54 & 0 \\
\hline
\text{Total}  & \frac12 n_\nu (3n_\nu+1)(n_u^2+n_d^2) + 5 n_e n_\nu n_u n_d & 465 \\
\bottomrule
\end{array}
\end{align*}
\caption[Tree-level matching for dimension-six $\Delta L=2$ operators in LEFT-SMEFT]{Dimension-six $\Delta L=2$ operators in LEFT.  There are also Hermitian conjugate operators, as indicated in the table heading.  The operators are divided into leptonic and semileptonic operators.  The second column is the number of operators for an arbitrary number of neutrino flavours $n_\nu$, charged lepton flavours $n_e$, $u$-type quark flavours $n_u$, and $d$-type quark flavours $n_d$, and the third column is the number in the SM LEFT with $n_\nu=3$, $n_e=3$, $n_u=2$, and $n_d=3$.  The last column is the tree-level matching coefficient in SMEFT. Table taken from \cite{Jenkins:2017jig}.}
\label{dim6l2}
\end{table}
%%%
% --- END TABLE
%

%
% --- START TABLE
%%%
\begin{table}[ht]
\renewcommand{\arraystretch}{1.2}
\small
\begin{align*}
\begin{adjustbox}{center}
\begin{array}[t]{c|c|c|c}
\multicolumn{4}{c}{\boldsymbol{\Delta B = \Delta L = 1 + {\rm h.c.}}} \\
\toprule
& \text{Number} &  \text{SM} &   \text{Matching} \\
\midrule\midrule
\op{udd}{S}{LL} & n_\nu n_u n_d^2 &  54 & - C_{\substack{ qqql \\ prst}} - C_{\substack{ qqql \\ rpst}} \\
\op{duu}{S}{LL}  & n_e n_d n_u^2  &  36 &  - C_{\substack{ qqql \\ prst}} - C_{\substack{ qqql \\ rpst}}  \\
%
%\hline
%
\op{uud}{S}{LR}  & \frac12 n_d n_u (n_u-1) n_e & 9 &  0 \\
\op{duu}{S}{LR}  &  n_e n_u^2 n_d &  36 &  -C_{\substack{ qque \\ prst}} - C_{\substack{ qque \\ rpst}} \\
%
%\hline
%
%
\op{uud}{S}{RL} & \frac12 n_d n_u (n_u-1) n_e & 9 &  0 \\
\op{duu}{S}{RL}  & n_e n_u^2 n_d & 36 &  C_{\substack{ duql \\ prst}} \\
\op{dud}{S}{RL} & n_\nu n_u n_d^2  & 54 &  -C_{\substack{ duql \\ prst}} \\
\op{ddu}{S}{RL}  & \frac12 n_d (n_d-1)n_u n_\nu  & 18  & 0 \\
%
%\hline
\op{duu}{S}{RR}  &  n_e n_d n_u^2   &  36 &  C_{\substack{ duue \\ prst}} \\
\hline
\text{Total} &  \frac52 n_d^2 n_\nu n_u+ 5 n_d n_e n_u^2 - n_d n_e n_u - \frac12 n_d n_\nu n_u 
 &   288 \\
\bottomrule
\end{array}
\end{adjustbox}
\end{align*}
\caption[Tree-level matching for dimension-six $\Delta B = \Delta L=1$ operators in LEFT-SMEFT]{Dimension-six $\Delta B = \Delta L=1$ operators in LEFT.  There are also Hermitian conjugate $\Delta B = \Delta L=-1$ operators, as indicated in the table heading.  The second column is the number of operators for an arbitrary number of neutrino flavours $n_\nu$, charged lepton flavours $n_e$, $u$-type quark flavours $n_u$, and $d$-type quark flavours $n_d$, and the third column is the number in the SM LEFT with $n_\nu=3$, $n_e=3$, $n_u=2$, and $n_d=3$.  The last column is the tree-level matching coefficient in SMEFT. Table taken from \cite{Jenkins:2017jig}.}
\label{dim6bl}
\end{table}
%%%
% --- END TABLE
%%%

%%%
% --- START TABLE
%%%
\begin{table}[ht]
\renewcommand{\arraystretch}{1.2}
\small
\begin{align*}
\begin{adjustbox}{center}
\begin{array}[t]{c|c|c|c}
\multicolumn{4}{c}{\boldsymbol{\Delta B = - \Delta L = 1 + {\rm h.c.}}} \\
\toprule
 & \text{Number} & \text{SM} &   \text{Matching} \\
\midrule\midrule
\op{ddd}{S}{LL} & \frac13 n_d (n_d^2-1) n_e & 24 & 0 \\
%
%\hline
%
\op{udd}{S}{LR}  & n_\nu n_u n_d^2  & 54 &  0 \\
\op{ddu}{S}{LR} & \frac12 n_d (n_d-1)n_u n_\nu  & 18 & 0 \\
\op{ddd}{S}{LR} & \frac12 n_d^2 (n_d-1) n_e & 27 &  0 \\
%
%\hline
%
%
\op{ddd}{S}{RL} & \frac12 n_d^2 (n_d-1) n_e & 27 & 0 \\
%
%\hline
%
%
\op{udd}{S}{RR}  & n_\nu n_u n_d^2 & 54 & 0 \\
\op{ddd}{S}{RR} & \frac13 n_d (n_d^2-1) n_e & 24 &  0 \\
\hline
\text{Total} &  \frac53 n_d^3 n_e +\frac52  n_\nu n_d^2 n_u - n_d^2 n_e -\frac12 n_d n_\nu n_u - \frac 23 n_d n_e    & 228 \\
\bottomrule
\end{array}
\end{adjustbox}
\end{align*}
\caption[Tree-level matching for dimension-six $\Delta B = -\Delta L=1$ operators in LEFT-SMEFT]{Dimension-six $\Delta B = -\Delta L=1$ operators in LEFT.  There are also Hermitian conjugate $\Delta B = -\Delta L=-1$ operators, as indicated in the table heading.  The second column is the number of operators for an arbitrary number of neutrino flavours $n_\nu$, charged lepton flavours $n_e$, $u$-type quark flavours $n_u$, and $d$-type quark flavours $n_d$, and the third column is the number in the SM LEFT with $n_\nu=3$, $n_e=3$, $n_u=2$, and $n_d=3$.  The last column is the tree-level matching coefficient in SMEFT. Table taken from \cite{Jenkins:2017jig}.}
\label{dim6bml}
\end{table}
%%%
% --- END TABLE
%%%

\backmatter
%\chapter{Glossary [if relevant]}
\bibliographystyle{JHEP}
\bibliography{UOS}
%\chapter{Bibliography}
%To use bibliography as well as the references section use the \texttt{multibbl} package.
%\chapter{Index [if relevant]}
\end{document}